\documentclass[twoside,11pt,preprint]{article}

\usepackage{jmlr2e_custom}

\usepackage{amsmath,amsfonts,amssymb,amsthm}
\usepackage{mathtools}
\usepackage{bm}
\usepackage{dsfont}
\usepackage{microtype}
\usepackage{etoolbox}
\usepackage{xparse}
\usepackage{needspace}
\usepackage{ragged2e}

\allowdisplaybreaks[2]
\usepackage[table,dvipsnames]{xcolor}
\usepackage{subcaption}
\usepackage{float}
\usepackage{longtable}
\usepackage{booktabs}
\usepackage{array}
\usepackage{makecell}
\usepackage{tabularx}
\usepackage{multirow}
\usepackage{enumitem}

\graphicspath{{./art/}}

\usepackage{xurl}
\hypersetup{
  colorlinks=true,
  linkcolor=Blue,
  citecolor=RoyalBlue,
  urlcolor=Blue,
  filecolor=Blue,
  breaklinks=true
}

\usepackage{lastpage}

\definecolor{CompHeader}{RGB}{38,83,128}
\definecolor{CompStripe}{RGB}{235,242,248}
\definecolor{CompGreen}{RGB}{24,130,76}
\definecolor{CompRed}{RGB}{190,45,45}
\definecolor{CompGray}{RGB}{110,110,110}
\definecolor{CompBlue}{RGB}{42,92,130}
\definecolor{CompBack}{RGB}{242,247,250}

\newcommand{\cmark}{%
  \textcolor{CompGreen}{\ensuremath{\bm{\checkmark}}}%
}

\newcommand{\xmark}{%
  \textcolor{CompRed}{\ensuremath{\bm{\times}}}%
}

\newcommand{\namark}{%
  \textcolor{CompGray}{\textemdash}%
}

\usepackage[ruled,vlined]{algorithm2e}

\SetKwInput{KwInput}{Input}
\SetKwInput{KwOutput}{Output}
\SetKwFor{For}{for}{do}{end}
\SetKwIF{If}{ElseIf}{Else}{if}{then}{else if}{else}{end}
\DontPrintSemicolon
\SetKwComment{Comment}{$\triangleright$\ }{}

\newcommand{\AlgBlockComment}[1]{%
  \BlankLine
  \noindent
  \textcolor{RoyalBlue}{%
    \(\blacktriangledown\)\ \textit{#1}%
  }\;
}

\makeatletter

\renewcommand{\algocf@captiontext}[2]{%
  #1\algocf@typo.\ \AlCapFnt{}#2%
}

\def\@algocf@capt@plain{top}

\renewcommand{\algocf@makecaption}[2]{%
  \addtolength{\hsize}{\algomargin}%
  \sbox\@tempboxa{\algocf@captiontext{#1}{#2}}%
  \ifdim\wd\@tempboxa>\hsize
    \hskip .5\algomargin
    \parbox[t]{\hsize}{%
      \algocf@captiontext{#1}{#2}%
    }%
  \else
    \global\@minipagefalse
    \hbox to\hsize{\box\@tempboxa}%
  \fi
  \addtolength{\hsize}{-\algomargin}%
}

\makeatother

\newcommand{\exprcell}[1]{%
  \begingroup
  \renewcommand{\arraystretch}{1.05}%
  \makecell[l]{\(\displaystyle #1\)}%
  \endgroup
}

\newcolumntype{H}{%
  >{\setbox0=\hbox\bgroup}c<{\egroup}%
}

\definecolor{ExprBack}{RGB}{255,241,238}

\NewDocumentCommand{\exprbox}{m g}{%
  \begingroup
  \setlength{\fboxsep}{3pt}%
  \colorbox{ExprBack}{%
    \begin{minipage}[t]{0.95\linewidth}
      \IfValueTF{#2}{%
        #1:\quad
        $\displaystyle
        \begin{aligned}[t]
          #2
        \end{aligned}$%
      }{%
        $\displaystyle
        \begin{aligned}[t]
          #1
        \end{aligned}$%
      }%
    \end{minipage}%
  }%
  \endgroup
}

\newcommand{\reprowsep}{%
  \addlinespace[2pt]
  \arrayrulecolor{black!12}
  \midrule
  \arrayrulecolor{black}
  \addlinespace[2pt]
}

\definecolor{RawBlue}{RGB}{28,82,155}
\definecolor{FinalGreen}{RGB}{24,120,72}

\newcommand{\rawtag}{%
  \textcolor{RawBlue}{\textbf{\textit{Raw}}}%
}

\newcommand{\finaltag}{%
  \textcolor{FinalGreen}{\textbf{\textit{Final}}}%
}

\newtheorem{theorem}{Theorem}
\newtheorem{lemma}{Lemma}
\newtheorem{proposition}{Proposition}
\newtheorem{corollary}{Corollary}
\newtheorem{definition}{Definition}
\newtheorem{remark}{Remark}

\newtheorem{assumption}{Assumption}

\renewcommand{\theassumption}{\arabic{assumption}}

\newcommand{\resetstatementcounters}{%
  \setcounter{theorem}{0}%
  \setcounter{lemma}{0}%
  \setcounter{proposition}{0}%
  \setcounter{corollary}{0}%
  \setcounter{definition}{0}%
  \setcounter{remark}{0}%
  \setcounter{example}{0}%
  \setcounter{assumption}{0}%
  \setcounter{aassumption}{0}%
  \setcounter{conjecture}{0}%
  \setcounter{axiom}{0}%
  \setcounter{condition}{0}%
}

\usepackage{listings}
\usepackage[most]{tcolorbox}
\tcbuselibrary{listings,breakable,skins}

\definecolor{codebg}{RGB}{248,249,252}
\definecolor{codeframe}{RGB}{190,198,215}
\definecolor{codetitlebg}{RGB}{232,238,248}
\definecolor{codekeyword}{RGB}{0,82,155}
\definecolor{codestring}{RGB}{163,21,21}
\definecolor{codecomment}{RGB}{90,120,90}
\definecolor{codeidentifier}{RGB}{45,45,45}

\lstdefinestyle{pysettings}{
  language=Python,
  basicstyle=\ttfamily\footnotesize\color{codeidentifier},
  columns=fullflexible,
  keepspaces=true,
  showstringspaces=false,
  breaklines=true,
  breakatwhitespace=false,
  tabsize=4,
  keywordstyle=\bfseries\color{codekeyword},
  stringstyle=\color{codestring},
  commentstyle=\itshape\color{codecomment},
  numbers=left,
  numberstyle=\scriptsize\color{black!45},
  numbersep=8pt,
  xleftmargin=1.2em,
  frame=none,
  emph={
    dsr_settings,
    estimator,
    program_,
    n_samples,
    batch_size,
    max_length,
    protected,
    random_state,
    function_set
  },
  emphstyle=\color{codekeyword}
}

\newtcblisting{pysettingsbox}[2][]{%
  enhanced,
  breakable,
  listing only,
  listing options={style=pysettings},
  colback=codebg,
  colframe=codeframe,
  coltitle=black,
  colbacktitle=codetitlebg,
  title={#2},
  fonttitle=\bfseries\footnotesize,
  attach boxed title to top left={
    xshift=1.2em,
    yshift=-2mm
  },
  boxed title style={
    colback=codetitlebg,
    colframe=codeframe,
    arc=2pt,
    boxrule=0.4pt
  },
  sharp corners,
  arc=2pt,
  boxrule=0.5pt,
  left=1em,
  right=0.8em,
  top=0.8em,
  bottom=0.7em,
  before skip=8pt,
  after skip=8pt,
  overlay first={
    \node[
      anchor=south east,
      font=\scriptsize\itshape\color{black!55}
    ] at (frame.south east) {continued on next page};
  },
  overlay middle={
    \node[
      anchor=north east,
      font=\scriptsize\itshape\color{black!55}
    ] at (frame.north east) {continued from previous page};
    \node[
      anchor=south east,
      font=\scriptsize\itshape\color{black!55}
    ] at (frame.south east) {continued on next page};
  },
  overlay last={
    \node[
      anchor=north east,
      font=\scriptsize\itshape\color{black!55}
    ] at (frame.north east) {continued from previous page};
  },
  #1
}

\definecolor{HierRed}{HTML}{8F2A22}
\newcommand{\hierbosss}{\textcolor{HierRed}{\texttt{HierBOSSS}}}

\newcommand{\operon}{\texttt{operon}}
\newcommand{\gplearn}{\texttt{gplearn}}
\newcommand{\pysr}{\texttt{PySR}}
\newcommand{\dsr}{\texttt{DSR}}
\newcommand{\qlattice}{\texttt{QLattice}}
\newcommand{\sisso}{\texttt{SISSO}}
\newcommand{\bms}{\texttt{BMS}}
\newcommand{\bsr}{\texttt{BSR}}

\newcommand{\aif}{\texttt{AI-Feynman}}
\newcommand{\sr}{\texttt{SR}}

\def\T{\top}

\colorlet{CrossRefColor}{BrickRed}
\colorlet{MainRefColor}{CrossRefColor}

\newcommand{\localnamedref}[2]{%
  \hyperref[#2]{#1~\ref*{#2}}%
}

\newcommand{\localeqnref}[1]{%
  \hyperref[#1]{(\ref*{#1})}%
}

\newcommand{\crossdoclink}[2]{%
  \begingroup
  \hypersetup{linkcolor=CrossRefColor}%
  \hyperref[#1]{%
    \textcolor{CrossRefColor}{#2}%
  }%
  \endgroup
}

\newcommand{\mainnamedref}[2]{%
  \crossdoclink{#2}{%
    #1~\ref*{#2} of the main manuscript%
  }%
}

\newcommand{\mainequationref}[1]{%
  \crossdoclink{#1}{(\ref*{#1}) of the main manuscript}
}

\newcommand{\appendixnamedref}[1]{%
  \crossdoclink{app-#1}{%
    Appendix~\ref*{app-#1}%
  }%
}

\NewDocumentCommand{\secref}{o g}{%
  \IfValueTF{#1}
    {\localnamedref{Section}{#1}}
    {\localnamedref{Section}{#2}}%
}

\NewDocumentCommand{\figref}{o g}{%
  \IfValueTF{#1}
    {\localnamedref{Figure}{#1}}
    {\localnamedref{Figure}{#2}}%
}

\NewDocumentCommand{\tabref}{o g}{%
  \IfValueTF{#1}
    {\localnamedref{Table}{#1}}
    {\localnamedref{Table}{#2}}%
}

\NewDocumentCommand{\algref}{o g}{%
  \IfValueTF{#1}
    {\localnamedref{Algorithm}{#1}}
    {\localnamedref{Algorithm}{#2}}%
}

\NewDocumentCommand{\eqnref}{o g}{%
  \IfValueTF{#1}
    {\localeqnref{#1}}
    {\localeqnref{#2}}%
}

\NewDocumentCommand{\appref}{o g}{%
  \IfValueTF{#1}
    {\appendixnamedref{#1}}
    {\appendixnamedref{#2}}%
}

\NewDocumentCommand{\suppref}{o g}{%
  \IfValueTF{#1}
    {\appendixnamedref{#1}}
    {\appendixnamedref{#2}}%
}

\NewDocumentCommand{\mainref}{o g}{%
  \IfValueTF{#1}
    {\mainnamedref{Section}{#1}}
    {\mainnamedref{Section}{#2}}%
}

\NewDocumentCommand{\maineqref}{o g}{%
  \IfValueTF{#1}
    {\mainequationref{#1}} 
    {\mainequationref{#2}}%
}

\NewDocumentCommand{\mainfigref}{o g}{%
  \IfValueTF{#1}
    {\mainnamedref{Figure}{#1}}
    {\mainnamedref{Figure}{#2}}%
}

\NewDocumentCommand{\maintabref}{o g}{%
  \IfValueTF{#1}
    {\mainnamedref{Table}{#1}}
    {\mainnamedref{Table}{#2}}%
}

\NewDocumentCommand{\mainalgref}{o g}{%
  \IfValueTF{#1}
    {\mainnamedref{Algorithm}{#1}}
    {\mainnamedref{Algorithm}{#2}}%
}

\NewDocumentCommand{\mainassref}{o g}{%
  \IfValueTF{#1}
    {\mainnamedref{Assumption}{#1}}
    {\mainnamedref{Assumption}{#2}}%
}

\makeatletter

\let\combined@originallabel\label

\newcommand{\usemainlabelaliases}{%
  \renewcommand{\label}[1]{%
    \combined@originallabel{##1}%
    \combined@originallabel{main-##1}%
  }%
}

\newcommand{\useappendixlabelaliases}{%
  \renewcommand{\label}[1]{%
    \combined@originallabel{##1}%
    \combined@originallabel{app-##1}%
    \combined@originallabel{supp-##1}%
  }%
}

\makeatother

\newcommand{\beginmainappendices}{%
  \inappendixtrue
  \setcounter{tocdepth}{3}%
  \appendix

  \setcounter{section}{0}%
  \setcounter{subsection}{0}%
  \setcounter{subsubsection}{0}%
  \setcounter{equation}{0}%
  \setcounter{figure}{0}%
  \setcounter{table}{0}%
  \setcounter{algocf}{0}%

  \resetstatementcounters

  \counterwithin*{equation}{section}%
  \counterwithin*{figure}{section}%
  \counterwithin*{table}{section}%
  \counterwithin*{algocf}{section}%

  \renewcommand{\thesection}{\Alph{section}}%
  \renewcommand{\thesubsection}{%
    \Alph{section}.\arabic{subsection}%
  }%
  \renewcommand{\thesubsubsection}{%
    \Alph{section}.\arabic{subsection}.\arabic{subsubsection}%
  }%

  \renewcommand{\theequation}{%
    \Alph{section}\arabic{equation}%
  }%
  \renewcommand{\thefigure}{%
    \Alph{section}\arabic{figure}%
  }%
  \renewcommand{\thetable}{%
    \Alph{section}\arabic{table}%
  }%
  \renewcommand{\thealgocf}{%
    \Alph{section}\arabic{algocf}%
  }%

  \def\theHsection{appsec.\Alph{section}}%
  \def\theHsubsection{%
    appsubsec.\Alph{section}.\arabic{subsection}%
  }%
  \def\theHsubsubsection{%
    appsubsubsec.\Alph{section}.\arabic{subsection}.%
    \arabic{subsubsection}%
  }%

  \def\theHequation{%
    appeq.\Alph{section}.\arabic{equation}%
  }%
  \def\theHfigure{%
    appfig.\Alph{section}.\arabic{figure}%
  }%
  \def\theHtable{%
    apptab.\Alph{section}.\arabic{table}%
  }%
  \def\theHalgocf{%
    appalgo.\Alph{section}.\arabic{algocf}%
  }%

  \def\theHtheorem{appthm.\arabic{theorem}}%
  \def\theHlemma{applem.\arabic{lemma}}%
  \def\theHproposition{appprop.\arabic{proposition}}%
  \def\theHcorollary{appcorr.\arabic{corollary}}%
  \def\theHdefinition{appdef.\arabic{definition}}%
  \def\theHremark{apprem.\arabic{remark}}%
  \def\theHexample{appex.\arabic{example}}%
  \def\theHassumption{appass.\arabic{assumption}}%
  \def\theHaassumption{appaass.\arabic{aassumption}}%
  \def\theHconjecture{appconj.\arabic{conjecture}}%
  \def\theHaxiom{appaxiom.\arabic{axiom}}%
  \def\theHcondition{appcond.\arabic{condition}}%
}

\makeatletter

\newcounter{suppdivision}
\renewcommand{\thesuppdivision}{\Roman{suppdivision}}

\newcommand{\suppdivision}[1]{%
  \clearpage
  \refstepcounter{suppdivision}%
  \addcontentsline{toc}{suppdivision}{%
    Part~\thesuppdivision:\space #1%
  }%
  \begin{center}
    {\Large\bfseries
      Part~\thesuppdivision:\space #1%
    }
  \end{center}
  \vspace{0.1pt}%
}

\newcommand{\l@suppdivision}[2]{%
  \addpenalty{-\@highpenalty}%
  \addvspace{1.2em}%
  \noindent
  \parbox{\textwidth}{%
    \centering
    \bfseries
    #1%
  }%
  \par
  \addvspace{0.35em}%
}

\renewcommand{\l@section}{%
  \@dottedtocline{1}{0em}{8.7em}%
}

\renewcommand{\l@subsection}{%
  \@dottedtocline{2}{2.5em}{3.2em}%
}

\renewcommand{\l@subsubsection}{%
  \@dottedtocline{3}{5.7em}{4.0em}%
}

\def\onlineappendix@sict#1#2#3#4#5#6[#7]#8{%
  \ifnum #2>\c@secnumdepth
    \def\@svsec{}%
  \else
    \refstepcounter{#1}%
    \edef\@svsec{\csname the#1\endcsname}%
  \fi
  \@tempskipa #5\relax
  \ifdim \@tempskipa>\z@
    \begingroup #6\relax
      \ifstrequal{#1}{section}{%
        \@hangfrom{%
          \hskip #3\relax
          Online Appendix~\@svsec.\hskip 0.5em%
        }{%
          \interlinepenalty\@M #8\par
        }%
      }{%
        \@hangfrom{%
          \hskip #3\relax
          \@svsec\hskip 0.1em%
        }{%
          \interlinepenalty\@M #8\par
        }%
      }%
    \endgroup
    \csname #1mark\endcsname{#7}%
    \ifstrequal{#1}{section}{%
      \addcontentsline{toc}{#1}{%
        \ifnum #2>\c@secnumdepth\else
          \protect\numberline{%
            Online Appendix~\csname the#1\endcsname.%
          }%
        \fi
        #7%
      }%
    }{%
      \addcontentsline{toc}{#1}{%
        \ifnum #2>\c@secnumdepth\else
          \protect\numberline{\csname the#1\endcsname}%
        \fi
        #7%
      }%
    }%
  \else
    \def\@svsechd{%
      #6\hskip #3\@svsec #8%
      \csname #1mark\endcsname{#7}%
      \addcontentsline{toc}{#1}{%
        \ifnum #2>\c@secnumdepth\else
          \protect\numberline{\csname the#1\endcsname}%
        \fi
        #7%
      }%
    }%
  \fi
  \@xsect{#5}%
}

\newcommand{\beginsupplement}{%
  \counterwithout*{equation}{section}%
  \counterwithout*{figure}{section}%
  \counterwithout*{table}{section}%
  \counterwithout*{algocf}{section}%

  \setcounter{section}{0}%
  \setcounter{subsection}{0}%
  \setcounter{subsubsection}{0}%
  \setcounter{equation}{0}%
  \setcounter{figure}{0}%
  \setcounter{table}{0}%
  \setcounter{algocf}{0}%
  \setcounter{suppdivision}{0}%

  \resetstatementcounters

  \setcounter{secnumdepth}{3}%
  \setcounter{tocdepth}{3}%

  \renewcommand{\thesection}{\arabic{section}}%
  \renewcommand{\thesubsection}{%
    \arabic{section}.\arabic{subsection}%
  }%
  \renewcommand{\thesubsubsection}{%
    \arabic{section}.\arabic{subsection}.%
    \arabic{subsubsection}%
  }%

  \let\@sict\onlineappendix@sict

  \renewcommand{\theequation}{S\arabic{equation}}%
  \renewcommand{\thefigure}{S\arabic{figure}}%
  \renewcommand{\thetable}{S\arabic{table}}%
  \renewcommand{\thealgocf}{S\arabic{algocf}}%

  \def\theHsection{suppsec.\arabic{section}}%
  \def\theHsubsection{%
    suppsubsec.\arabic{section}.\arabic{subsection}%
  }%
  \def\theHsubsubsection{%
    suppsubsubsec.\arabic{section}.\arabic{subsection}.%
    \arabic{subsubsection}%
  }%

  \def\theHequation{suppeq.\arabic{equation}}%
  \def\theHfigure{suppfig.\arabic{figure}}%
  \def\theHtable{supptab.\arabic{table}}%
  \def\theHalgocf{suppalgo.\arabic{algocf}}%

  \def\theHtheorem{suppthm.\arabic{theorem}}%
  \def\theHlemma{supplem.\arabic{lemma}}%
  \def\theHproposition{suppprop.\arabic{proposition}}%
  \def\theHcorollary{suppcorr.\arabic{corollary}}%
  \def\theHdefinition{suppdef.\arabic{definition}}%
  \def\theHremark{supprem.\arabic{remark}}%
  \def\theHexample{suppex.\arabic{example}}%
  \def\theHassumption{suppass.\arabic{assumption}}%
  \def\theHaassumption{suppaass.\arabic{aassumption}}%
  \def\theHconjecture{suppconj.\arabic{conjecture}}%
  \def\theHaxiom{suppaxiom.\arabic{axiom}}%
  \def\theHcondition{suppcond.\arabic{condition}}%

  \setlength{\parindent}{0pt}%
  \setlength{\parskip}{4pt plus 1pt minus 1pt}%

  \setlength{\abovedisplayskip}{8pt plus 3pt minus 5pt}%
  \setlength{\belowdisplayskip}{8pt plus 3pt minus 5pt}%
  \setlength{\abovedisplayshortskip}{4pt plus 2pt minus 3pt}%
  \setlength{\belowdisplayshortskip}{6pt plus 2pt minus 3pt}%

  \setcounter{topnumber}{4}%
  \setcounter{bottomnumber}{2}%
  \setcounter{totalnumber}{6}%

  \renewcommand{\topfraction}{0.92}%
  \renewcommand{\bottomfraction}{0.85}%
  \renewcommand{\textfraction}{0.06}%
  \renewcommand{\floatpagefraction}{0.75}%

  \setlength{\textfloatsep}{12pt plus 3pt minus 4pt}%
  \setlength{\floatsep}{10pt plus 3pt minus 3pt}%
  \setlength{\intextsep}{10pt plus 3pt minus 3pt}%
}

\makeatother

\makeatletter
\newif\ifinsupplement
\insupplementfalse
\newif\ifinappendix
\inappendixfalse
\let\originaladdcontentsline\addcontentsline
\renewcommand{\addcontentsline}[3]{%
  \def\tempcontentsfile{#1}%
  \def\standardtoc{toc}%
  \ifx\tempcontentsfile\standardtoc
    \ifinsupplement
      \originaladdcontentsline{stoc}{#2}{#3}%
    \else
      \ifinappendix
        \originaladdcontentsline{atoc}{#2}{#3}%
      \fi
    \fi
  \else
    \originaladdcontentsline{#1}{#2}{#3}%
  \fi
}
\newcommand{\supplementtableofcontents}{%
  \begin{center}
    {\Large\bfseries Contents}
  \end{center}
  \vspace{0.5em}
  \begingroup
    \@starttoc{stoc}%
  \endgroup
  \clearpage
}
\newcommand{\appendixtableofcontents}{%
  \begingroup
    \renewcommand{\l@section}{\@dottedtocline{1}{0em}{2.4em}}%
    \renewcommand{\l@subsection}{\@dottedtocline{2}{2.4em}{3.0em}}%
    \renewcommand{\l@subsubsection}{\@dottedtocline{3}{5.4em}{3.8em}}%
    \begin{center}
      {\Large\bfseries Contents}
    \end{center}
    \vspace{0.5em}
    \@starttoc{atoc}%
  \endgroup
  \clearpage
}
\makeatother

\usepackage[normalem]{ulem}
\usepackage[textsize=scriptsize]{todonotes}

\newif\ifreviewcomments
\reviewcommentstrue

\ifreviewcomments
  \makeatletter
  \@mparswitchfalse
  \makeatother

\else
  
\fi

\jmlrheading
  {XX}
  {20XX}
  {1--\pageref{LastPage}}
  {X/XX; Revised X/XX}
  {X/XX}
  {XX-0000}
  {Somjit Roy, Pritam Dey, Bani K. Mallick, and Debdeep Pati}

\ShortHeadings
  {\hierbosss}
  {Roy, Dey, Mallick, and Pati}

\firstpageno{1}

\begin{document}

\title{Probabilistic Symbolic Regression for Equation Discovery via
Operator-Induced and Regularized Symbolic Forests}

\author{\name
Somjit Roy
  \email{\href{mailto:sroy_123@tamu.edu}{sroy\_123@tamu.edu}} \\
\addr Department of Statistics\\
Texas A\&M University\\
College Station, TX 77843, USA
\AND
\name
Pritam Dey
  \email{\href{mailto:pritam.dey@tamu.edu}{pritam.dey@tamu.edu}} \\
\addr Department of Statistics\\
Texas A\&M University\\
College Station, TX 77843, USA
\AND
\name
Bani K. Mallick
  \email{\href{mailto:bmallick@stat.tamu.edu}
               {bmallick@stat.tamu.edu}} \\
\addr Department of Statistics\\
Texas A\&M University\\
College Station, TX 77843, USA
\AND
\name
Debdeep Pati
  \email{\href{mailto:dpati2@wisc.edu}{dpati2@wisc.edu}} \\
\addr Department of Statistics\\
University of Wisconsin--Madison\\
Madison, WI 53706, USA
}

\editor{My editor}

\maketitle

\usemainlabelaliases

\resetstatementcounters

\begin{abstract}
Symbolic regression has emerged as a powerful tool for artificial intelligence-driven scientific discovery by learning interpretable analytical expressions that reveal governing relationships directly from data. Existing methods, however, often rely on heuristic search, struggle to balance predictive accuracy with expression complexity in noisy settings, and offer limited characterization of symbolic uncertainty. Probabilistic approaches that address these challenges in a unified manner remain underexplored. We introduce a probabilistic symbolic regression framework that represents mathematical expressions as ensembles of symbolic trees. A regularizing prior over tree topology controls expression complexity, while an Occam’s window-based posterior summary captures uncertainty across multiple plausible symbolic models. Given the limited existing theoretical treatment of symbolic regression, we develop posterior concentration guarantees when symbolic expressions approximate the underlying relationship arbitrarily well, with a near-parametric rate when an exact finite formula exists. Additionally, we establish a sharp oracle concentration result under symbolic misspecification. Comparisons of our proposed framework with state-of-the-art competitors demonstrate superior predictive accuracy, optimal symbolic complexity, and stable structural recovery when learning benchmark scientific equations, together with the identification of scientifically interpretable descriptor formulas in a challenging materials discovery application.
\end{abstract}


\section{Introduction}
\label{sec:introduction}

Scientific machine learning offers a promising avenue toward increasingly autonomous scientific discovery. In this vision, learning algorithms contribute across the scientific workflow, from extracting structure in complex observations to formulating hypotheses and guiding subsequent experiments~\citep{Wang2023ScientificDiscoveryAI}. By integrating data-driven modeling with scientific knowledge, including physical constraints and mechanistic structure, scientific machine learning seeks models that are not only predictive but also scientifically meaningful~\citep{SciML-materials-1,Karniadakis2021PhysicsInformed}.

Symbolic regression (\sr) is particularly well suited to this objective. It learns analytical expressions directly from data, expressing discovered relationships in the mathematical language in which scientific laws and mechanistic hypotheses are commonly formulated~\citep{MakkeChawla2024SRReview}. Its outputs can therefore be inspected, tested, and refined using domain knowledge. By connecting flexible data-driven search with equation-based scientific reasoning, \sr\ provides a compelling route from prediction to interpretable structural discovery.

\subsection{The Symbolic Regression Problem}
\label{subsec:SR-problem}

Formally, consider an unknown input-output relation
$f:\mathfrak{X}\subset\mathbb{R}^{p}\to\mathbb{R}$ observed through data
$\mathcal{D}_n=\{(\bm{x}_i,y_i): i=1,\ldots,n\}$. Given primary features
$x_1,\ldots,x_p$ and a prescribed library of mathematical operators (such as $+$, $\times$, $\sin$, $\log$), \sr\ searches over a structured class of analytical expressions generated by composing these features and operators. In contrast to prediction-centric regression methods~\citep{LASSO-Tibshirani,RasmussenWilliams2006,BCART-Mallick,BCART,BART}, \sr\ seeks a compact expression $\hat f$ that approximates $f$ accurately while revealing the algebraic and compositional structure of the underlying relationship. Successful applications of \sr\ include sparse identification of nonlinear dynamical systems~\citep{SR-SciML-2}, materials discovery~\citep{SR-SciML-3}, and extraction of fundamental scientific laws~\citep{SR-SciML-1}.

Despite this promise, \sr\ poses several intertwined computational and statistical challenges~\citep{LaCava-NIPS}. The space of candidate expressions grows combinatorially with the number of features, operators, and expression complexity, making \sr\ NP-hard~\citep{virgolin2022symbolic}. Beyond computation, \sr\ must balance predictive accuracy with scientific interpretability. Noise further exacerbates this trade-off by promoting spurious variables, operators, and subexpressions, particularly in small-sample scientific datasets. In addition, symbolic representations are non-identifiable, since algebraically distinct expressions can represent the same function or yield nearly indistinguishable predictions over the observed domain. These issues make characterization of symbolic uncertainty essential. At the same time, formal statistical theory of \sr\ also remains largely scarce.

\subsection{Motivating Application}
\label{subsec:motivating-application}

Our investigation is strongly motivated by materials discovery. Since the launch of the Materials Genome Initiative~\citep{MaterialsGenomeInitiative2011}, materials informatics has integrated computation and high-throughput experimentation to accelerate the design of new materials~\citep{Green2017HighThroughput}. A central objective is to discover materials genes (descriptors), which are low-dimensional, physically interpretable combinations of structural, compositional, or electronic features governing a target material property~\citep{Ghiringhelli2015,Foppa2021MaterialsGenes}. In this setting, \sr\ searches for predictive and interpretable analytical descriptors that reveal important symbolic structure-property relationships.

Specifically, in \hyperref[sec:perovskites-data-study]{Section~\ref{sec:perovskites-data-study}} we study catalytic-descriptor discovery for oxide perovskites, a compositionally versatile class of materials central to oxygen electrocatalysis and renewable-energy conversion~\citep{Weng2020SimpleDescriptor}. Their catalytic activity is governed by strongly coupled geometric, chemical, and electronic effects, making the recovery of a compact structure-activity relation both scientifically consequential and statistically challenging. The dataset analyzed~\citep{Weng2020SimpleDescriptor} contains only a limited collection of experimentally characterized compounds, motivating \sr\ methods that can identify parsimonious activity descriptors for guiding new catalyst designs while accounting for uncertainty across competing symbolic explanations.

\subsection{Related Work}
\label{subsec:related-works}

The aforementioned challenges and scientific applications of \sr\ have fueled several complementary lines of work, spanning evolutionary search, modern machine learning, engineering-oriented screening methods, and probabilistic model-based approaches. Classical \sr\ algorithms such as $\gplearn$~\citep{stephens2016gplearn}, $\operon$~\citep{operon}, and $\pysr$~\citep{pysr} based on genetic programming~\citep{koza1992genetic}, explore symbolic expression spaces through stochastic mutation, crossover, and selection. Despite their flexibility, such methods can be computationally expensive, sensitive to initialization and search heuristics, and prone to overly complex output expressions~\citep{GP-bad}. Recent machine learning-based approaches, including deep symbolic regression ($\dsr$)~\citep{Deep-SR}, $\aif$~\citep{AI-Feynman}, and $\qlattice$~\citep{feyn-qlattice}, improve scalability, prediction, and search efficiency by incorporating neural architectures, physics-informed decompositions, or graph-based symbolic learning. However, due to the NP-hard nature of \sr, their empirical success can nevertheless depend on favorable regimes with sufficiently large samples, low noise, and strong structural regularity. In parallel, engineering-focused methods such as sure independence screening and sparsifying operator (\sisso)~\citep{Ghiringhelli2016LearningPD,SISSO} construct astronomically large symbolic dictionaries and use compressed sensing or penalized regression to identify sparse analytical expressions, with applications in phase stability analysis~\citep{SISSO-application-1} and catalysis~\citep{SISSO-application-2}. Related extensions~\citep{iBART} reduce the resulting search space by combining nonparametric feature selection with subsequent sparse symbolic estimation, but still rely on large candidate libraries and multi-stage heuristics. Collectively, these developments broaden the computational reach of \sr, but also highlight the need for model-based \sr\ frameworks that can jointly learn symbolic structure, control complexity, and account for uncertainty across multiple symbolic representations, particularly in noisy small- to moderate-sample scientific settings.

Existing probabilistic model-based formulations of \sr\ are built around expression tree representations~\citep{Bartlett,vasst}, where terminal nodes correspond to primary features and nonterminal nodes encode mathematical operators, as illustrated in \hyperref[fig:symbolic-tree-examples]{Figures~\ref{fig:symbolic-tree-examples}} and~\ref{fig:comparison}. Two prominent such \sr\ modules relevant to our work are Bayesian symbolic regression ($\bsr$)~\citep{BSR} and Bayesian Machine Scientist ($\bms$)~\citep{BMS}. $\bsr$ models the response as a linear combination of expression trees, but the outer linear coefficients are re-estimated by ordinary least squares at each proposed forest. Candidate forests are therefore evaluated using a likelihood at plug-in coefficient estimates, preventing posterior uncertainty propagation through the additive regression layer and potentially making structural comparisons sensitive to weakly identified or collinear tree outputs.
Moreover, operator choices in \bsr\ are governed by fixed weights shared across all trees, limiting both data-adaptive learning of operator preferences and the ability of different trees to independently explore distinct regions of the symbolic expression space. It also relies on small coefficient estimates to render redundant trees negligible, without explicitly selecting active symbolic components. These limitations can lead to unnecessarily complicated expressions even when the underlying relationship is simple~\citep{LaCava-NIPS}, as demonstrated in \hyperref[sec:learning-Feynman-equations]{Section~\ref{sec:learning-Feynman-equations}}. In contrast, \bms\ uses an approximate marginal posterior over a single closed-form expression. Its single-tree representation requires additive scientific structure to be encoded within one expression rather than as separately interpretable components, while its restriction against sampling repeated or equivalent expressions may hinder exploration of structurally related symbolic representations. Moreover, the marginal posterior approximation, computed after optimizing expression-specific continuous parameters, may be inaccurate in small-sample scientific settings. Furthermore, the expression prior of \bms\ is calibrated from operator frequencies in an external corpus and remains fixed during analysis, making symbolic preferences corpus-dependent rather than adaptive to the observed data. More broadly, although both \bsr\ and \bms\ are Bayesian \sr\ methods, their standard reported outputs center on a single representative fitted expression rather than an interpretable posterior set of competing symbolic structures. Consequently, neither framework directly summarizes structural uncertainty through multiple plausible equations that may provide complementary scientific explanations of the phenomenon underlying the data. \hyperref[tab:probabilistic-sr-comparison]{Table~\ref{tab:probabilistic-sr-comparison}} summarizes these methodological limitations and contrasts both formulations with our proposed framework, which we introduce next.

\begingroup
\scriptsize
\setlength{\tabcolsep}{2.8pt}
\renewcommand{\arraystretch}{1.10   }

\begin{table}[!ht]
\scriptsize
\centering
\caption{\centering
Methodological advances over existing probabilistic \sr\ methods.
}
\label{tab:probabilistic-sr-comparison}

\begin{tabular}{@{}
    >{\raggedright\arraybackslash}p{0.40\textwidth}
    >{\centering\arraybackslash}p{0.18\textwidth}
    >{\centering\arraybackslash}p{0.18\textwidth}
    >{\centering\arraybackslash}p{0.19\textwidth}
@{}}

\toprule
\toprule

\textbf{Methodological characteristic}
&
\makecell{
    \textbf{\texttt{BSR}}\\[-1pt]
    \tiny\citep{BSR}
}
&
\makecell{
    \textbf{\texttt{BMS}}\\[-1pt]
    \tiny\citep{BMS}
}
&
\makecell{
    \textbf{\hierbosss}\\[-1pt]
    \tiny (our framework)
}
\\

\midrule

\textbf{Symbolic representation}
&
Tree ensemble
&
Single tree
&
Tree ensemble
\\

\addlinespace[2pt]

\textbf{Regression parameter learning}
&
Least-squares plug-in
&
\namark
&
Full posterior
\\

\addlinespace[2pt]

\textbf{Operator/feature preferences}
&
\makecell{
    Fixed\\[-1pt]
    \tiny (user-specified)
}
&
\makecell{
    Fixed\\[-1pt]
    \tiny (corpus-based)
}
&
Data-adaptive
\\

\addlinespace[2pt]

\textbf{Tree-specific operator/feature weights}
&
\xmark
&
\namark
&
\cmark
\\

\addlinespace[2pt]

\textbf{Active symbolic component selection}
&
\xmark
&
\namark
&
\cmark
\\

\addlinespace[2pt]

\textbf{Symbolic uncertainty summary}
&
\xmark
&
\xmark
&
\cmark
\\

\bottomrule
\bottomrule

\end{tabular}
\end{table}

\endgroup

\subsection{Our Contributions}
\label{subsec:contributions}

To address the preceding gaps, we propose \hierbosss\ (\textcolor{HierRed}{\texttt{Hier}}archical \textcolor{HierRed}{\texttt{B}}ayesian \textcolor{HierRed}{\texttt{O}}perator-induced \textcolor{HierRed}{\texttt{S}}ymbolic regression forests for \textcolor{HierRed}{\texttt{S}}tructural discovery of \textcolor{HierRed}{\texttt{S}}ymbolic expressions), a probabilistic framework in which symbolic expressions are recursively represented as random operator-induced symbolic trees, as demonstrated in \hyperref[subsec:symbolic-tree-representation]{Section~\ref{subsec:symbolic-tree-representation}}. The \hierbosss\ model in \hyperref[subsec:symbolic-forest-component]{Section~\ref{subsec:symbolic-forest-component}} represents the unknown regression surface $f$ (under noise) as an affine combination of symbolic trees, allowing additive scientific structure while preserving the interpretability of individual symbolic components. In accordance with the Occam's razor principle~\citep{Occams-Razor-1}, the prior on tree topology in \hyperref[subsec:prior-specification]{Section~\ref{subsec:prior-specification}} regularizes expression complexity by favoring simpler representations both among algebraically equivalent formulas and among distinct symbolic functions with comparable predictive support. Unlike existing probabilistic \sr\ formulations, \hierbosss\ learns symbolic preferences data-adaptively through Dirichlet priors on tree-specific operator and feature assignment weights, and uses conjugate priors on the outer regression coefficients and noise variance, thereby providing full uncertainty propagation across all model unknowns.

A principal methodological highlight is an efficient posterior computation and symbolic model selection strategy, as developed in \hyperref[subsec:posterior-inference]{Sections~\ref{subsec:posterior-inference}} and~\ref{subsec:symbolic-model-selection-refinement}. By marginalizing the operator and feature assignment weights together with the regression and noise parameters, we obtain the joint marginal posterior over ensembles of symbolic trees (symbolic forests), which drives posterior sampling across the discrete symbolic expression space. Rather than returning a single best-fitted expression, as is common in many \sr\ methods, \hierbosss\ uses an Occam’s window-based~\citep{madigan1994model} posterior summary to retain multiple interpretable symbolic explanations with high posterior support. As emphasized above, such uncertainty-aware summarization is especially important in scientific regimes with noisy, small-sample or collinear datasets, where several formulas may be empirically competitive.

Given the scarcity of formal statistical guarantees for \sr, \hyperref[sec:posterior-contraction]{Section~\ref{sec:posterior-contraction}} underscores a key technical contribution of this work that establishes posterior concentration results for \hierbosss\ under mild regularity conditions on the symbolic function class and the prior. One of the main challenges is the derivation of a symbolic complexity scale that simultaneously controls the intertwined discrete and continuous complexities inherent to the symbolic expression space. When the true data-generating regression surface $f_0$ is expressible, or can be arbitrarily well approximated, by symbolic forests from the chosen operator library, the posterior concentrates around $f_0$ at a rate governed by the optimal trade-off between empirical symbolic approximation error and this complexity scale. Moreover, exact finite symbolic representability yields a near-parametric rate up to logarithmic factors. We also address the practically important setting of symbolic misspecification, where $f_0$ lies outside the \hierbosss-induced symbolic function class, and establish a sharp oracle concentration result~\citep{sharp-oracle} around the optimal population symbolic approximation error, even when this optimum is not attained by any symbolic model. In contrast to classical Bayesian misspecification theory~\citep{kleijn2006misspecification}, our analysis does not require a finite set of Kullback-Leibler minimizers or the verification of specialized testing separation conditions over convex covers, which are particularly cumbersome to establish for discrete and nonconvex symbolic model classes. Observe that, owing to the non-identifiability of symbolic representations, these guarantees are at the level of predictive functions rather than asserting recovery of a unique symbolic expression or a set of minimal-complexity symbolic representations.

Finally, \hyperref[sec:learning-Feynman-equations]{Sections~\ref{sec:learning-Feynman-equations}} and~\ref{sec:perovskites-data-study} provide strong empirical evidence for the effectiveness of \hierbosss\ by comparing it against several state-of-the-art \sr\ methods within \texttt{SRBench}~\citep{LaCava-NIPS} on a suite of Feynman equations~\citep{AI-Feynman} and the oxide perovskite catalyst dataset~\citep{Weng2020SimpleDescriptor}, respectively.
These studies involve moderate-dimensional feature sets, with $p=4, 5, 8$ for the Feynman equations and $p=8$ for the perovskite dataset. This range of $p$ is representative of scientific discovery settings, where domain knowledge guides feature specification before symbolic structure learning. 
On the Feynman benchmarks, \hierbosss\ consistently achieves a favorable balance between predictive accuracy, symbolic parsimony, and exact structural recovery, whereas competing methods typically fail to satisfy at least one of these desiderata. For the oxide perovskite catalyst data study, \hierbosss\ learns compact, scientifically interpretable descriptor expressions that recover important structure--activity relationships underlying oxygen evolution catalysis~\citep[Table 2]{Weng2020SimpleDescriptor}.
A \texttt{Python} implementation of \hierbosss\ is available at \href{https://github.com/Roy-SR-007/HierBOSSS}{\texttt{github.com/Roy-SR-007/HierBOSSS}}.

\section{The \texorpdfstring{\hierbosss}{HierBOSSS} Framework}
\label{sec:methodology}

Here we present the statistical formulation and methodological foundations of \hierbosss\ for learning interpretable symbolic relationships from noisy data. By representing candidate expressions as structured ensembles of symbolic trees, the resulting probabilistic framework provides the basis for the inferential and computational developments in the sequel.

\subsection{Symbolic Tree Representation of Scientific Expressions}
\label{subsec:symbolic-tree-representation}

\begin{figure}[H]
\centering

\begin{subfigure}{0.28\textwidth}
    \centering
    \includegraphics[width=0.35\linewidth]{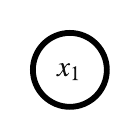}
    \caption{}
    \label{fig:tree-1}
\end{subfigure}
\hfill
\begin{subfigure}{0.28\textwidth}
    \centering
    \includegraphics[width=0.80\linewidth]{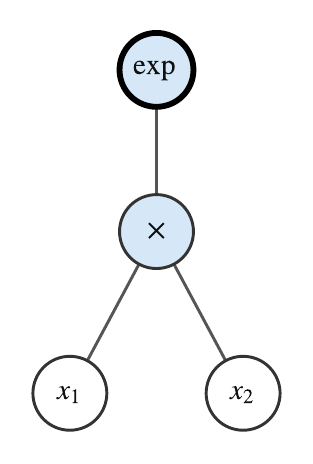}
    \caption{}
    \label{fig:tree-2}
\end{subfigure}
\hfill
\begin{subfigure}{0.28\textwidth}
    \centering
    \includegraphics[width=0.80\linewidth]{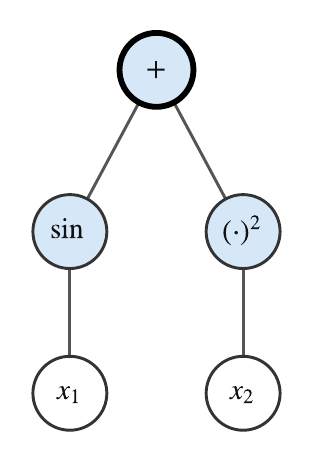}
    \caption{}
    \label{fig:tree-3}
\end{subfigure}
\caption{\centering{Symbolic tree representations: (a) a primitive expression ($\underline{x}_1$), (b) a unary transformation ($\underline{\exp}\{x_1 x_2\}$), and (c) binary combination of sub-expressions ($\sin x_1 \underline{+} x_{2}^{2}$).}}
\label{fig:symbolic-tree-examples}
\end{figure}

Scientific expressions or equations are typically built from primary features $x_1, \ldots, x_p$ and a library of continuous mathematical operators $\mathbb O = \mathbb O_u \cup \mathbb O_b$, where \(\mathbb O_u\) and \(\mathbb O_b\) denote unary and binary operator sets, respectively. For instance, $\mathbb O$ may contain operators such as $\exp$, $\sin$, $\mathrm{inv} \equiv 1/(\cdot)$, $\mathrm{neg} \equiv -(\cdot)$, $+$, and $\times$. For \hierbosss, we allow $\mathbb{O}$ to be user-specified. Starting from primitive features, more complex expressions are formed recursively by applying unary operators to one subexpression and binary operators to pairs of subexpressions, for example, $x_1$, $\exp\{x_1 x_2\}$, and $\sin(x_1) + x_2^2$, as illustrated in \hyperref[fig:symbolic-tree-examples]{Figure~\ref{fig:symbolic-tree-examples}}. This recursive construction described in \hyperref[alg:symbolic-tree-generation]{Algorithm~\ref{alg:symbolic-tree-generation}} naturally induces the following symbolic tree representation.

\begin{definition}[Symbolic tree]
\label{def:symbolic-tree}
A symbolic tree is a rooted tree that encodes an analytical expression, with terminal nodes labeled by primary features $x_1, \ldots, x_p$ and nonterminal nodes labeled by operators from $\mathbb O$. A node assigned a unary operator from $\mathbb O_u$ has one child, whereas a node assigned a binary operator from $\mathbb O_b$ has two children. The associated expression is obtained by evaluating the tree recursively from the terminal nodes to the root.
\end{definition}

Let $\mathbb{T}_{\mathbb O, p}$ denote the collection of all such symbolic trees generated recursively by initializing \hyperref[alg:symbolic-tree-generation]{Algorithm~\ref{alg:symbolic-tree-generation}} at depth $D=0$, using the operator library $\mathbb O$ and $p$ primary features $x_1,\ldots,x_p$. For any $T\in\mathbb T_{\mathbb O,p}$, let $\mathcal{N}(T) = \mathcal{N}_{\mathrm{op}}(T) \cup \mathcal{N}_{\mathrm{ft}}(T)$ be its node set, where $\mathcal N_{\mathrm{op}}(T)$ and $\mathcal N_{\mathrm{ft}}(T)$ are the sets of nonterminal and terminal nodes, respectively. Let $\zeta_0$ denote the root node of $T$. 
As a consequence of \hyperref[alg:symbolic-tree-generation]{Algorithm~\ref{alg:symbolic-tree-generation}}, node depths are assigned by setting $m_{\zeta_0} = 0$ and, for every child $\zeta'$ of a node $\zeta$, setting $m_{\zeta'} = m_{\zeta}+1$. The depth of $T$ is defined as $\mathrm{d}(T) := \max_{\zeta \in \mathcal{N}(T)} m_{\zeta}$ and $S(T) := |\mathcal N_{\mathrm{op}}(T)|$ defines the tree operator count.

\begin{algorithm}[H]
\caption{Recursive symbolic tree generation: \texttt{GrowSymbolicTree}($D$)}
\label{alg:symbolic-tree-generation}
\KwInput{
Operator set $\mathbb O = \{\psi_1, \ldots, \psi_{|\mathbb  O|}\}$, features $x_1,\ldots,x_p$, starting depth $D$.
}
\textbf{Initialize} a node $\zeta$ and set its depth as $m_\zeta= D$.\;

\textbf{Assign} split indicator $B_\zeta \in \{0, 1\}$.\Comment*{\footnotesize{Terminal if $B_{\zeta}=0$; split otherwise}}

\If{$B_\zeta=0$}{
\textbf{Assign} feature $x_{\ell}$ to $\zeta$ for $\ell\in\{1,\ldots,p\}$.\Comment*{\footnotesize{Feature assignment}}}

\Else{
\textbf{Assign} operator $\psi_o \in \mathbb O$ to $\zeta$.\Comment*{\footnotesize{Operator assignment}}
\textbf{Generate} left child subtree of $\zeta$ as \texttt{GrowSymbolicTree}($D + 1$).\;
\If{$\mathrm{arity}(\psi_o) = 2$}{
\textbf{Generate} right child subtree of $\zeta$ as \texttt{GrowSymbolicTree}($D + 1$).
}
}
\KwOutput{
A symbolic subtree $T$ rooted at $\zeta$.
}
\end{algorithm}
\begin{figure}[H]
    \centering
    \includegraphics[width=0.95\linewidth]{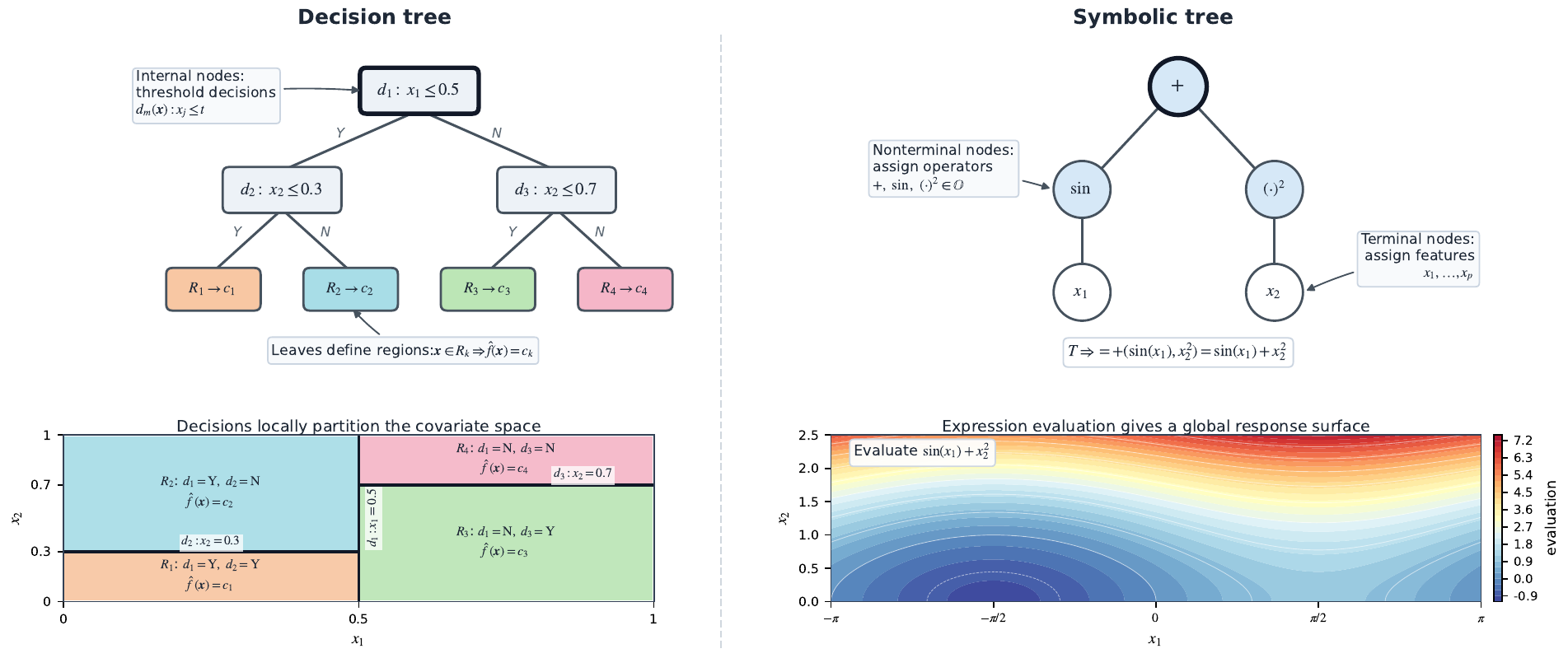}
    \caption{\centering{Local partitioning of covariate space using decision tree versus global operator-feature composition using symbolic tree.}}
    \label{fig:comparison}
\end{figure}

It is useful to distinguish the symbolic tree representation of \hierbosss\ from decision trees, as summarized in \hyperref[fig:comparison]{Figure~\ref{fig:comparison}}. Decision tree-based models~\citep{random-forests, BCART-Mallick, BCART, BART} recursively induce partitions of the covariate space through threshold rules, for example $x_1 \leq 0.5$ followed by additional splits on $x_2$, and assign a local prediction $c_k$ to each resulting region $R_k$. Thus, their output is determined by the region in which $\bm x$ falls. In contrast, a symbolic tree as in \hyperref[def:symbolic-tree]{Definition~\ref{def:symbolic-tree}} represents an analytical (scientific) expression, where primary features $x_1$ and $x_2$ are composed through the mathematical operators $\sin$, $^{2}$, and $+$ to give $\sin(x_1) + x_2^2$. The symbolic tree is therefore evaluated algebraically from terminal nodes to root, yielding a global smooth scientific response surface rather than a piecewise partition of the covariate space.

\subsection{Symbolic Forest Component}
\label{subsec:symbolic-forest-component}

Let $\mathcal{D}_n = \{(\bm x_i, y_i):i=1,\ldots, n\} \subset \mathfrak{X}\times \mathbb{R}$ denote the observed data, where $\mathfrak{X} \subset \mathbb{R}^{p}$. For any observed feature vector $\bm{x}\in\mathfrak X$, let 
$\mathrm{ev}[T](\bm{x})$ denote the real-valued evaluation of the symbolic expression represented by $T$. The corresponding symbolic dictionary is $\mathbb{G}_{\mathbb O, p} = \{\mathrm{ev}[T]: T\in \mathbb{T}_{\mathbb O, p}\}$.
\hierbosss\ models the symbolic relationship between the scalar response $y_i$ and observed primary feature vector $\bm x_i = (x_{i1},\ldots,x_{ip})^{\T}$ using a symbolic forest $\mathcal{T} = (T_1, \ldots, T_K) \in \mathbb{T}_{\mathbb O, p}^{K}$ of size $K\in \mathbb{N}$ as
\begin{align}
\label{eq:symbolic-forest-component}
y_i = \beta_0 + \sum_{j=1}^{K}\mathrm{ev}[T_j](\bm x_i)\beta_j + \epsilon_i,\quad i=1,\ldots, n,
\end{align}
where $\epsilon_i \sim \mathrm{N}(0, \sigma^2)$ is independent Gaussian noise, $\bm \beta = (\beta_0, \ldots, \beta_K)^{\T}\in \mathbb{R}^{K+1}$ is the outer model regression coefficient vector, and $\sigma^2 > 0$ is the model noise variance.
For fixed $K$, the symbolic forest component in \hyperref[eq:symbolic-forest-component]{\eqref{eq:symbolic-forest-component}} induces the following \hierbosss\ symbolic function class
\begin{equation}
\label{eq:BayeSymX-classes}
\mathcal{F}_K := \Big\{f_{\bm \beta, \mathcal T}:\mathfrak{X} \to \mathbb{R}\mid f_{\bm \beta, \mathcal T}(\bm x) = \beta_0 + \sum_{j=1}^{K}\mathrm{ev}[T_j](\bm x) \beta_j, \bm \beta\in \mathbb{R}^{K+1}, T_j\in \mathbb{T}_{\mathbb O, p}\Big\},
\end{equation}
consisting of affine (linear) combinations of $K$ symbolic expressions generated by $\mathbb O$ on $\mathfrak X$. 
The choice of $K$ represents a trade-off between exploration of the ensuing expression space and computational efficiency. In practice, we recommend choosing the symbolic forest size $K$ data-adaptively from a modest candidate set using held-out predictive accuracy and runtime diagnostics, as illustrated in \suppref{sec:ablation}. We complete the hierarchical Bayesian specification by placing priors on $(\bm \beta, \sigma^{2}, \mathcal T)$, thus inducing a prior over $\{f_{\bm \beta, \mathcal{T}} \in \mathcal F_K, \sigma^2 >0 \}$.

\subsection{Prior Specifications}
\label{subsec:prior-specification}

The outer regression parameters in $\bm \beta$ jointly with the model noise variance $\sigma^2$ are endowed with the conjugate Normal-Inverse-Gamma (NIG) prior distribution
\begin{align}
\label{eq:NIG-prior}
\mathrm{NIG}_{K+1}(\bm \beta, \sigma^2 \mid \bm \mu_{\beta}, \bm\Sigma_{\beta}, \nu, \lambda) = \mathrm{N}_{K+1}(\bm \beta \mid \bm \mu_{\beta}, \sigma^{2}\bm \Sigma_{\beta})\;\mathrm{IG}\Big(\sigma^{2}\mid \frac{\nu}{2}, \frac \lambda 2\Big),
\end{align}
where $\bm \mu_{\beta}\in \mathbb{R}^{K+1}$, $\bm \Sigma_{\beta} \succ 0$ (positive definite), and $\nu, \lambda >0$ are fixed hyperparameters of the $(K+1)$-variate Normal and Inverse-Gamma distributions with weakly informative default values $(\bm 0_{K+1}, 10\bm I_{K+1}, 0.05, 0.05)$.

\begin{figure}[H]
    \centering
    \includegraphics[width=0.95\linewidth]{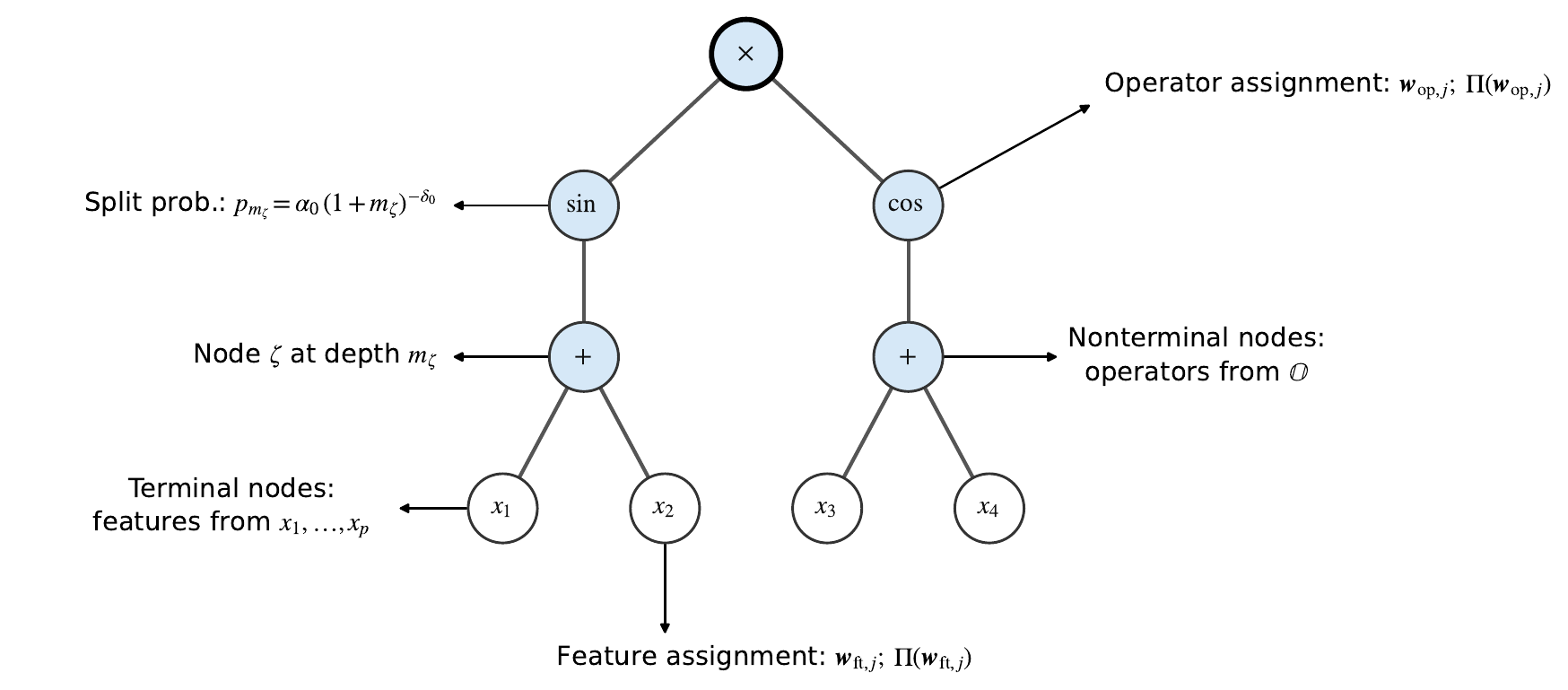}
    \caption{\centering{Symbolic tree prior over $T$, where $\mathrm{ev}[T](\bm x) = \sin(x_1+ x_2)\cos(x_3+x_4)$.}}
    \label{fig:symbolic-tree-prior}
\end{figure}

We next specify the prior on each symbolic tree $T_j$ independently for $j=1, \ldots, K$, as shown in \hyperref[fig:symbolic-tree-prior]{Figure \ref{fig:symbolic-tree-prior}}. Our symbolic tree prior mirrors the recursive formulation in line with \hyperref[def:symbolic-tree]{Definition~\ref{def:symbolic-tree}}. In \hyperref[alg:symbolic-tree-generation]{Algorithm~\ref{alg:symbolic-tree-generation}}, a node $\zeta \in \mathcal{N}(T_j)$ of depth $m_{\zeta}\in \{0, 1, \ldots\}$ splits with
\begin{align}
\label{eq:split-probability}
\mathsf{P}(B_{\zeta}=1) = p_{m_\zeta} = \alpha_0(1 + m_\zeta)^{-\delta_0}, \quad 0< \alpha_0 < 1, \quad \delta_0 > 0,
\end{align}
probability. Thus, nodes closer to the root are more likely to split, while splitting of deeper nodes is increasingly penalized. The hyperparameters $(\alpha_0,\delta_0)$ control the prior complexity of symbolic trees. Recommended defaults are $(\alpha_0,\delta_0)=(0.95,1.20)$ and $(\alpha_0,\delta_0)=(0.95,2.00)$ for experiments in \hyperref[sec:learning-Feynman-equations]{Sections \ref{sec:learning-Feynman-equations}} and \ref{sec:perovskites-data-study}, respectively, as they provided stable empirical performance across the applications.

Given that a node $\zeta$ is nonterminal, \hyperref[alg:symbolic-tree-generation]{Algorithm~\ref{alg:symbolic-tree-generation}} assigns an operator label drawn from the library $\mathbb O$ according to the tree-specific operator weight vector $\bm w_{\mathrm{op}, j} = (w_{\mathrm{op}, j, 1}, \ldots, w_{\mathrm{op}, j, |\mathbb O|}) \in \mathcal{S}_{|\mathbb O|}$, where $\mathcal S_m = \{(z_1, \ldots, z_m)^{\T}\in \mathbb{R}^{m}: \sum_{i=1}^{m}z_i = 1 \text{ and }z_i\geq 0\text{ for }i=1,\ldots,m\}$ represents the $(m-1)$-dimensional simplex. Similarly, \hyperref[alg:symbolic-tree-generation]{Algorithm~\ref{alg:symbolic-tree-generation}} assigns primary features $x_1, \ldots, x_p$ to terminal nodes according to the tree-specific feature weight vector $\bm w_{\mathrm{ft}, j} = (w_{\mathrm{ft}, j, 1}, \ldots, w_{\mathrm{ft}, j, p}) \in \mathcal{S}_{p}$. 
Let $\bm \xi_j = (\xi_{j, 1}, \ldots, \xi_{j, |\mathbb O|})$ and $\bm \varrho_{j} = (\varrho_{j, 1}, \ldots, \varrho_{j, p})$ denote the operator and feature count vectors in $T_j$, respectively. Also, let $\mathcal{N}_{\mathrm{op}}(T_j, m)$ and $\mathcal{N}_{\mathrm{ft}}(T_j, m)$ be the sets of nonterminal and terminal nodes of $T_j$ at depth $m$, respectively. Conditional on the weights, the prior over $T_j$ is
\begin{align}
\label{eq:symbolic-tree-prior-Tj}
\begin{split}
&\Pi(T_j\mid \bm w_{\mathrm{op}, j}, \bm w_{\mathrm{ft}, j}, \alpha_0, \delta_0)\\
&\qquad= \prod_{o=1}^{|\mathbb O|}(w_{\mathrm{op}, j, o})^{\xi_{j, o}}\prod_{h=1}^{p}(w_{\mathrm{ft}, j, h})^{\varrho_{j, h}}\prod_{m= 0}^{\infty} \left\{p_{m}^{|\mathcal N_{\mathrm{op}}(T_j, m)|}(1-p_m)^{|\mathcal{N}_{\mathrm{ft}}(T_j, m)|}\right\}.
\end{split}
\end{align}
%
We specify Dirichlet priors over the weight vectors for data-adaptive learning of operator and feature relevance
\begin{align}
\label{eq:weight-Dirichlet-priors}
\bm w_{\mathrm{op}, j} \sim \mathrm{Dir}(\bm \alpha_{\mathrm{op}}), \quad \bm w_{\mathrm{ft}, j} \sim \mathrm{Dir}(\bm \alpha_{\mathrm{ft}}),
\end{align}
independently. Here,  $\bm \alpha_{\mathrm{op}} = (\alpha_{\mathrm{op}, 1}, \ldots, \alpha_{\mathrm{op}, |\mathbb O|})$ and $\bm \alpha_{\mathrm{ft}} = (\alpha_{\mathrm{ft}, 1}, \ldots, \alpha_{\mathrm{ft}, p})$ are positive concentration hyperparameters, with the default choices $\alpha_{\mathrm{op}, o}=1$ and $\alpha_{\mathrm{ft}, h}=1$. These encode prior preferences over operators and features, unlike Dirichlet splitting priors for regression trees~\citep{Linero-Dirichlet-sparsity} which enforce sparsity in variable selection. 
With \hyperref[eq:symbolic-tree-prior-Tj]{\eqref{eq:symbolic-tree-prior-Tj}} and \hyperref[eq:weight-Dirichlet-priors]{\eqref{eq:weight-Dirichlet-priors}}, marginalizing out $\bm w_{\mathrm{op}, j}$ and $\bm w_{\mathrm{ft}, j}$, the symbolic tree prior over $T_j$ is
\begin{align}
\label{eq:symbolic-tree-prior-Tj-1}
\begin{split}
&\Pi_{\mathrm{tree}}(T_j\mid \alpha_0, \delta_0, \bm \alpha_{\mathrm{op}}, \bm \alpha_{\mathrm{ft}})
\\ & \qquad 
\propto \mathfrak{B}(\bm \alpha_{\mathrm{op}} + \bm \xi_j)\;\mathfrak{B}(\bm \alpha_{\mathrm{ft}} + \bm \varrho_j)\; \prod_{m= 0}^{\infty} \left\{p_{m}^{|\mathcal N_{\mathrm{op}}(T_j, m)|}(1-p_m)^{|\mathcal{N}_{\mathrm{ft}}(T_j, m)|}\right\},
\end{split}
\end{align}
where $\mathfrak{B}(\cdot)$ is the multivariate Beta function.
Consequently from \hyperref[eq:symbolic-tree-prior-Tj-1]{\eqref{eq:symbolic-tree-prior-Tj-1}}, the symbolic forest prior over $\mathcal T = (T_1, \ldots, T_K) \in \mathbb T_{\mathbb O, p}^{K}$ is
\begin{align}
\label{eq:symbolic-forest-prior}
\Pi_{\mathrm{forest}, K}(\mathcal T\mid  \alpha_0, \delta_0, \bm \alpha_{\mathrm{op}}, \bm \alpha_{\mathrm{ft}}) = \prod_{j=1}^{K}\Pi_{\mathrm{tree}}(T_j \mid  \alpha_0, \delta_0, \bm \alpha_{\mathrm{op}}, \bm \alpha_{\mathrm{ft}}).
\end{align}
Note that, the depth-dependent splitting rule in \hyperref[eq:split-probability]{\eqref{eq:split-probability}} imparts regularization over the symbolic tree depth $\mathrm{d}(T_j)$ and operator count $S(T_j)$, in the spirit of~\cite{BCART}. This phenomenon is formalized in \hyperref[prop:tree-forest-prior-tail-control]{Proposition~\ref{prop:tree-forest-prior-tail-control}} below, where the symbolic tree and forest priors in \hyperref[eq:symbolic-tree-prior-Tj-1]{\eqref{eq:symbolic-tree-prior-Tj-1}} and \hyperref[eq:symbolic-forest-prior]{\eqref{eq:symbolic-forest-prior}} assign exponentially small mass to overly complex symbolic structures.
\begin{proposition}[Tail control of symbolic tree and forest priors]
\label{prop:tree-forest-prior-tail-control}
For every $d, s\in \mathbb{N}$, the symbolic tree prior over $T\in \mathbb{T}_{\mathbb O, p}$ in \hyperref[eq:symbolic-tree-prior-Tj-1]{\eqref{eq:symbolic-tree-prior-Tj-1}} satisfies
\begin{equation*}
\begin{gathered}
\Pi_{\mathrm{tree}}(\mathrm{d}(T) \geq d\mid \alpha_0, \delta_0, \bm{\alpha}_{\mathrm{op}}, \bm{\alpha}_{\mathrm{ft}}) \leq C_1 \exp\{-c_1 d\log d\},\\
\Pi_{\mathrm{tree}}(S(T) > s\mid \alpha_0, \delta_0, \bm{\alpha}_{\mathrm{op}}, \bm{\alpha}_{\mathrm{ft}}) \leq C_2\exp\{-c_2 s\},
\end{gathered}
\end{equation*}
where $C_1, c_1, C_2, c_2$ are positive constants.
Consequently,  for every $s\in \mathbb{N}$, the symbolic forest prior over $\mathcal T=(T_1,\ldots,T_K) \in \mathbb{T}_{\mathbb O, p}^{K}$ in \hyperref[eq:symbolic-forest-prior]{\eqref{eq:symbolic-forest-prior}} satisfies
\begin{equation*}
\Pi_{\mathrm{forest}, K}\left(S(\mathcal T) > s\mid \alpha_0, \delta_0, \bm \alpha_{\mathrm{op}}, \bm\alpha_{\mathrm{ft}}\right) \leq C_2'\exp\{-c_2's\},
\end{equation*}
where $S(\mathcal T)=\sum_{j=1}^K S(T_j)$ is the symbolic forest operator count and $C_2', c_2'$ are positive constants.
\end{proposition}

\begin{proof}
See \suppref{sec:tail-control-proof}.
\end{proof}

In the context of \hierbosss, \hyperref[prop:tree-forest-prior-tail-control]{Proposition~\ref{prop:tree-forest-prior-tail-control}} implies that the prior regularization favors learning of interpretable, simple expressions in accordance with the Occam's razor principle~\citep{Occams-Razor-1}. This regularization operates in two complementary ways. Among algebraically equivalent symbolic tree structures, the prior in~\hyperref[eq:symbolic-tree-prior-Tj-1]{\eqref{eq:symbolic-tree-prior-Tj-1}} tends to favor parsimonious encodings of the same mathematical function. Across distinct symbolic functions with comparable predictive support, the prior allocates greater mass to those admitting less complex symbolic forms.

At the same time, we deliberately retain algebraically distinct symbolic tree structures as distinct symbolic models. The prior in~\hyperref[eq:symbolic-tree-prior-Tj-1]{\eqref{eq:symbolic-tree-prior-Tj-1}} may allocate different masses to trees $T$ and $T'$ even when $\mathrm{ev}[T] \equiv \mathrm{ev}[T']$. This is useful as posterior computation in \hyperref[subsec:posterior-inference]{Section \ref{subsec:posterior-inference}} is carried out over symbolic structures. Two functionally equivalent symbolic expressions can have substantially different structural neighborhoods. Excluding such equivalent representations~\citep{BMS} can therefore remove structurally useful intermediate states and restrict exploration of the discrete space of symbolic expressions. Conversely, explicitly quotienting by symbolic equivalence~\citep{eggsr} would require specifying and repeatedly validating a large collection of nonlinear algebraic rewrite rules, which is computationally expensive and generally incomplete. We therefore work on the full symbolic expression space, allowing efficient structural exploration while using posterior summaries as in \hyperref[subsec:symbolic-model-selection-refinement]{Section \ref{subsec:symbolic-model-selection-refinement}} to identify functionally equivalent or predictively similar symbolic explanations.

\subsection{Posterior Inference}
\label{subsec:posterior-inference}

Combining the likelihood of the \hierbosss\ model in \hyperref[eq:symbolic-forest-component]{\eqref{eq:symbolic-forest-component}} with the priors specified in \hyperref[eq:NIG-prior]{\eqref{eq:NIG-prior}} and \hyperref[eq:symbolic-forest-prior]{\eqref{eq:symbolic-forest-prior}} above, the \hierbosss-induced posterior distribution is
\begin{align}
\label{eq:BayeSymX-posterior}
\begin{split}
&\Pi(\mathcal T, \bm\beta, \sigma^2\mid \mathcal D_n, \alpha_0, \delta_0, \bm \alpha_{\mathrm{op}}, \bm\alpha_{\mathrm{ft}}, \bm\mu_{\beta}, \bm\Sigma_{\beta}, \nu, \lambda)\\
&\propto \mathrm{N}_n(\bm y\mid \mathcal E(\bm X; \mathcal T) \bm \beta, \sigma^2\bm I_n)\mathrm{NIG}_{K+1}(\bm \beta, \sigma^2 \mid \bm \mu_{\beta}, \bm\Sigma_{\beta}, \nu, \lambda) \Pi_{\mathrm{forest}, K}(\mathcal T\mid  \alpha_0, \delta_0, \bm \alpha_{\mathrm{op}}, \bm \alpha_{\mathrm{ft}}),
\end{split}
\end{align}
where $\bm{y} = (y_1, \ldots, y_n)^{\T} \in \mathbb R^{n}$ is the response vector, $\bm X = (\bm x_1, \ldots, \bm x_n)^{\T}\in \mathfrak{X}^{n}\subset \mathbb{R}^{n\times p}$ is the primary design matrix, and $\mathcal E (\bm X; \mathcal T) \in \mathbb{R}^{n\times (K+1)}$ is the symbolic design matrix with the $(i, j)$th entry being $\mathrm{ev}[T_{j-1}](\bm x_i)$ for $j >1$ and $1$ for $j=1$.

For an efficient Markov chain Monte Carlo (\texttt{MCMC}) posterior sampling, we first marginalize $(\bm\beta, \sigma^2)$ from \hyperref[eq:BayeSymX-posterior]{\eqref{eq:BayeSymX-posterior}} using the NIG conjugacy; see \suppref[sec:posterior-derivation]. This gives the joint marginal posterior ($\mathrm{JMP}$) distribution of $\mathcal{T}$ as
\begin{align}
\label{eq:JMP}
\begin{split}
\mathrm{JMP}(\mathcal T) &= \Pi(\mathcal T \mid \mathcal D_n, \alpha_0, \delta_0, \bm \alpha_{\mathrm{op}}, \bm\alpha_{\mathrm{ft}}, \bm\mu_{\beta}, \bm\Sigma_{\beta}, \nu, \lambda)\\
&\propto  \det(\bm \Sigma_{\beta}^{\star}(\mathcal T))^{\frac 12} (\lambda^\star(\mathcal T))^{-\frac{\nu^\star}{2}} \Pi_{\mathrm{forest}, K}(\mathcal T\mid  \alpha_0, \delta_0, \bm \alpha_{\mathrm{op}}, \bm \alpha_{\mathrm{ft}}),
\end{split}
\end{align}
where
\begin{align*}
\begin{gathered}
\nu^{\star} = \nu + n, \quad \lambda^{\star}(\mathcal T) = \lambda + \bm y^{\T}\bm{y} + \bm{\mu}_{\beta}^{\T}\bm\Sigma_{\beta}^{-1}\bm\mu_{\beta} - (\bm\mu_{\beta}^{\star}(\mathcal T))^{\T}(\bm\Sigma_\beta^\star(\mathcal T))^{-1}\bm\mu_\beta^\star(\mathcal T),\\
\bm\mu_\beta^\star(\mathcal T) = \bm\Sigma_\beta^\star(\mathcal T)(\bm\Sigma_\beta^{-1}\bm\mu_\beta + \mathcal{E}^{\T}(\bm X; \mathcal T) \bm y), \quad \bm\Sigma^{\star}_{\beta}(\mathcal T) = (\bm\Sigma_\beta^{-1} + \mathcal{E}^{\T}(\bm X; \mathcal T) \mathcal{E}(\bm X; \mathcal T))^{-1}.
\end{gathered}
\end{align*}

In \hyperref[alg:simple-BayeSymX-posterior-sampling]{Algorithm~\ref{alg:simple-BayeSymX-posterior-sampling}}, we sample from the posterior in \hyperref[eq:BayeSymX-posterior]{\eqref{eq:BayeSymX-posterior}} using a Metropolis-within-partially collapsed Gibbs algorithm~\citep{MH-1,vanDyk01062008}. At each iteration, the symbolic forest $\mathcal T$ is updated using the collapsed target $\mathrm{JMP}(\mathcal T)$ in \hyperref[eq:JMP]{\eqref{eq:JMP}}. Conditional on the updated forest, the outer regression coefficients in $\bm \beta$ and the model noise variance $\sigma^2$ are then sampled directly from their joint conjugate full conditional distribution. In practice, to facilitate exploration of the highly multimodal symbolic expression space, we run multiple instances of \hyperref[alg:simple-BayeSymX-posterior-sampling]{Algorithm~\ref{alg:simple-BayeSymX-posterior-sampling}} in parallel starting from randomly initialized symbolic forests.
\begin{algorithm}[!htp] \caption{Metropolis-within-partially collapsed Gibbs sampler for \hierbosss} 
\label{alg:simple-BayeSymX-posterior-sampling} 
\KwInput{ $\mathcal{D}_n$, $K$, $\mathbb O$, \texttt{max\_iter}, $\alpha_0$, $\delta_0$, $\bm \alpha_{\mathrm{op}}$, $\bm\alpha_{\mathrm{ft}}$, $\bm\mu_{\beta}$, $\bm\Sigma_{\beta}$, $\nu$, $\lambda$.}

\textbf{Initialize} $(\bm\beta^{(0)},(\sigma^2)^{(0)},\mathcal T^{(0)})$.\; 

\For{$\mathrm{iter}=1,\ldots,\;$\texttt{max\_iter}}
{ \textbf{Update} $\mathcal{T}^{(\mathrm{iter})}$ using the \texttt{MH} step (\hyperref[supple-alg:MH-step]{\textcolor{BrickRed}{Algorithm \ref*{supple-alg:MH-step}}}) targeting $\mathrm{JMP}(\mathcal T)$.\;

\textbf{Update}  $(\bm \beta ^{(\mathrm{iter})},(\sigma^2)^{(\mathrm{iter})}) \sim \mathrm{NIG}_{K+1}(\bm \beta,\sigma^2 \mid \bm\mu_{\beta}^{\star}(\mathcal T^{(\mathrm{iter})}), \bm\Sigma_{\beta}^{\star}(\mathcal T^{(\mathrm{iter})}), \nu^{\star}, \lambda^{\star}(\mathcal T^{(\mathrm{iter})}))$.} 

\KwOutput{$\{(\bm\beta ^{(\mathrm{iter})}, (\sigma^2)^{(\mathrm{iter})}, \mathcal{T}^{(\mathrm{iter})}) :\mathrm{iter}=1,\ldots,\texttt{max\_iter}\}$.} 
\end{algorithm}

\begin{figure}[!htp]
    \centering
    \includegraphics[width=\linewidth]{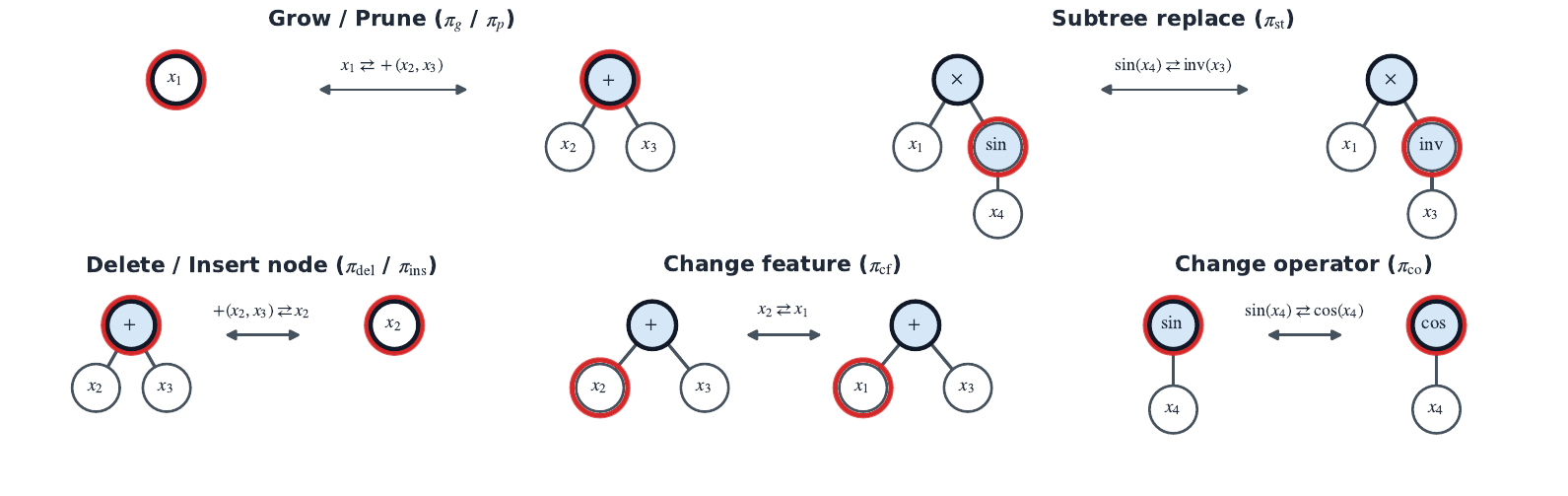}
    \caption{\centering{Proposal moves for symbolic tree update.}}
    \label{fig:proposal-moves}
\end{figure}

The Metropolis-Hastings (\texttt{MH}) update of $\mathcal T$ is performed tree-wise. For $j=1,\ldots, K$, the current symbolic tree $T_j$ is updated conditional on the remaining symbolic forest $\mathcal{T}_{-j} = (T_1, \ldots, T_{j-1}, T_{j+1}, \ldots, T_K) \in \mathbb{T}_{\mathbb O, p}^{K-1}$, using the target $\Pi(T_j \mid \mathcal{T}_{-j}, \mathcal D_n) \propto \mathrm{JMP}(\mathcal T)$. Candidate symbolic trees are proposed through one of the following seven local symbolic tree moves on the corresponding existing tree: growing a terminal node ($\mathrm{g}$), pruning a nonterminal node ($\mathrm{p}$), replacing a subtree ($\mathrm{st}$), deleting a nonterminal node ($\mathrm{del}$), inserting a nonterminal node between two existing nodes ($\mathrm{ins}$), changing the feature assigned to a terminal node ($\mathrm{cf}$), and changing the operator assigned to a nonterminal node while preserving arity ($\mathrm{co}$). These moves are selected according to $\bm\pi_{\mathrm{move}} = (\pi_{\mathrm{g}}, \pi_{\mathrm{p}}, \pi_{\mathrm{st}}, \pi_{\mathrm{del}}, \pi_{\mathrm{ins}}, \pi_{\mathrm{cf}}, \pi_{\mathrm{co}})\in \mathcal{S}_7$ and are illustrated in \hyperref[fig:proposal-moves]{Figure \ref{fig:proposal-moves}}. The \texttt{MH} step is detailed in \hyperref[supple-alg:MH-step]{\textcolor{BrickRed}{Algorithm~\ref*{supple-alg:MH-step}}} in \suppref[sec:posterior-sampler-details], with default configurations provided in \suppref[subsec:default-MH-config].

\subsection{Occam's Window Selection and Symbolic Model Refinement}
\label{subsec:symbolic-model-selection-refinement}

\hyperref[alg:simple-BayeSymX-posterior-sampling]{Algorithm~\ref{alg:simple-BayeSymX-posterior-sampling}} produces a collection of visited symbolic forests. Our goal is to identify a representative set of symbolic expressions that balances predictive power and scientific interpretability. This differs from standard Bayesian tree frameworks, where model selection is primarily driven by predictive strength only~\citep{BCART,BCART-Mallick,BART}. Furthermore, in \sr, the expression space is discrete and ultra-high-dimensional rendering model-visit frequency unreliable for model selection.
Hence, we construct a representative set by pooling the sampled symbolic forests across multiple parallel instances of \hyperref[alg:simple-BayeSymX-posterior-sampling]{Algorithm~\ref{alg:simple-BayeSymX-posterior-sampling}} and retaining the top $r\in\mathbb N$ forests with the largest $\mathrm{JMP}(\mathcal T)$ values in \hyperref[eq:JMP]{\eqref{eq:JMP}} as
\begin{equation}
\label{eq:Jr}
\text{Occam's window set}:\; \mathcal J_r = \{\mathcal{T}^{(1)}, \ldots, \mathcal{T}^{(r)}\}.
\end{equation}

Thus, $\mathcal{J}_r$ is a posterior summary based on the Occam's window principle~\citep{madigan1994model}, retaining high-scoring symbolic forests visited by the sampler while discarding low posterior probability alternatives. While a full characterization of posterior uncertainty through a formal Bayesian credible set is infeasible due to the NP-hard nature of \sr~\citep{virgolin2022symbolic}, a practical summary depending on the \texttt{MCMC} path and the chosen Occam's window cutoff $r$ provides principled insight into symbolic uncertainty by indicating whether the scientific data favors a single dominant expression or several plausible symbolic explanations.

The representative posterior set $\mathcal{J}_r$ in~\hyperref[eq:Jr]{\eqref{eq:Jr}} identifies the symbolic forests strongly supported by the \hierbosss-induced posterior in~\hyperref[eq:BayeSymX-posterior]{\eqref{eq:BayeSymX-posterior}}. However, the raw symbolic expressions (\rawtag) represented by the forests in~\hyperref[eq:Jr]{\eqref{eq:Jr}} may contain collinear or weakly contributing symbolic trees. We therefore apply a post-\texttt{MCMC} symbolic model refinement step to each forest in $\mathcal J_r$; see \hyperref[alg:post-mcmc-symbolic-model-refinement-single]{\textcolor{BrickRed}{Algorithm~\ref*{alg:post-mcmc-symbolic-model-refinement-single}}} of \suppref{sec:symbolic-model-refinement}. Specifically, the symbolic trees in a fixed high-$\mathrm{JMP}$ forest are treated as candidate components of an additive symbolic model, from which a smaller subset of informative tree-structured expressions is chosen using Bayesian information criterion (\texttt{BIC}).
This selected subset also provides an effective symbolic forest size ($K_{\mathrm{eff}}$), reflecting the number of components needed to capture the symbolic structure underlying the data. The outer regression coefficient vectors $\{\bm \beta^{(\ell)}: \ell=1, \ldots, r\}$ are re-estimated, and the resulting symbolic expressions are algebraically simplified using \texttt{SymPy}~\citep{meurer2017sympy}. This refinement yields final reported symbolic expressions (\finaltag) which retain the symbolic structure supported by the \hierbosss-induced posterior while improving scientific interpretability and parsimony.

\section{Posterior Concentration Guarantees for \texorpdfstring{\hierbosss}{HierBOSSS}}
\label{sec:posterior-contraction}

Let $\mathfrak{X} \subset \mathbb{R}^{p}$ be a compact subset and $L_2(\mathfrak{X})$ be the set of square-integrable functions on $\mathfrak{X}$. Then, for any $f\in L_2(\mathfrak X)$ and $\sigma^{2} > 0$, the conditional law and density of $\bm y\mid \bm X$ are
\begin{equation}
\label{eq:general-setup-1}
\mathbb{P}^{n}_{f, \sigma^2} = \otimes_{i=1}^{n}\mathrm{N}(f(\bm x_i), \sigma^2), \quad p_{f, \sigma^{2}}^{n} = \prod_{i=1}^{n}\mathrm{N}(y_i\mid f(\bm x_i), \sigma^2).
\end{equation}
Since $\mathfrak{X}$ is compact and $\mathbb O$ consists of continuous operators, for the \hierbosss\ model in \hyperref[subsec:symbolic-forest-component]{Section~\ref{subsec:symbolic-forest-component}}, $f \in \mathcal{F}_K$, as in \hyperref[eq:BayeSymX-classes]{\eqref{eq:BayeSymX-classes}}, with $\mathcal{F}_K \subset L_2(\mathfrak X)$. Here, we consider that $\mathcal{D}_n$ is generated following \hyperref[eq:general-setup-1]{\eqref{eq:general-setup-1}} with the truth $f= f_0 \in L_2(\mathfrak X)$  and $\sigma^{2}=\sigma_0^{2} > 0$. The corresponding true law and density of $\bm y \mid \bm X$ are then $\mathbb{P}_{0}^{n}$ and $p_0^{n }$, respectively. 

We measure posterior concentration in the empirical $L_2$ metric $d_n$ viz., $d_n^{2}(f, g) = n^{-1}\sum_{i=1}^{n}[f(\bm x_i) - g(\bm x_i)]^{2}$, where $f, g \in L_2(\mathfrak X)$. For a fixed symbolic forest operator count budget $S\in \mathbb{N}$, define the $S$-restricted \hierbosss\ function class as 
\begin{equation}
\label{eq:FKS}
\mathcal{F}_{K, S} := \Big\{f_{\bm \beta, \mathcal T} \in \mathcal{F}_K: S(\mathcal T) = \sum_{j=1}^{K}S(T_j) \leq S\Big\}.
\end{equation}
The corresponding empirical best symbolic approximation error in $\mathcal{F}_{K, S}$ is
\begin{equation}
\label{eq:aKSn}
a_{K, S, n}(f_0) = \inf_{f\in \mathcal F_{K, S}} d_n(f, f_0).
\end{equation}
Now, we impose the following general assumptions.
{
\renewcommand{\theassumption}{SE}
\begin{assumption}[Global symbolic evaluation envelope]
\label{ass:global-symbolic-evaluation-envelope}
For every $S\in \mathbb{N}$, there exist $U_{S}\geq 1$ and constants $\overline{C}_U > 1, c_U > 0$ such that
$$
\sup_{\substack{\bm x \in \mathfrak X,\;
T\in \mathbb{T}_{\mathbb O, p}:S(T)\leq S}} |\mathrm{ev}[T](\bm x)| \leq U_S \leq \overline{C}_U\exp\{c_U S\}.
$$
\end{assumption}
}
{
\renewcommand{\theassumption}{PR}
\begin{assumption}[Prior regularity]
\label{ass:prior-regularity}
The Dirichlet parameters $\bm \alpha_{\mathrm{op}}$ and $\bm \alpha_{\mathrm{ft}}$ satisfy, $0<a_{\mathrm{op}} < \alpha_{\mathrm{op}, o} < A_{\mathrm{op}}$ for all $o\in \{1, \ldots, |\mathbb O|\}$ and $0< a_{\mathrm{ft}}<\alpha_{\mathrm{ft}, h} < A_{\mathrm{ft}} < \infty$ for all $h\in \{1, \ldots, p\}$.
\end{assumption}
}

\hyperref[ass:global-symbolic-evaluation-envelope]{Assumption~\ref{ass:global-symbolic-evaluation-envelope}} controls the growth of evaluated symbolic expressions in $\mathbb{G}_{\mathbb O, p}$. This is needed because symbolic expressions may contain nonlinear operators such as $\{\times, ^2, \sqrt{\cdot}, \exp, \texttt{inv}\}$, whose evaluations can grow rapidly with symbolic tree complexity. \hyperref[ass:prior-regularity]{Assumption~\ref{ass:prior-regularity}} ensures that the symbolic tree prior in \hyperref[eq:symbolic-forest-prior]{\eqref{eq:symbolic-forest-prior}} assigns sufficient mass to all operators in $\mathbb O$ and features $x_1, \ldots, x_p$. Finally, the central quantity that governs the posterior concentration rates in \hyperref[theorem:posterior-contraction]{Theorems~\ref{theorem:posterior-contraction}} and~\ref{theorem:sieve-truncated-population-symbolic-oracle} is
\begin{align}
\label{eq:CKSn}
\begin{split}
\mathfrak{C}_{K, S, n} &= S[1 + \log(S + |\mathbb O|)] + [S + K][1 + \log(p+S + K)]\\
&\qquad + [K+1][\log \overline{C}_U + c_U S] + [K+2]\log n.
\end{split}
\end{align}
The quantity $\mathfrak{C}_{K, S, n}$ is the effective complexity scale of $\mathcal{F}_{K, S}$ in \hyperref[eq:FKS]{\eqref{eq:FKS}}. 
It encapsulates several fundamentally different sources of complexity on a common statistical scale namely, the combinatorially large number of admissible symbolic forest structures, the allocation of operators and features within those structures, the growth of nonlinear symbolic evaluations with the forest operator count budget, and the continuous resolution required for the outer regression coefficients and noise variance. At the same time, $\mathfrak{C}_{K, S, n}$ must be sufficiently sharp to control the metric entropy of symbolic sieves, guarantee adequate prior mass near suitable symbolic approximants, and ensure exponentially small prior mass outside the sieves. Thus, $\mathfrak{C}_{K, S, n}$ is the problem-specific complexity scale determining how symbolic expressivity is penalized statistically, and consequently, controls the posterior concentration rates in the sequel.
More specifically, the first two terms $S[1 + \log(S + |\mathbb O|)]$ and $[S + K][1 + \log(p+S + K)]$ in \hyperref[eq:CKSn]{\eqref{eq:CKSn}} correspond to the combinatorial complexity of symbolic forest structures, operator assignments to nonterminal nodes, and feature assignments to terminal nodes. The third term $[K+1][\log \overline{C}_U + c_U S]$ in \hyperref[eq:CKSn]{\eqref{eq:CKSn}} accounts for the growth in the magnitudes of the symbolic evaluations with increasing operator count budget, while the final term $[K+2]\log n$ in \hyperref[eq:CKSn]{\eqref{eq:CKSn}} is the statistical resolution cost of covering the outer regression coefficients in $\bm \beta$ and variance parameter $\sigma^2$. Before turning to the results, we clarify why the empirical metric $d_n$ is an appropriate metric for posterior concentration of \hierbosss. 

\begin{remark}[Empirical metric and symbolic non-identifiability]
\label{remark-d-n-identifiability}
Since symbolic expressions are non-identifiable, the likelihood of the \hierbosss\ model in~\hyperref[eq:symbolic-forest-component]{\eqref{eq:symbolic-forest-component}} depends on $f(\bm x) = \beta_0 + \sum_{j=1}^{K}\mathrm{ev}[T_j](\bm x)\beta_j$ only through its evaluation at the observed design points $\bm x_i$ for $i=1,\ldots, n$, and is agnostic to its symbolic form and out-of-sample behavior. The empirical metric $d_n$ exactly encodes this property, making it the natural metric for investigating likelihood-based inference in the present \sr\ setting. Consequently, ensuing guarantees are function-level statements and do not imply recovery of a unique expression or a collection of minimal-complexity symbolic representations. Indeed, it is easy to see that asymptotically non-negligible posterior mass persists around symbolically distinct approximations of $f_0$ with positive prior support.
\end{remark}

\subsection{Posterior Concentration Under Approximate Symbolic Realizability}
\label{subsec:well-specified-contraction}

We first consider the case where the truth $f_0$ can be approximated arbitrarily well by more and more complex symbolic forests in $\mathcal{F}_K$, as explicated in \hyperref[ass:well-specified]{Assumption~\ref{ass:well-specified}}.
{
\renewcommand{\theassumption}{ASR}
\begin{assumption}[Approximate Symbolic Realizability]
\label{ass:well-specified}
For every $S\in \mathbb{N}$, there exists $f_{(S)} \in \mathcal{F}_{K, S}$ such that, $\lim_{S\to \infty}\;d_n(f_{(S)}, f_0) = 0$ that is $ \lim_{S\to \infty}\;a_{K, S, n}(f_0) = 0$.
\end{assumption}
}

We demonstrate in \hyperref[theorem:posterior-contraction]{Theorem~\ref{theorem:posterior-contraction}} that the \hierbosss-induced posterior in \hyperref[eq:BayeSymX-posterior]{\eqref{eq:BayeSymX-posterior}} concentrates in shrinking neighborhoods of the truth $f_0$; see \hyperref[fig:concentration-diagram]{Figure~\ref{fig:concentration-diagram}}.

\begin{theorem}[Concentration under approximate symbolic realizability]
\label{theorem:posterior-contraction}
For fixed $K$, $p$, and $\mathbb O$, grant \hyperref[ass:global-symbolic-evaluation-envelope]{Assumptions~\ref{ass:global-symbolic-evaluation-envelope}}, \ref{ass:prior-regularity}, and~\ref{ass:well-specified}. Then there exists a constant $M>0$ such that
\begin{equation*}
\lim_{n\to \infty}\;\mathsf{E}_{\mathbb{P}_0^{n}}[\Pi(f_{\bm \beta, \mathcal T}\in \mathcal F_{K}: d_{n}(f_{\bm \beta, \mathcal T}, f_0) > M \mathsf r_{n, K} \mid \mathcal{D}_n )] = 0,
\end{equation*}
where the rate $\mathsf{r}_{n, K}$ is defined by
\begin{equation*}
n \mathsf r^{2}_{n, K} := \inf_{S\in \mathbb N}[\{\mathfrak{C}_{K, S, n} + na^{2}_{K, S, n}(f_0)\}
\{1 + \log(\mathfrak{C}_{K, S, n} + na^{2}_{K, S, n}(f_0) + K + p + |\mathbb O|)\}].
\end{equation*}
\end{theorem}

\begin{proof}
See \suppref{sec:posterior-contraction-supplement}.
\end{proof}

\hyperref[theorem:posterior-contraction]{Theorem~\ref{theorem:posterior-contraction}} solidifies that, for each $S \in \mathbb N$, the concentration rate $\mathsf{r}_{n, K}$ balances the empirical symbolic approximation error $a_{K, S, n}(f_0)$ in \hyperref[eq:aKSn]{\eqref{eq:aKSn}}  against the discrete and combinatorial metric complexity scale $\mathfrak{C}_{K, S, n}$ of the corresponding symbolic function class $\mathcal F_{K, S}$. Although the proof is organized within the general Bayesian nonparametric concentration framework of~\cite{ghosal-vdv}, its principal technical challenge lies in deriving and controlling the problem-specific complexity scale $\mathfrak{C}_{K,S,n}$, as discussed above.
We conclude by recording the important special case where the truth $f_0$ admits an exact finite symbolic representation.

\begin{corollary}[Concentration under exact symbolic realizability]
\label{corollary:exact-symbolic-realizability}
In addition to \hyperref[ass:global-symbolic-evaluation-envelope]{Assumptions~\ref{ass:global-symbolic-evaluation-envelope}}, \ref{ass:prior-regularity}, and \ref{ass:well-specified}, suppose that $f_0$ admits a finite complexity symbolic representation that is there exists $S_0\in \mathbb{N}$ such that, $f_0 \in \mathcal{F}_{K, S_0}$. Then, for every $S\geq S_0$, $a_{K, S, n}(f_0) = 0$. Consequently, for fixed $K$, $p$, and $\mathbb O$, the conclusion of \hyperref[theorem:posterior-contraction]{Theorem~\ref{theorem:posterior-contraction}} holds with
$$
n\mathsf r_{n, K}^{2} \leq \mathfrak{C}_{K, S_0, n}[1 + \log(\mathfrak{C}_{K, S_0, n} + K + p +|\mathbb O|)].
$$
\end{corollary}

\begin{proof}
See \suppref{sec:posterior-contraction-supplement}.
\end{proof}

Observe that from \hyperref[eq:CKSn]{\eqref{eq:CKSn}}, $\mathfrak{C}_{K, S_0, n} = \mathcal{O}(\log n)$, where $\mathcal O$ is the big-O notation. Hence, \hyperref[corollary:exact-symbolic-realizability]{Corollary \ref{corollary:exact-symbolic-realizability}} leads to a near-parametric posterior concentration rate that is $\mathsf r_{n, K} = \mathcal O(n^{-1/2}(\log n \log \log n)^{1/2})$.

\subsection{Posterior Concentration Under Symbolic Misspecification}
\label{subsec:symbolic-misspecification}

We next consider the case of symbolic misspecification, where the truth $f_0$ may not be exactly, or even approximately, representable by the fixed-$K$ \hierbosss\ symbolic function class $\mathcal{F}_K$. 
Here, we operate with a random design formulation, where $\bm x_1, \ldots, \bm x_n$ are independently sampled from a density $q_X$ supported on $\mathfrak{X}$. Thus, the joint law and density of $\mathcal{D}_n$ are $\mathbb{Q}^{n}_{f, \sigma^2}$ and $q^{n}_{f, \sigma^2}$, respectively. 

Define the population-level best symbolic approximation error in $\mathcal F_K$ as
\begin{equation}
\label{eq:aK}
a^{2}_K(f_0) := \inf_{f\in \mathcal F_K}\int_{\mathfrak X}|f(\bm x) - f_0(\bm x)|^{2}q_X(\bm x)d\bm x.
\end{equation}
\begin{figure}[!t]
    \centering
    \includegraphics[width=0.75\linewidth]{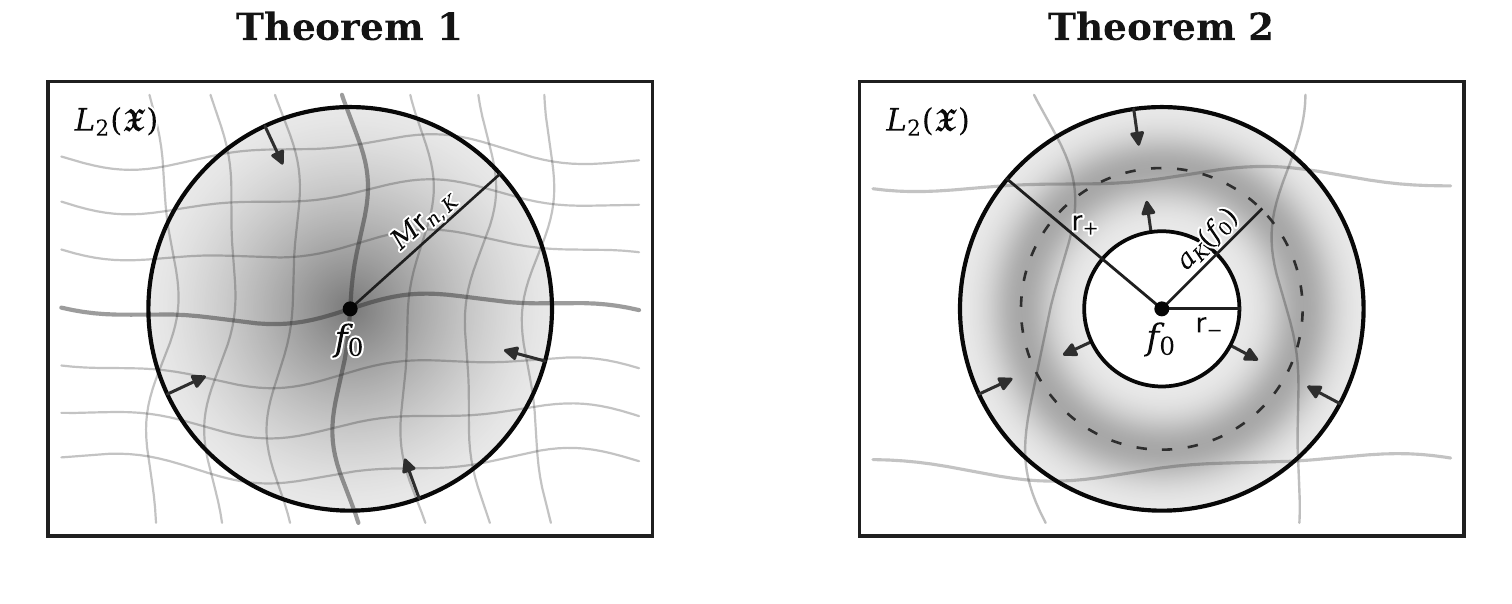}
    \caption{\hierbosss\ posterior concentration regions. Fibres denote subsets of $\mathcal{F}_K$ generated by fixed $\mathcal{T}\in \mathbb{T}_{\mathbb{O}, p}^{K}$ whose union is $\mathcal{F}_K \subset L_2(\mathfrak X)$. Fibre intersections denote multiple symbolic representations. In \hyperref[theorem:posterior-contraction]{Theorem~\ref{theorem:posterior-contraction}}, the posterior concentrates on symbolic functions in a shrinking ball with center $f_0\in \mathcal F_K$, while the concentration is on a shrinking annulus around $f_0\notin \mathcal{F}_K$ with radii $\mathsf{r}_{-} = [a_K^{2}(f_0) - M_{-}\mathsf{q}_{n, K}]^{1/2}$ and $r_{+} = [a_K^{2}(f_0) + M_{+}\mathsf{q}_{n, K}]^{1/2}$ for \hyperref[theorem:sieve-truncated-population-symbolic-oracle]{Theorem~\ref{theorem:sieve-truncated-population-symbolic-oracle}}.}
    \label{fig:concentration-diagram}
\end{figure}

In contrast to \hyperref[theorem:posterior-contraction]{Theorem~\ref{theorem:posterior-contraction}} (as visualized in \hyperref[fig:concentration-diagram]{Figure~\ref{fig:concentration-diagram}}), now the \hierbosss-induced posterior in \hyperref[eq:BayeSymX-posterior]{\eqref{eq:BayeSymX-posterior}} necessarily concentrates on regions of $\mathcal{F}_K$ whose squared empirical distance from $f_0$ is within a vanishing tolerance of the optimal error $a_K^{2}(f_0)$ in \hyperref[eq:aK]{\eqref{eq:aK}}. We work under the condition that $a_K^{2}(f_0)$ can be approached arbitrarily closely by symbolic representations in $\mathcal F_K$ whose outer regression coefficients remain uniformly controlled, as formalized in \hyperref[ass:bounded-coefficient-symbolic-approximability]{Assumption~\ref{ass:bounded-coefficient-symbolic-approximability}}.
{
\renewcommand{\theassumption}{BC}
\begin{assumption}[Bounded coefficient]
\label{ass:bounded-coefficient-symbolic-approximability}
There exists a constant $R_0 > 0$ such that $|f_0(\bm x)| \leq R_0$ uniformly for $\bm x \in \mathfrak{X}$.
Further, there exists a sequence $\{f_m=f_{\bm\beta_{m},\mathcal T_m}: m \in \mathbb N\} \subset \mathcal F_K$ such that $\lim_{m\to \infty}\;\mathsf{E}_{q_X}[|f_m(\bm x) - f_0(\bm x)|^2] = a_K^{2}(f_0)$ and $\lVert\bm\beta_{m}\rVert_\infty \leq C_\beta$ for all $m \in \mathbb N$ and some constant $C_{\beta} > 0$.
\end{assumption}
}

Consider constants $C_B$, $\underline v$, $\overline v$, and $C_D$, satisfying $C_{B} > C_{\beta}+1$, $0<\underline{v}<\sigma_0^{2}$, $\overline{v} > \sigma_0^{2} + a_{K}^{2}(f_0) + 1$, and $C_D > 2[R_0 + (a^{2}_K(f_0) + 1)^{1/2}]$, respectively. Let $\{S_n: n\in \mathbb{N}\}$ with $\lim_{n\to \infty}\;S_n = \infty$ be a deterministic sequence of sieve budgets such that $\rho_{n,K}$ defined as
\begin{align}
\label{eq:rho-nK}
\begin{gathered}
n\rho_{n,K}^2 := B_{n, K}^{4}\mathfrak C_{K,S_n,n}[1+ \log(\mathfrak C_{K,S_n,n}+K+p+|\mathbb O|)],
\\
B_{n, K}= 1 + R_0+ C_B(1 + K \overline{C}_U\exp\{c_U S_n\}),
\end{gathered}
\end{align}
satisfies $\lim_{n\to \infty}\;\rho_{n,K}=0$ and $\lim_{n\to \infty}\;n\rho_{n,K}^2 = \infty$.
Now define the functional and parametric sieves as
\begin{align}
\label{eq:sieve-misspecified}
\begin{gathered}
\mathcal F_n^D := \Big\{
f_{\bm\beta,\mathcal T}\in\mathcal F_K: S(\mathcal T)\leq S_n,  \lVert\bm\beta\rVert_\infty\leq C_B, d_n(f_{\bm\beta,\mathcal T},0) \leq C_D \Big\},
\\
\Theta_n^D := \Big\{(f, \sigma^2):f\in\mathcal F_n^D, \underline v \leq \sigma^2 \leq \overline v \Big\}.
\end{gathered}
\end{align}
Finally, let $\Pi_n^D(\cdot)$ be the observed design dependent truncation of the prior $\Pi(\cdot)$ in \hyperref[subsec:prior-specification]{Section~\ref{subsec:prior-specification}} to the sieve $\Theta_n^{D}$ in \hyperref[eq:sieve-misspecified]{\eqref{eq:sieve-misspecified}}.
We illustrate in \hyperref[theorem:sieve-truncated-population-symbolic-oracle]{Theorem~\ref{theorem:sieve-truncated-population-symbolic-oracle}} that the sieve truncated \hierbosss-induced posterior $\Pi_n^{D}(\cdot\mid \mathcal D_n)$ concentrates on a thin annulus centered at $f_0\notin \mathcal{F}_K$, with the inner and outer radii both shrinking toward $a_{K}(f_0)$; a pictorial representation is in \hyperref[fig:concentration-diagram]{Figure~\ref{fig:concentration-diagram}}.

\begin{theorem}[Concentration under symbolic misspecification]
\label{theorem:sieve-truncated-population-symbolic-oracle}
For fixed $K$, $p$, and $\mathbb{O}$, grant \hyperref[ass:global-symbolic-evaluation-envelope]{Assumptions~\ref{ass:global-symbolic-evaluation-envelope}},~\ref{ass:prior-regularity}, and~\ref{ass:bounded-coefficient-symbolic-approximability}. Then there exists a sequence $\xi_n \downarrow 0$ as $n\to \infty$ such that for $\mathsf q_{n, K} = \max\{\rho_{n,K}, \xi_n\}$ and sufficiently large constants $M_{+}, M_{-} > 0$
\begin{align*}
\lim_{n\to \infty}\;\mathsf E_{\mathbb Q_0^{n}}
[
\Pi_n^D(
f_{\bm\beta,\mathcal T}\in\mathcal F_n^D:
-M_{-}\mathsf q_{n,K} < d_n^2(f_{\bm\beta,\mathcal T},f_0) - a^{2}_{K}(f_0)
< M_{+}\mathsf q_{n,K}
\mid
\mathcal D_n)]
= 1,
\end{align*}
where $\mathbb{Q}_0^{n}$ is the joint law of $\mathcal{D}_n$ corresponding to the truth $f=f_0$ and $\sigma^2 = \sigma^{2}_0>0$.
\end{theorem}

\begin{proof}
See \suppref{sec:theorem-sieve-truncated-population-symbolic-oracle}.
\end{proof}

Some important upshots of \hyperref[theorem:sieve-truncated-population-symbolic-oracle]{Theorem~\ref{theorem:sieve-truncated-population-symbolic-oracle}} are in order. 
The result can be interpreted as a sharp oracle concentration statement~\citep{sharp-oracle,Fractional-Annals} that is $\Pi_{n}^{D}(\cdot \mid \mathcal D_n)$ concentrates around the population symbolic oracle error $a^{2}_{K}(f_0)$ in~\hyperref[eq:aK]{\eqref{eq:aK}} with leading constant $1$ and excess error $\mathcal{O}_{\mathbb{Q}_0^n}(\mathsf{q}_{n, K})$, as
$$
d_n^{2}(f_{\bm \beta, \mathcal T}, f_0) \leq 1\cdot a^{2}_{K}(f_0) + \mathcal{O}_{\mathbb{Q}_0^n}(\mathsf{q}_{n, K}).
$$
Here, $\mathsf{q}_{n, K} = \max\{\rho_{n, K}, \xi_n\}$ is an interplay between the symbolic complexity scale $\mathfrak{C}_{K, S_n, n}$, which determines $\rho_{n, K}$ in \hyperref[eq:rho-nK]{\eqref{eq:rho-nK}} and $\xi_n$, where $\xi_n$ captures the combined effect of how the population error of the bounded symbolic approximating sequence in \hyperref[ass:bounded-coefficient-symbolic-approximability]{Assumption~\ref{ass:bounded-coefficient-symbolic-approximability}} approaches $a_{K}^{2}(f_0)$, together with the  sampling fluctuation incurred by the random design setting. 
A further important feature is that \hyperref[theorem:sieve-truncated-population-symbolic-oracle]{Theorem~\ref{theorem:sieve-truncated-population-symbolic-oracle}} requires neither the existence of a finite set of Kullback-Leibler ($\mathrm{KL}$) minimizers nor the verification of the specialized testing entropy conditions used in classical Bayesian misspecification theory~\citep[Theorem 2.4]{kleijn2006misspecification}. The latter need shells surrounding the $\mathrm{KL}$ minimizers to be covered by convex sets over which a likelihood-ratio separation condition holds uniformly, thereby ensuring the existence of exponentially powerful tests. Such conditions are particularly cumbersome to verify for \sr\ since the class $\mathcal{F}_K$ is discrete and nonconvex. Moreover, $\mathcal{F}_K$ may admit no symbolic minimizer or infinitely many equivalent ones, as a sequence $\{f_m: m\in \mathbb{N}\}$ of symbolic functions may satisfy $\lim_{m\to \infty}\;\mathsf{E}_{q_X}[|f_m(\bm x) - f_0(\bm x)|^2] = a_K^{2}(f_0)$ without converging to any limiting symbolic representation. \hyperref[theorem:sieve-truncated-population-symbolic-oracle]{Theorem~\ref{theorem:sieve-truncated-population-symbolic-oracle}} instead establishes concentration around the oracle population symbolic approximation error under conditions formulated in terms of symbolic approximation and complexity, even when no pseudo-true symbolic model exists.

\section{Symbolic Recovery of Feynman Equations}
\label{sec:learning-Feynman-equations}

We investigate the ability of \hierbosss\ to recover interpretable scientific laws using the Feynman Symbolic Regression Database (\texttt{FSReD})~\citep{PMLB,AI-Feynman}, a benchmark containing over $100$ equations from the Feynman Lectures on Physics~\citep{feynman-1}, with $10^{5}$ observations available for each equation. We select the following $5$ representative physical laws having different levels of structural complexity namely the Coulomb's law~\hyperref[eq:feynman-cl]{\eqref{eq:feynman-cl}}, the Lorentz force on a moving charge in an electromagnetic field~\hyperref[eq:feynman-fce]{\eqref{eq:feynman-fce}}, the energy of a forced harmonic oscillator~\hyperref[eq:feynman-ehfo]{\eqref{eq:feynman-ehfo}}, harmonic oscillation with quadratic nonlinear correction~\hyperref[eq:feynman-hoqn]{\eqref{eq:feynman-hoqn}}, and the frictional force on a moving magnetic moment~\hyperref[eq:feynman-fmmm]{\eqref{eq:feynman-fmmm}}.

\begin{align}
&\mathrm{I\_12\_2}:\quad F
= \frac{q_1 q_2}{4\pi \epsilon r^{2}}, & p = 4
\tag{{CL}}
\label{eq:feynman-cl}
\\
&\mathrm{I\_12\_11}:\quad F
= q\left(E_f+vB\sin\theta\right), & p = 5
\tag{{FCE}} 
\label{eq:feynman-fce}
\\
&\mathrm{I\_24\_6}:\quad E_n = \frac{1}{4}m(\omega^2 + \omega_0^2)x^{2}, & p = 4
\tag{{EHFO}}
\label{eq:feynman-ehfo}
\\
&\mathrm{I\_50\_26}:\quad x=x_1[\cos(\omega t) + \alpha \cos^{2}(\omega t)], & p = 4
\tag{{HOQN}}
\label{eq:feynman-hoqn}
\\
&\mathrm{II\_36\_38}:\quad f = \frac{\mathrm{mom} H}{k_b T} + \frac{\mathrm{mom} \alpha}{\epsilon c^2 k_b T}M. & p = 8
\tag{{FMMM}}
\label{eq:feynman-fmmm}
\end{align}
A brief description on the scientific relevance of these equations is in \suppref{sec:experiment-protocol-evaluation-criteria}.

For each Feynman equation, we compare \hierbosss\ with state-of-the-art \sr\ methods from \texttt{SRBench}~\citep{LaCava-NIPS,Imai-SRBench}, including \gplearn~\citep{stephens2016gplearn}, \operon~\citep{operon}, \pysr~\citep{pysr}, \qlattice~\citep{feyn-qlattice}, \dsr~\citep{Deep-SR}, \sisso$++$~\citep{SISSO++}, \bms~\citep{BMS}, and \bsr~\citep{BSR}. We conduct $5$ independent repetitions for each equation. In every repetition, we randomly subsample $n=2000$ observations and partition them into a $90\%$ training set and a $10\%$ held-out test set. Both the original noiseless response and noisy regimes with two different levels ($\sigma$) of additive Gaussian noise are considered. For fair comparison, all methods were supplied with the common operator set $\mathbb{O} = \{+, \times, \mathrm{neg}, \mathrm{inv}, \sin, \cos, \exp, ^2, ^3\}$, held fixed across the experiments; except for \sisso$++$, which
is supplied a per-equation subset of $\mathbb{O}$, since its recursive construction of expressions grows combinatorially with the size of the operator
library. Full details of the experimental protocol, method-specific configurations, and evaluation criteria are given in~\suppref[sec:experiment-protocol-evaluation-criteria] and \suppref[sec:experimental-settings-competitors-Feynman].

The nominal symbolic forest size in \hierbosss\ is chosen to be $K=3$ for \hyperref[eq:feynman-fce]{\eqref{eq:feynman-fce}} and $K=4$ for the remaining Feynman equations following the ablation-based practical guideline in \suppref[sec:ablation], which balances predictive performance against computational cost. For comparability, the same $K$ values are also used in \bsr. Consequently, \hierbosss\ adapts the number of symbolic components in the \finaltag\ expression through the post-\texttt{MCMC} refinement step in \hyperref[subsec:symbolic-model-selection-refinement]{Section~\ref{subsec:symbolic-model-selection-refinement}}, which removes redundant and collinear tree structures to determine the effective forest size $K_{\mathrm{eff}} \leq K$ data-adaptively; see~\suppref[sec:effective-K-Feynman].

\begin{figure}[!t]
\centering

\begin{subfigure}[t]{0.325\textwidth}
    \centering
    \includegraphics[width=\linewidth]
    {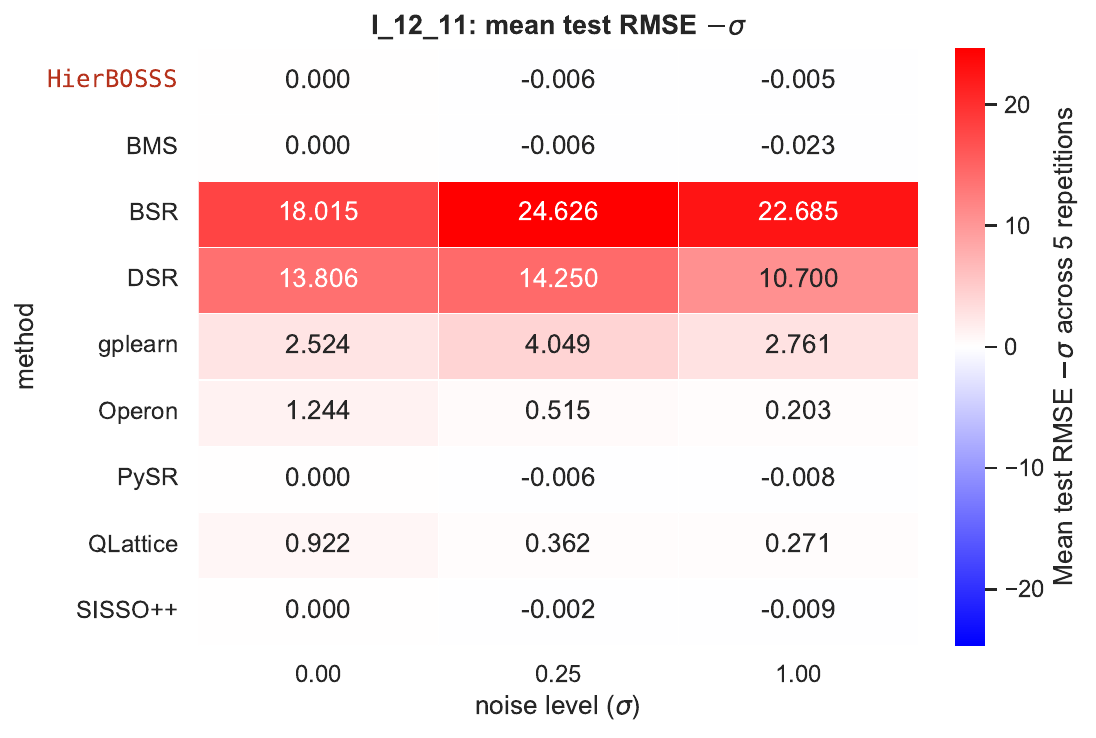}
    \caption{$\mathrm{I\_12\_11}$~\eqref{eq:feynman-fce}}
    \label{fig:feynman-rmse-fce}
\end{subfigure}
\hfill
\begin{subfigure}[t]{0.325\textwidth}
    \centering
    \includegraphics[width=\linewidth]
    {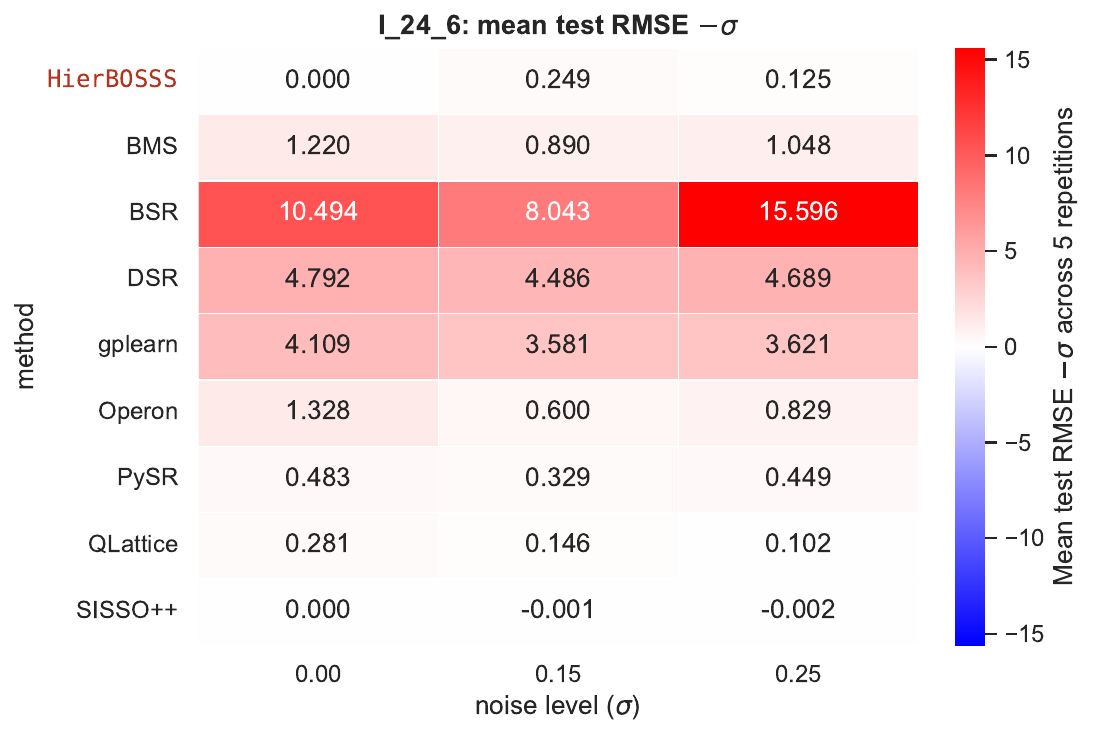}
    \caption{$\mathrm{I\_24\_6}$~\eqref{eq:feynman-ehfo}}
    \label{fig:feynman-rmse-ehfo}
\end{subfigure}
\hfill
\begin{subfigure}[t]{0.325\textwidth}
    \centering
    \includegraphics[width=\linewidth]
    {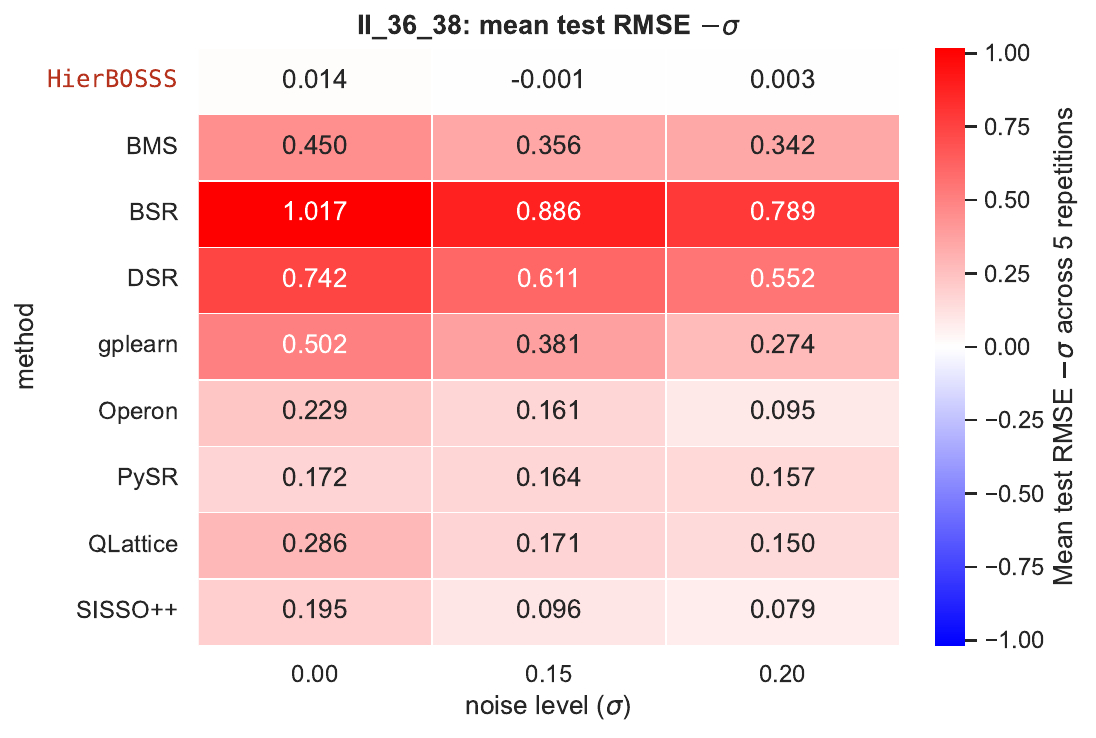}
    \caption{$\mathrm{II\_36\_38}$~\eqref{eq:feynman-fmmm}}
    \label{fig:feynman-rmse-fmmm}
\end{subfigure}

\caption{\centering{
Average (across repetitions) test $\texttt{RMSE}-\sigma$ for different noise levels.
}
}
\label{fig:feynman-rmse}
\end{figure}

We first assess predictive accuracy using the average test root mean squared error (\texttt{RMSE}) over the $5$ repetitions. \hyperref[fig:feynman-rmse]{Figure~\ref{fig:feynman-rmse}} presents representative results for the Feynman equations~\hyperref[eq:feynman-fce]{\eqref{eq:feynman-fce}},~\hyperref[eq:feynman-ehfo]{\eqref{eq:feynman-ehfo}}, and~\hyperref[eq:feynman-fmmm]{\eqref{eq:feynman-fmmm}}, which respectively exemplify low-, medium-, and high-complexity physical laws. With the exception of \bsr, \dsr, and \gplearn, most \sr\ methods achieve competitive predictive performance across noise levels, with only moderate deterioration as the noise level increases. Across all three structural complexity regimes, \hierbosss\ consistently outperforms the competing approaches. This advantage is particularly pronounced for the high-complexity equation~\hyperref[eq:feynman-fmmm]{\eqref{eq:feynman-fmmm}}, where \hierbosss\ improves upon the competing methods by orders of magnitude in test \texttt{RMSE}. Results for the remaining Feynman equations, together with the corresponding test coefficient of determination ($R^2$), show the same overall pattern and are reported in~\suppref[subsec:accuracy-runtime-Feynman].

\begin{figure}[H]
    \centering
    \includegraphics[width=0.80\linewidth]{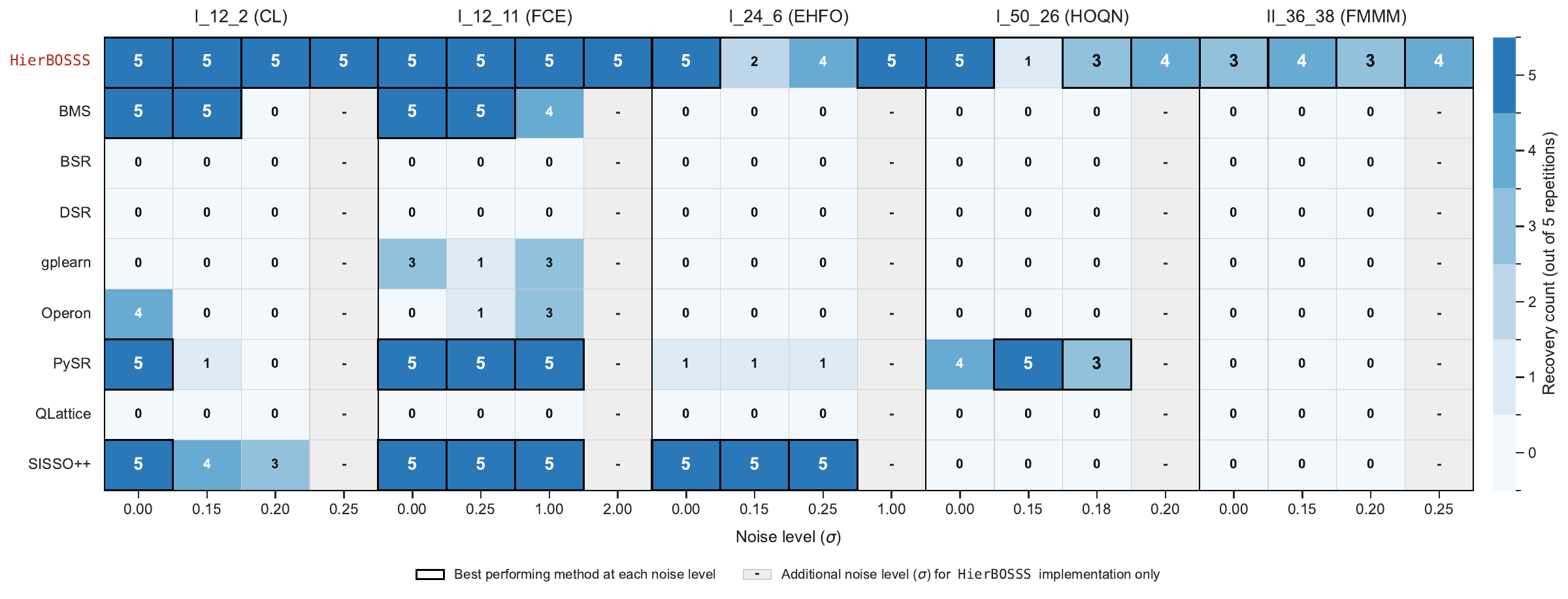}
    \caption{\centering{Structural recovery performance (out of $5$ repetitions) for different noise levels.}}
    \label{fig:Feynman-frequency-main}
\end{figure}
We next investigate whether strong predictive performance translates into recovery of the underlying Feynman equations. As shown in \hyperref[fig:Feynman-frequency-main]{Figure~\ref{fig:Feynman-frequency-main}}, this is generally not the case as several \sr\ methods (for example, \bms, \operon, \pysr, and \qlattice) achieve competitive accuracy, yet fail to consistently recover the correct symbolic structure, particularly with increasing noise and equation complexity levels. In contrast, \hierbosss\ exhibits robust structural recovery across all Feynman equations in~\hyperref[eq:feynman-cl]{\eqref{eq:feynman-cl}}-\hyperref[eq:feynman-fmmm]{\eqref{eq:feynman-fmmm}}, maintaining high recovery frequencies over $5$ repetitions and demonstrating stability to increasing noise. Furthermore, for each equation, we evaluate \hierbosss\ under an additional higher noise level, where it continues to reliably learn the underlying symbolic structure. The complete symbolic expressions learned by each \sr\ method across representative repetitions and all noise levels are recorded in~\suppref[sec:symbolic-expressions-Feynman] and \suppref{sec:symbolic-expressions-Feynman-BayeSymX-additional-noise-level}. Additionally, the Occam’s window set $\mathcal{J}_{r}$ with $r=10$, reported in~\suppref[sec:Occams-window-set-Feynman-BayeSymX], provides a posterior summary of uncertainty across the competing symbolic explanations learned by \hierbosss. Across representative repetitions and noise levels, these $\mathrm{JMP}$-ranked sets consistently place the ground-truth Feynman structure among the top-ranked candidate expressions, further demonstrating the stability of \hierbosss\ beyond selection of a single best symbolic model.

\begin{figure}[t]
    \centering
    \includegraphics[width=\linewidth]{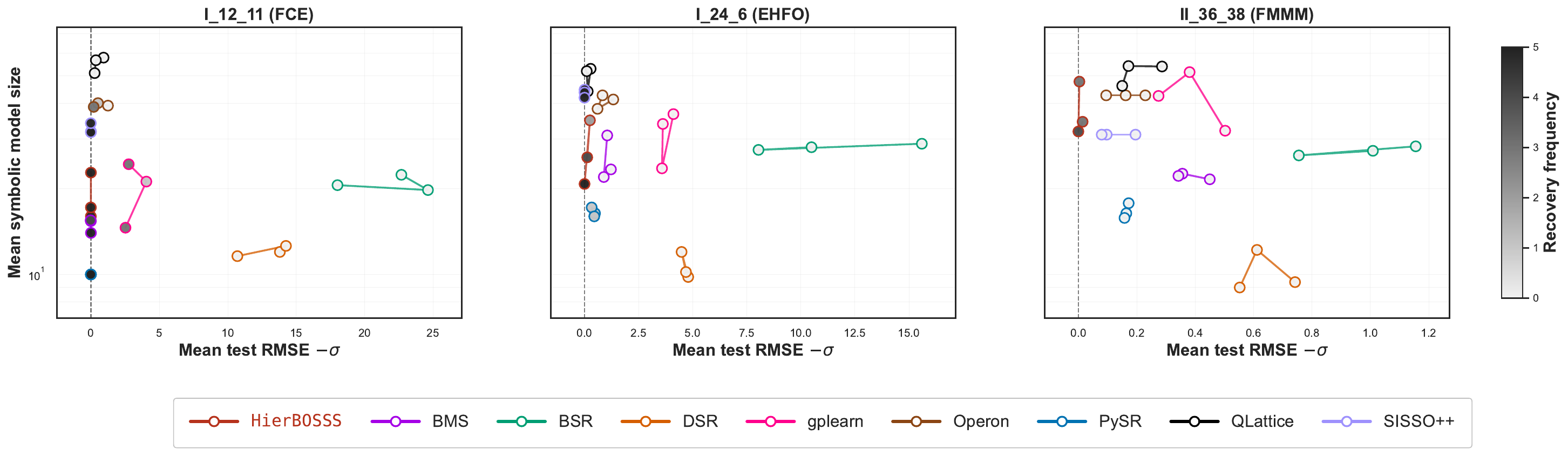}
    \caption{\centering{Accuracy-complexity-recovery trade-off across repetitions and noise levels.}}
    \label{fig:accuracy-complexity-recovery}
\end{figure}

Now, we study the symbolic complexity of the expressions learned by each method. Complexity is measured through the symbolic model size, defined as the total number of nodes in the learned expression tree, including mathematical operators, primitive features, and constants~\citep{LaCava-NIPS,Imai-SRBench}. The three-way comparison in \hyperref[fig:accuracy-complexity-recovery]{Figure~\ref{fig:accuracy-complexity-recovery}} summarizes the resulting accuracy–complexity–recovery trade-off for~\hyperref[eq:feynman-fce]{\eqref{eq:feynman-fce}}, \hyperref[eq:feynman-ehfo]{\eqref{eq:feynman-ehfo}}, and~\hyperref[eq:feynman-fmmm]{\eqref{eq:feynman-fmmm}}. Several competing methods attain low test \texttt{RMSE} only by producing substantially larger and unwieldy expressions, often without recovering the ground-truth structure (such as \operon\ and \qlattice). By comparison, \hierbosss\ consistently identifies compact expressions with near-noise-level prediction error and high recovery frequency across equation complexities and noise regimes. This behavior is a direct consequence of \hierbosss' probabilistic framework that combines a regularizing symbolic tree prior in~\hyperref[eq:symbolic-tree-prior-Tj-1]{\eqref{eq:symbolic-tree-prior-Tj-1}}, symbolic model selection based on $\mathrm{JMP}(\mathcal T)$, and the post-\texttt{MCMC} symbolic model refinement step in \hyperref[subsec:symbolic-model-selection-refinement]{Section~\ref{subsec:symbolic-model-selection-refinement}}. Together, these components yield a favorable balance among predictive accuracy, parsimony, and structural fidelity. Detailed complexity analyses and corresponding results for the remaining Feynman equations are provided in~\suppref[subsec:expression-complexity-Feynman].
\begin{figure}[H]
    \centering
    \includegraphics[width=\linewidth]{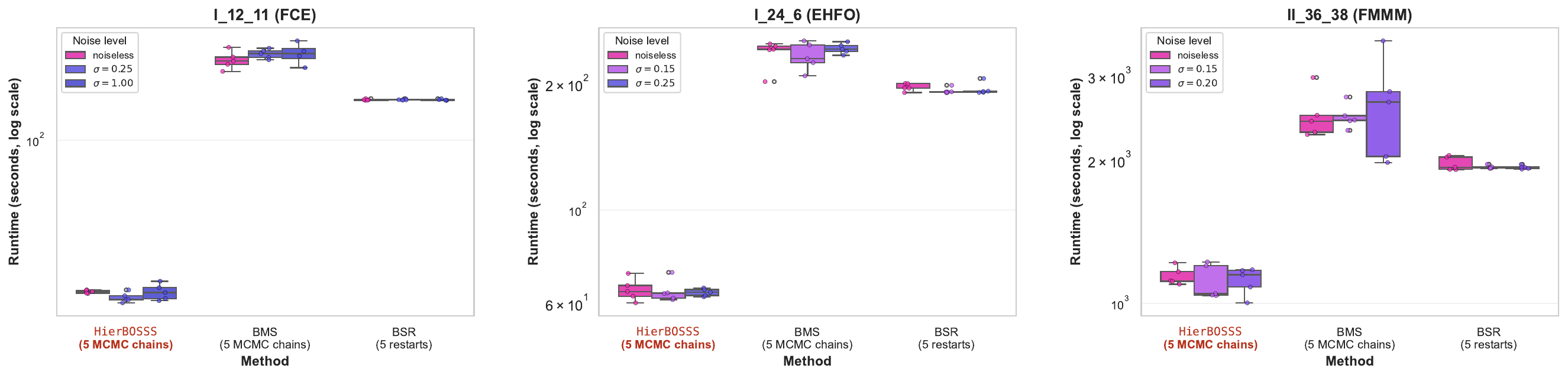}
    \caption{\centering{Runtimes (in $\log$-scale) of \hierbosss, \bms, and \bsr\ across repetitions and noise levels.}}
    \label{fig:feynman-Bayesian-runtime}
\end{figure}

\textbf{\emph{Summary}}.
Taken together, the preceding Feynman experiments reveal several recurring patterns across methods. \bsr, \dsr, and \gplearn\ generally exhibit weaker predictive accuracy and a tendency to return overly complex output symbolic expressions, while the performance of all competitors degrades markedly for the high-complexity law~\hyperref[eq:feynman-fmmm]{\eqref{eq:feynman-fmmm}}. Although \bms, \operon, \pysr, and \qlattice\ often predict accurately, their exact structural recovery is inconsistent, with \operon\ and \qlattice\ frequently producing particularly large expressions. \sisso$++$ achieves perfect recovery when the ground-truth law lies within its enumerated search space through the prescribed recursive depth parameter. However, more complex equations like \hyperref[eq:feynman-hoqn]{\eqref{eq:feynman-hoqn}} and \hyperref[eq:feynman-fmmm]{\eqref{eq:feynman-fmmm}} fall outside the explored search space under the present computational budget, while increasing the recursion depth requires exhaustive enumeration over an exponentially growing symbolic expression space. 
Finally, \hierbosss\ is considerably more computationally efficient than the competing Bayesian \sr\ modules \bms\ and \bsr, as illustrated in \hyperref[fig:feynman-Bayesian-runtime]{Figure~\ref{fig:feynman-Bayesian-runtime}}. Corresponding posterior convergence diagnostics for all Bayesian methods are reported in~\suppref[sec:trace-plots-Feynman].

\section{Descriptor Discovery for Oxide Perovskite Catalysts}
\label{sec:perovskites-data-study}

We consider the experimental oxide perovskite catalyst dataset~\citep{Weng2020SimpleDescriptor}, developed to uncover compact structure-activity relations governing the oxygen evolution reaction (\texttt{OER}) activity of perovskite catalysts. With general formula $ABO_{3}$, these materials offer substantial compositional flexibility. The $A$-site cation primarily influences lattice geometry and structural stability, whereas the transition metal $B$-site cation governs much of the electronic structure and surface chemistry underlying catalytic performance~\citep{Hwang2017Perovskites}. This flexibility makes oxide perovskites promising candidates for oxygen electrocatalysis and renewable-energy conversion, while also posing a challenging descriptor discovery problem. \sr\ is particularly well suited to this setting as it searches directly for compact analytical expressions (materials genes) linking catalytic activity to chemically and geometrically interpretable structure-activity relations, guiding the discovery of new perovskites.
\begin{figure}[H]
    \centering
    \includegraphics[width=0.45\linewidth]{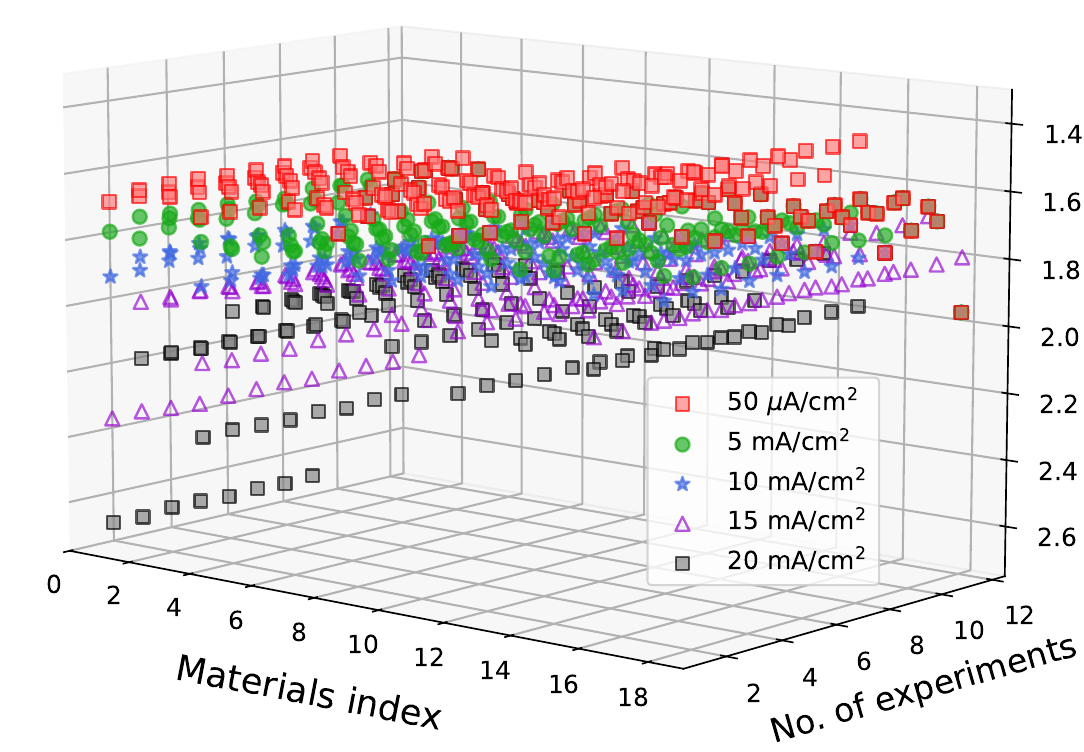}
    \caption{\centering{$V_{\mathrm{RHE}}$ landscape for the oxide perovskite catalyst dataset.}}
    \label{fig:VRHE_landscape_3D}
\end{figure}

The dataset comprises $18$ oxide perovskite catalysts synthesized and characterized under a common experimental protocol~\citep{Weng2020SimpleDescriptor}. For each material, $4$ independently prepared samples were measured $3$ times at each of $5$ prescribed current density ($\rho$) levels, yielding $n=1080$ observations. The response is the potential $V_{\mathrm{RHE}}$ measured in electron volts (eV) relative to the reversible hydrogen electrode, required to sustain the $\rho$ levels. As shown in \hyperref[fig:VRHE_landscape_3D]{Figure~\ref{fig:VRHE_landscape_3D}}, the measurements vary systematically across both catalyst composition and operating condition.
Each observation is described by $\rho$ and $7$ chemically motivated descriptors, $R_A$, $Q_A$, $N_d$, $\chi_A$, $\chi_B$, $\mu$, and $t$, which summarize the composition, electronic character, and lattice geometry of the catalyst. Further details are relegated to~\suppref[sec:oxide-perovskite-data-description].
\begin{figure}[H]
\centering
\includegraphics[width=0.624\linewidth]{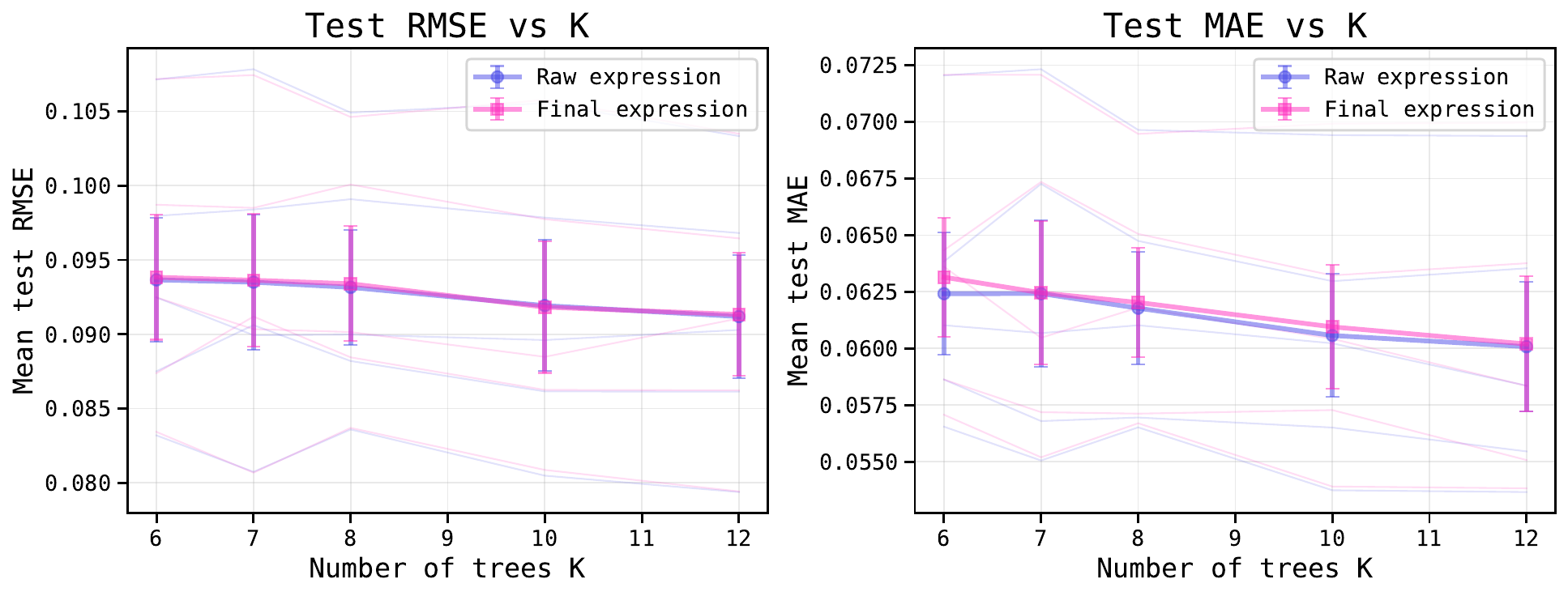}
\includegraphics[width=0.338\linewidth]{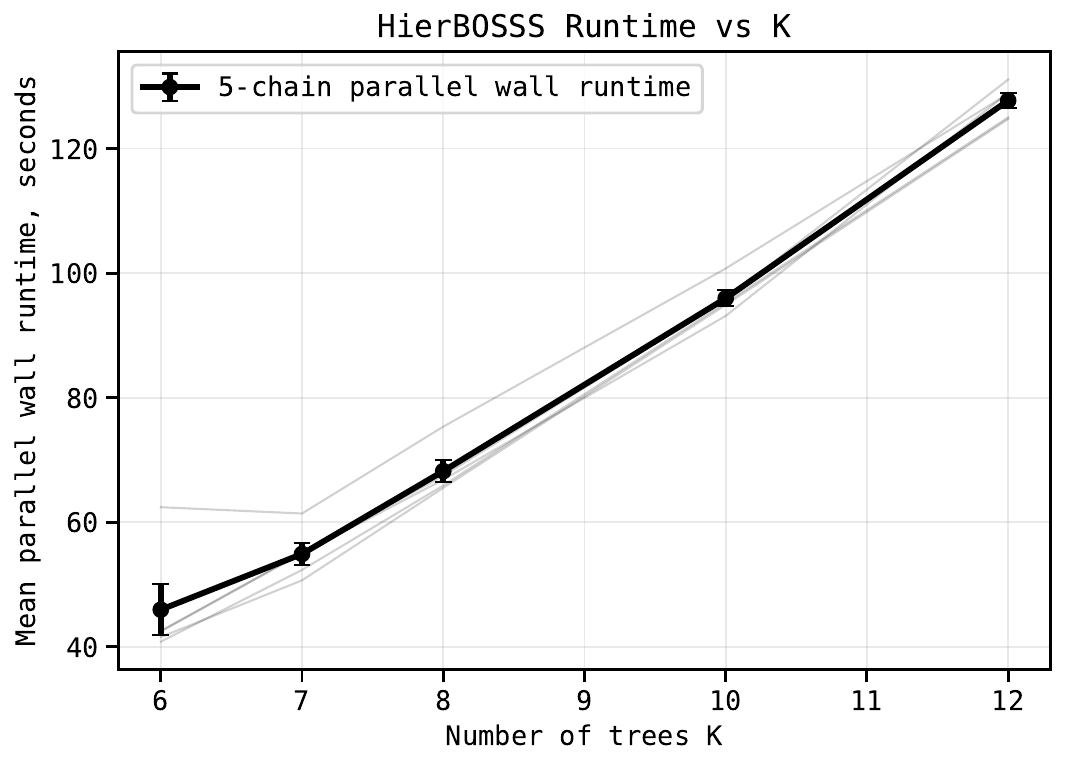}
\caption{\centering{
Ablation results of \hierbosss\ for the oxide perovskite catalyst dataset.
}
}
\label{fig:ablation-perovskite}
\end{figure} 
As in \hyperref[sec:learning-Feynman-equations]{Section~\ref{sec:learning-Feynman-equations}}, we first determine the nominal symbolic forest size $K$ in \hierbosss\ using the ablation-based practical diagnostic. The ablation results in \hyperref[fig:ablation-perovskite]{Figure~\ref{fig:ablation-perovskite}} indicate that predictive performance of \hierbosss\ stabilizes around $K=8$, beyond which larger $K$ incurs additional computational cost without significant accuracy gains; we therefore use $K=8$ for this analysis. We consider $5$ independent $90/10$ train-test splits of the dataset, fit the same suite of methods as in \hyperref[sec:learning-Feynman-equations]{Section~\ref{sec:learning-Feynman-equations}} on the training set, and evaluate held-out performance using test \texttt{RMSE} and mean absolute error (\texttt{MAE}). Across all methods, the common operator set $\mathbb{O} = \{+, \times, \mathrm{neg}, \mathrm{inv}, \sqrt{\cdot}\}$ has been used following~\cite{Weng2020SimpleDescriptor}. Complete experimental settings appear in~\suppref[sec:experimental-settings-perovskite].
\begin{figure}[H]
\centering
\includegraphics[width=0.65\linewidth]{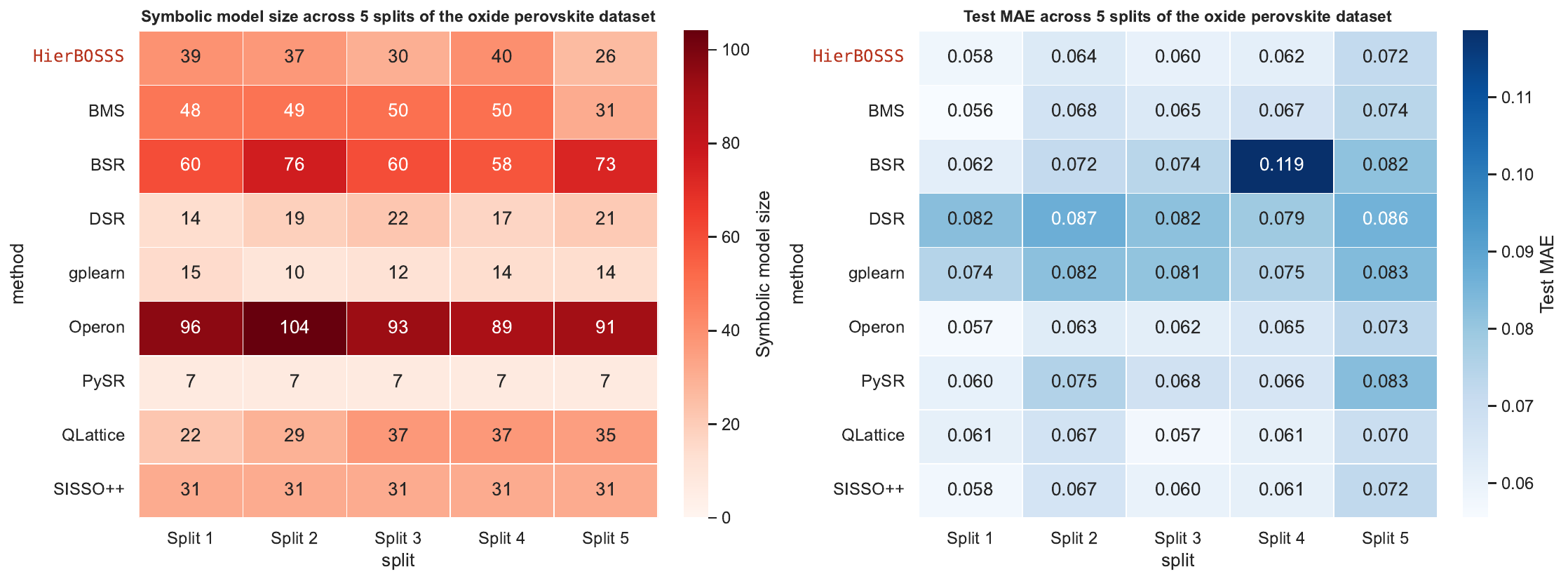}
\includegraphics[width=0.335\linewidth]{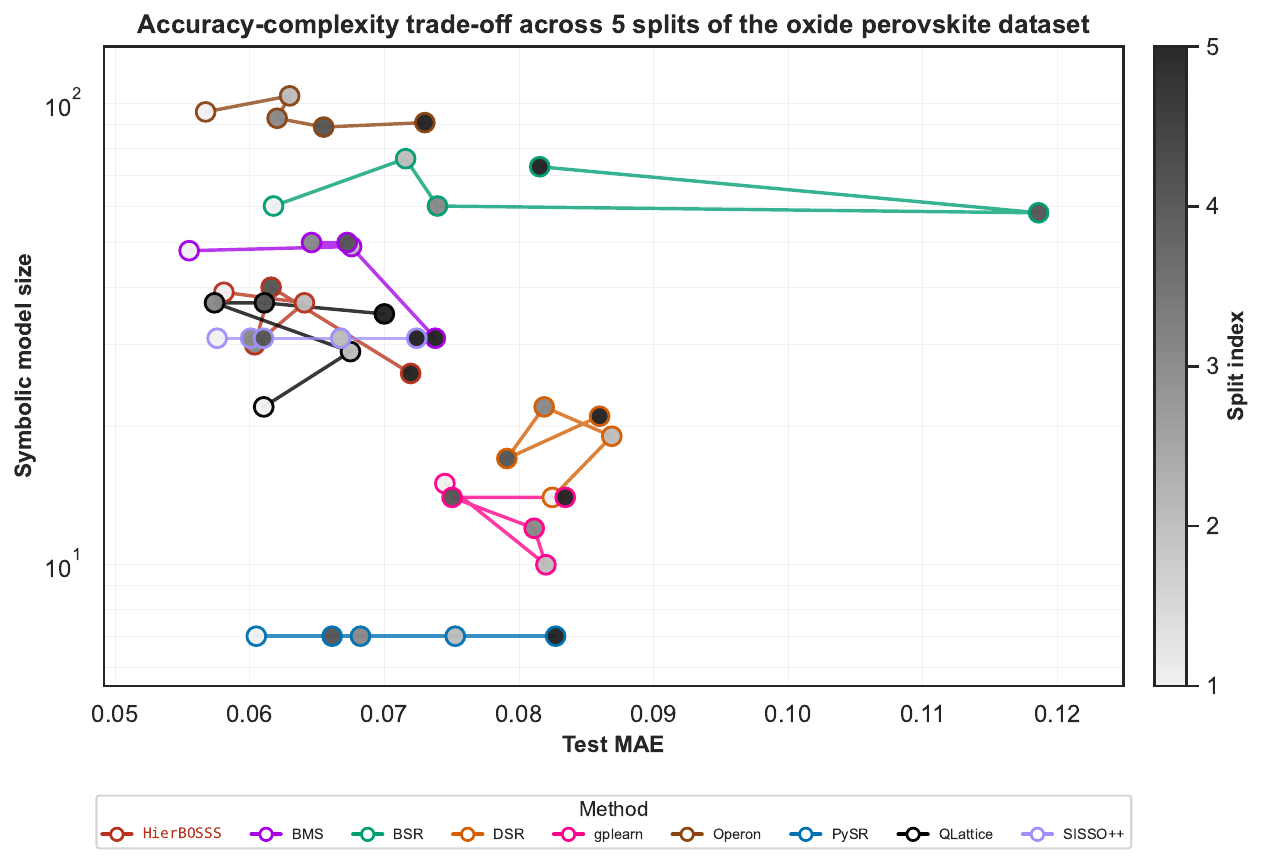}
\caption{\centering{
Symbolic model size, test \texttt{MAE}, and accuracy-complexity trade-off of \hierbosss\ and competitors, across $5$ splits of the oxide perovskite catalyst dataset.
}
}
\label{fig:accuracy-complexity-trade-off-perovskite}
\end{figure}

The split-wise descriptor learning behavior of each method is summarized in \hyperref[fig:accuracy-complexity-trade-off-perovskite]{Figure~\ref{fig:accuracy-complexity-trade-off-perovskite}}. The heatmaps highlight that low predictive error does not necessarily translate into compact symbolic descriptors. \operon\ consistently achieves competitive test \texttt{MAE}, but at the cost of the largest symbolic model sizes ($89$-$104$ nodes), while \bsr\ also produces relatively accurate descriptors yet requires substantially larger expression structures ($58$-$76$ nodes). On the contrary, \pysr\ and \gplearn\ return compact descriptors ($7$ and $10$-$15$ nodes, respectively), although their predictive performance is generally less competitive across the $5$ splits. \bms, \qlattice, and \sisso$++$ occupy intermediate regimes, balancing moderate accuracy with symbolic models of moderate complexity. In contrast, \hierbosss\ combines competitive predictive performance with descriptor expressions containing only $26$-$40$ nodes, markedly smaller than those of the strongest high-accuracy competitors. This favorable balance becomes particularly evident in the joint accuracy-complexity plot in \hyperref[fig:accuracy-complexity-trade-off-perovskite]{Figure~\ref{fig:accuracy-complexity-trade-off-perovskite}}, where \hierbosss\ repeatedly lies near the Pareto frontier across all splits, demonstrating that it achieves high predictive accuracy with symbolic parsimony. Extended complexity analysis, the descriptor expressions learned by all methods, computational comparisons highlighting the runtime advantage of \hierbosss\ over Bayesian competitors, and posterior diagnostics for the Bayesian \sr\ modules are provided in~\suppref[sec:symbolic-expressions-perovskites], \suppref[sec:accuracy-complexity-tradeoff-perovskites], \suppref[sec:runtime-complexity], and \suppref[sec:trace-plots-perovskites].

\begingroup
\scriptsize
\setlength{\tabcolsep}{2.2pt}
\renewcommand{\arraystretch}{1.22}

\begin{table}[t]
\scriptsize
\centering
\caption{\centering
{
Results on top-ranked $\mathrm{JMP}$ descriptor expressions learned by
\hierbosss\ across $5$ splits of the oxide perovskite catalyst dataset.
Descriptors highlighted in \textcolor{BrickRed}{\textbf{---}} were
identified as particularly
important for \texttt{OER} activity~\citep[Table 2]{Weng2020SimpleDescriptor}.
}
}
\label{tab:BayeSymX-top-expressions-across-splits}

\begin{tabular}{@{}c
>{\raggedright\arraybackslash}p{0.75\textwidth}
H c c c@{}}

\toprule
\toprule
Split
& \centering\finaltag\ learned descriptor expression
& Test RMSE
& Test \texttt{MAE}
& Model 
& {$K_{\mathrm{eff}}$}\\

& 
& (eV)
& (eV)
& size
&\\
\midrule


1 &
\exprcell{\exprbox{
\begin{aligned}[t]
& \tfrac{0.429}{\textcolor{BrickRed}{\mu}}
-0.004\textcolor{blue}{\rho}^{2}
-\tfrac{0.811}
 {\textcolor{BrickRed}{\sqrt{\chi_A}}-\sqrt{\textcolor{blue}{\rho}}}
-0.454\sqrt{\textcolor{blue}{\rho}}
+0.006N_d\textcolor{blue}{\rho}
+{0.339\textcolor{blue}{\rho}}
 \textcolor{BrickRed}{\tfrac{1}{\chi_B}}
+\tfrac{0.206\textcolor{BrickRed}{\mu}}
 {\textcolor{blue}{\rho}}
\end{aligned}
}}
& 0.084
& 0.058
& 39
& {7}\\

\reprowsep

2 &
\exprcell{\exprbox{
\begin{aligned}[t]
& 1.158
+0.292\textcolor{BrickRed}{\sqrt{Q_A}}
+\tfrac{0.530\textcolor{blue}{\rho}}{N_d}
+0.095\sqrt{N_d}\textcolor{blue}{\rho}
-0.068\textcolor{BrickRed}{\chi_B}\textcolor{blue}{\rho}
-0.073R_A\textcolor{blue}{\rho}
-0.073\textcolor{BrickRed}{\chi_A}\textcolor{blue}{\rho}
\end{aligned}
}}
& 0.087
& 0.064
& 37
& {5}\\

\reprowsep

3 &
\exprcell{\exprbox{
\begin{aligned}[t]
& 2.103\textcolor{BrickRed}{\sqrt{\mu}}
+0.282\textcolor{BrickRed}{\chi_A}
+{0.254\textcolor{blue}{\rho}}
 \textcolor{BrickRed}{\tfrac{1}{\chi_B}}
 \textcolor{BrickRed}{\tfrac{1}{\chi_B}}
+0.005N_d\textcolor{blue}{\rho}
-0.214\textcolor{BrickRed}{\mu}\textcolor{blue}{\rho}
\end{aligned}
}}
& 0.094
& 0.060
& 30
& {5}\\

\reprowsep

4 &
\exprcell{\exprbox{
\begin{aligned}[t]
& {0.720}\textcolor{BrickRed}{\tfrac{1}{t}}
+\tfrac{0.509}{\textcolor{BrickRed}{Q_A}}
+0.468\sqrt{\textcolor{BrickRed}{Q_A}+\textcolor{blue}{\rho}}
-0.169\sqrt{\textcolor{blue}{\rho}}
+{0.238\textcolor{blue}{\rho}}
 \textcolor{BrickRed}{\tfrac{1}{\chi_B}}
+0.035N_d\textcolor{blue}{\rho}
-0.148\sqrt{N_d}\textcolor{blue}{\rho}
\end{aligned}
}}
& 0.099
& 0.062
& 40
& {7}\\

\reprowsep

5 &
\exprcell{\exprbox{
\begin{aligned}[t]
& 0.371\textcolor{BrickRed}{t}
+{1.288}\textcolor{BrickRed}{\tfrac{1}{t}}
+\tfrac{0.334\textcolor{blue}{\rho}}{N_d}
+0.014N_d\textcolor{blue}{\rho}
-0.069\textcolor{BrickRed}{\chi_B}\textcolor{blue}{\rho}
\end{aligned}
}}
& 0.108
& 0.072
& 26
& {5}\\

\bottomrule
\bottomrule
\end{tabular}
\end{table}

\endgroup

\emph{\textbf{Scientific findings from \hierbosss\ results}}.
\hyperref[tab:BayeSymX-top-expressions-across-splits]{Table~\ref{tab:BayeSymX-top-expressions-across-splits}} presents the  top-ranked (by $\mathrm{JMP}$) descriptor expressions learned by \hierbosss\ across the $5$ splits of the dataset. Uniformly across the splits, the learned expressions consistently involve physically meaningful descriptor formulas involving tolerance factor $t$, octahedral factor $\mu$, electronegativities $\chi_A$ and $\chi_B$, and the $A$-site cation charge $Q_A$, all of which have been identified as important determinants of \texttt{OER} activity by~\cite{Weng2020SimpleDescriptor}. They also repeatedly include the prescribed current density $\rho$, reflecting the systematic dependence of $V_{\mathrm{RHE}}$ on the electrochemical operating condition observed in \hyperref[fig:VRHE_landscape_3D]{Figure~\ref{fig:VRHE_landscape_3D}}. Moreover, although the nominal symbolic forest size is chosen as $K=8$, the post-\texttt{MCMC} symbolic model refinement in \hyperref[subsec:symbolic-model-selection-refinement]{Section~\ref{subsec:symbolic-model-selection-refinement}} data-adaptively retains only $K_{\mathrm{eff}} = 5$ or $7$ additive components, balancing predictive performance with parsimony. Beyond the single top-ranked expression, \hierbosss\ also provides an Occam's window set of descriptors ($\mathcal{J}_r$, $r=10$) in \suppref[sec:Occams-window-set-perovskite-BayeSymX] as a posterior summary, representing uncertainty in symbolic structure while preserving multiple scientifically plausible structure-activity relationships. Collectively, these results demonstrate that \hierbosss\ combines predictive accuracy with compact, physically interpretable, and probabilistically supported descriptor discovery for oxide perovskite catalysts.


\section{Discussion}

The present work opens several promising methodological and theoretical directions for future research in probabilistic \sr. First, extending the symbolic tree representation in \hierbosss\ to infer internal constants would facilitate the recovery of scientific laws and equations involving unknown physical coefficients. While the current formulation can encode simple constants compositionally (for example, $\exp\{2x\} = [\exp\{x\}]^2$), it does not directly accommodate free constants within nonlinear expressions such as $\log(2x + 3)$. Second, as equivalent representations are ubiquitous in symbolic expression spaces, a promising albeit challenging theoretical direction is to characterize semantic equivalence classes of symbolic trees and forests as fibres of the corresponding evaluation homomorphism, and to study the induced kernel congruence and quotient algebra. This framework could also be extended to data-induced empirical equivalence and incorporated methodologically through structural data augmentation or equivalence-aware Monte Carlo tree search~\citep{eggsr}. Finally, a central question is to delineate the classes of target functions that admit arbitrarily accurate $K$-term approximations over the symbolic dictionary $\mathbb{G}_{\mathbb{O}, p}$, and to understand how this approximability depends on $\mathbb{O}$, $K$, and the inclusion of internal constants.


\section*{Appendices}
Appendices below include \hyperref[superdiv:methodological-details]{\textcolor{BrickRed}{Part~\ref*{superdiv:methodological-details}}}: additional methodological details of \hierbosss, \hyperref[superdiv:theoretical-proofs]{\textcolor{BrickRed}{Part \ref*{superdiv:theoretical-proofs}}}: proofs of theoretical results in \hyperref[sec:posterior-contraction]{Section \ref{sec:posterior-contraction}}, \hyperref[superdiv:feynman]{\textcolor{BrickRed}{Part~\ref*{superdiv:feynman}}}: additional empirical results for Feynman equations in \hyperref[sec:learning-Feynman-equations]{Section \ref{sec:learning-Feynman-equations}}, and \hyperref[superdiv:perovskites]{\textcolor{BrickRed}{Part~\ref*{superdiv:perovskites}}}: additional empirical results for the oxide perovskite catalyst data study in \hyperref[sec:perovskites-data-study]{Section \ref{sec:perovskites-data-study}}.

\clearpage
\begin{center}
{\Huge\bfseries
Appendices}  
\end{center}

\beginmainappendices

\useappendixlabelaliases

\appendixtableofcontents

\section{Notation}
\label{sec:notation}

For ease of reference, we collect below the notation in~\hyperref[tab:notation]{Table \ref{tab:notation}} used throughout the methodological developments, theoretical proofs, and empirical results. Bold lowercase and uppercase letters denote vectors and matrices, respectively. Calligraphic letters generally denote sets, classes, or collections, while blackboard-bold letters are used for number systems, tree spaces, and probability measures. Unless otherwise stated, all vectors are column vectors.

\begingroup
\scriptsize
\setlength{\tabcolsep}{5pt}
\renewcommand{\arraystretch}{1.16}

\begin{longtable}{
    >{\raggedright\arraybackslash}p{0.25\textwidth}
    >{\raggedright\arraybackslash}p{0.70\textwidth}
}
\caption{Frequently used notation in the main manuscript and online appendices.}
\label{tab:notation}\\

\toprule
\toprule
\textbf{Notation} & \textbf{Description} \\
\midrule
\endfirsthead

\multicolumn{2}{c}%
{{\tablename\ \thetable{} continued from the previous page}}\\
\toprule
\textbf{Notation} & \textbf{Description} \\
\midrule
\endhead

\midrule
\multicolumn{2}{r}{Continued on the next page}\\
\endfoot

\bottomrule
\bottomrule
\endlastfoot

\reprowsep
\multicolumn{2}{l}{\textcolor{BrickRed}{\textbf{\textit{Basic mathematical notation}}}}\\
\addlinespace[2pt]

\reprowsep

$\mathbb N$
&
Set of positive integers, $\{1,2,\ldots\}$.
\\[2pt]

$\mathbb R$
&
Set of real numbers.
\\[2pt]

$\mathbb R^{+}$
&
Set of strictly positive real numbers.
\\[2pt]

$\mathbb R^{n}$
&
The $n$-dimensional real Euclidean space.
\\[2pt]

$\mathbb R^{n\times K}$
&
Space of real-valued $n\times K$ matrices.
\\[2pt]

$|A|$
&
Cardinality of a finite set $A$.
\\[2pt]

$d!$
&
Factorial of the nonnegative integer $d$.
\\[2pt]

$\bm I_n$
&
The $n\times n$ identity matrix.
\\[2pt]

$\mathrm{det}(\bm A)$
&
Determinant of a square matrix $\bm A$.
\\[2pt]

$\bm A\succ 0$
&
The symmetric matrix $\bm A$ is positive definite.
\\[2pt]

$\|\bm z\|_1$
&
The $\ell_1$ norm, $\|\bm z\|_1=\sum_j|z_j|$.
\\[2pt]

$\|\bm z\|_2$
&
The Euclidean norm, $\|\bm z\|_2=(\sum_j z_j^2)^{1/2}$.
\\[2pt]

$\|\bm z\|_\infty$
&
The supremum norm, $\|\bm z\|_\infty=\max_j|z_j|$.
\\[2pt]

$\mathds{1}_{A}$
&
Indicator of the event or set $A$, equal to one when $A$ occurs and zero
otherwise.
\\[2pt]

$\mathcal O(a_n)$
&
A sequence whose magnitude is bounded above by a constant multiple of $a_n$
for all sufficiently large $n$.
\\[2pt]

$\mathfrak o(a_n)$
&
A sequence that is asymptotically negligible relative to $a_n$.
\\[2pt]

$\mathrm{Vol}(A)$
&
Volume of a measurable set $A$ with respect to the ambient measure.
\\[2pt]

$\mathrm{Vol}_{d}(A)$
&
$d$-dimensional volume of a measurable set $A\subseteq\mathbb R^d$.
\\[2pt]

$\mathrm{Leb}(A)$
&
Lebesgue measure of a measurable set $A\subseteq\mathbb R^d$.
\\[8pt]


\reprowsep

\multicolumn{2}{l}{\textcolor{BrickRed}{\textbf{\textit{Data, predictors, and regression functions}}}}\\
\addlinespace[2pt]

\reprowsep

$x$
&
A scalar-valued primary feature or a generic scalar argument.
\\[2pt]

$\bm x=(x_1,\ldots,x_p)^{\T}$
&
A generic $p$-dimensional feature vector.
\\[2pt]

$\mathfrak X\subset\mathbb R^p$
&
Predictor domain.
\\[2pt]

$\bm X=(\bm x_1,\ldots,\bm x_n)^{\T}$
&
The $n\times p$ design matrix.
\\[2pt]

$\bm y=(y_1,\ldots,y_n)^{\T}$
&
Observed response vector.
\\[2pt]

$f:\mathfrak X\rightarrow\mathbb R$
&
A generic regression function.
\\[2pt]

$f_0$
&
The true data-generating regression function.
\\[2pt]

$\mathrm{ev}[T](\bm x)$
&
Real-valued evaluation of the symbolic expression represented by the symbolic tree $T$ at
input $\bm x$.
\\[2pt]

$\mathbb G_{\mathbb O,p}$
&
Symbolic dictionary
$\{\mathrm{ev}[T]:T\in\mathbb T_{\mathbb O,p}\}$.
\\[2pt]

$f_{\bm\beta,\mathcal T}$
&
Regression function induced by the symbolic forest $\mathcal T$ and outer
regression coefficient vector $\bm\beta \in \mathbb{R}^{K+1}$.
\\[2pt]

$\mathcal F_K$
&
Class of regression functions representable as linear combinations of $K$
symbolic trees.
\\[2pt]

$\mathcal F_{K,S}$
&
Subset of $\mathcal F_K$ consisting of symbolic forests whose total number of
nonterminal nodes is at most, or equal to $S$.
\\[8pt]

\reprowsep
\multicolumn{2}{l}{\textcolor{BrickRed}{\textbf{\textit{Operators and symbolic trees}}}}\\
\addlinespace[2pt]

\reprowsep

$\mathbb O$
&
Prescribed library of admissible mathematical operators.
\\[2pt]

$\mathbb O_u$
&
Subset of unary operators in $\mathbb O$.
\\[2pt]

$\mathbb O_b$
&
Subset of binary operators in $\mathbb O$.
\\[2pt]

$o$
&
A generic operator belonging to $\mathbb O$.
\\[2pt]

$\mathrm{arity}(o)$
&
Number of arguments taken by operator $o\in \mathbb{O}$; in the present setting,
$\mathrm{arity}(o)\in\{1,2\}$.
\\[2pt]

$\mathbb T_{\mathbb O,p}$
&
Collection of all admissible symbolic trees generated from the
operator set $\mathbb O$ and primary features $x_1,\ldots,x_p$.
\\[2pt]

$\mathbb T_{\mathbb O,p}^{K}$
&
Cartesian product of $K$ symbolic tree spaces; equivalently, the space of
ordered symbolic forests containing $K$ symbolic trees.
\\[2pt]

$T$
&
A generic symbolic tree.
\\[2pt]

$T_j$
&
The $j$th symbolic tree in a forest, for $j=1,\ldots,K$.
\\[2pt]

$\mathcal T=(T_1,\ldots,T_K)$
&
An ordered symbolic forest containing $K$ symbolic trees.
\\[2pt]

$\mathrm{root}(T)$
&
Root node of tree $T$; this is also denoted by $\zeta_0$ where convenient.
\\[2pt]

$\mathcal N(T)$
&
Set of all nodes in tree $T$.
\\[2pt]

$\mathcal N_{\mathrm{op}}(T)$
&
Set of nonterminal, or operator, nodes in $T$.
\\[2pt]

$\mathcal N_{\mathrm{ft}}(T)$
&
Set of terminal, or primary feature, nodes in $T$.
\\[2pt]

$\mathcal N_{\mathrm{op}}(T,m)$
&
Set of nonterminal nodes of $T$ located at depth $m$.
\\[2pt]

$\mathcal N_{\mathrm{ft}}(T,m)$
&
Set of terminal nodes of $T$ located at depth $m$.
\\[2pt]

$m_\zeta$
&
Depth of node $\zeta$, with the root assigned depth zero.
\\[2pt]

$\mathrm d(T)$
&
Depth of tree $T$,
$\mathrm d(T)=\max_{\zeta\in\mathcal N(T)}m_\zeta$.
\\[2pt]

$S(T)$
&
Operator count of tree $T$, defined as the number of nonterminal nodes:
$S(T)=|\mathcal N_{\mathrm{op}}(T)|$.
\\[2pt]

$S(\mathcal T)$
&
Total forest operator count,
$S(\mathcal T)=\sum_{j=1}^{K}S(T_j)$.
\\[2pt]

$\mathcal S_m$
&
The $(m-1)$-dimensional simplex, $\mathcal S_m
=\{\bm z\in\mathbb R^m: z_i\geq0,\ \sum_{i=1}^{m}z_i=1\}$.
\\[2pt]

$N_{\mathcal T}(K,S)$
&
Number of admissible ordered symbolic forests containing $K$ trees and having maximum total operator count $S$.
\\[8pt]

\reprowsep

\multicolumn{2}{l}{\textcolor{BrickRed}{\textbf{\textit{Symbolic evaluation and model parameters}}}}\\
\addlinespace[2pt]

\reprowsep

$K$
&
Number of symbolic trees in the forest.
\\[2pt]

$\mathrm{ev}[T](\bm X)$
&
The $n$-dimensional evaluation vector, $\mathrm{ev}[T](\bm X)
=
\bigl(\mathrm{ev}[T](\bm x_1),\ldots,\mathrm{ev}[T](\bm x_n)\bigr)^{\T}$.
\\[2pt]

$\mathcal E(\bm X;\mathcal T)$
&
Symbolic design matrix associated with forest $\mathcal T$, $\mathcal E(\bm X;\mathcal T)
=[\bm 1_n,\,
\mathrm{ev}[T_1](\bm X),\ldots,
\mathrm{ev}[T_K](\bm X)]$.
\\[2pt]

$\bm\beta=(\beta_0,\ldots,\beta_K)^{\T}$
&
Outer regression coefficient vector, including the intercept $\beta_0$.
\\[2pt]

$\sigma^2$
&
Model and observation noise variance.
\\[2pt]

$\bm\mu_\beta,\bm\Sigma_\beta$
&
Prior location vector and prior scale matrix for $\bm\beta$ under the
Normal-Inverse-Gamma prior.
\\[2pt]

$\nu,\lambda$
&
Shape and scale hyperparameters of the Inverse-Gamma component of the
Normal-Inverse-Gamma prior.
\\[2pt]

$\bm\mu_\beta^\star(\mathcal T),
 \bm\Sigma_\beta^\star(\mathcal T)$
&
Conditional posterior location vector and scale matrix of $\bm\beta$ for a
fixed symbolic forest $\mathcal T$.
\\[2pt]

$\nu^\star,\lambda^\star(\mathcal T)$
&
Updated Inverse-Gamma hyperparameters for a fixed forest $\mathcal T$.
\\[8pt]

\reprowsep
\multicolumn{2}{l}{\textcolor{BrickRed}{\textbf{\textit{Probability distributions and operators}}}}\\
\addlinespace[2pt]

\reprowsep

$\mathrm N(\mu,\sigma^2)$
&
Univariate Gaussian distribution with mean $\mu$ and variance $\sigma^2$.
\\[2pt]

$\mathrm N_n(\bm\mu,\bm\Sigma),\;\mathrm N_n(\cdot \mid \bm\mu,\bm\Sigma)$
&
$n$-variate Gaussian density or distribution with mean vector $\bm\mu$ and
covariance matrix $\bm\Sigma$; when used as a density, the realized argument
is displayed before the conditioning bar.
\\[2pt]

$\mathrm{IG}(a,b),\; \mathrm{IG}(\cdot \mid a,b)$
&
Inverse-Gamma distribution with shape parameter $a$ and scale parameter $b$; when used as a density, the realized argument is displayed before the conditioning bar.
\\[2pt]

$\mathrm{NIG}_{K+1}(\bm\beta,\sigma^2
 \mid\bm\mu_\beta,\bm\Sigma_\beta,\nu,\lambda)$
&
Normal-Inverse-Gamma distribution defined by $\bm\beta\mid\sigma^2
\sim
\mathrm N_{K+1}
(\bm\mu_\beta,\sigma^2\bm\Sigma_\beta),\;
\sigma^2\sim
\mathrm{IG}(\nu/2,\lambda/2)$.
\\[2pt]

$\mathrm{G}(a,b),\; \mathrm{G}(\cdot \mid a,b)$
&
Gamma distribution with shape parameter $a$ and rate parameter $b$;
when used as a density, the realized argument is displayed before the
conditioning bar.
\\[2pt]

$\mathrm{Dir}(\bm\alpha)$
&
Dirichlet distribution with concentration vector $\bm\alpha$.
\\[2pt]

$\mathrm{Bernoulli}(q)$
&
Bernoulli distribution with success probability $q$.
\\[2pt]

$\mathrm{Uniform}(a, b)$
&
Uniform distribution over the interval $(a, b)$, where $a<b$.
\\[2pt]

$\Pi$
&
A generic prior probability measure.
\\[2pt]

$\Pi(\cdot\mid\mathcal D_n)$
&
Posterior probability measure conditional on the observed data $\mathcal D_n$.
\\[2pt]

$\Pi_{\mathrm{tree}}$
&
Marginal prior distribution on an individual symbolic tree.
\\[2pt]

$\Pi_{\mathrm{forest},K}$
&
Prior distribution on a symbolic forest containing $K$ trees.
\\[2pt]

$\mathsf{P}(A)$
&
Probability of event $A$.
\\[2pt]

$\mathsf E(X)$
&
Expectation of a random quantity $X$.
\\[2pt]

$\mathsf{V}(X)$
&
Variance of a random quantity $X$.
\\[2pt]

$\mathsf E_{\mathbb P}(X)$
&
Expectation of a random quantity $X$ under probability measure $\mathbb P$.
\\[2pt]

$\mathsf{V}_{\mathbb P}(X)$
&
Variance of a random quantity $X$ under probability measure $\mathbb P$.
\\[8pt]

\reprowsep

\multicolumn{2}{l}{\textcolor{BrickRed}{\textbf{\textit{Sampling laws and divergence quantities}}}}\\
\addlinespace[2pt]

\reprowsep

$\mathbb P^n_{f,\sigma^2}$
&
Fixed design joint sampling law.
\\[2pt]

$p^n_{f,\sigma^2}$
&
Density corresponding to $\mathbb P^n_{f,\sigma^2}$.
\\[2pt]

$\mathbb P_0^n$
&
True fixed design sampling law,
$\mathbb P_0^n\equiv\mathbb P^n_{f_0,\sigma_0^2}$.
\\[2pt]

$p_0^n$
&
Density corresponding to the true law $\mathbb P_0^n$.
\\[2pt]

$\mathbb Q^n_{f,\sigma^2}$
&
Joint sampling law under the random design formulation, including the
distribution of the predictors.
\\[2pt]

$q^n_{f,\sigma^2}$
&
Density corresponding to $\mathbb Q^n_{f,\sigma^2}$.
\\[2pt]

$q_X$
&
Density of the random predictor vector $\bm x$ on $\mathfrak X$.
\\[2pt]

$\mathrm{KL}(P\parallel Q)$
&
Kullback-Leibler divergence from probability measure $Q$ to
probability measure $P$.
\\[8pt]

\reprowsep

\multicolumn{2}{l}{\textcolor{BrickRed}{\textbf{\textit{Metrics, approximation, and covering numbers}}}}\\
\addlinespace[2pt]

\reprowsep

$d_n(f,g)$
&
Empirical $L_2$ distance, $
d_n(f,g)
=
[
\frac1n\sum_{i=1}^{n}
\bigl(f(\bm x_i)-g(\bm x_i)\bigr)^2
]^{1/2}.
$
\\[2pt]


$N(\varepsilon,\mathcal F,d)$
&
$\varepsilon$-covering number of the set $\mathcal F$ under metric $d$; that
is, the minimum number of $d$-balls of radius $\varepsilon$ needed to cover
$\mathcal F$.
\\[2pt]

$a_{K,S,n}(f_0)$
&
Best empirical approximation error of $f_0$ by a symbolic regression function
with $K$ trees and maximum total operator count $S$.
\\[2pt]

$a_K^{2}(f_0)$
&
Best population-level approximation error of $f_0$ over the class
$\mathcal F_K$.
\\

\end{longtable}
\endgroup
\newpage
\suppdivision{Details on the \texorpdfstring{\hierbosss}{HierBOSSS} Methodology}
\label{superdiv:methodological-details}


\section{Details on the \texorpdfstring{\hierbosss}{HierBOSSS} Posterior Sampling Algorithm}
\label{sec:posterior-details}

Here, we provide the computational details underlying posterior inference for \hierbosss, complementing \hyperref[subsec:posterior-inference]{\textcolor{BrickRed}{Section~\ref*{subsec:posterior-inference} of the main manuscript}} with the derivations and sampling mechanisms used to generate and update symbolic forest structures.

\subsection{Full Conditional of \texorpdfstring{$(\bm{\beta}, \sigma^2)$}{(beta, sigma2)} and \texorpdfstring{$\mathrm{JMP}$}{JMP} of \texorpdfstring{$\mathcal T$}{T}}
\label{sec:posterior-derivation}

\underline{\emph{Posterior full conditional distribution of $(\bm \beta, \sigma^2)$}}. From \hyperref[eq:BayeSymX-posterior]{\textcolor{BrickRed}{(\ref*{eq:BayeSymX-posterior}) of the main manuscript}}, the \hierbosss-induced posterior is
{
\begin{align}
\label{eq:BayeSymX-posterior-supp}
\begin{split}
&\Pi\left(\mathcal T, \bm\beta, \sigma^2\mid \mathcal D_n, \alpha_0, \delta_0, \bm \alpha_{\mathrm{op}}, \bm\alpha_{\mathrm{ft}}, \bm\mu_{\beta}, \bm\Sigma_{\beta}, \nu, \lambda\right)\\
&\propto \mathrm{N}_n\left(\bm y\mid \mathcal E(\bm X; \mathcal T) \bm \beta, \sigma^2\bm I_n\right)\mathrm{NIG}_{K+1}\left(\bm \beta, \sigma^2 \mid \bm \mu_{\beta}, \bm\Sigma_{\beta}, \nu, \lambda\right) \Pi_{\mathrm{forest}, K}\left(\mathcal T\mid  \alpha_0, \delta_0, \bm \alpha_{\mathrm{op}}, \bm \alpha_{\mathrm{ft}}\right).
\end{split}
\end{align}
}
From \hyperref[eq:BayeSymX-posterior-supp]{\eqref{eq:BayeSymX-posterior-supp}} and by the conjugacy of the Normal-Inverse-Gamma (NIG) prior, the joint posterior full conditional distribution over $(\bm \beta, \sigma^2)$ is
\begin{align}
\label{eq:beta-sigma2-full-condtional}
\begin{split}
&\Pi\left(\bm\beta, \sigma^2\mid \mathcal T, \mathcal D_n, \alpha_0, \delta_0, \bm \alpha_{\mathrm{op}}, \bm\alpha_{\mathrm{ft}}, \bm\mu_{\beta}, \bm\Sigma_{\beta}, \nu, \lambda\right)\\
& \propto \mathrm{N}_n\left(\bm y\mid \mathcal E(\bm X; \mathcal T) \bm \beta, \sigma^2\bm I_n\right)\mathrm{NIG}_{K+1}\left(\bm \beta, \sigma^2 \mid \bm \mu_{\beta}, \bm\Sigma_{\beta}, \nu, \lambda\right) \\
& = \mathrm{N}_n\left(\bm y\mid \mathcal E(\bm X; \mathcal T) \bm \beta, \sigma^2\bm I_n\right)\mathrm{N}_{K+1}\left(\bm \beta \mid \bm \mu_{\beta}, \sigma^{2}\bm \Sigma_{\beta}\right)\;\mathrm{IG}\left(\sigma^{2}\mid \nu/2, \lambda/2\right) \\
& \propto \mathrm{det}\left(\bm \Sigma_{\beta}^{\star}(\mathcal T)\right)^{\frac{1}{2}} \left(\lambda^\star(\mathcal T)\right)^{-\nu^\star/2} \times \mathrm{NIG}_{K+1}\left(\bm \beta, \sigma^2 \mid \bm\mu_\beta^\star(\mathcal T), \bm\Sigma_\beta^\star(\mathcal T), \nu^{\star}, \lambda^\star(\mathcal T)\right),
\end{split}
\end{align}
where
\begin{align}
\label{eq:NIG-parameters-supp}
\begin{gathered}
\nu^{\star} = \nu + n, \quad \lambda^{\star}(\mathcal T) = \lambda + \bm y^{\T}\bm{y} + \bm{\mu}_{\beta}^{\T}\bm\Sigma_{\beta}^{-1}\bm\mu_{\beta} - (\bm\mu_{\beta}^{\star}(\mathcal T))^{\T}(\bm\Sigma_\beta^\star(\mathcal T))^{-1}\bm\mu_\beta^\star(\mathcal T),\\
\bm\mu_\beta^\star(\mathcal T) = \bm\Sigma_\beta^\star(\mathcal T)\left(\bm\Sigma_\beta^{-1}\bm\mu_\beta + \mathcal{E}^{\T}(\bm X; \mathcal T) \bm y\right), \\
\bm\Sigma^{\star}_{\beta}(\mathcal T) = \left(\bm\Sigma_\beta^{-1} + \mathcal{E}^{\T}(\bm X; \mathcal T) \mathcal{E}(\bm X; \mathcal T)\right)^{-1}.
\end{gathered}
\end{align}
Therefore the joint posterior full conditional distribution over $(\bm \beta, \sigma^2)$ is
\begin{align}
\Pi\left(\bm\beta, \sigma^2\mid \mathcal T, \mathcal D_n, \alpha_0, \delta_0, \bm \alpha_{\mathrm{op}}, \bm\alpha_{\mathrm{ft}}, \bm\mu_{\beta}, \bm\Sigma_{\beta}, \nu, \lambda\right) = \mathrm{NIG}_{K+1}\left(\bm \beta, \sigma^2 \mid \bm\mu_\beta^\star(\mathcal T), \bm\Sigma_\beta^\star(\mathcal T), \nu^{\star}, \lambda^\star(\mathcal T)\right).
\end{align}

\underline{\emph{Joint marginal posterior ($\mathrm{JMP}$) distribution of $\mathcal{T}$}}.
Marginalizing over $\bm \beta$ and $\sigma^2$ in \hyperref[eq:BayeSymX-posterior-supp]{\eqref{eq:BayeSymX-posterior-supp}} above yields
\begin{align}
\label{eq:BayeSymX-posterior-marginalization-1}
\begin{split}
&\Pi\left(\mathcal T \mid \mathcal D_n, \alpha_0, \delta_0, \bm \alpha_{\mathrm{op}}, \bm\alpha_{\mathrm{ft}}, \bm\mu_{\beta}, \bm\Sigma_{\beta}, \nu, \lambda\right) \propto \Pi_{\mathrm{forest}, K}\left(\mathcal T\mid  \alpha_0, \delta_0, \bm \alpha_{\mathrm{op}}, \bm \alpha_{\mathrm{ft}}\right)
\\ & \qquad \times \int_{0}^{\infty} \int_{\mathbb{R}^{K + 1}} \mathrm{N}_n\left(\bm y\mid \mathcal E(\bm X; \mathcal T) \bm \beta, \sigma^2\bm I_n\right)\mathrm{NIG}_{K+1}\left(\bm \beta, \sigma^2 \mid \bm \mu_{\beta}, \bm\Sigma_{\beta}, \nu, \lambda\right) d\bm\beta d\sigma^2.
\end{split}
\end{align}
Using \hyperref[eq:beta-sigma2-full-condtional]{\eqref{eq:beta-sigma2-full-condtional}} in \hyperref[eq:BayeSymX-posterior-marginalization-1]{\eqref{eq:BayeSymX-posterior-marginalization-1}}, the integral has a closed form as below
\begin{align}
\label{eq:BayeSymX-posterior-marginalization-2}
\begin{split}
&\int_{0}^{\infty} \int_{\mathbb{R}^{K + 1}} \mathrm{N}_n\left(\bm y\mid \mathcal E(\bm X; \mathcal T) \bm \beta, \sigma^2\bm I_n\right)\mathrm{NIG}_{K+1}\left(\bm \beta, \sigma^2 \mid \bm \mu_{\beta}, \bm\Sigma_{\beta}, \nu, \lambda\right) d\bm\beta d\sigma^2 
\\ &
= \int_{0}^{\infty} \int_{\mathbb{R}^{K + 1}} \mathrm{N}_n\left(\bm y\mid \mathcal E(\bm X; \mathcal T) \bm \beta, \sigma^2\bm I_n\right)\mathrm{N}_{K+1}\left(\bm \beta \mid \bm \mu_{\beta}, \sigma^{2}\bm \Sigma_{\beta}\right)\;\mathrm{IG}\left(\sigma^{2}\mid \nu/2, \lambda/2\right) d\bm\beta d\sigma^2 \\
&\propto  \mathrm{det}\left(\bm \Sigma_{\beta}^{\star}(\mathcal T)\right)^{1/2} \left(\lambda^\star(\mathcal T)\right)^{-\nu^\star/2} \int_{0}^{\infty} \int_{\mathbb{R}^{K + 1}} \mathrm{NIG}_{K+1}\left(\bm \beta, \sigma^2 \mid \bm\mu_\beta^\star(\mathcal T), \bm\Sigma_\beta^\star(\mathcal T), \nu^{\star}, \lambda^\star(\mathcal T)\right) d\bm\beta d\sigma^2\\
&= \mathrm{det}\left(\bm \Sigma_{\beta}^{\star}(\mathcal T)\right)^{1/2} \left(\lambda^\star(\mathcal T)\right)^{-\nu^\star/2}.
\end{split}
\end{align}
Substituting \hyperref[eq:BayeSymX-posterior-marginalization-2]{\eqref{eq:BayeSymX-posterior-marginalization-2}} in \hyperref[eq:BayeSymX-posterior-marginalization-1]{\eqref{eq:BayeSymX-posterior-marginalization-1}} gives
\begin{align}
\label{supple-eq:JMP}
\boxed{
\begin{aligned}
\mathrm{JMP}(\mathcal T) &= \Pi\left(\mathcal T \mid \mathcal D_n, \alpha_0, \delta_0, \bm \alpha_{\mathrm{op}}, \bm\alpha_{\mathrm{ft}}, \bm\mu_{\beta}, \bm\Sigma_{\beta}, \nu, \lambda\right)
\\&
\propto  \mathrm{det}\left(\bm \Sigma_{\beta}^{\star}(\mathcal T)\right)^{\frac 12} \left(\lambda^\star(\mathcal T)\right)^{-\frac{\nu^{\star}}{2}} \Pi_{\mathrm{forest}, K}\left(\mathcal T\mid  \alpha_0, \delta_0, \bm \alpha_{\mathrm{op}}, \bm \alpha_{\mathrm{ft}}\right).
\end{aligned}
}
\end{align}

\subsection{Generative Algorithm for Symbolic Trees and Subtrees}
\label{subsec:symbolic-tree-generation}

\hyperref[supple-alg:symbolic-tree-generation]{Algorithm~\ref{supple-alg:symbolic-tree-generation}} is a detailed version of \hyperref[alg:symbolic-tree-generation]{\textcolor{BrickRed}{Algorithm~\ref*{alg:symbolic-tree-generation} of the main manuscript}} which implements the recursive symbolic tree generation procedure introduced in \hyperref[subsec:symbolic-tree-representation]{\textcolor{BrickRed}{Section~\ref*{subsec:symbolic-tree-representation} of the main manuscript}}. This procedure serves as a fundamental building block for the local symbolic tree proposal moves used in the \texttt{MH} updates detailed in \hyperref[sec:posterior-sampler-details]{Appendix~\ref{sec:posterior-sampler-details}}.

At each node, the depth-dependent splitting probability, as in \maineqref[eq:split-probability], determines whether the node is terminal or nonterminal. Terminal nodes are assigned feature labels, whereas nonterminal nodes are assigned operator labels and expanded by recursively generating child subtrees according to the operator arity. 

{\small
\begin{algorithm}[H]
\caption{Symbolic subtree generation: \texttt{SymSubTreeGen}($D$, $\bm w_{\mathrm{op}}$, $\bm w_{\mathrm{ft}}$, $\alpha_0,\delta_0$) }
\label{supple-alg:symbolic-tree-generation}

\KwInput{
Operator set $\mathbb O$, features $x_1,\ldots,x_p$, starting depth $D$, operator weight vector
$\bm w_{\mathrm{op}} = (w_{\mathrm{op},1},\ldots, w_{\mathrm{op},|\mathbb O|})$, feature weight vector $\bm w_{\mathrm{ft}} = (w_{\mathrm{ft},1},\ldots, w_{\mathrm{ft},p})$, tree split probability parameters $\alpha_0,\delta_0$.
}

\AlgBlockComment{Initialize current node}

\textbf{Initialize} a node $\zeta$ and set its depth as $m_\zeta= D$.\;

\AlgBlockComment{Decide whether the node is terminal}

\textbf{Draw} split indicator $B_\zeta\sim \mathrm{Bernoulli}(p_{m_{\zeta}} = \alpha_0(1 + m_{\zeta})^{-\delta_0})$.\;

\If{$B_\zeta=0$}{

\AlgBlockComment{Create a terminal feature node}

\textbf{Assign} feature $x_{\ell}$ to $\zeta$ for $\ell\in\{1,\ldots,p\}$ with probability
$w_{\mathrm{ft},\ell}$.\;
}

\Else{

\AlgBlockComment{Create a nonterminal operator node}

\textbf{Assign} operator $o \in \mathbb O$ to $\zeta$ with probability $w_{\mathrm{op},o}$.\;

\textbf{Generate} left child subtree of $\zeta$ as \texttt{SymSubTreeGen}($D + 1$, $\bm w_{\mathrm{op}}$, $\bm w_{\mathrm{ft}}$, $\alpha_0,\delta_0$).\;

\If{$\mathrm{arity}(o) = 2$}{

\AlgBlockComment{Nonterminal binary operator node}

\textbf{Generate} right child subtree of $\zeta$ as \texttt{SymSubTreeGen}($D + 1$, $\bm w_{\mathrm{op}}$, $\bm w_{\mathrm{ft}}$, $\alpha_0,\delta_0$).
}
}
\KwOutput{
A symbolic subtree $T$ rooted at $\zeta$.
}
\end{algorithm}
}

The probability mass function of a symbolic subtree $T$ generated using \hyperref[supple-alg:symbolic-tree-generation]{Algorithm~\ref{supple-alg:symbolic-tree-generation}} is
\begin{align}
\label{eq:symbolic-tree-pmf}
\mathcal{G}(T \mid D,\bm w_{\mathrm{op}},\bm w_{\mathrm{ft}}, \alpha_0, \delta_0) =\prod_{o=1}^{|\mathbb O|}(w_{\mathrm{op}, o})^{\xi_{o}}\prod_{h=1}^{p}(w_{\mathrm{ft}, h})^{\varrho_{h}}\prod_{d=D}^{\infty} \left\{p_{d}^{|\mathcal N_{\mathrm{op}}(T, d)|}(1-p_d)^{|\mathcal{N}_{\mathrm{ft}}(T, d)|}\right\},
\end{align}
with $\bm \xi = (\xi_{1}, \ldots, \xi_{|\mathbb O|})$ and $\bm \varrho = (\varrho_{ 1}, \ldots, \varrho_{p})$ denoting the operator and feature count vectors for $T$, respectively. 
Also, $\mathcal{N}_{\mathrm{op}}(T, d)$ and $\mathcal{N}_{\mathrm{ft}}(T, d)$ are the sets of nonterminal and terminal nodes of $T$ at depth $d$, respectively.

\subsection{Metropolis-Hastings Step for the Sampling of Symbolic Forest \texorpdfstring{$\mathcal T$}{T}}
\label{sec:posterior-sampler-details}

For updating the symbolic forest $\mathcal T$ at each iteration of \hyperref[alg:simple-BayeSymX-posterior-sampling]{\textcolor{BrickRed}{Algorithm \ref*{alg:simple-BayeSymX-posterior-sampling} of the main manuscript}}, we use a \texttt{MH} step treating $\mathrm{JMP}(\mathcal T)$ in~\hyperref[supple-eq:JMP]{\eqref{supple-eq:JMP}} as the target distribution.

\emph{\underline{General description of the \texttt{MH} step}}. Starting from an initial symbolic forest $\mathcal{T}^{(0)}$, we iteratively simulate the transitions $\mathcal{T}^{(\mathrm{iter} - 1)}$ to $\mathcal{T}^{(\mathrm{iter})}$ by updating each symbolic tree $T_j$, $j=1,\ldots, K$. At every iteration, we pass through a sequence of intermediate symbolic forests $\mathcal T^{(\mathrm{iter}-1)} \equiv \mathcal T^{(\mathrm{iter},0)} \to \mathcal T^{(\mathrm{iter},1)} \to \cdots \to \mathcal T^{(\mathrm{iter},K)} \equiv \mathcal T^{(\mathrm{iter})}$ defined in \hyperref[supple-alg:MH-step]{Algorithm \ref{supple-alg:MH-step}}.

{\small
\begin{algorithm}[H]
\caption{Metropolis-Hastings step: \texttt{MH}($\mathcal T^{(\mathrm{iter}-1)}$, $\bm\pi_{\mathrm{move}}$)}
\label{supple-alg:MH-step}

\KwInput{
Current symbolic forest
$\mathcal T^{(\mathrm{iter}-1)}
=
\left(T_1^{(\mathrm{iter}-1)},\ldots,T_K^{(\mathrm{iter}-1)}\right)$
and move probability vector
$\bm\pi_{\mathrm{move}}
=
\left(
\pi_{\mathrm{g}},
\pi_{\mathrm{p}},
\pi_{\mathrm{st}},
\pi_{\mathrm{del}},
\pi_{\mathrm{ins}},
\pi_{\mathrm{cf}},
\pi_{\mathrm{co}}
\right)
\in\mathcal S_7$.
}

\AlgBlockComment{Initialize the intermediate symbolic forest}

\textbf{Set} $\displaystyle
\mathcal T^{(\mathrm{iter},0)}
=
\mathcal T^{(\mathrm{iter}-1)}
=
\left(
T_1^{(\mathrm{iter}-1)},\ldots,T_K^{(\mathrm{iter}-1)}
\right)$.
\;

\AlgBlockComment{Sequentially update each symbolic tree}

\For{$j=1,\ldots,K$}{

\AlgBlockComment{Proposal tree generation}
\textbf{Generate} a candidate symbolic tree $T_j^\star$ with transition probability $q\left(T_j^{(\mathrm{iter}-1)}\to T_j^\star\right)$ by selecting one of the
seven local symbolic tree moves
according to $\bm\pi_{\mathrm{move}}$. The generation steps and corresponding transition probabilities for each move
type are discussed below.
\;

\textbf{Set} $\displaystyle
\mathcal T^{(\mathrm{iter},j),\star}
=
\left(
T_1^{(\mathrm{iter})},\ldots,T_{j-1}^{(\mathrm{iter})},
T_j^\star,
T_{j+1}^{(\mathrm{iter}-1)},\ldots,T_K^{(\mathrm{iter}-1)}
\right)$.
\;

\AlgBlockComment{Accept/Reject the proposed candidate tree}
\textbf{Set} $T_j^{(\mathrm{iter})}=T_j^\star$ with probability
\begin{align}
\label{eq:MH-ratio}
\mathrm{\texttt{MH}}
\left(T_j^{(\mathrm{iter}-1)}\to T_j^\star\right)
=
\min\left\{
1,\,
\frac{
\mathrm{JMP}\left(\mathcal T^{(\mathrm{iter},j),\star}\right)
}{
\mathrm{JMP}\left(\mathcal T^{(\mathrm{iter},j-1)}\right)
}
\frac{
q\left(T_j^\star\to T_j^{(\mathrm{iter}-1)}\right)
}{
q\left(T_j^{(\mathrm{iter}-1)}\to T_j^\star\right)
}
\right\},
\end{align}
otherwise set
$T_j^{(\mathrm{iter})}=T_j^{(\mathrm{iter}-1)}$.\;

\textbf{Set} $\displaystyle
\mathcal T^{(\mathrm{iter},j)}
=
\left(
T_1^{(\mathrm{iter})},\ldots,T_j^{(\mathrm{iter})},
T_{j+1}^{(\mathrm{iter}-1)},\ldots,T_K^{(\mathrm{iter}-1)}
\right)$.
\;
}

\AlgBlockComment{Complete the Metropolis-Hastings step}

\textbf{Set}
$\mathcal T^{(\mathrm{iter})}
=
\mathcal T^{(\mathrm{iter},K)}$.\;

\KwOutput{
Updated symbolic forest $\mathcal T^{(\mathrm{iter})}$.
}

\end{algorithm}
}

\emph{\underline{Generation of candidate symbolic trees and calculation of transition probabilities}}. Here, we describe in detail the generative step for a candidate proposal symbolic tree in \hyperref[supple-alg:MH-step]{Algorithm~\ref{supple-alg:MH-step}} along with the calculation of the respective transition probabilities. This is done by first choosing a valid symbolic move from $\{\mathrm p, \mathrm g, \mathrm{st}, \mathrm{del}, \mathrm{ins}, \mathrm{co}, \mathrm{cf}\}$, then choosing a valid node of the existing symbolic tree to perform that move and finally performing the move itself.

Given a current symbolic tree $T$, the grow and change feature moves can be performed only on any terminal node of $T$; the prune, delete node, and change operator moves can be only performed on any nonterminal node of $T$; and the insert node and subtree replacement moves can be performed on any node of $T$. 

Let $\mathcal A(T)\subseteq \{\mathrm p, \mathrm g, \mathrm{st}, \mathrm{del}, \mathrm{ins}, \mathrm{co}, \mathrm{cf}\}$ denote the set of move types which can be performed on at least one node of $T$, and for $m\in \mathcal A(T)$
let $\mathcal V_m(T)$ denote the set of valid nodes of $T$ on which move $m$ can be performed. A move $m$ is selected with probability proportional to its prescribed weight conditional on $\mathcal A(T)$, namely
\begin{align}
\label{eq:conditional-move-prob}
\mathsf{P}\left(m\mid \mathcal A(T)\right) = \frac{\pi_m}{\sum_{\ell \in \mathcal A(T)}\pi_{\ell}},\quad m\in \mathcal A(T),
\end{align}
and, then a node $\zeta \in\mathcal V_m(T)$ is selected uniformly. Conditional on $(m,\zeta)$, a local modification is proposed according to the corresponding move-specific transition kernel $q_m(T\to T^\star;\zeta)$. Thus, the transition probability is
\begin{align}
\label{eq:move-transition-prob}
q\left(T\to T^\star\right)
=
\mathsf{P}\left(m\mid \mathcal A(T)\right)\,
\frac{1}{|\mathcal V_m(T)|}\,
q_m\left(T\to T^\star;\zeta\right).
\end{align} 

Now, we describe the seven local symbolic tree moves along with their transition kernels $q_m(T\to T^\star;\zeta)$ as follows.

\begin{enumerate}
    \item \underline{\emph{Grow}}.
    The terminal node $\zeta \in \mathcal{N}_{\mathrm{ft}}(T)$ of the current symbolic tree $T$ is replaced by a newly generated non-primitive subtree starting at depth $m_{\zeta}$. This subtree is generated by assigning an operator $o^{\star} \in \mathbb O$ at $\zeta$ according to the operator proposal weight vector $\bm{w}_{\mathrm{op,prop}} \in \mathcal{S}_{|\mathbb O|}$. Subsequently a left child subtree $T^{1, \star}$ is generated if $o^{\star}$ is unary and left and right children subtrees ($T^{1, \star}$ and $T^{2, \star}$) are generated if $o^{\star}$ is binary according to \hyperref[supple-alg:symbolic-tree-generation]{Algorithm~\ref{supple-alg:symbolic-tree-generation}} starting from depth $m_\zeta + 1$, operator proposal weight vector $\bm{w}_{\mathrm{op,prop}} \in \mathcal{S}_{|\mathbb O|}$, feature proposal weight vector $\bm{w}_{\mathrm{ft, prop}}$, and split parameters $(\alpha_0, \delta_0)$. Hence, the grow move transition kernel is
    \begin{align}
    \label{eq:grow-transition-kernel}
    q_{\mathrm{g}}\left(T \to T^{\star}; \zeta\right) = 
    w_{\mathrm{op, prop}, o^\star} \prod_{h = 1}^{\mathrm{arity}(o^\star)}\mathcal{G}(T^{h, \star} \mid m_{\zeta} + 1,\bm w_{\mathrm{op, prop}},\bm w_{\mathrm{ft, prop}}, \alpha_0, \delta_0),
    \end{align}
    with $\mathcal{G}$ as defined in \hyperref[eq:symbolic-tree-pmf]{\eqref{eq:symbolic-tree-pmf}}. The grow move is illustrated in \hyperref[fig:grune]{Figure~\ref{fig:grune} ($x_1\to +(x_2, x_3)$)}, where the reverse of grow is a prune move.

    \item \underline{\emph{Prune}}.
    The nonterminal node $\zeta \in \mathcal{N}_{\mathrm{op}}(T)$ of the current symbolic tree $T$ is collapsed to a terminal node by deleting all its descendants and sampling a new feature $x_\ell$ according to the feature proposal weight vector $\bm{w}_{\mathrm{ft, prop}}\in \mathcal{S}_p$. Hence, the prune move transition kernel is
    \begin{equation}
    \label{eq:prune-transition-kernel}
    q_{\mathrm{p}}\left(T \to T^{\star}; \zeta\right) = w_{\mathrm{ft, prop}, \ell}.
    \end{equation}
    The prune move is illustrated in \hyperref[fig:grune]{Figure~\ref{fig:grune} ($+(x_2, x_3) \to x_1$)}, where the reverse of prune is a grow move.

    \begin{figure}[H]
        \centering
        \includegraphics[width=0.5\linewidth]{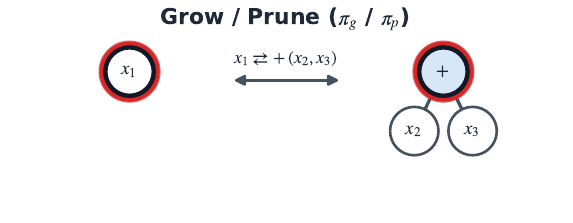}
        \caption{The grow and prune symbolic tree moves.}
        \label{fig:grune}
    \end{figure}

    \item \underline{\emph{Subtree replacement}}.
    For the node $\zeta\in \mathcal{N}(T)$ of the current symbolic tree $T$, 
    the subtree $T_{\zeta}$ of $T$ rooted at $\zeta$ is replaced by a new symbolic subtree $T_{\zeta}^{\star}$. This new subtree $T^{\star}_\zeta$ is generated according to \hyperref[supple-alg:symbolic-tree-generation]{Algorithm~\ref{supple-alg:symbolic-tree-generation}} starting from depth $m_\zeta$, operator proposal weight vector $\bm{w}_{\mathrm{op,prop}} \in \mathcal{S}_{|\mathbb O|}$, feature proposal weight vector $\bm{w}_{\mathrm{ft, prop}}$, and split parameters $(\alpha_0, \delta_0)$. Hence, the subtree replacement transition kernel is
    \begin{equation}
    \label{eq:subtree-replacement-transition-kernel}
    \begin{split}
    &q_{\mathrm{st}}\left(T \to T^{\star}; \zeta\right)= \mathcal{G}(T_\zeta^{\star} \mid m_\zeta,\bm w_{\mathrm{op, prop}},\bm w_{\mathrm{ft, prop}}, \alpha_0, \delta_0),
    \end{split}
    \end{equation}
    with $\mathcal{G}$ as defined in \hyperref[eq:symbolic-tree-pmf]{\eqref{eq:symbolic-tree-pmf}}. The subtree replacement move is illustrated in \hyperref[fig:subtree-repalcement]{Figure \ref{fig:subtree-repalcement}}, where the reverse of subtree replacement is another subtree replacement move.
    \begin{figure}[H]
        \centering
        \includegraphics[width=0.5\linewidth]{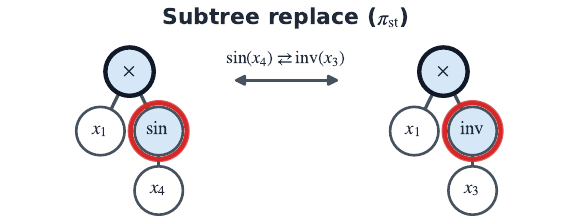}
        \caption{The subtree replacement symbolic tree move.}
        \label{fig:subtree-repalcement}
    \end{figure}

    \item \underline{\emph{Delete node}}.
    The nonterminal node $\zeta \in \mathcal{N}_{\mathrm{op}}(T)$ of the current symbolic tree $T$ is removed and one of its child subtrees is promoted to occupy its position. If the operator originally assigned to $\zeta$ is unary, its only child is promoted with probability one, and if the operator originally assigned to $\zeta$ is binary, one of the two children is promoted uniformly. Hence, the delete node transition kernel is
    \begin{equation}
    \label{eq:delete-node-transition-kernel}
    q_{\mathrm{del}}(T\to T^\star;\zeta)
    =
    \begin{cases}
    1, & \text{if operator at }\zeta\text{ is unary},\\
    \frac{1}{2}, & \text{if operator at }\zeta\text{ is binary}.
    \end{cases}
    \end{equation}
    The delete node move is illustrated in \hyperref[fig:delete-insert-node]{Figure~\ref{fig:delete-insert-node} ($+(x_2, x_3) \to x_2$)}, where the reverse of delete node is an insert node move.

    \item \underline{\emph{Insert node}}.
    For the node $\zeta \in \mathcal{N}(T)$ of the current symbolic tree $T$, let the symbolic subtree of $T$ rooted at $\zeta$ be $T_{\zeta}$. The node $\zeta$ is then assigned an operator $o^{\star} \in \mathbb O$ according to the proposal operator weight vector $\bm{w}_{\mathrm{op, prop}}$. If $o^{\star}$ is unary, then $T_{\zeta}$ is its child. Otherwise, for binary $o^{\star}$, $T_{\zeta}$ is its left child and its right child subtree $(T_\zeta)'$ is generated according to \hyperref[supple-alg:symbolic-tree-generation]{Algorithm~\ref{supple-alg:symbolic-tree-generation}} starting from depth $m_\zeta + 1$, operator proposal weight vector $\bm{w}_{\mathrm{op,prop}} \in \mathcal{S}_{|\mathbb O|}$, feature proposal weight vector $\bm{w}_{\mathrm{ft, prop}}$, and split parameters $(\alpha_0, \delta_0)$. Hence, the insert node transition kernel is
    \begin{equation}
    \label{eq:insert-node-transition-kernel}
    \begin{gathered}
    q_{\mathrm{ins}}\left(T\to T^{\star}; \zeta\right)
    = \begin{cases}
        w_{\mathrm{op, prop},o^{\star}}, & \text{if operator $o^{\star}$ at }\zeta\text{ is unary},\\
        w_{\mathrm{op, prop},o^{\star}}\cdot G', & \text{if operator $o^{\star}$ at }\zeta\text{ is binary},\\
    \end{cases}\\
    G' = \mathcal{G}((T_\zeta)' \mid m_\zeta + 1,\bm w_{\mathrm{op, prop}},\bm w_{\mathrm{ft, prop}}, \alpha_0, \delta_0),
    \end{gathered}
    \end{equation}
    with $\mathcal{G}$ as defined in \hyperref[eq:symbolic-tree-pmf]{\eqref{eq:symbolic-tree-pmf}}. The insert node move is illustrated in \hyperref[fig:delete-insert-node]{Figure~\ref{fig:delete-insert-node} ($x_2\to +(x_2, x_3)$)}, where the reverse of insert node is a delete node move.
    \begin{figure}[H]
        \centering
        \includegraphics[width=0.5\linewidth]{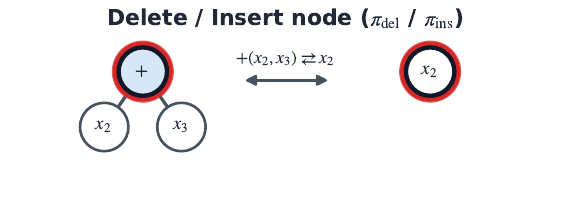}
        \caption{The delete node and insert node symbolic tree moves.}
        \label{fig:delete-insert-node}
    \end{figure}

    \item \underline{\emph{Change operator}}.
    The nonterminal node $\zeta\in \mathcal{N}_{\mathrm{op}}(T)$ of the current symbolic tree $T$ changes its operator from its current assignment $o\in \mathbb O$ to $o^{\star} \in \mathbb O$ according to the proposal operator weight vector $\bm w_{\mathrm{op, prop}}$ conditional on $\mathrm{arity}(o) = \mathrm{arity}(o^\star)$ and $o\neq o^{\star}$. Arity preservation ensures that the existing symbolic subtree topology is unchanged. Hence, the change operator transition kernel is
    \begin{equation}
    \label{eq:change-operator-transition-kernel}
    q_{\mathrm{co}}(T\to T^\star;\zeta) = \frac{w_{\mathrm{op, prop}, o^\star}}
    {\sum_{o'\in \mathbb O:\operatorname{arity}(o')=\operatorname{arity}(o),\,o'\neq o} w_{\mathrm{op, prop}, o'}}.
    \end{equation}
    The change operator move is illustrated in \hyperref[fig:change-operator]{Figure~\ref{fig:change-operator}}, where the reverse of change operator is another change operator move.
    \begin{figure}[H]
        \centering
        \includegraphics[width=0.5\linewidth]{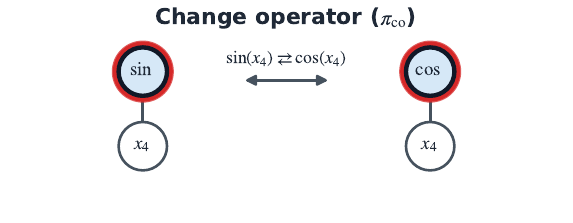}
        \caption{The change operator symbolic tree move.}
        \label{fig:change-operator}
    \end{figure}

    \item \underline{\emph{Change feature}}.
    The terminal node $\zeta \in \mathcal{N}_{\mathrm{ft}}(T)$ of the current symbolic tree $T$ changes its feature from its current assignment $x_a$ to a different feature $x_b$, where the new feature $x_b$ is sampled according to the proposal feature weight vector $\bm{w}_{\mathrm{ft, prop}}$ conditional on $x_a\neq x_b$. Hence, the change feature transition kernel is
    \begin{equation}
    \label{eq:change-feature-transition-kernel}
    q_{\mathrm{cf}}(T\to T^\star;\zeta) = \frac{w_{\mathrm{ft, prop}, b}}{\sum_{\ell\in \{1, \ldots, p\}\setminus \{a\}} w_{\mathrm{ft, prop}, \ell}}.
    \end{equation}
    The change feature move is illustrated in \hyperref[fig:change-feature]{Figure~\ref{fig:change-feature}}, where the reverse of change feature is another change feature move.
    \begin{figure}[H]
        \centering
        \includegraphics[width=0.5\linewidth]{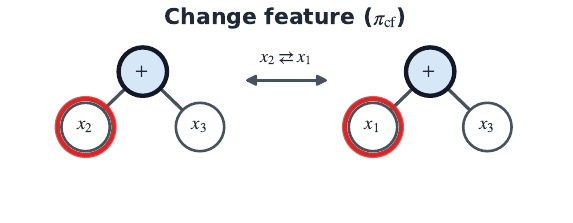}
        \caption{The change feature symbolic tree move.}
        \label{fig:change-feature}
    \end{figure}
\end{enumerate}

\subsection{Default Metropolis-Hastings Configurations}
\label{subsec:default-MH-config}

While the \texttt{MH} tuning parameters $\bm\pi_{\mathrm{move}}$, $\bm w_{\mathrm{op,prop}}$, and $\bm w_{\mathrm{ft,prop}}$ may be chosen to reflect the scientific context or application-specific prior knowledge, we recommend and use uniform proposal weights as broadly applicable defaults unless otherwise mentioned. Specifically, we set 
\begin{equation*}
\begin{gathered}
\bm\pi_{\mathrm{move}}=\left(\tfrac{1}{7},\ldots,\tfrac{1}{7}\right),\quad
\bm w_{\mathrm{op,prop}}=\left(\tfrac{1}{|\mathbb O|},\ldots,\tfrac{1}{|\mathbb O|}\right),\quad
\bm w_{\mathrm{ft,prop}}=\left(\tfrac{1}{p},\ldots,\tfrac{1}{p}\right),
\end{gathered}
\end{equation*}
so that all seven local symbolic tree moves receive equal initial weight, and all operators and features are proposed with equal probability. These uniform choices provide a simple, neutral specification when no operator, feature, or move type is preferred a priori.

\section{Post-\texorpdfstring{\texttt{MCMC}}{MCMC} Symbolic Model Refinement Algorithm}
\label{sec:symbolic-model-refinement}

{\small
\begin{algorithm}[H]
\caption{Post-\texttt{MCMC} symbolic model refinement for a single forest}
\label{alg:post-mcmc-symbolic-model-refinement-single}

\KwInput{
A symbolic forest $\mathcal T=(T_1,\ldots,T_K)\in \mathbb{T}_{\mathbb{O}, p}^{K}$, data $\mathcal D_n=\{(\bm x_i,y_i):i=1,\ldots,n\}$.
}

\AlgBlockComment{Construct symbolic design matrix}

\textbf{Evaluate} the symbolic design matrix $\mathcal{E}(\bm X; \mathcal T) \in \mathbb{R}^{n\times (K+1)}$ with $(i, j)$th entry being 
\begin{equation*}
(\mathcal{E}(\bm X; \mathcal T))_{i, j} = \begin{cases}
    \mathrm{ev}[T_{j-1}](\bm x_i), & j = 2, \ldots, K+1,\\
    1, & j=1.
\end{cases}
\end{equation*}

\AlgBlockComment{Subsets of symbolic components}

\For{each nonempty subset $\mathcal S\subseteq\{1,\ldots,K+1\}$}{

Let $\mathcal{E}_{\mathcal S}(\bm X; \mathcal T)$ be the submatrix of $\mathcal{E}(\bm X; \mathcal T)$ containing the columns indexed by $\mathcal S$.\;

\textbf{Estimate} the outer regression coefficient vector $\displaystyle \widehat{\bm \beta}_{\mathcal S} = \arg\min_{\bm b}\lVert \bm y - \mathcal{E}_{\mathcal{S}}(\bm X; \mathcal T)\bm b\rVert_{2}^{2}$.\;

\textbf{Compute} the residual sum of squares $\displaystyle \texttt{RSS}_{\mathcal S}
=
\|
\bm y-\mathcal{E}_{\mathcal S}(\bm X; \mathcal T)\widehat{\bm\beta}_{\mathcal S}
\|_2^2$.\;

\textbf{Compute} $\displaystyle \texttt{BIC}(\mathcal S) = n \log\left(\frac{\texttt{RSS}_{\mathcal{S}}}{n}\right) + |\mathcal S|\log n$.\;
}

\AlgBlockComment{Select refined symbolic structure}

\textbf{Select} the refined component subset $\displaystyle \widehat{\mathcal S} = \arg\min_{\mathcal S}\texttt{BIC}(\mathcal S)$.\;

\AlgBlockComment{Re-estimate outer regression coefficient vector}

\textbf{Re-estimate} the final outer regression coefficient vector $\bm \beta_{\widehat{\mathcal S}} = \bm \mu^{\star}_{\beta, \widehat{\mathcal S}}(\mathcal T)$ on $\mathcal{E}_{\widehat{\mathcal S}}(\bm X; \mathcal T)$ as
\begin{align*}
\begin{gathered}
\bm \mu^{\star}_{\beta, \widehat{\mathcal S}}(\mathcal T) = \bm \Sigma^{\star}_{\beta, \widehat{\mathcal S}}(\mathcal T)\left(\bm \Sigma_{\beta, \widehat{\mathcal S}}^{-1}\bm \mu_{\beta, \widehat{\mathcal S}} + \mathcal{E}^{\T}_{\widehat{\mathcal S}}(\bm X; \mathcal T)\bm y\right),\\
\bm \Sigma^{\star}_{\beta, \widehat{\mathcal S}}(\mathcal T) = \left(\bm \Sigma_{\bm \beta, \widehat{\mathcal S}}^{-1} + \mathcal{E}^{\T}_{\widehat{\mathcal S}}(\bm X; \mathcal T)\mathcal {E}_{\widehat{\mathcal S}}(\bm X; \mathcal T)\right)^{-1},
\end{gathered}
\end{align*}
where $\bm \mu_{\beta, \widehat{\mathcal S}}$ and $\bm \Sigma_{\beta, \mathcal{\widehat S}}$ represents the subvector and submatrix of $\bm \mu_{\beta}$ and $\bm \Sigma_{\beta}$ indexed by $\widehat{\mathcal S}$.\;

\AlgBlockComment{Form final reported expression}

\textbf{Set} all coefficients indexed by $\widehat{\mathcal S}^{c}$ to zero.\;

\textbf{Simplify} the retained symbolic expression algebraically using \texttt{SymPy}~\citep{meurer2017sympy}.\;

\KwOutput{
The \finaltag\ reported symbolic expression and the effective symbolic forest size $K_{\mathrm{eff}} = |\widehat{\mathcal S}\cap\{2,\ldots,K+1\}|$.
}
\end{algorithm}
}
\newpage
\suppdivision{Theoretical Guarantees for \texorpdfstring{\hierbosss}{HierBOSSS}}
\label{superdiv:theoretical-proofs}

\section{Details of \texorpdfstring{\hyperref[supple-prop:tree-forest-prior-tail-control]{Proposition~\ref{supple-prop:tree-forest-prior-tail-control}}}{Proposition 1}}
\label{sec:tail-control-proof}

\begin{proposition}[Tail control of symbolic tree and forest priors]
\label{supple-prop:tree-forest-prior-tail-control}
For every $d, s\in \mathbb{N}$, the symbolic tree prior over $T\in \mathbb{T}_{\mathbb O, p}$ in~\maineqref[eq:symbolic-tree-prior-Tj-1] satisfies
\begin{equation*}
\begin{gathered}
\Pi_{\mathrm{tree}}\left(\mathrm{d}(T) \geq d\mid \alpha_0, \delta_0, \bm{\alpha}_{\mathrm{op}}, \bm{\alpha}_{\mathrm{ft}}\right) \leq C_1 \exp\left\{-c_1 d\log d\right\},\\
\Pi_{\mathrm{tree}}\left(S(T) > s\mid \alpha_0, \delta_0, \bm{\alpha}_{\mathrm{op}}, \bm{\alpha}_{\mathrm{ft}}\right) \leq C_2\exp\left\{-c_2 s\right\},
\end{gathered}
\end{equation*}
where $C_1, c_1, C_2, c_2$ are positive constants.
Consequently,  for every $s\in \mathbb{N}$, the symbolic forest prior over $\mathcal T=(T_1,\ldots,T_K) \in \mathbb{T}_{\mathbb O, p}^{K}$ in~\maineqref[eq:symbolic-forest-prior] satisfies
\begin{equation*}
\Pi_{\mathrm{forest}, K}\left(S(\mathcal T) > s\mid \alpha_0, \delta_0, \bm \alpha_{\mathrm{op}}, \bm\alpha_{\mathrm{ft}}\right) \leq C_2'\exp\{-c_2's\},
\end{equation*}
where $S(\mathcal T)=\sum_{j=1}^K S(T_j)$ is the symbolic forest operator count and $C_2', c_2'$ are positive constants.
\end{proposition}

\subsection{Auxiliary Lemma for \texorpdfstring{\hyperref[supple-prop:tree-forest-prior-tail-control]{Proposition~\ref{supple-prop:tree-forest-prior-tail-control}}}{Proposition 1}}

\begin{lemma}[Subcritical Galton-Watson tail bound]
\label{lemma:subcritical-GW-exponential-tail}
Let $W$ denote the total progeny of a Galton-Watson process started from one individual, where each individual has offspring distribution $Z=2B$, where $B\sim \mathrm{Bernoulli}(q)$ and $0<q< 1/2$. Then there exist constants $C, c > 0$, depending only on $q$, such that
\begin{equation}
\label{eq:exponential-tail-subcricitical-GW}
\mathsf{P}(W> s) \leq C \exp\{-c s\},
\end{equation}
for $s\geq 1$.
\end{lemma}

\begin{proof}
Let $\phi(z) = \mathsf{E}(z^{Z}) = 1-q+qz^{2}$ be the probability generating function of the offspring distribution. Its mean is $\phi'(1) = \mathsf{E}(Z) = 2q < 1$. So the Galton-Watson process is subcritical. 

We first show that $W$ has a finite exponential moment. Define $\psi(z) := z^{-1}\phi(z)$ for $z> 0$. Since $\phi(1) = 1$, we have $\psi(1) = 1$. Moreover, $\psi'(z) = z^{-2}(z\phi'(z) - \phi(z))$
and therefore
\begin{equation}
\psi'(1) = \phi'(1) - \phi(1) = 2q-1 < 0.
\end{equation}
By continuity, there exist $A > 1$, sufficiently close to $1$ such that, $\psi(A) < 1$, or equivalently $\phi(A) < A$. Consequently, $A / \phi(A) > 1$. Choose $r$ such that, $1<r<\min\{A, A/\phi(A)\}$. Then $r\leq A$ and $r\phi(A) \leq A$.

Let $W_m$ denote the total number of individuals appearing up to an including generation $m$. Thus, $W_0 = 1$ and $W_m \uparrow W$ as $m\to \infty$. Define $F_{m}(r) := \mathsf{E}(r^{W_m})$. Now we show by induction that, $F_{m}(r) \leq A$ for every $m\geq 0$.

For $m=0$, $F_0(r) = r \leq A$. Suppose that, $F_{m}(r) \leq A$ for some $m \geq 0$. Conditional on the number $Z$ of children of the initial ancestor, the descendant subtrees are independent Galton-Watson processes with the same offspring law. Hence, $W_{m+1} = 1 + \sum_{j=1}^{Z}W_{m}^{(j)}$, where $W_{m}^{(1)}, \ldots, W_{m}^{(Z)}$ are independent copies of $W_{m}$, conditionally on $Z$. It follows that
\begin{equation}
F_{m+1}(r) = \mathsf{E}(r^{1 + \sum_{j=1}^{Z}W_{m}^{(j)}}) = r\mathsf{E}(F_m(r)^{Z}) = r\phi(F_m(r)).
\end{equation}
As $\phi$ is increasing on $[0, \infty)$, the induction hypothesis gives 
\begin{equation}
F_{m+1}(r) = r\phi(F_{m}(r)) \leq r\phi(A) \leq A.
\end{equation}
Thus, $F_{m}(r) \leq A$ for every $m\geq 0$.

Since $W_m \uparrow W$ as $m\to \infty$, the monotone convergence theorem yields
\begin{equation}
\mathsf{E}(r^{W}) = \lim_{m\to \infty}\mathsf{E}(r^{W_m}) = \lim_{m\to \infty}F_{m}(r) \leq A<\infty.
\end{equation}
Finally, by Markov's inequality for every $s\geq 1$
\begin{equation}
\mathsf{P}(W>s) = \mathsf{P}(r^{W} > r^{s}) \leq r^{-s}\mathsf{E}(r^{W}) \leq A r^{-s}.
\end{equation}
Taking $C = A$ and $c = \log r > 0$ proves
\begin{equation*}
\boxed{
\mathsf{P}(W>s) \leq C\exp\{-c s\},
}
\end{equation*}
for $s\geq 1$.
\end{proof}
 
\subsection{Proof of \texorpdfstring{\hyperref[supple-prop:tree-forest-prior-tail-control]{Proposition~\ref{supple-prop:tree-forest-prior-tail-control}}}{Proposition 1}}

\begin{proof}{}
Let $T\sim \Pi_{\mathrm{tree}}(\cdot \mid \alpha_0, \delta_0,\bm \alpha_{\mathrm{op}}, \bm \alpha_{\mathrm{ft}})$ as in \maineqref[eq:symbolic-tree-prior-Tj-1]. Fix an integer $d\geq 1$. Consider a particular root-to-depth-$d$ path. For this path to exist in the symbolic tree $T$, the node at depth $0$, the node at depth $1$, and so on up to the node at depth $d-1$, must all split. Since splitting decisions are independent across nodes conditional on their existence, the probability that this particular path survives to depth $d$ is
$$
\prod_{m=0}^{d-1}p_{m} = \prod_{m=0}^{d-1}\alpha_0(1 + m)^{-\delta_0} = \alpha_0^{d}\prod_{m=0}^{d-1}(1+m)^{-\delta_0} = \alpha^{d}_0(d!)^{-\delta_0}.
$$
There are at most $2^{d}$ possible binary root-to-depth-$d$ paths. Thus, by the union bound
\begin{align*}
\begin{split}
\Pi_{\mathrm{tree}}(\mathrm{d}(T) \geq d \mid \alpha_0, \delta_0, \bm\alpha_{\mathrm{op}}, \bm\alpha_{\mathrm{ft}}) &\leq (2\alpha_0)^{d}(d!)^{-\delta_0} \leq (2\alpha_0)^{d}\exp\{\delta_0 d\}d^{-\delta_0 d}\\
&\leq \exp\left\{d\log (2\alpha_0) + \delta_0 d - \delta_0 d\log d\right\}.
\end{split}
\end{align*}

Since, $-\delta_0 d\log d$ dominates the linear term $d\log (2\alpha_0) + \delta_0 d$ as $d\to \infty$, there exist constants $C_1, c_1>0$ such that
\begin{align*}
\boxed{\Pi_{\mathrm{tree}}(\mathrm{d}(T) \geq d\mid \alpha_0, \delta_0, \bm\alpha_{\mathrm{op}}, \bm\alpha_{\mathrm{ft}}) \leq C_1 \exp\left\{-c_1 d \log d\right\},}
\end{align*}
for $d\geq 1$. This proves the depth-tail bound.

Now we derive bounds on the tree and forest operator counts. Since $p_{m} = \alpha_0(1+m)^{-\delta_0} \stackrel{m\to \infty}{\longrightarrow} 0$, there exist an integer $m_0 \geq 0$ and a constant $q<1/2$ such that, $p_{m} \leq q< 1/2$ for every $m\geq m_0$.

For a symbolic tree node $v$ at depth $m$, let $X_v \in \{0, 1, 2\}$ denote its number of children and let $B_v$ denote the corresponding splitting indicator. Thus, $B_v\sim \mathrm{Bernoulli}(p_m)$ and $X_v\leq 2B_v$ almost surely: a terminal (nonsplitting) node has no children , while a nonterminal (splitting) node has either one child or two children. 

For $m\geq m_0$, couple $B_v$ with an independent random variable $I_v\sim \mathrm{Bernoulli}(q)$ so that, $B_v\leq I_v$ almost surely. For example, using independent $U_v\sim \mathrm{Uniform}(0, 1)$, one may set $B_v = \mathds{1}\{U_v\leq p_m\}$ and $I_v = \mathds{1}\{U_v \leq q\}$. Since $p_m \leq q$, this coupling gives $X_v \leq 2B_v \leq 2I_v$ almost surely. Therefore, every descendant subtree rooted at depth $m_0$ is stochastically dominated by a homogeneous Galton-Watson process whose offspring distribution is $Z = 2I$, where $I \sim \mathrm{Bernoulli}(q)$. This domination holds irrespective of the relative probabilities assigned to unary and binary operators.

The number of nodes lying strictly above depth $m_0$ is deterministically bounded by
$$
N_{< m_0} = \sum_{m=0}^{m_0-1}2^{m} = 2^{m_0}-1,
$$
with the convention $N_{<0} = 0$. There are at most $2^{m_0}$ nodes at depth $m_0$. Let $W_1, \ldots, W_{2^{m_0}}$ be independent copies of the total progeny $W$ of the dominating Galton-Watson process, with each $W_{\ell}$ including its initial ancestor. The total number of nodes in the symbolic tree $T$, and hence also its operator count $S(T)$, is stochastically bounded above by $N_{<m_0} + \sum_{\ell=1}^{2^{m_0}}W_{\ell}$.

By~\hyperref[lemma:subcritical-GW-exponential-tail]{Lemma~\ref{lemma:subcritical-GW-exponential-tail}}, $W$ has finite exponential moment that is there exist $\theta > 0$ such that, $M_{\theta} = \mathsf{E}[\exp\{\theta W\}] < \infty$. Consequently, for every $s>0$, Markov's inequality gives
\begin{equation}
\begin{split}
\mathsf{P}\left(N_{<m_0} + \sum_{\ell=1}^{2^{m_0}}W_{\ell} > s\right) &\leq \exp\{-\theta(s - N_{<m_0})\}\mathsf{E}\left[\exp\left\{\theta\sum_{\ell=1}^{2^{m_0}}W_{\ell}\right\}\right]\\
&= \exp\{\theta N_{<m_0}\}M_{\theta}^{2^{m_0}}\exp\{-\theta s\}.
\end{split}
\end{equation}

It follows that, there exist constants $C_2, c_2 > 0$ such that
$$
\boxed{\Pi_{\mathrm{tree}}(S(T) > s\mid \alpha_0, \delta_0, \bm\alpha_{\mathrm{op}}, \bm\alpha_{\mathrm{ft}}) \leq C_2\exp\{-c_2s\},}
$$
for every $s\in \mathbb{N}$.

Now recall that $S(\mathcal T) = \sum_{j=1}^{K} S(T_j)$. If $S(\mathcal T) > s$, then at least one tree must satisfy $S(T_j) > sK^{-1}$. Therefore, $\{S(\mathcal T) > s\} \subseteq \cup_{j=1}^{K}\{S(T_j) > sK^{-1}\}$  and by the union bound
\begin{align*}
\begin{gathered}
\Pi_{\mathrm{forest}, K}(S(\mathcal T) > s \mid \alpha_0, \delta_0, \bm \alpha_{\mathrm{op}}, \bm \alpha_{\mathrm{ft}}) \leq \sum_{j=1}^{K} \Pi_{\mathrm{tree}}(S(T_j) > s K^{-1}\mid \alpha_0, \delta_0, \bm\alpha_{\mathrm{op}}, \bm \alpha_{\mathrm{ft}})\\
\leq K C_2 \exp\{-c_2sK^{-1}\}\\
= C_2'\exp\{-c_2' s\},\\
\boxed{\Pi_{\mathrm{forest}, K}(S(\mathcal T) > s \mid \alpha_0, \delta_0, \bm\alpha_{\mathrm{op}}, \bm\alpha_{\mathrm{ft}})\leq C_2'\exp\{-c_2' s\},}
\end{gathered}
\end{align*}
for $s\in \mathbb N$.
\end{proof}

\newpage
\section{Details of \texorpdfstring{\hyperref[supple-theorem:posterior-contraction]{Theorem~\ref{supple-theorem:posterior-contraction}}}{Theorem 1} and \texorpdfstring{\hyperref[supple-corollary:exact-symbolic-realizability]{Corollary~\ref{supple-corollary:exact-symbolic-realizability}}}{Corollary 1}}
\label{sec:posterior-contraction-supplement}

\begin{theorem}[Concentration under approximate symbolic realizability]
\label{supple-theorem:posterior-contraction}
For fixed $K$, $p$, and $\mathbb O$, grant \hyperref[main-ass:global-symbolic-evaluation-envelope]{\textcolor{BrickRed}{Assumptions \ref*{main-ass:global-symbolic-evaluation-envelope}, \ref*{main-ass:prior-regularity}, and \ref*{main-ass:well-specified} of the main manuscript}}. Then there exists a constant $M>0$ such that
\begin{align}
\label{eq:contraction-2}
\lim_{n\to \infty}\;\mathsf{E}_{\mathbb{P}_0^{n}}\left[\Pi\left(f_{\bm \beta, \mathcal T}\in \mathcal F_{K}: d_{n}(f_{\bm \beta, \mathcal T}, f_0) > M \mathsf r_{n, K} \mid \mathcal{D}_n \right)\right] = 0,
\end{align}
where the symbolic complexity quantity $\mathfrak{C}_{K, S, n}$ and concentration rate $\mathsf{r}_{n, K}$ are
\begin{align}
\label{eq:contraction-1}
\begin{gathered}
\mathfrak{C}_{K, S, n} = S[1 + \log(S + |\mathbb O|)] + [S + K][1 + \log(p+S + K)]\\
\qquad + [K+1][\log \overline{C}_U + c_U S] + [K+2]\log n,\\
n \mathsf r^{2}_{n, K} = \inf_{S\in \mathbb N}\left[\left\{\mathfrak{C}_{K, S, n} + na^{2}_{K, S, n}(f_0)\right\}
\left\{1 + \log\left(\mathfrak{C}_{K, S, n} + na^{2}_{K, S, n}(f_0) + K + p + |\mathbb O|\right)\right\}\right].
\end{gathered}
\end{align}
\end{theorem}

\begin{corollary}[Concentration under exact symbolic realizability]
\label{supple-corollary:exact-symbolic-realizability}
In addition to \hyperref[main-ass:global-symbolic-evaluation-envelope]{\textcolor{BrickRed}{Assumptions \ref*{main-ass:global-symbolic-evaluation-envelope}, \ref*{main-ass:prior-regularity}, and \ref*{main-ass:well-specified} of the main manuscript}}, suppose that $f_0$ admits a finite complexity symbolic representation that is there exists $S_0\in \mathbb{N}$ such that, $f_0 \in \mathcal{F}_{K, S_0}$. Then, for every $S\geq S_0$, $a_{K, S, n}(f_0) = 0$. Consequently, for fixed $K$, $p$, and $\mathbb O$,~\hyperref[eq:contraction-2]{\eqref{eq:contraction-2}} holds with
\begin{equation*}
n\mathsf r_{n, K}^{2} \leq \mathfrak{C}_{K, S_0, n}\left[1 + \log\left(\mathfrak{C}_{K, S_0, n} + K + p +|\mathbb O|\right)\right].
\end{equation*}
\end{corollary}

\subsection{Auxiliary Lemmata for \texorpdfstring{\hyperref[supple-theorem:posterior-contraction]{Theorem~\ref{supple-theorem:posterior-contraction}}}{Theorem 1} and \texorpdfstring{\hyperref[supple-corollary:exact-symbolic-realizability]{Corollary~\ref{supple-corollary:exact-symbolic-realizability}}}{Corollary 1}}
\label{subsec:posterior-contraction-lemmata}

\begin{lemma}[Symbolic forest count]
\label{lemma:symbolic-forest-count}
Let $N_{\mathcal T}(K, S)$ denote the number of possible ordered symbolic forests $\mathcal{T} = (T_1, \ldots, T_K)$ $\in \mathbb{T}^{K}_{\mathbb O, p}$ with total symbolic forest operator count $S(\mathcal{T}) = \sum_{j=1}^{K}S(T_j) \leq S$. Then
\begin{align}
\label{eq:symbolic-forest-count}
\log N_{\mathcal T}(K, S) \leq (2S+K+1)\log 3 + S\log|\mathbb O| + (S+K)\log p.
\end{align}
\end{lemma}

\begin{proof}
We count the number of symbolic forests using the following steps.

\underline{\emph{Counting number of possible ordered forest shapes}}. Each symbolic tree is an ordered rooted tree with nodes of the following three structural types: (i) a terminal node, (ii) a unary nonterminal node with one child, and (iii) a binary nonterminal node with two children. For each tree $T_j$ in the symbolic forest $\mathcal T$, the number of terminal nodes is at most $1$ more than the number of nonterminal nodes that is $|\mathcal N_{\mathrm{ft}}(T_j)| \leq 1 + |\mathcal{N}_{\mathrm{op}}(T_j)|$. Indeed, under the recursive construction of a symbolic tree, replacing a terminal node by a unary operator leaves the number of terminal nodes unchanged, whereas replacing it by a binary operator increases the number of terminal nodes by exactly one. Since the tree begins with a single terminal node, the stated inequality follows. Therefore, the total number of terminal nodes in $\mathcal{T}$ is at most $S(\mathcal{T}) + K \leq S+K$ and the total number of nodes in $\mathcal{T}$ is at most $2S(\mathcal{T}) + K \leq 2S+K$.

Now encode the ordered forest shape by a preorder traversal of the $K$ roots. During this traversal, each node is assigned one of the following structural symbols: $0\to$ terminal, $1\to$ unary nonterminal, and $2\to$ binary nonterminal. The length of this structural code is at most $2S+K$ yielding at most $\sum_{m=0}^{2S+K}3^{m} \leq 3^{2S+K+1}$ possible unique ordered symbolic forest shapes.

\underline{\emph{Assignments of operators and features}}. There are at most $S$ nonterminal nodes. Each nonterminal node receives one operator from $\mathbb O$. Therefore, the number of operator assignments is at most $|\mathbb O|^{S}$. Similarly, for at most $S+K$ terminal nodes, where each terminal node receives one feature from $x_1, \ldots, x_p$, the number of feature assignments is at most $p^{S+K}$.

Combining the forest shape, operator, and feature assignments yields~\hyperref[eq:symbolic-forest-count]{\eqref{eq:symbolic-forest-count}}
\begin{align*}
\boxed{\log N_{\mathcal T}(K, S) \leq (2S+K+1)\log 3 + S\log|\mathbb O| + (S+K)\log p.}
\end{align*}
\end{proof}

\begin{lemma}[Metric entropy of symbolic forests]
\label{lemma:metric-entropy}
For $K, S\in \mathbb{N}$ and $B>0$, define
$$
\mathcal S(K, S, B) := \left\{f_{\bm \beta, \mathcal T}(\cdot) \in \mathcal{F}_K: S(\mathcal T) = \sum_{j=1}^{K}S(T_j) \leq S, \lVert \bm\beta \rVert_{\infty}\leq B\right\}.
$$
Under~\mainassref[ass:global-symbolic-evaluation-envelope], for every $0<\varepsilon < 1$
\begin{align}
\label{eq:metric-entropy}
\begin{aligned}
\log N(\varepsilon,\mathcal S(K,S,B),d_n)
&\le
(2S+K+1)\log 3 + S\log|\mathbb O| + (S+K)\log p\\
&\qquad + (K+1)\left[c_US + \log \overline{C}_U + \log\left(1 + \frac{2B(K+1)}{\varepsilon}\right)\right],
\end{aligned}
\end{align}
where $\overline{C}_U>1$ and $c_U > 0$ are envelope constants.
\end{lemma}

\begin{proof}
Fix a symbolic forest $\mathcal{T} = (T_1, \ldots, T_K)\in \mathbb{T}_{\mathbb O, p}^{K}$ with $S(\mathcal{T}) \leq S$. Let $\mathcal{S}(\mathcal {T}, B)$ denote the coefficient-indexed slice obtained at that particular fixed forest $\mathcal{T}$
\begin{align}
\label{eq:metric-entropy-1}
\mathcal{S}(\mathcal{T}, B) = \left\{f_{\bm \beta, \mathcal{T}}(\cdot) \in \mathcal{F}_K, \lVert \bm \beta\rVert_{\infty} \leq B\right\} \subset \mathcal{S}(K, S, B).
\end{align}
For a coefficient vector $\bm \beta \in [-B, B]^{K+1}$, define $f_{\bm \beta, \mathcal T}(\bm x) := \beta_0 + \sum_{j=1}^{K}\beta_{j}\mathrm{ev}[T_j](\bm x)$. Let $\bm \beta, \bm \beta' \in [-B, B]^{K+1}$. For each design point $\bm x_i \in \mathfrak{X} \subset \mathbb{R}^{p}$ compact
\begin{align}
\label{eq:metric-entropy-2}
f_{\bm \beta, \mathcal T}(\bm x_i) - f_{\bm \beta', \mathcal T}(\bm x_i) = (\beta_0 - \beta_0') + \sum_{j = 1}^K(\beta_j - \beta'_j)\mathrm{ev}[T_j](\bm x_i).
\end{align}
Define $\Phi_{i0}:=1$ and $\Phi_{ij} := \mathrm{ev}[T_j](\bm x_i)$, for all $i=1,\ldots, n$ and $j=1,\ldots,K$. By \mainassref[ass:global-symbolic-evaluation-envelope] and an application of triangle and Cauchy-Schwarz inequalities in~\hyperref[eq:metric-entropy-2]{\eqref{eq:metric-entropy-2}}, we have
\begin{align}
\label{eq:metric-entropy-3}
\begin{split}
\left|f_{\bm \beta, \mathcal T}(\bm x_i) - f_{\bm \beta', \mathcal T}(\bm x_i)\right| &\leq \left|\sum_{\ell=0}^{K}\Phi_{i\ell}(\beta_{\ell} -  \beta'_{\ell})\right| 
\leq \sum_{\ell=0}^{K}|\Phi_{i\ell}||\beta_{\ell} - \beta'_{\ell}|\\
&\leq U_S\sum_{\ell=0}^{K}|\beta_{\ell} - \beta'_{\ell}|
\leq U_{S}\sqrt{K+1}\lVert \bm \beta- \bm \beta'\rVert_{2}.
\end{split}
\end{align}
In~\hyperref[eq:metric-entropy-3]{\eqref{eq:metric-entropy-3}}, squaring, averaging over $i=1,\ldots,n$, and taking the square root yields
\begin{align}
\label{eq:metric-entropy-4}
d_{n}(f_{\bm \beta, \mathcal T}, f_{\bm \beta', \mathcal T}) \leq U_{S}\sqrt{K+1}\lVert \bm \beta- \bm \beta'\rVert_{2}.
\end{align}
Therefore, if the coefficient cube $[-B, B]^{K+1}$ is covered in Euclidean norm by balls of radius $r_{\varepsilon} = \varepsilon U^{-1}_S(K+1)^{-1/2}$, then using the display above in~\hyperref[eq:metric-entropy-4]{\eqref{eq:metric-entropy-4}}, the corresponding functions are covered in $d_n$-distance by balls of radius $\varepsilon$.

We construct a grid on the Euclidean cube $[-B, B]^{K+1}$ by dividing each coordinate interval $[-B, B]$ into subintervals of length $r_{\varepsilon}(K+1)^{-1/2}$. Thus, any point $\bm \beta \in [-B, B]^{K+1}$ lies at a distance at most $r_{\varepsilon}$ from one of the corresponding grid points. This yields that the Euclidean cube $[-B, B,]^{K+1}$ can be covered by at most $(1 + 2B\sqrt{K+1}r^{-1}_{\varepsilon})^{K+1}$ balls of radius $r_\varepsilon$. Hence, for each fixed symbolic forest $\mathcal{T}$, the coefficient-indexed slice $\mathcal{S}(\mathcal T, B)$ in~\hyperref[eq:metric-entropy-1]{\eqref{eq:metric-entropy-1}} can be covered by at most $(1 + 2BU_{S}(K+1)\varepsilon^{-1})^{K+1}$ balls of radius $\varepsilon$ in the $d_n$-metric. 

By~\hyperref[lemma:symbolic-forest-count]{Lemma~\ref{lemma:symbolic-forest-count}}, the number of possible discrete symbolic forests $\mathcal{T}\in \mathbb{T}^{K}_{\mathbb O, p}$ with total symbolic forest operator count $S(\mathcal{T})$ at most $S$ is at most $N_{\mathcal T}(K, S) \leq 3^{2S+K+1}|\mathbb O|^{S}p^{S+K}$. Taking the union of $\mathcal{S}(\mathcal{T}, B)$ over all discrete symbolic forests gives
\begin{align}
\label{eq:metric-entropy-5}
\begin{split}
    N(\varepsilon, \mathcal{S}(K, S, B), d_n) &\leq N_{\mathcal T}(K, S)\left(1 + \frac{2B U_S(K+1)}{\varepsilon}\right)^{K+1}\\
    &\leq 3^{2S+K+1}|\mathbb O|^{S}p^{S+K}\left(1 + \frac{2B U_S(K+1)}{\varepsilon}\right)^{K+1}.
\end{split}
\end{align}

By~\mainassref[ass:global-symbolic-evaluation-envelope]
$$
U_S\leq \overline C_U\exp\{c_U S\}\implies 1 + 2B(K+1)\varepsilon^{-1}U_S \leq 1 + 2B(K+1)\varepsilon^{-1}\overline C_U\exp\{c_U S\}.
$$
Since $\exp\{c_U S\} \geq 1$, $\overline C_U\exp\{c_U S\} \geq 1$, and $2B(K+1)\varepsilon^{-1}\overline C_U \exp\{c_U S\} \leq 2B(K+1)\varepsilon^{-1}\\\overline C_U\exp\{c_U S\}$, we have
\begin{align}
\label{eq:metric-entropy-6}
1 + \frac{2B(K+1)}{\varepsilon}U_S \leq \overline{C}_U\left(1 + \frac{2B(K+1)}{\varepsilon}\right)\exp\{c_U S\}.
\end{align}
Taking logarithm in the above display in~\hyperref[eq:metric-entropy-6]{\eqref{eq:metric-entropy-6}} and substituting in the bound in~\hyperref[eq:metric-entropy-5]{\eqref{eq:metric-entropy-5}} yields~\hyperref[eq:metric-entropy]{\eqref{eq:metric-entropy}}
\begin{align*}
\boxed{
\begin{aligned}
\log N(\varepsilon,\mathcal S(K,S,B),d_n)
&\le
(2S+K+1)\log 3 + S\log|\mathbb O| + (S+K)\log p\\
&\qquad + (K+1)\left[c_US + \log\overline{C}_U + \log\left(1 + \frac{2B(K+1)}{\varepsilon}\right)\right].
\end{aligned}
}
\end{align*}
{}
\end{proof}

\begin{lemma}[Dirichlet predictive lower bound]
\label{lemma:Dirichlet-predictive-lower-bound}
Let $d\in \mathbb{N}$ and $\bm \eta = (\eta_1, \ldots, \eta_d)$, where $0<a \leq \eta_\ell \leq A<\infty$ for all $\ell=1,\ldots, d$. Consider the $d$-tuple of nonnegative integers $\mathfrak{n} = (\mathfrak n_1, \ldots, \mathfrak n_{d})$ such that, $\sum_{\ell=1}^{d}\mathfrak{n}_{\ell} = M$. Then
\begin{align}
\label{eq:Dirichlet-predictive-lower-bound}
\frac{\mathfrak{B}(\bm \eta + \mathfrak{n})}{\mathfrak{B}(\bm \eta)} \geq \left(\frac{a}{Ad + M}\right)^{M},
\end{align}
where $\mathfrak{B}(\bm \eta) = \prod_{\ell=1}^{d}[\Gamma(\eta_{\ell})]/\Gamma(\sum_{\ell\in [d]}\eta_{\ell})$ is the multivariate Beta function.
\end{lemma}

\begin{proof}
For $d\in \mathbb{N}$, let $z_1, \ldots, z_M\in \{1, \ldots, d\}$ be any fixed sequence of labels whose count vector is $\mathfrak{n} = (\mathfrak n_1, \ldots, \mathfrak n_d)$. Thus, for each label $\ell\in \{1, \ldots, d\}$, $\mathfrak n_\ell = \# \{r\in \{1, \ldots, M\}:z_r = \ell\}$. Let $\bm w = (w_1, \ldots, w_d) \sim \mathrm{Dir}(\bm \eta)$. Conditional on $\bm w$, suppose the labels are drawn independently with $\mathsf{P}(z_r=\ell\mid \bm w) = w_{\ell}$. Then, the conditional probability of observing the fixed sequence $z_1, \ldots, z_M$ is $\prod_{r=1}^{M}w_{z_r}$. Since label $\ell$ appears $\mathfrak n_\ell$ times, we have $\prod_{r=1}^{M}w_{z_r} = \prod_{\ell=1}^{d}w_{\ell}^{\mathfrak n_{\ell}}$.

The Dirichlet density of $\bm w$ is $p(\bm w\mid \bm \eta) = \mathfrak{B}(\bm \eta)^{-1}\prod_{\ell=1}^{d}w_{\ell}^{\eta_{\ell}-1}$. Therefore, the integrated probability of the fixed sequence is
$$
\int_{\Delta_{d-1}}\prod_{\ell=1}^{d}w_{\ell}^{\mathfrak{n}_{\ell}}\frac{1}{\mathfrak{B}(\bm \eta)}\prod_{\ell=1}^{d}w_{\ell}^{\eta_{\ell}-1}d\bm w = \frac{1}{\mathfrak{B}(\bm \eta)}\int_{\Delta_{d-1}}\prod_{\ell=1}^{d}w_{\ell}^{\eta_{\ell} + \mathfrak{n}_{\ell} - 1}d\bm w  = \frac{\mathfrak{B}(\bm\eta + \mathfrak{n})}{\mathfrak{B}(\bm \eta)},
$$
where $\Delta_{d-1} = \{(z_1, \ldots, z_d)\in \mathbb{R}^{d}: z_j \geq 0, j\in \{1, \ldots, d\}, \sum_{j=1}^{d}z_j = 1\}$ is the $(d-1)$-dimensional simplex. Hence
\begin{align}
\label{eq:Dirichlet-LB-1}
\frac{\mathfrak{B}(\bm \eta + \mathfrak{n})}{\mathfrak{B}(\bm \eta)} = \int_{\Delta_{d-1}}\prod_{r=1}^{M}w_{z_r}d\mathrm{Dir}(\bm w; \bm\eta).
\end{align}
So it remains to obtain a lower bound of the right-hand-side in~\hyperref[eq:Dirichlet-LB-1]{\eqref{eq:Dirichlet-LB-1}}. We use the sequential predictive representation of the Dirichlet-multinomial model. For each $r=1,\ldots, M$, define $N_{\ell}(r-1) := \# \{q<r:z_q = \ell\}$. That is, $N_{\ell}(r-1)$ is the number of previous occurrences of label $\ell$ before step $r$.

After observing the first $r-1$ labels, the posterior distribution of $\bm w$ is
$$
\bm w\mid z_1, \ldots, z_{r-1} \sim \mathrm{Dir}(\eta_1 + N_1(r-1), \ldots, \eta_{d} + N_{d}(r-1)).
$$
Therefore, the posterior predictive probability of observing the required next label $z_r$ is
\begin{align}
\label{eq:Dirichlet-LB-2}
\mathsf{P}(z_r\mid z_1, \ldots, z_{r-1}) = \frac{\eta_{z_r} + N_{z_r}(r-1)}{\sum_{\ell=1}^{d}\eta_{\ell} + r-1}.
\end{align}
Now we bound the above probability in~\hyperref[eq:Dirichlet-LB-2]{\eqref{eq:Dirichlet-LB-2}} from below. For the numerator, $\eta_{z_r} + N_{z_r}(r-1) \geq \eta_{z_r} \geq a$. For the denominator, since $\eta_{\ell} \leq A$ for all $\ell =1, \ldots, d$, we have $\sum_{\ell=1}^{d}\eta_{\ell} + r-1 \leq Ad + M$. Thus
\begin{align*}
\boxed{
    \begin{aligned}
        &\mathsf{P}(z_1, \ldots, z_M) = \prod_{r=1}^{M}\mathsf{P}(z_r\mid z_1, \ldots, z_{r-1}) \geq \left(\frac{a}{Ad+M}\right)^{M}\\
        &\qquad \qquad \implies \frac{\mathfrak{B}(\bm\eta + \mathfrak{n})}{\mathfrak{B}(\bm \eta)} \geq \left(\frac{a}{Ad + M}\right)^{M},
    \end{aligned}
}
\end{align*}
yielding~\hyperref[eq:Dirichlet-predictive-lower-bound]{\eqref{eq:Dirichlet-predictive-lower-bound}}.
\end{proof}

\begin{lemma}[Prior mass of a fixed symbolic forest]
\label{lemma:prior-mass-forest}
Let $\mathcal{T}^{\dagger} = (T_1^{\dagger}, \ldots, T_{K}^{\dagger})\in \mathbb{T}_{\mathbb O, p}^{K}$ be a fixed ordered symbolic forest with $S(\mathcal T^{\dagger}) = \sum_{j=1}^{K}S(T_{j}^{\dagger}) \leq S$. Under~\mainassref[ass:prior-regularity], the prior mass over $\mathcal{T}^{\dagger}$ satisfies
\begin{align}
\label{eq:prior-mass-forest}
\begin{split}
&\Pi_{\mathrm{forest}, K}(\mathcal{T}^{\dagger}\mid \alpha_0, \delta_0, \bm \alpha_{\mathrm{op}}, \bm \alpha_{\mathrm{ft}})\\
&\qquad \geq \exp\left\{-C_4\left[S(1 + \log(S+|\mathbb O|)) + (S+K)(1 + \log(p+S+K))\right]\right\},
\end{split}
\end{align}
where $C_4>0$.
\end{lemma}

\begin{proof}
The prior mass over the fixed symbolic forest $\mathcal{T}^{\dagger}$ is
\begin{align}
\label{eq:prior-mass-1}
\begin{split}
    &\Pi_{\mathrm{forest}, K}(\mathcal{T}^{\dagger}\mid \alpha_0, \delta_0, \bm\alpha_{\mathrm{op}}, \bm\alpha_{\mathrm{ft}})\\
    &\qquad = \prod_{j=1}^{K}\left[\frac{\mathfrak{B}(\bm \alpha_{\mathrm{op}} + \bm \xi_{j})}{\mathfrak{B}(\bm \alpha_{\mathrm{op}})}\frac{\mathfrak{B}(\bm \alpha_{\mathrm{ft}} + \bm \varrho_{j})}{\mathfrak{B}(\bm \alpha_{\mathrm{ft}})}\prod_{m=0}^{\infty}\left\{p_{m}^{|\mathcal{N}_{\mathrm{op}}(T_j^{\dagger}, m)|}(1-p_m)^{|\mathcal{N}_{\mathrm{ft}}(T_j^{\dagger}, m)|}\right\}\right],
\end{split}
\end{align}
where $|\mathcal{N}_{\mathrm{op}}(T_j^{\dagger}, m)|$ is the number of nonterminal nodes of $T_{j}^{\dagger}$ at depth m, $|\mathcal{N}_{\mathrm{ft}}(T_j^{\dagger}, m)|$ is the number of terminal nodes of $T_{j}^{\dagger}$ at depth m, $\bm \xi_{j} = (\xi_{j, 1}, \ldots, \xi_{j, |\mathbb O|})$, and $\bm \varrho_{j} = (\varrho_{j, 1}, \ldots, \varrho_{j, p})$, for $j=1,\ldots, K$. Also, $\xi_{j, o}$ is the number of times the operator $o\in \mathbb O$ appears in the symbolic expression represented by $T_j^{\dagger}$ and $\varrho_{j, h}$ is the number of times the feature $x_h$ appears in the symbolic expression represented by $T_{j}^{\dagger}$. We obtain lower bounds for each of the three contributions on the right-hand-side of~\hyperref[eq:prior-mass-1]{\eqref{eq:prior-mass-1}}.

\underline{\emph{Lower bound for the split probability contributions}}. 
The total split probability contributions in \hyperref[eq:prior-mass-1]{\eqref{eq:prior-mass-1}} over all symbolic trees in $\mathcal{T}^{\dagger}$ is
\begin{align}
\label{eq:prior-mass-split-probability}
\begin{split}
\Pi_{\mathrm{split}}(\mathcal{T}^{\dagger}, \alpha_0, \delta_0) &= 
\prod_{j=1}^{K}\prod_{m= 0}^{\infty}\left\{p_{m}^{|\mathcal{N}_{\mathrm{op}}(T_j^{\dagger}, m)|}(1-p_m)^{|\mathcal{N}_{\mathrm{ft}}(T_j^{\dagger}, m)|}\right\} \\
&= \underbrace{\prod_{j=1}^{K}\prod_{m=0}^{\infty}p_{m}^{|\mathcal{N}_{\mathrm{op}}(T_j^{\dagger}, m)|}}_{\text{(i)}}
\underbrace{\prod_{j=1}^{K}\prod_{m=0}^{\infty}(1-p_m)^{|\mathcal{N}_{\mathrm{ft}}(T_j^{\dagger}, m)|}}_{\text{(ii)}}.
\end{split}
\end{align}
The number of terms in the product in (i) in~\hyperref[eq:prior-mass-split-probability]{\eqref{eq:prior-mass-split-probability}} is equal to the number of nonterminal nodes in the forest $\mathcal{T}^{\dagger}$, which is at most $S$. Recall, $p_m = \alpha_0(1+m)^{-\delta_0}$. The depth of any node in the forest $\mathcal{T}^{\dagger}$, can be at most $S$, yielding a lower bound $p_m \geq \alpha_0(1+S)^{-\delta_0}$. Combining the lower bound on $p_m$ and the upper bound on the number of terms in (i) in~\hyperref[eq:prior-mass-split-probability]{\eqref{eq:prior-mass-split-probability}}, yields
\begin{equation}
\label{eq:prior-mass-split-probability-lower-bound-1}
\prod_{j=1}^{K}\prod_{m=0}^{\infty}p_{m}^{|\mathcal{N}_{\mathrm{op}}(T_j^{\dagger}, m)|} \geq \exp\{S\log \alpha_0 - \delta_0 S\log(1+S)\}.
\end{equation}
Now, following the argument in proof of~\hyperref[lemma:symbolic-forest-count]{Lemma~\ref{lemma:symbolic-forest-count}}, the total number of terminal nodes in $\mathcal{T}^{\dagger}$ is at most $S(\mathcal{T}^{\dagger}) + K \leq S+K$. Thus, the number of terms in the product in (ii) in~\hyperref[eq:prior-mass-split-probability]{\eqref{eq:prior-mass-split-probability}} is equal to the number of terminal nodes in the forest $\mathcal{T}^{\dagger}$, which is at most $S + K$. Combining this with $1-p_m \geq 1 - \alpha_0$, yields the following lower bound
\begin{equation}
\label{eq:prior-mass-split-probability-lower-bound-2}
\prod_{j=1}^{K}\prod_{m=0}^{\infty}(1-p_{m})^{|\mathcal{N}_{\mathrm{ft}}(T_j^{\dagger}, m)|} \geq \exp\{(S + K)\log (1-\alpha_0)\}.
\end{equation}
Combining the lower bounds in \hyperref[eq:prior-mass-split-probability-lower-bound-1]{\eqref{eq:prior-mass-split-probability-lower-bound-1}} and \hyperref[eq:prior-mass-split-probability-lower-bound-2]{\eqref{eq:prior-mass-split-probability-lower-bound-2}}, and substituting in \hyperref[eq:prior-mass-split-probability]{\eqref{eq:prior-mass-split-probability}}
\begin{align}
\label{eq:prior-mass-split-probability-lower-bound}
\begin{split}
\Pi_{\mathrm{split}}(\mathcal{T}^{\dagger}, \alpha_0, \delta_0) 
\geq \exp\left\{-S\log \frac{1}{\alpha_0} - (S + K)\log \frac{1}{(1-\alpha_0)} - \delta_0 S\log(1+S)\right\}.
\end{split}   
\end{align}

\underline{\emph{Lower bound on the operator and feature assignment contributions}}. 
The total probability contributions of the operator and feature assignments in \hyperref[eq:prior-mass-1]{\eqref{eq:prior-mass-1}} over all symbolic trees in $\mathcal{T}^{\dagger}$ are respectively
\begin{align}
\label{eq:prior-mass-operators-features-1}
\begin{split}
\Pi_{\mathrm{op}}(\mathcal{T}^{\dagger}, \bm \alpha_{\mathrm{op}}) = 
\prod_{j=1}^{K} \frac{\mathfrak{B}(\bm \alpha_{\mathrm{op}} + \bm \xi_{j})}{\mathfrak{B}(\bm \alpha_{\mathrm{op}})}, \quad
\Pi_{\mathrm{ft}}(\mathcal{T}^{\dagger}, \bm \alpha_{\mathrm{ft}}) = 
\prod_{j=1}^{K} \frac{\mathfrak{B}(\bm \alpha_{\mathrm{ft}} + \bm \varrho_{j})}{\mathfrak{B}(\bm \alpha_{\mathrm{ft}})}.
\end{split}
\end{align}
By an application of~\hyperref[lemma:Dirichlet-predictive-lower-bound]{Lemma \ref{lemma:Dirichlet-predictive-lower-bound}}
\begin{align}
\label{eq:prior-mass-operators-features-2}  
\frac{\mathfrak{B}(\bm \alpha_{\mathrm{op}} + \bm \xi_{j})}{\mathfrak{B}(\bm \alpha_{\mathrm{op}})} \geq \left(\frac{a_{\mathrm{op}}}{A_{\mathrm{op}}|\mathbb O| + \lVert\bm \xi_j\rVert_{1}}\right)^{\lVert \bm \xi_j\rVert_{1}}, \quad \frac{\mathfrak{B}(\bm \alpha_{\mathrm{ft}} + \bm \varrho_{j})}{\mathfrak{B}(\bm \alpha_{\mathrm{ft}})} \geq \left(\frac{a_{\mathrm{ft}}}{A_{\mathrm{ft}}p + \lVert \bm\varrho_j\rVert_{1}}\right)^{\lVert\bm \varrho_j\rVert_{1}},
\end{align}
for each $j=1, \ldots, K$, where $\lVert\bm \xi_j\rVert_{1} = \sum_{o=1}^{|\mathbb O|} \xi_{j, o}$ and $\lVert \bm \varrho_j\rVert_{1} = \sum_{h=1}^{p} \varrho_{j, h}$, respectively. Now, we obtain the lower bounds of $\Pi_{\mathrm{op}}(\mathcal{T}^{\dagger}, \bm \alpha_{\mathrm{op}})$ and $\Pi_{\mathrm{ft}}(\mathcal{T}^{\dagger}, \bm \alpha_{\mathrm{ft}})$ by noting that $\sum_{j = 1}^{K} \lVert\bm \xi_j\rVert_{1}$ and $\sum_{j=1}^{K} \lVert\bm \varrho_j\rVert_{1}$ are the number of nonterminal and terminal nodes respectively, that is 
$\lVert\bm \xi_{j}\rVert_{1} \leq \sum_{j'=1}^{K} \lVert\bm \xi_{j'}\rVert_{1} \leq S$ and $\lVert\bm \varrho_{j}\rVert_{1} \leq \sum_{j'=1}^{K} \lVert\bm \varrho_{j'}\rVert_{1} \leq S + K$, for all $j = 1, \ldots, K$. Combining these with \hyperref[eq:prior-mass-operators-features-2]{\eqref{eq:prior-mass-operators-features-2}} and substituting back in \hyperref[eq:prior-mass-operators-features-1]{\eqref{eq:prior-mass-operators-features-1}}, we get
\begin{align}
\label{eq:prior-mass-operators-features-3}
\begin{split}
\Pi_{\mathrm{op}}(\mathcal{T}^{\dagger}, \bm\alpha_{\mathrm{op}}) 
&= \prod_{j=1}^{K} \frac{\mathfrak{B}(\bm \alpha_{\mathrm{op}} + \bm\xi_{j})}{\mathfrak{B}(\bm\alpha_{\mathrm{op}})} 
\geq \prod_{j=1}^{K} \left(\frac{a_{\mathrm{op}}}{A_{\mathrm{op}}|\mathbb O| + \lVert\bm \xi_j\rVert_{1}}\right)^{\lVert\bm \xi_j\rVert_{1}} 
\geq \prod_{j=1}^{K} \left(\frac{a_{\mathrm{op}}}{A_{\mathrm{op}}|\mathbb O| + S }\right)^{\lVert\bm \xi_j\rVert_{1}}
\\ &
\geq 
\left(\frac{a_{\mathrm{op}}}{A_{\mathrm{op}}|\mathbb O| + S}\right)^{S}
= \exp \left\{-S \log \frac{1}{a_{\mathrm{op}}} - S \log(A_{\mathrm{op}}|\mathbb O| + S)\right\},
\end{split}
\end{align}
and
\begin{align}
\label{eq:prior-mass-operators-features-4}
\begin{split}
\Pi_{\mathrm{ft}}(\mathcal{T}^{\dagger}, \bm \alpha_{\mathrm{ft}}) 
&= \prod_{j=1}^{K} \frac{\mathfrak{B}(\bm \alpha_{\mathrm{ft}} + \bm \xi_{j})}{\mathfrak{B}(\bm \alpha_{\mathrm{ft}})} 
\geq \prod_{j=1}^{K} \left(\frac{a_{\mathrm{ft}}}{A_{\mathrm{ft}}p + \lVert\bm \varrho_j\rVert_{1}}\right)^{\lVert\bm \varrho_j\rVert_{1}} 
\geq \prod_{j=1}^{K} \left(\frac{a_{\mathrm{ft}}}{A_{\mathrm{ft}}p + S + K }\right)^{\lVert\bm \varrho_j\rVert_{1}}
\\ &
\geq 
\left(\frac{a_{\mathrm{ft}}}{A_{\mathrm{ft}}p + S + K}\right)^{S + K}
= \exp \left\{-(S + K) \log \frac{1}{a_{\mathrm{ft}}} - (S + K) \log(A_{\mathrm{ft}}p + S + K)\right\}.
\end{split}
\end{align}

\underline{\emph{Final prior lower bound}}.
Combining \hyperref[eq:prior-mass-split-probability-lower-bound]{\eqref{eq:prior-mass-split-probability-lower-bound}}, \hyperref[eq:prior-mass-operators-features-3]{\eqref{eq:prior-mass-operators-features-3}}, and \hyperref[eq:prior-mass-operators-features-4]{\eqref{eq:prior-mass-operators-features-4}}, and substituting back in \hyperref[eq:prior-mass-1]{\eqref{eq:prior-mass-1}}, we get
\begin{align}
\label{eq:prior-mass-operators-features-5}
\begin{split}
&\log \Pi_{\mathrm{forest}, K}(\mathcal{T}^{\dagger}\mid \alpha_0, \delta_0, \bm\alpha_{\mathrm{op}}, \bm\alpha_{\mathrm{ft}})\\
&\qquad = 
\log \Pi_{\mathrm{split}}(\mathcal{T}^{\dagger}, \alpha_0, \delta_0) + 
\log \Pi_{\mathrm{op}}(\mathcal{T}^{\dagger}, \bm\alpha_{\mathrm{op}}) + \log \Pi_{\mathrm{ft}}(\mathcal{T}^{\dagger}, \bm\alpha_{\mathrm{ft}})\\ 
&\qquad\geq -S \log\left(\frac{1}{\alpha_0 a_{\mathrm{op}}}\right) - (S+K)\log \left(\frac{1}{(1-\alpha_0)a_{\mathrm{ft}}}\right) -\delta_0 S \log(1 + S) \\
&\qquad\qquad -S \log (A_{\mathrm{op}}|\mathbb O| + S) -(S + K) \log (A_{\mathrm{ft}}p + S + K).
\end{split}
\end{align}
Observe that for $C_3 > 0$
\begin{align*}
\begin{gathered}
-S\log(\alpha_0 a_{\mathrm{op}}) \leq C_3S \leq C_3S(1 + \log(S + |\mathbb O|)),\delta_0 S \log(1+S) \leq C_3S(1 + \log(S+ |\mathbb O|)),\\
S\log(A_{\mathrm{op}}|\mathbb O| + S) \leq C_3S(1 + \log(S+ |\mathbb O|)),\\
-(S+K)\log((1-\alpha_0)a_{\mathrm{ft}}) \leq C_3(S+K) \leq C_3(S+K)(1 + \log(p+S+K)),\\
(S+K)\log(A_{\mathrm{ft}}p + S+ K) \leq C_3(S+K)(1 + \log(p+S+K)).
\end{gathered}
\end{align*}
Using these in~\hyperref[eq:prior-mass-operators-features-5]{\eqref{eq:prior-mass-operators-features-5}} gives
\begin{align*}
\boxed{
\begin{aligned}
&\Pi_{\mathrm{forest}, K}(\mathcal{T}^{\dagger}\mid \alpha_0, \delta_0, \bm\alpha_{\mathrm{op}}, \bm\alpha_{\mathrm{ft}})\\
&\qquad \geq \exp\left\{-C_4\left[S(1 + \log(S+|\mathbb O|)) + (S+K)(1 + \log(p+S+K))\right]\right\},
\end{aligned}
}
\end{align*}
yielding~\hyperref[eq:prior-mass-forest]{\eqref{eq:prior-mass-forest}}.
\end{proof}

\begin{lemma}[Prior mass of a fixed symbolic model]
\label{lemma:prior-mass-symbolic-model}
Fix $(f^{\dagger}, (\sigma^\dagger)^2)\in \mathcal{F}_{K, S^{\dagger}} \times \mathbb{R}^{+}$, where $f^{\dagger}(x)\equiv f_{\bm \beta^{\dagger}, \mathcal{T}^{\dagger}}(x) = \beta_0^{\dagger} + \sum_{j=1}^{K}\beta_{j}^{\dagger}\mathrm{ev}[T_{j}^{\dagger}](\bm x)$. Under \hyperref[main-ass:global-symbolic-evaluation-envelope]{\textcolor{BrickRed}{Assumptions \ref*{main-ass:global-symbolic-evaluation-envelope} and \ref*{main-ass:prior-regularity} of the main manuscript}}, for sufficiently small $0<\varepsilon <1$
\begin{align}
\label{eq:prior-mass-symbolic-model}
\begin{split}
&\Pi(d_{n}(f_{\bm \beta, \mathcal{T}^{\dagger}}, f_{\bm \beta^{\dagger}, \mathcal{T}^{\dagger}})\leq \varepsilon, |\sigma^{2} - (\sigma^\dagger)^2| \leq \varepsilon)\\
&\qquad \geq \exp\Bigg\{-C_5\Bigg[S^{\dagger}(1 + \log(S^{\dagger} + |\mathbb O|)) + (S^{\dagger} + K)(1 + \log(p+S^{\dagger} + K))\\
&\qquad\qquad + (K+1)(\log \overline{C}_U + c_U S^{\dagger}) + (K+2)\log(1/\varepsilon)\Bigg]\Bigg\},
\end{split}
\end{align}
with $C_5>0$.
\end{lemma}

\begin{proof}
From~\hyperref[eq:metric-entropy-4]{\eqref{eq:metric-entropy-4}}, we have $d_{n}(f_{\bm \beta, \mathcal{T}^{\dagger}}, f_{\bm \beta^{\dagger}, \mathcal{T}^{\dagger}}) \leq U_{S^{\dagger}} \sqrt{K+1}\lVert \bm\beta-\bm\beta^{\dagger}\rVert_{2}$. Hence, if $\lVert \bm \beta-\bm \beta^{\dagger}\rVert_{2} \leq \varepsilon U^{-1}_{S^{\dagger}}(K+1)^{-1/2}$, then $d_{n}(f_{\bm \beta, \mathcal{T}^{\dagger}}, f_{\bm \beta^{\dagger}, \mathcal{T}^{\dagger}}) \leq \varepsilon$. Thus, the event
\begin{align}
\label{eq:prior-mass-symbolic-model-1}
\begin{split}
&\left\{\mathcal{T} = \mathcal{T}^{\dagger}, \lVert \bm \beta-\bm \beta^{\dagger}\rVert_{2}\leq \frac{\varepsilon}{U_{S^{\dagger}}\sqrt{K+1}}, |\sigma^{2} - (\sigma^\dagger)^2| \leq \varepsilon\right\}\\
&\qquad \subseteq \left\{d_{n}(f_{\bm \beta, \mathcal{T}^{\dagger}}, f_{\bm \beta^{\dagger}, \mathcal{T}^{\dagger}}) \leq \varepsilon, |\sigma^{2} - (\sigma^\dagger)^2| \leq \varepsilon\right\}.
\end{split}
\end{align}
Following~\hyperref[eq:prior-mass-symbolic-model-1]{\eqref{eq:prior-mass-symbolic-model-1}}

\begin{align}
\label{eq:prior-mass-symbolic-model-2}
\begin{split}
&\Pi(d_{n}(f_{\bm \beta, \mathcal{T}^{\dagger}}, f_{\bm \beta^{\dagger}, \mathcal{T}^{\dagger}})\leq \varepsilon, |\sigma^{2} - (\sigma^\dagger)^2| \leq \varepsilon) \\
&\geq \Pi\left(\mathcal{T} = \mathcal{T}^{\dagger}, \lVert \bm \beta-\bm \beta^{\dagger}\rVert_{2}\leq \frac{\varepsilon}{U_{S^{\dagger}}\sqrt{K+1}}, |\sigma^{2} - (\sigma^\dagger)^2| \leq \varepsilon\right)\\
&= \Pi_{\mathrm{forest}, K}\left(\mathcal{T}^{\dagger}\mid  \alpha_0, \delta_0, \bm \alpha_{\mathrm{op}}, \bm \alpha_{\mathrm{ft}}\right)\\
&\qquad \times \mathrm{NIG}_{K+1}\left( \lVert \bm\beta-\bm\beta^{\dagger}\rVert_{2}\leq \frac{\varepsilon}{U_{S^{\dagger}}\sqrt{K+1}}, |\sigma^{2} - (\sigma^\dagger)^2| \leq \varepsilon\;\Bigg|\; \bm\mu_{\beta}, \bm\Sigma_{\beta}, \nu, \lambda\right).
\end{split}
\end{align}

Now, we lower bound the NIG prior component on the right-hand-side of~\hyperref[eq:prior-mass-symbolic-model-2]{\eqref{eq:prior-mass-symbolic-model-2}}. The joint prior density of $(\bm \beta, \sigma^{2})$ is
$$
\mathrm{NIG}_{K+1}({\bm \beta}, \sigma^2\mid \bm \mu_{\beta}, \bm \Sigma_{\beta}, \nu, \lambda) \equiv \mathrm{N}_{K+1}\left(\bm \beta\mid \bm \mu_{\beta}, \sigma^{2}\bm \Sigma_{\beta}\right)\mathrm{IG}\left(\sigma^{2}\;\Bigg|\; \frac{\nu}{2}, \frac{\lambda}{2}\right),
$$
which is continuous and strictly positive at $(\bm \beta^{\dagger}, (\sigma^\dagger)^2)$ using $\bm \Sigma_{\beta}\succ 0$. Therefore, there exist constants $r_0 > 0$ and $c_{0} > 0$ such that, $\mathrm{NIG}_{K+1}(\bm \beta, \sigma^{2} \mid \bm\mu_{\beta}, \bm\Sigma_{\beta}, \nu, \lambda) \geq c_0$, whenever $\lVert \bm\beta-\bm \beta^{\dagger}\rVert_{2} \leq r_0$ and $|\sigma^{2}- (\sigma^\dagger)^2| \leq r_0$. Taking $\varepsilon > 0$ sufficiently small so that, $\varepsilon \leq r_0$ and $\varepsilon U^{-1}_{S^{\dagger}}(K+1)^{-1/2} \leq r_0$. Then
\begin{align}
\label{eq:prior-mass-symbolic-model-3}
\begin{split}
&\mathcal{B}_{\beta}(\varepsilon) \times \mathcal{I}_{\sigma}(\varepsilon)\\
&\qquad = \left\{\bm \beta \in \mathbb{R}^{K+1}:\lVert \bm \beta-\bm \beta^{\dagger}\rVert_{2} \leq \frac{\varepsilon}{U_{S^{\dagger}}\sqrt{K+1}}\right\} \times \left\{\sigma^{2} \in \mathbb{R}^{+}: |\sigma^{2} - (\sigma^\dagger)^2| \leq \varepsilon\right\}\\
&\qquad \qquad \subseteq \left\{(\bm \beta, \sigma^{2})\in \mathbb{R}^{K+1}\times \mathbb{R}^{+}: \mathrm{NIG}_{K+1}(\bm\beta, \sigma^{2} \mid \bm\mu_{\beta}, \bm\Sigma_{\beta}, \nu, \lambda) \geq c_0\right\}.
\end{split}
\end{align}
Therefore
\begin{align}
\label{eq:prior-mass-symbolic-model-4}
\mathrm{NIG}_{K+1}(\mathcal{B}_{\beta}(\varepsilon)\times \mathcal{I}_{\sigma}(\varepsilon) \mid \bm\mu_{\beta}, \bm\Sigma_{\beta}, \nu, \lambda) \geq c_0 \cdot \mathrm{Vol}_{K+1}(\mathcal{B}_{\beta}(\varepsilon))\cdot \mathrm{Leb}(\mathcal{I}_{\sigma}(\varepsilon)),
\end{align}
where $\mathrm{Vol}_{K+1}(\mathcal{B}_{\beta}(\varepsilon))$ and $\mathrm{Leb}(\mathcal{I}_{\sigma}(\varepsilon))$ are the Lebesgue measures of $\mathcal{B}_{\beta}(\varepsilon)$ and $\mathcal{I}_{\sigma}(\varepsilon)$, respectively. Trivially, $\mathrm{Leb}(\mathcal{I}_{\sigma}(\varepsilon)) = 2\varepsilon$ and $\mathrm{Vol}_{K+1}(\mathcal{B}_{\beta}(\varepsilon)) = v_{K+1}\varepsilon^{K+1}(U_{S^{\dagger}}\sqrt{K+1})^{-K-1}$, where $v_{K+1}$ is the volume of the unit ball in $\mathbb{R}^{K+1}$. Using these observations in~\hyperref[eq:prior-mass-symbolic-model-4]{\eqref{eq:prior-mass-symbolic-model-4}}, we get
\begin{align}
\label{eq:prior-mass-symbolic-model-5}
\mathrm{NIG}_{K+1}(\mathcal{B}_{\beta}(\varepsilon)\times \mathcal{I}_{\sigma}(\varepsilon) \mid \bm \mu_{\beta}, \bm \Sigma_{\beta}, \nu, \lambda) \geq c_0' \varepsilon^{K+2}U_{S^{\dagger}}^{-K-1}, \quad c_0' = 2c_0 v_{K+1}(K+1)^{-\frac{K+1}{2}}.
\end{align}
Following~\mainassref[ass:global-symbolic-evaluation-envelope], $U_{S^{\dagger}} \leq \overline C_{U}\exp\{c_U S^{\dagger}\}$, and from~\hyperref[eq:prior-mass-symbolic-model-5]{\eqref{eq:prior-mass-symbolic-model-5}} we have
\begin{align}
\label{eq:prior-mass-symbolic-model-6}
\begin{split}
\mathrm{NIG}_{K+1}(\mathcal{B}_{\beta}(\varepsilon)\times \mathcal{I}_{\sigma}(\varepsilon) \mid \bm\mu_{\beta}, \bm\Sigma_{\beta}, \nu, \lambda) &\geq c_0' \varepsilon^{K+2} \exp\left\{-(K+1)\left[\log \overline C_{U} + c_US^{\dagger}\right]\right\}.
\end{split}
\end{align}
Finally, combining~\hyperref[eq:prior-mass-symbolic-model-6]{\eqref{eq:prior-mass-symbolic-model-6}} with~\hyperref[eq:prior-mass-forest]{\eqref{eq:prior-mass-forest}} in~\hyperref[lemma:prior-mass-forest]{Lemma~\ref{lemma:prior-mass-forest}} and substituting in~\hyperref[eq:prior-mass-symbolic-model-2]{\eqref{eq:prior-mass-symbolic-model-2}}, we get
\begin{align*}
\boxed{
\begin{aligned}
&\Pi(d_{n}(f_{\bm \beta, \mathcal{T}^{\dagger}}, f_{\bm \beta^{\dagger}, \mathcal{T}^{\dagger}})\leq \varepsilon, |\sigma^{2} - (\sigma^\dagger)^2| \leq \varepsilon)\\
&\qquad \geq \exp\Bigg\{-C_5\Bigg[S^{\dagger}(1 + \log(S^{\dagger} + |\mathbb O|)) + (S^{\dagger} + K)(1 + \log(p+S^{\dagger} + K))\\
&\qquad\qquad  + (K+1)(\log \overline{C}_U + c_U S^{\dagger}) + (K+2)\log(1/\varepsilon)\Bigg]\Bigg\},
\end{aligned}
}
\end{align*}
yielding~\hyperref[eq:prior-mass-symbolic-model]{\eqref{eq:prior-mass-symbolic-model}}, where $C_5>0$.
\end{proof}

\begin{lemma}[Gaussian Kullback-Leibler ($\mathrm{KL}$) and variance control]
\label{lemma:KL-control}
For fixed design points $\bm x_1$, $\ldots, \bm x_n$, consider $\mathbb{P}^{n}_{f_\star, \sigma_\star^2} = \otimes_{i=1}^{n}\mathrm{N}(f_{\star}(\bm x_i), \sigma_\star^{2})$ and $\mathbb{P}^{n}_{f, \sigma^2} = \otimes_{i=1}^{n}\mathrm{N}(f(\bm x_i), \sigma^{2})$.
For sufficiently small $\varepsilon>0$, if $d_n(f, f_{\star}) \leq \varepsilon$ and $|\sigma^2-\sigma_\star^{2}| \leq \varepsilon$, then there exist constants $C_6, C_6'>0$ such that
\begin{align}
\label{eq:KL-control}
\mathrm{KL}\left( \mathbb P^{n}_{f_\star, \sigma_\star^2}\parallel \mathbb P^{n}_{f, \sigma^2}\right) \leq C_6 n\varepsilon^{2}, \quad \mathsf{V}_{\mathbb P^{n}_{f_{\star},\sigma_{\star}^{2}}}\left(\log\frac{p^{n}_{f, \sigma^2}}{p^{n}_{f_\star, \sigma_\star^2}}\right) \leq C_6' n \varepsilon^2,
\end{align}
where $\mathrm{KL}(P_1 \mid \mid P_2)$ denotes the Kullback-Leibler ($\mathrm{KL}$) divergence between probability measures $P_1$ and $P_2$.
\end{lemma}

\begin{proof}
The Kullback-Leibler ($\mathrm{KL}$) divergence between $\mathbb{P}^{n}_{f_\star, \sigma_{\star}^2}$ and $\mathbb{P}^{n}_{f, \sigma^2}$ follows from standard Gaussian $\mathrm{KL}$ formula

\begin{align}
\label{eq:KL-control-1}
\begin{split}
    \mathrm{KL}\left( \mathbb P^{n}_{f_\star, \sigma_\star^2}\parallel \mathbb P^{n}_{f, \sigma^2}\right) &= \sum_{i=1}^{n}\mathrm{KL}\left(\mathrm{N}(f_{\star}(\bm x_i), \sigma_{\star}^{2})\parallel \mathrm{N}(f(\bm x_i), \sigma^{2})\right)\\
    &= \frac{1}{2}\sum_{i=1}^{n}\left[\log\left(\frac{\sigma^2}{\sigma_{\star}^{2}}\right) + \frac{\sigma_{\star}^{2}}{\sigma^2} - 1 + \frac{(f(\bm x_i) - f_{\star}(\bm x_i))^2}{\sigma^{2}}\right]\\
    &= \frac{1}{2\sigma^{2}}\sum_{i=1}^{n}(f(\bm x_i) - f_{\star}(\bm x_i))^2 + \frac{n}{2}\left(\log\left(\frac{\sigma^2}{\sigma_{\star}^{2}}\right) + \frac{\sigma_{\star}^{2}}{\sigma^2} - 1\right)\\
    &= \frac{n d_n^{2}(f, f_{\star})}{2\sigma^2} + \frac{n}{2}\left(\log\left(\frac{\sigma^2}{\sigma_{\star}^{2}}\right) + \frac{\sigma_{\star}^{2}}{\sigma^2} - 1\right).
\end{split}
\end{align}
Now if $d_{n}(f, f_{\star}) \leq \varepsilon$, then $(2\sigma^2)^{-1}nd_{n}^{2}(f, f_{\star}) \leq (2\sigma^2)^{-1}n\varepsilon^{2}$, which upper bounds the first term in the last equality of~\hyperref[eq:KL-control-1]{\eqref{eq:KL-control-1}}. Define $h(r) := \log r + r^{-1} - 1$, for $r>0$. Note that, $h(1) = 0 = h'(1)$. Therefore, by Taylor's theorem, $h(r) \leq \mathfrak{C}(r-1)^{2}$. Taking $r = \sigma^{2}/\sigma_{\star}^{2}$ and using $|\sigma^{2} - \sigma_{\star}^{2}| \leq \varepsilon$, we have
\begin{align}
\label{eq:KL-control-2}
\frac{n}{2}\left(\log\left(\frac{\sigma^2}{\sigma_{\star}^{2}}\right) + \frac{\sigma_{\star}^{2}}{\sigma^2} - 1\right) \leq \frac{n}{2}\mathfrak{C}\sigma_{\star}^{-4}\varepsilon^{2}.
\end{align}
Using~\hyperref[eq:KL-control-2]{\eqref{eq:KL-control-2}} and $(2\sigma^2)^{-1}nd_{n}^{2}(f, f_{\star}) \leq (2\sigma^2)^{-1}n\varepsilon^{2}$ in~\hyperref[eq:KL-control-1]{\eqref{eq:KL-control-1}} for sufficiently small $\varepsilon \leq \sigma_{\star}/2$, we get
$$
\boxed{\mathrm{KL}\left( \mathbb P^{n}_{f_\star, \sigma_\star^2}\parallel \mathbb P^{n}_{f_{\beta, \mathcal T}, \sigma^2}\right) \leq C_6 n\varepsilon^{2},}
$$
which proves the $\mathrm{KL}$ bound in~\hyperref[eq:KL-control]{\eqref{eq:KL-control}}.

Now we bound $\mathsf{V}_{\mathbb P^{n}_{f_{\star},\sigma_{\star}^{2}}}\left(\log\left({p^{n}_{f, \sigma^2}}/{p^{n}_{f_\star, \sigma_\star^2}}\right)\right)$. Working under $\mathbb{P}^{n}_{f_{\star}, \sigma^{2}_{\star}}$ that is $y_{i} = f_{\star}(\bm x_i) + \sigma_{\star}Z_i$ with $Z_{i} \sim \mathrm{N}(0, 1)$ independently and identically for all $i=1,\ldots, n$. Write $\Delta_{i} = f_{\star}(\bm x_i) - f(\bm x_i)$. The per-observation $\log$-likelihood ratio is
\begin{align}
\label{eq:KL-control-3}
\mathcal{L}_i = \log\frac{\mathrm{N}(y_i\mid f_{\star}(\bm x_i), \sigma_{\star}^{2})}{\mathrm{N}(y_i\mid f(\bm x_i), \sigma^2)} = -\frac{1}{2}\log \frac{\sigma_{\star}^{2}}{\sigma^2} - \frac{(y_i - f_{\star}(\bm x_i))^2}{2\sigma_{\star}^2} + \frac{(y_i - f(\bm x_i))^2}{2\sigma^2}.
\end{align}
Substituting $y_i = f_{\star}(\bm x_i) + \sigma_{\star}Z_i$ in~\hyperref[eq:KL-control-3]{\eqref{eq:KL-control-3}}, we obtain
\begin{align}
\label{eq:KL-control-4}
\mathcal{L}_i = -\frac{1}{2}\log \frac{\sigma_{\star}^{2}}{\sigma^2} + \frac{\Delta_i^2}{2\sigma^2} + \frac{\sigma_{\star}}{\sigma^2}\Delta_iZ_i + \frac{1}{2}\left(\frac{\sigma_{\star}^{2}}{\sigma^2} - 1\right)Z_i^2.
\end{align}
Note that, $\mathsf{E}(Z_i) = 0 = \mathsf{E}(Z_i^2-1)$, $\mathsf{V}(Z_i) = 1$, $\mathsf{V}(Z_i^2-1) = 2$, and $\mathrm{cov}(Z_i, Z_i^2-1) = \mathsf{E}(Z_i^3) - \mathsf{E}(Z_i) = 0$. Write the last term in \hyperref[eq:KL-control-4]{\eqref{eq:KL-control-4}} above as $b(Z_i^2-1) + b$, where $b =(\sigma_{\star}^{2}/\sigma^2-1)/2$. Then
\begin{align}
\label{eq:KL-control-5}
\mathcal{L} = \sum_{i=1}^{n}\mathcal{L}_i = \mathfrak{C}' + \sum_{i=1}^{n}\frac{\sigma_{\star}\Delta_i}{\sigma^2}Z_i + b\sum_{i=1}^{n}(Z_i^2-1), \quad \mathfrak{C}' = \sum_{i=1}^{n}\frac{\Delta_i^2}{2\sigma^2} - \frac{n}{2}\log \frac{\sigma_{\star}^{2}}{\sigma^2} + nb.
\end{align}
By independence across observations and $\mathrm{cov}(Z_i, Z_i^{2}-1) = 0$, from~\hyperref[eq:KL-control-5]{\eqref{eq:KL-control-5}} we have
\begin{align}
\label{eq:KL-control-6}
    \mathsf{V}_{\mathbb P^{n}_{f_{\star}, \sigma_{\star}^2}}(\mathcal{L}) = \sum_{i=1}^{n}\frac{\sigma_{\star}^2\Delta_i^2}{\sigma^{4}}\mathsf{V}(Z_i) + b^{2}\sum_{i=1}^{n}\mathsf{V}(Z_i^{2}-1) = \frac{\sigma_{\star}^2}{\sigma^4}\sum_{i=1}^{n}\Delta_i^2 + 2nb^2.
\end{align}
Observe that, $\sum_{i=1}^{n}\Delta_i^{2} = nd_{n}^{2}(f, f_{\star}) \leq n\varepsilon^{2}$. Also for $\varepsilon \leq \sigma_{\star}^{2}/2$ and using $|\sigma^2 - \sigma_{\star}^{2}| \leq \varepsilon$, we have $2n b^{2} = n(\sigma_{\star}^{2}/\sigma^2-1)^2/2 \leq 2n\varepsilon^{2}/\sigma_{\star}^{4}$ and $\sigma_{\star}^{2}/\sigma^{4} \leq 4/\sigma_{\star}^{2}$. Combining these observations with~\hyperref[eq:KL-control-6]{\eqref{eq:KL-control-6}}, we get
\begin{align*}
\begin{split}
\boxed{\mathsf{V}_{\mathbb P^{n}_{f_{\star},\sigma_{\star}^{2}}}\left(\log\frac{p^{n}_{f, \sigma^2}}{p^{n}_{f_\star, \sigma_\star^2}}\right) \leq \frac{4}{\sigma_{\star}^{2}}n\varepsilon^2 + \frac{2}{\sigma_{\star}^{4}}n\varepsilon^2 = C_{6}'n\varepsilon^{2},}
\end{split}
\end{align*}
yielding the variance bound in~\hyperref[eq:KL-control]{\eqref{eq:KL-control}}.
\end{proof}

\begin{lemma}[Kullback-Leibler ($\mathrm{KL}$) prior mass]
\label{lemma:KL-prior-mass}
Consider that, $\mathcal{D}_n$ follows $y_i = f_0(\bm x_i) + \epsilon_i$, where $\epsilon_i \sim \mathrm{N}(0, \sigma_0^2)$ independently for $i=1,\ldots, n$ and $\sigma_0^2 > 0$. Hence, the true law is $\mathbb{P}_{0}^{n} \equiv \mathbb{P}^{n}_{f_0, \sigma_0^2} = \otimes_{i=1}^{n}\mathrm{N}(f_0(\bm x_i), \sigma_0^{2})$. Define the Kullback-Leibler ($\mathrm{KL}$) neighborhood for sufficiently large $C_{\mathrm{KL}}>0$
\begin{align}
\label{eq:KL-neighborhood}
\begin{split}
\mathcal{B}_{\mathrm{KL}, n} &:= \Bigg\{(f_{\bm \beta, \mathcal{T}}, \sigma^2)\in \mathcal{F}_K\times \mathbb{R}^{+}: \mathrm{KL}\left(\mathbb{P}_{0}^{n}\parallel \mathbb{P}_{f_{\bm \beta, \mathcal T}, \sigma^2}^{n}\right) \leq C_{\mathrm{KL}}n\varepsilon_{n, K}^{2},
\\ &\qquad \qquad \qquad \qquad \qquad \qquad  
\mathsf{V}_{\mathbb{P}^{n}_{0}}\left(\log \frac{p_0^{n}}{p^{n}_{f_{\bm \beta, \mathcal T}, \sigma^2}}\right) \leq C_{\mathrm{KL}}n\varepsilon_{n, K}^{2}\Bigg\},
\end{split}
\end{align}
Then under \hyperref[main-ass:global-symbolic-evaluation-envelope]{\textcolor{BrickRed}{Assumptions \ref*{main-ass:global-symbolic-evaluation-envelope} and \ref*{main-ass:prior-regularity} of the main manuscript}}, there exist $C_7>0$ such that
\begin{align}
\label{eq:KL-prior-mass-1}
\Pi(\mathcal{B}_{\mathrm{KL}, n}) \geq \exp\left\{-C_7 n\varepsilon_{n, K}^{2}\right\},
\end{align}
where $\varepsilon_{n, K}^{2} = \inf_{S\in \mathbb{N}}\left\{a^{2}_{K, S, n}(f_0) + n^{-1}\mathfrak{C}_{K, S, n}\right\}$ with $\mathfrak{C}_{K, S, n}$ defined in \hyperref[supple-theorem:posterior-contraction]{Theorem~\ref{supple-theorem:posterior-contraction}}.
\end{lemma}

\begin{proof}
Let $S^{\dagger}\in \mathbb{N}$ be an approximately optimal symbolic forest operator count satisfying $a^{2}_{K, S^{\dagger}, n}(f_0) + n^{-1}\mathfrak{C}_{K, S^{\dagger}, n} \leq 2\varepsilon_{n, K}^{2}$. 
Therefore, as $a^{2}_{K, S^{\dagger}, n}(f_0) \leq 2\varepsilon_{n, K}^{2}$,
there exist $f^{\dagger}(\bm x)\equiv f_{\bm \beta^{\dagger}, \mathcal{T}^{\dagger}}(\bm x) = \beta_0^{\dagger} + \sum_{j=1}^{K}\beta_{j}^{\dagger}\mathrm{ev}[ T_{j}^{\dagger}](\bm x)\in \mathcal{F}_{K, S^{\dagger}}$ such that, $d_{n}(f^{\dagger}, f_0)\leq \sqrt{2}\varepsilon_{n, K}$.
Using \hyperref[lemma:prior-mass-symbolic-model]{Lemma~\ref{lemma:prior-mass-symbolic-model}} with $\varepsilon = \varepsilon_{n, K}$ and $\sigma^{\dagger} = \sigma_0$, we have
\begin{align}
\label{eq:KL-prior-mass-2}
\begin{split}
&\Pi(d_{n}(f_{\bm \beta, \mathcal{T}}, f_{\bm \beta^{\dagger}, \mathcal{T}^{\dagger}})\leq  \varepsilon_{n, K}, |\sigma^{2} - \sigma_0^2| \leq  \varepsilon_{n, K})\\
&\qquad \geq \Pi(d_{n}(f_{\bm \beta, \mathcal{T}^{\dagger}}, f_{\bm \beta^{\dagger}, \mathcal{T}^{\dagger}})\leq  \varepsilon_{n, K}, |\sigma^{2} - \sigma_0^2| \leq  \varepsilon_{n, K})\\
&\qquad \geq \exp\left\{-C_5\left[\mathcal{A}(K, S^{\dagger}, |\mathbb O|, p) + (K+2)\log(1/\varepsilon_{n, K})\right]\right\},
\end{split}
\end{align}
where
\begin{align}
\label{eq:KL-prior-mass-3}
\begin{split}
\mathcal{A}(K, S^{\dagger}, |\mathbb O|, p) &= S^{\dagger}[1 + \log(S^{\dagger} + |\mathbb O|)] + [S^{\dagger} + K][1 + \log(p+S^{\dagger} + K)]\\
&\qquad + [K+1][\log \overline{C}_U+ c_U S^{\dagger}].
\end{split}
\end{align} 
Note that, $\mathfrak{C}_{K, S^{\dagger}, n} = \mathcal{A}(K, S^{\dagger}, |\mathbb O|, p) + (K+2)\log n$. Therefore
\begin{align}
\label{eq:KL-prior-mass-3.1}
\mathcal{A}(K, S^\dagger, |\mathbb O|, p) \leq \mathfrak{C}_{K, S^{\dagger}, n} \leq 2n\varepsilon_{n, K}^{2}.
\end{align}
It remains to control $(K+2)\log(1/\varepsilon_{n, K})$ in~\hyperref[eq:KL-prior-mass-2]{\eqref{eq:KL-prior-mass-2}}.

Again, $(K+2)\log n \leq \mathfrak{C}_{K, S^{\dagger}, n} \leq 2n\varepsilon_{n, K}^{2}$ which yields $\varepsilon_{n, K} \geq (2n)^{-1/2}[(K+2)\log n]^{1/2}$. This implies that for large $n$, $\log(1/\varepsilon_{n, K}) \leq [\log 2 + \log n - \log(K+2) - \log \log n]/2 \leq \log n$, leading to
\begin{align}
\label{eq:KL-prior-mass-3.2}
(K+2)\log(1/\varepsilon_{n, K}) \leq (K+2)\log n \leq \mathfrak{C}_{K, S^{\dagger}, n} \leq 2n\varepsilon_{n, K}^{2}.
\end{align}
Therefore, using the bounds on $\mathcal{A}(K, S^{\dagger}, |\mathbb O|, p)$ in~\hyperref[eq:KL-prior-mass-3.1]{\eqref{eq:KL-prior-mass-3.1}} and $(K+2)\log(1/\varepsilon_{n, K})$ in~\hyperref[eq:KL-prior-mass-3.2]{\eqref{eq:KL-prior-mass-3.2}}, from~\hyperref[eq:KL-prior-mass-2]{\eqref{eq:KL-prior-mass-2}} we get
\begin{align}
\label{eq:KL-prior-mass-4}
\begin{split}
&\Pi(d_{n}(f_{\bm \beta, \mathcal{T}}, f_{\bm \beta^{\dagger}, \mathcal{T}^{\dagger}})\leq \varepsilon_{n, K}, |\sigma^{2} - \sigma_0^2| \leq \varepsilon_{n, K})\geq \exp\left\{-4C_5n\varepsilon_{n, K}^{2}\right\}.
\end{split}
\end{align}
Now define $\mathcal E_n := \left\{(f_{\bm \beta, \mathcal T}, \sigma^2)\in \mathcal{F}_K\times \mathbb{R}^{+}:d_{n}(f_{\bm \beta, \mathcal{T}}, f_{\bm \beta^{\dagger}, \mathcal{T}^{\dagger}})\leq \varepsilon_{n, K}, |\sigma^{2} - \sigma_0^2| \leq \varepsilon_{n, K}\right\}$. On $\mathcal{E}_n$, using the triangle inequality, we have
\begin{align*}
\begin{split}
d_{n}(f_{\bm \beta, \mathcal T}, f_0) &\leq d_{n}(f_{\bm \beta, \mathcal T}, f_{\bm \beta^{\dagger}, \mathcal{T}^{\dagger}}) + d_n(f_{\bm \beta^{\dagger}, \mathcal{T}^{\dagger}}, f_0) \leq \varepsilon_{n, K} + \sqrt{2}\varepsilon_{n, K}.
\end{split}
\end{align*} 
For $n$ sufficiently large, there exist $c_7 > 0$ such that $d_{n}(f_{\bm \beta, \mathcal T},f_0) \leq c_7\varepsilon_{n, K}$ and $|\sigma^{2} - \sigma_0^{2}| \leq c_7\varepsilon_{n, K}$. Therefore using \hyperref[lemma:KL-control]{Lemma~\ref{lemma:KL-control}}
\begin{align}
\label{eq:KL-prior-mass-5}
\mathrm{KL}\left( \mathbb P^{n}_0\parallel \mathbb P^{n}_{f_{\bm \beta, \mathcal T}, \sigma^2}\right) \leq C_{\mathrm{KL}}n\varepsilon_{n, K}^{2}, \quad \mathsf{V}_{\mathbb P^{n}_{0}}\left(\log\frac{p^{n}_{f_{\bm \beta, \mathcal T}, \sigma^2}}{p^{n}_{0}}\right) \leq C_{\mathrm{KL}} n \varepsilon_{n, K}^2,
\end{align}
where $C_{\mathrm{KL}} = \max\{C_6(c_7)^2, C_6'(c_7)^2\} > 0$. From~\hyperref[eq:KL-prior-mass-5]{\eqref{eq:KL-prior-mass-5}}, it follows that $\mathcal{E}_n \subseteq \mathcal{B}_{\mathrm{KL}, n}$. Hence using~\hyperref[eq:KL-prior-mass-4]{\eqref{eq:KL-prior-mass-4}}
\begin{align*}
\boxed{\Pi(\mathcal{B}_{\mathrm{KL}, n}) \geq \exp\left\{-C_7 n\varepsilon_{n, K}^{2}\right\},}
\end{align*}
yielding~\hyperref[eq:KL-prior-mass-1]{\eqref{eq:KL-prior-mass-1}}, where $C_7 = 4C_5>0$.
\end{proof}

\begin{lemma}[Prior mass on sieve complement]
\label{lemma:sieve-complement}
For constants $L_S, L_{\beta}, L_{\ell}, L_u > 0$ and $n\varepsilon_{n, K}^{2} \stackrel{n \to \infty}{\longrightarrow}\infty$,
define the sieve
\begin{align}
\label{eq:sieve}
\begin{split}
\Theta_{n} &:= \Bigg\{(f_{\bm \beta, \mathcal T}, \sigma^2)\in \mathcal{F}_K\times \mathbb{R}^{+}: S(\mathcal T) \leq L_Sn\varepsilon_{n, K}^{2}, \lVert \bm \beta\rVert_{\infty} \leq \exp\{L_{\beta}n\varepsilon_{n, K}^{2}\},\\
&\qquad \qquad \qquad \qquad \qquad  -L_{\ell}n\varepsilon_{n, K}^{2} \leq \log \sigma^{2} \leq L_u n \varepsilon_{n, K}^{2}\Bigg\},
\end{split}
\end{align}
where $\varepsilon_{n, K}$ is defined in \hyperref[lemma:KL-prior-mass]{Lemma~\ref{lemma:KL-prior-mass}}.
Since $\bm \Sigma_{\beta}\succ 0$, there exist a constant $C_{\Theta}>0$ such that for sufficiently large $n$
\begin{align}
\label{eq:sieve-complement-bound}
\Pi(\Theta_n^{c}) \leq \exp\{-C_{\Theta}n\varepsilon_{n, K}^{2}\}.
\end{align}
\end{lemma}

\begin{proof}
Note that, the complement of the sieve $\Theta_n$ in~\hyperref[eq:sieve]{\eqref{eq:sieve}} is
\begin{align}
\label{eq:sieve-complement-1}
\begin{split}
\Theta_n^{c} &\subseteq \{S(\mathcal T) > L_S n\varepsilon_{n, K}^{2}\} \cup \{\lVert \bm\beta \rVert_{\infty} > \exp\{L_{\beta}n\varepsilon^{2}_{n, K}\}\}\\
&\qquad \cup \{\log \sigma^{2} < -L_{\ell}n\varepsilon_{n, K}^{2}\}\cup \{\log \sigma^{2} > L_{u}n\varepsilon_{n, K}^{2}\}.
\end{split}
\end{align}
From~\hyperref[eq:sieve-complement-1]{\eqref{eq:sieve-complement-1}} and union bound, it follows that
\begin{align}
\label{eq:sieve-complement-2}
\begin{split}
\Pi(\Theta_n^{c})& \leq  \Pi_{\mathrm{forest},K}(S(\mathcal T) > L_S n\varepsilon_{n, K}^{2}\mid \alpha_0, \delta_0, \bm \alpha_{\mathrm{op}}, \bm \alpha_{\mathrm{ft}})\\
&\qquad + \mathrm{NIG}_{K+1}(\lVert\bm\beta \rVert_{\infty} > \exp\{L_{\beta}n\varepsilon^{2}_{n, K}\}, \sigma^2>0\mid \bm\mu_{\beta}, \bm\Sigma_{\beta}, \nu, \lambda)\\
&\qquad + \mathrm{IG}(\log \sigma^{2} < -L_{\ell}n\varepsilon_{n, K}^{2}\mid \nu, \lambda)
+ \mathrm{IG}(\log \sigma^{2} > L_{u}n\varepsilon_{n, K}^{2}\mid \nu, \lambda).
\end{split}
\end{align}
We bound each term in the right-hand-side of~\hyperref[eq:sieve-complement-2]{\eqref{eq:sieve-complement-2}} above separately, as follows.

\underline{\emph{Symbolic forest operator count tail}}. 
From~\hyperref[supple-prop:tree-forest-prior-tail-control]{Proposition~\ref{supple-prop:tree-forest-prior-tail-control}}, for fixed $K$, constants $C_2', c_2'>0$, and $s = L_Sn\varepsilon_{n, K}^{2}$, we have
\begin{align}
\label{eq:sieve-complement-3}
\Pi_{\mathrm{forest}, K}(S(\mathcal T) > L_Sn\varepsilon_{n, K}^{2} \mid \alpha_0, \delta_0, \bm\alpha_{\mathrm{op}}, \bm\alpha_{\mathrm{ft}}) \leq C_2' \exp\{-c_2'L_Sn\varepsilon_{n, K}^{2}\}.
\end{align}

\underline{\emph{Upper tail of the Inverse-Gamma variance parameter}}. 
Under the NIG prior, $\sigma^{2}\sim \mathrm{IG}(\nu/2, \lambda/2)$ with density proportional to $p(\sigma^2) \propto (\sigma^{2})^{-\nu/2-1}\exp\{-\lambda/(2\sigma^{2})\}$, for $\sigma^{2} > 0$. Note that, \\$\exp\{L_u n\varepsilon_{n, K}^2\} \geq 1$, hence
\begin{align}
\label{eq:sieve-complement-4}
\begin{split}
\mathrm{IG}(\sigma^{2} > \exp\{L_{u}n\varepsilon_{n, K}^{2}\}\mid \nu, \lambda) &\leq C_8\int_{\exp\{L_u n\varepsilon_{n, K}^{2}\}}^{\infty}s^{-\nu/2-1}ds 
\leq C_8'\exp\{-\nu L_un\varepsilon_{n, K}^{2}/2\},
\end{split}
\end{align}
for $C_8, C_8' > 0$.

\underline{\emph{Lower tail of the Inverse-Gamma variance parameter}}. 
Note that, $\sigma^{-2} \sim \mathrm{G}(\nu/2, \lambda/2)$. The Gamma distribution has an exponentially decaying upper tail. Therefore
\begin{align}
\label{eq:sieve-complement-5}
\begin{split}
\mathrm{IG}(\sigma^{-2} > \exp\{L_\ell n\varepsilon_{n, K}^{2}\}\mid  \nu, \lambda) &\leq C_9\int_{\exp\{L_{\ell}n\varepsilon_{n, K}^{2}\}}^{\infty} s^{\nu/2-1}\exp\{-\lambda s/2\}ds\\
&\leq C_9'\exp\{-\lambda\exp\{L_{\ell}n\varepsilon_{n, K}^{2}\}/4\},
\end{split}
\end{align}
for $C_9, C_9'>0$.

\underline{\emph{Coefficient tail}}. 
Since $\bm \beta\in \mathbb{R}^{K+1}$ and $K$ is fixed, by union bound
\begin{align} \label{eq:sieve-complement-6} 
&\mathrm{NIG}_{K+1}(\lVert\bm\beta\rVert_{\infty} > \exp\{L_{\beta}n\varepsilon_{n, K}^{2}\}, \sigma^2>0\mid \bm\mu_{\beta}, \bm\Sigma_{\beta}, \nu, \lambda) \nonumber\\ 
&\leq \sum_{\ell =0}^{K}\mathrm{NIG}_{K+1}(|\beta_{\ell}| > \exp\{L_{\beta}n\varepsilon_{n, K}^{2}\}, \sigma^2>0\mid \bm\mu_{\beta}, \bm\Sigma_{\beta}, \nu, \lambda) \nonumber\\ 
&\leq \sum_{\ell=0}^{K}\Bigg[\mathrm{NIG}_{K+1}(|\beta_\ell| > \exp\{L_{\beta}n\varepsilon_{n, K}^{2}\}, \sigma^{2}\leq \exp\{L_{\beta}n\varepsilon_{n, K}^{2}\}\mid \bm\mu_{\beta}, \bm\Sigma_{\beta}, \nu, \lambda) \nonumber\\ 
&\qquad + \mathrm{NIG}_{K+1}(|\beta_\ell| > \exp\{L_{\beta}n\varepsilon_{n, K}^{2}\}, \sigma^{2} > \exp\{L_{\beta}n\varepsilon_{n, K}^{2}\}\mid \bm\mu_{\beta}, \bm\Sigma_{\beta}, \nu, \lambda)\Bigg]\nonumber \displaybreak[4]\\ 
&\leq \sum_{\ell=0}^{K}\mathrm{NIG}_{K+1}(|\beta_\ell| > \exp\{L_{\beta}n\varepsilon_{n, K}^{2}\}, \sigma^{2}\leq \exp\{L_{\beta}n\varepsilon_{n, K}^{2}\}\mid \bm\mu_{\beta}, \bm\Sigma_{\beta}, \nu, \lambda) \nonumber\\ 
&\qquad + \sum_{\ell=0}^{K}C_8'\exp\{-\nu L_{\beta}n\varepsilon_{n, K}^{2}/2\},\text{ follows from~\hyperref[eq:sieve-complement-4]{\eqref{eq:sieve-complement-4}}}.
\end{align}
Next we bound $\mathrm{NIG}_{K+1}(|\beta_\ell| > \exp\{L_{\beta}n\varepsilon_{n, K}^{2}\}, \sigma^{2}\leq \exp\{L_{\beta}n\varepsilon_{n, K}^{2}\}\mid \bm\mu_{\beta}, \bm\Sigma_{\beta}, \nu, \lambda)$ in~\hyperref[eq:sieve-complement-6]{\eqref{eq:sieve-complement-6}}. Under the NIG prior, $\bm \beta \mid \sigma^{2} \sim \mathrm{N}_{K+1}(\bm \mu_\beta, \sigma^2\bm \Sigma_{\beta})$ that is $\beta_{\ell}\mid \sigma^{2}=s \sim \Pi(\beta_{\ell}\mid \sigma^{2}=s) \equiv \mathrm{N}(\mu_{\beta, \ell}, s\Sigma_{\beta, \ell\ell})$ for $\ell=0, \ldots, K$, where $\Sigma_{\beta, \ell\ell} > 0$ as $\bm{\Sigma}_{\beta}\succ 0$. Conditioning on $\sigma^2$
\begin{align}
\label{eq:sieve-complement-7}
\begin{split}
&\mathrm{NIG}_{K+1}(|\beta_\ell| > \exp\{L_{\beta}n\varepsilon_{n, K}^{2}\}, \sigma^{2}\leq \exp\{L_{\beta}n\varepsilon_{n, K}^{2}\}\mid \bm\mu_{\beta}, \bm\Sigma_{\beta}, \nu, \lambda)\\
&=\int_{0}^{\exp\{L_{\beta}n\varepsilon_{n, K}^{2}\}}\Pi(|\beta_\ell| > \exp\{L_{\beta}n\varepsilon_{n, K}^{2}\}\mid \sigma^{2}=s)d(\mathrm{IG}(s\mid \nu/2, \lambda/2))\\
&\leq \sup_{0<s\leq \exp\{L_{\beta}n\varepsilon_{n, K}^{2}\}}\Pi(|\beta_{\ell}| > \exp\{L_{\beta}n\varepsilon_{n, K}^{2}\}\mid \sigma^{2}=s).
\end{split}
\end{align}
Note that, $\beta_{\ell} - \mu_{\beta, \ell} = \sqrt{s \Sigma_{\beta, \ell\ell}}Z$ where $Z\sim \mathrm{N}(0, 1)$. For all sufficiently large $n$, $\exp\{L_{\beta}n\varepsilon_{n, K}^{2}\} > 2|\mu_{\beta, \ell}|$. Hence, $|\beta_{\ell}| > \exp\{L_{\beta}n\varepsilon^{2}_{n, K}\} \implies |\beta_{\ell}-\mu_{\beta, \ell}| > \exp\{L_{\beta}n\varepsilon_{n, K}^{2}\} - |\mu_{\beta, \ell}| \geq \exp\{L_{\beta}n\varepsilon_{n, K}^{2}\}/2$. Using this observation, for $0<s \leq \exp\{L_{\beta}n\varepsilon_{n, K}^{2}\}$
\begin{align}
\label{eq:sieve-complement-8}
\begin{split}
\Pi(|\beta_\ell| > \exp\{L_{\beta}n\varepsilon_{n, K}^{2}\}\mid \sigma^{2}=s) &\leq \Pi(|\beta_{\ell} - \mu_{\beta, \ell}| > \exp\{L_{\beta}n\varepsilon_{n, K}^{2}\}/2\mid \sigma^{2}=s)\\
&= \Pi(|Z| > \exp\{L_{\beta}n\varepsilon_{n, K}^{2}\}/(2\sqrt{s\Sigma_{\beta, \ell\ell}})).
\end{split}
\end{align}
Since $s\leq \exp\{L_{\beta}n\varepsilon^{2}_{n, K}\}$, $\exp\{L_{\beta}n\varepsilon_{n, K}^{2}\}/(2\sqrt{s\Sigma_{\beta, \ell\ell}}) \geq \sqrt{\exp\{L_{\beta}n\varepsilon_{n, K}^{2}\}}/(2\sqrt{\Sigma_{\beta, \ell\ell}})$. Combining this with the standard Gaussian tail bound, $\int_{t}^{\infty}\sqrt{2/\pi}\exp\{-v^{2}/2\}dv \leq 2\exp\{-t^{2}/2\}$ for $t>0$, from~\hyperref[eq:sieve-complement-8]{\eqref{eq:sieve-complement-8}} we get
\begin{align}
\label{eq:sieve-complement-9}
\begin{split}
\Pi(|\beta_\ell| > \exp\{L_{\beta}n\varepsilon_{n, K}^{2}\}\mid \sigma^{2}=s) &\leq 2\exp\{-\exp\{L_{\beta}n\varepsilon_{n, K}^{2}\}/(8\Sigma_{\beta, \ell\ell})\}.
\end{split}
\end{align}
Now define $\overline{\Sigma}_{\beta}:=\max_{\ell=0,\ldots,K}\Sigma_{\beta, \ell\ell}$. Since $K$ is fixed and $\bm{\Sigma}_{\beta}\succ 0$, $0<\overline{\Sigma}_{\beta}<\infty$. Thus, uniformly over $\ell = 0, \ldots, K$
$$
\exp\{-\exp\{L_\beta n\varepsilon^{2}_{n, K}\}/(8\Sigma_{\beta, \ell\ell})\} \leq \exp\{-\exp\{L_\beta n\varepsilon^{2}_{n, K}\}/(8\overline{\Sigma}_{\beta})\}.
$$ 
Therefore, from~\hyperref[eq:sieve-complement-9]{\eqref{eq:sieve-complement-9}} and~\hyperref[eq:sieve-complement-7]{\eqref{eq:sieve-complement-7}}
\begin{align}
\label{eq:sieve-complement-10}
\begin{split}
&\mathrm{NIG}_{K+1}(|\beta_\ell| > \exp\{L_{\beta}n\varepsilon_{n, K}^{2}\}, \sigma^{2}\leq \exp\{L_{\beta}n\varepsilon_{n, K}^{2}\}\mid \bm\mu_{\beta}, \bm\Sigma_{\beta}, \nu, \lambda) \\
&\qquad \leq  2\exp\{-\exp\{L_{\beta}n\varepsilon_{n, K}^{2}\}/(8\overline{\Sigma}_{\beta})\}.
\end{split}
\end{align}
Plugging~\hyperref[eq:sieve-complement-10]{\eqref{eq:sieve-complement-10}} in~\hyperref[eq:sieve-complement-6]{\eqref{eq:sieve-complement-6}} yields
\begin{align}
\label{eq:sieve-complement-11}
\begin{split}
&\mathrm{NIG}_{K+1}(\lVert \bm\beta\rVert_{\infty} > \exp\{L_{\beta}n\varepsilon_{n, K}^{2}\}\mid \bm\mu_{\beta}, \bm\Sigma_{\beta}, \nu, \lambda)\\
&\qquad\leq (K+1)\left[C_8'\exp\{-\nu L_{\beta} n\varepsilon_{n, K}^{2}/2\} + 2\exp\{-\exp\{L_{\beta}n\varepsilon_{n, K}^{2}\}/(8\overline{\Sigma}_{\beta})\}\right].
\end{split}
\end{align}

\underline{\emph{Combining the bounds}}.
Combining the preceding bounds in~\hyperref[eq:sieve-complement-3]{\eqref{eq:sieve-complement-3}},~\hyperref[eq:sieve-complement-4]{\eqref{eq:sieve-complement-4}},~\hyperref[eq:sieve-complement-5]{\eqref{eq:sieve-complement-5}}, and~\hyperref[eq:sieve-complement-11]{\eqref{eq:sieve-complement-11}}, the prior mass on the sieve complement $\Theta_{n}^{c}$ is
\begin{align}
\label{eq:sieve-complement}
\begin{split}
\Pi(\Theta_{n}^{c}) &\leq C_2'\exp\{-c_2' L_Sn\varepsilon^{2}_{n, K}\} + C_8'\exp\left\{-\frac{\nu L_u}{2}n\varepsilon_{n, K}^{2}\right\} + C_9'\exp\left\{-\frac{\lambda}{4}\exp\{L_{\ell}n\varepsilon_{n, K}^{2}\}\right\}\\
&\qquad + (K+1)C_8'\exp\left\{-\frac{\nu L_{\beta}}{2}n\varepsilon_{n, K}^{2}\right\} + 2(K+1)\exp\left\{-\frac{\exp\{L_{\beta}n\varepsilon^{2}_{n, K}\}}{8\overline{\Sigma}_{\beta}}\right\}.
\end{split}
\end{align}

Choose $C_{10}$ as $0< C_{10} < \min\{c_2'L_S, \nu L_{u}/2, \nu L_{\beta}/2 \}$. Also, $C_9'\exp\{-\lambda \exp\{L_{\ell}n\varepsilon_{n, K}^{2}\}/4\} = \mathfrak{o}(\exp\{-C_{10}n\varepsilon_{n, K}^{2}\})$ and $2(K+1)\exp\{-\exp\{L_{\beta}n\varepsilon^{2}_{n, K}\}/(8\overline{\Sigma}_{\beta})\} = \mathfrak{o}(\exp\{-C_{10}n\varepsilon_{n, K}^{2}\})$, since\\ $n\varepsilon^{2}_{n, K} \stackrel{n \to \infty}{\longrightarrow}\infty$. Moreover, by the choice of $C_{10}$, we have
\begin{align*}
C_2'\exp\{-c_2'L_S n\varepsilon_{n, K}^{2}\} &+ C_{8}'\exp\{-\nu L_u n\varepsilon_{n, K}^{2}/2\} \\
&\qquad + (K+1)C_8'\exp\{-\nu L_{\beta} n\varepsilon^{2}_{n, K}/2\} \leq C_{10}'\exp\{-C_{10}n\varepsilon_{n, K}^{2}\},    
\end{align*}
for some constant $C_{10}'>0$. Hence, for all sufficiently large $n$, $\Pi(\Theta_n^{c}) \leq C_{10}''\exp\{-C_{10}n\varepsilon_{n, K}^{2}\}$, for some constant $C_{10}'' > 0$. Since $n\varepsilon_{n, K}^{2}\stackrel{n \to \infty}{\longrightarrow} \infty$, $C_{10}'' \exp\{-C_{10}n\varepsilon_{n, K}^{2}\} \leq \exp\{-C_{10}n\varepsilon_{n, K}^{2}/2\} = \exp\{-C_{\Theta}n\varepsilon_{n, K}^{2}\}$, where $C_{\Theta}=C_{10}/2$. Therefore
$$
\boxed{\Pi(\Theta_n^{c}) \leq \exp\{-C_{\Theta}n\varepsilon_{n, K}^{2}\},}
$$
which yields the sieve complement bound in~\hyperref[eq:sieve-complement-bound]{\eqref{eq:sieve-complement-bound}}.
\end{proof}

\begin{lemma}[Univariate Gaussian Hellinger bound]
\label{lemma:Gaussian-Hellinger-bound}
Let $\mu_1, \mu_2 \in \mathbb{R}$ and $\sigma_1^{2}, \sigma_2^{2} > 0$. Then, the squared Hellinger distance between $p_{\mu_1, \sigma_1^2}(y)\equiv \mathrm{N}(y\mid \mu_1, \sigma_1^2)$ and $p_{\mu_2, \sigma_2^2}(y) \equiv \mathrm{N}(y\mid \mu_2, \sigma_2^2)$ is bounded by
\begin{align}
\label{eq:Hellinger-univariate-normal}
\frac{1}{2}\int_{\mathbb{R}}\left(\sqrt{p_{{\mu_1, \sigma_1^2}}(y)} - \sqrt{p_{\mu_2, \sigma_2^2}(y)}\right)^{2} dy\leq \frac{1}{2}\left[\frac{(\mu_1-\mu_2)^2}{L} + \frac{(\sigma_1^{2} - \sigma_2^{2})^2}{L^{2}}\right],
\end{align}
where $0 < L \leq \min\{\sigma_1^{2}, \sigma^{2}_{2}\}$.
\end{lemma}

\begin{proof}
For univariate Gaussian distributions, the squared Hellinger distance is
\begin{align}
\label{eq:Hellinger-1}
\frac{1}{2}\int_{\mathbb{R}}\left(\sqrt{p_{{\mu_1, \sigma_1^2}}(y)} - \sqrt{p_{\mu_2, \sigma_2^2}(y)}\right)^{2} dy = 1 - \left(\frac{2\sigma_1 \sigma_2}{\sigma_1^{2} + \sigma_2^{2}}\right)^{\frac{1}{2}}\exp\left\{-\frac{(\mu_1 - \mu_2)^2}{4(\sigma_1^2 + \sigma_2^2)}\right\}.
\end{align}
Using, $1-ab \leq 1 - a + 1-b$ for $a,b \in [0, 1]$, we get
\begin{align}
\label{eq:Hellinger-2}
\frac{1}{2}\int_{\mathbb{R}}\left(\sqrt{p_{{\mu_1, \sigma_1^2}}(y)} - \sqrt{p_{\mu_2, \sigma_2^2}(y)}\right)^{2} dy \leq 1 - \left(\frac{2\sigma_1\sigma_2}{\sigma_1^2 + \sigma_2^2}\right)^{\frac{1}{2}} + 1 - \exp\left\{-\frac{(\mu_1 - \mu_2)^2}{4(\sigma_1^2 + \sigma_2^2)}\right\}.
\end{align}

\underline{\emph{Bounding $1 - \exp\{-(\mu_1-\mu_2)^2/(4(\sigma_1^2 + \sigma_2^2))\}$ in~\hyperref[eq:Hellinger-2]{\eqref{eq:Hellinger-2}}}}.
Note that, $$1 - \exp\{-(\mu_1-\mu_2)^2/(4(\sigma_1^2 + \sigma_2^2))\} \leq (\mu_1-\mu_2)^2/(4(\sigma_1^2 + \sigma_2^2)) \leq (\mu_1 - \mu_2)^2/(8L),$$ using $1-\exp\{-x\} \leq x$ and $L\leq \min\{\sigma_1^2, \sigma_2^2\}$. 

\underline{\emph{Bounding $1 - (2\sigma_1\sigma_2/(\sigma_1^2 + \sigma_2^2))^{1/2}$ in~\hyperref[eq:Hellinger-2]{\eqref{eq:Hellinger-2}}}}. 
Let $x = 2\sigma_1\sigma_2 / (\sigma_1^2 + \sigma_2^2)$. By AM-GM inequality, $0 < x\leq 1$. Then, $1-x^{1/2} \leq 1-x$. So it is enough to bound $1-x$. Now, 
$$1-x = 
\frac{(\sigma_1 - \sigma_2)^2}{\sigma_1^{2} + \sigma_2^2} 
= \frac{(\sigma_1^{2} - \sigma_2^{2})^2}{(\sigma_1^{2} + \sigma_2^{2})(\sigma_1 + \sigma_2)^2} 
\leq \frac{(\sigma_1^2-\sigma_2^2)^{2}}{(\sigma_1^2 + \sigma_2^2)\sigma_1^2}
\leq \frac{(\sigma_1^2 - \sigma_2^2)^2}{2L^{2}},$$ 
where the last inequality follows from $L\leq \min\{\sigma_1^2, \sigma_2^2\}$.

Combining the above two bounds gives
$$
\boxed{\frac{1}{2}\int_{\mathbb{R}}\left(\sqrt{p_{{\mu_1, \sigma_1^2}}(y)} - \sqrt{p_{\mu_2, \sigma_2^2}(y)}\right)^{2} dy\leq \frac{1}{2}\left[\frac{(\mu_1-\mu_2)^2}{L} + \frac{(\sigma_1^{2} - \sigma_2^{2})^2}{L^{2}}\right],}
$$
yielding~\hyperref[eq:Hellinger-univariate-normal]{\eqref{eq:Hellinger-univariate-normal}}.
\end{proof}

\begin{lemma}[Hellinger metric entropy]
\label{lemma:Hellinger-metric-entropy}
Define the average squared Hellinger distance
\begin{align}
\label{eq:Hellinger-distance}
h_{n}^{2}\left(\mathbb{P}^{n}_{f, \sigma^{2}}\parallel \mathbb{P}^{n}_{g, \tau^2}\right) := \frac{1}{n}\sum_{i=1}^{n}\int_{\mathbb{R}}\left(\sqrt{p_{{f(\bm x_i), \sigma^2}}(y_i)} - \sqrt{p_{g(\bm x_i), \tau^2}(y_i)}\right)^{2} dy_i,
\end{align}
where $p_{\mu, s^{2}}(y) \equiv \mathrm{N}(y\mid \mu, s^{2})$. With $\mathsf r_{n, K}$ in~\hyperref[supple-theorem:posterior-contraction]{Theorem~\ref{supple-theorem:posterior-contraction}} and $\Theta_n$ in \hyperref[lemma:sieve-complement]{Lemma~\ref{lemma:sieve-complement}}, there exist a constant $C_{11} > 0$ such that, under \mainassref[ass:global-symbolic-evaluation-envelope], for all sufficiently large $n$
\begin{align}
\label{eq:Hellinger-entropy}
\log N(\mathsf r_{n, K}, \Theta_n, h_n) \leq C_{11} n \mathsf r_{n, K}^{2}.
\end{align}
\end{lemma}

\begin{proof}
For $(f, \sigma^2), (g, \tau^2) \in \Theta_n$, using~\hyperref[lemma:Gaussian-Hellinger-bound]{Lemma~\ref{lemma:Gaussian-Hellinger-bound}} we have
\begin{align}
\label{eq:Hellinger-entropy-1}
\begin{split}
h_{n}^{2}\left(\mathbb{P}^{n}_{f, \sigma^{2}}\parallel \mathbb{P}^{n}_{g, \tau^2}\right) &\leq \frac{1}{n}\sum_{i=1}^{n}\left[\frac{(f(x_i) - g(x_i))^2}{\ell_n} + \frac{(\sigma^2 - \tau^2)^2}{\ell_n^2}\right]
\leq \left[\frac{d_n^{2}(f, g)}{\ell_n} + \frac{(\sigma^2 - \tau^2)^2}{\ell_n^2}\right],
\end{split}
\end{align}
where $\ell_n = \exp\{-L_{\ell}n\varepsilon_{n, K}^{2}\}$ with $\varepsilon_{n, K}$ as defined in~\hyperref[lemma:KL-prior-mass]{Lemma~\ref{lemma:KL-prior-mass}}. Using $\sqrt{a+b} \leq \sqrt{a} + \sqrt{b}$ for $a, b>0$, from~\hyperref[eq:Hellinger-entropy-1]{\eqref{eq:Hellinger-entropy-1}} we have
\begin{align}
\label{eq:Hellinger-entropy-2}
h_{n}\left(\mathbb{P}^{n}_{f, \sigma^{2}}\parallel \mathbb{P}^{n}_{g, \tau^2}\right) &\leq \left[\frac{d_n(f, g)}{\sqrt{\ell_n}} + \frac{|\sigma^2-\tau^2|}{\ell_n}\right].
\end{align}
Hence from~\hyperref[eq:Hellinger-entropy-2]{\eqref{eq:Hellinger-entropy-2}}, an $h_n$-cover of $\Theta_n$ with radius $\mathsf r_{n, K}$ is obtained by covering the function part in $d_n$-radius, $\eta_{f, n, K} = c_{11} \mathsf r_{n, K}\sqrt{\ell_n}$, and the variance interval in Euclidean radius, $\eta_{\sigma, n, K} = c_{11}\mathsf r_{n, K}\ell_n$, for a sufficiently small constant $c_{11} > 0$. Now define the function part of the sieve
\begin{align}
\label{eq:Hellinger-entropy-3}
\begin{split}
\mathcal{S}_n &:= \left\{f_{\bm \beta, \mathcal T}\in \mathcal{F}_K: S(\mathcal T) \leq L_{S}n\varepsilon_{n, K}^{2}, \lVert \bm \beta\rVert_{\infty}\leq \exp\{L_{\beta}n\varepsilon_{n, K}^{2}\}\right\}\\
&\equiv \mathcal{S}(K, L_{S}n\varepsilon_{n, K}^{2}, \exp\{L_{\beta}n\varepsilon_{n, K}^{2}\}),
\end{split}
\end{align}
Invoking \hyperref[lemma:metric-entropy]{Lemma~\ref{lemma:metric-entropy}} yields
\begin{align}
\label{eq:Hellinger-entropy-4}
\begin{split}
&\log N(\eta_{f, n, K},\mathcal S_n,d_n)\\
&\qquad \le
(2L_Sn\varepsilon_{n, K}^2+K+1)\log 3 + L_S n\varepsilon_{n, K}^{2}\log|\mathbb O| + (L_S n\varepsilon_{n, K}^{2}+K)\log p\\
&\qquad + (K+1)\left[c_U L_S n\varepsilon_{n, K}^{2} + \log\overline{C}_U + \log\left(1 + \frac{2\exp\{L_{\beta}n\varepsilon_{n, K}^{2}\}(K+1)}{\eta_{f, n, K}}\right)\right].
\end{split}
\end{align}
Now we bound each contribution on the right-hand-side of~\hyperref[eq:Hellinger-entropy-4]{\eqref{eq:Hellinger-entropy-4}} as follows:
\begin{align}
\label{eq:Hellinger-entropy-5}
\begin{gathered}
(2L_Sn\varepsilon_{n, K}^2+K+1)\log 3 \leq (2L_Sn \mathsf{r}_{n, K}^2+K+1)\log 3 = \mathcal{O}(n \mathsf r_{n, K}^{2}),\\
L_S n\varepsilon_{n, K}^{2}\log|\mathbb O| \leq L_S n \mathsf r_{n, K}^{2}\log|\mathbb O| = \mathcal{O}(n \mathsf r_{n, K}^{2}),\\
(L_S n\varepsilon_{n, K}^{2}+K)\log p \leq (L_S n \mathsf r_{n, K}^{2}+K)\log p = \mathcal{O}(n \mathsf r_{n, K}^{2}),\\
(K+1)(c_U L_S n\varepsilon_{n, K}^{2} + \log\overline{C}_U) \leq (K+1)(c_U L_S n \mathsf r_{n, K}^{2} + \log\overline{C}_U) \leq \mathcal{O}(n \mathsf r_{n, K}^{2}).
\end{gathered}
\end{align}
It remains to obtain an upper bound of $(K+1)\log(1 + (2\exp\{L_{\beta}n\varepsilon_{n, K}^{2}\}(K+1))/\eta_{f, K, n})$.
\begin{align}
\label{eq:Hellinger-entropy-6}
\begin{split}
&(K+1) \log\left(1 + \frac{2\exp\{L_{\beta}n\varepsilon_{n, K}^{2}\}(K+1)}{\eta_{f, n, K}}\right)
\leq (K+1) \log\left( \frac{4\exp\{L_{\beta}n\varepsilon_{n, K}^{2}\}(K+1)}{c_{11}\mathsf r_{n, K}\sqrt{\exp\{-L_{\ell}n\varepsilon_{n, K}^{2}\}}}\right)\\
&\qquad \leq (K+1)\left[\log \frac{4(K+1)}{c_{11}} + \log\frac{1}{\mathsf r_{n, K}} + L_{\beta}n\varepsilon_{n, K}^{2} + \frac{L_{\ell}}{2}n\varepsilon_{n, K}^{2}\right]\\
&\qquad \leq (K+1)\left[\log \frac{4(K+1)}{c_{11}} + \log\frac{1}{\varepsilon_{n, K}} + L_{\beta}n \mathsf r_{n, K}^{2} + \frac{L_{\ell}}{2}n \mathsf r_{n, K}^{2}\right]\\
&\qquad = \mathcal{O}(n \mathsf r_{n, K}^{2}),
\end{split}
\end{align}
where the last equality in the above display follows as $\log(\varepsilon_{n, K}) \geq \log(n^{-1}(K+2)\log n) = \log(K+2) + \log\log n - \log n \implies \log(\varepsilon_{n, K}^{-1}) \leq  -\log(K+2) - \log\log n + \log n = \mathcal{O}(\log n) = \mathfrak{o}(n\varepsilon_{n, K}^{2}) = \mathfrak{o}(n \mathsf r_{n, K}^{2})$. Hence, using~\hyperref[eq:Hellinger-entropy-5]{\eqref{eq:Hellinger-entropy-5}} and~\hyperref[eq:Hellinger-entropy-6]{\eqref{eq:Hellinger-entropy-6}} in~\hyperref[eq:Hellinger-entropy-4]{\eqref{eq:Hellinger-entropy-4}}, we have
\begin{align}
\label{eq:Hellinger-entropy-7}
\log N(\eta_{f, n, K},\mathcal S_n,d_n) = \mathcal{O}(n \mathsf r_{n, K}^{2}).
\end{align}
On the sieve $\Theta_n$, $-L_{\ell} n\varepsilon_{n, K}^{2} \leq \log \sigma^{2} \leq L_{u} n\varepsilon_{n, K}^{2}$. Hence, the interval \\$[\exp\{-L_{\ell}n\varepsilon_{n, K}^{2}\},\exp\{L_{u}n\varepsilon_{n, K}^{2}\}]$ can be covered by at most $N_{\sigma}$ balls of radius $\eta_{\sigma, n, K}$, where
$$
N_{\sigma} \leq 1 + \frac{\exp\{L_{u}n\varepsilon_{n, K}^{2}\}}{c_{11} \mathsf r_{n, K}\exp\{-L_{\ell}n\varepsilon_{n, K}^{2}\}} \leq \frac{2\exp\{(L_{u}+L_{\ell})n \mathsf r_{n, K}^{2}\}}{c_{11}\varepsilon_{n, K}}.
$$
Therefore, $\log N_{\sigma} \leq \log(2/c_{11}) + (L_{u} +L_{\ell})n\varepsilon_{n, K}^{2} + \log(\varepsilon_{n, K}^{-1}) = \mathcal{O}(n \mathsf r_{n, K}^{2})$. Since $\Theta_n = \mathcal{S}_n \times [\exp\{-L_{\ell}n\varepsilon_{n, K}^{2}\},\exp\{L_{u}n\varepsilon_{n, K}^{2}\}]$, using the upper bound on $\log N_{\sigma}$ and~\hyperref[eq:Hellinger-entropy-7]{\eqref{eq:Hellinger-entropy-7}}
\begin{align*}
\boxed{
\log N(\mathsf r_{n, K}, \Theta_n, h_{n}) \leq \log N(\eta_{f, n, K},\mathcal S_n,d_n) + \log N_{\sigma}\leq C_{11} n \mathsf r_{n, K}^2,
}
\end{align*}
for some constant $C_{11}>0$, thus yielding~\hyperref[eq:Hellinger-entropy]{\eqref{eq:Hellinger-entropy}}.
\end{proof}

\begin{lemma}[Tests for $d_n$-separated alternatives on the sieve]
\label{lemma:tests}
For $M>0$, define the sieve restricted alternative set
\begin{align}
\label{eq:alternative-set}
A_{n}(M) := \left\{(f_{\bm \beta, \mathcal{T}}, \sigma^2)\in \Theta_n: d_{n}(f_{\bm \beta, \mathcal{T}}, f_0) > M \mathsf r_{n, K}\right\},
\end{align}
where $\mathsf r_{n, K}$ and $\Theta_n$ are as in \hyperref[supple-theorem:posterior-contraction]{Theorem~\ref{supple-theorem:posterior-contraction}} and \hyperref[lemma:sieve-complement]{Lemma~\ref{lemma:sieve-complement}}, respectively. Then under \hyperref[main-ass:global-symbolic-evaluation-envelope]{\textcolor{BrickRed}{Assumptions \ref*{main-ass:global-symbolic-evaluation-envelope} and \ref*{main-ass:well-specified} of the main manuscript}}, there exist tests $\phi_n$ such that
\begin{align}
\label{eq:tests-existence}
\begin{gathered}
\mathsf{E}_{\mathbb{P}^{n}_{0}}(\phi_n) \leq \exp\{C_{12} n \mathsf r_{n, K}^{2} - C_{13}M^{2}n \mathsf r_{n, K}^{2}\} + \exp\{C_{11} n \mathsf r_{n, K}^{2} - C_{13}''' n\},\\
\sup_{(f_{\bm \beta, \mathcal T}, \sigma^2)\in A_{n}(M)}\mathsf{E}_{\mathbb{P}^{n}_{f_{\bm \beta, \mathcal T}, \sigma^2}}(1-\phi_n) \leq \exp\{-C_{13}'M^{2}n \mathsf r_{n, K}^{2}\} + \exp\{-C_{13}''' n\},
\end{gathered}
\end{align}
for constants $C_{11}, C_{12}, C_{13}, C_{13}', C_{13}''' > 0$. Consequently, if $M$ is chosen sufficiently large then, $\mathsf{E}_{\mathbb{P}_0^{n}}(\phi_n) \stackrel{n\to \infty}{\longrightarrow} 0$ and $\sup_{(f_{\bm \beta, \mathcal T}, \sigma^2)\in A_{n}(M)}\mathsf{E}_{\mathbb{P}^{n}_{f_{\bm \beta, \mathcal T}, \sigma^2}}(1-\phi_n) \stackrel{n\to \infty}{\longrightarrow} 0$.
\end{lemma}

\begin{proof}
For convenience, we write $f = f_{\bm \beta, \mathcal T}$. Split the alternative set in~\hyperref[eq:alternative-set]{\eqref{eq:alternative-set}} into two parts, $A_{n}(M) = A_{n, 1}(M) \cup A_{n, 2}(M)$, where
\begin{align}
\label{eq:tests-1}
\begin{gathered}
A_{n, 1}(M) = A_{n}(M) \cap \left\{\frac{\sigma_0^2}{2} \leq \sigma^2\leq 2\sigma_0^2\right\},\; A_{n, 2}(M) = A_{n}(M) \cap \left\{\sigma^{2} < \frac{\sigma^{2}_0}{2}\cup\sigma^{2} > 2\sigma_0^{2}\right\}.
\end{gathered}
\end{align}
We construct tests separately for $A_{n, 1}(M)$ and $A_{n, 2}(M)$ in~\hyperref[eq:tests-1]{\eqref{eq:tests-1}}, respectively.

\underline{\emph{Variance-near alternatives}}.
Let the function part of the sieve be, $\mathcal{S}_n$, as in~\hyperref[eq:Hellinger-entropy-3]{\eqref{eq:Hellinger-entropy-3}}. With $\varepsilon_{n, K}$ as defined in \hyperref[lemma:KL-prior-mass]{Lemma~\ref{lemma:KL-prior-mass}}, for sufficiently large $n$, $c_{11}\sqrt{\exp\{-L_{\ell}n\varepsilon_{n, K}^{2}\}} \leq M/8$. Therefore, using the proof of \hyperref[lemma:Hellinger-metric-entropy]{Lemma~\ref{lemma:Hellinger-metric-entropy}}, $\log N(M \mathsf r_{n, K}/8, \mathcal S_n, d_n) \leq \log N(\eta_{f, n, K}, \mathcal S_{n}, d_n) \leq C_{12} n \mathsf r_{n, K}^{2}$, for constant $C_{12}>0$ and $\eta_{f, n, K} = c_{11} \mathsf r_{n, K}\sqrt{\exp\{-L_{\ell}n\varepsilon_{n, K}^{2}\}}$. Thus, there exist $N_{M, n} = N(M \mathsf r_{n, K}/8, \mathcal{S}_n, d_n)$ many functions, $\{f_1, \ldots, f_{N_{M, n}}\}$, such that the $d_n$-balls of radius $M \mathsf r_{n, K}/8$ cover $\mathcal S_n$. That is, for every $f\in \mathcal{S}_n$, there exists $j\in \{1, \ldots, N_{M, n}\}$ such that, $d_{n}(f, f_j) \leq M \mathsf r_{n, K}/8$. Let
$$
\mathcal{G}_{n}(M) = \left\{f_{\bm \beta, \mathcal{T}}\in \mathcal{F}_K: d_{n}(f_{\bm \beta, \mathcal{T}}, f_0) > M \mathsf r_{n, K} \right\}.
$$
Now let $J_n$ be the set of indices whose covering balls intersect $\mathcal{G}_n(M)$. Thus, for each $j\in J_n$, there exist some $f\in \mathcal{G}_{n}(M)$ such that, $d_{n}(f, f_j) \leq M \mathsf r_{n, K}/8$. For $f\in \mathcal{G}_{n}(M)$, we have $d_{n}(f, f_0) > M \mathsf r_{n, K}$. Therefore, by triangle inequality
\begin{align}
\label{eq:tests-2}
d_{n}(f_j, f_0) \geq d_n(f, f_0) - d_n(f, f_j) > M \mathsf r_{n, K} - \frac{M \mathsf r_{n, K}}{8} = \frac{7M}{8} \mathsf r_{n, K}.
\end{align}
For each $j\in J_n$, define $\Delta_{j, i}:= f_{j}(\bm x_i) - f_0(\bm x_i)$, $i=1,\ldots, n$ and hence $\lVert \Delta_{j}\rVert_{2}^{2} = \sum_{i=1}^{n}\Delta_{j, i}^{2} = n d_n^{2}(f_j, f_0)$. Finally, define the test
\begin{align}
\label{eq:tests-3}
\phi_{n, j} := \mathds{1}\left\{T_j > 0\right\}, \quad T_j = \sum_{i=1}^{n}\Delta_{j, i}\left[y_i - \frac{f_{0}(\bm x_i) + f_{j}(\bm x_i)}{2}\right].
\end{align}
We now bound the type-I and local type-II errors of $\phi_{n, j}$ in~\hyperref[eq:tests-3]{\eqref{eq:tests-3}}.

\underline{\emph{Type-I error for $\phi_{n, j}$ in~\eqref{eq:tests-3}}}.
Under true law $\mathbb{P}_{0}^{n}$, $y_i = f_0(\bm x_i) + \epsilon_i$, where $\epsilon_i \sim \mathrm{N}(0, \sigma_0^{2})$ independently. Hence, $y_i - (f_0(\bm x_i) + f_{j}(\bm x_i))/2 = \epsilon_i - \Delta_{j, i}/2$. Thus under $\mathbb{P}_{0}^{n}$, the distribution of $T_j$ is, $T_{j} \sim \mathrm{N}(-\lVert \Delta_{j}\rVert_{2}^{2}/2, \sigma_0^{2}\lVert \Delta_j\rVert_{2}^{2})$. Hence
\begin{align}
\label{eq:tests-4}
\mathsf{E}_{\mathbb{P}_{0}^{n}}(\phi_{n, j}) = \mathbb{P}_{0}^{n}(T_j>0) = \mathsf{P}\left(Z>\frac{\lVert \Delta_j\rVert_{2}}{2\sigma_0}\right) \leq \exp\left\{-\frac{\lVert \Delta_j\rVert_{2}^{2}}{8\sigma_0^2}\right\},
\end{align}
where $Z\sim \mathrm{N}(0, 1)$ and the last inequality above follows from the Gaussian tail bound, $\mathsf{P}(Z>x) \leq \exp\{-x^{2}/2\}$. Using~\hyperref[eq:tests-2]{\eqref{eq:tests-2}}, we have $\lVert \Delta_j\rVert_{2}^{2} = nd_n^{2}(f_j, f_0) \geq (64)^{-1}49M^{2}n \mathsf r_{n, K}^{2}$. Thus, using~\hyperref[eq:tests-4]{\eqref{eq:tests-4}}
\begin{align}
\label{eq:tests-5}
\mathsf{E}_{\mathbb{P}_0^{n}}(\phi_{n, j}) \leq \exp\left\{-C_{13}M^{2} n \mathsf r_{n, K}^{2}\right\},
\end{align}
where $C_{13}={49}/{(512\sigma_0^{2})}$.

\underline{\emph{Local type-II error for $\phi_{n, j}$ in~\hyperref[eq:tests-3]{\eqref{eq:tests-3}}}}.
Now consider an alternative $(f, \sigma^2) \in A_{n, 1}(M)$ satisfying $d_{n}(f, f_j) \leq M \mathsf r_{n, K}/8$. Under $\mathbb{P}_{f, \sigma^2}^{n}$, $y_i = f(\bm x_i) + \sigma Z_i$, where $Z_i\sim \mathrm{N}(0, 1)$ independently. Then
\begin{align}
\label{eq:tests-6}
T_j = \sum_{i=1}^{n} \Delta_{j, i}\left[f(\bm x_i) - \frac{f_0(\bm x_i) + f_{j}(\bm x_i)}{2}\right] + \sigma\sum_{i=1}^{n}\Delta_{j, i}Z_i.
\end{align}
The mean of $T_{j}$ under $\mathbb{P}^{n}_{f, \sigma^2}$ is
\begin{align}
\label{eq:tests-7}
m_{j}(f) = \sum_{i=1}^{n}\Delta_{j, i}\left[f(\bm x_i) - \frac{f_0(\bm x_i) + f_{j}(\bm x_i)}{2}\right].
\end{align}
Write $f(\bm x_i) - (f_0(\bm x_i) + f_{j}(\bm x_i))/2 = \Delta_{j, i}/2 + (f(\bm x_i) - f_{j}(\bm x_i))$. Therefore, from~\hyperref[eq:tests-7]{\eqref{eq:tests-7}} we have, $m_{j}(f) = \lVert \Delta_j\rVert_{2}^{2}/2 + \sum_{i=1}^{n}\Delta_{j, i}(f(\bm x_i) - f_{j}(\bm x_i))$. By Cauchy-Schwarz inequality
\begin{align}
\label{eq:tests-8}
\left|\sum_{i=1}^{n} \Delta_{j, i}(f(\bm x_i) - f_j(\bm x_i))\right| \leq \lVert \Delta_j\rVert_{2}\sqrt{n}d_{n}(f, f_j).
\end{align}
Since $d_{n}(f, f_j) \leq M \mathsf r_{n, K}/8$ and from~\hyperref[eq:tests-2]{\eqref{eq:tests-2}}, $d_{n}(f_j, f_0) \geq 7 M \mathsf r_{n, K}/8$, we have $d_{n}(f, f_j) \leq d_{n}(f_j, f_0)/7$. The mean function $m_{j}(f)$ satisfies, $m_{j}(f) \geq 5\lVert \Delta_j\rVert_{2}^{2}/14$. Moreover, $\mathsf{V}_{\mathbb{P}^{n}_{f, \sigma^2}}(T_j) = \sigma^{2}\lVert \Delta_j\rVert_{2}^{2}$. On $A_{n, 1}(M)$, $\sigma^{2} \leq 2\sigma_0^2 \implies \mathsf{V}_{\mathbb{P}^{n}_{f, \sigma^2}}(T_j)\leq 2\sigma_0^2 \lVert \Delta_j\rVert_{2}^{2}$. The type-II error for this local ball is, $\mathsf{E}_{\mathbb{P}^{n}_{f, \sigma^2}}(1-\phi_{n, j}) = \mathbb{P}^{n}_{f, \sigma^2}(T_j\leq 0)$. Since $T_j$ is Gaussian with mean, $m_{j}(f) \geq 5\lVert \Delta_j\rVert_{2}^{2}/14$, and variance, $\mathsf{V}_{\mathbb{P}^{n}_{f, \sigma^2}}(T_j)\leq 2\sigma_0^2 \lVert \Delta_j\rVert_{2}^{2}$, using the one-sided Gaussian tail bound, we obtain
\begin{align}
\label{eq:tests-9}
\mathbb{P}_{f, \sigma^2}^{n}(T_j\leq 0) \leq \exp\left\{-\frac{5^{2}}{4\cdot 14^{2}\sigma_0^{2}}\|\Delta_j\|_{2}^{2}\right\}
\end{align}
Using again, $\lVert \Delta_j\rVert_{2}^{2} \geq (64)^{-1}49M^{2}n \mathsf r_{n, K}^{2}$, from~\hyperref[eq:tests-9]{\eqref{eq:tests-9}} we get
\begin{align}
\label{eq:tests-10}
\sup_{\substack{(f, \sigma^2)\in A_{n, 1}(M)\\ d_{n}(f, f_j) \leq M \mathsf r_{n, K}/8}} \mathsf{E}_{\mathbb{P}^{n}_{f, \sigma^2}}(1-\phi_{n, j}) \leq \exp\left\{-C_{13}' M^{2} n\mathsf r_{n, K}^{2}\right\},
\end{align}
where $C_{13}' = 25/(1024 \sigma_0^{2})$.

\underline{\emph{Combining the variance-near tests}}.
Define $\phi_{n}^{(1)} := \max_{j\in J_n}\phi_{n, j}$. Then from~\hyperref[eq:tests-5]{\eqref{eq:tests-5}} and $\log N_{M, n} \leq C_{12} n \mathsf r_{n, K}^{2}$
\begin{align}
\label{eq:tests-11}
\begin{split}
\mathsf{E}_{\mathbb{P}^{n}_{0}}(\phi^{(1)}_{n}) &\leq \sum_{j\in J_n} \mathsf{E}_{\mathbb{P}^{n}_{0}}(\phi_{n, j}) \leq N_{M, n}\exp\{-C_{13} M^{2} n \mathsf r_{n, K}^{2}\} \\
&\leq \exp\{C_{12}n \mathsf r_{n, K}^{2} -C_{13} M^{2} n \mathsf r_{n, K}^{2}\}.
\end{split}
\end{align}
Also, for every $(f, \sigma^2) \in A_{n, 1}(M)$, there exists $j\in J_n$ with $d_{n}(f, f_j) \leq M \mathsf r_{n, K}/8$. Therefore from~\hyperref[eq:tests-10]{\eqref{eq:tests-10}}
\begin{align}
\label{eq:tests-12}
\sup_{(f, \sigma^2)\in A_{n, 1}(M)}\mathsf{E}_{\mathbb{P}^{n}_{f, \sigma^2}}(1 - \phi^{(1)}_{n}) \leq \exp\{-C_{13}'M^{2} n \mathsf r_{n, K}^{2}\}.
\end{align}

\underline{\emph{Variance-far alternatives}}.
Now consider $A_{n, 2}(M)$ in~\hyperref[eq:tests-1]{\eqref{eq:tests-1}}. We show that all such alternatives are separated from the truth in averaged squared Hellinger distance by a fixed positive constant. From~\hyperref[eq:Hellinger-1]{\eqref{eq:Hellinger-1}} in \hyperref[lemma:Gaussian-Hellinger-bound]{Lemma~\ref{lemma:Gaussian-Hellinger-bound}}
\begin{align}
\label{eq:tests-13}
\int_{\mathbb{R}}\left(\sqrt{p_{{\mu_1, \sigma_0^2}}(y)} - \sqrt{p_{\mu_2, \sigma^2}(y)}\right)^{2} dy = 1 - \left(\frac{2\sigma_0 \sigma}{\sigma_0^{2} + \sigma^{2}}\right)^{\frac{1}{2}}\exp\left\{-\frac{(\mu_1 - \mu_2)^2}{4(\sigma_0^2 + \sigma^2)}\right\}.
\end{align}
For fixed $\sigma^{2}>0$,~\hyperref[eq:tests-13]{\eqref{eq:tests-13}} is minimized when $\mu_1 = \mu_2$ yielding
\begin{align}
\label{eq:tests-14}
\int_{\mathbb{R}}\left(\sqrt{p_{{\mu_1, \sigma_0^2}}(y)} - \sqrt{p_{\mu_2, \sigma^2}(y)}\right)^{2} dy \geq 1 - \left(\frac{2\sigma_0 \sigma}{\sigma_0^{2} + \sigma^{2}}\right)^{\frac{1}{2}}.
\end{align}
For $\sigma^{2} < \sigma_0^{2}/2$ or $\sigma^{2} > 2\sigma_0^2$, from~\hyperref[eq:tests-14]{\eqref{eq:tests-14}}
\begin{align}
\label{eq:tests-15}
\int_{\mathbb{R}}\left(\sqrt{p_{{\mu_1, \sigma_0^2}}(y)} - \sqrt{p_{\mu_2, \sigma^2}(y)}\right)^{2} dy \geq 1 - \left(\frac{2\sigma_0 \sigma}{\sigma_0^{2} + \sigma^{2}}\right)^{\frac{1}{2}} \geq 1 - \left(\frac{8}{9}\right)^{\frac{1}{4}}.
\end{align}
Therefore~\hyperref[eq:tests-15]{\eqref{eq:tests-15}} implies, $h_{n}^{2}\left(\mathbb{P}^{n}_{0}\parallel \mathbb{P}^{n}_{f, \sigma^2}\right) \geq 1 - (8/9)^{1/4} = C_{13}''$, for all $(f, \sigma^2)\in A_{n, 2}(M)$. By \hyperref[lemma:Hellinger-metric-entropy]{Lemma~\ref{lemma:Hellinger-metric-entropy}}, $\log N(\mathsf r_{n, K}, \Theta_n, h_n) \leq C_{11}n \mathsf r_{n, K}^{2}$. Hence $A_{n, 2}(M)\subseteq \Theta_n$ can be covered by at most $\tilde{N}_{M, n} \leq \exp\{C_{11}n \mathsf r_{n, K}^{2}\}$ balls of $h_n$-radius $\mathsf r_{n, K}$ centered at $\{(\tilde{f}_1, \tilde{\sigma}_1^2), \ldots, (\tilde{f}_{\tilde{N}_{M, n}}, \tilde{\sigma}^{2}_{\tilde{N}_{M, n}})\}$. This leads to, $A_{n, 2}(M) \subseteq \cup_{j=1}^{\tilde{N}_{M, n}} \mathcal{B}_{h_n}((\tilde{f}_{j}, \tilde{\sigma}^{2}_{j}), \mathsf r_{n, K})$, where $\mathcal{B}_{h_n}((\tilde{f}_{j}, \tilde{\sigma}^{2}_{j}), \mathsf r_{n, K}) = \{(f, \sigma^{2})\in \Theta_n: h_{n}(\mathbb{P}^{n}_{f, \sigma^{2}}\parallel \mathbb{P}^{n}_{\tilde f_j, \tilde\sigma_j^{2}}) \leq \mathsf r_{n, K}\}$. Let $\tilde{J}_{n}$ denote the indices whose covering balls intersect $A_{n, 2}{(M)}$. For each $j\in \tilde{J}_{n}$, there exist some $(f, \sigma^2) \in A_{n, 2}{(M)}$ such that, $h_{n}(\mathbb{P}^{n}_{f, \sigma^2}\parallel \mathbb{P}^{n}_{\tilde{f}_{j}, \tilde{\sigma}_{j}^{2}}) \leq \mathsf r_{n, K}$. An application of triangle inequality gives
\begin{align}
\label{eq:tests-15.1}
h_{n}\left(\mathbb{P}^{n}_{\tilde{f}_j, \tilde{\sigma}^{2}_j}\parallel \mathbb{P}^{n}_0\right) \geq h_n\left(\mathbb{P}^{n}_{f, \sigma^2}\parallel \mathbb{P}^{n}_{0}\right) - h_n\left(\mathbb{P}^{n}_{f, \sigma^2}\parallel \mathbb{P}^{n}_{\tilde{f}_j, \tilde{\sigma}_j^{2}}\right) \geq \sqrt{C_{13}''} - \mathsf r_{n, K}.
\end{align}
By \mainassref[ass:well-specified], since $\mathsf r_{n, K} \stackrel{n\to \infty}{\longrightarrow} 0$, for all sufficiently large $n$, $\mathsf r_{n, K} < \sqrt{C_{13}''}/8$. So from~\hyperref[eq:tests-15.1]{\eqref{eq:tests-15.1}}, $h_{n}(\mathbb{P}^{n}_{\tilde f_j, \tilde{\sigma}_j^2}\parallel \mathbb{P}_0^{n}) \geq 7\sqrt{C_{13}''}/8$. Moreover, for every $(f, \sigma^2)\in \mathcal{B}_{h_n}((\tilde{f}_{j}, \tilde{\sigma}^{2}_{j}), \mathsf r_{n, K})$
\begin{align}
\label{eq:tests-15.2}
\begin{split}
h_{n}\left(\mathbb{P}^{n}_{f, \sigma^2}\parallel \mathbb{P}^{n}_{0}\right) &\geq h_{n}\left(\mathbb{P}^{n}_{\tilde{f}_j, \tilde\sigma^{2}_j}\parallel \mathbb{P}^{n}_{0}\right) - h_{n}\left(\mathbb{P}^{n}_{f, \sigma^2}\parallel \mathbb{P}^{n}_{\tilde f_j, \tilde\sigma^{2}_j}\right) \geq \frac{7}{8}\sqrt{C_{13}''} - \mathsf r_{n, K} \geq \frac{3}{4}\sqrt{C_{13}''},
\end{split}
\end{align}
again for sufficiently large $n$. Therefore, for each $j\in \tilde{J}_n$
\begin{align}
\label{eq:tests-15.3}
\inf_{(f, \sigma^2)\in \mathcal{B}_{h_n}\left((\tilde f_j, \tilde\sigma^{2}_j), \mathsf r_{n, K}\right)} h_{n}\left(\mathbb{P}^{n}_{f, \sigma^2}\parallel \mathbb{P}^{n}_{0}\right) \geq \frac{3}{4}\sqrt{C_{13}''}.
\end{align}

\underline{\emph{Type-I and type-II error bounds for tests $\tilde{\phi}_{n, j}$}}.
By the standard Hellinger testing lemma for product measures~\citep[Lemma 2]{ghosal-vdv}, there exist tests $\tilde{\phi}_{n, j}$ such that
\begin{align}
\label{eq:tests-16}
\begin{gathered}
\mathsf{E}_{\mathbb{P}^{n}_0}(\tilde{\phi}_{n, j}) \leq \exp\left\{-\frac{1}{2}n h_{n}^{2}\left(\mathbb{P}^{n}_{0}\parallel \mathbb{P}^{n}_{f, \sigma^2}\right)\right\} \leq \exp\left\{-{C_{13}'''n}\right\},\\
\sup_{(f, \sigma^2)\in \mathcal{B}_{h_n}\left((\tilde f_j, \tilde\sigma^{2}_j), \mathsf r_{n, K}\right)}\mathsf{E}_{\mathbb{P}^{n}_{f, \sigma^2}}(1-\tilde{\phi}_{n, j}) \leq \exp\left\{-\frac{1}{2}n h_{n}^{2}\left(\mathbb{P}^{n}_{0}\parallel \mathbb{P}^{n}_{f, \sigma^2}\right)\right\} \leq \exp\left\{-{C_{13}'''n}\right\},
\end{gathered}
\end{align}
where $C_{13}''' = 9C_{13}''/32$.

\underline{\emph{Combining the variance-far tests}}.
Define $\phi_{n}^{(2)} := \max_{j \in \tilde{J}_n} \tilde \phi_{n, j}$. Then, by the union bound and $\tilde{N}_{M, n} \leq \exp\{C_{11}n \mathsf r_{n, K}^{2}\}$, the type-I error is
\begin{align}
\label{eq:tests-17}
\mathsf{E}_{\mathbb{P}_0^{n}}(\phi^{(2)}_{n}) \leq \exp\{C_{11}n \mathsf r_{n, K}^{2} - C_{13}''' n\}.
\end{align}
For the type-II error, take any $(f, \sigma^2) \in A_{n, 2}(M)$. Then there exist $j\in \tilde{J}_n$ such that, $(f, \sigma^2) \in \mathcal{B}_{h_n}\left((\tilde f_j, \tilde\sigma^{2}_j), \mathsf r_{n, K}\right)$ and
\begin{align}
\label{eq:tests-18}
\sup_{(f, \sigma^2)\in A_{n, 2}(M)}\mathsf{E}_{\mathbb{P}^{n}_{f, \sigma^2}}(1-\phi^{(2)}_n) \leq \exp\{-C_{13}''' n\}.
\end{align}

\underline{\emph{Combining variance-far and variance-near tests}}.
Define $\phi_n := \max\{\phi_{n}^{(1)}, \phi_{n}^{(2)}\}$. Then, $\mathsf{E}_{\mathbb{P}_0^{n}}(\phi_n) \leq \mathsf{E}_{\mathbb{P}_0^{n}}(\phi_n^{(1)}) + \mathsf{E}_{\mathbb{P}_0^{n}}(\phi_n^{(2)})$. From~\hyperref[eq:tests-11]{\eqref{eq:tests-11}} and~\hyperref[eq:tests-17]{\eqref{eq:tests-17}}
\begin{align}
\label{eq:tests-19}
\boxed{\mathsf{E}_{\mathbb{P}^{n}_{0}}(\phi_n) \leq \exp\{C_{12} n \mathsf r_{n, K}^{2} - C_{13}M^{2}n \mathsf r_{n, K}^{2}\} + \exp\{C_{11} n \mathsf r_{n, K}^{2} - C_{13}''' n\}.}
\end{align}
For $M>0$ sufficiently large, $C_{13}M^{2} > C_{12}$, so $\exp\{C_{12} n \mathsf r_{n, K}^{2} - C_{13}M^{2}n \mathsf r_{n, K}^{2}\} \stackrel{n\to \infty}{\longrightarrow} 0$ as $\mathsf r_{n, K}\stackrel{n\to \infty}{\longrightarrow} 0$ (by \mainassref[ass:well-specified]) and $n \mathsf r_{n, K}^{2} \stackrel{n\to\infty}{\longrightarrow} \infty$, yielding 
$$
\boxed{\mathsf{E}_{\mathbb{P}^{n}_{0}}(\phi_n) \stackrel{n\to\infty}{\longrightarrow} 0.}
$$

For type-II error, if $(f, \sigma^2) \in A_{n}(M)$, then $(f, \sigma^2) \in A_{n, 1}(M)$ or $(f, \sigma^2)\in A_{n, 2}(M)$. Therefore $\sup_{(f, \sigma^2)\in A_{n}(M)}\mathsf{E}_{\mathbb{P}^{n}_{f, \sigma^2}}(1-\phi_n) \leq \max\{\sup_{(f, \sigma^2)\in A_{n, 1}(M)}\mathsf{E}_{\mathbb{P}^{n}_{f, \sigma^2}}(1-\phi_n^{(1)}), \sup_{(f, \sigma^2)\in A_{n, 2}(M)}\mathsf{E}_{\mathbb{P}^{n}_{f, \sigma^2}}(1-\phi_n^{(2)})\}$ and
\begin{align}
\label{eq:tests-20}
\boxed{
\begin{aligned}
\sup_{(f, \sigma^2)\in A_{n}(M)}\mathsf{E}_{\mathbb{P}^{n}_{f, \sigma^2}}(1-\phi_n) \leq \exp\{-C_{13}'M^{2}n \mathsf r_{n, K}^{2}\} + \exp\{-C_{13}''' n\},
\end{aligned}
}
\end{align}
which follows from~\hyperref[eq:tests-12]{\eqref{eq:tests-12}} and~\hyperref[eq:tests-18]{\eqref{eq:tests-18}}. Since $n \mathsf r_{n, K}^{2} \stackrel{ n\to \infty}{\longrightarrow} \infty$
$$
\boxed{\sup_{(f, \sigma^2)\in A_{n}(M)}\mathsf{E}_{\mathbb{P}^{n}_{f, \sigma^2}}(1-\phi_n)\stackrel{n\to\infty}{\longrightarrow} 0.}
$$
{}
\end{proof}


\subsection{Proof of \texorpdfstring{\hyperref[supple-theorem:posterior-contraction]{Theorem~\ref{supple-theorem:posterior-contraction}}}{Theorem 1}}

\begin{proof}{}
The proof proceeds by aggregating the preceding ingredients: Kullback-Leibler ($\mathrm{KL}$) prior concentration, sieve complement control, metric entropy condition, and tests; following~\cite{ghosal-vdv}.
Define the likelihood ratio
\begin{align}
\label{eq:contraction-proof-1}
\mathcal{R}_n(f_{\bm \beta, \mathcal T}, \sigma^{2}) := \frac{p^{n}_{f_{\bm \beta, \mathcal T}, \sigma^{2}}(y_{1}, \ldots, y_n)}{p^{n}_{0}(y_1, \ldots, y_n)}.
\end{align}
The posterior probability of a measurable set $A \in \mathcal{F}_K \times \mathbb{R}^{+}$ is
\begin{align}
\label{eq:contraction-proof-2}
\Pi(A\mid \mathcal{D}_n) = \frac{\int_A \mathcal{R}_n(f_{\bm \beta, \mathcal T}, \sigma^2)d\Pi(f_{\bm \beta, \mathcal T}, \sigma^2)}{\mathcal Z_n}, \quad \mathcal{Z}_n = \int_{\mathcal{F}_K \times \mathbb{R}^{+}} \mathcal{R}_n(f_{\bm \beta, \mathcal T}, \sigma^2)d\Pi(f_{\bm \beta, \mathcal T}, \sigma^2).
\end{align}
Let $A_{n}(M) = \{(f_{\bm \beta, \mathcal{T}}, \sigma^2)\in \Theta_n: d_{n}(f_{\bm \beta, \mathcal{T}}, f_0) > M \mathsf r_{n, K}\}$ be as in~\hyperref[eq:alternative-set]{\eqref{eq:alternative-set}}.

\underline{\emph{High-probability lower bound of $\mathcal Z_n$}}.
By~\hyperref[lemma:KL-prior-mass]{Lemma~\ref{lemma:KL-prior-mass}}, there exist a Kullback-Leibler ($\mathrm{KL}$) neighborhood $\mathcal{B}_{\mathrm{KL}, n}$ as in~\hyperref[eq:KL-neighborhood]{\eqref{eq:KL-neighborhood}} such that, $\Pi(\mathcal{B}_{\mathrm{KL}, n}) \geq \exp\{-C_7 n\varepsilon_{n, K}^{2}\}$, and for all $(f_{\bm \beta, \mathcal T}, \sigma^2)\in \mathcal{B}_{\mathrm{KL}, n}$, $\mathrm{KL}(\mathbb{P}_{0}^{n}\parallel \mathbb{P}_{f_{\bm \beta, \mathcal T}, \sigma^2}^{n}) \leq C_{\mathrm{KL}}n\varepsilon_{n, K}^{2}$ and $\mathsf{V}_{\mathbb{P}^{n}_{0}}(\log {p_0^{n}}/{p^{n}_{f_{\bm \beta, \mathcal T}, \sigma^2}}) \leq C_{\mathrm{KL}}n\varepsilon_{n, K}^{2}$.

Define the normalized prior restricted to the $\mathrm{KL}$ ball
$$
\Pi_{\mathcal{B}_{\mathrm{KL}, n}}(\cdot) := \frac{\Pi(\cdot \cap \mathcal{B}_{\mathrm{KL}, n})}{\Pi(\mathcal B_{\mathrm{KL}, n})}.
$$
Then
\begin{align}
\label{eq:contraction-proof-3}
\begin{split}
\mathcal{Z}_n &\geq \int_{\mathcal{B}_{\mathrm{KL}, n}} \mathcal{R}_n(f_{\bm \beta, \mathcal T}, \sigma^{2})d\Pi(f_{\bm \beta, \mathcal T}, \sigma^2)
= \Pi(\mathcal{B}_{\mathrm{KL}, n})\int_{\mathcal B_{\mathrm{KL}, n}} \mathcal{R}_n(f_{\bm \beta, \mathcal T}, \sigma^{2}) d\Pi_{\mathcal B_{\mathrm{KL}, n}}(f_{\bm \beta, \mathcal T}, \sigma^2).
\end{split}
\end{align}
By Jensen's inequality
\begin{align}
\label{eq:contraction-proof-4}
\log \int_{\mathcal B_{\mathrm{KL}, n}} \mathcal{R}_n(f_{\bm \beta, \mathcal T}, \sigma^{2}) d\Pi_{\mathcal B_{\mathrm{KL}, n}}(f_{\bm \beta, \mathcal T}, \sigma^2) \geq \int_{\mathcal B_{\mathrm{KL}, n}} \log \mathcal{R}_n(f_{\bm \beta, \mathcal T}, \sigma^{2}) d\Pi_{\mathcal B_{\mathrm{KL}, n}}(f_{\bm \beta, \mathcal T}, \sigma^2).
\end{align}
Using~\hyperref[eq:contraction-proof-4]{\eqref{eq:contraction-proof-4}} in~\hyperref[eq:contraction-proof-3]{\eqref{eq:contraction-proof-3}}
\begin{align}
\label{eq:contraction-proof-5}
\log \mathcal Z_n \geq \log \Pi(\mathcal B_{\mathrm{KL}, n}) + \underbrace{\int_{\mathcal B_{\mathrm{KL}, n}} \log \mathcal{R}_n(f_{\bm \beta, \mathcal T}, \sigma^{2}) d\Pi_{\mathcal B_{\mathrm{KL}, n}}(f_{\bm \beta, \mathcal T}, \sigma^2)}_{-\tilde{\mathcal{Z}}_n}.
\end{align}
Now $\mathsf{E}_{\mathbb{P}^{n}_0}(\tilde{\mathcal{Z}}_n) = \int_{\mathcal{B}_{\mathrm{KL}, n}}\mathrm{KL}(\mathbb{P}_{0}^{n}\parallel \mathbb{P}_{f_{\bm \beta, \mathcal T}, \sigma^2}^{n})d\Pi_{\mathcal B_{\mathrm{KL}, n}}(f_{\bm \beta, \mathcal T}, \sigma^2) \leq C_{\mathrm{KL}}n\varepsilon_{n, K}^{2}$. Also, using Cauchy-Schwarz inequality for covariances
\begin{align*}
\begin{split}
\mathsf{V}_{\mathbb{P}^{n}_0}(\tilde{\mathcal{Z}}_n) \leq \int_{\mathcal{B}_{\mathrm{KL}, n}} \int_{\mathcal{B}_{\mathrm{KL}, n}}& \left[\mathsf{V}_{\mathbb{P}^{n}_{0}}\left(\log \frac{p_0^{n}}{p^{n}_{f_{\bm \beta, \mathcal T}, \sigma^2}}\right) \mathsf{V}_{\mathbb{P}^{n}_{0}}\left(\log \frac{p_0^{n}}{p^{n}_{f_{\bm \beta', \mathcal T'}, (\sigma')^2}}\right)\right]^{1/2}\\
&d\Pi_{\mathcal B_{\mathrm{KL}, n}}(f_{\bm \beta, \mathcal T}, \sigma^2) d\Pi_{\mathcal B_{\mathrm{KL}, n}}(f_{\bm \beta', \mathcal T'}, (\sigma')^2) \leq C_{\mathrm{KL}}n \varepsilon_{n, K}^{2}.
\end{split}
\end{align*}
Therefore, for any fixed $\delta>0$, Chebyshev's inequality gives, $\mathbb{P}^{n}_0(\tilde{\mathcal{Z}}_n > (C_{\mathrm{KL}} + \delta)n\varepsilon_{n, K}^{2}) \leq C_{\mathrm{KL}}/(\delta^2 n \varepsilon_{n, K}^{2}) \stackrel{n\to\infty}{\longrightarrow} 0$, as $n\varepsilon_{n, K}^{2} \stackrel{n\to\infty}{\longrightarrow} \infty$. Hence, with $\mathbb{P}_0^{n}$-probability at least $1 - C_{\mathrm{KL}}/(\delta^2n\varepsilon_{n, K}^{2})$, $\tilde{\mathcal{Z}}_n \leq (C_{\mathrm{KL}} + \delta)n\varepsilon_{n, K}^{2}$. On the event, $\mathcal{E}_{\tilde{\mathcal Z}_n} = \{\tilde{\mathcal{Z}}_n \leq (C_{\mathrm{KL}} + \delta)n\varepsilon_{n, K}^{2}\}$, from~\hyperref[eq:contraction-proof-5]{\eqref{eq:contraction-proof-5}}, we have $\log \mathcal{Z}_n \geq -C_{7} n\varepsilon_{n, K}^{2} - (C_{\mathrm{KL}} + \delta)n\varepsilon_{n, K}^{2} = -(C_7 + C_{\mathrm{KL}} + \delta)n\varepsilon_{n, K}^2$. Thus, for any $C_{\mathcal{Z}} > C_{7} + C_{\mathrm{KL}}$, choose $\delta>0$ so that, $C_{\mathcal Z} = C_{7} + C_{\mathrm{KL}} + \delta$. Then, for $\mathcal{E}_{\mathcal{Z}_n} = \{\log \mathcal{Z}_n \geq -C_{\mathcal{Z}}n\varepsilon^2_{n, K}\}$
\begin{align}
\label{eq:EZnc}
\mathbb{P}_{0}^{n}(\mathcal{E}_{\mathcal Z_n}^{c}) \stackrel{n\to\infty}{\longrightarrow} 0.
\end{align}

\underline{\emph{Decomposition into sieve and sieve complement}}.
Let $\Theta_n$ be the sieve as in~\hyperref[eq:sieve]{\eqref{eq:sieve}} in \hyperref[eq:sieve-complement]{Lemma~\ref{lemma:sieve-complement}}. Decompose $A_{n}(M)$ as, $A_{n}(M) = \{A_{n}(M) \cap \Theta_n\} \cup \{A_{n}(M) \cap \Theta_{n}^{c}\}$. Therefore
\begin{align}
\label{eq:contraction-proof-6}
\Pi(A_{n}(M)\mid \mathcal{D}_n) \leq \Pi(A_{n}(M)\cap \Theta_n\mid \mathcal{D}_n) + \Pi(\Theta_n^{c}\mid \mathcal{D}_n).
\end{align}
We treat the two terms on the right-hand-side of~\hyperref[eq:contraction-proof-6]{\eqref{eq:contraction-proof-6}} separately.

\underline{\emph{Control of the posterior mass inside the sieve}}.
By \hyperref[lemma:tests]{Lemma~\ref{lemma:tests}}, for $M>0$ sufficiently large, there exist tests $\phi_n$ such that
\begin{align}
\label{eq:contraction-proof-7}
\begin{gathered}
\mathsf{E}_{\mathbb{P}^{n}_{0}}(\phi_n) \stackrel{n\to\infty}{\longrightarrow} 0,\\
\sup_{(f_{\bm \beta, \mathcal T}, \sigma^2) \in A_{n}(M)\cap \Theta_n}\mathsf{E}_{\mathbb{P}^{n}_{f_{\bm \beta, \mathcal T}, \sigma^2}}(1-\phi_n) \leq \exp\{-C_{13}'M^{2}n \mathsf r_{n, K}^{2}\} + \exp\{-C_{13}''' n\},
\end{gathered}
\end{align}
for $C_{13}', C_{13}''' > 0$. On the event $\mathcal{E}_{\mathcal Z_n}$, $\mathcal{Z}_n^{-1} \leq \exp\{C_{\mathcal{Z}}n\varepsilon_{n, K}^{2}\}$
\begin{align}
\label{eq:contraction-proof-8}
\begin{split}
&\Pi(A_{n}(M) \cap \Theta_n\mid \mathcal D_n)\\
&\qquad = \mathcal{Z}_n^{-1}\int_{A_{n}(M)\cap \Theta_n}\mathcal{R}_n(f_{\bm \beta, \mathcal T}, \sigma^{2})d\Pi(f_{\bm \beta, \mathcal T}, \sigma^2)\\
&\qquad \leq \phi_n + (1-\phi_n)\mathcal{Z}_n^{-1}\int_{A_{n}(M)\cap \Theta_n}\mathcal{R}_n(f_{\bm \beta, \mathcal T}, \sigma^{2})d\Pi(f_{\bm \beta, \mathcal T}, \sigma^2)\\
&\qquad \leq \phi_n + \exp\{C_{\mathcal Z}n\varepsilon_{n, K}^{2}\}\int_{A_{n}(M)\cap \Theta_n}(1-\phi_n)\mathcal{R}_n(f_{\bm \beta, \mathcal T}, \sigma^{2})d\Pi(f_{\bm \beta, \mathcal T}, \sigma^2).
\end{split}
\end{align}
Taking $\mathbb{P}_0^{n}$-expectation in~\hyperref[eq:contraction-proof-8]{\eqref{eq:contraction-proof-8}}
\begin{align}
\label{eq:contraction-proof-9}
\begin{split}
&\mathsf{E}_{\mathbb{P}_0^{n}}\left[\Pi(A_{n}(M) \cap \Theta_n\mid \mathcal D_n)\mathds{1}_{\mathcal{E}_{\mathcal{Z}_n}}\right] \\
&\qquad \leq \mathsf{E}_{\mathbb{P}_0^{n}}(\phi_n) + \exp\{C_{\mathcal Z}n\varepsilon_{n, K}^{2}\}\mathsf{E}_{\mathbb{P}_0^{n}}\int_{A_{n}(M)\cap \Theta_n}(1-\phi_n)\mathcal{R}_n(f_{\bm \beta, \mathcal T}, \sigma^{2})d\Pi(f_{\bm \beta, \mathcal T}, \sigma^2)\\
&\qquad \leq \mathsf{E}_{\mathbb{P}_0^{n}}(\phi_n) + \exp\{C_{\mathcal Z}n\varepsilon_{n, K}^{2}\}\int_{A_{n}(M)\cap \Theta_n}\mathsf{E}_{\mathbb{P}_0^{n}}\left[(1-\phi_n)\mathcal{R}_n(f_{\bm \beta, \mathcal T}, \sigma^{2})\right]d\Pi(f_{\bm \beta, \mathcal T}, \sigma^2),
\end{split}
\end{align}
where the last inequality above follows from Fubini's theorem. For fixed $(f_{\bm \beta, \mathcal T}, \sigma^2)$, \\$\mathsf{E}_{\mathbb{P}_0^{n}}\left[(1-\phi_n)\mathcal{R}_n(f_{\bm\beta, \mathcal T}, \sigma^{2})\right] = \mathsf{E}_{\mathbb{P}_{f_{\bm \beta, \mathcal T}, \sigma^2}^{n}}\left[(1-\phi_n)\right]$. Using this in~\hyperref[eq:contraction-proof-9]{\eqref{eq:contraction-proof-9}} yields
\begin{align}
\label{eq:contraction-proof-10}
\begin{split}
&\mathsf{E}_{\mathbb{P}_0^{n}}\left[\Pi(A_{n}(M) \cap \Theta_n\mid \mathcal D_n)\mathds{1}_{\mathcal{E}_{\mathcal{Z}_n}}\right]\\
&\qquad \leq \mathsf{E}_{\mathbb{P}^{n}_{0}}(\phi_n) + \exp\{C_{\mathcal Z}n\varepsilon_{n, K}^{2}\}\sup_{(f_{\bm \beta, \mathcal{T}}, \sigma^2)\in A_{n}(M)\cap \Theta_n}\mathsf{E}_{\mathbb{P}_{f_{\bm \beta, \mathcal T}, \sigma^2}^{n}}(1-\phi_n)\\
&\qquad \leq \mathsf{E}_{\mathbb{P}^{n}_{0}}(\phi_n) + \exp\{C_{\mathcal Z}n\varepsilon_{n, K}^{2}\}\left[\exp\left\{-C_{13}'M^{2} n\mathsf r_{n, K}^{2}\right\} + \exp\left\{-C_{13}''' n\right\}\right],
\end{split}
\end{align}
where the last inequality above follows from~\hyperref[eq:contraction-proof-7]{\eqref{eq:contraction-proof-7}}. Observe that, as $n \varepsilon_{n, K}^{2} \leq n \mathsf r_{n, K}^{2}$, $\exp\{C_{\mathcal Z}n \varepsilon_{n, K}^{2} - C_{13}'M^{2} n \mathsf r_{n, K}^{2}\} \leq \exp\{C_{\mathcal Z}n \varepsilon_{n, K}^{2} - C_{13}'M^{2} n\varepsilon_{n, K}^{2}\} \stackrel{n\to\infty}{\longrightarrow} 0$, for $M$ sufficiently large so that $C_{13}'M^{2} > C_{\mathcal Z}$. Also since $\mathsf r_{n, K}\stackrel{n\to\infty}{\longrightarrow} 0$ by \mainassref[ass:well-specified], $n \mathsf r_{n, K}^{2} = \mathfrak{o}(n)$, and $n\varepsilon_{n, K}^{2} = \mathfrak{o}(n)$, thus yielding $\exp\{C_{\mathcal{Z}}n \varepsilon_{n, K}^{2} - C_{13}''' n\} \stackrel{n\to\infty}{\longrightarrow} 0$. Using these observations along with $\mathsf{E}_{\mathbb{P}^{n}_{0}}(\phi_n) \stackrel{n\to\infty}{\longrightarrow} 0$ and $\mathbb{P}_0^{n}(\mathcal{E}_{\mathcal{Z}_n}^{c}) \stackrel{n\to\infty}{\longrightarrow} 0$, from~\hyperref[eq:contraction-proof-10]{\eqref{eq:contraction-proof-10}} we get, $\Pi(A_{n}(M)\cap \Theta_n \mid \mathcal{D}_n) \stackrel{n\to\infty}{\longrightarrow} 0$ in $\mathbb{P}_0^{n}$-probability.

\underline{\emph{Control of the posterior mass of the sieve complement}}.
By \hyperref[lemma:sieve-complement]{Lemma~\ref{lemma:sieve-complement}}, $\Pi(\Theta_n^{c}) \leq \exp\{-C_{\Theta} n\varepsilon_{n, K}^{2}\}$, for $C_{\Theta} > 0$. Choose $C_{\Theta} > C_{\mathcal Z}$. On $\mathcal{E}_{\mathcal Z_n}$
\begin{align}
\label{eq:contraction-proof-11}
\begin{split}
\Pi(\Theta_n^{c}\mid \mathcal{D}_n) &= \mathcal{Z}_n^{-1}\int_{\Theta_n^{c}} \mathcal{R}_n(f_{\bm \beta, \mathcal T}, \sigma^{2})d\Pi(f_{\bm \beta, \mathcal T}, \sigma^2)\\
&\leq \exp\{C_{\mathcal Z}n\varepsilon_{n, K}^{2}\}\int_{\Theta_n^{c}} \mathcal{R}_n(f_{\bm \beta, \mathcal T}, \sigma^{2})d\Pi(f_{\bm \beta, \mathcal T}, \sigma^2).
\end{split}
\end{align}
Taking $\mathbb{P}_0^{n}$-expectation in~\hyperref[eq:contraction-proof-11]{\eqref{eq:contraction-proof-11}} and applying Fubini's theorem
\begin{align}
\label{eq:contraction-proof-12}
\begin{split}
\mathsf{E}_{\mathbb{P}^{n}_0}\left[\Pi(\Theta_n^{c}\mid \mathcal{D}_n)\mathds{1}_{\mathcal{E}_{\mathcal{Z}_n}}\right] &\leq \exp\{C_{\mathcal Z}n\varepsilon_{n, K}^{2}\}\int_{\Theta_n^c}\mathsf{E}_{\mathbb{P}^{n}_0}[\mathcal{R}_n(f_{\bm \beta, \mathcal T}, \sigma^{2})]d\Pi({f_{\bm \beta, \mathcal T}, \sigma^2})\\
&\leq \exp\{C_{\mathcal{Z}} n\varepsilon_{n, K}^{2}\}\Pi(\Theta_n^{c}),
\end{split}
\end{align}
where the last inequality above follows from $\mathsf{E}_{\mathbb{P}^{n}_0}[\mathcal{R}_n(f_{\bm \beta, \mathcal T}, \sigma^{2})]=1$ for every $(f_{\bm \beta, \mathcal T}, \sigma^2)$. Further, using the sieve complement bound, from~\hyperref[eq:contraction-proof-12]{\eqref{eq:contraction-proof-12}} we get
\begin{align}
\label{eq:contraction-proof-13}
\mathsf{E}_{\mathbb{P}^{n}_0}\left[\Pi(\Theta_n^{c}\mid \mathcal{D}_n)\mathds{1}_{\mathcal{E}_{\mathcal{Z}_n}}\right] \leq \exp\{-(C_{\Theta} -C_{\mathcal Z}) n\varepsilon_{n, K}^{2}\} \stackrel{n\to\infty}{\longrightarrow} 0,
\end{align}
as we have chosen $C_{\Theta} > C_{\mathcal Z}$ and also $n \varepsilon_{n, K}^{2} \stackrel{n\to \infty}{\longrightarrow}\infty$. Therefore, from~\hyperref[eq:contraction-proof-13]{\eqref{eq:contraction-proof-13}} we conclude that, $\Pi(\Theta_n^{c}\mid \mathcal D_n) \stackrel{n\to\infty}{\longrightarrow} 0$ in $\mathbb{P}_0^{n}$-probability.

\underline{\emph{Combining the bounds}}.
Recall, $\mathcal{E}_{\mathcal{Z}_n} = \{\log \mathcal{Z}_n \geq -C_{\mathcal{Z}}n\varepsilon^2_{n, K}\}$. Since posterior probabilities are bounded by $1$, we have
\begin{align}
\label{eq:contraction-proof-14}
\begin{split}
\mathsf E_{\mathbb P_0^{n}}\left[\Pi(A_n(M)\mid \mathcal D_n)\right]
&=
\mathsf E_{\mathbb P_0^{n}}\left[
\Pi(A_n(M)\mid \mathcal D_n)\mathds 1_{\mathcal E_{\mathcal Z_n}}
\right] 
+
\mathsf E_{\mathbb P_0^{n}}\left[
\Pi(A_n(M)\mid \mathcal D_n)\mathds 1_{\mathcal E_{\mathcal Z_n}^{c}}
\right] 
\\ &
\leq
\mathsf E_{\mathbb P_0^{n}}\left[
\Pi(A_n(M)\mid \mathcal D_n)\mathds 1_{\mathcal E_{\mathcal Z_n}}
\right]
+
\mathbb P_0^{n}(\mathcal E_{\mathcal Z_n}^{c}).
\end{split}
\end{align}
Furthermore note that, after multiplying $\mathds{1}_{\mathcal{E}_{\mathcal{Z}_n}}$ and taking $\mathbb{P}_0^{n}$-expectation over~\hyperref[eq:contraction-proof-6]{\eqref{eq:contraction-proof-6}} yields
\begin{align}
\label{eq:contraction-proof-15}
\begin{split}
&\mathsf E_{\mathbb P_0^{n}}\left[
\Pi(A_n(M)\mid \mathcal D_n)\mathds 1_{\mathcal E_{\mathcal Z_n}}
\right] 
\leq
\mathsf E_{\mathbb P_0^{n}}\left[
\Pi(A_n(M)\cap \Theta_n\mid \mathcal D_n)\mathds 1_{\mathcal E_{\mathcal Z_n}}
\right]
+
\mathsf E_{\mathbb P_0^{n}}\left[
\Pi(\Theta_n^c\mid \mathcal D_n)\mathds 1_{\mathcal E_{\mathcal Z_n}}
\right].
\end{split}
\end{align}
Using~\hyperref[eq:contraction-proof-13]{\eqref{eq:contraction-proof-13}} and arguments following~\hyperref[eq:contraction-proof-10]{\eqref{eq:contraction-proof-10}}, from~\hyperref[eq:contraction-proof-15]{\eqref{eq:contraction-proof-15}} we conclude that, \\
$\mathsf E_{\mathbb P_0^{n}}\left[
\Pi(A_n(M)\mid \mathcal D_n)\mathds 1_{\mathcal E_{\mathcal Z_n}}
\right] \stackrel{n\to\infty}{\longrightarrow} 0$. Finally, using $\mathbb{P}_0^{n}(\mathcal{E}_{\mathcal Z_n}^{c}) \stackrel{n\to\infty}{\longrightarrow} 0$ from~\hyperref[eq:EZnc]{\eqref{eq:EZnc}} and~\hyperref[eq:contraction-proof-15]{\eqref{eq:contraction-proof-15}} in~\hyperref[eq:contraction-proof-14]{\eqref{eq:contraction-proof-14}}, we have
$$
\boxed{\lim_{n\to \infty}\;\mathsf{E}_{\mathbb{P}_0^{n}}\left[\Pi\left(f_{\bm\beta, \mathcal T}\in \mathcal F_{K}: d_{n}(f_{\bm\beta, \mathcal T}, f_0) > M \mathsf r_{n, K} \mid \mathcal{D}_n \right)\right] = 0.}
$$
{}
\end{proof}

\subsection{Proof of \texorpdfstring{\hyperref[supple-corollary:exact-symbolic-realizability]{Corollary~\ref{supple-corollary:exact-symbolic-realizability}}}{Corollary 1}}

\begin{proof}{}
Since $f_0 \in \mathcal{F}_{K, S_0} \subseteq \mathcal{F}_{K, S}$ for $S_0\in \mathbb{N}$ for all $S\geq S_0$, it implies that, $a_{K, S, n}(f_0) = 0$ for all $S\geq S_0$. Taking $S=S_0$, for fixed $K$, $p$, and $\mathbb O$, \hyperref[supple-theorem:posterior-contraction]{Theorem~\ref{supple-theorem:posterior-contraction}} yields
$$
\boxed{n\mathsf r_{n, K}^{2} \leq \mathfrak{C}_{K, S_0, n}\left[1 + \log\left(\mathfrak{C}_{K, S_0, n} + K + p +|\mathbb O|\right)\right].}
$$
{}
\end{proof}

\newpage
\section{Details of \texorpdfstring{\hyperref[supple-theorem:sieve-truncated-population-symbolic-oracle]{Theorem~\ref{supple-theorem:sieve-truncated-population-symbolic-oracle}}}{Theorem 2}}
\label{sec:theorem-sieve-truncated-population-symbolic-oracle}

We first recall the general setting and important notations required for~\hyperref[supple-theorem:sieve-truncated-population-symbolic-oracle]{Theorem~\ref{supple-theorem:sieve-truncated-population-symbolic-oracle}}. In addition to the conditional law $\mathbb P^{n}_{f, \sigma^2}$ defined earlier for~\hyperref[supple-theorem:posterior-contraction]{Theorem~\ref{supple-theorem:posterior-contraction}}, here we operate under a random design setting that is there exist a density $q_X$ over $\mathfrak X$ such that $\bm x_i \sim q_X$ independently for $i = 1, \ldots, n$. In the sequel, we represent the joint law over $\mathcal D_n$ by $\mathbb Q^{n}_{f, \sigma^2}$ with corresponding density $q_{f, \sigma^2}^{n}$.

Consider the constants $C_B$, $\underline v$, $\overline v$, and $C_D$, satisfying  
$$
C_B > C_{\beta} + 1,
\quad
0<\underline v<\sigma_0^2,
\quad
\overline v>\sigma_0^2+a_K^2(f_0)+1,
\quad
C_D > 2\left\{R_0+\sqrt{a_K^2(f_0)+1}\right\}.
$$
Let $S_n \stackrel{n \to \infty}{\longrightarrow} \infty$ be a deterministic sequence of sieve budgets chosen such that $\rho_{n,K}$ defined as
\begin{align}
n\rho_{n,K}^2 := B^{4}_{n, K}\mathfrak C_{K,S_n,n}\left[1+ \log\bigl(\mathfrak C_{K,S_n,n}+K+p+|\mathbb O|\bigr)\right],
\end{align}
satisfies $\rho_{n,K} \stackrel{n \to \infty}{\longrightarrow} 0$ and $n\rho_{n,K}^2 \stackrel{n \to \infty}{\longrightarrow} \infty$, where $B_{n, K} = 1 + R_0 + C_B(1 + K \overline{C}_U\exp\{c_U S_n\})$.
We define the functional and parametric sieves as
\begin{align}
\label{supple-eq:sieve-misspecified}
\begin{gathered}
\mathcal F_n^D := \left\{
f_{\bm\beta,\mathcal T}\in\mathcal F_K: S(\mathcal T)\leq S_n,  \lVert\bm\beta\rVert_\infty\leq C_B, d_n(f_{\bm\beta,\mathcal T},0) \leq C_D \right\},
\\
\Theta_n^D := \left\{(f, \sigma^2):f\in\mathcal F_n^D, \underline v \leq \sigma^2 \leq \overline v \right\}.
\end{gathered}
\end{align}
Finally, let $\Pi_n^D(\cdot)$ be the observed design dependent truncation of the prior $\Pi(\cdot)$ to the sieve $\Theta_n^{D}$ in~\eqnref[supple-eq:sieve-misspecified].


\begin{theorem}[Concentration under symbolic misspecification]
\label{supple-theorem:sieve-truncated-population-symbolic-oracle}
For fixed $K$, $p$, and $\mathbb{O}$, grant \hyperref[main-ass:global-symbolic-evaluation-envelope]{\textcolor{BrickRed}{Assumptions \ref*{main-ass:global-symbolic-evaluation-envelope}, \ref*{main-ass:prior-regularity}, and \ref*{main-ass:bounded-coefficient-symbolic-approximability} of the main manuscript}}. Then there exists a sequence $\xi_n \downarrow 0$ as $n\to \infty$ such that for $\mathsf q_{n, K} = \max\{\rho_{n,K}, \xi_n\}$ and sufficiently large constants $M_{+}, M_{-} > 0$
\begin{align}
\label{eq:posterior-concentration-mispecified}
\begin{split}
\lim_{n\to \infty}\;\mathsf E_{\mathbb Q_0^{n}}
\left[
\Pi_n^D
\left(
f_{\bm\beta,\mathcal T}\in\mathcal F_n^D:
-M_{-}\mathsf q_{n,K} < d_n^2(f_{\bm\beta,\mathcal T},f_0) - a^{2}_{K}(f_0)
< M_{+}\mathsf q_{n,K}
\mid
\mathcal D_n
\right)
\right]= 1,
\end{split}
\end{align}
where $\mathbb{Q}_0^{n}$ is the joint law of $\mathcal{D}_n$ corresponding to the truth $f=f_0$ and $\sigma^2 = \sigma^{2}_0>0$.
\end{theorem}

\subsection{Auxiliary Lemma for \texorpdfstring{\hyperref[supple-theorem:sieve-truncated-population-symbolic-oracle]{Theorem~\ref{supple-theorem:sieve-truncated-population-symbolic-oracle}}}{Theorem 2}}
\label{subsec:aux-lemma-misspec}

\begin{lemma}[Empirical realization of bounded population symbolic approximants]
\label{lemma:empirical-realization-bounded-population-approximants}
Suppose\\$S_n \stackrel{n \to \infty}{\longrightarrow} \infty$ be any deterministic sequence of sieve budgets. Under \mainassref[ass:bounded-coefficient-symbolic-approximability], there exist an index sequence \(m_n \stackrel{n \to \infty}{\longrightarrow} \infty\) and a deterministic sequence \(\xi_n\downarrow0\) as $n\to \infty$ such that
\begin{align}
\label{eq:empirical-realization-bounded-population-approximants-1}
S(\mathcal T_{m_n})\leq S_n\quad\text{and}\quad\mathbb Q_0^{n} \left(d_n^2(f_{m_n},f_0) \leq
a_K^2(f_0)+\xi_n\right) \stackrel{n\to \infty}{\longrightarrow} 1.
\end{align}
\end{lemma}

\begin{proof}
Consider a fixed positive deterministic sequence $\varepsilon_j \downarrow 0$ as $j \to \infty$. For instance, one may choose $\varepsilon_j = j^{-1}$, however the rest of the proof only requires that the sequence $\{\varepsilon_j\}_{j\geq 1}$ is a positive sequence decreasing to zero. 

By \mainassref[ass:bounded-coefficient-symbolic-approximability], for every $j \in \mathbb N$, one can choose $\tilde m_j \in \mathbb N$ such that $\tilde m_j \stackrel{j \to \infty}{\longrightarrow} \infty$ and 
$\mathsf{E}_{q_X}[|f_{\tilde{m}_j}(\bm x) - f_0(\bm x)|^2] \leq a_K^2(f_0)+\varepsilon_j$.
By the weak law of large numbers, for each fixed $j \in \mathbb N$, there exists $N_j^{(1)} \in \mathbb N$ such that, for all $n \geq N_j^{(1)}$
\begin{align}
\label{eq:empirical-realization-bounded-population-approximants-2}
\mathbb Q_0^{n}
\left(d_n^2(f_{\tilde m_j},f_0)
\leq
\mathsf{E}_{q_X}[|f_{\tilde{m}_j}(\bm x) - f_0(\bm x)|^2]+\varepsilon_j
\right)
\geq 1-\varepsilon_j.
\end{align}
Since $S_n \stackrel{n \to \infty}{\longrightarrow} \infty$ and $S(\mathcal T_{\tilde{m}_j})<\infty$ for each fixed $j$, there exists $N_j^{(2)}$ such that, for all $n \geq N_j^{(2)}$, $S(\mathcal T_{\tilde m_j})\leq S_n$. Choose an increasing deterministic sequence $N_j \geq \max\{N_j^{(1)},N_j^{(2)},N_{j-1}+1\}$, with $N_0 = 0$ and define the index sequence $m_n$ piecewise as $m_n = \tilde m_j$ whenever $N_j\le n<N_{j+1}$.
Then \(m_n \stackrel{n \to \infty}{\longrightarrow} \infty\). Moreover, for $N_j\le n<N_{j+1}$, we have
$$
\boxed{
S(\mathcal T_{m_n})=S(\mathcal T_{\tilde m_j})\leq S_n.
}
$$

On the event $\{d_n^2(f_{\tilde{m}_j},f_0) \leq \mathsf{E}_{q_X}[|f_{\tilde{m}_j}(\bm x) - f_0(\bm x)|^2]+\varepsilon_j\}$, which has probability at least $1-\varepsilon_j$ by~\hyperref[eq:empirical-realization-bounded-population-approximants-2]{\eqref{eq:empirical-realization-bounded-population-approximants-2}}, and since $\mathsf{E}_{q_X}[|f_{\tilde{m}_j}(\bm x) - f_0(\bm x)|^2] \leq a_K^2(f_0)+\varepsilon_j$, we obtain
\begin{align}
\label{eq:empirical-realization-bounded-population-approximants-3}
d_n^2(f_{m_n},f_0) = d_n^2(f_{\tilde m_j},f_0) \leq \mathsf{E}_{q_X}[|f_{\tilde{m}_j}(\bm x) - f_0(\bm x)|^2]+\varepsilon_j \leq a_K^2(f_0)+ 2\varepsilon_j.
\end{align}
Therefore, for $N_j \leq n < N_{j + 1}$, $d_n^2(f_{m_n},f_0) \leq a_K^2(f_0)+2\varepsilon_j$ with probability at least $1 - \varepsilon_j$. Define $\xi_n := 2\varepsilon_j$, whenever $N_j \leq n < N_{j + 1}$. Then $\xi_n \downarrow 0$ as $n\to \infty$ and
$$
\boxed{
\mathbb Q_0^{n}
\left(
d_n^2(f_{m_n},f_0)
\le
a_K^2(f_0)+\xi_n
\right)
\stackrel{n \to \infty}{\longrightarrow} 1,
}
$$
which together with $S(\mathcal T_{m_n}) \leq S_n$ yields \eqnref[eq:empirical-realization-bounded-population-approximants-1].
\end{proof}

\subsection{Proof of \texorpdfstring{\hyperref[supple-theorem:sieve-truncated-population-symbolic-oracle]{Theorem~\ref{supple-theorem:sieve-truncated-population-symbolic-oracle}}}{Theorem 2}}
\label{subsec:proof-supple-theorem:sieve-truncated-population-symbolic-oracle}

We show
\begin{align}
\label{eq:to-show-outer}
\lim_{n\to \infty}\;\mathsf E_{\mathbb{Q}_0^{n}}
\left[
\Pi_n^D
\left(
f_{\bm\beta,\mathcal T}\in\mathcal F_n^D:
d_n^2(f_{\bm\beta,\mathcal T},f_0)
>
a_K^2(f_0)+M_{+}\mathsf q_{n,K}
\mid
\mathcal D_n
\right)
\right]
= 0,\\
\label{eq:to-show-inner}
\lim_{n\to \infty}\;\mathsf E_{\mathbb{Q}_0^{n}}
\left[
\Pi_n^D
\left(
f_{\bm\beta,\mathcal T}\in\mathcal F_n^D:
d_n^2(f_{\bm\beta,\mathcal T},f_0)
<
a_K^2(f_0)-M_{-}\mathsf q_{n,K}
\mid
\mathcal D_n
\right)
\right]
= 0,
\end{align}
in \hyperref[subsubsec:outer-radius]{Sections \ref{subsubsec:outer-radius}} and \ref{subsubsec:inner-radius}, respectively, which collectively with the union bound proves~\eqnref[eq:posterior-concentration-mispecified] in~\hyperref[supple-theorem:sieve-truncated-population-symbolic-oracle]{Theorem~\ref{supple-theorem:sieve-truncated-population-symbolic-oracle}}.

\subsubsection{Proof of \texorpdfstring{\hyperref[supple-theorem:sieve-truncated-population-symbolic-oracle]{Theorem~\ref{supple-theorem:sieve-truncated-population-symbolic-oracle}}}{Theorem 2}: Outer Radius}
\label{subsubsec:outer-radius}

\begin{proof}{}

\emph{\underline{Construction of a bounded symbolic approximate projection}}. For the deterministic sequence of sieve budgets $S_n \stackrel{n \to \infty}{\longrightarrow} \infty$, by \hyperref[lemma:empirical-realization-bounded-population-approximants]{Lemma~\ref{lemma:empirical-realization-bounded-population-approximants}}, there exist an index sequence \(m_n \stackrel{n \to \infty}{\longrightarrow} \infty\) and a deterministic sequence $\xi_n\downarrow 0$ as $n\to \infty$ such that $S(\mathcal T_{m_n}) \leq S_n$ and for the event $\Omega_{\mathfrak X, n}$ defined by
\begin{align}
\label{eq:posterior-mispecified-proof-1}
\Omega_{\mathfrak{X}, n} := \{d_n^2(f_{m_n},f_0) \leq
a_K^2(f_0)+\xi_n\},
\end{align}
we have $\mathbb Q_0^{n} \left(\Omega_{\mathfrak X, n}\right) \stackrel{n\to \infty}{\longrightarrow} 1$.

Now $\xi_n \downarrow 0$ as $n\to \infty$ and $\rho_{n,K} \stackrel{n \to \infty}{\longrightarrow} 0$ imply $\mathsf{q}_{n,K} \stackrel{n \to \infty}{\longrightarrow} 0$. Also, $n\mathsf q_{n,K}^2 = \max\{n\rho_{n,K}^2, n\xi_n^2\} \stackrel{n \to \infty}{\longrightarrow} \infty$.

Choose $f_n^\dagger = f_{\bm \beta_n^\dagger, \mathcal T_n^\dagger}$, where $\bm \beta_n^\dagger = \bm \beta_{m_n}$ and $\mathcal T_n^\dagger = \mathcal T_{m_n}$. On the high probability event $\Omega_{\mathfrak X,n}$ in \eqnref[eq:posterior-mispecified-proof-1]
\begin{align}
\label{eq:posterior-mispecified-proof-2}
S(\mathcal T_n^\dagger)\leq S_n, \quad
\lVert \bm\beta_n^\dagger\rVert_\infty \leq C_\beta, \quad
d_n^2(f_n^\dagger,f_0) \leq a_K^2(f_0)+\xi_n \leq a_K^2(f_0)+\mathsf q_{n, K}.
\end{align}
By \mainassref[ass:bounded-coefficient-symbolic-approximability], $d_n^2(f_0,0) = n^{-1}\sum_{i=1}^n |f_0(\bm x_i)|^2 \leq n^{-1}\sum_{i=1}^n R_0^2 = R_0^2$ and using \eqnref[eq:posterior-mispecified-proof-2]
\begin{align}
\label{eq:posterior-mispecified-proof-3}
d_n(f_n^\dagger,0) \leq
d_n(f_n^\dagger,f_0)+d_n(f_0,0) \leq
\sqrt{a_K^2(f_0)+ \mathsf q_{n,K}}+R_0 \leq C_D/2,
\end{align}
for all sufficiently large $n$ since $\mathsf q_{n,K} \stackrel{n \to \infty}{\longrightarrow} 0$. Further, define $v_n^\dagger := (\sigma_n^\dagger)^2 = \sigma_0^2+d_n^2(f_n^\dagger,f_0)$. Since $d_n^2(f_n^\dagger,f_0) \leq a_K^2(f_0)+\mathsf q_{n,K}$, and $\mathsf q_{n,K} \stackrel{n \to \infty}{\longrightarrow} 0$, the choice of $\underline v$ and $\overline v$ implies
$v_n^\dagger\in(\underline v,\overline v)$ for all sufficiently large $n$.

In the rest of the proof, we work on $\Omega_{\mathfrak X,n}$ and handle the $\mathbb Q_0^{n}$-probability tending to zero event $\Omega_{\mathfrak X,n}^c$ at the end.

\emph{\underline{Construction and containment of a local denominator block}}.
Let $S_n^{\dagger} = S(\mathcal T^\dagger_n)$ and $\delta_{\beta, n} = c(K+1)^{-1}U^{-1}_{S_n^\dagger} \mathsf{q}_{n, K}^{2}$, where $c>0$ is a sufficiently small constant.
Define the local block
\begin{align}
\label{eq:posterior-mispecified-proof-4}
\mathcal C_n^\dagger(c)
:=
\left\{
\mathcal T=\mathcal T_n^\dagger,\ 
\|\bm\beta-\bm\beta_n^\dagger\|_\infty\le \delta_{\beta,n},\
|\sigma^2-v_n^\dagger|\le c\mathsf q_{n,K}
\right\}.
\end{align}

We first show that $\mathcal C_n^\dagger(c)\subseteq\Theta_n^D$. For every element of \(\mathcal C_n^\dagger(c)\), \mainassref[ass:global-symbolic-evaluation-envelope] gives
\begin{align}
\label{eq:posterior-mispecified-proof-5}
d_n(f_{\bm\beta,\mathcal T_n^\dagger},f_n^\dagger)
\le
(K+1)U_{S_n^\dagger}
\|\bm\beta-\bm\beta_n^\dagger\|_\infty
\le
c \mathsf{q}_{n,K}^2.
\end{align}
By~\hyperref[eq:posterior-mispecified-proof-2]{\eqref{eq:posterior-mispecified-proof-2}}, $S(\mathcal T_n^\dagger)\le S_n$. Also, by~\hyperref[eq:posterior-mispecified-proof-2]{\eqref{eq:posterior-mispecified-proof-2}} and~\hyperref[eq:posterior-mispecified-proof-5]{\eqref{eq:posterior-mispecified-proof-5}}, $\|\bm \beta\|_{\infty} \leq \|\bm \beta_n^{\dagger}\|_{\infty} + \|\bm \beta - \bm \beta_n^{\dagger}\|_{\infty}\leq C_{\beta} + \delta_{\beta, n} \leq C_{B}$, for all sufficiently large \(n\), since \(C_B > C_\beta+1\) and \(\delta_{\beta,n} \stackrel{n \to \infty}{\longrightarrow} 0\).
Further, using triangle inequality along with~\hyperref[eq:posterior-mispecified-proof-3]{\eqref{eq:posterior-mispecified-proof-3}} and~\hyperref[eq:posterior-mispecified-proof-5]{\eqref{eq:posterior-mispecified-proof-5}} we have, $d_n(f_{\bm \beta, \mathcal{T}_n^\dagger}, 0) \leq d_n(f_{\bm \beta, \mathcal{T}_n^\dagger}, f_n^{\dagger}) + d_n(f_n^\dagger, 0) \leq c\mathsf{q}_{n, K}^{2} + C_D/2 \leq C_D$, for all sufficiently large $n$.
Finally, because $|\sigma^2 - v_n^{\dagger}| \leq c\mathsf{q}_{n, K}$ as in~\hyperref[eq:posterior-mispecified-proof-4]{\eqref{eq:posterior-mispecified-proof-4}} and $v_n^{\dagger}\in (\underline v, \overline v)$ for all large $n$, we have $\underline{v} \leq \sigma^2 \leq \overline{v}$. Combining these observations, we conclude that $\mathcal C_n^\dagger(c)\subseteq\Theta_n^D$.

\emph{\underline{Local prior mass lower bound}}.
Since $\mathcal C_n^\dagger(c)\subseteq\Theta_n^D$, we have $\Pi_{n}^{D}(\mathcal{C}_n^{\dagger}(c)) = \Pi(\mathcal C_n^{\dagger}(c)) / \Pi(\Theta_n^D) \geq \Pi(\mathcal{C}_n^{\dagger}(c))$. We now lower bound the original prior mass of \(\mathcal C_n^\dagger(c)\)that is $\Pi(\mathcal{C}_n^{\dagger}(c))$.

Note that
\begin{align}
\label{eq:posterior-mispecified-proof-6}
\begin{split}
\Pi\left(\mathcal C_n^\dagger(c)\right) &= \Pi_{\mathrm{forest}, K}\left(\mathcal{T}_n^{\dagger}\mid \alpha_0, \delta_0, \bm \alpha_{\mathrm{op}}, \bm \alpha_{\mathrm{ft}}\right)\\
&\qquad \times \mathrm{NIG}_{K+1}\left(\lVert \bm \beta -\bm \beta_n^{\dagger}\rVert_{\infty}\leq \delta_{\beta, n}, |\sigma^2 - v_n^{\dagger}| \leq c \mathsf{q}_{n, K}\mid \bm \mu_{\beta}, \bm\Sigma_{\beta}, \nu, \lambda\right).
\end{split}
\end{align}
By \hyperref[lemma:prior-mass-forest]{Lemma~\ref{lemma:prior-mass-forest}}, the prior mass of $\mathcal{T}_n^{\dagger}$ in~\hyperref[eq:posterior-mispecified-proof-6]{\eqref{eq:posterior-mispecified-proof-6}} is
\begin{align}
\label{eq:posterior-mispecified-proof-7}
\begin{split}
&\Pi_{\mathrm{forest}, K}\left(\mathcal T=\mathcal T_n^\dagger\mid \alpha_0, \delta_0, \bm \alpha_{\mathrm{op}}, \bm \alpha_{\mathrm{ft}}\right)\\
&\qquad \geq \exp\left\{-C_4\left[S_n^\dagger(1 + \log(S_n^\dagger+|\mathbb O|)) + (S_n^\dagger+K)(1 + \log(p+S_n^\dagger+K))\right]\right\}\\
&\qquad \geq \exp\left\{-C_4\left[S_n (1 + \log(S_n+|\mathbb O|)) + (S_n+K)(1 + \log(p+S_n+K))\right]\right\}\\
&\qquad \geq \exp\left\{-C_4 \mathfrak{C}_{K, S_n, n} \right\}
\geq \exp\left\{-C_4 n\rho_{n, K}^2 \right\} 
\geq \exp\left\{-C_4 n\mathsf q_{n, K}^2 \right\},
\end{split}
\end{align}
since $S_n^\dagger \leq  S_n$ and $\mathfrak{C}_{K, S_n, n} \leq n \rho^{2}_{n, K} \leq n \mathsf{q}_{n, K}^{2}$, where $C_4 > 0$. Now, we consider the prior mass of the outer regression coefficient and model variance parts in~\hyperref[eq:posterior-mispecified-proof-6]{\eqref{eq:posterior-mispecified-proof-6}}. Since $\{\lVert \bm \beta\rVert_{\infty} \leq C_B, \sigma^2 \in [\underline v, \overline v]\}$ is compact, the continuous $\mathrm{NIG}$ prior density attains an infimum $m_{\beta, \sigma^2} > 0$ on it. Further, using preceding arguments $\{\lVert \bm \beta -\bm \beta_n^{\dagger}\rVert_{\infty}\leq \delta_{\beta, n}, |\sigma^2 - v_n^{\dagger}| \leq c \mathsf{q}_{n, K}\} \subset \{\lVert \bm \beta\rVert_{\infty} \leq C_B, \sigma^2 \in [\underline v, \overline v]\}$, the same lower bound holds on $\{\lVert \bm \beta -\bm \beta_n^{\dagger}\rVert_{\infty}\leq \delta_{\beta, n}, |\sigma^2 - v_n^{\dagger}| \leq c \mathsf{q}_{n, K}\}$. Then
\begin{align}
\label{eq:posterior-mispecified-proof-8}
\begin{split}
&\mathrm{NIG}_{K+1}\left(\lVert \bm \beta -\bm \beta_n^{\dagger}\rVert_{\infty}\leq \delta_{\beta, n}, |\sigma^2 - v_n^{\dagger}| \leq c \mathsf{q}_{n, K}\mid \bm \mu_{\beta}, \bm\Sigma_{\beta}, \nu, \lambda\right)\\
&\quad \geq m_{\beta, \sigma^2} \times \mathrm{Vol}\left(\{\lVert \bm \beta -\bm \beta_n^{\dagger}\rVert_{\infty}\leq \delta_{\beta, n}, |\sigma^2 - v_n^{\dagger}| \leq c \mathsf{q}_{n, K}\}\right)\\
&\quad= m_{\beta, \sigma^{2}} 2c \mathsf{q}_{n, K}(2\delta_{\beta, n})^{K+1}\\
&\quad= m_{\beta, \sigma^2} 2^{K+2} c\mathsf{q}_{n, K}\left(c (K+1)^{-1}U_{S_n^\dagger}^{-1}\mathsf{q}^{2}_{n, K}\right)^{K+1}\\
&\quad \geq C^{\star} \mathsf{q}_{n, K}^{2K+3} (\overline{C}_U)^{-(K+1)}\exp\left\{-(K+1)c_U S_n^{\dagger}\right\}\\
&\quad = C^{\star \star} \exp\left\{-c^{\star}\left[S_n^{\dagger} + \log(1 / \mathsf{q}_{n, K})\right]\right\},
\end{split}
\end{align}
where the penultimate inequality follows from~\mainassref[ass:global-symbolic-evaluation-envelope], with $C^{\star\star} = C^{\star}(\overline{C}_U)^{-(K+1)}$, $C^{\star} =m_{\beta, \sigma^{2}} (2c)^{K+2}(K+1)^{-(K+1)}$, and $c^{\star} = \max\{(K+1) c_U, 2K+3\}$. Using, $S_n^\dagger \leq S_n \leq \mathfrak{C}_{K, S_n, n} \leq n \rho_{n, K}^{2} \leq n \mathsf{q}_{n, K}^{2}$ and $\log(1/\mathsf{q}_{n, K}) \leq \log(1/\rho_{n, K}) \leq \log n/2 - \log(n \rho_{n, K}^{2})/2 \leq \log n /2 \leq \mathfrak{C}_{K, S_n, n}\leq n \rho_{n, K}^{2} \leq n\mathsf{q}_{n, K}^{2}$, from~\hyperref[eq:posterior-mispecified-proof-8]{\eqref{eq:posterior-mispecified-proof-8}} we obtain
\begin{align}
\label{eq:posterior-mispecified-proof-9}
\mathrm{NIG}_{K+1}\left(\lVert \bm \beta -\bm \beta_n^{\dagger}\rVert_{\infty}\leq \delta_{\beta, n}, |\sigma^2 - v_n^{\dagger}| \leq c \mathsf{q}_{n, K}\mid \bm \mu_{\beta}, \bm\Sigma_{\beta}, \nu, \lambda\right) \geq \exp\left\{-\tilde{C} n \mathsf{q}_{n, K}^{2}\right\},
\end{align}
for some constant $\tilde{C} > 0$. Using~\hyperref[eq:posterior-mispecified-proof-7]{\eqref{eq:posterior-mispecified-proof-7}} and~\hyperref[eq:posterior-mispecified-proof-9]{\eqref{eq:posterior-mispecified-proof-9}} in~\hyperref[eq:posterior-mispecified-proof-6]{\eqref{eq:posterior-mispecified-proof-6}}, we conclude that
\begin{align}
\label{eq:posterior-mispecified-proof-10}
\Pi_n^D\left(\mathcal C_n^\dagger(c)\right)
\ge
\exp\left\{-C^{\dagger} n \mathsf q_{n,K}^2\right\},
\end{align}
for some $C^{\dagger} = C_4 + \tilde{C} > 0$.

\emph{\underline{Expectation and variance bounds of the $\log$-likelihood ratio on the local block $\mathcal{C}_n^{\dagger}(c)$}}.
Here, we use the proof strategy of \hyperref[lemma:KL-control]{Lemma~\ref{lemma:KL-control}}. From~\hyperref[eq:posterior-mispecified-proof-5]{\eqref{eq:posterior-mispecified-proof-5}}, $d_n(f_{\bm \beta, \mathcal{T}_n^{\dagger}}, f^{\dagger}_n) \leq c \mathsf{q}_{n, K}^{2}$. Therefore
\begin{align}
\label{eq:posterior-mispecified-proof-11}
\begin{split}
|d_n^{2}(f_{\bm \beta, \mathcal T_n^{\dagger}}, f_0) - d_n^{2}(f_n^{\dagger}, f_0)| &\leq 2 d_n(f_n^{\dagger}, f_0)d_n(f_{\bm \beta, \mathcal{T}_n^{\dagger}}, f_n^{\dagger}) + d_n^{2}(f_{\bm\beta, \mathcal{T}_n^{\dagger}}, f_n^{\dagger})\\
&\leq 2 d_n(f_n^{\dagger}, f_0) c \mathsf{q}^{2}_{n, K} + c^{2}\mathsf{q}^{4}_{n, K}\\
&\leq 2c \mathsf{q}_{n, K}^{2}(a^{2}_K(f_0) + \mathsf{q}_{n, K})^{\frac 12} + c^{2}\mathsf{q}^{4}_{n, K}\\
&\leq C^{\dagger \dagger} \mathsf{q}_{n, K}^{2},
\end{split}
\end{align}
for some $C^{\dagger \dagger} > 0$. Also inside $\mathcal{C}_n^{\dagger}(c)$, $|\sigma^{2} - v_n^{\dagger}| \leq c \mathsf{q}_{n, K}$. For Gaussian likelihoods
\begin{align}
\label{eq:posterior-mispecified-proof-12}
\begin{split}
\mathsf E_{\mathbb{Q}_0^{n}}\left(
\log
\frac{
q_{f_n^\dagger,v_n^\dagger}^{n}
}{
q_{f_{\bm \beta, \mathcal{T}_n^{\dagger}},\sigma^2}^{n}
}\right)
&=
\frac n2
\left[
\log\left(\frac {\sigma^2}{v_n^\dagger}\right)
+
\frac{\sigma_0^2+d_n^{2}(f_{\bm \beta, \mathcal{T}_n^{\dagger}}, f_0)}{\sigma^2}
-
\frac{\sigma_0^2+d_n^{2}(f_n^{\dagger}, f_0)}{v_n^\dagger}
\right]\\
&= \frac n2
\left[
\log\left(\frac {\sigma^2}{v_n^\dagger}\right)
+
\frac{\sigma_0^2+d_n^{2}(f_{\bm \beta, \mathcal{T}_n^{\dagger}}, f_0)}{\sigma^2}
-
1
\right]\\
&\leq \tilde{C}^{\dagger \dagger} n \mathsf{q}_{n, K}^{2},
\end{split}
\end{align}
where the last inequality follows from the same arguments as in the proof of \hyperref[lemma:KL-control]{Lemma~\ref{lemma:KL-control}} for some $\tilde{C}^{\dagger \dagger} > 0$. Once again, using the proof of \hyperref[lemma:KL-control]{Lemma~\ref{lemma:KL-control}}, for some $\tilde{\tilde{C}}^{\dagger \dagger} > 0$
\begin{align}
\label{eq:posterior-mispecified-proof-13}
\mathsf{V}_{\mathbb Q_0^{n}}
\left(
\log
\frac{
q_{f_n^\dagger,v_n^\dagger}^{n}
}{
q_{f_{\bm \beta, \mathcal{T}_n^{\dagger}},\sigma^2}^{n}
}
\right)
\le
\tilde{\tilde{C}}^{\dagger \dagger} n \mathsf{q}_{n,K}^2.
\end{align}

\emph{\underline{Denominator lower bound}}.
For $(f, \sigma^2) \in \Theta_n^{D}$, define the likelihood ratio relative to the approximating center $(f_n^\dagger, v_n^{\dagger})$ by
\begin{align}
\label{eq:posterior-mispecified-proof-14}
\begin{split}
\tilde{\mathcal{R}}_n\left(f, \sigma^2\right) := \frac{q_{f, \sigma^2}^{n}(\mathcal D_n)}{q_{f_n^\dagger, v_n^\dagger}^{n}(\mathcal{D}_n)}.
\end{split}
\end{align}
The sieve-truncated posterior denominator is
\begin{align}
\label{eq:posterior-mispecified-proof-15}
\begin{split}
\tilde{\mathcal{Z}}_n = \int_{\Theta_n^D} \tilde{\mathcal R}_n(f, \sigma^2) d\Pi_n^{D}\left(f, \sigma^2\right).
\end{split}
\end{align}
Since $\mathcal C_n^\dagger(c)\subseteq\Theta_n^D$, we have
\begin{align}
\label{eq:posterior-mispecified-proof-16}
\tilde{\mathcal{Z}_n} \geq \Pi_{n}^{D}\left(\mathcal{C}_n^{\dagger}(c)\right)\int_{\mathcal{C}_n^{\dagger}(c)}\tilde{\mathcal{R}}_n(f, \sigma^2)d\Pi_{\mathcal{C}_n^{\dagger}(c)}(f, \sigma^2),
\end{align}
where \(\Pi_{\mathcal C_n^\dagger}\) denotes the normalized restriction of \(\Pi_n^D(\cdot)\) to \(\mathcal C_n^\dagger(c)\).
By Jensen's inequality
\begin{align}
\label{eq:posterior-mispecified-proof-17}
\log
\int_{\mathcal C_n^\dagger(c)}
\tilde{\mathcal{R}}_n(f,\sigma^2)
\,d\Pi_{\mathcal C_n^\dagger}(f,\sigma^2)
\ge
\int_{\mathcal C_n^\dagger(c)}
\log \tilde{\mathcal{R}}_n(f,\sigma^2)
\,d\Pi_{\mathcal C_n^\dagger}(f,\sigma^2).
\end{align}
Using~\hyperref[eq:posterior-mispecified-proof-17]{\eqref{eq:posterior-mispecified-proof-17}} in~\hyperref[eq:posterior-mispecified-proof-16]{\eqref{eq:posterior-mispecified-proof-16}}, we get
\begin{align}
\label{eq:posterior-mispecified-proof-18}
\begin{split}
\log \tilde{\mathcal{Z}}_n &\geq \log \Pi_n^{D}\left (\mathcal{C}_n^{\dagger}(c)\right ) + \int_{\mathcal C_n^\dagger(c)}
\log \tilde{\mathcal{R}}_n(f,\sigma^2)
\,d\Pi_{\mathcal C_n^\dagger}(f,\sigma^2)\\
&\geq -C^{\dagger} n \mathsf{q}_{n, K}^{2} + \int_{\mathcal C_n^\dagger(c)}
\log \tilde{\mathcal{R}}_n(f,\sigma^2)
\,d\Pi_{\mathcal C_n^\dagger}(f,\sigma^2),
\end{split}
\end{align}
where the last inequality above follows from~\hyperref[eq:posterior-mispecified-proof-10]{\eqref{eq:posterior-mispecified-proof-10}}. Note that, $\mathsf{E}_{\mathbb{Q}_0^{n}}(-\log \tilde{\mathcal R}_n(f, \sigma^2)) \leq \tilde{C}^{\dagger \dagger} n \mathsf{q}_{n, K}^{2}$ from~\hyperref[eq:posterior-mispecified-proof-12]{\eqref{eq:posterior-mispecified-proof-12}} as well as $\mathsf{V}_{\mathbb{Q}_0^{n}}(-\log \mathcal{\tilde{R}}_n(f, \sigma^2)) \leq \tilde{\tilde{C}}^{\dagger\dagger} n \mathsf{q}_{n, K}^{2}$ from~\hyperref[eq:posterior-mispecified-proof-13]{\eqref{eq:posterior-mispecified-proof-13}}. Following the arguments in the proof of~\hyperref[supple-theorem:posterior-contraction]{Theorem~\ref{supple-theorem:posterior-contraction}} (while obtaining a high-probability lower bound of the posterior denominator $\mathcal{Z}_n$), by Chebyshev's inequality, with $\mathcal{E}_{\tilde{\mathcal{Z}}} = \{\tilde{\mathcal{Z}}_n \geq \exp\{-C_{\tilde{\mathcal Z}}n \mathsf{q}_{n, K}^{2}\}\}$ for some $C_{\tilde{\mathcal{Z}}} > 0$, we have
\begin{align}
\label{eq:posterior-mispecified-proof-19}
\mathbb{Q}_0^{n}\left(\Omega_{\mathfrak X, n} \setminus \mathcal{E}_{\tilde{\mathcal Z}_n} \right) \stackrel{n \to \infty}{\longrightarrow} 0.
\end{align}

\emph{\underline{Metric entropy and empirical process bound}}.
Define $G_n(f) := 2n^{-1}\sum_{i=1}^{n}\epsilon_i\{f(\bm x_i) - f_n^{\dagger}(\bm x_i)\}$, where $\epsilon_i \sim \mathrm{N}(0, \sigma_0^{2})$ independently for $i=1,\ldots, n$ and $\sigma_0^2>0$. Conditional on the design matrix $\bm X$, observe that $G_{n}(f)$ is a centered Gaussian process with canonical metric $d_{G}(f, g) = (\mathsf{E}_{\mathbb{Q}_0^{n}}[G_n(f) - G_n(g)]^2)^{1/2} = 2\sigma_0n^{-1/2}d_n(f, g)$.

Now, we bound the metric entropy of $\mathcal{F}_n^{D}$ in~\hyperref[supple-eq:sieve-misspecified]{\eqref{supple-eq:sieve-misspecified}}. The number of symbolic forests with total operator count at most $S_n$ that is $N_{\mathcal{T}}(K, S_n)$ is bounded as
\begin{align}
\label{eq:posterior-mispecified-proof-20}
N_{\mathcal{T}}(K, S_n) \leq \exp\left\{(2S_n + K +1)\log 3 + S_n\log|\mathbb O| + (S_n+K)\log p\right\},
\end{align}
by \hyperref[lemma:symbolic-forest-count]{Lemma~\ref{lemma:symbolic-forest-count}}. For a fixed symbolic forest $\mathcal T \in \mathbb{T}_{\mathbb O, p}^{K}$ with $S(\mathcal{T}) \leq S_n$ and for $\bm \beta, \bm \beta' \in \mathbb{R}^{K+1}$, \mainassref[ass:global-symbolic-evaluation-envelope] gives
\begin{align}
\label{eq:posterior-mispecified-proof-21}
d_{n}(f_{\bm \beta, \mathcal T}, f_{\bm \beta', \mathcal{T}}) \leq (K+1) U_{S_n}\lVert \bm \beta - \bm \beta'\rVert_{\infty}.
\end{align}
Therefore, the coefficient cube $\{\|\bm \beta\|_{\infty} \leq C_{B}\}$ can be covered by at most $(1 + (2C_B(K+1)U_{S_n})/u)^{K+1}$ balls of $d_n$-radius $u$. Combining this with~\hyperref[eq:posterior-mispecified-proof-20]{\eqref{eq:posterior-mispecified-proof-20}} yields
\begin{align}
\label{eq:posterior-mispecified-proof-22}
\begin{split}
\log N(u, \mathcal F_n^{D}, d_n) &\leq (2S_n + K +1)\log 3 + S_n \log|\mathbb O| + (S_n+K)\log p
\\ &\qquad 
+ (K+1)\log \left(1 + \frac{2C_B(K+1)U_{S_n}}{u}\right)\\
&\leq (2S_n + K +1)\log 3 + S_n \log|\mathbb O| + (S_n+K)\log p
\\&\qquad 
+ (K+1)\log\left(1 + \frac{2C_B(K+1)\overline{C}_U\exp\{c_US_n\}}{u}\right)\\
&\leq \mathcal{C}^{\star}\mathfrak{C}_{K, S_n, n} + \mathcal{C}^{\star \star}(K+1)\log\left(\frac{\mathcal{C}^{\star\star\star}}{u}\right),
\end{split}
\end{align}
for large enough constants $\mathcal{C}^{\star}, \mathcal C^{\star \star}, \mathcal{C}^{\star \star \star} > 0$ and where $\mathfrak C_{K, S_n, n}$ absorbs all $S_n$ dependent terms. Additionally the constant $\mathcal{C}^{\star \star \star}$ is chosen to be larger than $3C_D/2$. The penultimate inequality in the above display follows from~\mainassref[ass:global-symbolic-evaluation-envelope]. Furthermore, the last inequality in the above display follows as $u$ can be chosen less than the $d_n$-diameter of $\mathcal F_n^{D}$ which can be bounded as
\begin{align}
\label{eq:posterior-mispecified-proof-23}
d_n(f, f_n^{\dagger}) \leq d_n(f, 0) + d_{n}(f_n^{\dagger}, 0) \leq \frac{3C_D}{2},
\end{align}
following from triangle inequality, definition of $\mathcal{F}_n^D$ in~\hyperref[supple-eq:sieve-misspecified]{\eqref{supple-eq:sieve-misspecified}}, and~\hyperref[eq:posterior-mispecified-proof-3]{\eqref{eq:posterior-mispecified-proof-3}}. Therefore from~\hyperref[eq:posterior-mispecified-proof-22]{\eqref{eq:posterior-mispecified-proof-22}}
\begin{align}
\label{eq:posterior-mispecified-proof-24}
\begin{split}
&\int_{0}^{3C_D/2}\sqrt{\log N(u, \mathcal{F}_n^D, d_n)}\;du
\leq \frac{3C_D}{2}\sqrt{\mathcal{C}^{\star}\mathfrak{C}_{K, S_n, n}} + \sqrt{\mathcal{C}^{\star \star}(K+1)}\int_{0}^{3C_D/2}\sqrt{\log \frac{\mathcal{C}^{\star\star\star}}{u}}\;du.
\end{split}
\end{align}
The integral on the right hand side of \eqnref[eq:posterior-mispecified-proof-24] is a constant depending on $C_D$ and $\mathcal C^{\star\star\star}$ as $\mathcal C^{\star\star\star}$ was chosen greater than $3C_D/2$. Hence
\begin{align}
\label{eq:posterior-mispecified-proof-25}
\int_{0}^{3C_D/2}\sqrt{\log N(u, \mathcal{F}_n^D, d_n)}\;du \leq \tilde{\mathcal{C}}\sqrt{\mathfrak{C}_{K, S_n, n}} \leq \tilde{\mathcal C}\sqrt{n}\rho_{n, K} \leq \tilde{\mathcal C}\sqrt{n}\mathsf{q}_{n, K},
\end{align}
for some constant $\tilde{\mathcal{C}} > 0$. Using Dudley's entropy bound~\citep{gine-nickl2016,vershynin2018} with some $\tilde{\tilde{\mathcal{C}}} > 0$ and~\hyperref[eq:posterior-mispecified-proof-25]{\eqref{eq:posterior-mispecified-proof-25}}
\begin{align}
\label{eq:posterior-mispecified-proof-26}
\mathsf{E}_{\mathbb{Q}_0^{n}}\left(\sup_{f\in \mathcal F_n^D}|G_n(f)|\right) \leq \frac{\tilde{\tilde{\mathcal{C}}}}{\sqrt{n}}\int_{0}^{3C_D/2}\sqrt{\log N(u, \mathcal{F}_n^D, d_n)}\;du \leq \tilde{\tilde{\mathcal{C}}}^{\star}\mathsf{q}_{n, K},
\end{align}
where $\tilde{\tilde{\mathcal{C}}}^{\star} = \tilde{\mathcal{C}} \tilde{\tilde{\mathcal{C}}} > 0$. Also, $\sup_{f\in \mathcal{F}_n^{D}}\mathsf{V}_{\mathbb{Q}_0^{n}}(G_n(f)) = \sup_{f\in \mathcal{F}_n^{D}}4\sigma_0^{2}d_n^{2}(f, f_n^{\dagger})/n \leq 9\sigma_0^{2}C_D^{2}/n$ from~\hyperref[eq:posterior-mispecified-proof-23]{\eqref{eq:posterior-mispecified-proof-23}}. Hence, an application of Borell-TIS inequality~\citep{adler-taylor2007} gives
\begin{align}
\label{eq:posterior-mispecified-proof-27}
\sup_{f\in\mathcal F_n^D} |G_n(f)| = \mathcal{O}_{\mathbb{Q}_0^{n}}(\mathsf q_{n,K}).  
\end{align}

\emph{\underline{Residual separation}}.
Define $\hat{Q}_n(f) := n^{-1}\sum_{i=1}^{n}\{y_i - f(\bm x_i)\}^2$. For every $f:\mathfrak{X}\to \mathbb{R}$
\begin{align}
\label{eq:posterior-mispecified-proof-28}
\hat{Q}_n(f) -\hat{Q}_n(f_n^{\dagger}) = d_n^{2}(f, f_0) - d_n^{2}(f_n^{\dagger}, f_0) - G_n(f).
\end{align}
Let $A_n^{+}(M_{+}) = \{f_{\bm \beta, \mathcal T}\in \mathcal{F}_n^{D}: d_{n}^{2}(f_{\bm \beta, \mathcal{T}}, f_0) > a^{2}_K(f_0) + M_{+} \mathsf{q}_{n, K}\}$, where $M_{+}>0$. For $f_{\bm \beta, \mathcal T} \in A_n^{+}(M_{+})$, $d_n^{2}(f_{\bm \beta, \mathcal T}, f_0) - d_n^{2}(f_n^{\dagger}, f_0) > (M_{+}-1)\mathsf{q}_{n, K}$, using~\hyperref[eq:posterior-mispecified-proof-2]{\eqref{eq:posterior-mispecified-proof-2}}. By~\hyperref[eq:posterior-mispecified-proof-27]{\eqref{eq:posterior-mispecified-proof-27}}, $\sup_{f_{\bm \beta, \mathcal{T}}\in\mathcal F_n^D} |G_n(f_{\bm \beta, \mathcal{T}})| = \mathcal{O}_{\mathbb{Q}_0^{n}}(\mathsf q_{n,K})$. Thus, for sufficiently large $M_{+}$, there exist $c_M > 0$ such that
\begin{align}
\label{eq:posterior-mispecified-proof-29}
\mathbb{Q}_0^{n}\left(\left\{\inf_{f_{\bm \beta, \mathcal{T}}\in A_{n}^{+}(M_{+})}\left\{\hat{Q}_n(f_{\bm \beta, \mathcal{T}}) - \hat{Q}_n(f_n^{\dagger})\right\} \geq 2c_M \mathsf{q}_{n, K}\right\}\cap \Omega_{\mathfrak X, n}\right) \stackrel{n\to \infty}{\longrightarrow} 1.
\end{align}
Next
\begin{align}
\label{eq:posterior-mispecified-proof-30}
\begin{split}
\hat{Q}_n(f_n^{\dagger}) - v_n^{\dagger} &= \frac2n\sum_{i=1}^n
\{f_0(\bm x_i)-f_n^\dagger(\bm x_i)\}\epsilon_i
+
\left(
\frac1n\sum_{i=1}^n\epsilon_i^2-\sigma_0^2
\right).
\end{split}
\end{align}
In \eqnref[eq:posterior-mispecified-proof-30], the quantity $2 n^{-1}\sum_{i=1}^n \{f_0(\bm x_i)-f_n^\dagger(\bm x_i)\}\epsilon_i \sim \mathrm{N}(0, 4n^{-1}\sigma_0^2d^2_n(f_0, f_n^\dagger))$, where $d_n(f_0, f_n^\dagger) \leq d_n(f_0, 0) + d_n(f_n^{\dagger}, 0) \leq 3C_D/2$ by an application of triangle inequality which implies $2 n^{-1}\sum_{i=1}^n \{f_0(\bm x_i)-f_n^\dagger(\bm x_i)\}\epsilon_i = \mathcal{O}_{\mathbb{Q}_0^{n}}(n^{-\frac{1}{2}})$. Also $n^{-1}\sum_{i=1}^n\epsilon_i^2-\sigma_0^2 = \mathcal{O}_{\mathbb{Q}_0^{n}}(n^{-\frac{1}{2}})$. As $n \mathsf{q}_{n, K}^{2} \stackrel{n\to \infty}{\longrightarrow} \infty$, we have $\hat{Q}_n(f_n^{\dagger}) - v_n^{\dagger} =  \mathcal{O}_{\mathbb{Q}_0^{n}}(n^{-\frac{1}{2}}) = \mathfrak{o}_{\mathbb{Q}_0^{n}}(\mathsf{q}_{n, K})$, and further combining this with~\hyperref[eq:posterior-mispecified-proof-29]{\eqref{eq:posterior-mispecified-proof-29}} gives
\begin{align}
\label{eq:posterior-mispecified-proof-31}
\mathbb{Q}_0^{n}\left(\left\{\inf_{f_{\bm \beta, \mathcal T}\in A_n^{+}(M_{+})} \hat Q_n(f_{\bm \beta, \mathcal T}) \ge v_n^\dagger+c_M \mathsf q_{n,K}\right\}\cap \Omega_{\mathfrak X, n}\right) \stackrel{n\to \infty}{\longrightarrow} 1.
\end{align}

\emph{\underline{Numerator bound}}.
For any $f:\mathfrak{X} \to \mathbb{R}$ and any \(\sigma^2>0\), the Gaussian likelihood satisfies the profile bound
\begin{align}
\label{eq:posterior-mispecified-proof-32}
  q_{f,\sigma^2}^{n}(y_1,\ldots,y_n \mid \bm X) \leq (2\pi\hat Q_n(f))^{-\frac{n}{2}}\exp\left\{-\frac{n}{2}\right\}.  
\end{align}
Therefore, uniformly over $f_{\bm \beta, \mathcal{T}}\in A_n^{+}(M_{+})$ and \(\sigma^2\in[\underline v,\overline v]\)
\begin{align}
\label{eq:posterior-mispecified-proof-33}
\log \left(\frac{q_{f_{\bm \beta, \mathcal T},\sigma^2}^{n}(\mathcal D_n)}{q_{f_n^\dagger,v_n^\dagger}^{n}(\mathcal D_n)}\right)
\leq \frac n2
\log
\left(
\frac{v_n^\dagger}{\hat Q_n(f_{\bm \beta, \mathcal T})}
\right)
-\frac n2 + \frac{n\hat Q_n(f_n^\dagger)}{2v_n^\dagger}.
\end{align}
We consider each terms on the right-hand-side of~\eqnref[eq:posterior-mispecified-proof-33] above. By \eqnref[eq:posterior-mispecified-proof-31], $\hat{Q}_n(f_{\bm \beta, \mathcal{T}}) \geq v_n^\dagger + c_M\mathsf q_{n, K}$ uniformly over \(f_{\bm \beta, \mathcal T}\in A_n^{+}(M_{+})\) with high $\mathbb{Q}_0^{n}$-probability. Using this, we have $\log({v_n^\dagger}/{\hat Q_n(f_{\bm \beta, \mathcal T})} ) \leq -\log(1 + c_M (v_n^\dagger)^{-1} \mathsf q_{n,K}) \leq -c_M (2v_n^\dagger)^{-1} \mathsf q_{n,K} \leq -c_M (2 \overline v)^{-1} \mathsf q_{n,K}$ for sufficiently large $n$ (follows from $\log(1+a) > a/2$ for $a$ sufficiently small and since $\mathsf q_{n, K} \stackrel{n\to \infty}{\longrightarrow} 0$). Moreover, from $\hat{Q}_n(f_n^{\dagger}) = v_n^{\dagger} + \mathfrak{o}_{\mathbb{Q}_0^{n}}(\mathsf{q}_{n, K})$ it follows that $-n/2 + {n\hat Q_n(f_n^\dagger)}/(2v_n^\dagger) = \mathfrak{o}_{\mathbb{Q}_0^{n}}(n \mathsf{q}_{n,K})$. Hence, combining these observations and using~\eqnref[eq:posterior-mispecified-proof-33], we have
\begin{align}
\label{eq:posterior-mispecified-proof-34}
\mathbb{Q}^{n}_0\left(\left\{\sup_{\substack{f_{\bm \beta, \mathcal T}\in A_n^{+}(M_{+})\\ \sigma^2\in[\underline v,\overline v]}}
\log\left(
\frac{
q_{f_{\bm \beta, \mathcal T},\sigma^2}^{n}(\mathcal D_n)
}{
q_{f_n^\dagger,v_n^\dagger}^{n}(\mathcal D_n)
}\right)
\le
-c_M n \mathsf{q}_{n,K}\right\}\cap \Omega_{\mathfrak X, n}\right) \stackrel{n\to \infty}{\longrightarrow} 1.
\end{align}
Consequently from~\eqnref[eq:posterior-mispecified-proof-34]
\begin{align}
\label{eq:posterior-mispecified-proof-35}
\mathbb{Q}^{n}_0\left(\left\{\int_{A_n^{+}(M_{+})} \tilde{\mathcal{R}}_n(f_{\bm \beta, \mathcal T},\sigma^2)
\,d\Pi_n^D(f_{\bm \beta, \mathcal T},\sigma^2)
\le
\exp\{-c_M n \mathsf{q}_{n,K}\}\right\}\cap \Omega_{\mathfrak X, n}\right)  \stackrel{n\to \infty}{\longrightarrow} 1.
\end{align}

\emph{\underline{Final posterior bound}}.
The truncated-sieve posterior on $A_n^{+}(M_{+})$ is
\begin{align}
\label{eq:posterior-mispecified-proof-36}
\Pi_n^{D}(A_n^{+}(M_{+})\mid \mathcal D_n) = \frac{\int_{A_n^{+}(M_{+})} \tilde{\mathcal{R}}_n(f_{\bm \beta, \mathcal T},\sigma^2)
\,d\Pi_n^D(f_{\bm \beta, \mathcal T},\sigma^2)}{\tilde{\mathcal{Z}}_n}.
\end{align}
Combining the numerator bound from~\eqnref[eq:posterior-mispecified-proof-35] with the denominator lower bound from~\eqnref[eq:posterior-mispecified-proof-19], we obtain using~\eqnref[eq:posterior-mispecified-proof-36]
\begin{align}
\label{eq:posterior-mispecified-proof-37}
\mathbb{Q}_0^{n}\left(\left\{\Pi_n^{D}(A_n^{+}(M_{+})\mid \mathcal D_n) \leq \exp\left\{C_{\tilde{\mathcal{Z}}}n \mathsf{q}_{n, K}^{2} - c_M n \mathsf{q}_{n, K}\right\}\right\} \cap \Omega_{\mathfrak X, n}\right) \stackrel{n\to \infty}{\longrightarrow} 1.
\end{align}
Note that, $n \mathsf{q}_{n, K}^{2} = \mathfrak{o}(n \mathsf{q}_{n, K})$ as $\mathsf{q}_{n, K} \stackrel{n\to \infty}{\longrightarrow} 0$. Also, $n\mathsf{q}_{n, K}^{2} \stackrel{n\to \infty}{\longrightarrow} \infty$. Therefore
\begin{align}
\label{eq:posterior-mispecified-proof-38}
\exp\left\{C_{\tilde{\mathcal{Z}}}n \mathsf{q}_{n, K}^{2} - c_M n \mathsf{q}_{n, K}\right\} \stackrel{n\to \infty}{\longrightarrow} 0.
\end{align}
Finally
\begin{align}
\label{eq:posterior-mispecified-proof-39}
\begin{split}
\mathsf E_{\mathbb{Q}_0^{n}}\left[\Pi_n^{D}(A_n^{+}(M_+)\mid \mathcal D_n)\right]
&=
\mathsf E_{\mathbb{Q}_0^{n}}\left[
\Pi_n^{D}(A_n^{+}(M_{+})\mid \mathcal D_n)\mathds 1_{\Omega_{\mathfrak{X}, n}}
\right] 
+\mathsf E_{\mathbb{Q}_0^{n}}\left[
\Pi_n^{D}(A_n^{+}(M_{+})\mid \mathcal D_n)1_{\Omega_{\mathfrak{X}, n}^{c}}
\right] \\
&\leq
\mathsf E_{\mathbb{Q}_0^{n}}\left[
\Pi_{n}^{D}(A_n^{+}(M_+)\mid \mathcal D_n)\mathds 1_{\Omega_{\mathfrak{X}, n}}
\right]
+
\mathbb{Q}_0^{n}(\Omega_{\mathfrak{X}, n}^{c}),
\end{split}
\end{align}
where the last inequality in the above display follows from the fact that posterior probability is bounded by one on $\Omega_{\mathfrak X, n}^c$. Since $\mathbb{Q}_0^{n} \left(\Omega_{\mathfrak X, n}\right) \stackrel{n\to \infty}{\longrightarrow} 1$, we have $\mathbb{Q}_0^{n} \left(\Omega_{\mathfrak X, n}^{c}\right) \stackrel{n\to \infty}{\longrightarrow} 0$. Further~\eqnref[eq:posterior-mispecified-proof-37] and ~\eqnref[eq:posterior-mispecified-proof-38] imply $\mathsf E_{\mathbb{Q}_0^{n}}\left[
\Pi_{n}^{D}(A_n^{+}(M_{+})\mid \mathcal D_n)\mathds 1_{\Omega_{\mathfrak{X}, n}}
\right] \stackrel{n\to \infty}{\longrightarrow} 0$ by bounded convergence in probability. Combining these with~\eqnref[eq:posterior-mispecified-proof-39] we obtain~\eqnref[eq:to-show-outer]
$$
\boxed{
\lim_{n\to \infty}\;\mathsf E_{\mathbb{Q}_0^{n}}
\left[
\Pi_n^D
\left(
f_{\bm\beta,\mathcal T}\in\mathcal F_n^D:
d_n^2(f_{\bm\beta,\mathcal T},f_0)
>
a_K^2(f_0)+M_{+}\mathsf q_{n,K}
\mid
\mathcal D_n
\right)
\right]
= 0.
}
$$
{}
\end{proof}

\subsubsection{Proof of \texorpdfstring{\hyperref[supple-theorem:sieve-truncated-population-symbolic-oracle]{Theorem~\ref{supple-theorem:sieve-truncated-population-symbolic-oracle}}}{Theorem 2}: Inner Radius}
\label{subsubsec:inner-radius}

\begin{proof}{}
\emph{\underline{Deterministic enlarged sieve and squared error class}}.
Let $\tilde{\mathcal{F}}_n = \{f_{\bm \beta, \mathcal T}\in \mathcal{F}_K: S(\mathcal T) \leq S_n, \lVert \bm \beta \rVert_{\infty} \leq C_{B}\} \supseteq \mathcal{F}_{n}^{D}$, where $\mathcal{F}_n^{D}$ is as in~\hyperref[supple-eq:sieve-misspecified]{\eqref{supple-eq:sieve-misspecified}}. For $f\in \tilde{\mathcal{F}_n}$, define $h_{f}(\bm x) := |f(\bm x) - f_0(\bm x)|^2$. Let $\mathcal{H}_n = \{h_f: f\in \tilde{\mathcal{F}}_n\}$. By~\mainassref[ass:global-symbolic-evaluation-envelope], for every $f_{\bm \beta, \mathcal T} \in \tilde{\mathcal{F}_n}$, $|f_{\bm \beta, \mathcal T}(\bm x)| \leq C_{B}(1 + K \overline{C}_U\exp\{c_U S_n\})$. Combining this with~\mainassref[ass:bounded-coefficient-symbolic-approximability], we obtain
\begin{equation}
\label{eq:inner-radius-1}
|f_{\bm \beta, \mathcal T}(\bm x) - f_0(\bm x)| \leq B_{n, K}.
\end{equation}
Thus, from~\eqnref[eq:inner-radius-1], $0\leq h_f(\bm x) \leq B^{2}_{n, K}$ uniformly over $h_{f} \in \mathcal{H}_n$ and $x\in \mathfrak{X}$.

\emph{\underline{Lipschitz transfer from symbolic functions to squared error functions}}.
Next, for any $f, g\in \tilde{\mathcal{F}}_n$
\begin{equation}
\label{eq:inner-radius-2}
|h_f(\bm x) - h_g(\bm x)| = \left||f(\bm x) - f_0(\bm x)|^2 - |g(\bm x) - f_0(\bm x)|^2\right| \leq 2 B_{n, K}|f(\bm x) - g(\bm x)|,
\end{equation}
using $|a^{2} - b^{2}| = |a+b| |a-b|$. From~\eqnref[eq:inner-radius-2], it follows that, $\|h_f - h_g\|_{\infty} \leq 2 B_{n, K} \|f-g\|_{\infty}$.

\emph{\underline{Covering the squared error class}}.
For a fixed symbolic forest $\mathcal{T} \in \mathbb{T}_{\mathbb O, p}^{K}$ with $S(\mathcal T) \leq S_n$,~\mainassref[ass:global-symbolic-evaluation-envelope] gives $\|f_{\bm \beta, \mathcal T} - f_{\bm \beta', \mathcal T}\|_{\infty} \leq (K+1)\overline{C}_{U}\exp\{c_U S_n\}\lVert \bm \beta - \bm \beta'\rVert_{\infty}$. Hence, the coefficient cube $\{\|\bm \beta\|_{\infty} \leq C_{B}\}$ can be covered so as to induce an $\varepsilon$-cover of $\mathcal H_n$ in $\|\cdot \|_{\infty}$ with at most
\begin{equation}
\label{eq:inner-radius-3}
\left(1 + \frac{4 C_B(K+1)\overline{C}_U B_{n, K}\exp\{c_U S_n\}}{\varepsilon}\right)^{K+1}
\end{equation}
points. By~\hyperref[lemma:symbolic-forest-count]{Lemma~\ref{lemma:symbolic-forest-count}}
\begin{equation}
\label{eq:inner-radius-4}
N_{\mathcal T}(K, S_n) \leq \exp\{(2S_n +K+1)\log 3 + S_n \log|\mathbb O| + (S_n+K)\log p\}.
\end{equation}
Combining~\eqnref[eq:inner-radius-3] with~\eqnref[eq:inner-radius-4] and using the definition of $\mathfrak{C}_{K, S_n, n}$, there exist a constant $C>0$ such that, for $0<\varepsilon < B_{n, K}^{2}$
\begin{equation}
\label{eq:inner-radius-5}
\log N(\varepsilon, \mathcal H_n, \|\cdot \|_{\infty}) \leq C \mathfrak{C}_{K, S_n, n} + C(K+1)\log\left(\frac{C B_{n, K}}{\varepsilon}\right).
\end{equation}

\emph{\underline{Finite net approximation}}.
Define $V_n := \mathfrak{C}_{K, S_n, n}[1 + \log(\mathfrak{C}_{K, S_n, n} + K + p + |\mathbb O|)]$ and for any real-valued function $g$ on $\mathfrak{X}$, let $P_n g = n^{-1}\sum_{i=1}^{n}g(\bm x_i)$ and $P g = \int_{\mathfrak{X}} g(\bm x) q_X(\bm x) d\bm x$. Choose $\varepsilon_n = n^{-1}B^{2}_{n, K}$. Let $\mathcal{H}_{n, \varepsilon_n}$ be an $\varepsilon_n$-net of $\mathcal{H}_n$ in $\|\cdot \|_{\infty}$. By~\eqnref[eq:inner-radius-5],
\begin{equation}
\label{eq:inner-radius-6}
\begin{split}
\log |\mathcal H_{n, \varepsilon_n}| &\leq C \mathfrak C_{K, S_n, n} + C(K+1)\log \frac{nCB_{n,K}}{B_{n,K}^2}
\leq C \mathfrak C_{K, S_n, n} + C(K+1)\log \frac{nC}{B_{n,K}}
\leq CV_n,
\end{split}
\end{equation}
where the penultimate inequality in~\eqnref[eq:inner-radius-6] follows as $B_{n,K} \geq 1$.
Now fix $h \in \mathcal H_n$. Let $h^{\triangle} \in \mathcal H_{n, \varepsilon_n}$ such that, $\|h-h^{\triangle}\|_{\infty} \leq \varepsilon_n$. Then $|(P_n - P)h| \leq |(P_n - P)h^{\triangle}| + |(P_n - P)(h-h^{\triangle})|$. Note that, $|(P_n - P)(h-h^{\triangle})| \leq |P_n(h-h^{\triangle})| + |P(h-h^{\triangle})| \leq 2 \varepsilon_n$. Therefore, $\sup_{h \in \mathcal H_n} |(P_n - P)h| \leq \max_{h^{\triangle} \in \mathcal H_{n, \varepsilon_n}} |(P_n - P)h^{\triangle}| + 2\varepsilon_n$. 

\emph{\underline{Uniform concentration over the finite net}}.
For fixed $h^{\triangle}$, since $0\leq h^{\triangle} \leq B^{2}_{n, K}$, Hoeffding's inequality yields, for every $t>0$, $\mathbb{Q}_0^{n}(|(P_n-P)h^{\triangle}| > t) \leq 2\exp\{-2 n t^{2}/B_{n, K}^{4}\}$. Taking the union bound over the net $\mathcal{H}_{n, \varepsilon_n}$, we have
\begin{equation}
\label{eq:inner-radius-7}
\begin{split}
\mathbb{Q}_0^{n}\left(\max_{h^{\triangle}\in \mathcal{H}_{n, \varepsilon_n}}|(P_n -P) h^{\triangle}| > t\right) \leq 2 |\mathcal{H}_{n, \varepsilon_n}|\exp\left\{-\frac{2n t^2}{B^{4}_{n, K}}\right\}\leq 2\exp\left\{C V_n - \frac{2n t^{2}}{B^{4}_{n, K}}\right\}.
\end{split}
\end{equation}
Now, choose $t = A B^{2}_{n, K}\sqrt{V_n / n}$, where $A > \sqrt{C/2} > 0$. Hence, from~\eqnref[eq:inner-radius-7], we have
\begin{equation}
\label{eq:inner-radius-8}
\begin{split}
&\mathbb{Q}_0^{n}\left(\max_{h^{\triangle}\in \mathcal{H}_{n, \varepsilon_n}}|(P_n -P) h^{\triangle}| > A B^{2}_{n, K}\sqrt{\frac{V_n}{n}}\right) \leq 2\exp\left\{-(2A^2-C) V_n\right\}\\
&\implies \mathbb{Q}_0^{n}\left(\max_{h^{\triangle}\in \mathcal{H}_{n, \varepsilon_n}}|(P_n -P) h^{\triangle}| > A\rho_{n, K}\right) \leq 2\exp\left\{-(2A^2-C) V_n\right\}.
\end{split}
\end{equation}
Note that, $2\varepsilon_n =  2n^{-1}B_{n, K}^{2}= \mathfrak{o}(B_{n, K}^{2}\sqrt{V_n/n}) = \mathfrak{o}(\rho_{n, K})$. Combining this with~\eqnref[eq:inner-radius-8], we obtain
\begin{equation}
\label{eq:inner-radius-9}
\mathbb{Q}_0^{n}\left(\sup_{h\in \mathcal{H}_{n}}|(P_n -P) h| > (A+1)\rho_{n, K}\right) \leq 2\exp\left\{-(2A^2-C) V_n\right\}.
\end{equation}

\emph{\underline{Uniform empirical population risk comparison}}.
Since $2A^{2} > C$, the right-hand-side of~\eqnref[eq:inner-radius-9] tends to zero, as $V_n \stackrel{n \to \infty}{\longrightarrow} \infty$. Hence, for the event
\begin{equation}
\label{eq:inner-radius-10}
\mathcal{E}_{\mathrm{in}, n} = \left\{\sup_{f\in \tilde{\mathcal{F}}_n} \left|d_n^{2}(f, f_0) - \mathsf{E}_{q_X}|f(\bm x) - f_0(\bm x)|^2\right| \leq (A+1)\rho_{n, K}\right\},
\end{equation}
we have $\mathbb{Q}_0^{n}(\mathcal{E}_{\mathrm{in}, n}) \stackrel{n\to \infty}{\longrightarrow} 1$. On $\mathcal{E}_{\mathrm{in}, n}$, for every $f\in \tilde{\mathcal{F}}_n$, $d_n^{2}(f, f_0) \geq \mathsf{E}_{q_X}|f(\bm x) - f_0(\bm x)|^{2} - (A+1)\rho_{n, K}$. Since $\tilde{\mathcal{F}}_n \subseteq \mathcal{F}_K$, the definition of $a^{2}_{K}(f_0)$ in~\maineqref[eq:aK] implies $d_n^{2}(f, f_0) \geq a_K^{2}(f_0) - (A+1)\rho_{n, K}$ uniformly over $f\in \tilde{\mathcal{F}}_n$. Also, since $\mathcal{F}_n^{D} \subseteq \tilde{\mathcal{F}}_n$, on $\mathcal{E}_{\mathrm{in}, n}$
\begin{equation}
\label{eq:inner-radius-11}
\inf_{f\in \mathcal{F}_n^{D}} d_n^{2}(f, f_0) \geq a_K^{2}(f_0) - (A+1)\rho_{n, K}.
\end{equation}

\emph{\underline{Posterior mass on the inner bad set}}.
Let $A_n^{-}(M_{-}) = \{f_{\bm \beta, \mathcal T}\in \mathcal{F}_n^{D}: d_n^{2}(f_{\bm \beta, \mathcal T}, f_0) < a_{K}^{2}(f_0) - M_{-}\mathsf{q}_{n, K}\}$. Since $\rho_{n, K}\leq \mathsf{q}_{n, K}$, choose $M_{-} > A+1$, then on $\mathcal{E}_{\mathrm{in}, n}$ by~\eqnref[eq:inner-radius-11], we have $A_{n}^{-}(M_{-}) = \emptyset$. Therefore
\begin{equation}
\label{eq:inner-radius-12}
\lim_{n\to \infty}\mathsf{E}_{\mathbb{Q}_0^{n}}\left[\Pi_{n}^{D}\left(A^{-}_{n}(M_{-})\mid \mathcal{D}_n\right)\right] \leq \lim_{n\to \infty}\mathbb{Q}_0^{n}(\mathcal{E}_{\mathrm{in}, n}^{c}) = 0,
\end{equation}
yields~\eqnref[eq:to-show-inner]
$$
\boxed{
\lim_{n\to \infty}\;\mathsf E_{\mathbb{Q}_0^{n}}
\left[
\Pi_n^D
\left(
f_{\bm\beta,\mathcal T}\in\mathcal F_n^D:
d_n^2(f_{\bm\beta,\mathcal T},f_0)
<
a_K^2(f_0)-M_{-}\mathsf q_{n,K}
\mid
\mathcal D_n
\right)
\right]
= 0.
}
$$
{}
\end{proof}
\newpage
\suppdivision{Empirical Results for Feynman Equations}
\label{superdiv:feynman}

\begin{tcolorbox}[
  enhanced,
  colback=CompBack,
  colframe=CompBlue,
  boxrule=0pt,
  arc=1pt,
  left=8pt,
  right=8pt,
  top=6pt,
  bottom=6pt,
  before skip=8pt,
  after skip=10pt,
  borderline west={2.5pt}{0pt}{CompBlue}
]
\small
\textcolor{CompBlue}{\textbf{Extended results.}}
A broader set of results for the Feynman equations is available \href{https://github.com/Roy-SR-007/HierBOSSS/blob/main/Feynman_Equations_Results.pdf}{\texttt{here}}.
\end{tcolorbox}

\section{Description of Feynman Equations, Experimental Protocol, and Evaluation Criteria}
\label{sec:experiment-protocol-evaluation-criteria}

We consider learning the following $5$ Feynman equations~\citep{AI-Feynman} as in \mainref[sec:learning-Feynman-equations], with details provided in~\hyperref[tab:feynman-equations-variable-interpretation]{Table~\ref{tab:feynman-equations-variable-interpretation}}. 
\begin{align}
&\mathrm{I\_12\_2}:\quad F
= \frac{q_1 q_2}{4\pi \epsilon r^{2}}, & \textcolor{BrickRed}{p = 4}
\tag{{CL}}
\label{supple-eq:feynman-cl}
\\
&\mathrm{I\_12\_11}:\quad F
= q\left(E_f+vB\sin\theta\right), & \textcolor{BrickRed}{p = 5}
\tag{{FCE}} 
\label{supple-eq:feynman-fce}
\\
&\mathrm{I\_24\_6}:\quad E_n = \frac{1}{4}m(\omega^2 + \omega_0^2)x^{2}, & \textcolor{BrickRed}{p = 4}
\tag{{EHFO}}
\label{supple-eq:feynman-ehfo}
\\
&\mathrm{I\_50\_26}:\quad x=x_1[\cos(\omega t) + \alpha \cos^{2}(\omega t)], & \textcolor{BrickRed}{p = 4}
\tag{{HOQN}}
\label{supple-eq:feynman-hoqn}
\\
&\mathrm{II\_36\_38}:\quad f = \frac{\mathrm{mom} H}{k_b T} + \frac{\mathrm{mom} \alpha}{\epsilon c^2 k_b T}M. & \textcolor{BrickRed}{p = 8}
\tag{{FMMM}}
\label{supple-eq:feynman-fmmm}
\end{align}

\begingroup
\scriptsize
\setlength{\tabcolsep}{3.4pt}
\renewcommand{\arraystretch}{1.18}

\begin{longtable}{@{}p{0.13\textwidth} p{0.20\textwidth} p{0.25\textwidth} p{0.35\textwidth}@{}}
\caption{Details on Feynman equations.}
\label{tab:feynman-equations-variable-interpretation}\\

\toprule
\toprule
\textbf{Equation} 
& \textbf{Response variable} 
& \textbf{Predictors} 
& \textbf{Scientific interpretation} \\
\midrule
\endfirsthead

\caption[]{Details on Feynman equations \emph{(continued)}.}\\
\toprule
\toprule
\textbf{Equation} 
& \textbf{Output variable} 
& \textbf{Input variables} 
& \textbf{Scientific interpretation} \\
\midrule
\endhead

\midrule
\multicolumn{4}{r}{\emph{Continued on next page}}\\
\endfoot

\bottomrule
\bottomrule
\endlastfoot

$\mathrm{I\_12\_2}$~\eqnref[supple-eq:feynman-cl] 
& $F$: electrostatic force 
& electric charges ($q_1, q_2$); permittivity ($\epsilon$); distance between charges ($r$) 
& Coulomb's law describing the magnitude of the electrostatic force between two point charges. The force is proportional to the product of the charges and inversely proportional to the squared separation distance, with the permittivity controlling the strength of interaction in the medium. \\[4pt]

\reprowsep

$\mathrm{I\_12\_11}$~\eqnref[supple-eq:feynman-fce]
& $F$: Lorentz force 
& particle charge ($q$); electric field ($E_f$); magnetic field ($B$); particle velocity ($v$); angle between velocity and magnetic field ($\theta$) 
& Force acting on a charged particle moving through combined electric and magnetic fields. The term $qE_f$ corresponds to the electric contribution, while $qvB\sin\theta$ captures the magnetic force perpendicular to the velocity-field plane. \\[4pt]

\reprowsep

$\mathrm{I\_24\_6}$~\eqnref[supple-eq:feynman-ehfo]
& $E_n$: harmonic oscillator energy 
& mass ($m$); angular frequency ($\omega$); natural angular frequency ($\omega_0$); displacement ($x$) 
& Energy associated with a harmonic oscillator-like system. The energy scales with mass, squared displacement, and the combined squared frequencies, representing stored mechanical energy in oscillatory motion. \\[4pt]

\reprowsep

$\mathrm{I\_50\_26}$~\eqnref[supple-eq:feynman-hoqn]
& $x$: position or displacement 
& amplitude/initial displacement scale ($x_1$); angular frequency ($\omega$); time ($t$); nonlinear correction parameter ($\alpha$) 
& Oscillatory displacement with a nonlinear correction term. The first term represents standard cosine oscillation, while $\alpha\cos^2(\omega t)$ introduces a second-order harmonic correction to the motion. \\[4pt]

\reprowsep

$\mathrm{II\_36\_38}$~\eqnref[supple-eq:feynman-fmmm]
& $f$: magnetic response 
& magnetic moment ($\mathrm{mom}$); magnetic field ($H$); Boltzmann constant ($k_b$); temperature ($T$); polarizability parameter ($\alpha$); permittivity ($\epsilon$); speed of light ($c$); mass-like parameter ($M$)
& Magnetic response governed by the balance between field--moment interaction and thermal energy. The first term captures the normalized contribution of the applied magnetic field, while the second term adds a correction involving polarizability, electromagnetic constants, and the material parameter $M$. \\

\end{longtable}
\endgroup

The data sets corresponding to the aforementioned Feynman equations are obtained from the Penn Machine Learning Benchmark (\texttt{PMLB}; \href{https://github.com/EpistasisLab/pmlb}{\texttt{github.com/EpistasisLab/pmlb}})~\citep{PMLB}. The data set associated with each equation contains $10^{5}$ observations.

As mentioned in~\mainref[sec:learning-Feynman-equations], the goal of this data study is to showcase the performance on \hierbosss\ against state-of-the-art competing \sr\ modules: deep symbolic regression (\dsr)~\citep{Deep-SR}, \qlattice~\citep{feyn-qlattice}, \sisso$++$~\citep{SISSO++}, \gplearn~\citep{stephens2016gplearn}, \operon~\citep{operon}, \pysr~\citep{pysr}, Bayesian Machine Scientist (\bms)~\citep{BMS} and Bayesian Symbolic Regression (\bsr)~\citep{BSR}. 

For each experimental run, we randomly subsample $n=2000$ observations from the corresponding Feynman dataset and partition them into a $90\%$ training set and a $10\%$ held-out test set. We consider both the noiseless setting, in which the original target values are used, and several noisy settings (to mimic experimental or observational noise). Specifically, for a prescribed noise standard deviation $\sigma\geq 0$, the
response is generated as
\begin{equation}
\label{eq:feynman-noise-generation}
    y_i^{(\sigma)} = y_i+\varepsilon_i,
    \qquad
    \varepsilon_i \overset{\mathrm{iid}}{\sim} \mathrm{N}(0,\sigma^2),
    \qquad 
    i=1,\ldots,n,
\end{equation}
where $y_i$ denotes the original target value. Thus, $\sigma=0$ corresponds to the noiseless setting. The equation-specific noise levels used in the experiments are summarized in
\hyperref[tab:feynman-noise-levels]{Table~\ref{tab:feynman-noise-levels}}. The first three noise levels for each equation are used to compare \hierbosss\ with the competing methods, whereas the final, higher noise setting is considered only for \hierbosss\ to further examine its performance under increased response contamination.
\begin{table}[H]
\centering
\small
\setlength{\tabcolsep}{6pt}
\renewcommand{\arraystretch}{1.18}

\caption{Noise levels used in each Feynman equation experiment.}
\label{tab:feynman-noise-levels}

\begin{tabular}{
    @{}
    >{\raggedright\arraybackslash}p{0.20\textwidth}
    >{\raggedright\arraybackslash}p{0.28\textwidth}
    >{\centering\arraybackslash}p{0.22\textwidth}
    >{\centering\arraybackslash}p{0.22\textwidth}
    @{}
}
\toprule
\toprule
Feynman equation
& Expression
& Noise levels for \hierbosss\ and competitors
& Additional higher noise level for \hierbosss\ only \\
\midrule

$\displaystyle
\mathrm{I\_12\_2}$~\eqnref[supple-eq:feynman-cl]
&
$F=\frac{q_1q_2}{4\pi\epsilon r^2}$
&
$\sigma\in\{0,\,0.15,\,0.20\}$
&
$\sigma=0.25$
\\[3pt]

$\displaystyle
\mathrm{I\_12\_11}$~\eqnref[supple-eq:feynman-fce]
&
$
F=q\left(E_f+vB\sin\theta\right)
$
&
$\sigma\in\{0,\,0.25,\,1.00\}$
&
$\sigma=2.00$
\\[3pt]

$\displaystyle
\mathrm{I\_24\_6}$~\eqnref[supple-eq:feynman-ehfo]
&
$
E_n=\frac{1}{4}m\left(\omega^2+\omega_0^2\right)x^2
$
&
$\sigma\in\{0,\,0.15,\,0.25\}$
&
$\sigma=1.00$
\\[3pt]

$\displaystyle
\mathrm{I\_50\_26}$~\eqnref[supple-eq:feynman-hoqn]
&
$
x=x_1\left[\cos(\omega t)+\alpha\cos^2(\omega t)\right]
$
&
$\sigma\in\{0,\,0.15,\,0.18\}$
&
$\sigma=0.20$
\\[3pt]

$\displaystyle
\mathrm{II\_36\_38}$~\eqnref[supple-eq:feynman-fmmm]
&
$
f=
\frac{\mathrm{mom}\,H}{k_bT}
+
\frac{\mathrm{mom}\,\alpha}{\epsilon c^2k_bT}M
$
&
$\sigma\in\{0,\,0.15,\,0.20\}$
&
$\sigma=0.25$
\\
\bottomrule
\bottomrule
\end{tabular}
\end{table}
For each Feynman equation and noise level, we conduct $5$ independent repetitions, with the random subsampling, train-test partition, and noise realization generated independently in each repetition. Predictive performance is evaluated on the held-out test set using the root mean squared error (\texttt{RMSE}) and coefficient of determination ($R^2$). These metrics are
defined as
\begin{equation}
\label{eq:feynman-predictive-metrics}
\begin{split}
\texttt{RMSE} = \left[\frac{1}{n_{\mathrm{test}}}
    \sum_{i=1}^{n_{\mathrm{test}}}
    \left(y_i^{(\sigma)}-\widehat{y}_i^{(\sigma)}\right)^2
    \right]^{1/2},
    \qquad
    R^2 = 1-
    \frac{
        \sum_{i=1}^{n_{\mathrm{test}}}
        \left(y_i^{(\sigma)}-\widehat{y}_i^{(\sigma)}\right)^2
    }{
        \sum_{i=1}^{n_{\mathrm{test}}}
        \left(y_i^{(\sigma)}-\overline{y}_{\mathrm{test}}^{(\sigma)}\right)^2
    },
\end{split}
\end{equation}
where $y_i^{(\sigma)}$ and $\widehat{y}_i^{(\sigma)}$ are the observed and predicted responses for the $i$th test observation at noise level $\sigma$, respectively, $n_{\mathrm{test}} = 200$ is the size of the $10\%$ held-out dataset, and $\overline{y}_{\mathrm{test}}^{(\sigma)} = n_{\mathrm{test}}^{-1} \sum_{i=1}^{n_{\mathrm{test}}}y_i^{(\sigma)}$.

\newpage
\section{Experimental Settings of \texorpdfstring{\hierbosss}{HierBOSSS} and Competitors}
\label{sec:experimental-settings-competitors-Feynman}

\begin{tcolorbox}[
  enhanced,
  colback=CompBack,
  colframe=CompBlue,
  boxrule=0pt,
  arc=1pt,
  left=8pt,
  right=8pt,
  top=6pt,
  bottom=6pt,
  before skip=8pt,
  after skip=10pt,
  borderline west={2.5pt}{0pt}{CompBlue}
]
\small
\textcolor{CompBlue}{\textbf{Computational environment.}}
All Feynman equation experiments were implemented in \texttt{Python} and executed on a MacBook Air equipped with an Apple M2 processor and 8GB of unified memory.
\end{tcolorbox}

\subsection{\texorpdfstring{\hierbosss}{HierBOSSS}}
\label{subsec:BayeSymX-Feynman-settings}

We evaluate \hierbosss\ on the Feynman equations in $\mathrm{I\_12\_2}$~\eqnref[supple-eq:feynman-cl]-$\mathrm{II\_36\_38}$~\eqnref[supple-eq:feynman-fmmm] using the experimental protocol in~\hyperref[sec:experiment-protocol-evaluation-criteria]{Appendix \ref{sec:experiment-protocol-evaluation-criteria}}. The operator set $\mathbb O$ in~\mainref[sec:methodology] is kept common across every Feynman equation. The set consists of two binary operators and seven unary operators, as given below. In the numerical implementation, the inverse operator is protected using a threshold $10^{-8}$, while the exponential operator is evaluated after clipping its input to $[-20,20]$ for numerical stability.

\begin{pysettingsbox}{\hierbosss\ operator set}
opset = [
    "add",  # (x, y) -> x + y
    "mul",  # (x, y) -> x*y
    "neg",  # x -> -x
    "inv",  # x -> 1/x, protected near zero
    "sin",  # x -> sin(x)
    "cos",  # x -> cos(x)
    "exp",  # x -> exp(x), with input clipped to [-20, 20]
    "sq",   # x -> x^2
    "cu",   # x -> x^3
]
\end{pysettingsbox}

The prior hyperparameters in~\mainref[subsec:prior-specification] are set uniformly across all Feynman equations. The Dirichlet concentration parameters for both operator and feature assignments are set to one that is $\bm \alpha_{\mathrm{op}} = (1, \ldots, 1)$ and $\bm{\alpha}_{\mathrm{ft}} = (1, \ldots, 1)$. The depth-dependent splitting probability is controlled by $(\alpha_0,\delta_0)=(0.95,1.20)$. For the outer regression coefficient vector $\bm \beta$ with an intercept, we set the Gaussian prior mean vector to $\bm \mu_{\beta} = \bm 0$, use Gaussian prior covariance matrix $\bm \Sigma_{\beta} = 10\bm I_{K+1}$, and assign the hyperparameters of the Inverse-Gamma prior on the model noise variance $\sigma^{2}$ to $\nu = 0.05 = \lambda$.

\begin{pysettingsbox}{\hierbosss\ prior configuration}
HierBOSSS_prior = {
    # operator and feature Dirichlet weight prior concentration
    "alpha_op": "ones(len(opset))",
    "alpha_ft": "ones(number_of_features)",

    # split probability
    "alpha_0": 0.95,
    "delta_0": 1.20,

    # Gaussian prior parameters of outer regression coefficient vector
    "add_intercept": True,
    "beta0": "zeros(K + 1)",
    "V0": "10 * I_(K + 1)",

    # Inverse-Gamma prior parameters of model noise variance
    "a0": 0.05,
    "b0": 0.05,
    "sigma2_prior": "Inverse-Gamma(a0 / 2, b0 / 2)",
}
\end{pysettingsbox}
For each Feynman equation, the symbolic forest size $K$ and number of Markov chain Monte Carlo (\texttt{MCMC}) iterations are chosen according to the symbolic complexity of the underlying target law, as given below. However, one can choose a suitable nominal $K$ value balancing the expressiveness and computational efficiency for learning the underlying Feynman equations; see the practical guideline provided in~\hyperref[sec:ablation]{Appendix \ref{sec:ablation}}.
\begin{pysettingsbox}{\hierbosss\ equation-specific $K$ and \texttt{MCMC} iterations}
# maxiter -> number of MCMC iterations within each parallel chain of HierBOSSS
# K -> symbolic forest size

HierBOSSS_feynman_settings = {
    "I_12_2": {  "K": 4,  "maxiter": 2000,},            # CL
    "I_12_11": { "K": 3,  "maxiter": 2000,},            # FCE
    "I_24_6": {  "K": 4,  "maxiter": 2000,},            # EHFO   
    "I_50_26": { "K": 4,  "maxiter": 2000,},            # HOQN
    "II_36_38": {"K": 4,  "maxiter": 20000,},           # FMMM
}
\end{pysettingsbox}

Each repetition is fit using $5$ independent parallel \texttt{MCMC} chains of \hierbosss. The final symbolic model selection is performed by pooling sampled forests across chains and ranking them by their joint marginal posterior over $\mathcal T$, $\mathrm{JMP}(\mathcal T)$ in \hyperref[eq:JMP]{\textcolor{BrickRed}{(\ref*{eq:JMP}) of the main manuscript}}.

\begin{pysettingsbox}{\hierbosss\ parallel chain configuration}
parallel_mcmc_settings = {
    "nchains": 5,               # number of parallel MCMC chains
    "burnin": 0,
    "thin": 1,
    "n_jobs": "nchains",
    "chain_seed_rule": "...",
    "target": "JMP(T)",
}
\end{pysettingsbox}

As described in~\mainref[subsec:posterior-inference], the Metropolis-Hastings (\texttt{MH}) kernel uses seven local symbolic tree proposal moves: grow ($\mathrm{g}$), prune ($\mathrm{p}$), change feature ($\mathrm{cf}$), change operator ($\mathrm{co}$), subtree replacement ($\mathrm{st}$), delete node ($\mathrm{del}$), and insert node ($\mathrm{ins}$). Unless a move is unavailable for the current tree state, all moves are assigned equal proposal weight through the move weight vector $\bm \pi_{\mathrm{move}} = \left(\pi_{\mathrm{g}}, \pi_{\mathrm{p}}, \pi_{\mathrm{st}}, \pi_{\mathrm{del}}, \pi_{\mathrm{ins}}, \pi_{\mathrm{cf}}, \pi_{\mathrm{co}}\right)$ as chosen in \hyperref[subsec:default-MH-config]{Appendix \ref{subsec:default-MH-config}}. At each tree update, the sampler first identifies the set of valid moves and valid node addresses, normalizes the proposal weights over the available moves, samples a move type, and then samples a valid node address uniformly for that move.

\begin{pysettingsbox}{\hierbosss\ local symbolic tree proposal moves}
# local symbolic tree proposal move weights

move_weights = {
    "grow": 1.0,                # g
    "prune": 1.0,               # p
    "change_feature": 1.0,      # cf
    "change_operator": 1.0,     # co    
    "subtree_replace": 1.0,     # st
    "delete_node": 1.0,         # del             
    "insert_node": 1.0,         # ins    
}
\end{pysettingsbox}

The initial symbolic forests and proposal-generated subtrees use uniform operator and feature weights that is uniform choices for $(\bm{w}_{\mathrm{op, init}}, \bm{w}_{\mathrm{op, prop}}, \bm{w}_{\mathrm{ft, init}}, \bm{w}_{\mathrm{ft, prop}})$; see \hyperref[subsec:default-MH-config]{Appendix \ref{subsec:default-MH-config}}. These weights are used only for initialization and proposal generation; the posterior target itself uses the Dirichlet-collapsed symbolic forest prior $\Pi_{\mathrm{forest}, K}(\mathcal T\mid \alpha_0, \delta_0, \bm{\alpha}_{\mathrm{op}}, \bm \alpha_{\mathrm{ft}})$ in~\maineqref[eq:symbolic-forest-prior]. Thus, operator and feature preferences are learned through the marginal prior contribution in the joint marginal posterior rather than fixed at the proposal level. 

\begin{pysettingsbox}{\hierbosss\ initialization and proposal weights}
# initialization of operator and feature weight vectors with uniform choice
# proposal operator and feature weight vectors with uniform choice

initialization_and_proposal = {
    "wts_init_op": "ones(len(opset))",
    "wts_init_ft": "ones(number_of_features)",
    "wts_prop_op": "wts_init_op",
    "wts_prop_ft": "wts_init_ft",
    "proposal_normalization": "normalize over available moves and valid labels",
}
\end{pysettingsbox}

After sampling, all retained symbolic forests from the parallel \texttt{MCMC} chains are pooled and ranked by their $\log$ joint marginal posterior over $\mathcal T$ that is $\log\mathrm{JMP}(\mathcal T)$. For each repetition, the top $r=10$ symbolic forests are retained as the representative posterior set that is the Occam's window set $\mathcal{J}_r$ in~\maineqref[eq:Jr] with $r=10$. For each retained forest, the posterior mean of the outer regression coefficient vector is computed under the conjugate Gaussian layer (see posterior mean vector $\bm{\mu}_{\beta}^{\star}(\mathcal{T})$ of $\bm \beta$ in~\mainref[subsec:posterior-inference]), yielding the corresponding raw symbolic expression (\rawtag).

\begin{pysettingsbox}{\hierbosss\ posterior ranking configuration and obtaing \rawtag\ expressions}
posterior_ranking = {
    "ranking_score": "log_JMP",
    "pooling": "all retained forests from all parallel chains",
    "top_r": 10,
    "raw_expression_coefficients": "posterior mean under NIG layer",
    "raw_expression_digits": 6, # rounding off decimals in raw expression
}
\end{pysettingsbox}

Finally, each top-ranked symbolic forest is passed through the post-\texttt{MCMC} symbolic model refinement algorithm of \hierbosss\ in~\hyperref[alg:post-mcmc-symbolic-model-refinement-single]{Algorithm \ref{alg:post-mcmc-symbolic-model-refinement-single}}.
The final reported expression corresponds to the \finaltag\ expression after this refinement step with effective symbolic forest size $K_{\mathrm{eff}} \leq K$.

\begin{pysettingsbox}{\hierbosss\ post-\texttt{MCMC} symbolic model refinement to obtain \finaltag\ expressions}
post_mcmc_refinement = {
    "candidate_models": "top_r forests ranked by log_JMP",
    "criterion": "BIC",
    "coefficient_refit": "posterior mean on selected symbolic design",
    "simplification": "SymPy",
    "final_expression_digits": 2, # rounding off decimals in final expression
    "coefficient_tolerance": 1e-4,
    "effective_forest_size": "K_eff = number of retained non-intercept trees",}
\end{pysettingsbox}

\subsection{\texorpdfstring{\dsr}{DSR}}
\label{subsec:DSR-Feynman-settings}

We use the \texttt{dso.DeepSymbolicRegressor} implementation\footnote{\texttt{dso.DeepSymbolicRegressor} implementation: \href{https://github.com/dso-org/deep-symbolic-optimization}{\texttt{github.com/dso-org/deep-symbolic-optimization}}} of deep symbolic regression (\dsr)~\citep{Deep-SR} for learning the $5$ Feynman equations in $\mathrm{I\_12\_2}$~\eqnref[supple-eq:feynman-cl]-$\mathrm{II\_36\_38}$~\eqnref[supple-eq:feynman-fmmm] using the experimental protocol in~\hyperref[sec:experiment-protocol-evaluation-criteria]{Appendix \ref{sec:experiment-protocol-evaluation-criteria}}. 

\dsr\ parameterizes a policy over symbolic expressions using an autoregressive recurrent neural network (\texttt{RNN}) which sequentially samples symbolic programs from a prescribed function set. Across all $5$ Feynman equations, the operator set is configured as follows.

\begin{pysettingsbox}{\dsr\ operator set}
function_set = [
    "add",      # (x, y) -> x + y
    "mul",      # (x, y) -> x*y
    "neg",      # x -> -x
    "inv",      # x -> 1/x
    "sin",      # x -> sin(x)
    "cos",      # x -> cos(x)
    "exp",      # x -> exp(x)
    "n2",       # x -> x^2
    "n3",       # x -> x^3
]
\end{pysettingsbox}

The regression task is run with protected operators enabled. For all equations, the the following \dsr\ configuration is maintained.

\begin{pysettingsbox}{\dsr\ settings}
dsr_settings = {
    "n_samples": 200000,  # total symbolic program sampling budget 
    "batch_size": 500,    # samples per RNN policy-gradient update
    "max_length": 20,     # maximum length of each sampled expression  
}
\end{pysettingsbox}

These settings correspond to approximately \(400\) \texttt{RNN} policy-gradient updates, unless early stopping occurs. During training, \dsr\ retains the best-reward symbolic program encountered over the search. The final recovered expression is extracted from the fitted estimator through

\begin{pysettingsbox}{Extracting the final \dsr\ expression}
estimator.program_
\end{pysettingsbox}
which stores the best symbolic program returned by \texttt{dso.DeepSymbolicRegressor}.

\subsection{\texorpdfstring{\qlattice}{QLattice}}
\label{subsec:QLattice-Feynman-settings}

We use the \texttt{feyn.QLattice} implementation (\href{https://docs.abzu.ai/}{\texttt{docs.abzu.ai/}}) of \qlattice~\citep{feyn-qlattice} for learning the $5$ Feynman equations in $\mathrm{I\_12\_2}$~\eqnref[supple-eq:feynman-cl]-$\mathrm{II\_36\_38}$~\eqnref[supple-eq:feynman-fmmm] using the experimental protocol in~\hyperref[sec:experiment-protocol-evaluation-criteria]{Appendix \ref{sec:experiment-protocol-evaluation-criteria}}.

\qlattice\ constructs a lattice-based search space over symbolic interactions among input variables and primitive functions, and iteratively generates and scores candidate symbolic models. Candidate formulas are generated through the \texttt{auto\_run} routine, ranked according to a specified model selection criterion, and the top-ranked model returned by \qlattice\ is used as the learned expression. The operator set across all $5$ Feynman equations is configured using the primitive functions available in the \texttt{feyn} implementation as follows.

\begin{pysettingsbox}{\qlattice\ operator set}
function_names = [
    "inverse",  # z -> 1/z
    "linear",   # z -> a*z + b
    "squared",  # z -> z^2
    "add",      # (z1, z2) -> z1 + z2
    "multiply", # (z1, z2) -> z1 * z2
    "exp",      # z -> exp(z)
    "sin",      # z -> sin(z)
    "cos",      # z-> cos(z)
]
\end{pysettingsbox}

For all equations, the \qlattice\ search is run with the following configuration.

\begin{pysettingsbox}{\qlattice\ settings}
qlattice_settings = {
    "n_epochs": 25,         # number of search epochs
    "max_complexity": 20,   # complexity of symbolic models
    "criterion": "bic",     # Bayesian information criterion
}
\end{pysettingsbox}

The final recovered expression is extracted from the top-ranked Bayesian information criterion (\texttt{BIC}) model using \texttt{Python}'s \texttt{SymPy} package~\citep{meurer2017sympy} as

\begin{pysettingsbox}{Extracting the final \qlattice\ expression}
# signif -> significant digits in fitted numerical coefficients

best_model = models[0] # top-ranked BIC model
expression = best_model.sympify(signif=6)
\end{pysettingsbox}

which converts the selected \qlattice\ model into a symbolic expression.

\subsection{\texorpdfstring{\sisso$++$}{SISSO++}}
\label{subsec:SISSO++-Feynman-settings}

For sure independence screening and sparsifying operator (\sisso)~\citep{SISSO}, we use the \texttt{SISSORegressor} implementation\footnote{\texttt{SISSORegressor} implementation: \href{https://sissopp_developers.gitlab.io/sissopp/}{\texttt{sissopp\_developers.gitlab.io/sissopp/}}} from \sisso$++$~\citep{SISSO++} for learning the $5$ Feynman equations in $\mathrm{I\_12\_2}$~\eqnref[supple-eq:feynman-cl]-$\mathrm{II\_36\_38}$~\eqnref[supple-eq:feynman-fmmm] using the experimental protocol in~\hyperref[sec:experiment-protocol-evaluation-criteria]{Appendix \ref{sec:experiment-protocol-evaluation-criteria}}.

\sisso$++$ constructs a large symbolic descriptor space (complete enumeration) by recursively applying a prescribed set of mathematical operators to the primary input variables $x_1,\ldots,x_p$. It then performs sure independence screening, \texttt{SIS}, followed by sparse descriptor selection through the sparsifying operator, \texttt{SO}.

For all equations, the \sisso$++$ regression task is run with the following common configuration.

\begin{pysettingsbox}{\sisso$++$ settings}
sisso_settings = {
    "n_dim": 3,           # number of selected descriptors
    "max_rung": 3,        # maximum recursive feature-construction depth
    "n_sis_select": 50,   # candidate descriptors retained by SIS
    "n_residual": 20,     # number of residual-based selections during SO
}
\end{pysettingsbox}

For \sisso$++$, we use problem-specific operator sets rather than the common collection of operators across all equations. This is because of the combinatorial growth of the \sisso$++$ descriptor space: as \(\texttt{max\_rung}\) increases, recursively applying a large operator library to the primary variables can generate an extremely large number of candidate descriptors, making the \texttt{SIS} and \texttt{SO} steps computationally expensive. Therefore, for each Feynman equation, we restrict the operator set to the algebraic and functional components required to represent the corresponding ground-truth expression. This provides a computationally feasible and structurally informed benchmark for \sisso$++$, while still allowing the method to search over a rich nonlinear descriptor space generated from the specified operators.

\begin{pysettingsbox}{\sisso$++$ operator set}
# add: +, mult: *, div: /, sq: ^2, sin: sin(), cos: cos(), 
sisso_operator_sets = {
    "I_12_2":   ["add", "div", "sq"],          # CL
    "I_12_11":  ["add", "mult", "sin"],        # FCE
    "I_24_6":   ["add", "mult", "sq"],         # EHFO
    "I_50_26":  ["add", "mult", "cos", "sq"],  # HOQN
    "II_36_38": ["add", "mult", "div", "sq"],  # FMMM 
}
\end{pysettingsbox}

The final recovered expression is taken as the sparse linear combination of selected symbolic descriptors returned by \texttt{SISSORegressor}.

\subsection{\texorpdfstring{\gplearn}{gplearn}}
\label{subsec:gplearn-Feynman-settings}

We use the \texttt{gplearn.genetic.SymbolicRegressor} (\href{https://gplearn.readthedocs.io/}{\texttt{gplearn.readthedocs.io/}}) implementation  of \gplearn~\citep{stephens2016gplearn} for learning the $5$ Feynman equations in $\mathrm{I\_12\_2}$~\eqnref[supple-eq:feynman-cl]-$\mathrm{II\_36\_38}$~\eqnref[supple-eq:feynman-fmmm] using the experimental protocol in~\hyperref[sec:experiment-protocol-evaluation-criteria]{Appendix \ref{sec:experiment-protocol-evaluation-criteria}}.

\gplearn\ evolves symbolic expressions using genetic programming~\citep{koza1992genetic}. Starting from an initial population of randomly generated symbolic programs, it repeatedly applies evolutionary operations such as selection, crossover, mutation, and reproduction to improve predictive fit. In our experiments, symbolic programs are scored using \texttt{RMSE}, and a parsimony penalty is included to discourage overly complex expressions.

Across all $5$ Feynman equations, the operator set is configured as follows.

\begin{pysettingsbox}{\gplearn\ operator set}
function_set = [
    "add",      # (x, y) -> x + y
    "mul",      # (x, y) -> x*y
    "neg",      # x -> -x
    "inv",      # x -> 1/x
    "sin",      # x -> sin(x)
    "cos",      # x -> cos(x)
    "exp",      # x -> exp(x)
    "sq",       # x -> x^2
    "cu",       # x -> x^3
]
\end{pysettingsbox}

The unary functions \(\mathrm{neg}\), \(\mathrm{inv}\), \(\sin\), \(\cos\), \(\exp\), \({}^2\), and \({}^3\) are supplied through custom wrapper \texttt{gplearn.functions.make\_function}. The inverse operator is protected near zero, and the exponential operator is clipped for numerical stability.

For all equations, the following \gplearn\ configurations are maintained.

\begin{pysettingsbox}{\gplearn\ settings}
gplearn_settings = {
    "metric": "rmse",  # fitness metric to select symbolic expression
    "generations": 20,  # number of evolutionary generations
    "population_size": 2000,  # tournament selection pressure used to choose parents
    "tournament_size": 20,  # tournament selection pressure used to choose parents 
    "parsimony_coefficient": 1e-4,  # expression complexity during fitness evaluation

    # gplearn default settings implemented below
    "stopping_criteria": 0.00,  # early stopping threshold for RMSE
    "p_crossover": 0.9,  # subtree crossover probability
    "p_subtree_mutation": 0.01,  # subtree mutation probability
    "p_hoist_mutation": 0.01,  # hoist mutation probability
    "p_point_mutation": 0.01,  # point mutation probability
    "max_samples": 1.0,  # fraction of training samples per program
}
\end{pysettingsbox}

The final recovered expression is extracted from the fitted estimator through

\begin{pysettingsbox}{Extracting the final \gplearn\ expression}
model._program
\end{pysettingsbox}
which stores the best symbolic program retained by \texttt{gplearn}.

\subsection{\texorpdfstring{\operon}{operon}}
\label{subsec:operon-Feynman-settings}

We use the \texttt{pyoperon.sklearn.SymbolicRegressor} implementation\footnote{\texttt{pyoperon.sklearn.SymbolicRegressor} implementation: \href{https://github.com/heal-research/operon}{\texttt{github.com/heal-research/operon}}} of \operon~\citep{operon} for learning the $5$ Feynman equations in $\mathrm{I\_12\_2}$~\eqnref[supple-eq:feynman-cl]-$\mathrm{II\_36\_38}$~\eqnref[supple-eq:feynman-fmmm] using the experimental protocol in~\hyperref[sec:experiment-protocol-evaluation-criteria]{Appendix \ref{sec:experiment-protocol-evaluation-criteria}}.

\operon\ is a genetic programming-based \sr\ method that evolves expression trees using evolutionary search. Starting from a population of candidate symbolic expressions, \operon\ iteratively applies variation and selection operators to improve predictive fit while searching over a prescribed set of primitive symbols. In our implementation, the symbolic search is performed through the \texttt{SymbolicRegressor} interface from the \texttt{Python} package \texttt{pyoperon}.

Across all $5$ Feynman equations, the primitive symbol set is configured as follows.

\begin{pysettingsbox}{\operon\ operator set}
allowed_symbols = [
    "add",       # (x, y) -> x + y
    "mul",       # (x, y) -> x*y
    "sub",       # (x, y) -> x - y
    "div",       # (x, y) -> x/y
    "sin",       # x -> sin(x)
    "cos",       # x -> cos(x)
    "exp",       # x -> exp(x)
    "square",    # x -> x^2
    "variable",  # input variable leaf nodes
    "constant",  # numerical constant leaf nodes
]
\end{pysettingsbox}

For all equations, \operon\ is implemented with the following configuration.

\begin{pysettingsbox}{\operon\ settings}
operon_settings = {
    "generations": 2000,  # number of evolutionary generations
    "population_size": 1000,  # number of candidate expressions in the population
    "max_length": 20,  # maximum length of the evolved expressions
}
\end{pysettingsbox}

The \operon\ estimator is initialized as follows.

\begin{pysettingsbox}{\operon\ module initialization}
model = SymbolicRegressor(
    allowed_symbols=allowed_symbols,
    generations=2000,
    population_size=1000,
    max_length=20,
)
\end{pysettingsbox}

The final recovered expression is extracted from the fitted estimator using the available \operon\ model-string interface namely,

\begin{pysettingsbox}{Extracting the final \operon\ expression}
model.get_model_string(model.model_)
\end{pysettingsbox}
whenever the fitted model is stored in \texttt{model\_}; otherwise, the corresponding best available expression attribute returned by the fitted \texttt{SymbolicRegressor} is used.

\subsection{\texorpdfstring{\pysr}{PySR}}
\label{subsec:PySR-Feynman-settings}

We use the \texttt{pysr.PySRRegressor} implementation\footnote{\texttt{pysr.PySRRegressor} implementation: \href{https://github.com/MilesCranmer/PySR}{\texttt{github.com/MilesCranmer/PySR}}} of \pysr~\citep{pysr} for learning the $5$ Feynman equations in $\mathrm{I\_12\_2}$~\eqnref[supple-eq:feynman-cl]-$\mathrm{II\_36\_38}$~\eqnref[supple-eq:feynman-fmmm] using the experimental protocol in~\hyperref[sec:experiment-protocol-evaluation-criteria]{Appendix \ref{sec:experiment-protocol-evaluation-criteria}}.

\pysr\ performs evolutionary \sr\ by searching over expression trees composed of user-specified binary and unary operators. The search maintains multiple evolving populations of candidate expressions and optimizes them using mutation, crossover, selection, and constant optimization. In our experiments, the symbolic search is performed through the \texttt{PySRRegressor} interface, with squared-error loss and deterministic serial execution.

Across all $5$ Feynman equations, the operator set is configured as follows.

\begin{pysettingsbox}{\pysr\ operator set}
binary_operators = [
    "+",       # (x, y) -> x + y
    "*",       # (x, y) -> x*y
]

unary_operators = [
    "neg(x) = -x",          # x -> -x
    "inv(x) = 1 / x",       # x -> 1/x
    "sin",                  # x -> sin(x)
    "cos",                  # x -> cos(x)
    "exp",                  # x -> exp(x)
    "sq(x) = x * x",        # x -> x^2
    "cu(x) = x * x * x",    # x -> x^3
]
\end{pysettingsbox}

For all equations, \pysr\ is run with the following configuration.

\begin{pysettingsbox}{\pysr\ settings}
pysr_settings = {
    "niterations": 200, # number of PySR evolutionary iterations
    "populations": 10, # number of populations evolved during the search
    "population_size": 50, # number of candidate expressions in each population
    "maxsize": 20, # maximum expression size
    "elementwise_loss": "loss(x, y) = (x - y)^2", # squared-error loss
    "model_selection": "best", # for symbolic expression selection
    "deterministic": True, # to ensure reproducibility
    "parallelism": "serial", # avoid variation owing to parallel execution
}
\end{pysettingsbox}

The final recovered expression is extracted from the best \pysr\ model using

\begin{pysettingsbox}{Extracting the best \pysr\ model}
model.get_best()["equation"]
\end{pysettingsbox}
whenever available; otherwise, the symbolic expression returned by
\begin{pysettingsbox}{Extracting the final \texttt{SymPy}-supported \pysr\ expression}
model.sympy()
\end{pysettingsbox}
is used.

\subsection{\texorpdfstring{\bms}{BMS}}
\label{subsec:BMS-Feynman-settings}

We use the \texttt{BMSRegressor} implementation from the \texttt{AutoRA} module\footnote{\texttt{AutoRA} \texttt{BMSRegressor} implementation: \href{https://autoresearch.github.io/autora/user-guide/theorists/bms/}{\texttt{autoresearch.github.io/autora/user-guide/theorists/bms/}}}~\citep{AutoRA} of Bayesian machine scientist (\bms)~\citep{BMS} for learning the $5$ Feynman equations in $\mathrm{I\_12\_2}$~\eqnref[supple-eq:feynman-cl]-$\mathrm{II\_36\_38}$~\eqnref[supple-eq:feynman-fmmm] using the experimental protocol in~\hyperref[sec:experiment-protocol-evaluation-criteria]{Appendix \ref{sec:experiment-protocol-evaluation-criteria}}.

Across all $5$ Feynman equations, the \bms\ operator set is configured using the primitive operators supported by the \texttt{BMSRegressor} interface. The common operator configuration is passed through the \texttt{ops} dictionary and a uniform prior weight is assigned to each included operator.

\begin{pysettingsbox}{\bms\ operator dictionary and weights}
bms_ops = {
    "+": 2,     # add: (x, y) -> x + y
    "*": 2,     # mul: (x, y) -> x * y
    "/": 2,     # div: (x, y) -> x/y
    "-": 1,     # neg: x -> -x
    "sin": 1,   # sin: x -> sin(x)
    "cos": 1,   # cos: x -> cos(x)
    "exp": 1,   # exp: x -> exp(x)
    "pow2": 1,  # sq: x -> x^2
    "pow3": 1,  # cu: x -> x^3
}

bms_prior = {
    "Nopi_+": 1.0,
    "Nopi_*": 1.0,
    "Nopi_/": 1.0,
    "Nopi_-": 1.0,
    "Nopi_sin": 1.0,
    "Nopi_cos": 1.0,
    "Nopi_exp": 1.0,
    "Nopi_pow2": 1.0,
    "Nopi_pow3": 1.0,
}
\end{pysettingsbox}

For all equations except $\mathrm{II\_36\_38}$~\eqnref[supple-eq:feynman-fmmm], each \bms\ chain is run for \(\texttt{epochs}=2000\) \texttt{MCMC} steps. For $\mathrm{II\_36\_38}$~\eqnref[supple-eq:feynman-fmmm], which is the most structurally complex Feynman equation considered here, we use \(\texttt{epochs}=20000\). \bms\ is run using \(5\) independent parallel \texttt{MCMC} chains.

\begin{pysettingsbox}{\bms\ settings}
bms_settings = {
    "n_parallel_chains": 5,
    "epochs": 2000, # 20000 for II_36_38 (FMMM)
    "parallel_chains": True,
}
\end{pysettingsbox}

After all chains are completed, the chain with the lowest in-sample \texttt{RMSE} (computed on a $90\%$ train split) is selected, and its symbolic expression is used as the final recovered expression.

\begin{pysettingsbox}{Extracting the final \bms\ expression across parallel chains}
best_chain = min(
    successful_chains,
    key=lambda chain: chain["chain_train_rmse"],
)

final_expression = best_chain["expression"]
\end{pysettingsbox}

\subsection{\texorpdfstring{\bsr}{BSR}}
\label{subsec:BSR-Feynman-settings}

We use the \texttt{BSRRegressor} implementation from the \texttt{AutoRA} module\footnote{\texttt{AutoRA} \texttt{BSRRegressor} implementation: \href{https://autoresearch.github.io/autora/user-guide/theorists/bsr/}{\texttt{autoresearch.github.io/autora/user-guide/theorists/bsr/}}} of Bayesian symbolic regression (\bsr)~\citep{BSR} for learning the $5$ Feynman equations in $\mathrm{I\_12\_2}$~\eqnref[supple-eq:feynman-cl]-$\mathrm{II\_36\_38}$~\eqnref[supple-eq:feynman-fmmm] using the experimental protocol in~\hyperref[sec:experiment-protocol-evaluation-criteria]{Appendix \ref{sec:experiment-protocol-evaluation-criteria}}.

Across all $5$ Feynman equations, the \bsr\ operator set is configured as follows.

\begin{pysettingsbox}{\bsr\ operator set}
bsr_operator_set = [
    "add",   # (x, y) -> x + y
    "mul",   # (x, y) -> x*y
    "neg",   # x -> -x
    "inv",   # x -> 1/x
    "sin",   # x -> sin(x)
    "cos",   # x -> cos(x)
    "exp",   # x -> exp(x)
    "sq",    # x -> x^2
    "cu",    # x -> x^3
]
\end{pysettingsbox}

The aforementioned operator set is converted to the corresponding \bsr\ prior dictionary.
\newline
\begin{pysettingsbox}{\bsr\ operator prior dictionary}
bsr_operator_prior = {
    "+": 1.0,
    "*": 1.0,
    "neg": 1.0,
    "inv": 1.0,
    "sin": 1.0,
    "cos": 1.0,
    "exp": 1.0,
    "pow2": 1.0,
    "pow3": 1.0,
}
\end{pysettingsbox}

For all equations except $\mathrm{II\_36\_38}$~\eqnref[supple-eq:feynman-fmmm], we set \(\texttt{itr\_num}=2000\), while for $\mathrm{II\_36\_38}$ we set \(\texttt{itr\_num}=20000\). We use \(5\) independent restarts for each \bsr\ fit. The main \bsr\ configuration is given below. Note that, the split probability parameters are maintained at their default specifications.

\begin{pysettingsbox}{\bsr\ settings}
# split probability parameters maintained at default settings

bsr_settings = {
    "n_restarts": 5,
    "itr_num": 2000, # 20000 for II_36_38 (FMMM)
    "prior_name": "restricted_ops",
}
\end{pysettingsbox}

The number of symbolic trees \(K\) is chosen to match the corresponding \hierbosss\ configuration described in~\hyperref[subsec:BayeSymX-Feynman-settings]{Appendix \ref{subsec:BayeSymX-Feynman-settings}}.
The equation-specific tree configurations are given below.

\begin{pysettingsbox}{\bsr\ $K$ configuration}
bsr_tree_num = {
    "I_12_2": 4,
    "I_12_11": 3,
    "I_24_6": 4,
    "I_50_26": 4,
    "II_36_38": 4,
}
\end{pysettingsbox}

Thus, for a given equation, the \bsr\ estimator is initialized as follow.

\begin{pysettingsbox}{\bsr\ module initialization}
model = BSRRegressor(
    tree_num=bsr_tree_num[equation_name],
    itr_num=2000, # 20000 for II_36_38 (FMMM)
)
\end{pysettingsbox}

After all restarts are completed, the final recovered expression is taken from the successful restart with the lowest in-sample \texttt{RMSE} (computed on a $90\%$ train set).

\begin{pysettingsbox}{Extracting the final \bsr\ expression across restarts}
best_restart = min(
    successful_restarts,
    key=lambda restart: restart["train_rmse"],
)

final_expression = best_restart["expression"]
\end{pysettingsbox}

The final expression is extracted as the additive symbolic model returned by \texttt{BSRRegressor}, together with the fitted coefficients of the selected symbolic trees.

\newpage
\section{Symbolic Expressions Learned by \texorpdfstring{\hierbosss}{HierBOSSS} and Competitors for Learning the Feynman Equations}
\label{sec:symbolic-expressions-Feynman}


\subsection{For \texorpdfstring{$\mathrm{I\_12\_2}$~\eqnref[supple-eq:feynman-cl]}{I.12.2}}

\begingroup
\footnotesize
\setlength{\tabcolsep}{2.8pt}
\renewcommand{\arraystretch}{1.20}

\begin{longtable}{@{}>{\centering\arraybackslash}p{0.105\textwidth} >{\raggedright\arraybackslash}p{0.585\textwidth} >{\centering\arraybackslash}p{0.090\textwidth} >{\centering\arraybackslash}p{0.080\textwidth} >{\centering\arraybackslash}p{0.090\textwidth}@{}}
\caption{Best repetition results for the Feynman equation $\mathrm{I\_12\_2}$~\eqnref[supple-eq:feynman-cl]: $F = \tfrac{q_1 q_2}{4\pi \epsilon r^2}$, across methods and noise levels. For each method and noise level, the displayed expression is selected from the repetition with the lowest train \texttt{RMSE} among $5$ independent repetitions. Test \texttt{RMSE} and $R^2$ are computed on the held-out $10\%$ test set.}
\label{tab:i-12-2-best-train-rmse-all-methods-results}\\

\toprule
\toprule
Method
& Learned expression
& Test \texttt{RMSE}
& Test $R^2$
& Runtime (s) \\
\midrule
\endfirsthead

\caption[]{\emph{(Continued)}. Best repetition results for the Feynman equation $\mathrm{I\_12\_2}$~\eqnref[supple-eq:feynman-cl]: $F = \tfrac{q_1 q_2}{4\pi \epsilon r^2}$, across methods and noise levels.}\\
\toprule
\toprule
Method
& Learned expression
& Test \texttt{RMSE}
& Test $R^2$
& Runtime (s) \\
\midrule
\endhead

\midrule
\multicolumn{5}{r}{\emph{Continued on next page}}\\
\endfoot

\bottomrule
\bottomrule
\endlastfoot

\addlinespace[3pt]
\multicolumn{5}{@{}l}{\textcolor{BrickRed}{\textit{Noiseless setting}}}\\
\addlinespace[2pt]

\rowcolor{ExprBack}\hierbosss
& \exprcell{
\exprbox{\finaltag}{
& 0.040\,\frac{2q_1q_2}{\epsilon r^2}
}
}
& $9.738{\times}10^{-7}$ & 1.000 & 47.098 \\

\reprowsep

\dsr
& \exprcell{
\displaystyle
\begin{gathered}
q_1
\Big\{
q_1+\epsilon\exp\!\big[
r\exp\{\sin(q_2^{-2})\}
\big]
\Big\}^{-1}
\end{gathered}
}
& 0.023 & 0.923 & 85.300 \\

\reprowsep

\qlattice
& \exprcell{
\displaystyle
\begin{gathered}
-4.30{\times}10^{-5}q_{1}
+3.97{\times}10^{-6} \\
{}+0.258(-0.249q_{1}-1.32{\times}10^{-4})
(-0.071q_{2}-0.307) \\
{}\times(0.214q_{2}+8.66{\times}10^{-4}) \\
{}\times
\Big\{
(0.088-0.585r)(-0.195\epsilon-5.70{\times}10^{-4}) \\
{}\times(-0.340q_{2}-1.473)(-0.311r-0.061)
\Big\}^{-1}
\end{gathered}
}
& 0.000 & 1.000 & 92.951 \\

\reprowsep

\sisso$++$
& \exprcell{
\displaystyle
\begin{gathered}
-4.642{\times}10^{-17}
+1.430{\times}10^{-22}\,
\epsilon q_1q_2^2(q_2/r)^2 \\
{}+2.206{\times}10^{-19}\,
\frac{\epsilon q_2(q_1/r)}{r^2(r/q_1)}
+0.080\,\frac{q_1q_2}{\epsilon r^2}
\end{gathered}
}
& 0.000 & 1.000 & 14.700 \\

\reprowsep

\gplearn
& \exprcell{
\displaystyle
\begin{gathered}
q_{1}^{2}q_{2}
\Big\{
q_{1}\big[
\epsilon\{r+\sin(q_{1}^{-1})-0.509\} \\
{}+r+\sin(e^{-\epsilon})-1.018
\big]+3
\Big\}^{-2}
\end{gathered}
}
& 0.013 & 0.982 & 17.639 \\

\reprowsep

\operon
& \exprcell{
\displaystyle
0.080\,q_{1}q_{2}(\epsilon r^{2})^{-1}
}
& 0.000 & 1.000 & 18.776 \\

\reprowsep

\pysr
& \exprcell{
\displaystyle
0.080\,q_{1}q_{2}(\epsilon r^{2})^{-1}
}
& 0.000 & 1.000 & 71.351 \\

\reprowsep

\bms
& \exprcell{
\displaystyle
-\frac{q_{1}q_{2}}{a_{0}\epsilon r^{2}}
-\frac{q_{2}\sin(a_{0})}{a_{0}^{2}}
}
& 0.000 & 1.000 & 231.897 \\

\reprowsep

\bsr
& \exprcell{
\displaystyle
\begin{gathered}
2.105-2.008\cos\{\exp(-\epsilon)\}
+0.076\cos(r) \\
{}-0.014\cos(\epsilon)-0.004q_{1}r
\end{gathered}
}
& 0.079 & 0.325 & 188.710 \\

\addlinespace[4pt]
\midrule

\addlinespace[3pt]
\multicolumn{5}{@{}l}{\textcolor{BrickRed}{\textit{Noise level}: $\sigma=0.15$}}\\
\addlinespace[2pt]

\rowcolor{ExprBack}\hierbosss
& \exprcell{
\exprbox{\finaltag}{
& 0.084\,\frac{q_1q_2}{\epsilon r^2}
}
}
& 0.144 & 0.235 & 48.198 \\

\reprowsep

\dsr
& \exprcell{
\displaystyle
\epsilon q_1\exp\{-(\epsilon+r)\}
}
& 0.148 & 0.185 & 95.300 \\

\reprowsep

\qlattice
& \exprcell{
\displaystyle
0.373\exp(-0.457\epsilon+0.346q_{1}+0.261q_{2}-1.107r)
+0.002
}
& 0.144 & 0.226 & 84.850 \\

\reprowsep

\sisso$++$
& \exprcell{
\displaystyle
\begin{gathered}
-0.004
+c_1\,\frac{r^5}{\epsilon q_2^2}
-0.009\,\frac{\epsilon r}{q_1^4} +0.079\,\frac{q_1q_2}{\epsilon r^2}
\end{gathered}
}
& 0.148 & 0.238 & 17.756 \\

\reprowsep

\gplearn
& \exprcell{
\displaystyle
\exp\!\bigg[
\frac{r\{-\epsilon q_{2}^{2}-q_{1}(q_{2}+2)\}}
{q_{1}q_{2}^{2}}
\bigg]
}
& 0.142 & 0.270 & 16.168 \\

\reprowsep

\operon
& \exprcell{
\displaystyle
-0.007
}
& 0.145 & 0.222 & 26.672 \\

\reprowsep

\pysr
& \exprcell{
\displaystyle
0.082\,q_{1}q_{2}(\epsilon r^{2})^{-1}
}
& 0.156 & 0.274 & 69.974 \\

\reprowsep

\bms
& \exprcell{
\displaystyle
\sin\!\bigg\{
\frac{q_{1}q_{2}}{a_{0}\epsilon r^{2}}
\bigg\}
}
& 0.144 & 0.232 & 349.022 \\

\reprowsep

\bsr
& \exprcell{
\displaystyle
\begin{gathered}
-0.308-0.009\cos(\epsilon^{2})
-0.435\exp(-r) \\
{}-0.008\cos\{\sin(r)\}
+0.271\exp(r^{-1})
\end{gathered}
}
& 0.153 & 0.062 & 188.704 \\

\addlinespace[4pt]
\midrule

\addlinespace[3pt]
\multicolumn{5}{@{}l}{\textcolor{BrickRed}{\textit{Noise level}: $\sigma=0.20$}}\\
\addlinespace[2pt]

\rowcolor{ExprBack}\hierbosss
& \exprcell{
\exprbox{\finaltag}{
& 0.082\,\frac{q_1q_2}{\epsilon r^2}
}
}
& 0.211 & 0.129 & 55.818 \\

\reprowsep

\dsr
& \exprcell{
\displaystyle
\sin\!\Big[
\epsilon\{r(\epsilon+q_2^{-1})\}^{-2}
\Big]
}
& 0.213 & 0.111 & 119.900 \\

\reprowsep

\qlattice
& \exprcell{
\displaystyle
\begin{gathered}
-0.277(1.244-0.209\epsilon)(0.193q_{2}+0.023) \\
{}\times
\{0.180q_{1}-0.921r-0.050\}^{-1}
+0.003
\end{gathered}
}
& 0.208 & 0.152 & 66.408 \\

\reprowsep

\sisso$++$
& \exprcell{
\displaystyle
\begin{gathered}
0.003
+c_1\,\frac{r^4q_1}{\epsilon q_2^2}
+0.019\,\frac{q_1q_2}{r^3} +0.077\,\frac{q_1q_2}{\epsilon^2 r^2}
\end{gathered}
}
& 0.198 & 0.144 & 17.098 \\

\reprowsep

\gplearn
& \exprcell{
\displaystyle
q_{1}^{2}\{\epsilon(q_{1}^{2}r^{2}+q_{1}r^{2}+1)\}^{-1}
}
& 0.202 & 0.173 & 16.173 \\

\reprowsep

\operon
& \exprcell{
\displaystyle
\begin{gathered}
-0.074e^{-\sin(16.325\epsilon/r)}(\epsilon r^{2})^{-1}
\Big[
0.284\epsilon r^{2}e^{\sin(16.325\epsilon/r)}
\sin(8.519q_{1}r) \\
{}-0.501\epsilon r^{2}e^{\sin(16.325\epsilon/r)}
+0.284\epsilon r^{2} \\
{}-q_{1}q_{2}e^{\sin(16.325\epsilon/r)}
\Big]
\end{gathered}
}
& 0.199 & 0.195 & 26.691 \\

\reprowsep

\pysr
& \exprcell{
\displaystyle
0.178\,r^{-1}
}
& 0.218 & 0.072 & 70.455 \\

\reprowsep

\bms
& \exprcell{
\displaystyle
\sin\!\bigg\{
a_{0}\{q_{1}q_{2}+\cos(a_{0}q_{2})\}
(\epsilon r^{2})^{-1}
\bigg\}
}
& 0.196 & 0.221 & 406.794 \\

\reprowsep

\bsr
& \exprcell{
\displaystyle
\begin{gathered}
0.114-0.034\cos\{\cos(q_{1})\}
-0.027\cos^{2}\{\sin(q_{2}^{2})\} \\
{}+5.686{\times}10^{-4}q_{1}^{3}
-0.057\sin^{2}\{\cos(\epsilon)+q_{2}\}
\end{gathered}
}
& 0.225 & 0.010 & 190.897 \\

\addlinespace[4pt]
\end{longtable}
\endgroup

\subsection{For \texorpdfstring{$\mathrm{I\_12\_11}$~\eqnref[supple-eq:feynman-fce]}{I.12.11}}

\begingroup
\footnotesize
\setlength{\tabcolsep}{2.8pt}
\renewcommand{\arraystretch}{1.20}

\begin{longtable}{@{}>{\centering\arraybackslash}p{0.105\textwidth} >{\raggedright\arraybackslash}p{0.585\textwidth} >{\centering\arraybackslash}p{0.090\textwidth} >{\centering\arraybackslash}p{0.080\textwidth} >{\centering\arraybackslash}p{0.090\textwidth}@{}}
\caption{Best repetition results for the Feynman equation $\mathrm{I\_12\_11}$~\eqnref[supple-eq:feynman-fce]: $F = q(E_f + Bv \sin \theta)$, across methods and noise levels. For each method and noise level, the displayed expression is selected from the repetition with the lowest train \texttt{RMSE} among $5$ independent repetitions. Test \texttt{RMSE} and $R^2$ are computed on the held-out $10\%$ test set.}
\label{tab:i-12-11-best-train-rmse-all-methods-results}\\

\toprule
\toprule
Method
& Learned expression
& Test \texttt{RMSE}
& Test $R^2$
& Runtime (s) \\
\midrule
\endfirsthead

\caption[]{\emph{(Continued)}. Best repetition results for the Feynman equation $\mathrm{I\_12\_11}$~\eqnref[supple-eq:feynman-fce]: $F = q(E_f + Bv \sin \theta)$, across methods and noise levels.}\\
\toprule
\toprule
Method
& Learned expression
& Test \texttt{RMSE}
& Test $R^2$
& Runtime (s) \\
\midrule
\endhead

\midrule
\multicolumn{5}{r}{\emph{Continued on next page}}\\
\endfoot

\bottomrule
\bottomrule
\endlastfoot

\addlinespace[3pt]
\multicolumn{5}{@{}l}{\textcolor{BrickRed}{\textit{Noiseless setting}}}\\
\addlinespace[2pt]

\rowcolor{ExprBack}\hierbosss
& \exprcell{
\exprbox{\finaltag}{
& E_{f} q + B q v \sin(\theta)
}
}
& $8.641{\times}10^{-8}$ & 1.000 & 15.484 \\

\reprowsep

\dsr
& \exprcell{
\displaystyle
\begin{gathered}
q+\sin(\theta)v
\Big\{
B\big[(\exp(\theta)+E_f)^{-1}+q\big]+B
\Big\}+B
\end{gathered}
}
& 9.812 & 0.829 & 44.400 \\

\reprowsep

\qlattice
& \exprcell{
\displaystyle
\begin{gathered}
-48.303(0.126q+5.576{\times}10^{-4}) \\
{}\times
\Big[
-0.163E_f
+(0.165-0.393\theta)(1.908-0.326\theta) \\
{}\times(4.395-1.400\theta)(-0.625B-0.002)
\Big] \\
{}\times(-4.409{\times}10^{-5}E_f-1.210)(0.156v+0.002) \\
{}-0.062
\end{gathered}
}
& 0.522 & 1.000 & 96.089 \\

\reprowsep

\sisso$++$
& \exprcell{
\displaystyle
\begin{gathered}
-9.001{\times}10^{-16}
-1.354{\times}10^{-16}
(v+B+\sin B)(\sin v+v) \\
{}+E_fq+Bvq\sin(\theta)
\end{gathered}
}
& 0.000 & 1.000 & 42.554 \\

\reprowsep

\gplearn
& \exprcell{
\displaystyle
q\{E_f+Bv\sin(\theta)\}
}
& 0.000 & 1.000 & 15.034 \\

\reprowsep

\operon
& \exprcell{
\displaystyle
\begin{gathered}
0.028
+0.496
\Big[
2.020qE_f
+1.822B\sin(\theta)\{1.115qv\}
\Big]
\end{gathered}
}
& 0.046 & 1.000 & 22.940 \\

\reprowsep

\pysr
& \exprcell{
\displaystyle
q\{E_f+Bv\sin(\theta)\}
}
& 0.000 & 1.000 & 136.217 \\

\reprowsep

\bms
& \exprcell{
\displaystyle
qE_f+Bqv\sin(\theta)
}
& 0.000 & 1.000 & 267.921 \\

\reprowsep

\bsr
& \exprcell{
\displaystyle
33.811-0.067(2\theta)^{3}-1.468B+0.055e^{B}
}
& 17.040 & 0.551 & 167.938 \\

\addlinespace[4pt]
\midrule

\addlinespace[3pt]
\multicolumn{5}{@{}l}{\textcolor{BrickRed}{\textit{Noise level}: $\sigma=0.25$}}\\
\addlinespace[2pt]

\rowcolor{ExprBack}\hierbosss
& \exprcell{
\exprbox{\finaltag}{
& 1.001\,E_{f} q + B q v \sin(\theta)
}
}
& 0.238 & 1.000 & 15.472 \\

\reprowsep

\dsr
& \exprcell{
\displaystyle
\begin{gathered}
Bqv
\Big[
\{\exp(B)v+\theta\}^{-1}+\sin^{3}(\theta)
\Big]
+\theta
\end{gathered}
}
& 11.847 & 0.800 & 42.100 \\

\reprowsep

\qlattice
& \exprcell{
\displaystyle
\begin{gathered}
-37.575(0.002-0.258B)(1.145q+0.005) \\
{}\times(0.178\theta-0.558)(0.243\theta-1.422) \\
{}\times(-0.326v-0.001)(0.003B-0.843\theta+0.339) \\
{}-37.575(0.053-1.318q)(0.156-4.545{\times}10^{-4}v) \\
{}\times(0.132E_f-0.009)+0.246
\end{gathered}
}
& 0.441 & 1.000 & 91.071 \\

\reprowsep

\sisso$++$
& \exprcell{
\displaystyle
\begin{gathered}
0.019
+0.998\,E_fq
+0.994\,Bvq\sin(\theta) \\
{}+0.004\,q\sin(\theta)(vB+v)
\end{gathered}
}
& 0.248 & 1.000 & 42.689 \\

\reprowsep

\gplearn
& \exprcell{
\displaystyle
q\{E_f+Bv\sin(\theta)\}
}
& 0.260 & 1.000 & 15.842 \\

\reprowsep

\operon
& \exprcell{
\displaystyle
\begin{gathered}
-0.002
+0.279
\Big[
1.790qE_f+1.790qE_f \\
{}+0.012E_f
+3.598qBv\sin(\theta)
\Big]
\end{gathered}
}
& 0.231 & 1.000 & 23.109 \\

\reprowsep

\pysr
& \exprcell{
\displaystyle
q\{E_f+Bv\sin(\theta)\}
}
& 0.237 & 1.000 & 119.976 \\

\reprowsep

\bms
& \exprcell{
\displaystyle
qE_f+Bqv\sin(\theta)
}
& 0.237 & 1.000 & 294.344 \\

\reprowsep

\bsr
& \exprcell{
\displaystyle
\begin{gathered}
19.906
-0.970(v+q\theta) \\
{}+19.892\sin\{\sin(e^{v})+\theta\}
-1.615\cos(E_f)
\end{gathered}
}
& 19.449 & 0.410 & 167.556 \\

\addlinespace[4pt]
\midrule

\addlinespace[3pt]
\multicolumn{5}{@{}l}{\textcolor{BrickRed}{\textit{Noise level}: $\sigma=1.00$}}\\
\addlinespace[2pt]

\rowcolor{ExprBack}\hierbosss
& \exprcell{
\exprbox{\finaltag}{
& 1.005\,E_{f} q + 0.999\,B q v \sin(\theta) \\
&\quad
- 0.110\,\sin{(e^{q} )}
}
}
& 1.061 & 0.998 & 17.241 \\

\reprowsep

\dsr
& \exprcell{
\displaystyle
v\{E_f+Bq\sin(\theta)\}
}
& 5.420 & 0.951 & 42.000 \\

\reprowsep

\qlattice
& \exprcell{
\displaystyle
\begin{gathered}
-29.274(3.813{\times}10^{-4}-0.085q) \\
{}\times
\Big[
0.406E_f
+(9.843{\times}10^{-4}-0.419B)(0.281-0.683\theta) \\
{}\times(0.951-0.303\theta)(1.904-0.325\theta) \\
{}\times(1.872v-0.002)-0.009
\Big]
-0.152
\end{gathered}
}
& 1.165 & 0.998 & 96.701 \\

\reprowsep

\sisso$++$
& \exprcell{
\displaystyle
\begin{gathered}
0.076
+0.991\,E_fq
+0.987\,Bvq\sin(\theta) \\
{}+0.008\,v\sin(\theta)(Bq+E_f+q)
\end{gathered}
}
& 0.994 & 0.998 & 43.563 \\

\reprowsep

\gplearn
& \exprcell{
\displaystyle
\begin{gathered}
qE_f+v
\Big[
qB\sin(\theta)
+\exp\{-(0.569^{-1})^{3}\} \\
{}+\exp\{-\exp(0.569^{-1})\}
+\exp\{-(0.569^{-1})^{3}\}
\Big]
\end{gathered}
}
& 1.052 & 0.998 & 16.375 \\

\reprowsep

\operon
& \exprcell{
\displaystyle
\begin{gathered}
0.056
-180.230(0.806q)
\Big[
2.276(0.073B)(0.073v)\sin(-0.995\theta) \\
{}-0.007E_f
-(0.073B)(0.073v)\sin(-0.988\theta)
\Big]
\end{gathered}
}
& 0.985 & 0.999 & 24.154 \\

\reprowsep

\pysr
& \exprcell{
\displaystyle
q\{E_f+Bv\sin(\theta)\}
}
& 1.056 & 0.998 & 117.745 \\

\reprowsep

\bms
& \exprcell{
\displaystyle
qE_f+Bqv\sin(\theta)
}
& 0.978 & 0.999 & 285.738 \\

\reprowsep

\bsr
& \exprcell{
\displaystyle
-5.198+2.083(B+\theta)+29.568\sin(\theta)+3.449\sin(B)
}
& 17.252 & 0.604 & 164.415 \\

\addlinespace[4pt]
\end{longtable}
\endgroup

\subsection{For \texorpdfstring{$\mathrm{I\_24\_6}$~\eqnref[supple-eq:feynman-ehfo]}{I.24.6}}

\begingroup
\footnotesize
\setlength{\tabcolsep}{2.8pt}
\renewcommand{\arraystretch}{1.20}

\begin{longtable}{@{}>{\centering\arraybackslash}p{0.105\textwidth} >{\raggedright\arraybackslash}p{0.585\textwidth} >{\centering\arraybackslash}p{0.090\textwidth} >{\centering\arraybackslash}p{0.080\textwidth} >{\centering\arraybackslash}p{0.090\textwidth}@{}}
\caption{Best repetition results for the Feynman equation $\mathrm{I\_24\_6}$~\eqnref[supple-eq:feynman-ehfo]: $E_n = \tfrac{1}{4} m (\omega^2 + \omega_0^2)x^{2}$, across methods and noise levels. For each method and noise level, the displayed expression is selected from the repetition with the lowest train \texttt{RMSE} among $5$ independent repetitions. Test \texttt{RMSE} and $R^2$ are computed on the held-out $10\%$ test set.}
\label{tab:i-24-6-best-train-rmse-all-methods-results}\\

\toprule
\toprule
Method
& Learned expression
& Test \texttt{RMSE}
& Test $R^2$
& Runtime (s) \\
\midrule
\endfirsthead

\caption[]{\emph{(Continued)}. Best repetition results for the Feynman equation $\mathrm{I\_24\_6}$~\eqnref[supple-eq:feynman-ehfo]: $E_n = \tfrac{1}{4} m (\omega^2 + \omega_0^2)x^{2}$, across methods and noise levels.}\\
\toprule
\toprule
Method
& Learned expression
& Test \texttt{RMSE}
& Test $R^2$
& Runtime (s) \\
\midrule
\endhead

\midrule
\multicolumn{5}{r}{\emph{Continued on next page}}\\
\endfoot

\bottomrule
\bottomrule
\endlastfoot

\addlinespace[3pt]
\multicolumn{5}{@{}l}{\textcolor{BrickRed}{\textit{Noiseless setting}}}\\
\addlinespace[2pt]

\rowcolor{ExprBack}\hierbosss
& \exprcell{
\exprbox{\finaltag}{
& 0.248\,m\omega^{2}x^{2}
+0.248\,m\omega_{0}^{2}x^{2}
}
}
& $2.547{\times}10^{-8}$ & 1.000 & 60.097 \\

\reprowsep

\dsr
& \exprcell{
\displaystyle
x\{x+m\omega\omega_0\}
}
& 4.721 & 0.899 & 30.600 \\

\reprowsep

\qlattice
& \exprcell{
\displaystyle
\begin{gathered}
0.039m
-11.667(0.969-2.099x) \\
{}\times
\Big[
0.250(0.469m+0.018)(-0.667x-0.544) \\
{}\times
\{-0.475\omega_0
+0.565(2.908-0.676\omega_0)(-0.080\omega-0.296)
\\
\times(0.710\omega-2.618)e^{0.449\omega_0}
-1.202\}
{}-0.018
\Big]
+0.288
\end{gathered}
}
& 0.271 & 1.000 & 95.968 \\

\reprowsep

\sisso$++$
& \exprcell{
\displaystyle
\begin{gathered}
-5.394{\times}10^{-16}
-7.822{\times}10^{-18}(x\omega)(xm)(\omega_0+\omega)(\omega_0m) \\
{}-5.118{\times}10^{-18}(x\omega)\omega^2(x+\omega+x^2)
+0.250\,mx^2(\omega_0^2+\omega^2)
\end{gathered}
}
& 0.000 & 1.000 & 20.426 \\

\reprowsep

\gplearn
& \exprcell{
\displaystyle
\begin{gathered}
e^{-1/\omega_0}
\Big[
e^{-2/\omega_0}
\{mx^2(\omega+1)e^{-1/\omega_0}+x(0.842+m\omega x)\} \\
{}+x(\omega_0^{-1}+m\omega x)
\Big]
\end{gathered}
}
& 2.154 & 0.979 & 17.269 \\

\reprowsep

\operon
& \exprcell{
\displaystyle
\begin{gathered}
0.485
-2.089
\Big[
e^{1.002\omega}(0.707)(-0.290x)e^{0.300\omega_0}
+\sin\{e^{0.477\omega_0}\}
\Big] \\
{}\times
\{(0.434m)(2.718x)e^{-0.581\omega}\}
\end{gathered}
}
& 0.704 & 0.998 & 29.436 \\

\reprowsep

\pysr
& \exprcell{
\displaystyle
0.250\,mx^2(\omega^2+\omega_0^2)
}
& 0.000 & 1.000 & 66.755 \\

\reprowsep

\bms
& \exprcell{
\displaystyle
mx^2\{\omega+\omega_0+2a_0\}
}
& 1.171 & 0.993 & 241.607 \\

\reprowsep

\bsr
& \exprcell{
\displaystyle
77.459
-88.506\cos(e^{-\omega_0})
-8.088\cos(x)
-12.942\cos(\omega_0)
+5.242mx
}
& 6.964 & 0.764 & 195.174 \\

\addlinespace[4pt]
\midrule

\addlinespace[3pt]
\multicolumn{5}{@{}l}{\textcolor{BrickRed}{\textit{Noise level}: $\sigma=0.15$}}\\
\addlinespace[2pt]

\rowcolor{ExprBack}\hierbosss
& \exprcell{
\exprbox{\finaltag}{
& 0.250\,m\omega^{2}x^{2}
+0.250\,m\omega_{0}^{2}x^{2}
}
}
& 0.143 & 1.000 & 61.153 \\

\reprowsep

\dsr
& \exprcell{
\displaystyle
\omega+\{-\omega_0+\omega+m\omega_0\}x^{2}
}
& 4.385 & 0.904 & 32.100 \\

\reprowsep

\qlattice
& \exprcell{
\displaystyle
\begin{gathered}
2.459(-6.498{\times}10^{-5}m-1)^2
(0.122m+1.847{\times}10^{-4}) \\
{}\times(-0.008\omega+1.825x-0.955) \\
{}\times
\Big[
1.256\omega_0
+0.720(\omega-0.068)^2
-2.947
+(0.547-0.123\omega_0)^{-1}
\Big]\\
\times
e^{0.295x}
-0.017
\end{gathered}
}
& 0.265 & 1.000 & 94.493 \\

\reprowsep

\sisso$++$
& \exprcell{
\displaystyle
\begin{gathered}
0.017
+0.250\,mx^2(\omega_0^2+\omega^2)
+0.001(x\omega_0)m(x+\omega)^2 \\
{}-0.003(x\omega_0)m(x\omega+x+\omega)
\end{gathered}
}
& 0.149 & 1.000 & 19.172 \\

\reprowsep

\gplearn
& \exprcell{
\displaystyle
\begin{gathered}
\sin(0.678^2)\cos(\sin\omega)
\Big[
m\omega_0x^2
\cos\{\cos[\sin(m\sin\omega\cos(\sin\omega))]\}
+\omega x^2
\Big] \\
{}+m\omega_0x^2\cos(\sin\omega)
\end{gathered}
}
& 2.428 & 0.974 & 16.442 \\

\reprowsep

\operon
& \exprcell{
\displaystyle
\begin{gathered}
0.011
-0.089
\Big[
-0.066\omega_0
+\{(-1.043\omega_0-1.101\omega)(-1.911x)\}^{2} \\
{}+\{(-1.047\omega_0+0.990\omega)(-2.012x)\}^{2}
\Big](-0.333m)
\end{gathered}
}
& 0.143 & 1.000 & 22.110 \\

\reprowsep

\pysr
& \exprcell{
\displaystyle
0.250\,mx^2(\omega_0^2+\omega^2)
}
& 0.143 & 1.000 & 63.275 \\

\reprowsep

\bms
& \exprcell{
\displaystyle
\begin{gathered}
\{(\omega+\omega_0)^2+a_0m\omega_0\}
m\sin(a_0)x^2 \\
{}+\{a_0(a_0+\omega^2)+\cos(\omega_0)\}^{2}
\end{gathered}
}
& 0.566 & 0.998 & 253.143 \\

\reprowsep

\bsr
& \exprcell{
\displaystyle
\begin{gathered}
-30.878
+0.729\cos(\omega)
-5.018m^{-1}
-6.903(\omega+\omega_0)
\\
+8.028(x+m+\omega\omega_0+x)
\end{gathered}
}
& 5.786 & 0.850 & 190.559 \\

\addlinespace[4pt]
\midrule

\addlinespace[3pt]
\multicolumn{5}{@{}l}{\textcolor{BrickRed}{\textit{Noise level}: $\sigma=0.25$}}\\
\addlinespace[2pt]

\rowcolor{ExprBack}\hierbosss
& \exprcell{
\exprbox{\finaltag}{
& 0.250\,m\omega^{2}x^{2}
+0.250\,m(\omega_{0}^{2}x^{2}+1)\\
&\quad
-0.250\,m
}
}
& 0.264 & 1.000 & 64.655 \\

\reprowsep

\dsr
& \exprcell{
\displaystyle
x\{m\omega\omega_0+x\}
}
& 5.300 & 0.869 & 34.700 \\

\reprowsep

\qlattice
& \exprcell{
\displaystyle
\begin{gathered}
20.732(0.382-1.114x)(0.386m-0.001)(-0.394x-0.199) \\
{}\times
\Big[
(-0.191\omega-0.095)
\{-0.375(-0.002m-0.554)(-0.027\omega-2.015)\\
\times
(0.809\omega-1.322)-0.441\}
{}+0.069(-\omega_0-0.015)^2
\Big]
+0.108
\end{gathered}
}
& 0.294 & 1.000 & 95.126 \\

\reprowsep

\sisso$++$
& \exprcell{
\displaystyle
\begin{gathered}
-0.012
-2.034{\times}10^{-4}x^4\omega^4
+3.247{\times}10^{-4}(x\omega)(x+\omega)x^2\omega^2 \\
{}+0.250\,mx^2(\omega_0^2+\omega^2)
\end{gathered}
}
& 0.249 & 1.000 & 19.302 \\

\reprowsep

\gplearn
& \exprcell{
\displaystyle
\begin{gathered}
\cos(x+\omega)+2\omega^2+m\omega_0x^2
+e^{-e^x} \\
{}+2\{\sin(-0.286)\}^{-1}
+\exp\{\cos(mx)\}
\end{gathered}
}
& 3.523 & 0.945 & 17.446 \\

\reprowsep

\operon
& \exprcell{
\displaystyle
\begin{gathered}
0.248
-0.281
\Big[
\{(0.485x-1.914x)-3.145\omega\}(0.485x)(0.707m)(1.875\omega) \\
{}-\{(-0.809\omega_0)(0.784x)\}^{2}(2.190m)
+2.074\omega(2.190m)
\Big]
\end{gathered}
}
& 0.838 & 0.996 & 24.293 \\

\reprowsep

\pysr
& \exprcell{
\displaystyle
0.250\,mx^2(\omega^2+\omega_0^2)
}
& 0.243 & 1.000 & 57.047 \\

\reprowsep

\bms
& \exprcell{
\displaystyle
\begin{gathered}
a_0\Big[
\{a_0x(\exp[\sin(\sin a_0)]^3+m)(\omega+a_0+\omega_0)\}^{2}
\\
+x\cos\{\omega_0+\omega_0\sin(\omega)\}
\Big]
\end{gathered}
}
& 0.901 & 0.996 & 246.667 \\

\reprowsep

\bsr
& \exprcell{
\displaystyle
21.838
-32.250x^{-1}
+1.481\omega_0^2
+0.577\exp\{\cos(xm)+x\}
+0.269\omega_0^3
}
& 8.730 & 0.660 & 191.737 \\

\addlinespace[4pt]
\end{longtable}
\endgroup

\subsection{For \texorpdfstring{$\mathrm{I\_50\_26}$~\eqnref[supple-eq:feynman-hoqn]}{I.50.26}}

\begingroup
\footnotesize
\setlength{\tabcolsep}{2.8pt}
\renewcommand{\arraystretch}{1.20}

\begin{longtable}{@{}>{\centering\arraybackslash}p{0.105\textwidth} >{\raggedright\arraybackslash}p{0.585\textwidth} >{\centering\arraybackslash}p{0.090\textwidth} >{\centering\arraybackslash}p{0.080\textwidth} >{\centering\arraybackslash}p{0.090\textwidth}@{}}
\caption{Best repetition results for the Feynman equation $\mathrm{I\_50\_26}$~\eqnref[supple-eq:feynman-hoqn]: $x=x_1[\cos(\omega t)+\alpha\cos^2(\omega t)]$, across methods and noise levels. For each method and noise level, the displayed expression is selected from the repetition with the lowest train \texttt{RMSE} among $5$ independent repetitions. Test \texttt{RMSE} and $R^2$ are computed on the held-out $10\%$ test set.}
\label{tab:i-50-26-best-train-rmse-all-methods-results}\\

\toprule
\toprule
Method
& Learned expression
& Test \texttt{RMSE}
& Test $R^2$
& Runtime (s) \\
\midrule
\endfirsthead

\caption[]{\emph{(Continued)}. Best repetition results for the Feynman equation $\mathrm{I\_50\_26}$~\eqnref[supple-eq:feynman-hoqn]: $x=x_1[\cos(\omega t)+\alpha\cos^2(\omega t)]$, across methods and noise levels.}\\
\toprule
\toprule
Method
& Learned expression
& Test \texttt{RMSE}
& Test $R^2$
& Runtime (s) \\
\midrule
\endhead

\midrule
\multicolumn{5}{r}{\emph{Continued on next page}}\\
\endfoot

\bottomrule
\bottomrule
\endlastfoot

\addlinespace[3pt]
\multicolumn{5}{@{}l}{\textcolor{BrickRed}{\textit{Noiseless setting}}}\\
\addlinespace[2pt]

\rowcolor{ExprBack}\hierbosss
& \exprcell{
\exprbox{\finaltag}{
& x_1\{-0.996\,\alpha\sin^2(\omega t)
+\alpha+\cos(\omega t)\}
}
}
& $7.586{\times}10^{-6}$ & 1.000 & 59.610 \\

\reprowsep

\dsr
& \exprcell{
\displaystyle
x_1\exp[\cos(\omega t)]
}
& 1.514 & 0.400 & 43.800 \\

\reprowsep

\qlattice
& \exprcell{
\displaystyle
\begin{gathered}
-0.330t+0.313x_1
+1.256(0.105-0.211x_1)\\
\times(3.152-2.507\alpha)
{}-0.186+\frac{1.256}{\textcolor{BrickRed}{A_2}},\\
\textcolor{BrickRed}{A_2}=43.822\textcolor{BrickRed}{B_2}^2+0.199,\\
\textcolor{BrickRed}{B_2}=0.353\omega+0.162t-0.404(0.505\omega+0.743)\\
\times
e^{-0.403t}-1
\end{gathered}
}
& 1.058 & 0.714 & 85.057 \\

\reprowsep

\sisso$++$
& \exprcell{
\displaystyle
\begin{gathered}
-1.698
+0.034(\alpha+t+\omega+x_1)(\alpha x_1+\alpha) \\
{}+3.251\cos^{2}(t\omega)
+0.196\cos(t\omega)(tx_1\omega)
\end{gathered}
}
& 0.585 & 0.903 & 26.002 \\

\reprowsep

\gplearn
& \exprcell{
\displaystyle
\begin{gathered}
\{\alpha\cos^2(\omega t)
+x_1[0.145\alpha x_1\cos^2(\omega t) \\
{}+0.145\omega\{0.145t
+0.145\cos[0.145\omega t\{\alpha x_1\cos^2(\omega t)+\sin(\omega t)\}]\}\\
{}\times\{x_1\cos^2(\omega t)+\cos^2(t)\}\cos(\omega t)
+0.145t+0.021x_1\\
+0.290\cos(\omega t)]\}\cos\{\sin(\omega t)\}+\cos(\omega t)
\end{gathered}
}
& 0.326 & 0.973 & 20.960 \\

\reprowsep

\operon
& \exprcell{
\displaystyle
\begin{gathered}
-0.972
\Big[
0.248\omega t(-0.693\alpha-0.910x_1)+1.537
\Big] \\
{}\times
\Big[
0.224\alpha+0.118x_1-\sin(1.767\omega t)
\Big]
+0.372
\end{gathered}
}
& 0.580 & 0.914 & 22.804 \\

\reprowsep

\pysr
& \exprcell{
\displaystyle
x_1\{\cos(\omega t)+\alpha\cos^2(\omega t)\}
}
& 0.000 & 1.000 & 100.024 \\

\reprowsep

\bms
& \exprcell{
\displaystyle
-\omega^2a_0^2(x_1+a_0-x_1t)
\{\alpha a_0+\alpha\cos[\sin(\omega t)]\}
}
& 0.827 & 0.829 & 268.458 \\

\reprowsep

\bsr
& \exprcell{
\displaystyle
35.478-37.326\cos(e^{-t})-0.388\cos\alpha
-2.534\cos t+0.457\alpha x_1
}
& 1.727 & 0.239 & 190.812 \\

\addlinespace[4pt]
\midrule

\addlinespace[3pt]
\multicolumn{5}{@{}l}{\textcolor{BrickRed}{\textit{Noise level}: $\sigma=0.15$}}\\
\addlinespace[2pt]

\rowcolor{ExprBack}\hierbosss
& \exprcell{
\exprbox{\finaltag}{
& x_1\{\alpha\cos^2(\omega t)+\cos(\omega t)\}
}
}
& 0.146 & 0.993 & 63.595 \\

\reprowsep

\dsr
& \exprcell{
\displaystyle
\{\sin[\exp\{\exp(t)^{-1}\}]+\cos(\omega t)\}^{3}
}
& 1.526 & 0.290 & 46.400 \\

\reprowsep

\qlattice
& \exprcell{
\displaystyle
\begin{gathered}
4.614(0.098\alpha-1)^2\textcolor{BrickRed}{A_5}^2+0.023,\\
\textcolor{BrickRed}{A_5}=-0.386\alpha-0.265x_1+0.849-e^{-45.106\textcolor{BrickRed}{B_5}^2},\\
\textcolor{BrickRed}{B_5}=0.418\omega
+0.268(1.747-0.661t)(-0.178\alpha\\
-2.005)-1
\end{gathered}
}
& 0.956 & 0.771 & 79.372 \\

\reprowsep

\sisso$++$
& \exprcell{
\displaystyle
\begin{gathered}
-1.491
+0.051(\alpha+t+t+\omega)(\alpha x_1+\cos t) \\
{}+3.242\cos^{2}(t\omega)
+0.200\cos(t\omega)(t\omega x_1)
\end{gathered}
}
& 0.600 & 0.908 & 27.228 \\

\reprowsep

\gplearn
& \exprcell{
\displaystyle
\begin{gathered}
0.391x_1\{2\alpha\cos^2(\omega t)-\sin(0.868x_1-\cos(\omega t))
+\cos^4(\omega t)\\
+3\cos(\omega t)+\cos(\cos x_1)\}
\end{gathered}
}
& 0.324 & 0.976 & 19.991 \\

\reprowsep

\operon
& \exprcell{
\displaystyle
\begin{gathered}
0.169x_1
\Big[
5.934\alpha+
\frac{5.908}{\cos(1.006\omega t)}
\Big]
\cos^2(1.003\omega t)
-0.011
\end{gathered}
}
& 0.143 & 0.995 & 26.387 \\

\reprowsep

\pysr
& \exprcell{
\displaystyle
x_1\{\cos(\omega t)+\alpha\cos^2(\omega t)\}
}
& 0.142 & 0.995 & 99.897 \\

\reprowsep

\bms
& \exprcell{
\displaystyle
\begin{gathered}
\omega\Big[
tx_1\{\cos(a_0^3)+\sin(2a_0)\}
\{\cos(\omega t)+a_0+\cos(2\omega t)+a_0^3+\alpha\}\\
{}+\alpha\exp\{-\exp[a_0+\sin(a_0)+\exp(t+a_0)]\}
\Big]
\end{gathered}
}
& 0.641 & 0.873 & 326.618 \\

\reprowsep

\bsr
& \exprcell{
\displaystyle
-2.954-0.051\cos\omega-0.922x_1^{-1}
-0.002(\omega+t)+0.502(x_1+\omega t+2\alpha)
}
& 1.622 & 0.340 & 190.397 \\

\addlinespace[4pt]
\midrule

\addlinespace[3pt]
\multicolumn{5}{@{}l}{\textcolor{BrickRed}{\textit{Noise level}: $\sigma=0.18$}}\\
\addlinespace[2pt]

\rowcolor{ExprBack}\hierbosss
& \exprcell{
\exprbox{\finaltag}{
& \alpha x_1\cos^2(\omega t)
+x_1\cos(\omega t)
+0.023\,\sin(\omega)
}
}
& 0.189 & 0.992 & 59.533 \\

\reprowsep

\dsr
& \exprcell{
\displaystyle
x_1\exp[\cos(t\omega)]
}
& 1.482 & 0.429 & 45.000 \\

\reprowsep

\qlattice
& \exprcell{
\displaystyle
\begin{gathered}
0.994\alpha-0.044t+0.728x_1
+1.085e^{\textcolor{BrickRed}{A_4B_4}}-2.569,\\
\textcolor{BrickRed}{A_4}=-0.269\omega-2.001,\\
\textcolor{BrickRed}{B_4}=0.076\omega+100.427(1-0.200\omega-0.196t)^2-0.810
\end{gathered}
}
& 0.980 & 0.793 & 95.180 \\

\reprowsep

\sisso$++$
& \exprcell{
\displaystyle
\begin{gathered}
-1.491
+0.051(\alpha+t+t+\omega)(\alpha x_1+\cos t) \\
{}+3.242\cos^{2}(t\omega)
+0.200\cos(t\omega)(tx_1\omega)
\end{gathered}
}
& 0.607 & 0.907 & 25.341 \\

\reprowsep

\gplearn
& \exprcell{
\displaystyle
\begin{gathered}
x_1\Big\{\alpha+\sin\Big[(\alpha+\cos\omega)\cos^3(\omega t)\\
{}+\sin^3\{\sin^3[(\alpha+\cos(\omega+0.528x_1))
\sin^2((\alpha+\cos(\omega+0.528x_1))\\
{}\times\cos^2(\omega t)-\omega^{-3})-t^{-3}]
+\cos^2x_1\cos(\omega t)\}\Big]\Big\}\cos^2(\omega t)\\
{}+\cos(\omega t)
\end{gathered}
}
& 0.533 & 0.919 & 20.623 \\

\reprowsep

\operon
& \exprcell{
\displaystyle
\begin{gathered}
-0.774x_1
\Big[
1.383\alpha
-\frac{1.481\alpha}{\cos\{\cos(1.003\omega t)\}}
+0.354
\Big] \\
{}+1.954\cos(1.001\omega t)+0.534
\end{gathered}
}
& 0.447 & 0.948 & 25.782 \\

\reprowsep

\pysr
& \exprcell{
\displaystyle
x_1\{\cos(\omega t)+\alpha\cos^2(\omega t)\}
}
& 0.190 & 0.992 & 99.745 \\

\reprowsep

\bms
& \exprcell{
\displaystyle
\begin{gathered}
\Big[
a_0\cos(e^{\sin t})
+\{\cos(\omega t)+a_0\}e^{\cos(a_0x_1)}\\
{}\times x_1(a_0+a_0\sin a_0)a_0
\{[\cos(-\omega\sin a_0)+\cos a_0]\\
\times 
[\sin x_1+\alpha]+\omega\}
+a_0
\Big]^2
\end{gathered}
}
& 0.521 & 0.942 & 317.026 \\

\reprowsep

\bsr
& \exprcell{
\displaystyle
0.116-1.696\alpha^{-1}
+0.965t^2+0.037e^{\cos(\alpha x_1)+\alpha}
-0.227t^3
}
& 1.770 & 0.186 & 191.171 \\

\addlinespace[4pt]
\end{longtable}
\endgroup

\subsection{For \texorpdfstring{$\mathrm{II\_36\_38}$~\eqnref[supple-eq:feynman-fmmm]}{II.36.38}}

\begingroup
\footnotesize
\setlength{\tabcolsep}{2.8pt}
\renewcommand{\arraystretch}{1.20}

\begin{longtable}{@{}>{\centering\arraybackslash}p{0.105\textwidth} >{\raggedright\arraybackslash}p{0.585\textwidth} >{\centering\arraybackslash}p{0.090\textwidth} >{\centering\arraybackslash}p{0.080\textwidth} >{\centering\arraybackslash}p{0.090\textwidth}@{}}
\caption{Best repetition results for the Feynman equation $\mathrm{II\_36\_38}$~\eqnref[supple-eq:feynman-fmmm]: $f = \tfrac{\mathrm{mom}H}{k_b T} + \tfrac{\mathrm{mom}\alpha}{\epsilon c^{2} k_b T}M$, across methods and noise levels. For each method and noise level, the displayed expression is selected from the repetition with the lowest train \texttt{RMSE} among $5$ independent repetitions. Test \texttt{RMSE} and $R^2$ are computed on the held-out $10\%$ test set.}
\label{tab:ii-36-38-best-train-rmse-all-methods-results}\\

\toprule
\toprule
Method
& Learned expression
& Test \texttt{RMSE}
& Test $R^2$
& Runtime (s) \\
\midrule
\endfirsthead

\caption[]{\emph{(Continued)}. Best repetition results for the Feynman equation $\mathrm{II\_36\_38}$~\eqnref[supple-eq:feynman-fmmm]: $f = \tfrac{\mathrm{mom}H}{k_b T} + \tfrac{\mathrm{mom}\alpha}{\epsilon c^{2} k_b T}M$, across methods and noise levels.}\\
\toprule
\toprule
Method
& Learned expression
& Test \texttt{RMSE}
& Test $R^2$
& Runtime (s) \\
\midrule
\endhead

\midrule
\multicolumn{5}{r}{\emph{Continued on next page}}\\
\endfoot

\bottomrule
\bottomrule
\endlastfoot

\addlinespace[3pt]
\multicolumn{5}{@{}l}{\textcolor{BrickRed}{\textit{Noiseless setting}}}\\
\addlinespace[2pt]

\rowcolor{ExprBack}\hierbosss
& \exprcell{
\exprbox{\finaltag}{
& \frac{1.0 H \mathrm{mom}}{T k_b} + \frac{1.0 M \alpha \mathrm{mom}}{T c^{2} \epsilon k_b}
}
}
& $4.782{\times}10^{-7}$ & 1.000 & 1116.153 \\

\reprowsep

\dsr
& \exprcell{
\displaystyle
\mathrm{mom}\{T^{-1}+\sin[(k_b^2)^{-1}]\}
}
& 0.650 & 0.635 & 41.300 \\

\reprowsep

\qlattice
& \exprcell{
\displaystyle
\begin{gathered}
0.207\alpha
+0.257\Big[
0.648H
+(0.366T-2.270)(0.065\epsilon-0.701)\\
{}\times
\{-0.869T-0.578c-1.064k_b+0.897\mathrm{mom}
+(0.868-0.162\epsilon)\\
\times (0.483M-0.091)+2.964\}
{}+0.238
\Big]^2
+0.132
\end{gathered}
}
& 0.262 & 0.955 & 89.144 \\

\reprowsep

\sisso$++$
& \exprcell{
\begin{aligned}[t]
& -0.459 \\
&\quad + (0.368)\,\{(\alpha\,\mathrm{mom})/(c\epsilon)\} \\
&\quad + (1.400)\,\{(M/k_b)/(cT)\} \\
&\quad + (0.993)\,\{(H\,\mathrm{mom})/(Tk_b)\}
\end{aligned}
}
& 0.173 & 0.971 & 17.246 \\

\reprowsep

\gplearn
& \exprcell{
\displaystyle
\frac{H\,\mathrm{mom}}{Tk_b^2}+\frac{H}{Tck_b}+\frac{\mathrm{mom}}{Tc}
}
& 0.324 & 0.904 & 19.868 \\

\reprowsep

\operon
& \exprcell{
\displaystyle
\begin{gathered}
-0.413
-0.671
\Bigg[
\left\{
\frac{-1.390\mathrm{mom}}{-0.522c}
+1.389\mathrm{mom}
\right\}\\
{}\times
\frac{1.599H+0.608M}
{\cos(0.236\alpha)(5.930T)(-0.597k_b)}
-\frac{-0.639}{-0.568\epsilon}
\Bigg]
\end{gathered}
}
& 0.285 & 0.917 & 24.610 \\

\reprowsep

\pysr
& \exprcell{
\displaystyle
\frac{\mathrm{mom}}{Tk_b}
\left\{
H+\frac{M\alpha}{\epsilon c^2}
\right\}
}
& 0.000 & 1.000 & 73.380 \\

\reprowsep

\bms
& \exprcell{
\displaystyle
\exp^{3}\!\left(
a_0\{H+\cos a_0+\sin(c+\epsilon)+\mathrm{mom}-Tk_b+\alpha^{2}a_0\}
\right)
}
& 0.394 & 0.858 & 2485.010 \\

\reprowsep

\bsr
& \exprcell{
\displaystyle
\begin{gathered}
4.692
-1.532\{k_b+\exp[\cos(k_b)]\}
+0.017\exp(\alpha) \\
{}+0.552H
-0.157\sin(\mathrm{mom}^{2})
\end{gathered}
} 
& 0.922 & 0.224 & 1912.210 \\

\addlinespace[4pt]
\midrule

\addlinespace[3pt]
\multicolumn{5}{@{}l}{\textcolor{BrickRed}{\textit{Noise level}: $\sigma=0.15$}}\\
\addlinespace[2pt]

\rowcolor{ExprBack}\hierbosss
& \exprcell{
\exprbox{\finaltag}{
& \frac{1.000 H \mathrm{mom}}{T k_b} + \frac{0.990 M \alpha \mathrm{mom}}{T c^{2} \epsilon k_b}
}
}
& 0.144 & 1.000 & 1039.296 \\

\reprowsep

\dsr
& \exprcell{
\displaystyle
\{\mathrm{mom}^{2}+\sin(\mathrm{mom})\}(k_bT)^{-1}
}
& 0.634 & 0.653 & 50.600 \\

\reprowsep

\qlattice
& \exprcell{
\displaystyle
\begin{gathered}
-0.195k_b
-1.809
\Big[
-0.715H
+(0.124-0.546\alpha)(1.655-0.579c)\\
\times (0.553M-0.110)
+0.207
\Big]\\
{}\times
(-0.680\epsilon+2.558\mathrm{mom}+0.449)
\exp(-0.653T)\exp(-0.537k_b)
+0.649
\end{gathered}
}
& 0.347 & 0.914 & 112.151 \\

\reprowsep

\sisso$++$
& \exprcell{
\begin{aligned}[t]
& -0.466 \\
&\quad + (0.372)\,\{(\alpha\,\mathrm{mom})/(c\epsilon)\} \\
&\quad + (1.418)\,\{(M/c)/(Tk_b)\} \\
&\quad + (0.993)\,\{(H\,\mathrm{mom})/(Tk_b)\}
\end{aligned}
}
& 0.229 & 0.951 & 17.099 \\

\reprowsep

\gplearn
& \exprcell{
\displaystyle
\begin{gathered}
c^{-1}
+T^{-1}\Bigg[
\frac{H(\mathrm{mom}-0.859)}{k_b}\\
{}+
T^{-1}\left\{
\frac{H\,\mathrm{mom}\cos(k_b^{-1})}{k_b^2}
-0.859+\frac{2}{k_b}
\right\}
\Bigg]
\end{gathered}
}
& 0.590 & 0.752 & 18.665 \\

\reprowsep

\operon
& \exprcell{
\displaystyle
\begin{gathered}
-0.001
+63.119
\\
\times \frac{
(0.930H)(0.862\mathrm{mom})
}{
2.718k_b
\left[
2.008c+
\frac{1.078H-1.772\alpha}{1.367c}
+1.333\epsilon+0.910H
\right]
(1.772T)
}
\end{gathered}
}
& 0.299 & 0.930 & 24.575 \\

\reprowsep

\pysr
& \exprcell{
\displaystyle
\frac{\mathrm{mom}}{Tk_b}
\left\{
H+\frac{M\{\cos(\epsilon)+1.423\}}{c^2}
\right\}
}
& 0.246 & 0.948 & 71.924 \\

\reprowsep

\bms
& \exprcell{
\displaystyle
\begin{gathered}
\{\mathrm{mom}(\alpha+H)+\cos c+M\}\\
{}\times
\exp\{a_0(T+a_0+k_b)\}+a_0
\end{gathered}
}
& 0.399 & 0.862 & 2417.780 \\

\reprowsep

\bsr
& \exprcell{
\displaystyle
\begin{gathered}
6.922
-1.301\cos(H)
-2.606\,\mathrm{mom}^{-1} \\
{}-0.467(H+c)
-0.243(T+\alpha+\epsilon k_b+M)
\end{gathered}
}
& 0.707 & 0.577 & 1924.410 \\

\addlinespace[4pt]
\midrule

\addlinespace[3pt]
\multicolumn{5}{@{}l}{\textcolor{BrickRed}{\textit{Noise level}: $\sigma=0.20$}}\\
\addlinespace[2pt]

\rowcolor{ExprBack}\hierbosss
& \exprcell{
\exprbox{\finaltag}{
& \frac{1.000 H \mathrm{mom}}{T k_b} + \frac{1.3 M \mathrm{mom} \left(2 \alpha + \epsilon\right)}{c^{2} k_b \left(T \epsilon + 1\right)^{2}} \\
&\quad + \frac{0.0043 \alpha^{2} \left(M \mathrm{mom} + M \cos{\left(k_b \right)} + 2\right)^{2}}{c^{2} \epsilon} \\
&\quad + \frac{0.077}{c}
}
}
& 0.202 & 1.000 & 1177.871 \\

\reprowsep

\dsr
& \exprcell{
\displaystyle
(\alpha+c+\mathrm{mom}^{2})(T+k_bT)^{-1}
}
& 0.677 & 0.648 & 57.800 \\

\reprowsep

\qlattice
& \exprcell{
\displaystyle
\begin{gathered}
28.951\exp\!\Big[
0.145M-0.641T
+(0.847-0.110H)\\
\times (-0.293\epsilon-1.027k_b+0.894\mathrm{mom}-3.160)\\
{}+(-0.171\alpha-0.065)(0.718c-2.117)
\Big]
+0.156
\end{gathered}
}
& 0.344 & 0.915 & 121.915 \\

\reprowsep

\sisso$++$
& \exprcell{
\begin{aligned}[t]
& -0.469 \\
&\quad + (0.373)\,\{(\alpha\,\mathrm{mom})/(c\epsilon)\} \\
&\quad + (1.424)\,\{(M/c)/(Tk_b)\} \\
&\quad + (0.994)\,\{(H\,\mathrm{mom})/(Tk_b)\}
\end{aligned}
}
& 0.263 & 0.936 & 16.868 \\

\reprowsep

\gplearn
& \exprcell{
\displaystyle
\begin{gathered}
c^{-1}\Bigg[
k_b e^{-k_b}
\left\{
e^{\mathrm{mom}/T}
+\frac{\mathrm{mom}(M+0.175)\sin(\epsilon^{-1})}{k_b^2}
\right\}\\
{}+
\frac{\mathrm{mom}\{-\cos H+c^{-1}\}\sin(\epsilon^{-1})}{k_b^2}
\Bigg]
+\frac{-\cos H+\mathrm{mom}/k_b}{T}
\end{gathered}
}
& 0.346 & 0.901 & 22.410 \\

\reprowsep

\operon
& \exprcell{
\displaystyle
\begin{gathered}
-0.015
+2.075
\frac{
\frac{\sin(-0.413H)}{0.184T}
}{
\frac{1.862k_b}{0.434\mathrm{mom}}
\left[
\sin\{\sin(-0.509\epsilon)\}
-\frac{2.186M+2.135\alpha}{(1.507\epsilon)(2.350c)}
\right]^{-1}
}
\end{gathered}
}
& 0.260 & 0.951 & 27.554 \\

\reprowsep

\pysr
& \exprcell{
\displaystyle
\frac{\mathrm{mom}}{Tk_b}
\left\{
H+\frac{\alpha+M-\epsilon}{c^2}
\right\}
}
& 0.228 & 0.957 & 110.782 \\

\reprowsep

\bms
& \exprcell{
\displaystyle
(H-c+\mathrm{mom})
\exp\!\left[
\cos\{a_0^{3}(k_b+a_0+T)\}
\right]
}
& 0.488 & 0.814 & 2647.510 \\

\reprowsep

\bsr
& \exprcell{
\displaystyle
\begin{gathered}
1.530
+3.028\,T^{-1}
-0.907\,k_b^2 \\
{}+0.023\exp\{\cos(M\mathrm{mom})+M\}
+0.218\,k_b^3
\end{gathered}
}
& 0.773 & 0.532 & 1933.380 \\

\addlinespace[4pt]
\end{longtable}
\endgroup

\newpage
\section{Symbolic Expressions Learned by \texorpdfstring{\hierbosss}{HierBOSSS} for the Feynman Equations at an Additional Noise Level}
\label{sec:symbolic-expressions-Feynman-BayeSymX-additional-noise-level}

\begingroup
\footnotesize
\setlength{\tabcolsep}{1.2pt}
\renewcommand{\arraystretch}{1.18}

\begin{longtable}{
    @{}
    c
    >{\centering\arraybackslash}p{0.25\textwidth}
    >{\raggedright\arraybackslash}p{0.42\textwidth}
    c
    c
    c
    @{}
}

\caption{Best repetition results of \hierbosss\ for the Feynman equations in $\mathrm{I\_12\_2}$~\eqnref[supple-eq:feynman-cl]-$\mathrm{II\_36\_38}$~\eqnref[supple-eq:feynman-fmmm], at an additional higher noise level. The displayed expression is selected from the repetition with the lowest train \texttt{RMSE} among $5$ independent repetitions. Test \texttt{RMSE} and $R^2$ are computed on the held-out $10\%$ test set.}
\label{tab:BayeSymX-best-highest-noise}
\\

\toprule
\toprule
{Noise level}
& {Feynman equation}
& {Learned expression}
& {Test \texttt{RMSE}}
& {Test $R^2$}
& {Runtime (s)} \\
\midrule
\endfirsthead

\caption[]{\textit{(Continued)}. Best repetition results of \hierbosss\ for the Feynman equations in $\mathrm{I\_12\_2}$~\eqnref[supple-eq:feynman-cl]-$\mathrm{II\_36\_38}$~\eqnref[supple-eq:feynman-fmmm], at an additional higher noise level.}
\\
\toprule
\toprule
{Noise level}
& {Feynman equation}
& {Learned expression}
& {Test \texttt{RMSE}}
& {Test $R^2$}
& {Runtime (s)} \\
\midrule
\endhead

\midrule
\multicolumn{6}{r}{\textit{Continued on next page}}\\
\endfoot

\bottomrule
\bottomrule
\endlastfoot

$\sigma=0.25$
&
\begin{minipage}[t]{\linewidth}
\centering
$\mathrm{I\_12\_2}$~\eqnref[supple-eq:feynman-cl]
\end{minipage}
&
\exprcell{\exprbox{\finaltag}{
& 
-0.089
+0.081\,\frac{q_1q_2}{\epsilon r^2}
+0.160\,\frac{1}{r}
}}
&
0.245
&
$-0.007$
&
58.167
\\[4pt]

\reprowsep

$\sigma=2.00$
&
\begin{minipage}[t]{\linewidth}
\centering
$\mathrm{I\_12\_11}$~\eqnref[supple-eq:feynman-fce]
\end{minipage}
&
\exprcell{\exprbox{\finaltag}{
& 
0.994\,E_fq
+1.004\,Bqv\sin(\theta)
}}
&
1.958
&
0.994
&
15.997
\\[4pt]

\reprowsep

$\sigma=1.00$
&
\begin{minipage}[t]{\linewidth}
\centering
$\mathrm{I\_24\_6}$~\eqnref[supple-eq:feynman-ehfo]
\end{minipage}
&
\exprcell{\exprbox{\finaltag}{
& 
0.250\,m\omega^2x^2
+0.250\,m\omega_0^2x^2
}}
&
0.974
&
0.995
&
67.711
\\[4pt]

\reprowsep

$\sigma=0.20$
&
\begin{minipage}[t]{\linewidth}
\centering
$\mathrm{I\_50\_26}$~\eqnref[supple-eq:feynman-hoqn]
\end{minipage}
&
\exprcell{\exprbox{\finaltag}{
& 
x_1\left\{
\alpha\cos^2(\omega t)
+\cos(\omega t)
\right\}
}}
&
0.195
&
0.990
&
68.162
\\[4pt]

\reprowsep

$\sigma=0.25$
&
\begin{minipage}[t]{\linewidth}
\centering
$\mathrm{II\_36\_38}$~\eqnref[supple-eq:feynman-fmmm]
\end{minipage}
&
\exprcell{\exprbox{\finaltag}{
& 
\frac{H\,\mathrm{mom}}{T k_b}
+
\frac{M\alpha\,\mathrm{mom}}
     {T c^2\epsilon k_b}
}}
&
0.211
&
1.000
&
1078.759
\\

\end{longtable}
\endgroup

\newpage
\section{Occam's Window Set \texorpdfstring{($\mathcal J_r, r=10$)}{Jr} Learned by \texorpdfstring{\hierbosss}{HierBOSSS} for the Feynman Equations}
\label{sec:Occams-window-set-Feynman-BayeSymX}


\subsection{For \texorpdfstring{$\mathrm{I\_12\_2}$~\eqnref[supple-eq:feynman-cl]}{I.12.2}}

\begingroup
\footnotesize
\setlength{\tabcolsep}{2.0pt}
\renewcommand{\arraystretch}{1.22}

\begin{longtable}{@{}c >{\raggedright\arraybackslash}p{0.50\textwidth} c c H@{}}
\caption{The Occam's window set ($\mathcal J_r, r=10$) of symbolic expressions ranked by $\mathrm{JMP}$ learned by \hierbosss\ for the Feynman equation $\mathrm{I\_12\_2}$~\eqnref[supple-eq:feynman-cl]: 
$F = \tfrac{q_1q_2}{4\pi\epsilon r^2}$, across noise levels. Test \texttt{RMSE} and $R^2$ are computed on the held-out $10\%$ test set.}
\label{tab:BayeSymX-i-12-2-rep1-top10-distinct-results}\\

\toprule
\toprule
Rank 
& Learned expression 
& Test \texttt{RMSE} 
& Test $R^2$ 
& $\log\mathrm{JMP}$ \\
\midrule
\endfirsthead

\caption[]{\emph{(Continued)}. The Occam's window set ($\mathcal J_r, r=10$) of symbolic expressions ranked by $\mathrm{JMP}$ learned by \hierbosss\ for the Feynman equation $\mathrm{I\_12\_2}$~\eqnref[supple-eq:feynman-cl]: 
$F = \tfrac{q_1q_2}{4\pi\epsilon r^2}$, across noise levels.}\\
\toprule
\toprule
Rank 
& Learned expression 
& Test \texttt{RMSE} 
& Test $R^2$ 
& $\log\mathrm{JMP}$ \\
\midrule
\endhead

\midrule
\multicolumn{5}{r}{\emph{Continued on next page}}\\
\endfoot

\bottomrule
\bottomrule
\endlastfoot

\addlinespace[3pt]
\multicolumn{5}{@{}l}{\textcolor{BrickRed}{\textit{Noiseless setting}, Repetition $1$}}\\
\addlinespace[2pt]

1--10

& \exprcell{\exprbox{\finaltag}{
& 0.080\,\frac{q_1q_2}{\epsilon r^2}
}}
& $5.503{\times}10^{-6}$ 
& 1.000 
& --- \\

\addlinespace[4pt]
\midrule
\addlinespace[2pt]
\multicolumn{5}{@{}l}{\textcolor{BrickRed}{\textit{Noise level}: $\sigma=0.15$, Repetition $1$}}\\
\addlinespace[2pt]

1--10

& \exprcell{\exprbox{\finaltag}{
& 0.084\,\frac{q_1q_2}{\epsilon r^2}
}}
& 0.144 
& 0.235 
& --- \\

\addlinespace[4pt]
\midrule
\addlinespace[2pt]
\multicolumn{5}{@{}l}{\textcolor{BrickRed}{\textit{Noise level}: $\sigma=0.20$, Repetition $1$}}\\
\addlinespace[2pt]

1--10

& \exprcell{\exprbox{\finaltag}{
& -0.045
+0.110\,\frac{1}{r}+0.085\,\frac{q_1q_2}{\epsilon r^{2}}
}}
& 0.198 
& 0.205 
& --- \\

\addlinespace[4pt]
\midrule
\addlinespace[2pt]
\multicolumn{5}{@{}l}{\textcolor{BrickRed}{\textit{Noise level}: $\sigma=0.25$, Repetition $1$}}\\
\addlinespace[2pt]

1

& \exprcell{\exprbox{\finaltag}{
& -0.067
+0.082\,\frac{q_1q_2}{\epsilon r^{2}}+0.180\,\frac{1}{r}
}}
& 0.236 
& 0.108 
& --- \\

\addlinespace[2pt]
\reprowsep
\addlinespace[2pt]

2--10

& \exprcell{\exprbox{\finaltag}{
& -0.064
-0.033\,\sin(q_1)+0.130\,\frac{q_2}{\epsilon r}
+0.180\,\frac{1}{r}
}}
& 0.240 
& 0.074 
& --- \\

\end{longtable}
\endgroup


\subsection{For \texorpdfstring{$\mathrm{I\_12\_11}$~\eqnref[supple-eq:feynman-fce]}{I.12.11}}

\begingroup
\footnotesize
\setlength{\tabcolsep}{2.0pt}
\renewcommand{\arraystretch}{1.22}

\begin{longtable}{@{}c >{\raggedright\arraybackslash}p{0.50\textwidth} c c H@{}}
\caption{The Occam's window set ($\mathcal J_r, r=10$) of symbolic expressions ranked by $\mathrm{JMP}$ learned by \hierbosss\ for the Feynman equation $\mathrm{I\_12\_11}$~\eqnref[supple-eq:feynman-fce]: 
$F = q(E_f + Bv \sin \theta)$, across noise levels. Test \texttt{RMSE} and $R^2$ are computed on the held-out $10\%$ test set.}
\label{tab:BayeSymX-i-12-11-rep1-top10-distinct-results}\\

\toprule
\toprule
Rank 
& Learned expression 
& Test \texttt{RMSE} 
& Test $R^2$ 
& $\log\mathrm{JMP}$ \\
\midrule
\endfirsthead

\caption[]{\emph{(Continued)}. The Occam's window set ($\mathcal J_r, r=10$) of symbolic expressions ranked by $\mathrm{JMP}$ learned by \hierbosss\ for the Feynman equation $\mathrm{I\_12\_11}$~\eqnref[supple-eq:feynman-fce]: 
$F = q(E_f + Bv \sin \theta)$, across noise levels.}\\
\toprule
\toprule
Rank 
& Learned expression 
& Test \texttt{RMSE} 
& Test $R^2$ 
& $\log\mathrm{JMP}$ \\
\midrule
\endhead

\midrule
\multicolumn{5}{r}{\emph{Continued on next page}}\\
\endfoot

\bottomrule
\bottomrule
\endlastfoot

\addlinespace[2pt]
\multicolumn{5}{@{}l}{\textcolor{BrickRed}{\textit{Noiseless setting}, Repetition $1$}}\\
\addlinespace[2pt]

1--10
& \exprcell{\exprbox{\finaltag}{
& E_{f} q + B q v \sin(\theta)
}}
& $8.641{\times}10^{-8}$ 
& 1.000 
& --- \\

\addlinespace[2pt]
\midrule
\addlinespace[2pt]
\multicolumn{5}{@{}l}{\textcolor{BrickRed}{\textit{Noise level}: $\sigma=0.25$, Repetition $1$}}\\
\addlinespace[2pt]

1--10
& \exprcell{\exprbox{\finaltag}{
& 1.001\,E_{f} q + B q v \sin(\theta)
}}
& 0.238 
& 1.000 
& --- \\

\addlinespace[2pt]
\midrule
\addlinespace[2pt]
\multicolumn{5}{@{}l}{\textcolor{BrickRed}{\textit{Noise level}: $\sigma=1.00$, Repetition $1$}}\\
\addlinespace[2pt]

1--10
& \exprcell{\exprbox{\finaltag}{
& 1.003\,E_{f} q + 0.999\,B q v \sin(\theta)
}}
& 0.978 
& 0.999 
& --- \\

\addlinespace[2pt]
\midrule
\addlinespace[2pt]
\multicolumn{5}{@{}l}{\textcolor{BrickRed}{\textit{Noise level}: $\sigma=2.00$, Repetition $1$}}\\
\addlinespace[2pt]

1--10
& \exprcell{\exprbox{\finaltag}{
& 1.005\,E_{f} q + B q v \sin(\theta)
}}
& 1.846 
& 0.995 
& --- \\

\end{longtable}
\endgroup


\subsection{For \texorpdfstring{$\mathrm{I\_24\_6}$~\eqnref[supple-eq:feynman-ehfo]}{I.24.6}}

\begingroup
\footnotesize
\setlength{\tabcolsep}{2.0pt}
\renewcommand{\arraystretch}{1.22}

\begin{longtable}{@{}c >{\raggedright\arraybackslash}p{0.50\textwidth} c c H@{}}
\caption{The Occam's window set ($\mathcal J_r, r=10$) of symbolic expressions ranked by $\mathrm{JMP}$ learned by \hierbosss\ for the Feynman equation $\mathrm{I\_24\_6}$~\eqnref[supple-eq:feynman-ehfo]: 
$E_n = \tfrac{1}{4}m(\omega^{2}+\omega_0^{2})x^{2}$, across noise levels. Test \texttt{RMSE} and $R^2$ are computed on the held-out $10\%$ test set.}
\label{tab:BayeSymX-i-24-6-rep1-top10-distinct-results}\\

\toprule
\toprule
Rank 
& Learned expression 
& Test \texttt{RMSE} 
& Test $R^2$ 
& $\log\mathrm{JMP}$ \\
\midrule
\endfirsthead

\caption[]{\emph{(Continued)}. The Occam's window set ($\mathcal J_r, r=10$) of symbolic expressions ranked by $\mathrm{JMP}$ learned by \hierbosss\ for the Feynman equation $\mathrm{I\_24\_6}$~\eqnref[supple-eq:feynman-ehfo]: 
$E_n = \tfrac{1}{4}m(\omega^{2}+\omega_0^{2})x^{2}$, across noise levels.}\\
\toprule
\toprule
Rank 
& Learned expression 
& Test \texttt{RMSE} 
& Test $R^2$ 
& $\log\mathrm{JMP}$ \\
\midrule
\endhead

\midrule
\multicolumn{5}{r}{\emph{Continued on next page}}\\
\endfoot

\bottomrule
\bottomrule
\endlastfoot

\addlinespace[3pt]
\multicolumn{5}{@{}l}{\textcolor{BrickRed}{\textit{Noiseless setting}, Repetition $1$}}\\
\addlinespace[2pt]

1--10
& \exprcell{\exprbox{\finaltag}{
& 0.248\,m\omega^{2}x^{2}
+0.248\,m\omega_{0}^{2}x^{2}
}}
& $2.595{\times}10^{-8}$ 
& 1.000 
& --- \\

\addlinespace[4pt]
\midrule
\addlinespace[2pt]
\multicolumn{5}{@{}l}{\textcolor{BrickRed}{\textit{Noise level}: $\sigma=0.15$, Repetition $1$}}\\
\addlinespace[2pt]

1--10
& \exprcell{\exprbox{\finaltag}{
& 0.250\,m\omega^{2}x^{2}
+0.250\,m\omega_{0}^{2}x^{2}
}}
& 0.143 
& 1.000 
& --- \\

\addlinespace[4pt]
\midrule
\addlinespace[2pt]
\multicolumn{5}{@{}l}{\textcolor{BrickRed}{\textit{Noise level}: $\sigma=0.25$, Repetition $1$}}\\
\addlinespace[2pt]

1--10
& \exprcell{\exprbox{\finaltag}{
& 0.250\,m\omega^{2}x^{2}
+0.250\,m\omega_{0}^{2}x^{2}
}}
& 0.245 
& 1.000 
& --- \\

\addlinespace[4pt]
\midrule
\addlinespace[2pt]
\multicolumn{5}{@{}l}{\textcolor{BrickRed}{\textit{Noise level}: $\sigma=1.00$, Repetition $1$}}\\
\addlinespace[2pt]

1--10
& \exprcell{\exprbox{\finaltag}{
& 0.250\,mx^{2}(\omega^{2}+\omega_{0}^{2})
}}
& 0.922 
& 0.995 
& --- \\

\end{longtable}
\endgroup


\subsection{For \texorpdfstring{$\mathrm{I\_50\_26}$~\eqnref[supple-eq:feynman-hoqn]}{I.50.26}}

\begingroup
\footnotesize
\setlength{\tabcolsep}{2.0pt}
\renewcommand{\arraystretch}{1.22}

\begin{longtable}{@{}c >{\raggedright\arraybackslash}p{0.50\textwidth} c c H@{}}
\caption{The Occam's window set ($\mathcal J_r, r=10$) of symbolic expressions ranked by $\mathrm{JMP}$ learned by \hierbosss\ for the Feynman equation $\mathrm{I\_50\_26}$~\eqnref[supple-eq:feynman-hoqn]: 
$x=x_1[\cos(\omega t)+\alpha\cos^2(\omega t)]$, across noise levels. Test \texttt{RMSE} and $R^2$ are computed on the held-out $10\%$ test set.}
\label{tab:BayeSymX-i-50-26-top10-rep1-rep2-results}\\

\toprule
\toprule
Rank 
& Learned expression 
& Test \texttt{RMSE} 
& Test $R^2$ 
& $\log\mathrm{JMP}$ \\
\midrule
\endfirsthead

\caption[]{\emph{(Continued)}. The Occam's window set ($\mathcal J_r, r=10$) of symbolic expressions ranked by $\mathrm{JMP}$ learned by \hierbosss\ for the Feynman equation $\mathrm{I\_50\_26}$~\eqnref[supple-eq:feynman-hoqn]: 
$x=x_1[\cos(\omega t)+\alpha\cos^2(\omega t)]$, across noise levels.}\\
\toprule
\toprule
Rank 
& Learned expression 
& Test \texttt{RMSE} 
& Test $R^2$ 
& $\log\mathrm{JMP}$ \\
\midrule
\endhead

\midrule
\multicolumn{5}{r}{\emph{Continued on next page}}\\
\endfoot

\bottomrule
\bottomrule
\endlastfoot

\addlinespace[3pt]
\multicolumn{5}{@{}l}{\textcolor{BrickRed}{\textit{Noiseless setting}, Repetition $1$}}\\
\addlinespace[2pt]

1--10
& \exprcell{\exprbox{\finaltag}{
& x_1\{\alpha\cos^2(\omega t)+\cos(\omega t)\}
}}
& $1.354{\times}10^{-5}$ 
& 1.000 
& --- \\

\addlinespace[2pt]
\midrule
\addlinespace[2pt]
\multicolumn{5}{@{}l}{\textcolor{BrickRed}{\textit{Noise level}: $\sigma=0.15$, Repetition $2$}}\\
\addlinespace[2pt]

1--10
& \exprcell{\exprbox{\finaltag}{
& x_1\{\alpha\cos^2(\omega t)+\cos(\omega t)\}
}}
& 0.146 
& 0.993 
& --- \\

\addlinespace[2pt]
\midrule
\addlinespace[2pt]
\multicolumn{5}{@{}l}{\textcolor{BrickRed}{\textit{Noise level}: $\sigma=0.18$, Repetition $2$}}\\
\addlinespace[2pt]

1--10
& \exprcell{\exprbox{\finaltag}{
& x_1\{\alpha\cos^2(\omega t)+\cos(\omega t)\}
}}
& 0.172 
& 0.994 
& --- \\

\addlinespace[2pt]
\midrule
\addlinespace[2pt]
\multicolumn{5}{@{}l}{\textcolor{BrickRed}{\textit{Noise level}: $\sigma=0.20$, Repetition $2$}}\\
\addlinespace[2pt]

1--10
& \exprcell{\exprbox{\finaltag}{
& x_1\{\alpha\cos^2(\omega t)+\cos(\omega t)\}
}}
& 0.211 
& 0.983 
& --- \\

\end{longtable}
\endgroup

\subsection{For \texorpdfstring{$\mathrm{II\_36\_38}$~\eqnref[supple-eq:feynman-fmmm]}{II.36.38}}

\begingroup
\footnotesize
\setlength{\tabcolsep}{2.0pt}
\renewcommand{\arraystretch}{1.22}

\begin{longtable}{@{}c >{\raggedright\arraybackslash}p{0.50\textwidth} c c H@{}}
\caption{The Occam's window set ($\mathcal J_r, r=10$) of symbolic expressions ranked by $\mathrm{JMP}$ learned by \hierbosss\ for the Feynman equation $\mathrm{II\_36\_38}$~\eqnref[supple-eq:feynman-fmmm]: 
$f = \tfrac{\mathrm{mom} H}{k_b T} + \tfrac{\mathrm{mom} \alpha}{\epsilon c^2 k_b T}M$, across noise levels. Test \texttt{RMSE} and $R^2$ are computed on the held-out $10\%$ test set.}
\label{tab:BayeSymX-ii-36-38-top10-rep1-except-noise020-rep2-results}\\

\toprule
\toprule
Rank 
& Learned expression 
& Test \texttt{RMSE} 
& Test $R^2$ 
& $\log\mathrm{JMP}$ \\
\midrule
\endfirsthead

\caption[]{\emph{(Continued)}. The Occam's window set ($\mathcal J_r, r=10$) of symbolic expressions ranked by $\mathrm{JMP}$ learned by \hierbosss\ for the Feynman equation $\mathrm{II\_36\_38}$~\eqnref[supple-eq:feynman-fmmm]: 
$f = \tfrac{\mathrm{mom} H}{k_b T} + \tfrac{\mathrm{mom} \alpha}{\epsilon c^2 k_b T}H$, across noise levels.}\\
\toprule
\toprule
Rank 
& Learned expression 
& Test \texttt{RMSE} 
& Test $R^2$ 
& $\log\mathrm{JMP}$ \\
\midrule
\endhead

\midrule
\multicolumn{5}{r}{\emph{Continued on next page}}\\
\endfoot

\bottomrule
\bottomrule
\endlastfoot

\addlinespace[3pt]
\multicolumn{5}{@{}l}{\textcolor{BrickRed}{\textit{Noiseless setting}, Repetition $1$}}\\
\addlinespace[2pt]

1--2
& \exprcell{\exprbox{\finaltag}{
& \frac{1.000 H \mathrm{mom}}{T k_b}
+ \frac{1.000 M\alpha \mathrm{mom}}{T c^{2}\epsilon k_b}
}}
& $4.782{\times}10^{-7}$ 
& 1.000 
& --- \\

\addlinespace[2pt]
\reprowsep
\addlinespace[2pt]

3--10
& \exprcell{\exprbox{\finaltag}{
& \frac{1.000 H \mathrm{mom}}{T k_b}
+ \frac{1.005 M\alpha \mathrm{mom}}{T c^{2}\epsilon k_b}
}}
& $3.377{\times}10^{-7}$ 
& 1.000 
& --- \\

\addlinespace[4pt]
\midrule
\addlinespace[2pt]
\multicolumn{5}{@{}l}{\textcolor{BrickRed}{\textit{Noise level}: $\sigma=0.15$, Repetition $1$}}\\
\addlinespace[2pt]

1--10
& \exprcell{\exprbox{\finaltag}{
& \frac{1.000 H \mathrm{mom}}{T k_b}
+ \frac{0.990 M\alpha \mathrm{mom}}{T c^{2}\epsilon k_b}
}}
& 0.144 
& 0.985 
& --- \\

\addlinespace[4pt]
\midrule
\addlinespace[2pt]
\multicolumn{5}{@{}l}{\textcolor{BrickRed}{\textit{Noise level}: $\sigma=0.20$, Repetition $2$}}\\
\addlinespace[2pt]

1--10
& \exprcell{\exprbox{\finaltag}{
& \frac{0.990 H \mathrm{mom}}{T k_b}
+ \frac{1.000 M\alpha \mathrm{mom}}{T c^{2}\epsilon k_b}
}}
& 0.193 
& 0.969 
& --- \\

\addlinespace[4pt]
\midrule
\addlinespace[2pt]
\multicolumn{5}{@{}l}{\textcolor{BrickRed}{\textit{Noise level}: $\sigma=0.25$, Repetition $1$}}\\
\addlinespace[2pt]

1--10
& \exprcell{\exprbox{\finaltag}{
& \frac{1.000 H \mathrm{mom}}{T k_b}
+ \frac{0.990 M\alpha \mathrm{mom}}{T c^{2}\epsilon k_b}
}}
& 0.231 
& 0.951 
& --- \\

\end{longtable}
\endgroup

\newpage


\section{Effective Symbolic Forest Size of \texorpdfstring{\hierbosss}{HierBOSSS} for the Feynman Equations}
\label{sec:effective-K-Feynman}

\begin{figure}[H]
    \centering
    \includegraphics[width=\linewidth]{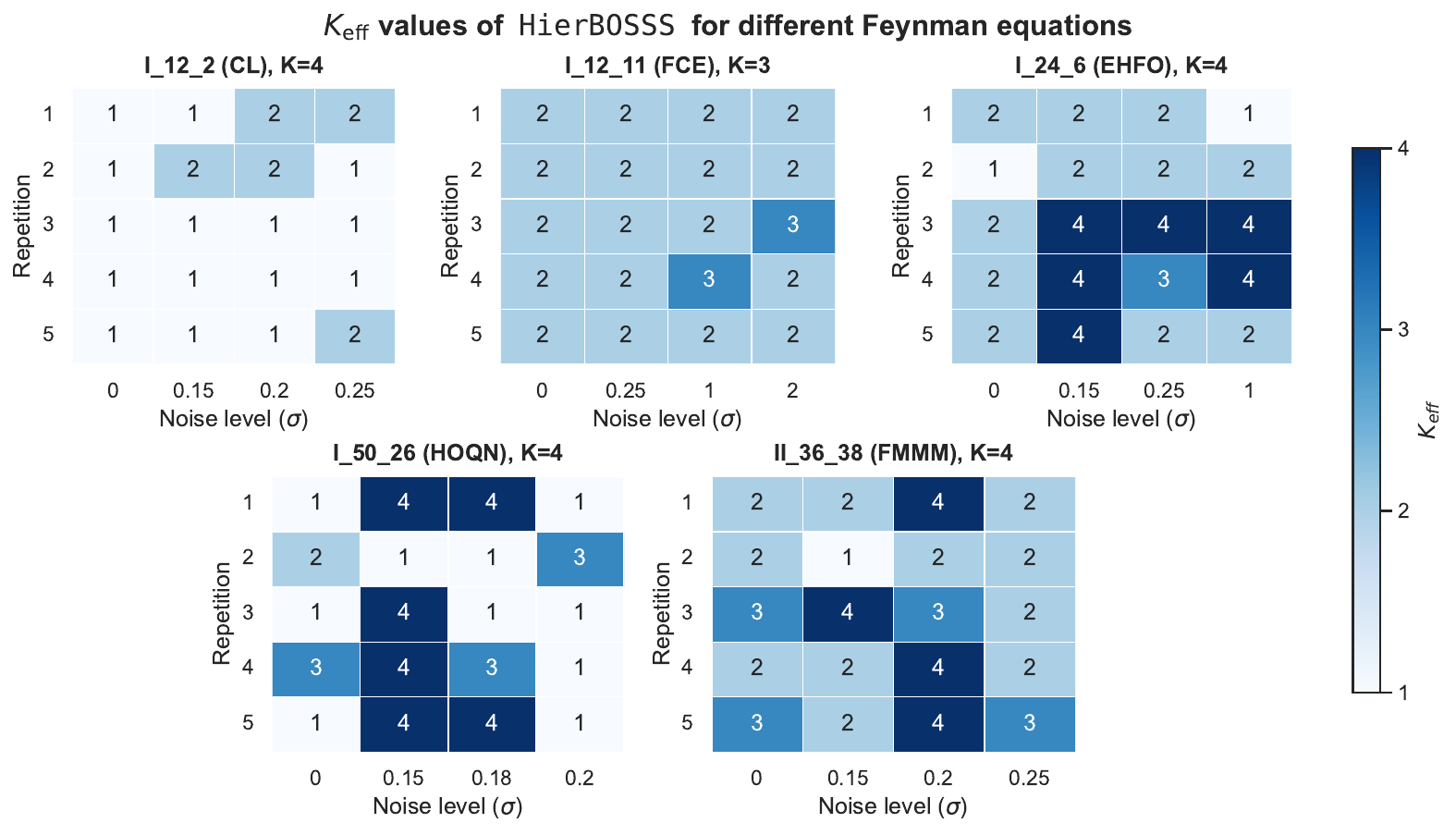}
    \caption{Effective symbolic forest size ($K_{\mathrm{eff}}$) of \hierbosss\ for the Feynman equations in $\mathrm{I\_12\_2}$~\eqnref[supple-eq:feynman-cl]-$\mathrm{II\_36\_38}$~\eqnref[supple-eq:feynman-fmmm], across $5$ independent repetitions and different noise levels.}
    \label{fig:Keff_heatmap}
\end{figure}

\hyperref[fig:Keff_heatmap]{Figure~\ref{fig:Keff_heatmap}} displays the effective symbolic forest size $K_{\mathrm{eff}}$ selected by the post-\texttt{MCMC} symbolic model refinement of \hierbosss\ in~\hyperref[alg:post-mcmc-symbolic-model-refinement-single]{Algorithm \ref{alg:post-mcmc-symbolic-model-refinement-single}}. Since $K_{\mathrm{eff}}$ is determined by the \texttt{BIC}-based subset selection step, it measures how many symbolic tree components are retained in the \finaltag\ expression after removing collinear and weakly contributing tree-structured symbolic expressions. The results show that \hierbosss\ often selects effective forests smaller than the nominal value of $K$, particularly for the simpler Feynman equations such as $\mathrm{I\_12\_2}$~\eqnref[supple-eq:feynman-cl] and $\mathrm{I\_12\_11}$~\eqnref[supple-eq:feynman-fce]. In contrast, structurally richer equations such as $\mathrm{I\_24\_6}$~\eqnref[supple-eq:feynman-ehfo], $\mathrm{II\_36\_38}$~\eqnref[supple-eq:feynman-fmmm], and $\mathrm{I\_50\_26}$~\eqnref[supple-eq:feynman-hoqn] sometimes retain $3$ or $4$ components (closer to the nominal $K$ used), especially under increased noise levels. This illustrates that the post-\texttt{MCMC} symbolic model refinement provides a data-adaptive notion of symbolic complexity, balancing parsimony with the need to preserve multiple meaningful additive components capturing the symbolic structure underlying the data.

\newpage
\section{Expression Catching, Complexity, Accuracy, and Runtime Summaries of \texorpdfstring{\hierbosss}{HierBOSSS} and Competitors for Learning Feynman Equations}
\label{sec:expression-catching-complexity-accuracy-runtime-Feynman}

\subsection{Structural Recovery}
\label{subsec:structural-recovery-Feynman}

\begin{figure}[!htp]
    \centering
    \includegraphics[width=\linewidth]{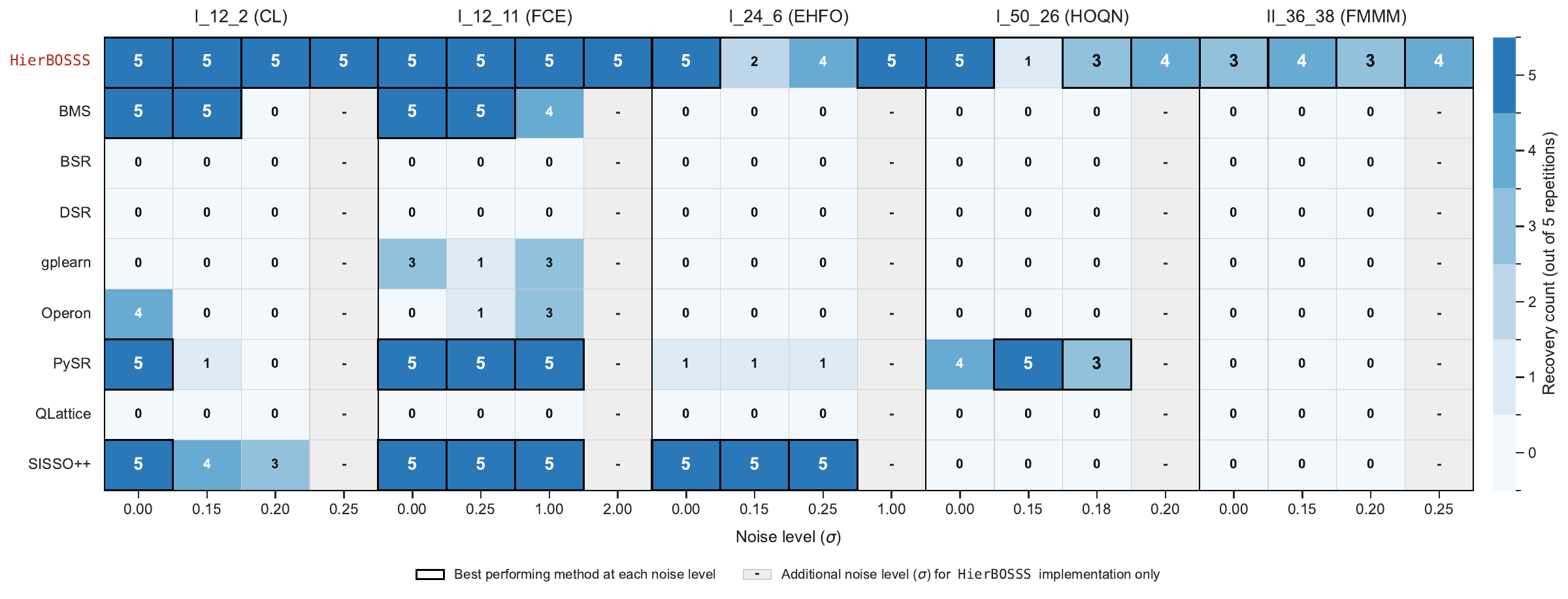}
    \caption{Structural recovery performance (out of $5$ independent repetitions) of \hierbosss\ and competitors across Feynman equations in $\mathrm{I\_12\_2}$~\eqnref[supple-eq:feynman-cl]-$\mathrm{II\_36\_38}$~\eqnref[supple-eq:feynman-fmmm] and different noise levels.}
    \label{fig:Feynman-frequency}
\end{figure}


\subsection{Expression Complexity}
\label{subsec:expression-complexity-Feynman}

We evaluate the complexity of the symbolic expressions learned by \hierbosss\ and the competing \sr\ methods, for the Feynman equations in $\mathrm{I\_12\_2}$~\eqnref[supple-eq:feynman-cl]-$\mathrm{II\_36\_38}$~\eqnref[supple-eq:feynman-fmmm]. Predictive accuracy alone does not fully characterize the quality of a learned scientific expression, since an accurate expression may still be difficult to interpret if it contains redundant operators, unnecessary constants, deeply nested transformation, or unwieldy algebraic structure. This behavior is observed for several competitors (see~\hyperref[sec:symbolic-expressions-Feynman]{Appendix \ref{sec:symbolic-expressions-Feynman}}), where good numerical fit is sometimes achieved through complex nested output expressions rather than compact, interpretable scientific laws. We therefore examine expression complexity jointly with structural recovery and held-out predictive error. Experimental configurations for \hierbosss\ and the competing methods are kept the same as those described in~\hyperref[sec:experimental-settings-competitors-Feynman]{Appendix \ref{sec:experimental-settings-competitors-Feynman}}.

For a learned expression, we calculate its symbolic model size as the total number of nodes appearing in its expression tree~\citep{Imai-SRBench,LaCava-NIPS}. Nodes include numerical constants, symbolic constants, variables or features, and mathematical operators. This gives a common symbolic model size measure across heterogeneous \sr\ methods. For the competing methods, symbolic model size is computed from the expression returned (learned) by the corresponding method. For \hierbosss, symbolic model size is computed from the \finaltag\ expression provided by the post-\texttt{MCMC} symbolic model refinement in \hyperref[alg:post-mcmc-symbolic-model-refinement-single]{Algorithm \ref{alg:post-mcmc-symbolic-model-refinement-single}}. For each Feynman equation in $\mathrm{I\_12\_2}$~\eqnref[supple-eq:feynman-cl]-$\mathrm{II\_36\_38}$~\eqnref[supple-eq:feynman-fmmm], noise level, and method, we average the symbolic model sizes across the $5$ independent repetitions. This average is reported together with the structural recovery frequency (out of the same $5$ repetitions) and the mean (across the same $5$ repetitions) test error gap that is mean test $\texttt{RMSE} - \sigma$.

\begin{figure}[H]
\centering

\begin{subfigure}[t]{0.88\textwidth}
\centering
\includegraphics[width=\linewidth]{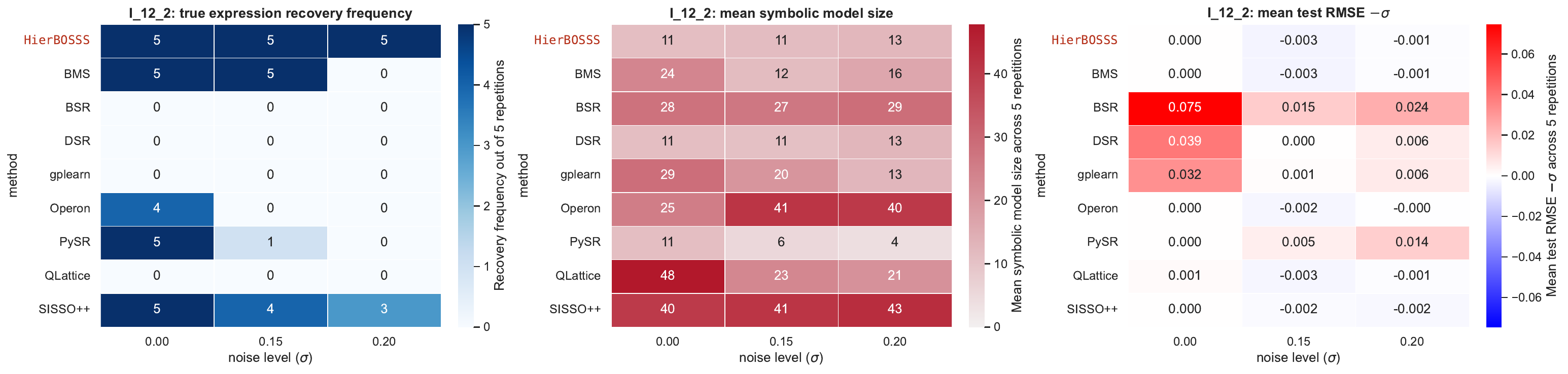}
\caption{$\mathrm{I\_12\_2}$~\eqnref[supple-eq:feynman-cl]}
\label{fig:diag-complexity-i-12-2}
\end{subfigure}

\begin{subfigure}[t]{0.88\textwidth}
\centering
\includegraphics[width=\linewidth]{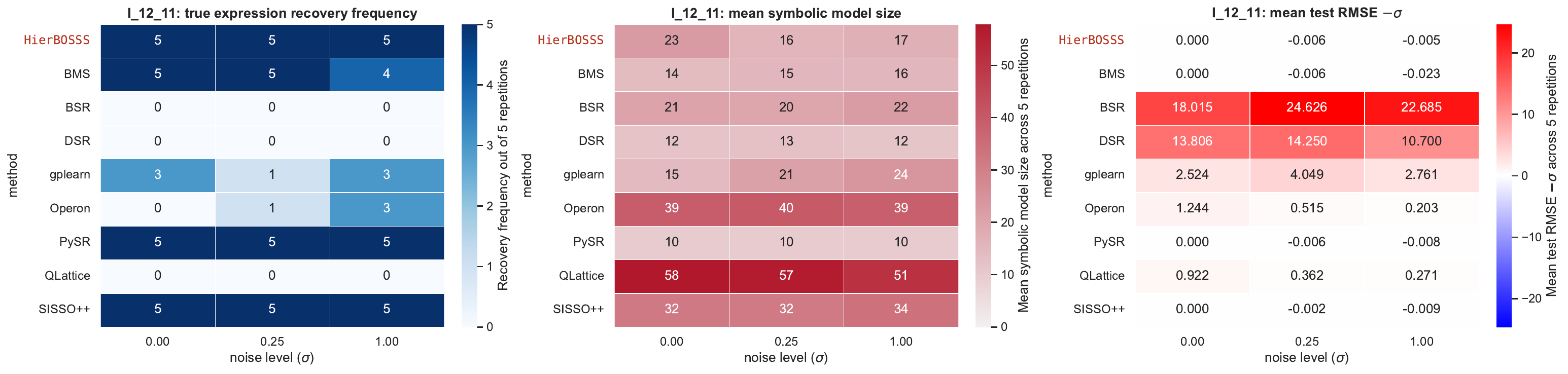}
\caption{$\mathrm{I\_12\_11}$~\eqnref[supple-eq:feynman-fce]}
\label{fig:diag-complexity-i-12-11}
\end{subfigure}

\begin{subfigure}[t]{0.88\textwidth}
\centering
\includegraphics[width=\linewidth]{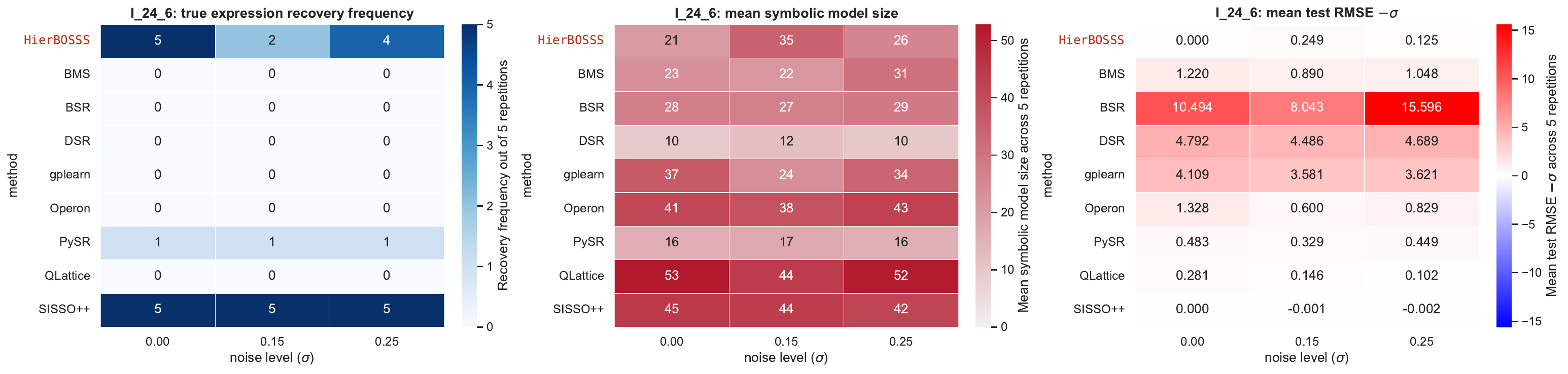}
\caption{$\mathrm{I\_24\_6}$~\eqnref[supple-eq:feynman-ehfo]}
\label{fig:diag-complexity-i-24-6}
\end{subfigure}

\begin{subfigure}[t]{0.88\textwidth}
\centering
\includegraphics[width=\linewidth]{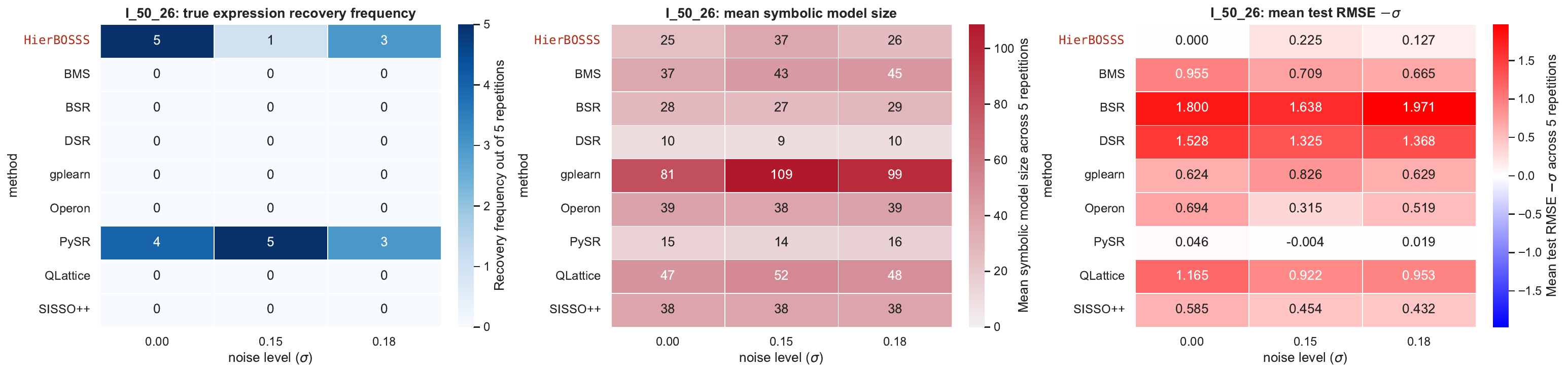}
\caption{$\mathrm{I\_50\_26}$~\eqnref[supple-eq:feynman-hoqn]}
\label{fig:diag-complexity-i-50-26}
\end{subfigure}

\begin{subfigure}[t]{0.88\textwidth}
\centering
\includegraphics[width=\linewidth]{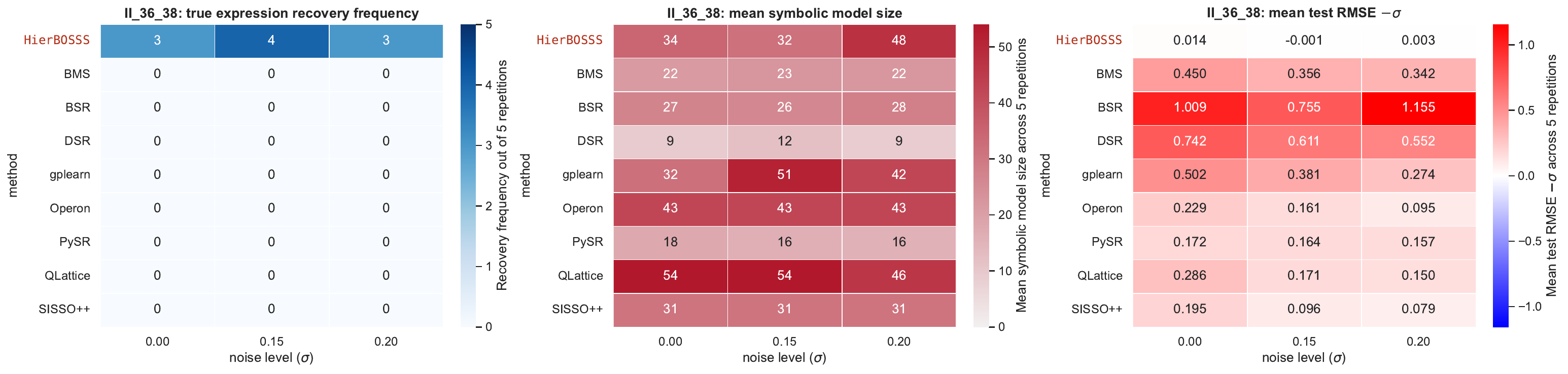}
\caption{$\mathrm{II\_36\_38}$~\eqnref[supple-eq:feynman-fmmm]}
\label{fig:diag-complexity-ii-36-38}
\end{subfigure}

\caption{Expression recovery, symbolic model size, and predictive error gap of \hierbosss\ and competitors for learning the Feynman equations in $\mathrm{I\_12\_2}$~\eqnref[supple-eq:feynman-cl]-$\mathrm{II\_36\_38}$~\eqnref[supple-eq:feynman-fmmm]. For each equation, the three panels report structurally correct recovery frequency, mean symbolic model size, and mean test \texttt{RMSE}$-\sigma$, respectively, across $5$ independent repetitions.}
\label{fig:feynman-expression-complexity-diagnostics}
\end{figure}

\hyperref[fig:feynman-expression-complexity-diagnostics]{Figure~\ref{fig:feynman-expression-complexity-diagnostics}} summarizes these three diagnostics for each of the Feynman equations in $\mathrm{I\_12\_2}$~\eqnref[supple-eq:feynman-cl]-$\mathrm{II\_36\_38}$~\eqnref[supple-eq:feynman-fmmm]. These heatmaps distinguish methods that recover the true symbolic structure of the underlying Feynman equations from methods that achieve reasonable prediction through larger or only partially correct expressions.

\begin{figure}[H]
\centering

\begin{subfigure}[t]{0.48\textwidth}
\centering
\includegraphics[width=\linewidth]{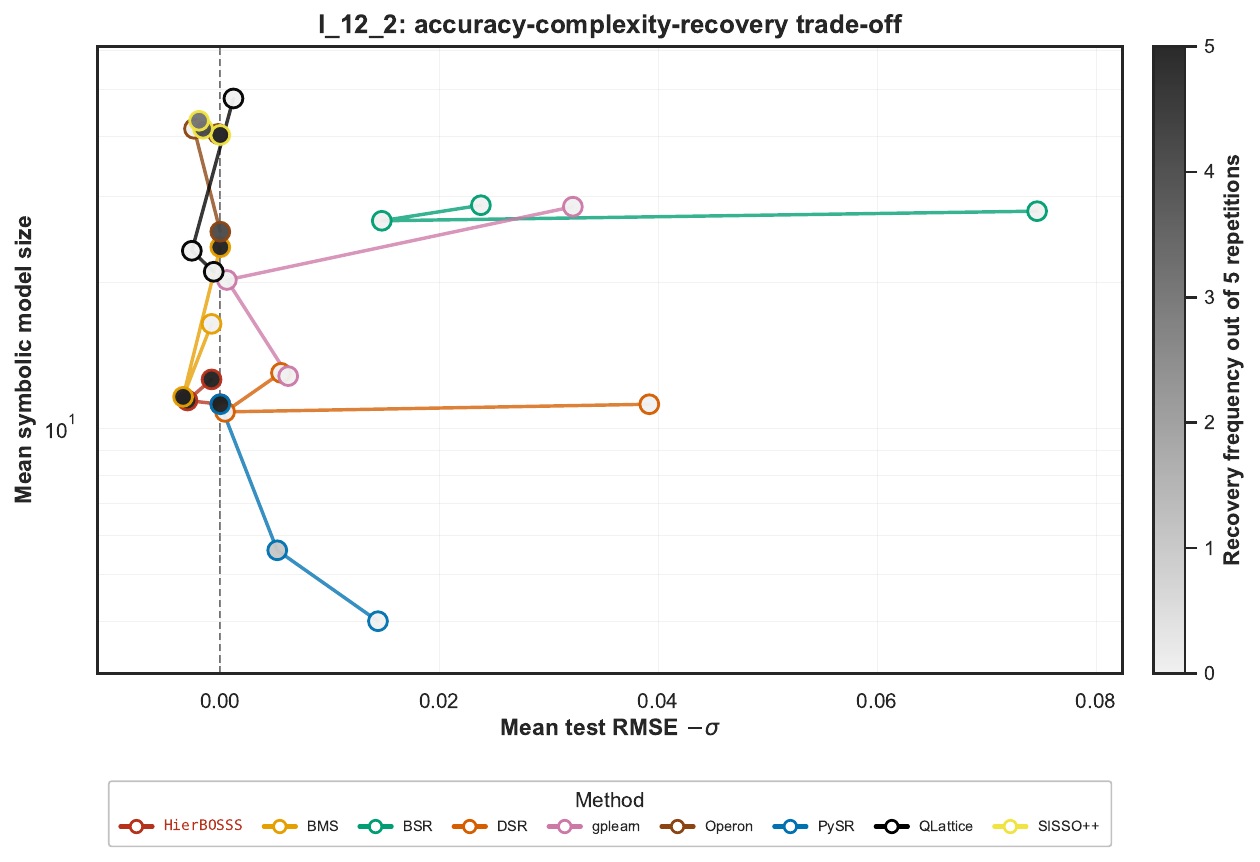}
\caption{$\mathrm{I\_12\_2}$~\eqnref[supple-eq:feynman-cl]}
\label{fig:tradeoff-i-12-2}
\end{subfigure}
\hfill
\begin{subfigure}[t]{0.48\textwidth}
\centering
\includegraphics[width=\linewidth]{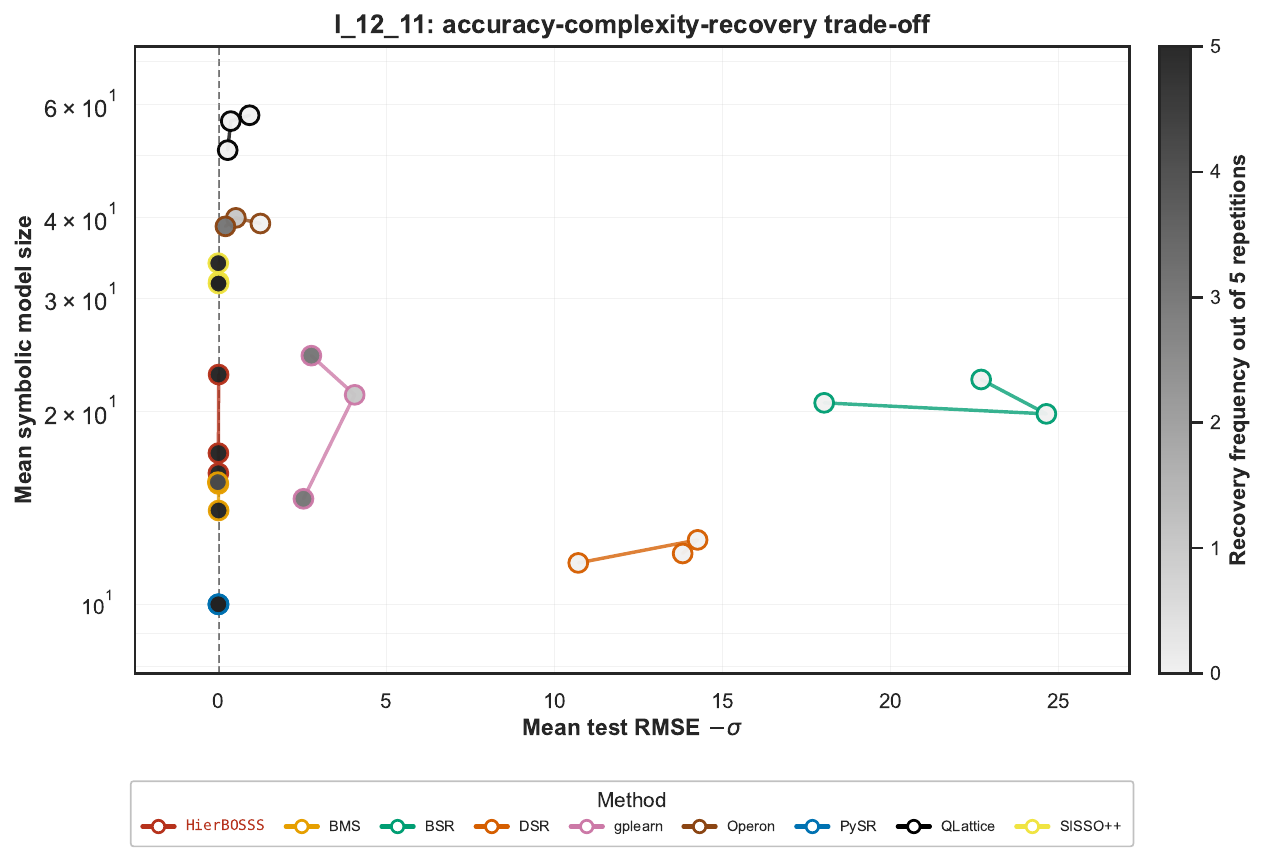}
\caption{$\mathrm{I\_12\_11}$~\eqnref[supple-eq:feynman-fce]}
\label{fig:tradeoff-i-12-11}
\end{subfigure}

\medskip

\begin{subfigure}[t]{0.48\textwidth}
\centering
\includegraphics[width=\linewidth]{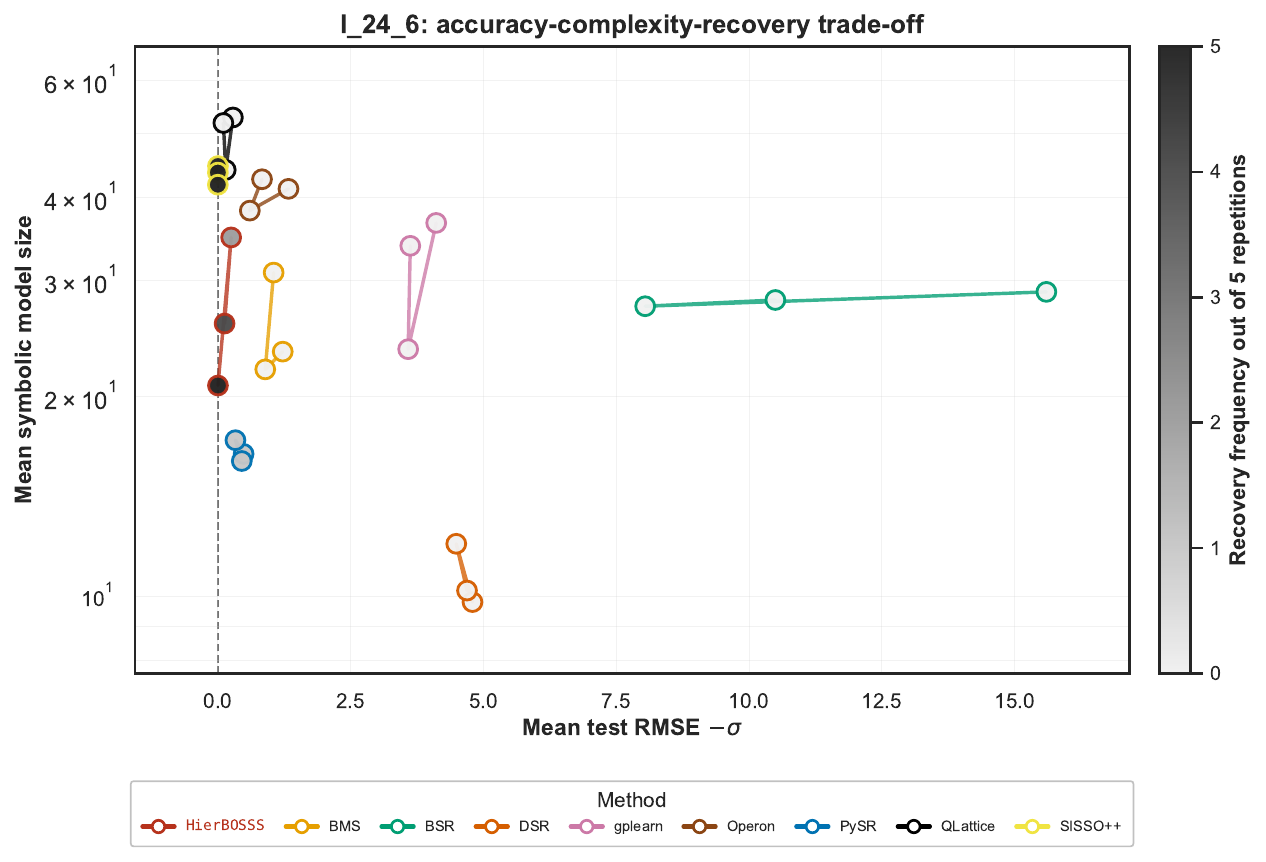}
\caption{$\mathrm{I\_24\_6}$~\eqnref[supple-eq:feynman-ehfo]}
\label{fig:tradeoff-i-24-6}
\end{subfigure}
\hfill
\begin{subfigure}[t]{0.48\textwidth}
\centering
\includegraphics[width=\linewidth]{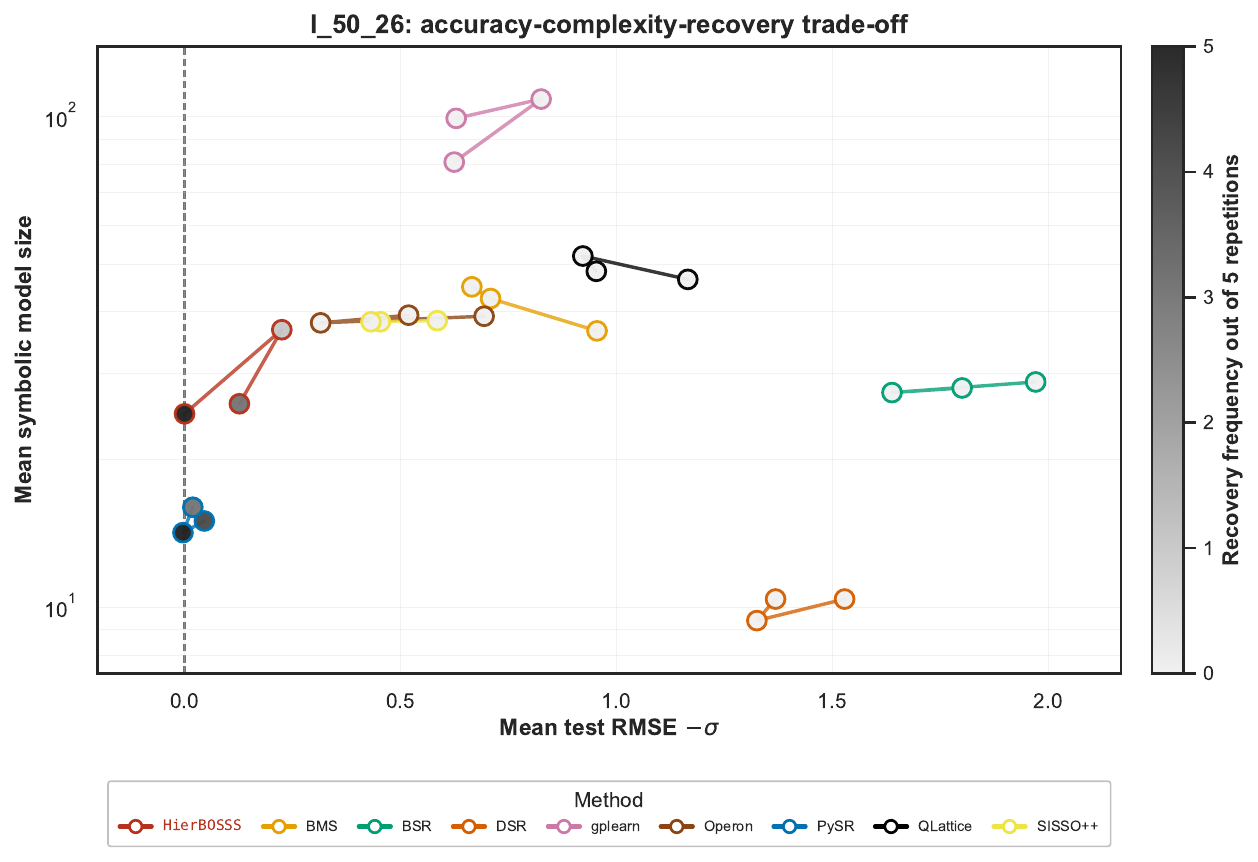}
\caption{$\mathrm{I\_50\_26}$~\eqnref[supple-eq:feynman-hoqn]}
\label{fig:tradeoff-i-50-26}
\end{subfigure}

\medskip

\begin{subfigure}[t]{0.48\textwidth}
\centering
\includegraphics[width=\linewidth]{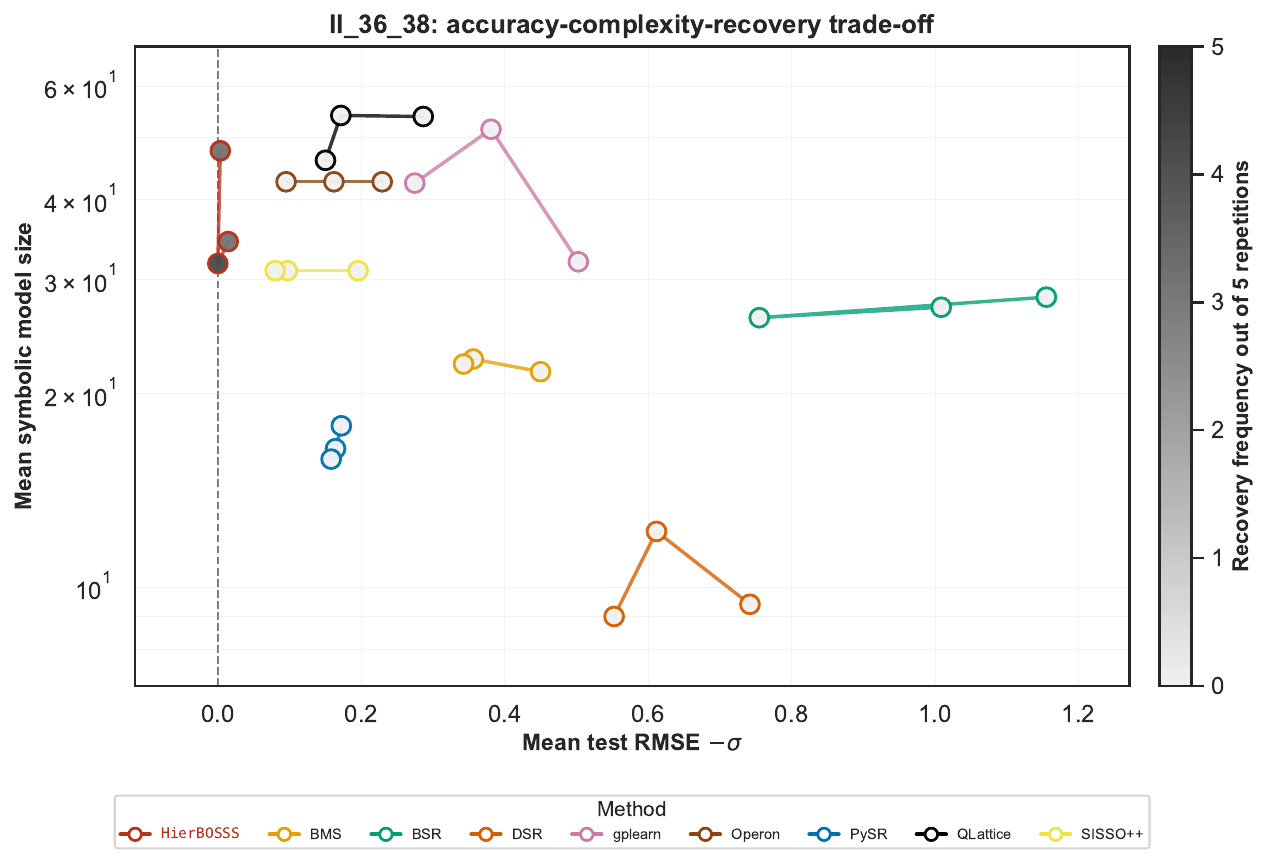}
\caption{$\mathrm{II\_36\_38}$~\eqnref[supple-eq:feynman-fmmm]}
\label{fig:tradeoff-ii-36-38}
\end{subfigure}

\caption{Accuracy-complexity-recovery trade-off plots of \hierbosss\ and competitors for the Feynman equations in $\mathrm{I\_12\_2}$~\eqnref[supple-eq:feynman-cl]-$\mathrm{II\_36\_38}$~\eqnref[supple-eq:feynman-fmmm]. Each point corresponds to a method at a given noise level. The horizontal axis reports mean test $\texttt{RMSE}-\sigma$, the vertical axis reports mean symbolic model size, and point shading indicates structurally correct recovery frequency, across $5$ independent repetitions.}
\label{fig:feynman-accuracy-complexity-recovery-tradeoff}
\end{figure}

To complement the preceding diagnostic heatmaps, we also provide accuracy-complexity-recovery trade-off plots in \hyperref[fig:feynman-accuracy-complexity-recovery-tradeoff]{Figure~\ref{fig:feynman-accuracy-complexity-recovery-tradeoff}}. Each point represents one method at one noise level, with the horizontal axis measuring the mean test-error gap namely, mean test $\texttt{RMSE}-\sigma$, the vertical axis measuring the mean symbolic model size, and the point shading indicating the structurally correct recovery frequency, across $5$ independent repetitions. Hence, the most desirable region is near the vertical zero-error reference line, with small model size and darker point shading. Such points correspond to expressions that are compact, predictive near the noise floor, and structurally correct.

Across the Feynman equations in $\mathrm{I\_12\_2}$~\eqnref[supple-eq:feynman-cl]-$\mathrm{II\_36\_38}$~\eqnref[supple-eq:feynman-fmmm], \hierbosss\ generally exhibits a favorable balance of the accuracy-complexity-recovery trade-off. For the simpler equations $\mathrm{I\_12\_2}$~\eqnref[supple-eq:feynman-cl] and $\mathrm{I\_12\_11}$~\eqnref[supple-eq:feynman-fce], \hierbosss\ recovers the true symbolic structure across all displayed noise levels while maintaining compact model sizes and test errors close to the noise level. Some competitors, such as \bms, \pysr, and \sisso$++$, can also perform well for selected equations and noise levels, but their behavior is less uniform across the Feynman benchmark. Other methods often either fail to recover the true structure, produce substantially larger expressions, or exhibit larger test error gaps.

For more challenging equations, such as $\mathrm{I\_24\_6}$~\eqnref[supple-eq:feynman-ehfo], $\mathrm{I\_50\_26}$~\eqnref[supple-eq:feynman-hoqn], and $\mathrm{II\_36\_38}$~\eqnref[supple-eq:feynman-fmmm], the distinction between prediction and structural discovery becomes more pronounced. Several competing \sr\ methods (including \dsr, \bms, \pysr, \operon, and \sisso$++$) obtain reasonable prediction errors or compact expressions in some settings, but this does not necessarily imply recovery of the underlying Feynman equation. In particular, a method may lie close to the zero-error line while still having low recovery frequency, indicating that it has found a predictive approximation rather than the correct symbolic structure. Conversely, some methods return large expressions with many redundant or nested components, making the resulting formulas less interpretable even when predictive performance is acceptable.

The behavior of \hierbosss\ is consistent. The depth-dependent splitting rule favors parsimonious symbolic tree structures, encoding the Occam's razor principle~\citep{Occams-Razor-1} directly into the symbolic forest prior over $\mathcal T$. In addition, post-\texttt{MCMC} symbolic model refinement in~\hyperref[alg:post-mcmc-symbolic-model-refinement-single]{Algorithm \ref{alg:post-mcmc-symbolic-model-refinement-single}} applies an Occam's window-based~\citep{madigan1994model} symbolic model selection using the joint marginal posterior of the symbolic forest, $\mathrm{JMP}(\mathcal{T})$, retaining high-posterior expressions while discarding unnecessary symbolic components. This complete probabilistic \sr\ framework of \hierbosss\ allows it balance expressiveness, predictive accuracy, structural recovery, and symbolic parsimony.

\newpage

\newpage

\subsection{Accuracy (Test \texorpdfstring{\texttt{RMSE}}{RMSE} and \texorpdfstring{$R^{2}$}{R2}) and Compute Runtimes}
\label{subsec:accuracy-runtime-Feynman}

\begingroup
\scriptsize
\setlength{\tabcolsep}{1.8pt}
\renewcommand{\arraystretch}{1.15}

\newcommand{\msd}[2]{%
    \makecell[c]{$#1$\\[-1pt]$({#2})$}%
}

\begin{longtable}{@{}
>{\raggedright\arraybackslash}p{0.095\textwidth}
*{9}{>{\centering\arraybackslash}p{0.1\textwidth}}
@{}}
\caption{Summary of test \texttt{RMSE}, test $R^2$, and runtime in seconds for \hierbosss\ and competitors across the Feynman equations in $\mathrm{I\_12\_2}$~\eqnref[supple-eq:feynman-cl]-$\mathrm{II\_36\_38}$~\eqnref[supple-eq:feynman-fmmm] and different noise levels. Each entry reports the mean over $5$ repetitions,
with the corresponding standard deviation shown in parentheses underneath. Test \texttt{RMSE} and $R^2$ are computed on the held-out $10\%$ test set.}
\label{tab:feynman-combined-performance-summary}\\

\toprule
\toprule
\endfirsthead

\caption[]{\emph{(Continued)}. Summary of test \texttt{RMSE}, test $R^2$, and runtime in seconds for \hierbosss\ and competitors across the Feynman equations in $\mathrm{I\_12\_2}$~\eqnref[supple-eq:feynman-cl]-$\mathrm{II\_36\_38}$~\eqnref[supple-eq:feynman-fmmm]
and different noise levels. Each entry reports the mean over $5$ repetitions, with the corresponding standard deviation shown in parentheses underneath.}\\

\toprule
\endhead

\midrule
\multicolumn{10}{r}{\emph{Continued on next page}}\\
\endfoot

\bottomrule
\bottomrule
\endlastfoot


\addlinespace[3pt]
\multicolumn{10}{@{}l}{$\mathrm{I\_12\_2}$~\eqnref[supple-eq:feynman-cl]}\\
\addlinespace[2pt]

\multirow{2}{*}{{Method}}
& \multicolumn{3}{c}{\textcolor{BrickRed}{\emph{Noiseless setting}}}
& \multicolumn{3}{c}{\textcolor{BrickRed}{\emph{Noise level}: $\sigma=0.15$}}
& \multicolumn{3}{c}{\textcolor{BrickRed}{\emph{Noise level}: $\sigma=0.20$}} \\
\cmidrule(lr){2-4}
\cmidrule(lr){5-7}
\cmidrule(lr){8-10}

& {Test \texttt{RMSE}}
& {Test $R^2$}
& {Runtime (s)}
& {Test \texttt{RMSE}}
& {Test $R^2$}
& {Runtime (s)}
& {Test \texttt{RMSE}}
& {Test $R^2$}
& {Runtime (s)} \\
\midrule

\rowcolor{ExprBack}
\hierbosss
& \msd{0.000}{0.000}
& \msd{1.000}{0.000}
& \msd{51.390}{6.365}
& \msd{0.147}{0.006}
& \msd{0.243}{0.060}
& \msd{56.357}{8.312}
& \msd{0.199}{0.007}
& \msd{0.196}{0.073}
& \msd{57.842}{1.986} \\

\reprowsep

\bms
& \msd{0.000}{0.000}
& \msd{1.000}{0.000}
& \msd{210.905}{14.077}
& \msd{0.147}{0.006}
& \msd{0.246}{0.063}
& \msd{361.300}{6.939}
& \msd{0.199}{0.007}
& \msd{0.196}{0.073}
& \msd{399.125}{15.148} \\

\reprowsep

\bsr
& \msd{0.075}{0.005}
& \msd{0.210}{0.096}
& \msd{188.939}{0.420}
& \msd{0.165}{0.008}
& \msd{0.053}{0.081}
& \msd{189.356}{1.034}
& \msd{0.224}{0.015}
& \msd{-0.011}{0.078}
& \msd{189.955}{2.828} \\

\reprowsep

\dsr
& \msd{0.039}{0.011}
& \msd{0.776}{0.092}
& \msd{76.420}{15.339}
& \msd{0.150}{0.005}
& \msd{0.206}{0.060}
& \msd{107.400}{7.225}
& \msd{0.205}{0.006}
& \msd{0.145}{0.058}
& \msd{123.560}{24.557} \\

\reprowsep

\gplearn
& \msd{0.032}{0.013}
& \msd{0.827}{0.113}
& \msd{17.349}{1.267}
& \msd{0.151}{0.005}
& \msd{0.202}{0.080}
& \msd{15.700}{0.688}
& \msd{0.206}{0.008}
& \msd{0.140}{0.047}
& \msd{15.695}{0.783} \\

\reprowsep

\operon
& \msd{0.000}{0.000}
& \msd{1.000}{0.000}
& \msd{19.513}{0.976}
& \msd{0.148}{0.007}
& \msd{0.235}{0.059}
& \msd{24.173}{1.647}
& \msd{0.200}{0.007}
& \msd{0.190}{0.074}
& \msd{25.032}{2.637} \\

\reprowsep

\pysr
& \msd{0.000}{0.000}
& \msd{1.000}{0.000}
& \msd{69.953}{4.908}
& \msd{0.155}{0.003}
& \msd{0.152}{0.085}
& \msd{72.777}{6.282}
& \msd{0.214}{0.013}
& \msd{0.073}{0.033}
& \msd{81.711}{15.020} \\

\reprowsep

\qlattice
& \msd{0.001}{0.001}
& \msd{1.000}{0.000}
& \msd{93.850}{1.890}
& \msd{0.147}{0.006}
& \msd{0.234}{0.068}
& \msd{83.086}{2.549}
& \msd{0.199}{0.005}
& \msd{0.194}{0.077}
& \msd{73.609}{4.531} \\

\reprowsep

\sisso$++$
& \msd{0.000}{0.000}
& \msd{1.000}{0.000}
& \msd{16.639}{1.111}
& \msd{0.148}{0.002}
& \msd{0.249}{0.030}
& \msd{17.552}{0.298}
& \msd{0.198}{0.003}
& \msd{0.155}{0.024}
& \msd{17.585}{0.505} \\

\addlinespace[4pt]
\midrule


\addlinespace[3pt]
\multicolumn{10}{@{}l}{$\mathrm{I\_12\_11}$~\eqnref[supple-eq:feynman-fce]}\\
\addlinespace[2pt]

\multirow{2}{*}{{Method}}
& \multicolumn{3}{c}{\textcolor{BrickRed}{\emph{Noiseless setting}}}
& \multicolumn{3}{c}{\textcolor{BrickRed}{\emph{Noise level}: $\sigma=0.25$}}
& \multicolumn{3}{c}{\textcolor{BrickRed}{\emph{Noise level}: $\sigma=1.00$}} \\
\cmidrule(lr){2-4}
\cmidrule(lr){5-7}
\cmidrule(lr){8-10}

& {Test \texttt{RMSE}}
& {Test $R^2$}
& {Runtime (s)}
& {Test \texttt{RMSE}}
& {Test $R^2$}
& {Runtime (s)}
& {Test \texttt{RMSE}}
& {Test $R^2$}
& {Runtime (s)} \\
\midrule

\rowcolor{ExprBack}
\hierbosss
& \msd{0.000}{0.000}
& \msd{1.000}{0.000}
& \msd{15.138}{0.313}
& \msd{0.244}{0.011}
& \msd{1.000}{0.000}
& \msd{14.082}{0.891}
& \msd{0.995}{0.038}
& \msd{0.999}{0.001}
& \msd{15.103}{1.520} \\

\reprowsep

\bms
& \msd{0.000}{0.000}
& \msd{1.000}{0.000}
& \msd{272.269}{30.810}
& \msd{0.244}{0.011}
& \msd{1.000}{0.000}
& \msd{293.312}{17.077}
& \msd{0.977}{0.014}
& \msd{0.999}{0.001}
& \msd{295.094}{40.632} \\

\reprowsep

\bsr
& \msd{18.015}{1.724}
& \msd{0.499}{0.047}
& \msd{165.651}{1.424}
& \msd{24.876}{3.416}
& \msd{0.090}{0.223}
& \msd{165.657}{1.107}
& \msd{23.685}{4.119}
& \msd{0.111}{0.329}
& \msd{165.410}{1.454} \\

\reprowsep

\dsr
& \msd{13.805}{4.072}
& \msd{0.691}{0.166}
& \msd{48.720}{8.014}
& \msd{14.500}{1.855}
& \msd{0.687}{0.092}
& \msd{42.600}{0.911}
& \msd{11.700}{4.319}
& \msd{0.774}{0.139}
& \msd{42.500}{0.283} \\

\reprowsep

\gplearn
& \msd{2.524}{3.543}
& \msd{0.976}{0.037}
& \msd{16.342}{1.109}
& \msd{4.298}{3.260}
& \msd{0.961}{0.041}
& \msd{16.877}{0.777}
& \msd{3.760}{3.785}
& \msd{0.962}{0.050}
& \msd{16.751}{1.640} \\

\reprowsep

\operon
& \msd{1.244}{0.868}
& \msd{0.997}{0.004}
& \msd{26.616}{2.265}
& \msd{0.765}{0.435}
& \msd{0.999}{0.001}
& \msd{24.694}{1.669}
& \msd{1.203}{0.304}
& \msd{0.998}{0.001}
& \msd{24.779}{1.212} \\

\reprowsep

\pysr
& \msd{0.000}{0.000}
& \msd{1.000}{0.000}
& \msd{127.246}{8.985}
& \msd{0.244}{0.011}
& \msd{1.000}{0.000}
& \msd{121.428}{9.605}
& \msd{0.992}{0.038}
& \msd{0.999}{0.001}
& \msd{123.313}{13.838} \\

\reprowsep

\qlattice
& \msd{0.922}{0.451}
& \msd{0.999}{0.001}
& \msd{93.075}{2.206}
& \msd{0.612}{0.285}
& \msd{1.000}{0.001}
& \msd{92.105}{1.032}
& \msd{1.271}{0.236}
& \msd{0.998}{0.001}
& \msd{93.531}{4.214} \\

\reprowsep

\sisso$++$
& \msd{0.000}{0.000}
& \msd{1.000}{0.000}
& \msd{43.668}{1.667}
& \msd{0.248}{0.003}
& \msd{1.000}{0.000}
& \msd{42.753}{0.650}
& \msd{0.991}{0.011}
& \msd{0.998}{0.000}
& \msd{43.789}{1.669} \\

\addlinespace[4pt]
\midrule


\addlinespace[3pt]
\multicolumn{10}{@{}l}{$\mathrm{I\_24\_6}$~\eqnref[supple-eq:feynman-ehfo]}\\
\addlinespace[2pt]

\multirow{2}{*}{{Method}}
& \multicolumn{3}{c}{\textcolor{BrickRed}{\emph{Noiseless setting}}}
& \multicolumn{3}{c}{\textcolor{BrickRed}{\emph{Noise level}: $\sigma=0.15$}}
& \multicolumn{3}{c}{\textcolor{BrickRed}{\emph{Noise level}: $\sigma=0.25$}} \\
\cmidrule(lr){2-4}
\cmidrule(lr){5-7}
\cmidrule(lr){8-10}

& {Test \texttt{RMSE}}
& {Test $R^2$}
& {Runtime (s)}
& {Test \texttt{RMSE}}
& {Test $R^2$}
& {Runtime (s)}
& {Test \texttt{RMSE}}
& {Test $R^2$}
& {Runtime (s)} \\
\midrule

\rowcolor{ExprBack}
\hierbosss
& \msd{0.000}{0.000}
& \msd{1.000}{0.000}
& \msd{64.626}{4.014}
& \msd{0.399}{0.236}
& \msd{0.999}{0.001}
& \msd{64.124}{3.986}
& \msd{0.375}{0.282}
& \msd{0.999}{0.002}
& \msd{63.645}{1.326} \\

\reprowsep

\bms
& \msd{1.219}{0.090}
& \msd{0.993}{0.001}
& \msd{235.951}{19.032}
& \msd{1.040}{0.325}
& \msd{0.994}{0.003}
& \msd{232.742}{17.890}
& \msd{1.298}{0.370}
& \msd{0.991}{0.005}
& \msd{242.652}{6.945} \\

\reprowsep

\bsr
& \msd{10.494}{2.528}
& \msd{0.459}{0.240}
& \msd{196.386}{4.105}
& \msd{8.193}{1.748}
& \msd{0.657}{0.147}
& \msd{192.569}{3.252}
& \msd{15.846}{5.789}
& \msd{-0.397}{0.980}
& \msd{194.237}{6.542} \\

\reprowsep

\dsr
& \msd{4.792}{0.087}
& \msd{0.892}{0.007}
& \msd{31.800}{1.131}
& \msd{4.637}{0.227}
& \msd{0.894}{0.013}
& \msd{32.720}{1.410}
& \msd{4.939}{0.258}
& \msd{0.880}{0.013}
& \msd{31.640}{2.182} \\

\reprowsep

\gplearn
& \msd{4.109}{1.872}
& \msd{0.909}{0.080}
& \msd{16.797}{2.877}
& \msd{3.731}{1.025}
& \msd{0.927}{0.040}
& \msd{14.813}{0.947}
& \msd{3.871}{0.612}
& \msd{0.925}{0.024}
& \msd{15.715}{1.215} \\

\reprowsep

\operon
& \msd{1.328}{0.536}
& \msd{0.991}{0.007}
& \msd{26.912}{2.841}
& \msd{0.750}{0.613}
& \msd{0.996}{0.005}
& \msd{23.552}{3.040}
& \msd{1.080}{0.387}
& \msd{0.994}{0.005}
& \msd{23.836}{1.224} \\

\reprowsep

\pysr
& \msd{0.482}{0.377}
& \msd{0.998}{0.002}
& \msd{69.940}{10.127}
& \msd{0.479}{0.261}
& \msd{0.999}{0.001}
& \msd{76.673}{8.929}
& \msd{0.699}{0.274}
& \msd{0.997}{0.002}
& \msd{67.912}{9.139} \\

\reprowsep

\qlattice
& \msd{0.281}{0.012}
& \msd{1.000}{0.000}
& \msd{95.689}{2.282}
& \msd{0.296}{0.036}
& \msd{1.000}{0.000}
& \msd{94.665}{1.604}
& \msd{0.352}{0.066}
& \msd{0.999}{0.001}
& \msd{94.928}{2.089} \\

\reprowsep

\sisso$++$
& \msd{0.000}{0.000}
& \msd{1.000}{0.000}
& \msd{19.535}{0.596}
& \msd{0.149}{0.002}
& \msd{1.000}{0.000}
& \msd{19.015}{0.130}
& \msd{0.248}{0.003}
& \msd{1.000}{0.000}
& \msd{19.163}{0.330} \\

\addlinespace[4pt]
\midrule


\addlinespace[3pt]
\multicolumn{10}{@{}l}{$\mathrm{I\_50\_26}$~\eqnref[supple-eq:feynman-hoqn]}\\
\addlinespace[2pt]

\multirow{2}{*}{{Method}}
& \multicolumn{3}{c}{\textcolor{BrickRed}{\emph{Noiseless setting}}}
& \multicolumn{3}{c}{\textcolor{BrickRed}{\emph{Noise level}: $\sigma=0.15$}}
& \multicolumn{3}{c}{\textcolor{BrickRed}{\emph{Noise level}: $\sigma=0.18$}} \\
\cmidrule(lr){2-4}
\cmidrule(lr){5-7}
\cmidrule(lr){8-10}

& {Test \texttt{RMSE}}
& {Test $R^2$}
& {Runtime (s)}
& {Test \texttt{RMSE}}
& {Test $R^2$}
& {Runtime (s)}
& {Test \texttt{RMSE}}
& {Test $R^2$}
& {Runtime (s)} \\
\midrule

\rowcolor{ExprBack}
\hierbosss
& \msd{0.000}{0.000}
& \msd{1.000}{0.000}
& \msd{60.498}{1.581}
& \msd{0.375}{0.167}
& \msd{0.959}{0.033}
& \msd{63.543}{3.170}
& \msd{0.307}{0.178}
& \msd{0.968}{0.036}
& \msd{62.391}{2.588} \\

\reprowsep

\bms
& \msd{0.955}{0.139}
& \msd{0.758}{0.078}
& \msd{274.056}{27.935}
& \msd{0.859}{0.239}
& \msd{0.792}{0.151}
& \msd{294.648}{29.925}
& \msd{0.845}{0.277}
& \msd{0.804}{0.150}
& \msd{300.061}{24.232} \\

\reprowsep

\bsr
& \msd{1.800}{0.093}
& \msd{0.159}{0.063}
& \msd{191.599}{1.395}
& \msd{1.788}{0.162}
& \msd{0.180}{0.106}
& \msd{191.651}{1.057}
& \msd{2.151}{0.316}
& \msd{-0.112}{0.320}
& \msd{193.930}{5.809} \\

\reprowsep

\dsr
& \msd{1.528}{0.151}
& \msd{0.390}{0.120}
& \msd{44.880}{1.018}
& \msd{1.475}{0.122}
& \msd{0.434}{0.122}
& \msd{45.760}{1.031}
& \msd{1.548}{0.170}
& \msd{0.433}{0.085}
& \msd{46.000}{1.304} \\

\reprowsep

\gplearn
& \msd{0.624}{0.391}
& \msd{0.869}{0.162}
& \msd{19.290}{1.877}
& \msd{0.975}{0.585}
& \msd{0.716}{0.258}
& \msd{20.649}{5.276}
& \msd{0.809}{0.198}
& \msd{0.838}{0.083}
& \msd{20.935}{2.589} \\

\reprowsep

\operon
& \msd{0.694}{0.217}
& \msd{0.867}{0.088}
& \msd{24.758}{3.164}
& \msd{0.465}{0.297}
& \msd{0.923}{0.074}
& \msd{26.250}{1.756}
& \msd{0.699}{0.357}
& \msd{0.865}{0.140}
& \msd{28.398}{1.562} \\

\reprowsep

\pysr
& \msd{0.046}{0.102}
& \msd{0.997}{0.007}
& \msd{108.801}{15.951}
& \msd{0.146}{0.006}
& \msd{0.994}{0.001}
& \msd{103.283}{3.208}
& \msd{0.199}{0.029}
& \msd{0.990}{0.004}
& \msd{100.307}{11.418} \\

\reprowsep

\qlattice
& \msd{1.165}{0.134}
& \msd{0.643}{0.087}
& \msd{85.273}{1.281}
& \msd{1.072}{0.097}
& \msd{0.704}{0.050}
& \msd{86.189}{7.616}
& \msd{1.133}{0.139}
& \msd{0.691}{0.083}
& \msd{85.358}{5.604} \\

\reprowsep

\sisso$++$
& \msd{0.585}{0.015}
& \msd{0.908}{0.003}
& \msd{26.372}{0.552}
& \msd{0.603}{0.012}
& \msd{0.903}{0.004}
& \msd{26.119}{0.638}
& \msd{0.612}{0.012}
& \msd{0.900}{0.005}
& \msd{25.999}{0.704} \\

\addlinespace[4pt]
\midrule


\addlinespace[3pt]
\multicolumn{10}{@{}l}{$\mathrm{II\_36\_38}$~\eqnref[supple-eq:feynman-fmmm]}\\
\addlinespace[2pt]

\multirow{2}{*}{{Method}}
& \multicolumn{3}{c}{\textcolor{BrickRed}{\emph{Noiseless setting}}}
& \multicolumn{3}{c}{\textcolor{BrickRed}{\emph{Noise level}: $\sigma=0.15$}}
& \multicolumn{3}{c}{\textcolor{BrickRed}{\emph{Noise level}: $\sigma=0.20$}} \\
\cmidrule(lr){2-4}
\cmidrule(lr){5-7}
\cmidrule(lr){8-10}

& {Test \texttt{RMSE}}
& {Test $R^2$}
& {Runtime (s)}
& {Test \texttt{RMSE}}
& {Test $R^2$}
& {Runtime (s)}
& {Test \texttt{RMSE}}
& {Test $R^2$}
& {Runtime (s)} \\
\midrule

\rowcolor{ExprBack}
\hierbosss
& \msd{0.014}{0.019}
& \msd{1.000}{0.001}
& \msd{1142.192}{49.481}
& \msd{0.150}{0.010}
& \msd{1.000}{0.001}
& \msd{1110.915}{92.768}
& \msd{0.203}{0.011}
& \msd{1.000}{0.000}
& \msd{1117.321}{74.164} \\

\reprowsep

\bms
& \msd{0.450}{0.047}
& \msd{0.826}{0.040}
& \msd{2486.476}{290.751}
& \msd{0.506}{0.075}
& \msd{0.791}{0.052}
& \msd{2469.174}{149.699}
& \msd{0.542}{0.096}
& \msd{0.770}{0.073}
& \msd{2601.064}{644.774} \\

\reprowsep

\bsr
& \msd{1.009}{0.134}
& \msd{0.137}{0.128}
& \msd{1965.850}{66.240}
& \msd{0.905}{0.120}
& \msd{0.332}{0.138}
& \msd{1933.160}{16.120}
& \msd{1.355}{0.538}
& \msd{-0.616}{1.231}
& \msd{1931.850}{17.040}\\

\reprowsep

\dsr
& \msd{0.742}{0.082}
& \msd{0.530}{0.081}
& \msd{41.420}{0.638}
& \msd{0.761}{0.104}
& \msd{0.529}{0.099}
& \msd{59.320}{14.941}
& \msd{0.752}{0.069}
& \msd{0.560}{0.085}
& \msd{65.980}{11.217} \\

\reprowsep

\gplearn
& \msd{0.502}{0.113}
& \msd{0.776}{0.096}
& \msd{19.391}{1.055}
& \msd{0.531}{0.086}
& \msd{0.768}{0.064}
& \msd{20.437}{2.257}
& \msd{0.474}{0.087}
& \msd{0.851}{0.031}
& \msd{21.182}{4.144} \\

\reprowsep

\operon
& \msd{0.229}{0.035}
& \msd{0.953}{0.022}
& \msd{27.523}{2.473}
& \msd{0.311}{0.038}
& \msd{0.921}{0.020}
& \msd{26.457}{2.915}
& \msd{0.295}{0.042}
& \msd{0.932}{0.020}
& \msd{29.408}{5.562} \\

\reprowsep

\pysr
& \msd{0.172}{0.110}
& \msd{0.964}{0.036}
& \msd{77.939}{11.421}
& \msd{0.314}{0.061}
& \msd{0.920}{0.026}
& \msd{90.886}{25.039}
& \msd{0.357}{0.098}
& \msd{0.896}{0.053}
& \msd{118.457}{59.492} \\

\reprowsep

\qlattice
& \msd{0.286}{0.035}
& \msd{0.927}{0.029}
& \msd{113.685}{16.482}
& \msd{0.321}{0.023}
& \msd{0.917}{0.007}
& \msd{110.192}{5.903}
& \msd{0.350}{0.048}
& \msd{0.905}{0.026}
& \msd{120.602}{28.463} \\

\reprowsep

\sisso$++$
& \msd{0.195}{0.023}
& \msd{0.968}{0.005}
& \msd{17.127}{0.111}
& \msd{0.246}{0.015}
& \msd{0.951}{0.005}
& \msd{16.914}{0.131}
& \msd{0.279}{0.013}
& \msd{0.937}{0.006}
& \msd{16.825}{0.169} \\

\end{longtable}
\endgroup

\begin{figure}[H]
    \centering
    \includegraphics[width=0.88\linewidth]{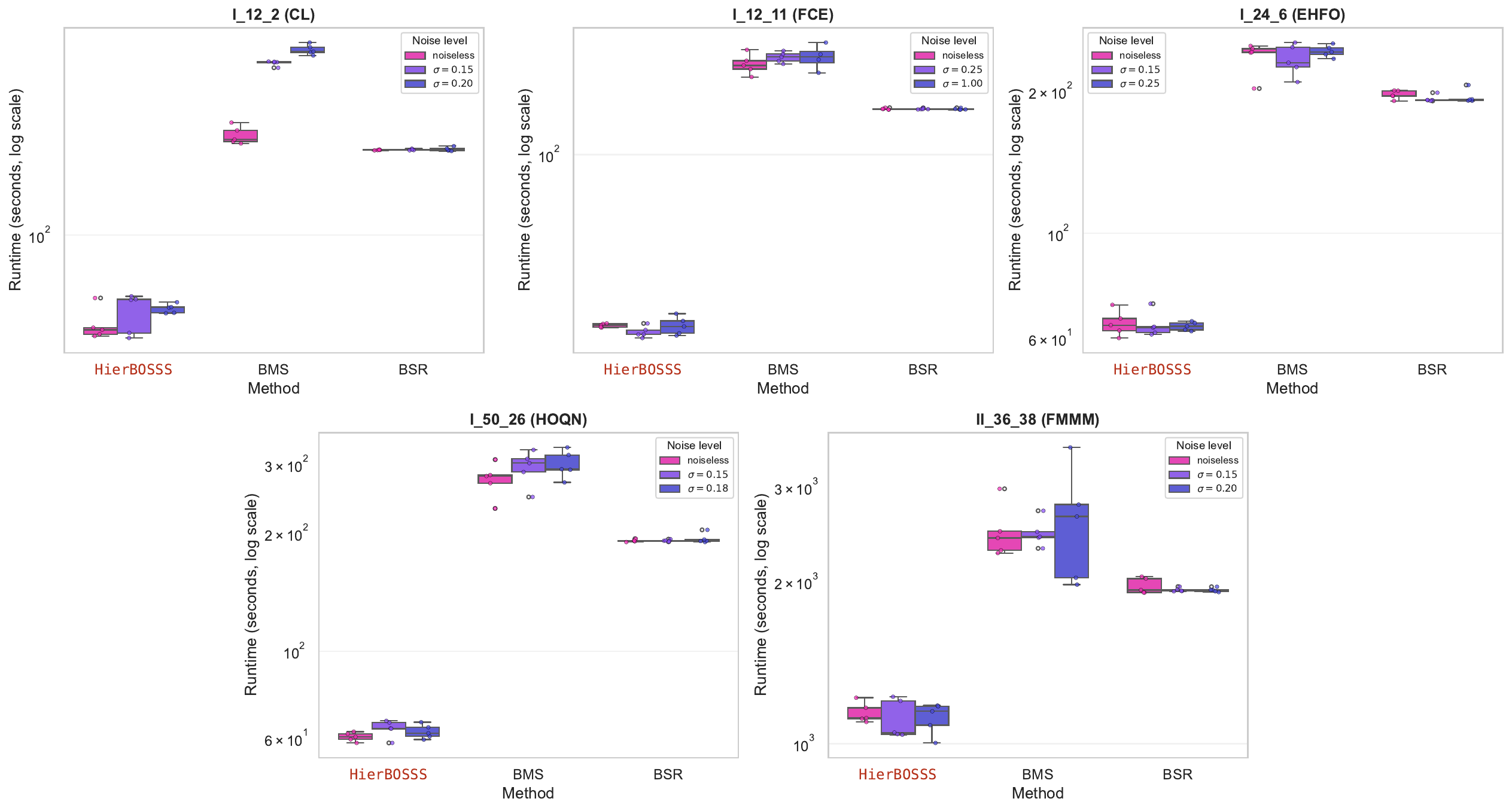}
    \caption{Compute runtimes (in $\log$-scale) of \hierbosss, \bms\ ($5$ parallel \texttt{MCMC} chains), and \bsr\ ($5$ restarts) for the Feynman equations in $\mathrm{I\_12\_2}$~\eqnref[supple-eq:feynman-cl]-$\mathrm{II\_36\_38}$~\eqnref[supple-eq:feynman-fmmm], across different noise levels and $5$ independent repetitions. Implementation details of \hierbosss\ and the Bayesian competitors for each Feynman equation follow from \hyperref[subsec:BayeSymX-Feynman-settings]{Appendices \ref{subsec:BayeSymX-Feynman-settings}}, \hyperref[subsec:BMS-Feynman-settings]{\ref{subsec:BMS-Feynman-settings}}, and \hyperref[subsec:BSR-Feynman-settings]{\ref{subsec:BSR-Feynman-settings}}.}
    \label{fig:runtime_Bayesian_SR}
\end{figure}

\newpage
\section{Ablation Study for \texorpdfstring{\hierbosss}{HierBOSSS} for the Feynman Equations and A Practical Diagnostic for Choosing the Symbolic Forest Size}
\label{sec:ablation}

We conduct an ablation study to assess the sensitivity of \hierbosss\ to the nominal number of symbolic trees \(K\) in the symbolic forest component \(\mathcal T\). At the same time, this study provides a practical diagnostic for choosing \(K\). The study is performed on two representative Feynman equations, \(\mathrm{I\_12\_2}\)~\eqnref[supple-eq:feynman-cl] and \(\mathrm{I\_12\_11}\)~\eqnref[supple-eq:feynman-fce], under a fixed noise level \(\sigma=0.25\). For each equation, we vary \(K\) over \(\{2,3,4,5,6,8,10,12\}\) and run \hierbosss\ with \(5\) independent repetitions under the experimental settings described in~\hyperref[subsec:BayeSymX-Feynman-settings]{Appendix \ref{subsec:BayeSymX-Feynman-settings}}.

\begin{figure}[H]
\centering
\includegraphics[width=0.80\textwidth]{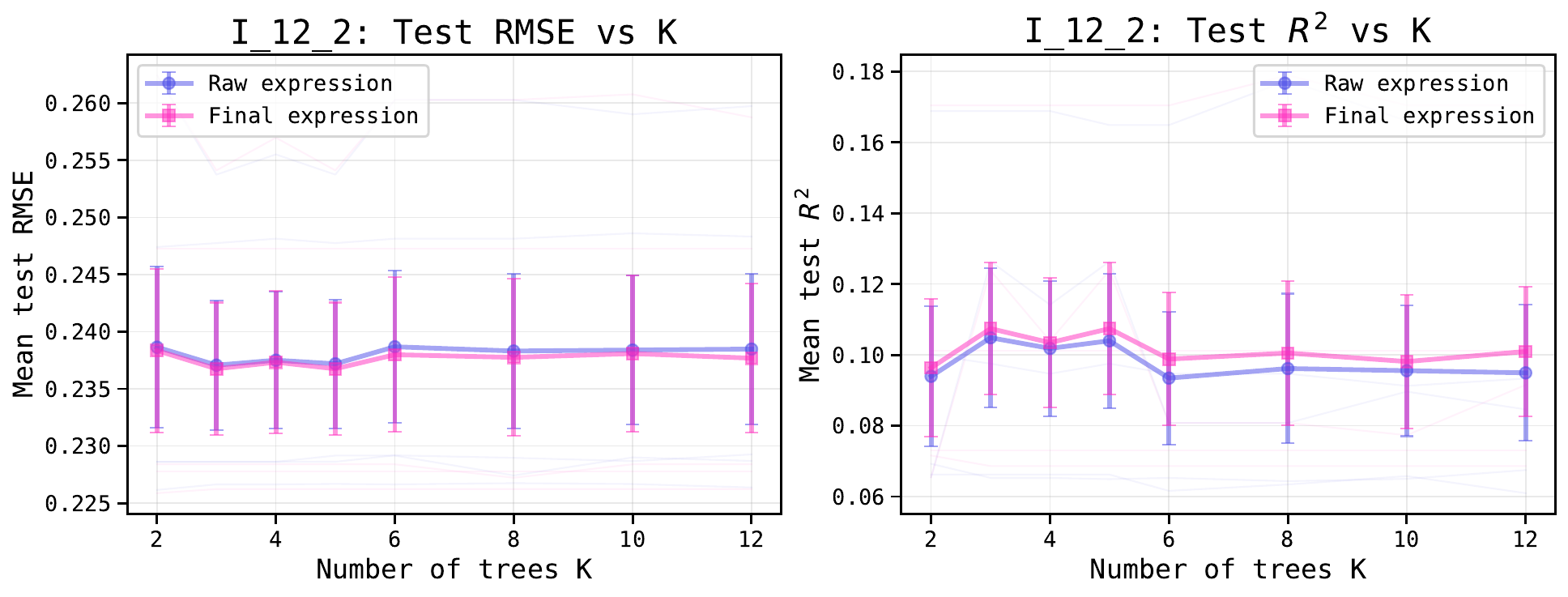}
\caption{Average (over $5$ independent repetitions) test \texttt{RMSE} and $R^2$ of \hierbosss\ across different values of $K$ for learning $\mathrm{I\_12\_2}$~\eqnref[supple-eq:feynman-cl] under noise level $\sigma=0.25$.}
\label{fig:k-ablation-i-12-2-rmse-r2}
\end{figure}

\begin{figure}[H]
\centering
\includegraphics[width=0.80\textwidth]{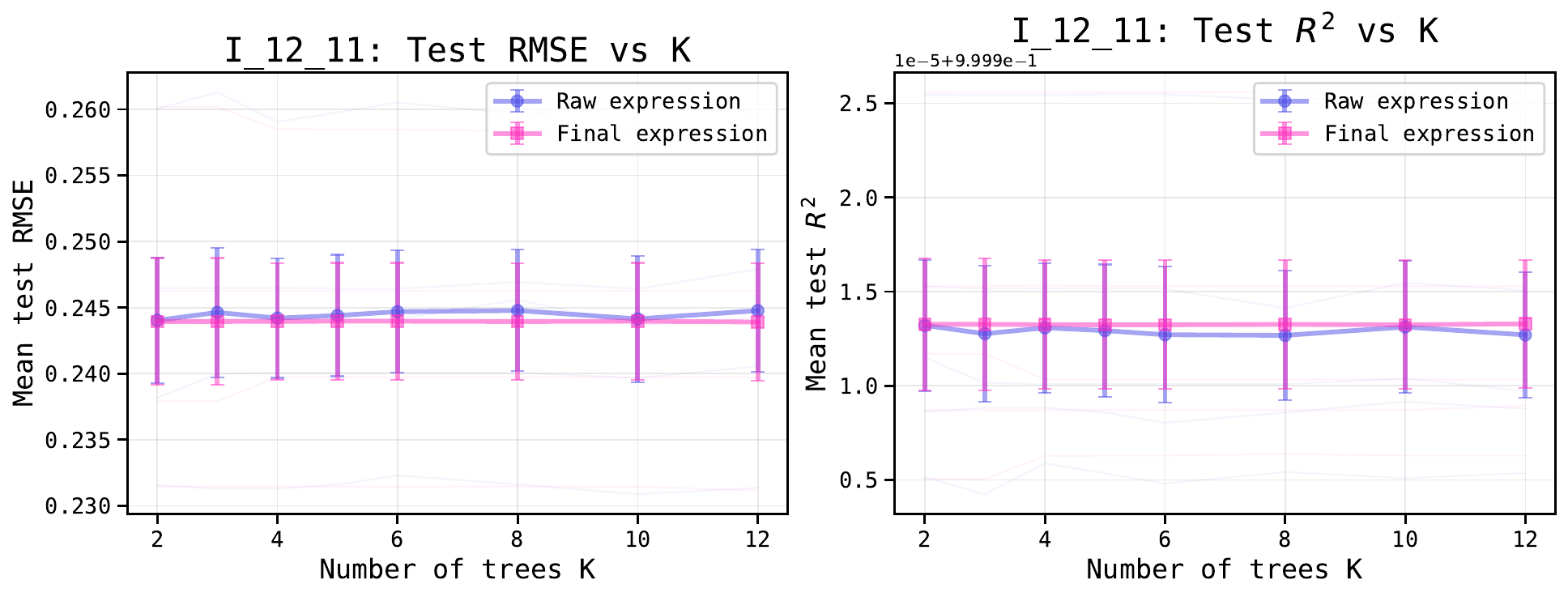}
\caption{Average (over $5$ independent repetitions) test \texttt{RMSE} and $R^2$ of \hierbosss\ across different values of $K$ for learning $\mathrm{I\_12\_11}$~\eqnref[supple-eq:feynman-fce] under noise level $\sigma=0.25$.}
\label{fig:k-ablation-i-12-11-rmse-r2}
\end{figure}

\hyperref[fig:k-ablation-i-12-2-rmse-r2]{Figures~\ref{fig:k-ablation-i-12-2-rmse-r2}} and~\ref{fig:k-ablation-i-12-11-rmse-r2} summarize the effect of \(K\) on the predictive accuracy of \hierbosss. Across the considered values of \(K\), both the \rawtag\ and \finaltag\ expressions exhibit stable average test \texttt{RMSE} and test \(R^2\) over \(5\) repetitions. For \(\mathrm{I\_12\_2}\)~\eqnref[supple-eq:feynman-cl], increasing \(K\) beyond small to moderate values yields little change in predictive performance. A similar pattern is observed for \(\mathrm{I\_12\_11}\)~\eqnref[supple-eq:feynman-fce], where the test \texttt{RMSE} and test \(R^2\) remain highly stable across the full range of \(K\). These results indicate that, once \(K\) is large enough to represent the main additive symbolic components, additional trees do not materially improve predictive performance.


\begin{figure}[!htp]
\centering

\begin{subfigure}[t]{0.45\textwidth}
\centering
\includegraphics[width=\linewidth]{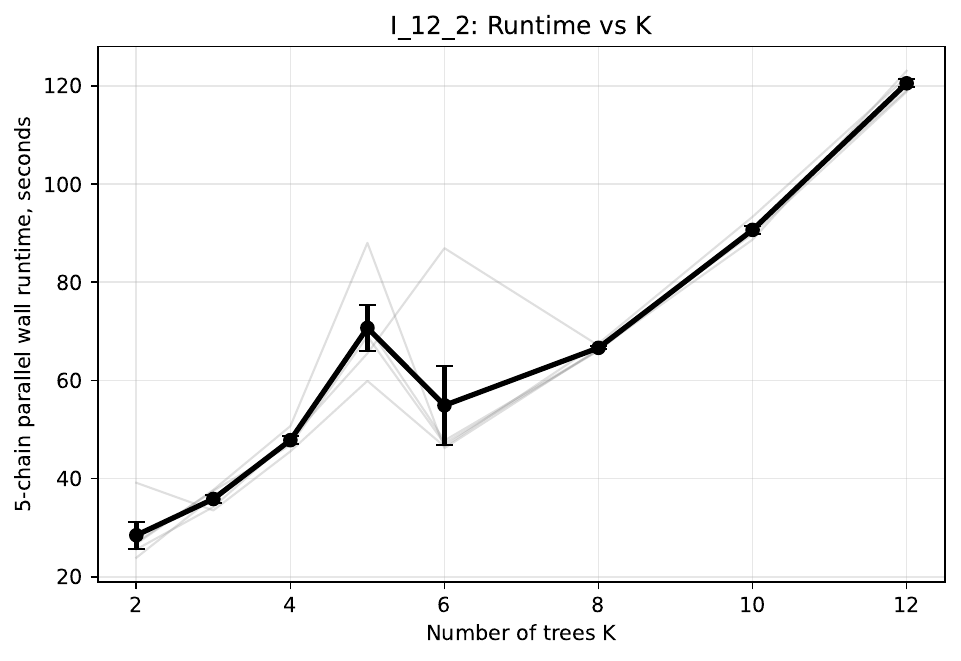}
\caption{$\mathrm{I\_12\_2}$~\eqnref[supple-eq:feynman-cl]}
\label{fig:k-ablation-i-12-2-runtime}
\end{subfigure}
\hfill
\begin{subfigure}[t]{0.45\textwidth}
\centering
\includegraphics[width=\linewidth]{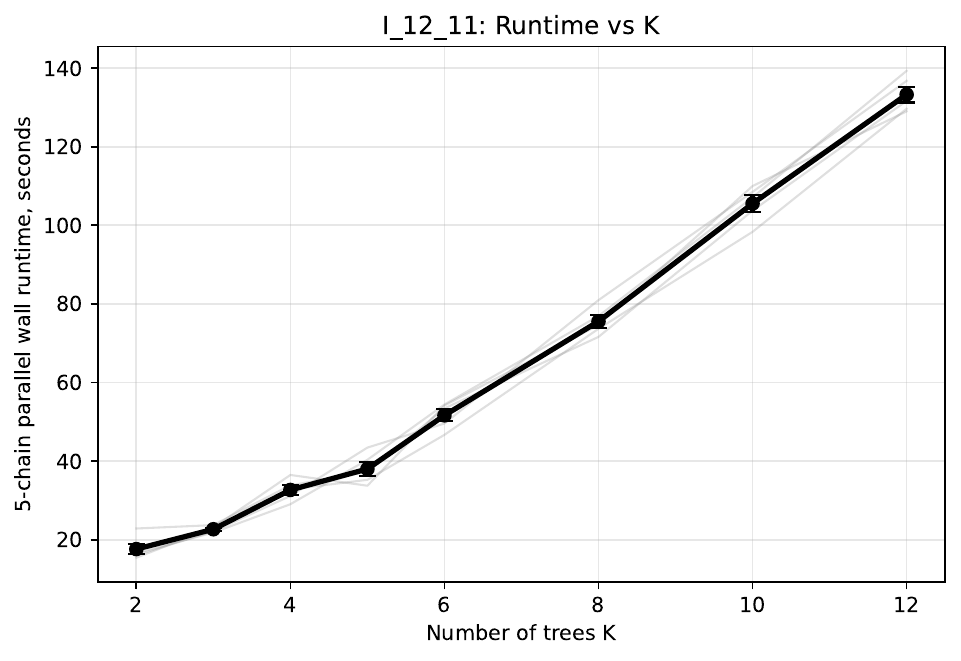}
\caption{$\mathrm{I\_12\_11}$~\eqnref[supple-eq:feynman-fce]}
\label{fig:k-ablation-i-12-11-runtime}
\end{subfigure}

\caption{Runtime behavior of \hierbosss\ across different values of $K$ under noise level $\sigma=0.25$ for $\mathrm{I\_12\_2}$~\eqnref[supple-eq:feynman-cl] and $\mathrm{I\_12\_11}$~\eqnref[supple-eq:feynman-fce]. Each panel reports the wall-clock runtime for $5$ parallel chains across $5$ independent repetitions, with the mean trend shown in black and repetition-level variability shown in gray.}
\label{fig:k-ablation-runtime}
\end{figure}

The runtime behavior is shown in \hyperref[fig:k-ablation-runtime]{Figure~\ref{fig:k-ablation-runtime}}. As expected, the computational cost increases with \(K\). This increase is moderate for smaller values of \(K\), but becomes more pronounced for larger symbolic forest sizes. Thus, although larger \(K\) expands the symbolic expression search space, it also introduces additional computational overhead without commensurate gains in predictive accuracy for these Feynman equations.


Taken together, the preceding ablation study suggests a principled strategy for choosing \(K\): select the smallest or a moderate value of \(K\) after which predictive accuracy stabilizes, while avoiding unnecessarily large forests that increase runtime. In this sense, \(K\) serves as a nominal expressivity parameter controlling the breadth of symbolic exploration, rather than the final number of symbolic components retained in the reported expression. The post-\texttt{MCMC} symbolic model refinement in \hyperref[alg:post-mcmc-symbolic-model-refinement-single]{Algorithm \ref{alg:post-mcmc-symbolic-model-refinement-single}} subsequently determines the effective forest size \(K_{\mathrm{eff}}\) in a data-adaptive manner by removing redundant or collinear symbolic tree structures. Therefore, a moderate nominal \(K\) provides sufficient exploratory flexibility, while the \finaltag\ expression remains parsimonious through the refinement step.

\newpage
\section{Posterior Diagnostics of the Bayesian \texorpdfstring{\sr}{SR} Methods for Learning Feynman Equations}
\label{sec:trace-plots-Feynman}

We provide trace plots to examine the stochastic search and posterior convergence behavior of \hierbosss, \bms, and \bsr, while learning the Feynman equations in $\mathrm{I\_12\_2}$~\eqnref[supple-eq:feynman-cl]-$\mathrm{II\_36\_38}$~\eqnref[supple-eq:feynman-fmmm]. For each equation, we show one representative repetition at the corresponding fixed noise level ($\sigma$), where \hierbosss\ and \bms\ use $5$ parallel chains and \bsr\ is executed for $5$ restarts. All other experimental configurations follow from~\hyperref[subsec:BayeSymX-Feynman-settings]{Appendices \ref{subsec:BayeSymX-Feynman-settings}}, \hyperref[subsec:BMS-Feynman-settings]{\ref{subsec:BMS-Feynman-settings}}, and \hyperref[subsec:BSR-Feynman-settings]{\ref{subsec:BSR-Feynman-settings}}, respectively. These traces visualize the stability of the posterior search and exploration of the Bayesian \sr\ methods over the symbolic expression space. Since the plotted quantity differs across methods, the traces should be interpreted method-wise rather than as direct numerical comparisons across methods.

\begin{figure}[H]
\centering

\begin{subfigure}[t]{0.41\textwidth}
\centering
\includegraphics[width=\linewidth]{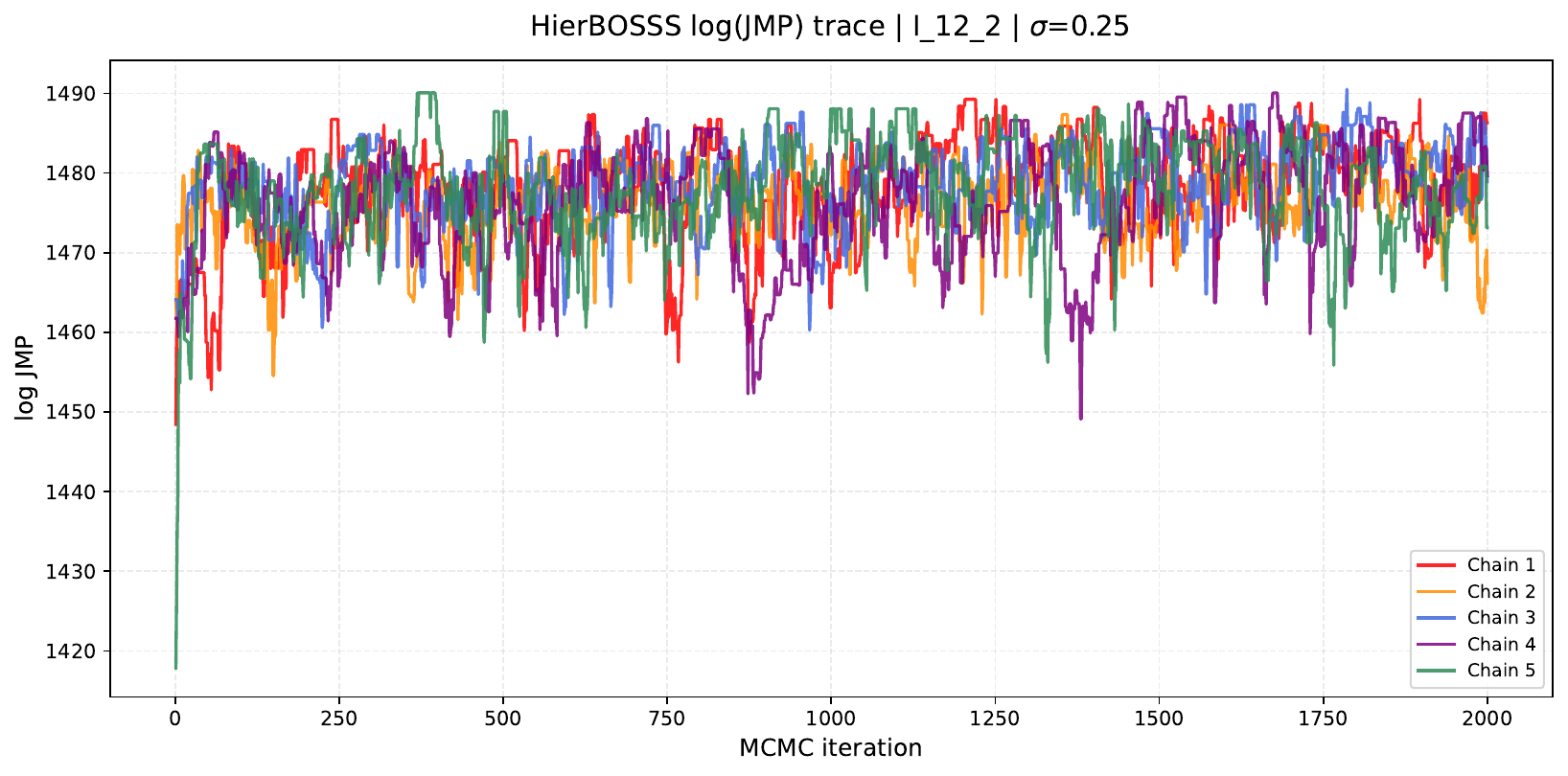}
\caption{$\mathrm{I\_12\_2}$~\eqnref[supple-eq:feynman-cl]}
\end{subfigure}
\hfill
\begin{subfigure}[t]{0.41\textwidth}
\centering
\includegraphics[width=\linewidth]{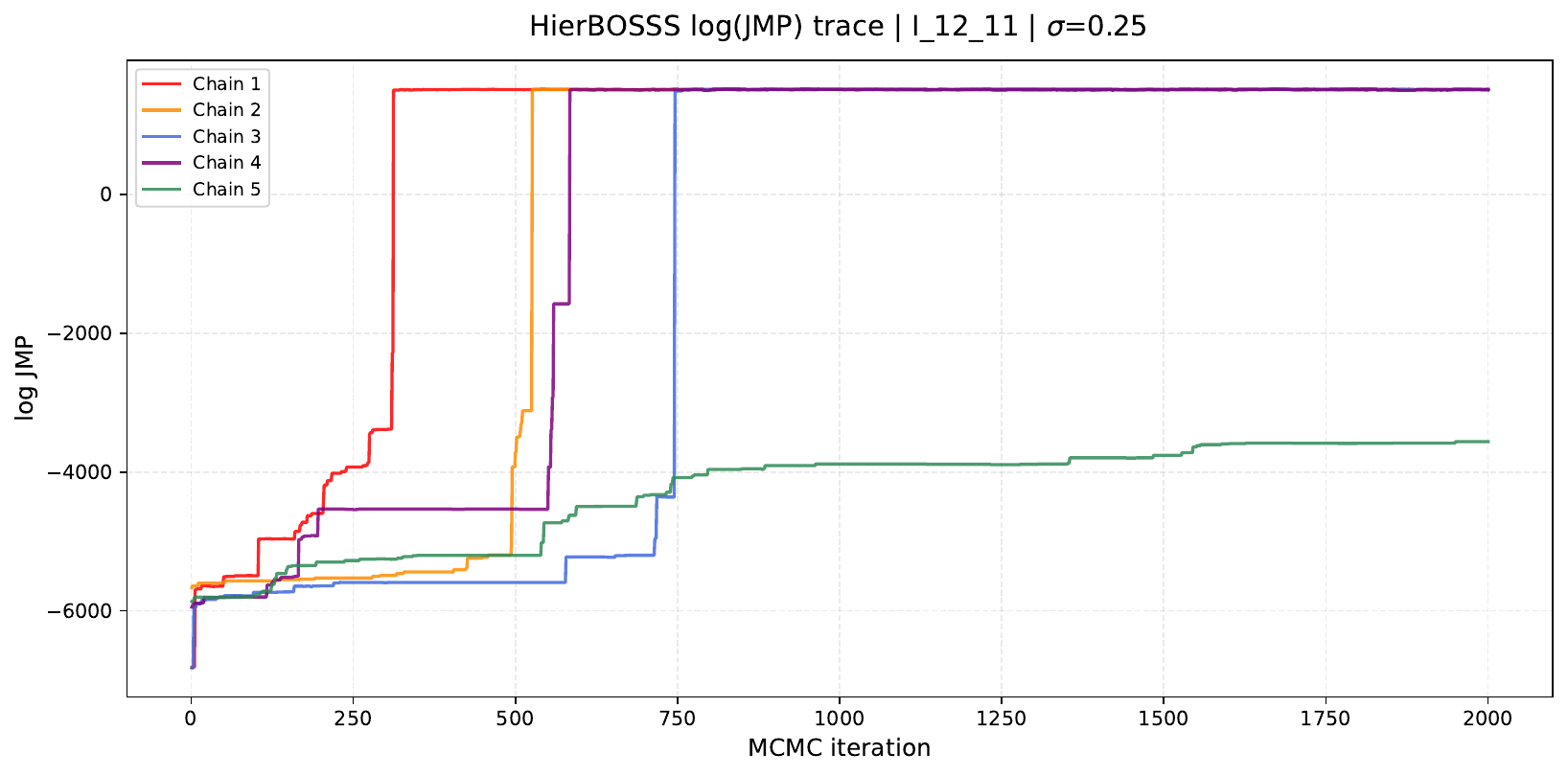}
\caption{$\mathrm{I\_12\_11}$~\eqnref[supple-eq:feynman-fce]}
\end{subfigure}
\hfill
\begin{subfigure}[t]{0.41\textwidth}
\centering
\includegraphics[width=\linewidth]{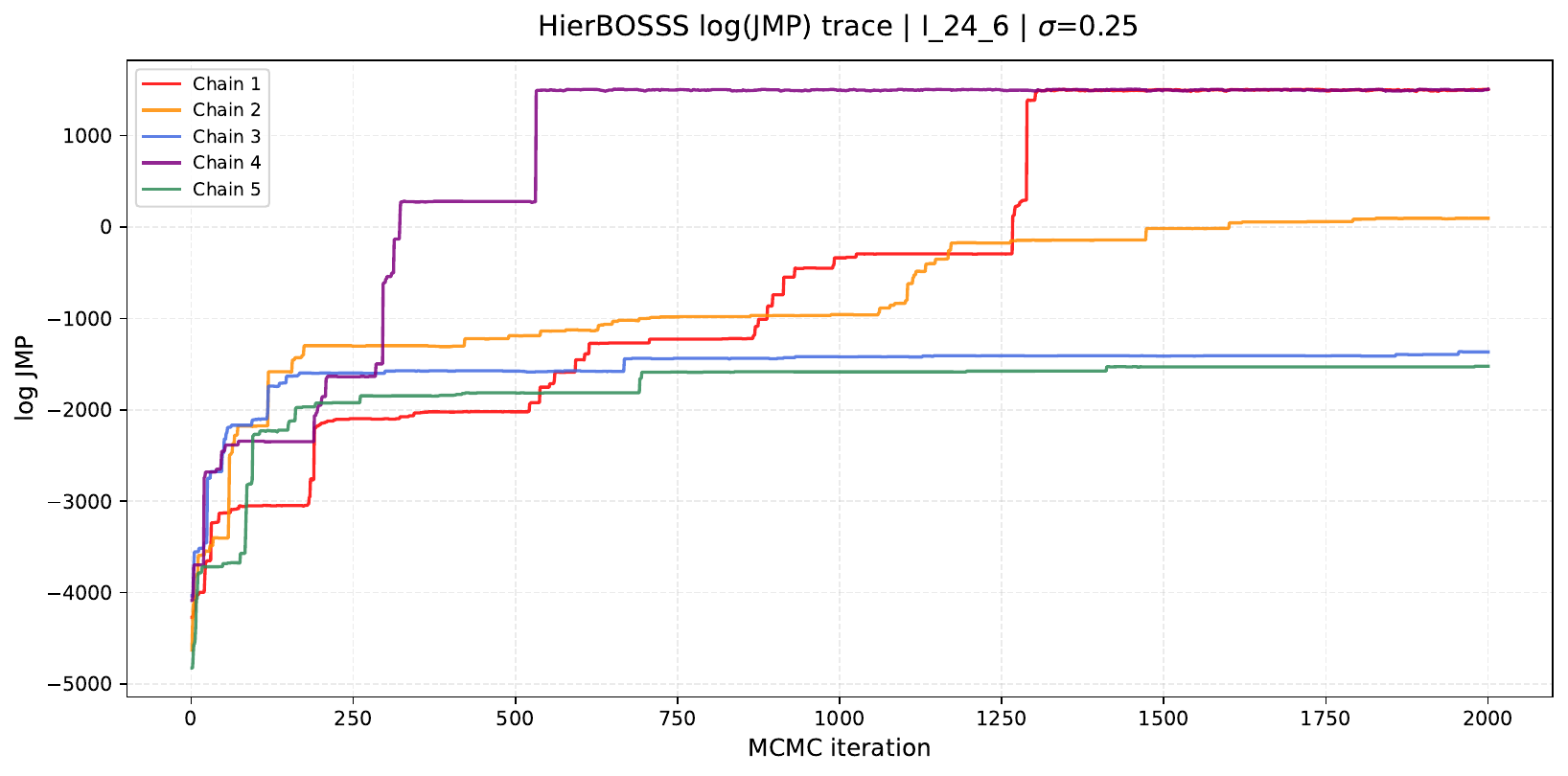}
\caption{$\mathrm{I\_24\_6}$~\eqnref[supple-eq:feynman-ehfo]}
\end{subfigure}

\medskip

\begin{subfigure}[t]{0.41\textwidth}
\centering
\includegraphics[width=\linewidth]{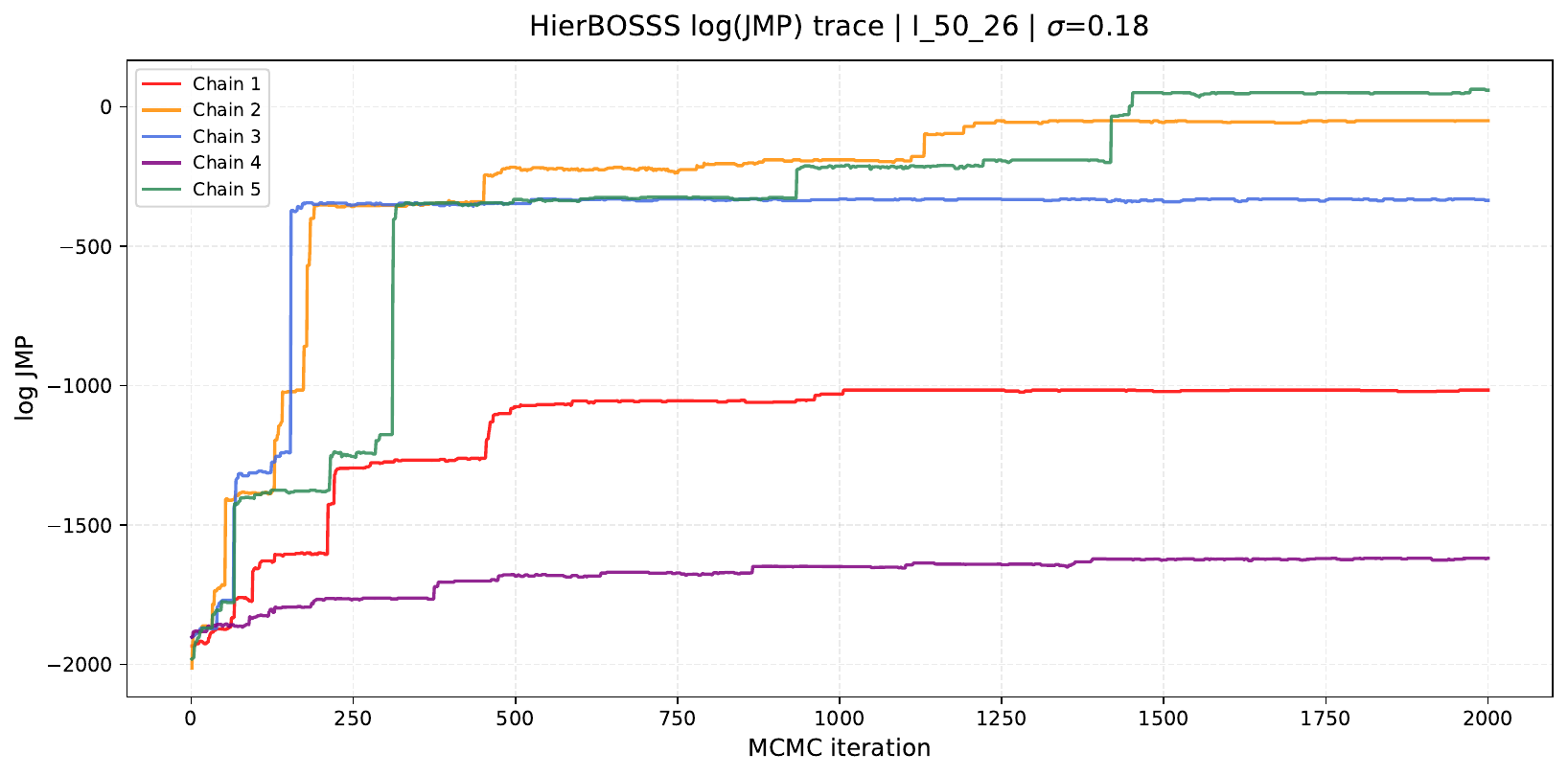}
\caption{$\mathrm{I\_50\_26}$~\eqnref[supple-eq:feynman-hoqn]}
\end{subfigure}
\hspace{0.06\textwidth}
\begin{subfigure}[t]{0.41\textwidth}
\centering
\includegraphics[width=\linewidth]{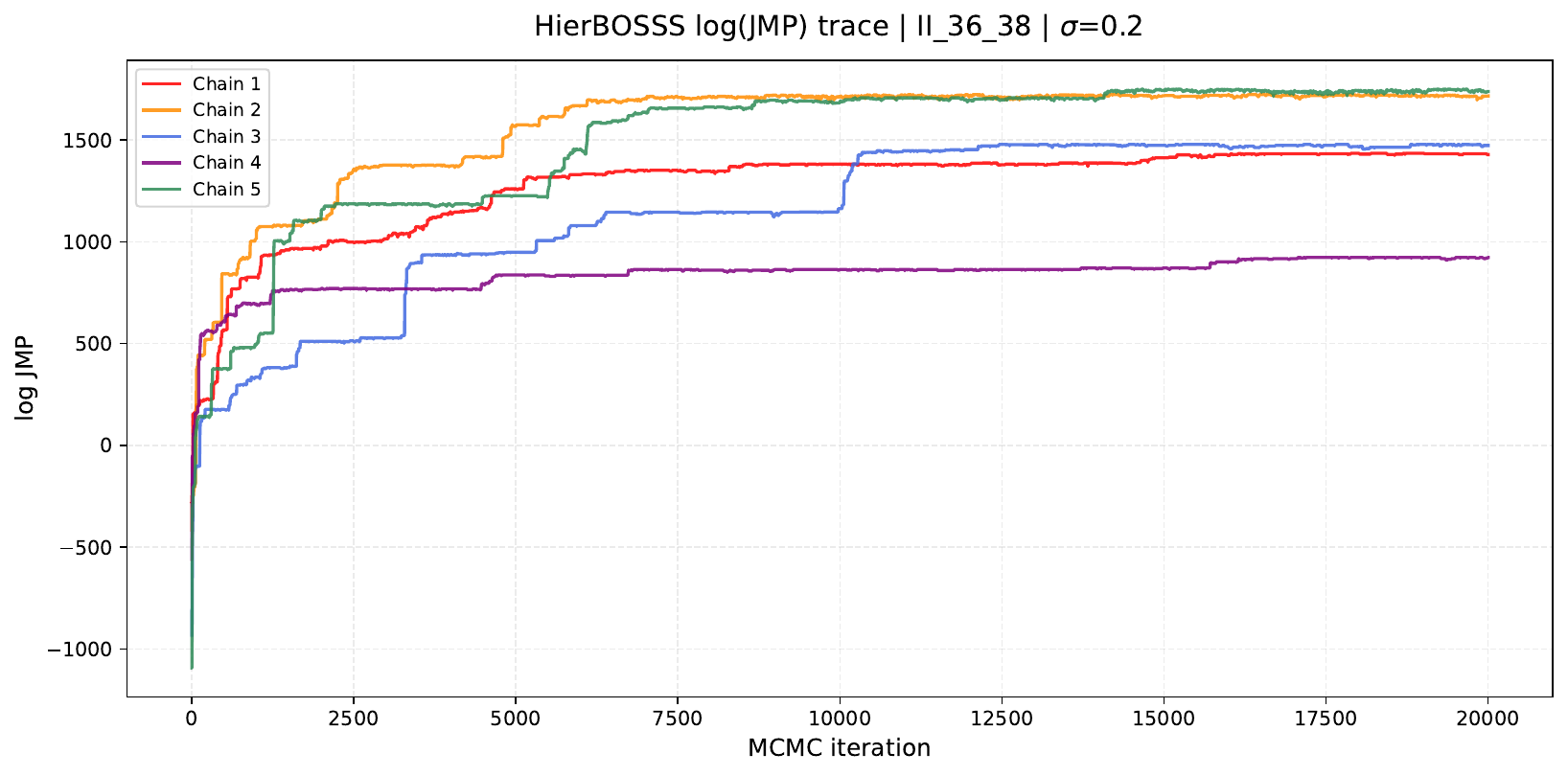}
\caption{$\mathrm{II\_36\_38}$~\eqnref[supple-eq:feynman-fmmm]}
\end{subfigure}

\caption{Trace plots for \hierbosss\ across the Feynman equations in $\mathrm{I\_12\_2}$~\eqnref[supple-eq:feynman-cl]--$\mathrm{II\_36\_38}$~\eqnref[supple-eq:feynman-fmmm].}
\label{fig:trace-BayeSymX-feynman}
\end{figure}

For \hierbosss, the traces in~\hyperref[fig:trace-BayeSymX-feynman]{Figure~\ref{fig:trace-BayeSymX-feynman}} show the evolution of the $\log$ joint marginal posterior value of $\mathcal{T}$, denoted by $\log(\mathrm{JMP})$, across \texttt{MCMC} iterations. These traces indicate how the sampler explores the posterior distribution over symbolic forests and whether the chains move toward regions of high posterior support. Stable traces across chains suggest proper mixing and that the posterior search repeatedly identifies comparable high-scoring symbolic structures.

\newpage

\begin{figure}[H]
\centering

\begin{subfigure}[t]{0.41\textwidth}
\centering
\includegraphics[width=\linewidth]{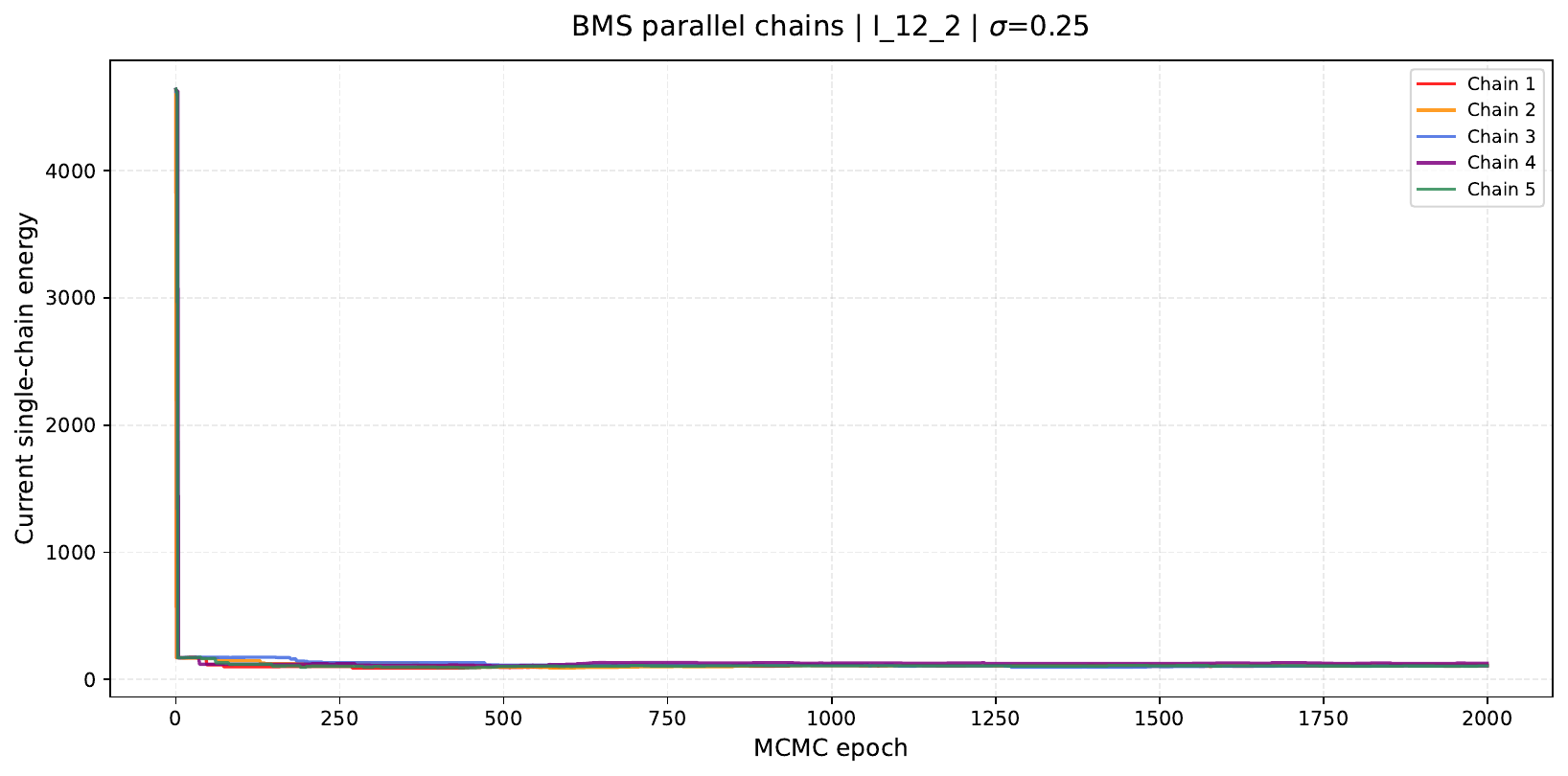}
\caption{$\mathrm{I\_12\_2}$~\eqnref[supple-eq:feynman-cl]}
\end{subfigure}
\hfill
\begin{subfigure}[t]{0.41\textwidth}
\centering
\includegraphics[width=\linewidth]{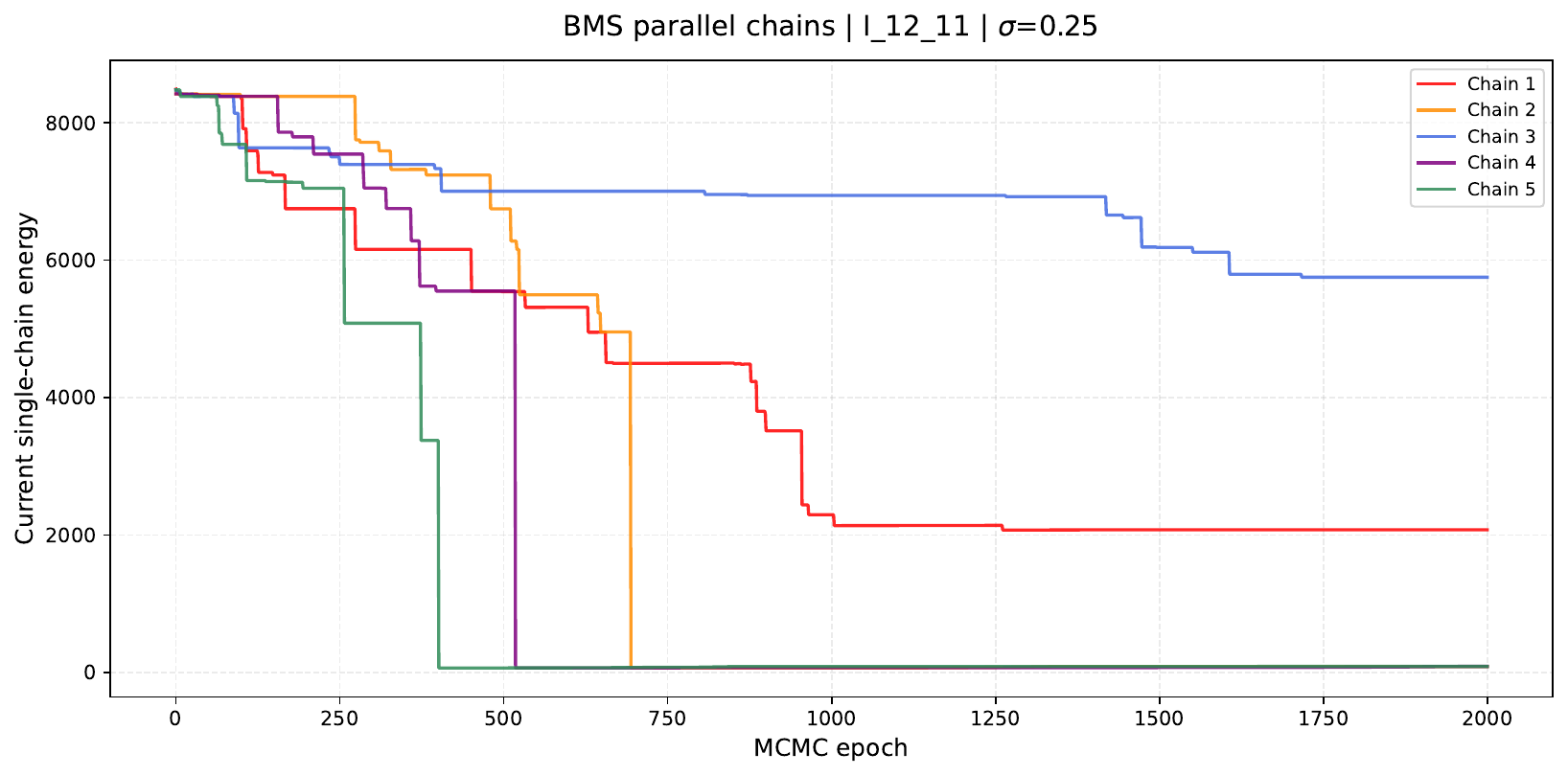}
\caption{$\mathrm{I\_12\_11}$~\eqnref[supple-eq:feynman-fce]}
\end{subfigure}
\hfill
\begin{subfigure}[t]{0.41\textwidth}
\centering
\includegraphics[width=\linewidth]{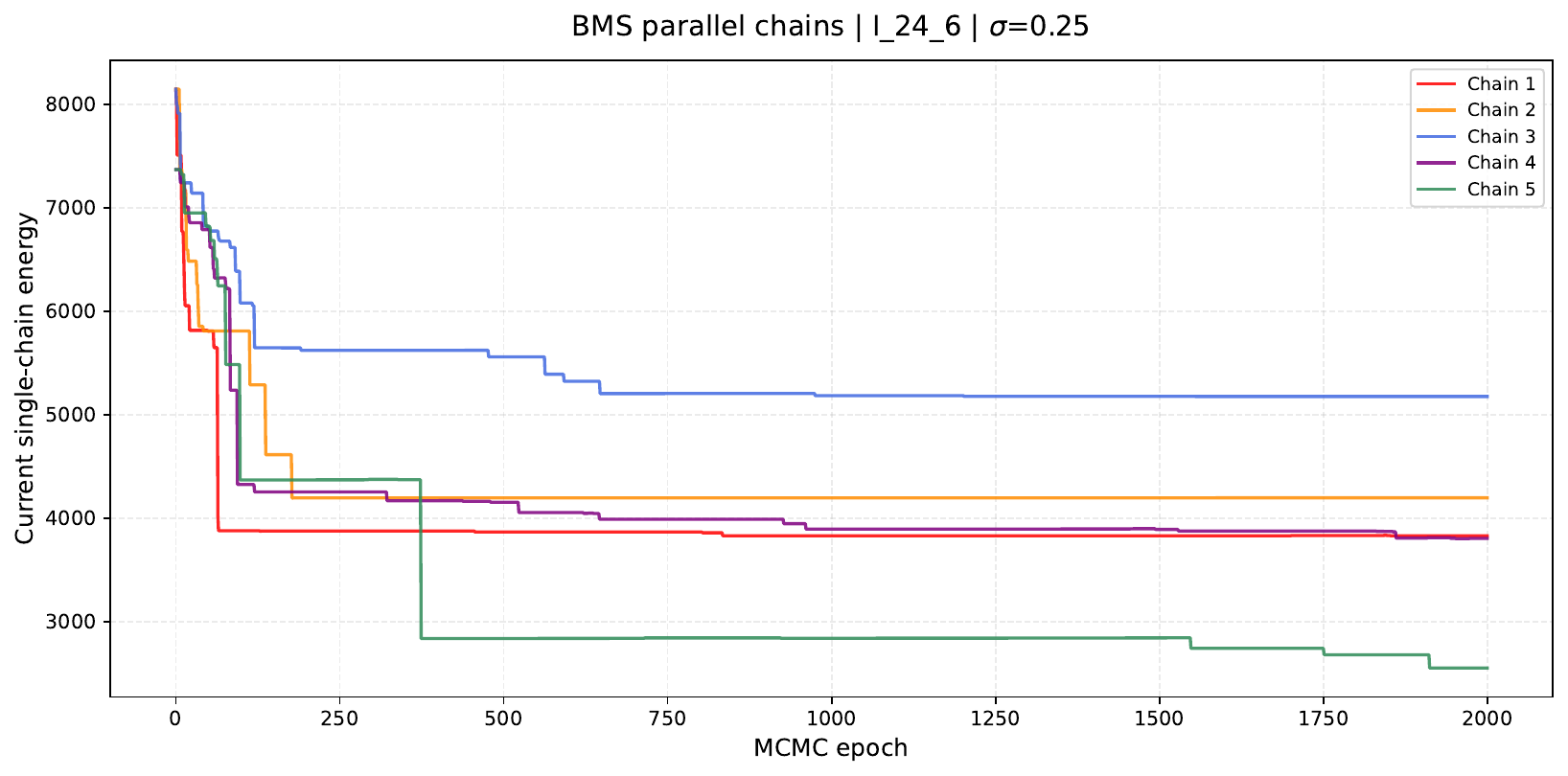}
\caption{$\mathrm{I\_24\_6}$~\eqnref[supple-eq:feynman-ehfo]}
\end{subfigure}

\medskip

\begin{subfigure}[t]{0.41\textwidth}
\centering
\includegraphics[width=\linewidth]{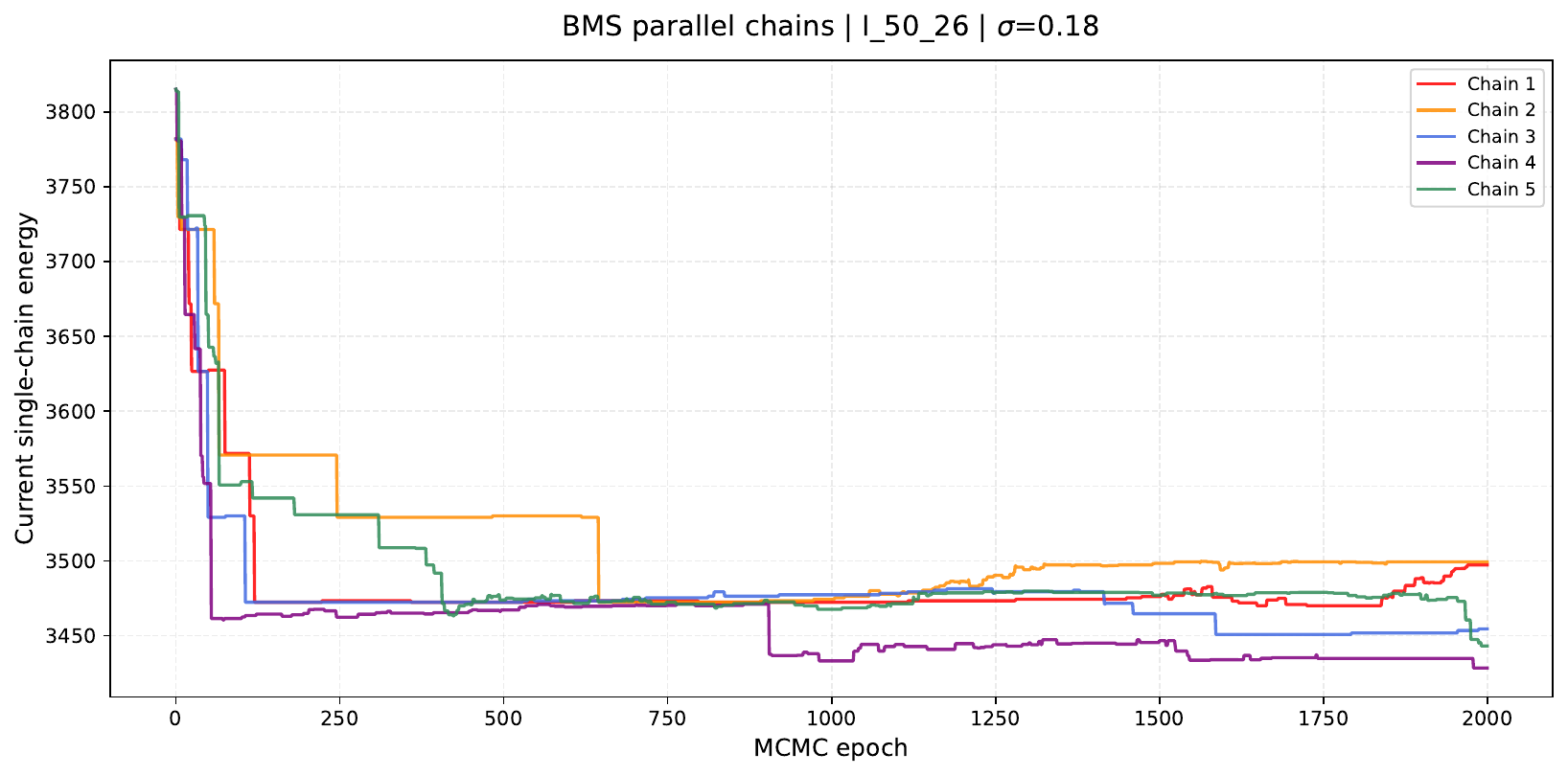}
\caption{$\mathrm{I\_50\_26}$~\eqnref[supple-eq:feynman-hoqn]}
\end{subfigure}
\hspace{0.06\textwidth}
\begin{subfigure}[t]{0.41\textwidth}
\centering
\includegraphics[width=\linewidth]{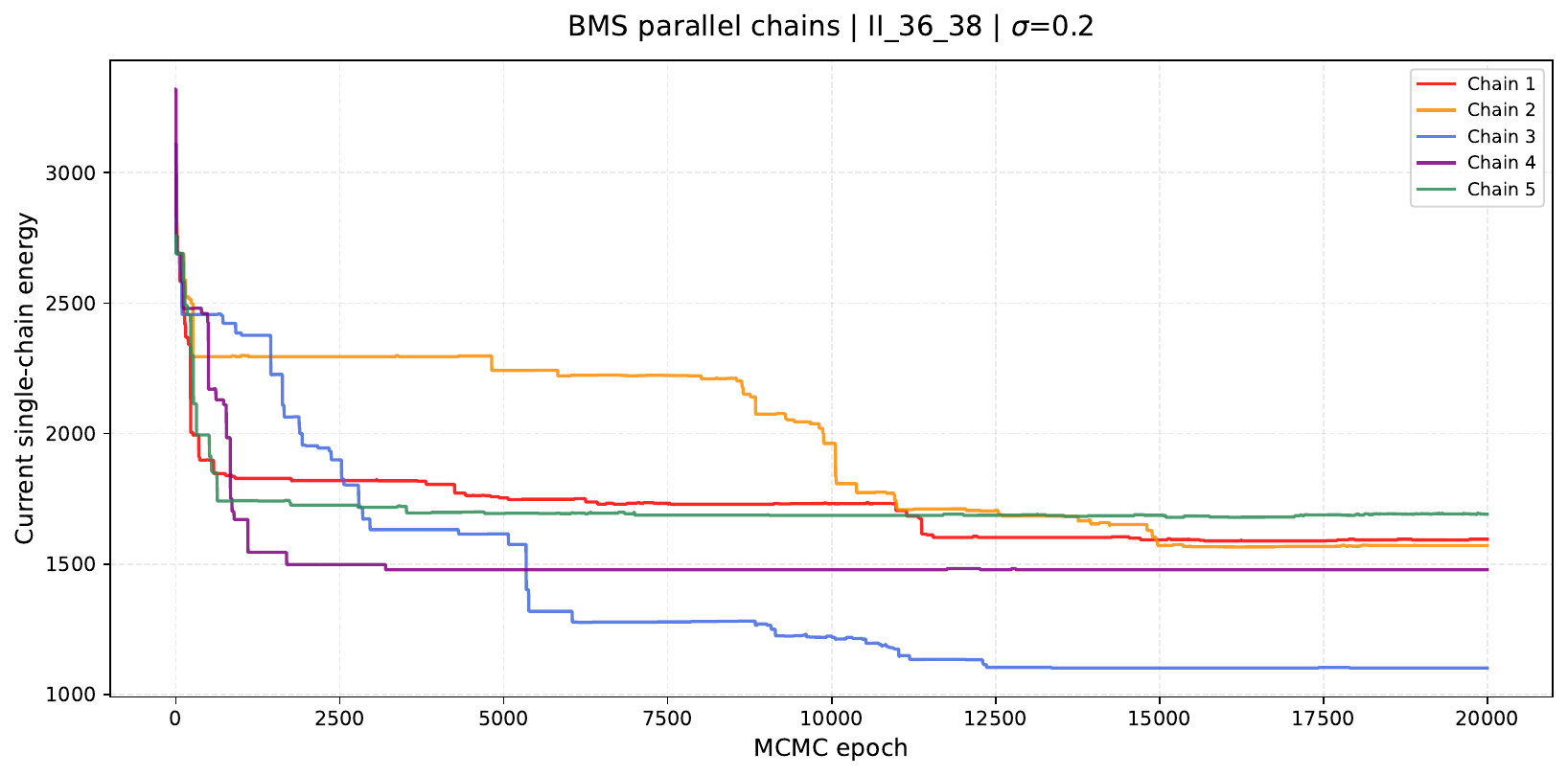}
\caption{$\mathrm{II\_36\_38}$~\eqnref[supple-eq:feynman-fmmm]}
\end{subfigure}

\caption{Trace plots for \bms\ across the Feynman equations in $\mathrm{I\_12\_2}$~\eqnref[supple-eq:feynman-cl]--$\mathrm{II\_36\_38}$~\eqnref[supple-eq:feynman-fmmm].}
\label{fig:trace-bms-feynman}
\end{figure}

For \bms, the traces in~\hyperref[fig:trace-bms-feynman]{Figure~\ref{fig:trace-bms-feynman}} report the current single-chain energy, or the description length~\citep{BMS}, over \texttt{MCMC} epochs. This energy combines the data-fit term and the prior or symbolic complexity penalty. Lower values indicate models with better posterior support. Therefore, downward movements correspond to improvements in fit, parsimony, or both, while flat regions indicate that the chain remains in similar symbolic states.

\begin{figure}[H]
\centering

\begin{subfigure}[t]{0.41\textwidth}
\centering
\includegraphics[width=\linewidth]{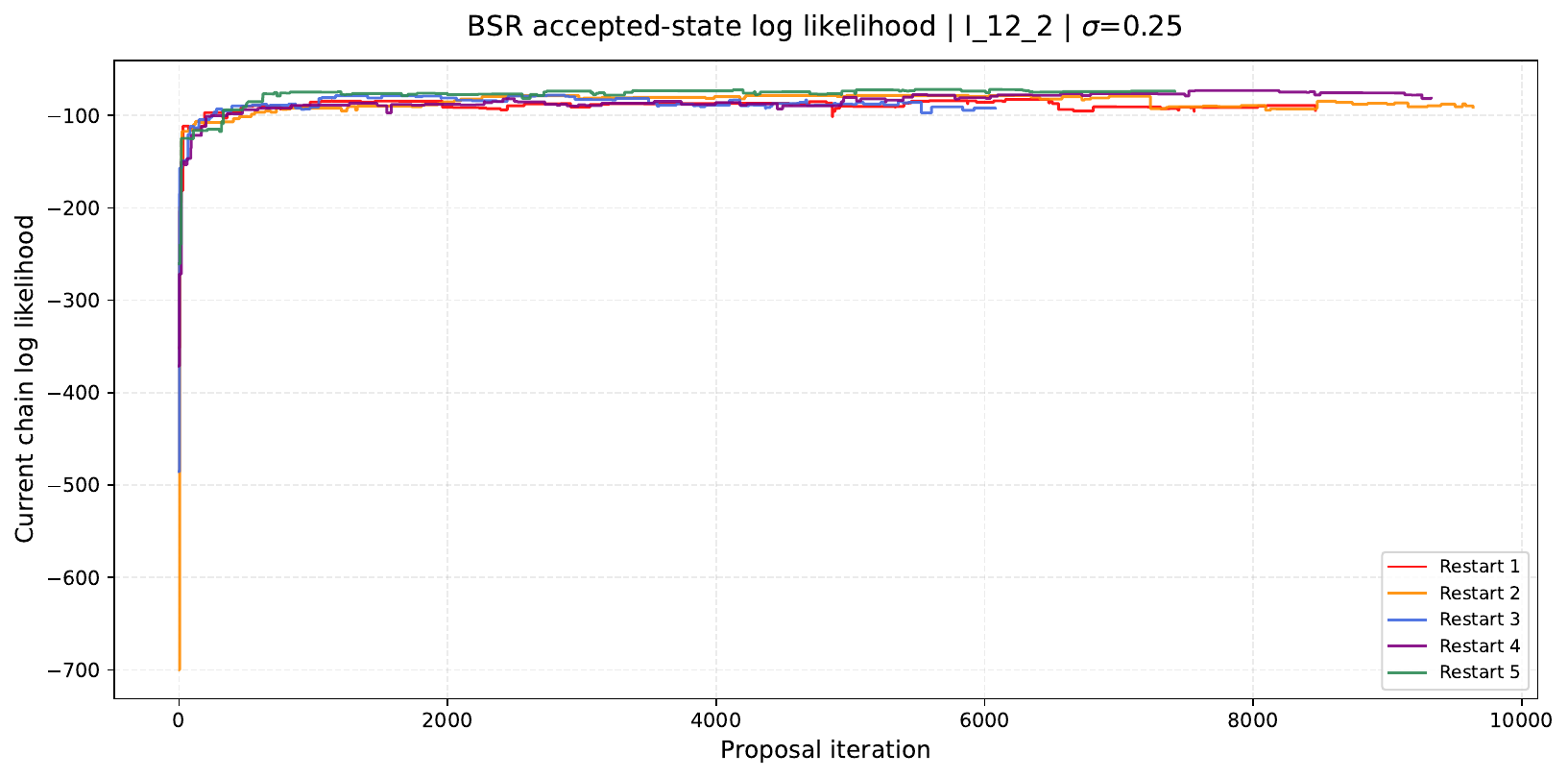}
\caption{$\mathrm{I\_12\_2}$~\eqnref[supple-eq:feynman-cl]}
\end{subfigure}
\hfill
\begin{subfigure}[t]{0.41\textwidth}
\centering
\includegraphics[width=\linewidth]{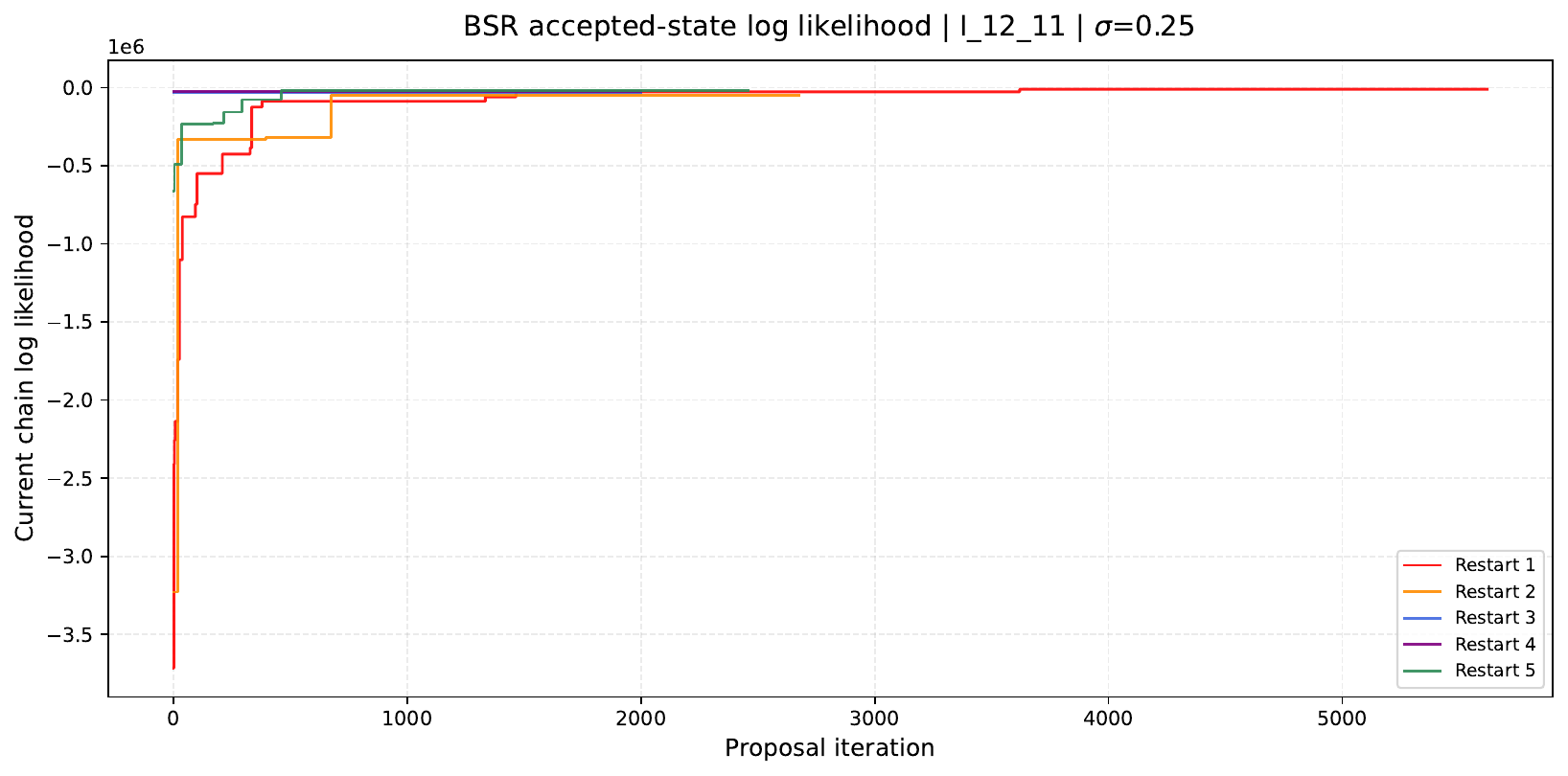}
\caption{$\mathrm{I\_12\_11}$~\eqnref[supple-eq:feynman-fce]}
\end{subfigure}
\hfill
\begin{subfigure}[t]{0.41\textwidth}
\centering
\includegraphics[width=\linewidth]{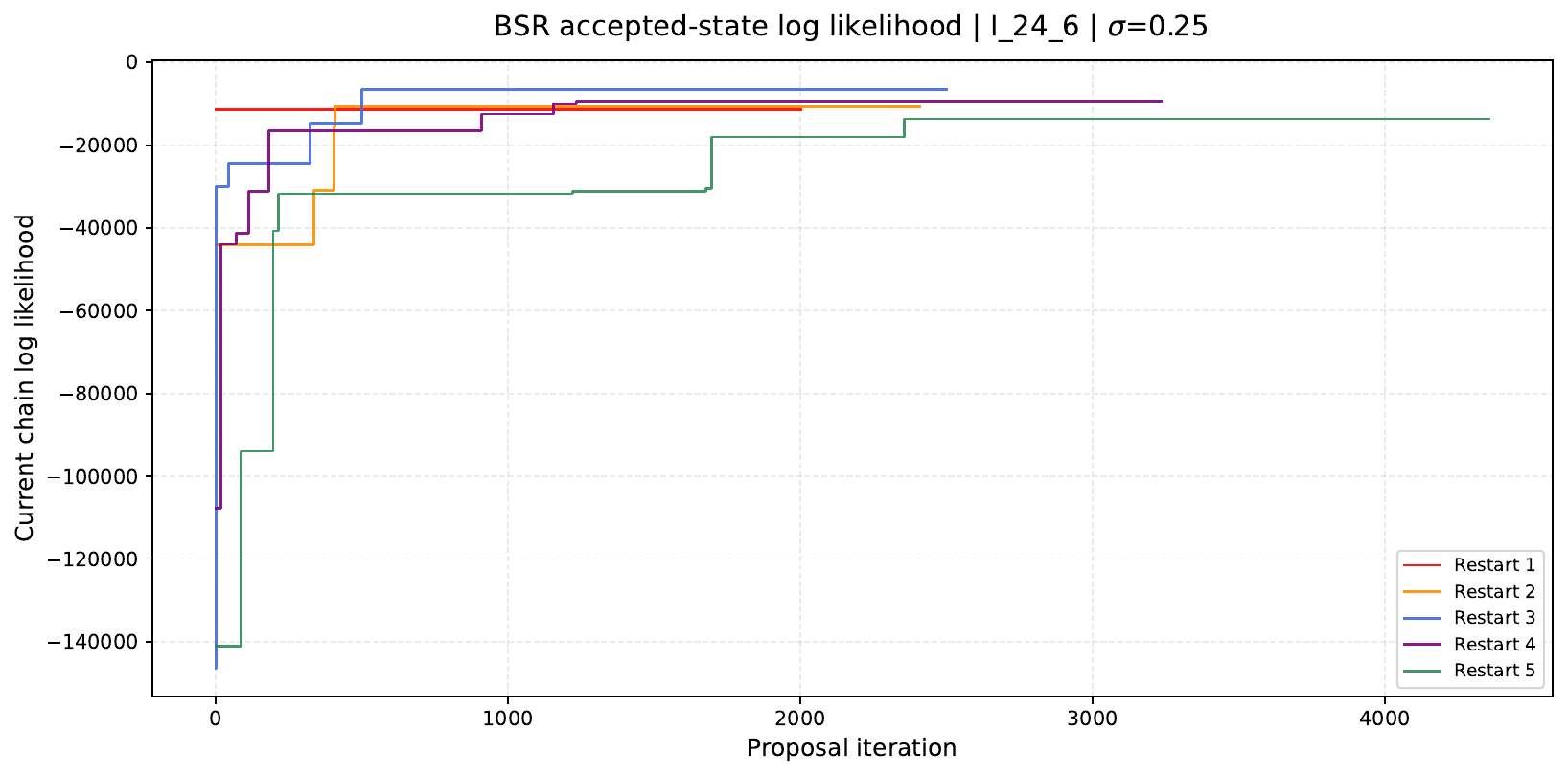}
\caption{$\mathrm{I\_24\_6}$~\eqnref[supple-eq:feynman-ehfo]}
\end{subfigure}

\medskip

\begin{subfigure}[t]{0.41\textwidth}
\centering
\includegraphics[width=\linewidth]{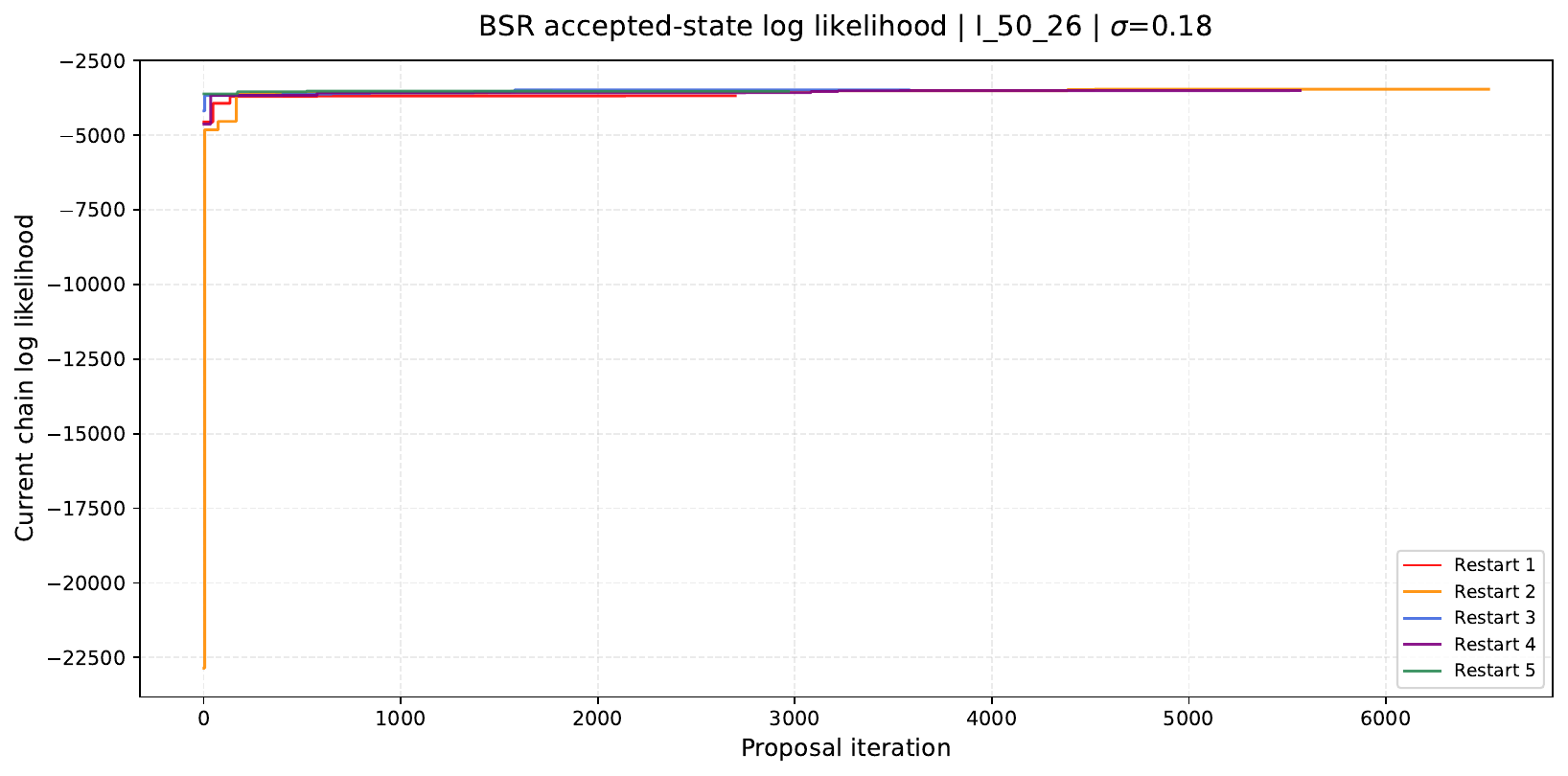}
\caption{$\mathrm{I\_50\_26}$~\eqnref[supple-eq:feynman-hoqn]}
\end{subfigure}
\hspace{0.06\textwidth}
\begin{subfigure}[t]{0.41\textwidth}
\centering
\includegraphics[width=\linewidth]{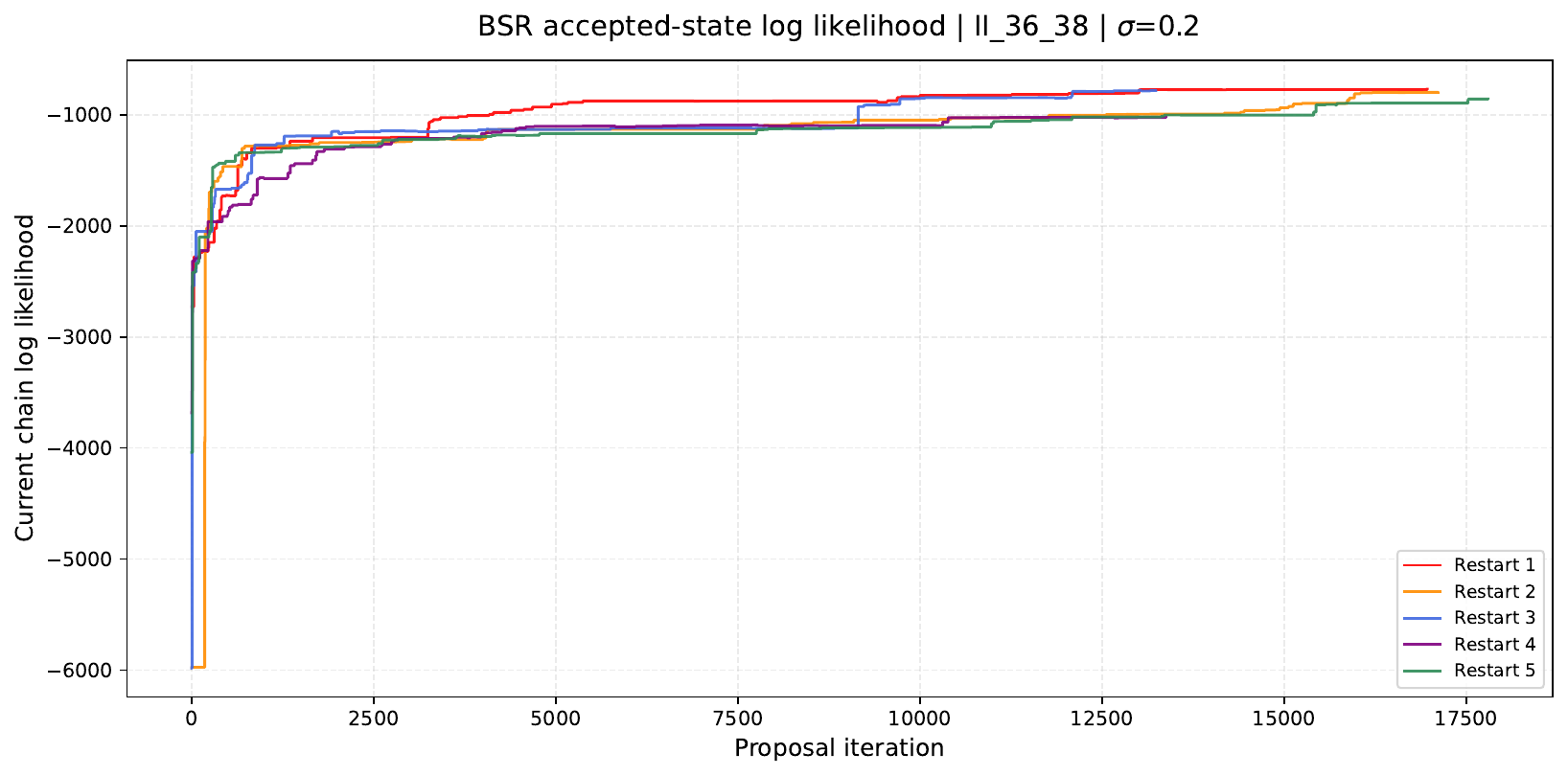}
\caption{$\mathrm{II\_36\_38}$~\eqnref[supple-eq:feynman-fmmm]}
\end{subfigure}

\caption{Trace plots for \bsr\ across the Feynman equations in $\mathrm{I\_12\_2}$~\eqnref[supple-eq:feynman-cl]--$\mathrm{II\_36\_38}$~\eqnref[supple-eq:feynman-fmmm].}
\label{fig:trace-bsr-feynman}
\end{figure}

\newpage
For \bsr, the traces in \hyperref[fig:trace-bsr-feynman]{Figure \ref{fig:trace-bsr-feynman}} show the accepted-state $\log$ likelihood across proposal iterations~\citep{BSR}. Since the trace updates only when a proposed symbolic modification is accepted, flat segments correspond to rejected proposals, while jumps indicate accepted moves. Upward jumps typically reflect improved data fit, although occasional downward moves may occur due to the exploratory nature of the underlying \texttt{MCMC} routine.

\newpage
\suppdivision{Empirical Results for Oxide Perovskite Catalyst Data Study}
\label{superdiv:perovskites}

\begin{tcolorbox}[
  enhanced,
  colback=CompBack,
  colframe=CompBlue,
  boxrule=0pt,
  arc=1pt,
  left=8pt,
  right=8pt,
  top=6pt,
  bottom=6pt,
  before skip=8pt,
  after skip=10pt,
  borderline west={2.5pt}{0pt}{CompBlue}
]
\small
\textcolor{CompBlue}{\textbf{Extended results.}}
A broader set of results for the oxide perovskite dataset is available \href{https://github.com/Roy-SR-007/HierBOSSS/blob/main/Oxide_Perovskite_Results.pdf}{\texttt{here}}.
\end{tcolorbox}



\section{Description of the Oxide Perovskite Data, Experimental Protocol, and Evaluation Criteria}
\label{sec:oxide-perovskite-data-description}

We analyze the experimental dataset of~\cite{Weng2020SimpleDescriptor}, comprising $18$ conventional oxide perovskites synthesized and characterized under a common experimental protocol. The response variable is the potential $V_{\mathrm{RHE}}$ (in electron volts (eV)), measured relative to the reversible hydrogen electrode, required for a catalyst to sustain a prescribed current density ($\rho$). At a fixed $\rho$, a smaller $V_{\mathrm{RHE}}$ indicates that less applied potential is needed to drive the oxygen evolution reaction and therefore corresponds to greater catalytic activity.

\begin{figure}[H]
    \centering
    \includegraphics[width=0.6\linewidth]{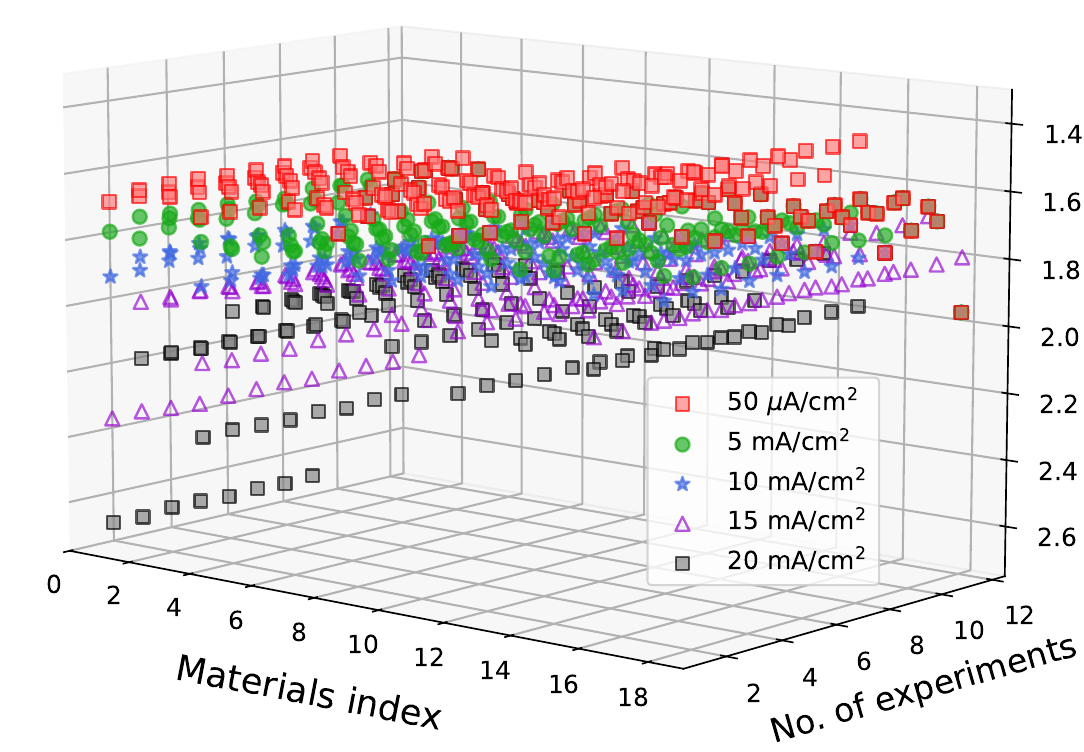}
    \caption{Experimental $V_{\mathrm{RHE}}$ measurements for the $18$ oxide perovskites across $12$ experimental replicates and $5$ prescribed current density ($\rho$) levels.}
    \label{fig:VRHE_landscape}
\end{figure}

For each material, $4$ independently prepared samples were measured $3$ times at each of $5$ $\rho$ levels: $0.05$, $5$, $10$, $15$, and $20$ $\mathrm{mA/cm}^{2}$. This yields $n = 18\times 4\times 3\times 5 = 1080$ observations. The resulting activity landscape is shown in \hyperref[fig:VRHE_landscape]{Figure~\ref{fig:VRHE_landscape}}, where measurements are organized by material, experimental replicate, and $\rho$ level. The landscape shows clear variation in $V_{\mathrm{RHE}}$ across both material compositions and operating conditions, together with noticeable within-material experimental variability. Larger $\rho$ values generally require larger applied potentials, while differences across materials reflect variation in intrinsic catalytic activity.

Thus, the dataset has response variable $V_{\mathrm{RHE}}$ and $p=8$ predictors: the prescribed $\rho$, and $7$ chemically motivated materials descriptors $R_A$, $Q_A$, $N_d$, $\chi_A$, $\chi_B$, $\mu$, and $t$, which are summarized in \hyperref[tab:perovskite-variable-interpretation]{Table~\ref{tab:perovskite-variable-interpretation}} leading to a \sr\ problem with $\textcolor{BrickRed}{(n, p) = (1080, 8)}$.

\begingroup
\scriptsize
\setlength{\tabcolsep}{3.4pt}
\renewcommand{\arraystretch}{1.18}

\begin{longtable}{@{}p{0.11\textwidth} p{0.12\textwidth}
                        p{0.34\textwidth} p{0.39\textwidth}@{}}
\caption{Response and input variables in the oxide perovskite data study.}
\label{tab:perovskite-variable-interpretation}\\

\toprule
\toprule
\textbf{Role}
& \textbf{Variable}
& \textbf{Description}
& \textbf{Scientific interpretation} \\
\midrule
\endfirsthead

\caption[]{Response and input variables in the oxide perovskite data study \emph{(continued)}.}\\
\toprule
\toprule
\textbf{Role}
& \textbf{Variable}
& \textbf{Description}
& \textbf{Scientific interpretation} \\
\midrule
\endhead

\midrule
\multicolumn{4}{r}{\emph{Continued on next page}}\\
\endfoot

\bottomrule
\bottomrule
\endlastfoot

Output
& $V_{\mathrm{RHE}}$
& Potential relative to the reversible hydrogen electrode required to sustain a prescribed current density.
& Measures \texttt{OER} performance. At a fixed current density, a smaller value indicates a smaller required applied potential and hence greater catalytic activity. \\[4pt]

\reprowsep

Input
& $\rho$
& Prescribed current density, taking values $0.05$, $5$, $10$, $15$, and $20\ \mathrm{mA}/\mathrm{cm}^{2}$.
& Defines the electrochemical operating condition at which $V_{\mathrm{RHE}}$ is measured. Larger current densities generally require greater applied potential to sustain a higher reaction rate. \\[4pt]

\reprowsep

Input
& $R_A$
& Composition-weighted ionic radius of the $A$-site cations, measured in \AA.
& Characterizes the effective size of the cations occupying the larger crystallographic site and influences lattice expansion, octahedral tilting, structural distortion, and the tolerance factor $t$. \\[4pt]

\reprowsep

Input
& $Q_A$
& Composition-weighted formal charge of the $A$-site cations.
& Contributes to charge neutrality and indirectly constrains the oxidation states and electronic configurations of the catalytically active $B$-site transition metals. \\[4pt]

\reprowsep

Input
& $N_d$
& Composition-weighted number of $d$ electrons associated with the transition metal $B$-site cations.
& Summarizes the $B$-site electronic configuration, which influences metal--oxygen covalency, orbital occupancy, adsorption energetics, and electron transfer during the \texttt{OER}. \\[4pt]

\reprowsep

Input
& $\chi_A$
& Composition-weighted electronegativity of the $A$-site cations.
& Describes the tendency of the $A$-site species to attract electronic charge and can indirectly affect bonding, charge redistribution, and $B$-site oxidation states. \\[4pt]

\reprowsep

Input
& $\chi_B$
& Composition-weighted electronegativity of the transition metal $B$-site cations.
& Reflects the electronic character of the catalytically active site and its interaction with oxygen ligands and adsorbed \texttt{OER} intermediates. \\[4pt]

\reprowsep

Input
& $t$
& Goldschmidt tolerance factor,
$\displaystyle t=(R_A+R_O)/[\sqrt{2}(R_B+R_O)]$.
& Measures ionic size compatibility with the ideal $ABO_{3}$ lattice. Deviations from the ideal geometric range are associated with octahedral rotation, lattice distortion, and changes in structural stability. \\[4pt]

\reprowsep

Input
& $\mu$
& Octahedral factor,
$\displaystyle \mu=R_B/R_O$.
& Measures the size compatibility of the $B$-site cation with its surrounding oxygen octahedron and is related to the formation and distortion of the $BO_{6}$ units central to perovskite structure and catalysis. \\

\end{longtable}
\endgroup
We perform $5$ independent random $90/10$ train-test splits of the oxide perovskite dataset. For each split, \hierbosss\ and the same suite of competing methods, as in \hyperref[superdiv:feynman]{Part \ref{superdiv:feynman}} for the Feynman equations, are fitted to the corresponding $90\%$ training set. Predictive performance is evaluated on both the training set and the held-out $10\%$ test set using root mean squared error (\texttt{RMSE}) and mean absolute error (\texttt{MAE}). For a sample of size $n$, these are defined as
\begin{equation}
\label{eq:perovskite-evaluation-metrics}
\texttt{RMSE}
=\left[\frac{1}{n}
\sum_{i=1}^{n}
\left(
V_{\mathrm{RHE},i} -
\widehat{V}_{\mathrm{RHE},i}
\right)^2\right]^{1/2},
\qquad
\texttt{MAE}
=\frac{1}{n}
\sum_{i=1}^{n}
\left|
V_{\mathrm{RHE},i} -
\widehat{V}_{\mathrm{RHE},i}
\right|.
\end{equation}
The results are summarized across the $5$ independent splits of the dataset to assess both predictive accuracy and variability across data partitions.

\newpage


\section{Experimental Settings of \texorpdfstring{\hierbosss}{HierBOSSS} and Competitors}
\label{sec:experimental-settings-perovskite}

\begin{tcolorbox}[
  enhanced,
  colback=CompBack,
  colframe=CompBlue,
  boxrule=0pt,
  arc=1pt,
  left=8pt,
  right=8pt,
  top=6pt,
  bottom=6pt,
  before skip=8pt,
  after skip=10pt,
  borderline west={2.5pt}{0pt}{CompBlue}
]
\small
\textcolor{CompBlue}{\textbf{Computational environment.}}
All experiments related to the oxide perovskite catalyst dataset were implemented in \texttt{Python} and executed on a MacBook Air equipped with an Apple M2 processor and 8GB of unified memory.
\end{tcolorbox}


\subsection{\texorpdfstring{\hierbosss}{HierBOSSS}}
\label{subsec:bayesymx-experimental-settings-perovskite}

We evaluate \hierbosss\ for discovering symbolic descriptors of catalytic activity from the oxide perovskite dataset of
\citet{Weng2020SimpleDescriptor} using the experimental protocol described in \hyperref[sec:oxide-perovskite-data-description]{Appendix \ref{sec:oxide-perovskite-data-description}}. The operator set $\mathbb O$ is kept fixed across all $5$ data splits as those used in~\cite{Weng2020SimpleDescriptor}.

\begin{pysettingsbox}{\hierbosss\ operator set}
opset = [
    "add",   # (x, y) -> x + y
    "mul",   # (x, y) -> x*y
    "neg",   # x -> -x
    "inv",   # x -> 1/x, protected near zero
    "sqrt",  # x -> sqrt(abs(x))
]
\end{pysettingsbox}

The prior hyperparameter specifications are shared across all data splits and these are kept at the same values as in the settings for the Feynman equations (see \hyperref[subsec:BayeSymX-Feynman-settings]{Appendix \ref{subsec:BayeSymX-Feynman-settings}}) except for the depth-dependent splitting probability parameters which are chosen to be $(\alpha_0,\delta_0)=(0.95,2.00)$.

We set the nominal symbolic forest size to $K=8$. The choice $K=8$ is guided by the practical diagnostic provided through the ablation study in
\hyperref[sec:perovskite-ablation]{Appendix \ref{sec:perovskite-ablation}},
which shows that predictive performance stabilizes around this value
while computational cost continues to increase for larger symbolic
forests. Each \texttt{MCMC} chain is run for $2000$ iterations.

\begin{pysettingsbox}{\hierbosss\ symbolic forest and \texttt{MCMC} configuration}
HierBOSSS_perovskite_settings = {"K": 8,  "maxiter": 2000,}
\end{pysettingsbox}

For each split of the oxide perovskite dataset, we run $5$ independent parallel \texttt{MCMC} chains of \hierbosss\ while preserving the same \texttt{MH} kernel and proposal parameters, posterior ranking scheme, and post-\texttt{MCMC} refinement steps as for the Feynman equation experiments; see \hyperref[subsec:BayeSymX-Feynman-settings]{Appendix \ref{subsec:BayeSymX-Feynman-settings}}.


\subsection{\texorpdfstring{\dsr}{DSR}}
\label{subsec:dsr-experimental-settings-perovskite}

As in the case of the Feynman equation experiments (see \hyperref[subsec:DSR-Feynman-settings]{Appendix \ref{subsec:DSR-Feynman-settings}}), we use the same implementation of deep symbolic regression (\dsr)~\citep{Deep-SR} for discovering symbolic descriptors of catalytic activity from the oxide
perovskite dataset of~\cite{Weng2020SimpleDescriptor} using the experimental protocol described in \hyperref[sec:oxide-perovskite-data-description]{Appendix \ref{sec:oxide-perovskite-data-description}}. Following~\cite{Weng2020SimpleDescriptor}, the operator set for all 5 splits is configured as follows.
\begin{pysettingsbox}{\dsr\ operator set}
function_set = [
    "add",      # (x, y) -> x + y
    "sub",      # (x, y) -> x - y
    "mul",      # (x, y) -> x*y
    "div",      # (x, y) -> x/y
    "sqrt",     # x -> sqrt(x)
]
\end{pysettingsbox}
All other configurations of \dsr\ are chosen same as in the case of the Feynman equation experiments; see \hyperref[subsec:DSR-Feynman-settings]{Appendix \ref{subsec:DSR-Feynman-settings}}.


\subsection{\texorpdfstring{\qlattice}{QLattice}}
\label{subsec:qlattice-experimental-settings-perovskite}

As in the case of the Feynman equation experiments (see \hyperref[subsec:QLattice-Feynman-settings]{Appendix \ref{subsec:QLattice-Feynman-settings}}), we use the same implementation of \qlattice~\citep{feyn-qlattice} for discovering symbolic descriptors of catalytic activity from the oxide
perovskite dataset of~\cite{Weng2020SimpleDescriptor} using the experimental protocol described in \hyperref[sec:oxide-perovskite-data-description]{Appendix \ref{sec:oxide-perovskite-data-description}}. Following~\cite{Weng2020SimpleDescriptor}, the primitive functions for all 5 splits are chosen as follows.
\begin{pysettingsbox}{\qlattice\ operator set}
function_names = [
    "inverse",  # z -> 1/z
    "linear",   # z -> a*z + b
    "sqrt",     # z -> sqrt(z)
    "add",      # (z1, z2) -> z1 + z2
    "multiply", # (z1, z2) -> z1*z2
]
\end{pysettingsbox}
All other configurations of \qlattice\ are chosen same as in the case of the Feynman equation experiments; see \hyperref[subsec:QLattice-Feynman-settings]{Appendix \ref{subsec:QLattice-Feynman-settings}}.


\subsection{\texorpdfstring{\sisso$++$}{SISSO++}}
\label{subsec:sisso++-experimental-settings-perovskite}

As in the case of the Feynman equation experiments (see \hyperref[subsec:SISSO++-Feynman-settings]{Appendix \ref{subsec:SISSO++-Feynman-settings}}), we use the same \sisso$++$~\citep{SISSO++} implementation of the sure independence screening and sparsifying operator (\sisso)~\citep{SISSO} for discovering symbolic descriptors of catalytic activity from the oxide
perovskite dataset of~\cite{Weng2020SimpleDescriptor} using the experimental protocol described in \hyperref[sec:oxide-perovskite-data-description]{Appendix \ref{sec:oxide-perovskite-data-description}}. Following~\cite{Weng2020SimpleDescriptor}, the common operators across all 5 splits are chosen as follows.

\begin{pysettingsbox}{\sisso$++$ operator set}
sisso_operators = [
    "add",        # add: x + y
    "sub",        # sub: x - y
    "mult",       # mult: x * y
    "div",        # div: x / y
    "sqrt",       # sqrt: sqrt(x)
]
\end{pysettingsbox}
All other configurations of \sisso$++$ are chosen same as in the case of the Feynman equation experiments; see \hyperref[subsec:SISSO++-Feynman-settings]{Appendix \ref{subsec:SISSO++-Feynman-settings}}.

\subsection{\texorpdfstring{\gplearn}{gplearn}}

\label{subsec:gplearn-experimental-settings-perovskite}
As in the case of the Feynman equation experiments (see \hyperref[subsec:gplearn-Feynman-settings]{Appendix \ref{subsec:gplearn-Feynman-settings}}), we use the same implementation of \gplearn~\citep{stephens2016gplearn} for discovering symbolic descriptors of catalytic activity from the oxide
perovskite dataset of~\cite{Weng2020SimpleDescriptor} using the experimental protocol described in \hyperref[sec:oxide-perovskite-data-description]{Appendix \ref{sec:oxide-perovskite-data-description}}. Following~\cite{Weng2020SimpleDescriptor}, the operator set for all 5 splits is configured as follows.
\begin{pysettingsbox}{\gplearn\ operator set}
function_set = [
    "add",   # (x, y) -> x + y
    "sub",   # (x, y) -> x - y
    "mul",   # (x, y) -> x * y
    "div",   # protected division
    "sqrt",  # protected square root
]
\end{pysettingsbox}
The protected implementations of $\mathrm{div}$ and $\mathrm{sqrt}$ are adopted to avoid numerical instabilities during symbolic evolution, following the default \texttt{gplearn} implementation. We maintain similar \gplearn\ settings as adopted in~\cite{Weng2020SimpleDescriptor} across all $5$ splits of the dataset.

\begin{pysettingsbox}{\gplearn\ settings}
# The settings closely follow those adopted in 
# Weng et al. (2020, Nature Communications)

gplearn_settings = {
    "population_size": 5000,          # symbolic programs per generation
    "generations": 20,                # maximum number of generations
    "stopping_criteria": 0.01,        # early stopping threshold for MAE
    "p_crossover": 0.5,               # subtree crossover probability
    "p_subtree_mutation": 1 / 6,      # subtree mutation probability
    "p_hoist_mutation": 1 / 6,        # hoist mutation probability
    "p_point_mutation": 1 / 6,        # point mutation probability
    "max_samples": 0.95,              # fraction of training samples per program
    "tournament_size": 20,            # tournament selection size
    "parsimony_coefficient": 5e-4,    # complexity penalty
    "metric": "mean absolute error",  # fitness criterion, MAE
    "constant_range": (-1, 1),        # range of ephemeral constants
}
\end{pysettingsbox}

The final recovered symbolic expression is extracted from the fitted estimator through

\begin{pysettingsbox}{Extracting the final \gplearn\ expression}
model._program
\end{pysettingsbox}

which stores the best symbolic program obtained after the final generation.


\subsection{\texorpdfstring{\operon}{operon}}
\label{subsec:operon-experimental-settings-perovskite}

As in the case of the Feynman equation experiments (see \hyperref[subsec:operon-Feynman-settings]{Appendix \ref{subsec:operon-Feynman-settings}}), we use the same implementation of \operon~\citep{operon} for discovering symbolic descriptors of catalytic activity from the oxide
perovskite dataset of~\cite{Weng2020SimpleDescriptor} using the experimental protocol described in \hyperref[sec:oxide-perovskite-data-description]{Appendix \ref{sec:oxide-perovskite-data-description}}. Following~\cite{Weng2020SimpleDescriptor}, the primitive symbol set for all 5 splits is configured as follows.

\begin{pysettingsbox}{\operon\ operator set}
allowed_symbols = [
    "add",       # (x, y) -> x + y
    "sub",       # (x, y) -> x - y
    "mul",       # (x, y) -> x*y
    "div",       # (x, y) -> x/y
    "sqrt",      # x -> sqrt(x)
    "constant",  # numerical-constant leaf nodes
    "variable",  # input-variable leaf nodes
]
\end{pysettingsbox}

All other configurations of \operon\ are chosen same as in the case of the Feynman equation experiments; see \hyperref[subsec:operon-Feynman-settings]{Appendix \ref{subsec:operon-Feynman-settings}}.


\subsection{\texorpdfstring{\pysr}{PySR}}
\label{subsec:pysr-experimental-settings-perovskite}

As in the case of the Feynman equation experiments (see \hyperref[subsec:PySR-Feynman-settings]{Appendix \ref{subsec:PySR-Feynman-settings}}), we use the same implementation of \pysr~\citep{pysr} for discovering symbolic descriptors of catalytic activity from the oxide
perovskite dataset of~\cite{Weng2020SimpleDescriptor} using the experimental protocol described in \hyperref[sec:oxide-perovskite-data-description]{Appendix \ref{sec:oxide-perovskite-data-description}}. Following~\cite{Weng2020SimpleDescriptor}, the operator set for all 5 splits is configured as follows.

\begin{pysettingsbox}{\pysr\ operator set}
binary_operators = [
    "+",       # (x, y) -> x + y
    "-",       # (x, y) -> x - y
    "*",       # (x, y) -> x*y
    "/",       # (x, y) -> x/y
]

unary_operators = [
    "sqrt",    # x -> sqrt(x)
]
\end{pysettingsbox}

For all $5$ splits of the dataset, \pysr\ is run with the following common settings.

\begin{pysettingsbox}{\pysr\ settings}
pysr_settings = {
    "niterations": 500,    # number of evolutionary search iterations
    "populations": 31,     # number of populations evolved during the search
    "population_size": 27, # number of candidate expressions in each population
    "maxsize": 20,         # maximum permitted expression tree size
    "elementwise_loss": "loss(x, y) = (x - y)^2", # squared-error loss
    "model_selection": "best",  # rule used to select the final expression
    "deterministic": True,
    "parallelism": "serial",
}
\end{pysettingsbox}

All other configurations of \pysr\ are chosen same as in the case of the Feynman equation experiments; see \hyperref[subsec:PySR-Feynman-settings]{Appendix \ref{subsec:PySR-Feynman-settings}}.


\subsection{\texorpdfstring{\bms}{BMS}}
\label{subsec:bms-experimental-settings-perovskite}

As in the case of the Feynman equation experiments (see \hyperref[subsec:BMS-Feynman-settings]{Appendix \ref{subsec:BMS-Feynman-settings}}), we use the same implementation of Bayesian Machine Scientist (\bms)~\citep{BMS} for discovering symbolic descriptors of catalytic activity from the oxide
perovskite dataset of~\cite{Weng2020SimpleDescriptor} using the experimental protocol described in \hyperref[sec:oxide-perovskite-data-description]{Appendix \ref{sec:oxide-perovskite-data-description}}. Following~\cite{Weng2020SimpleDescriptor}, the operator set for all 5 splits is configured as follows.

\begin{pysettingsbox}{\bms\ operator dictionary and prior}
bms_ops = {
    "+": 2,    # add(x, y) -> x + y
    "-": 1,    # sub(x, y) -> x - y
    "*": 2,    # mul(x, y) -> x * y
    "/": 2,    # div(x, y) -> x / y
    "psqrt": 1,  # sqrt(x)
}

bms_prior = {
    "Nopi_+": 1.0,
    "Nopi_-": 1.0,
    "Nopi_*": 1.0,
    "Nopi_/": 1.0,
    "Nopi_psqrt": 1.0,
}
\end{pysettingsbox}

All other configurations of \bms\ are chosen same as in the case of the Feynman equation experiments; see \hyperref[subsec:BMS-Feynman-settings]{Appendix \ref{subsec:BMS-Feynman-settings}}.


\subsection{\texorpdfstring{\bsr}{BSR}}
\label{subsec:bsr-experimental-settings-perovskite}

As in the case of the Feynman equation experiments (see \hyperref[subsec:BSR-Feynman-settings]{Appendix \ref{subsec:BSR-Feynman-settings}}), we use the same implementation of Bayesian Symbolic Regression (\bsr)~\citep{BSR} for discovering symbolic descriptors of catalytic activity from the oxide
perovskite dataset of~\cite{Weng2020SimpleDescriptor} using the experimental protocol described in \hyperref[sec:oxide-perovskite-data-description]{Appendix \ref{sec:oxide-perovskite-data-description}}. Following~\cite{Weng2020SimpleDescriptor}, the operator set for all 5 splits is configured as follows.

\begin{pysettingsbox}{\bsr\ operator set}
bsr_operator_set = [
    "add",      # (x, y) -> x + y
    "sub",      # (x, y) -> x - y
    "mul",      # (x, y) -> x*y
    "inv",      # x -> 1/x
    "psqrt",    # x -> sqrt(abs(x))
]
\end{pysettingsbox}

The aforementioned operator set is converted into the following prior dictionary, with equal prior weights assigned to all included operators.

\begin{pysettingsbox}{\bsr\ operator prior dictionary}
bsr_operator_prior = {
    "+": 1.0,
    "-": 1.0,
    "*": 1.0,
    "inv": 1.0,
    "psqrt": 1.0,
}
\end{pysettingsbox}

For each train/test split, the $5$ restarts employ $K=8$ symbolic trees similar to the \hierbosss\ configuration described in~\hyperref[subsec:bayesymx-experimental-settings-perovskite]{Appendix \ref{subsec:bayesymx-experimental-settings-perovskite}}. All other configurations of \bsr\ are chosen same as in the case of the Feynman equation experiments; see \hyperref[subsec:BSR-Feynman-settings]{Appendix \ref{subsec:BSR-Feynman-settings}}.
%
\newpage
\section{Ablation Study for \texorpdfstring{\hierbosss}{HierBOSSS}: Practical Guideline for Choosing the Nominal Symbolic Forest Size}
\label{sec:perovskite-ablation}

We conduct an ablation study: (i) to assess the sensitivity of \hierbosss\ to the nominal number of symbolic trees $K$ in the symbolic forest component $\mathcal T$ and (ii) as a practical guideline for selecting an appropriate nominal symbolic forest size that balances predictive performance with computational efficiency.

Following the experimental settings of \hierbosss\ described in~\hyperref[subsec:bayesymx-experimental-settings-perovskite]{Appendix~\ref{subsec:bayesymx-experimental-settings-perovskite}}, we consider symbolic forest sizes $K\in \{6, 7, 8, 10, 12\}$ and perform $5$ independent \(90/10\) train/test splits of the oxide perovskite dataset. We report the predictive performance (mean test \texttt{RMSE} and \texttt{MAE} across $5$ splits computed on the $10\%$ held-out test set) of both the \rawtag\ and \finaltag\ expressions of \hierbosss\ together with the mean wall-clock runtime (over $5$ splits) of the $5$ parallel \texttt{MCMC} chains.

\begin{figure}[!htp]
\centering
\includegraphics[width=\textwidth]{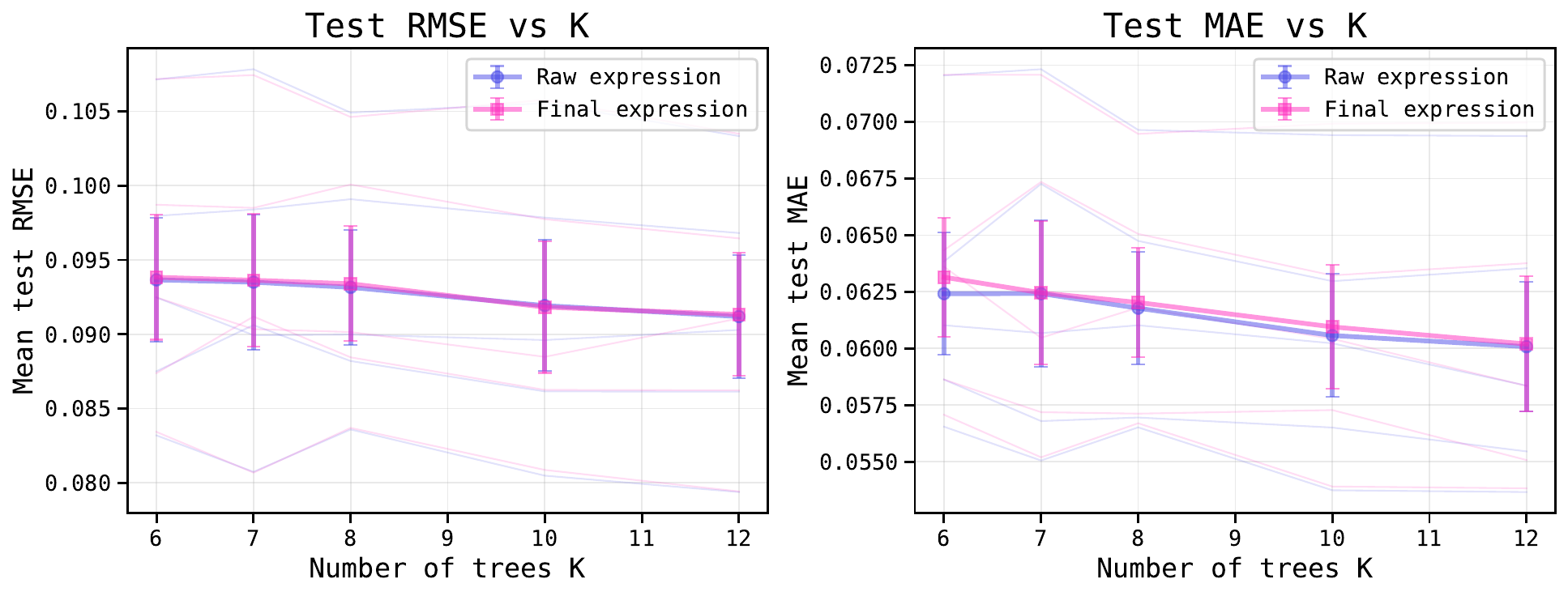}
\caption{Average (over $5$ splits) test \texttt{RMSE} and test \texttt{MAE} of \hierbosss\ across different values of $K$ for the oxide perovskite dataset.}
\label{fig:perovskite-k-ablation-performance}
\end{figure}

\hyperref[fig:perovskite-k-ablation-performance]{Figure~\ref{fig:perovskite-k-ablation-performance}} summarizes the predictive performance of \hierbosss\ across different symbolic forest sizes. The average test \texttt{RMSE} and test \texttt{MAE} remain stable over the considered values of \(K\), with only marginal improvements as the symbolic forest becomes larger. Furthermore, the predictive performance of the \rawtag\ and \finaltag\ expressions is nearly indistinguishable, indicating that the post-\texttt{MCMC} symbolic model refinement step in \hierbosss\ successfully removes redundant symbolic components while preserving predictive accuracy.

\begin{figure}[H]
\centering
\includegraphics[width=0.60\textwidth]{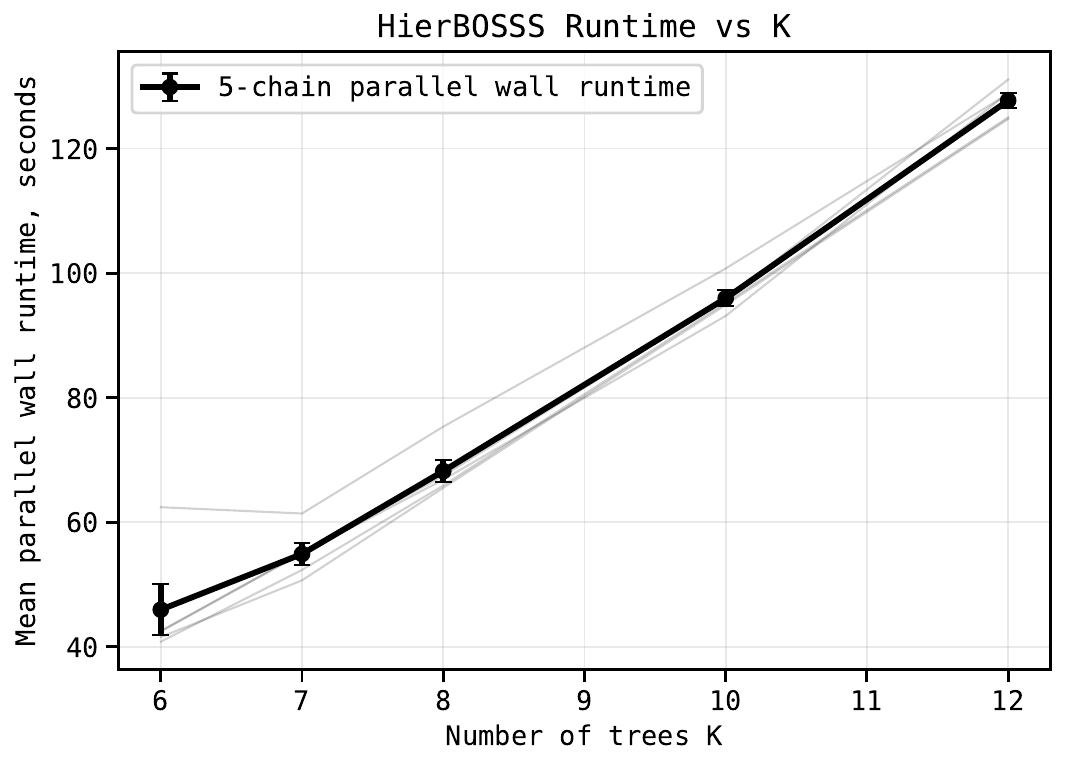}
\caption{Runtime behavior of \hierbosss\ across different values of $K$ for the oxide perovskite dataset. The wall-clock runtime for $5$ parallel chains across $5$ splits has been reported, with the mean trend in black and split-level variability shown in gray.}
\label{fig:perovskite-k-ablation-runtime}
\end{figure}

The computational cost associated with increasing the symbolic forest size is shown in \hyperref[fig:perovskite-k-ablation-runtime]{Figure~\ref{fig:perovskite-k-ablation-runtime}}. As expected, the mean parallel wall-clock runtime of running $5$ parallel \texttt{MCMC} chains of \hierbosss\ increases approximately monotonically with \(K\), reflecting exploration of the larger symbolic search space. Although larger symbolic forests usually increase computational cost, they provide little additional predictive benefit once moderate values of \(K\) are reached.

Taken together, the preceding ablation results suggest a practical strategy for choosing the nominal symbolic forest size: select the smallest moderate value of \(K\) after which predictive accuracy stabilizes while avoiding unnecessarily large symbolic forests that incur additional computational cost. In the oxide perovskite data study, the predictive metrics stabilize around \(K\approx8\), whereas the runtime continues to increase steadily for larger values of \(K\). Consequently, a moderate symbolic forest provides sufficient representational flexibility for discovering informative symbolic descriptors, while the post-\texttt{MCMC} symbolic model refinement in \hyperref[alg:post-mcmc-symbolic-model-refinement-single]{Algorithm \ref{alg:post-mcmc-symbolic-model-refinement-single}} data-adaptively removes redundant symbolic trees to produce a parsimonious final symbolic model and determines the effective forest size $K_{\mathrm{eff}}$. Guided by this trade-off, we set $K=8$ for the reported \hierbosss\ analysis of the oxide perovskite dataset.

\newpage
\section{Descriptors Learned by \texorpdfstring{\hierbosss}{HierBOSSS} and Competitors for the Oxide Perovskite Catalyst Data Study}
\label{sec:symbolic-expressions-perovskites}


\begingroup
\tiny
\setlength{\tabcolsep}{2.8pt}
\renewcommand{\arraystretch}{1.20}

\begin{longtable}{@{}
>{\centering\arraybackslash}p{0.10\textwidth}
>{\raggedright\arraybackslash}p{0.56\textwidth}
>{\centering\arraybackslash}p{0.085\textwidth}
>{\centering\arraybackslash}p{0.085\textwidth}
>{\centering\arraybackslash}p{0.085\textwidth}
@{}}
\caption{Results for split 1 of the oxide perovskite dataset. Test \texttt{RMSE} and \texttt{MAE} are computed on the corresponding $10\%$ held-out test set. Descriptors highlighted in \textcolor{BrickRed}{\textbf{---}} were identified as particularly important for \texttt{OER} activity~\citep[Table 2]{Weng2020SimpleDescriptor}. For \bms, $a_0$ denotes the fitted constant. For \hierbosss, the effective symbolic forest size is denoted by \textcolor{RoyalBlue}{$K_{\mathrm{eff}}$}.}
\label{tab:perovskite-results-split-1}\\

\toprule
\toprule
Method
& Learned descriptor expression
& Test \texttt{RMSE} (eV)
& Test \texttt{MAE} (eV)
& Runtime (s) \\
\midrule
\endfirsthead

\caption[]{\emph{(Continued)}. Results for split 1 of the oxide perovskite dataset.}\\
\toprule
\toprule
Method
& Learned descriptor expression
& Test \texttt{RMSE} (eV)
& Test \texttt{MAE} (eV)
& Runtime (s) \\
\midrule
\endhead

\midrule
\multicolumn{5}{r}{\emph{Continued on next page}}\\
\endfoot

\bottomrule
\bottomrule
\endlastfoot

\rowcolor{ExprBack}\hierbosss
& \exprcell{\exprbox{\finaltag}{
\begin{aligned}[t]
& \frac{0.429}{\textcolor{BrickRed}{\mu}}
-0.004\textcolor{blue}{\rho}^{2}
-\frac{0.811}{\textcolor{BrickRed}{\sqrt{\chi_A}}-\sqrt{\textcolor{blue}{\rho}}}
-0.454\sqrt{\textcolor{blue}{\rho}}
+0.006N_d\textcolor{blue}{\rho}
+{0.339\textcolor{blue}{\rho}}\textcolor{BrickRed}{\frac{1}{\chi_B}}
+\frac{0.206\textcolor{BrickRed}{\mu}}{\textcolor{blue}{\rho}}
\end{aligned}
}}
& 0.084
& 0.058
& 89.277 \\

& \exprcell{\exprbox{\textcolor{RoyalBlue}{$K_{\mathrm{eff}}$}}{
7
}}
&
&
& \\

\reprowsep

\dsr
& \exprcell{
\displaystyle
\sqrt{
\textcolor{BrickRed}{\mu}+
\sqrt{
\frac{
\textcolor{BrickRed}{Q_A}
\bigl(\textcolor{BrickRed}{Q_A}+\sqrt{\textcolor{blue}{\rho}}\bigr)
}{
\textcolor{BrickRed}{\chi_A\chi_B}
}
}
}
}
& 0.133
& 0.082
& 32.031 \\

\reprowsep

\qlattice
& \exprcell{
\displaystyle
\begin{aligned}[t]
&0.669\,
(0.074\,\textcolor{blue}{\rho}+0.359)
\Big(
0.066\,N_d-1.018
+\frac{1}{1.196\,\textcolor{BrickRed}{\chi_B}-1.121}
\Big) +1.569
\end{aligned}
}
& 0.084
& 0.061
& 84.321 \\

\reprowsep

\sisso$++$
& \exprcell{
\displaystyle
\begin{aligned}[t]
&1.237
-0.075
\Big[
\textcolor{blue}{\rho}\,\textcolor{BrickRed}{t}
-(\textcolor{blue}{\rho}-\textcolor{BrickRed}{\chi_A})
\Big]
+1.507
\left(
\textcolor{BrickRed}{\frac{\mu}{t}}
-\frac{\textcolor{BrickRed}{\mu}}{N_d}
\right)
+0.006
\frac{\textcolor{blue}{\rho}\,\textcolor{BrickRed}{\chi_A}}{\textcolor{BrickRed}{\chi_B}-R_A}
\end{aligned}
}
& 0.083
& 0.058
& 3.971 \\

\reprowsep

\gplearn
& \exprcell{
\displaystyle
\sqrt{
\sqrt{\textcolor{BrickRed}{Q_A}}\,
\sqrt{
\sqrt{
N_d+\frac{\textcolor{blue}{\rho}}{\textcolor{BrickRed}{\chi_A}^{3}}
}
}
}
}
& 0.140
& 0.074
& 43.128 \\

\reprowsep

\operon
& \exprcell{
\displaystyle
\begin{aligned}[t]
&1.736
+0.007
\left[
\frac{1.670\,\textcolor{BrickRed}{Q_A}}
     {0.382\,\textcolor{blue}{\rho}\textcolor{BrickRed}{\mu}
      -1.097\,\textcolor{BrickRed}{\chi_B}}
+
\frac{\textcolor{RoyalBlue}{\mathcal A_1}}{1.772\,\textcolor{BrickRed}{\chi_B}}
\right],\\[2pt]
\textcolor{RoyalBlue}{\mathcal A_1}
&=
\frac{
\dfrac{1.134\,\textcolor{BrickRed}{\chi_B}}
      {1.687\,\textcolor{BrickRed}{\mu}+1.921}
-
0.200\,\textcolor{BrickRed}{t}\textcolor{BrickRed}{Q_A}
}{0.007}
+
\frac{1.213\,\textcolor{blue}{\rho} N_d}
     {1.221-1.097\,\textcolor{BrickRed}{\chi_B}}
+
\frac{\textcolor{RoyalBlue}{\mathcal B_1}}
     {\sqrt{1.558}-1.097\,\textcolor{BrickRed}{\chi_B}},\\[2pt]
\textcolor{RoyalBlue}{\mathcal B_1}
&=
(0.382\,\textcolor{blue}{\rho}\textcolor{BrickRed}{\chi_A})^2
(0.790\,\textcolor{BrickRed}{Q_A}
 +1.097\,\textcolor{BrickRed}{\chi_B})
(0.166\,\textcolor{BrickRed}{\chi_A})
+1.389\,\textcolor{blue}{\rho}
\end{aligned}
}
& 0.082
& 0.057
& 27.062 \\

\reprowsep

\pysr
& \exprcell{
\displaystyle
0.015\,\textcolor{blue}{\rho}
+
4.011\,\textcolor{BrickRed}{\mu}
}
& 0.111
& 0.060
& 156.604 \\

\reprowsep

\bms
& \exprcell{
\displaystyle
\begin{aligned}[t]
&
\textcolor{BrickRed}{\mu}
\sqrt{\left|
a_0+
\textcolor{BrickRed}{\mu}
\left[
\frac{\textcolor{BrickRed}{\mu}}
{\textcolor{blue}{\rho}\left(\sqrt{|\textcolor{RoyalBlue}{\mathcal A_1}|}+\textcolor{RoyalBlue}{\mathcal B_1}\right)}
+
\frac{\textcolor{blue}{\rho}}{\textcolor{BrickRed}{\chi_A}\sqrt{|\textcolor{BrickRed}{\chi_B}|}}
\right]
\right|},\\[2pt]
\textcolor{RoyalBlue}{\mathcal A_1}
&=
\frac{\textcolor{BrickRed}{t}}
{N_d(-\textcolor{blue}{\rho})}\,\textcolor{BrickRed}{\chi_B},
\qquad
\textcolor{RoyalBlue}{\mathcal B_1}
=
\left(\textcolor{blue}{\rho}-R_A N_d\right)
\frac{
(\textcolor{BrickRed}{Q_A}+\textcolor{BrickRed}{\chi_A})/(-\textcolor{blue}{\rho})
}{
\textcolor{blue}{\rho}+\sqrt{|\textcolor{BrickRed}{\chi_B}|}+\textcolor{BrickRed}{\mu}
}
\end{aligned}
}
& 0.077
& 0.056
& 352.474 \\

\reprowsep

\bsr
& \exprcell{
\displaystyle
\begin{aligned}[t]
&-0.684
-0.497(\textcolor{blue}{\rho}-\textcolor{BrickRed}{Q_A})
+1.713R_A N_d-0.708\left(\textcolor{BrickRed}{\chi_B}+\frac{1}{N_d}\right)\\
&-0.496\left\{\frac{1}{1/(\textcolor{blue}{\rho}-\textcolor{BrickRed}{Q_A})}\right\}
-0.049(\textcolor{BrickRed}{Q_A}
+\textcolor{BrickRed}{\mu}-\textcolor{BrickRed}{\mu})
+\frac{1.563}{\textcolor{BrickRed}{\chi_A}}\\
&-0.714\left(
\textcolor{BrickRed}{Q_A}
+R_A^2N_d-\textcolor{BrickRed}{\mu}
\right)+1.008(\textcolor{blue}{\rho}-N_d-R_A)
\end{aligned}
}
& 0.104
& 0.062
& 525.408 \\
\end{longtable}
\endgroup

\begingroup
\tiny
\setlength{\tabcolsep}{2.8pt}
\renewcommand{\arraystretch}{1.20}

\begin{longtable}{@{}
>{\centering\arraybackslash}p{0.10\textwidth}
>{\raggedright\arraybackslash}p{0.56\textwidth}
>{\centering\arraybackslash}p{0.085\textwidth}
>{\centering\arraybackslash}p{0.085\textwidth}
>{\centering\arraybackslash}p{0.085\textwidth}
@{}}
\caption{Results for split 2 of the oxide perovskite dataset. Test \texttt{RMSE} and \texttt{MAE} are computed on the corresponding $10\%$ held-out test set. Descriptors highlighted in \textcolor{BrickRed}{\textbf{---}} were identified as particularly important for \texttt{OER} activity~\citep[Table 2]{Weng2020SimpleDescriptor}. For \bms, $a_0$ denotes the fitted constant. For \hierbosss, the effective symbolic forest size is denoted by \textcolor{RoyalBlue}{$K_{\mathrm{eff}}$}.}
\label{tab:perovskite-results-split-2}\\

\toprule
\toprule
Method
& Learned descriptor expression
& Test \texttt{RMSE} (eV)
& Test \texttt{MAE} (eV)
& Runtime (s) \\
\midrule
\endfirsthead

\caption[]{\emph{(Continued)}. Results for split 2 of the oxide perovskite dataset.}\\
\toprule
\toprule
Method
& Learned descriptor expression
& Test \texttt{RMSE} (eV)
& Test \texttt{MAE} (eV)
& Runtime (s) \\
\midrule
\endhead

\midrule
\multicolumn{5}{r}{\emph{Continued on next page}}\\
\endfoot

\bottomrule
\bottomrule
\endlastfoot

\rowcolor{ExprBack}\hierbosss
& \exprcell{\exprbox{\finaltag}{
\begin{aligned}[t]
& 1.158
+0.292\textcolor{BrickRed}{\sqrt{Q_A}}
+\frac{0.530\textcolor{blue}{\rho}}{N_d}
+0.095\sqrt{N_d}\textcolor{blue}{\rho}
-0.068\textcolor{BrickRed}{\chi_B}\textcolor{blue}{\rho}
-0.073R_A\textcolor{blue}{\rho}
-0.073\textcolor{BrickRed}{\chi_A}\textcolor{blue}{\rho}
\end{aligned}
}}
& 0.087
& 0.064
& 66.506 \\

& \exprcell{\exprbox{\textcolor{RoyalBlue}{$K_{\mathrm{eff}}$}}{
5
}}
&
&
& \\

\reprowsep

\dsr
& \exprcell{
\displaystyle
\sqrt{
\textcolor{BrickRed}{\mu}
\left[
\sqrt{R_A+\textcolor{blue}{\rho}}
+
\textcolor{BrickRed}{\mu}
\sqrt{R_A+\sqrt{\textcolor{BrickRed}{\mu}}}
+
R_A
\right]
+
\textcolor{BrickRed}{\chi_A}
}
}
& 0.128
& 0.087
& 31.175 \\

\reprowsep

\qlattice
& \exprcell{
\displaystyle
\begin{aligned}[t]
&0.030\,\textcolor{blue}{\rho}
-0.360\,
(1.365\,\textcolor{BrickRed}{\chi_B}-2.858)
(7.387\,\textcolor{BrickRed}{\chi_B}-11.859)
(-0.076\,\textcolor{blue}\rho+16.542\,\textcolor{BrickRed}{\mu}\\
&-6.813)+1.659
\end{aligned}
}
& 0.092
& 0.067
& 79.871 \\

\reprowsep

\sisso$++$
& \exprcell{
\displaystyle
\begin{aligned}[t]
&1.429
+0.003
(\textcolor{blue}{\rho}+N_d)(N_d-\textcolor{BrickRed}{Q_A})
+0.067
\frac{\textcolor{blue}{\rho}+\textcolor{BrickRed}{\chi_B}}{\textcolor{BrickRed}{\chi_B}-\textcolor{BrickRed}{\chi_A}}
-0.034
\frac{\textcolor{blue}{\rho}\,\textcolor{BrickRed}{\chi_B}}{\textcolor{BrickRed}{\chi_B}-\textcolor{BrickRed}{\chi_A}}
\end{aligned}
}
& 0.090
& 0.067
& 4.989 \\

\reprowsep

\gplearn
& \exprcell{
\displaystyle
\sqrt{
\sqrt{
\sqrt{
\textcolor{BrickRed}{Q_A}+\textcolor{blue}{\rho}
}
}
+
\frac{\textcolor{BrickRed}{\chi_A}}{0.803}
}
}
& 0.125
& 0.082
& 41.212 \\

\reprowsep

\operon
& \exprcell{
\displaystyle
\begin{aligned}[t]
&
1.485+0.001\,\textcolor{RoyalBlue}{\mathcal A_2},\\[2pt]
\textcolor{RoyalBlue}{\mathcal A_2}
&=
-14.338\,\textcolor{BrickRed}{t}N_d
+7.516\,\textcolor{BrickRed}{Q_A}
+1.107\,\textcolor{blue}{\rho}
+8.138\,\textcolor{BrickRed}{Q_A}N_d
+\sqrt{4.453\,\textcolor{blue}{\rho}}\,(0.923\,N_d)\\
&\;
+
\frac{2.718\,N_d+2.223\,\textcolor{blue}{\rho}}
     {(0.274\,\textcolor{BrickRed}{\chi_B})^4(0.923\,\textcolor{BrickRed}{\chi_B})}
\\
&-
\frac{1.414\,\textcolor{BrickRed}{Q_A}}
     {(0.274\,R_A)^3} \left(\frac{-1.155\,\textcolor{BrickRed}{\mu}
      (2.718\,\textcolor{BrickRed}{Q_A}+0.854\,\textcolor{blue}{\rho})}
     {0.923\,\textcolor{BrickRed}{\chi_B}}
+1.618\,N_d \right)^{-1}
\end{aligned}
}
& 0.088
& 0.063
& 29.585 \\

\reprowsep

\pysr
& \exprcell{
\displaystyle
0.016\,\textcolor{blue}{\rho}
+
3.998\,\textcolor{BrickRed}{\mu}
}
& 0.109
& 0.075
& 157.797 \\

\reprowsep

\bms
& \exprcell{
\displaystyle
\begin{aligned}[t]
&
-a_0\textcolor{BrickRed}{\mu}
\sqrt{\left|
\sqrt{\left|
\sqrt{\left|\frac{
R_A\sqrt{|a_0|}\,\textcolor{BrickRed}{\chi_B}
}{
\textcolor{BrickRed}{\mu}
}
+
\frac{
(\textcolor{blue}{\rho}+a_0)
}{
(\textcolor{BrickRed}{\chi_B}-a_0\textcolor{BrickRed}{\chi_A})R_A
} \frac{
\dfrac{\textcolor{blue}{\rho}}{N_d(\textcolor{BrickRed}{\chi_B}+\textcolor{BrickRed}{\mu})}
}{
(\textcolor{BrickRed}{\chi_B}+a_0)R_A+2\textcolor{BrickRed}{t}
}
+\textcolor{blue}{\rho}\right|}
\right|}
\right|}
\end{aligned}
}
& 0.092
& 0.068
& 429.338 \\

\reprowsep

\bsr
& \exprcell{
\displaystyle
\begin{aligned}[t]
&-1.913
+0.023\,\textcolor{blue}{\rho}\textcolor{BrickRed}{\mu}
+2.450\left(\frac{1}{\textcolor{BrickRed}{\chi_A}}+\textcolor{BrickRed}{\chi_A}\right)-0.163\left(
\textcolor{BrickRed}{\mu}N_d+\textcolor{BrickRed}{\chi_B}+R_A
\right)\\
&-0.627\left(
\frac{1}{R_A+N_d}+\textcolor{BrickRed}{\chi_A}
\right)-0.001\left[
\frac{1}{\textcolor{blue}{\rho}}+\textcolor{BrickRed}{\chi_A}
+\frac{1}{R_A}
\left(
\textcolor{BrickRed}{t}
-\frac{1}{\textcolor{BrickRed}{t}}
+\textcolor{BrickRed}{\chi_B}
\right)
\right]\\
&+0.281(N_d-\textcolor{BrickRed}{\mu})
-0.109\,\textcolor{BrickRed}{t}^{\,2}\textcolor{BrickRed}{\chi_B}N_d+0.002\,\textcolor{BrickRed}{Q_A}
(\textcolor{BrickRed}{\mu}+\textcolor{blue}{\rho})
\end{aligned}
}
& 0.102
& 0.072
& 510.201 \\
\end{longtable}
\endgroup


\begingroup
\tiny
\setlength{\tabcolsep}{2.8pt}
\renewcommand{\arraystretch}{1.20}

\begin{longtable}{@{}
>{\centering\arraybackslash}p{0.10\textwidth}
>{\raggedright\arraybackslash}p{0.56\textwidth}
>{\centering\arraybackslash}p{0.085\textwidth}
>{\centering\arraybackslash}p{0.085\textwidth}
>{\centering\arraybackslash}p{0.085\textwidth}
@{}}
\caption{Results for split 3 of the oxide perovskite dataset. Test \texttt{RMSE} and \texttt{MAE} are computed on the corresponding $10\%$ held-out test set. Descriptors highlighted in \textcolor{BrickRed}{\textbf{---}} were identified as particularly important for \texttt{OER} activity~\citep[Table 2]{Weng2020SimpleDescriptor}. For \bms, $a_0$ denotes the fitted constant. For \hierbosss, the effective symbolic forest size is denoted by \textcolor{RoyalBlue}{$K_{\mathrm{eff}}$}.}
\label{tab:perovskite-results-split-3}\\

\toprule
\toprule
Method
& Learned descriptor expression
& Test \texttt{RMSE} (eV)
& Test \texttt{MAE} (eV)
& Runtime (s) \\
\midrule
\endfirsthead

\caption[]{\emph{(Continued)}. Results for split 3 of the oxide perovskite dataset.}\\
\toprule
\toprule
Method
& Learned descriptor expression
& Test \texttt{RMSE} (eV)
& Test \texttt{MAE} (eV)
& Runtime (s) \\
\midrule
\endhead

\midrule
\multicolumn{5}{r}{\emph{Continued on next page}}\\
\endfoot

\bottomrule
\bottomrule
\endlastfoot

\rowcolor{ExprBack}\hierbosss
& \exprcell{\exprbox{\finaltag}{
\begin{aligned}[t]
& 2.103\textcolor{BrickRed}{\sqrt{\mu}}
+0.282\textcolor{BrickRed}{\chi_A}
+{0.254\textcolor{blue}{\rho}}\textcolor{BrickRed}{\frac{1}{\chi_B}}\textcolor{BrickRed}{\frac{1}{\chi_B}}
+0.005N_d\textcolor{blue}{\rho}
-0.214\textcolor{BrickRed}{\mu}\textcolor{blue}{\rho}
\end{aligned}
}}
& 0.094
& 0.060
& 69.089 \\

& \exprcell{\exprbox{\textcolor{RoyalBlue}{$K_{\mathrm{eff}}$}}{
5
}}
&
&
& \\

\reprowsep

\dsr
& \exprcell{
\displaystyle
\sqrt{
\sqrt{
\sqrt{
\textcolor{BrickRed}{t}+
\frac{
\textcolor{BrickRed}{t}
\textcolor{BrickRed}{Q_A}^{\,2}
\bigl(
1+\textcolor{blue}{\rho}+\textcolor{BrickRed}{Q_A}
+\textcolor{BrickRed}{\chi_B}+\textcolor{BrickRed}{t}
\bigr)
}{
\textcolor{BrickRed}{\chi_A}
}
}
}
}
}
& 0.139
& 0.082
& 33.378 \\

\reprowsep

\qlattice
& \exprcell{
\displaystyle
\begin{aligned}[t]
&-1.139\,\textcolor{BrickRed}{t}
+0.096\,
(17.657-8.738\,\textcolor{BrickRed}{\chi_B})
(0.073\,\textcolor{blue}{\rho}-0.031)\\
&\times
\Big[
1.051\,\textcolor{BrickRed}{\chi_B}
+(0.385\,N_d-2.099)
(0.823\,N_d-3.891)
-0.754
\Big]
+2.811
\end{aligned}
}
& 0.090
& 0.057
& 85.091 \\

\reprowsep

\sisso$++$
& \exprcell{
\displaystyle
\begin{aligned}[t]
&1.534
+0.205
\frac{\textcolor{blue}{\rho}/N_d}{\textcolor{BrickRed}{\chi_B}-\textcolor{BrickRed}{\chi_A}}
+0.025
(\textcolor{blue}{\rho}+\textcolor{BrickRed}{Q_A})
\frac{\textcolor{BrickRed}{Q_A}}{\textcolor{BrickRed}{\chi_B}}
-0.266
\frac{\textcolor{blue}{\rho}\,\textcolor{BrickRed}{Q_A}}
{\textcolor{BrickRed}{\chi_B} N_d}
\end{aligned}
}
& 0.095
& 0.060
& 6.363 \\

\reprowsep

\gplearn
& \exprcell{
\displaystyle
\sqrt{
\sqrt{
\sqrt{2\textcolor{blue}{\rho}}
+
2\textcolor{BrickRed}{Q_A}
+
0.523
}
}
}
& 0.150
& 0.081
& 42.081 \\

\reprowsep

\operon
& \exprcell{
\displaystyle
\begin{aligned}[t]
&
1.459+0.000\,\textcolor{RoyalBlue}{\mathcal A_3},\\[2pt]
\textcolor{RoyalBlue}{\mathcal A_3}
&=
\frac{
\dfrac{
\dfrac{
1.446\,\textcolor{blue}{\rho}+3.142\,\textcolor{BrickRed}{Q_A}
}{
0.102/\!\left(1.952\,\textcolor{BrickRed}{\mu}\right)
\big/\sqrt{0.536\,\textcolor{BrickRed}{\mu}}
}
}{
0.176\,\textcolor{BrickRed}{\chi_B}
}
}{
\dfrac{4.008}{
\dfrac{\sqrt{0.498\,\textcolor{BrickRed}{\mu}}}
      {0.571\,\textcolor{BrickRed}{\chi_B}(-0.131\,\textcolor{BrickRed}{\chi_B})}
+1.367\,\textcolor{BrickRed}{\mu}
}
}
\frac{1}{-0.070\,\textcolor{BrickRed}{\chi_B}}\\
&\quad
+
\frac{
\dfrac{0.975}
     {(0.497\,\textcolor{BrickRed}{\chi_B})(0.542\,\textcolor{BrickRed}{\chi_B})(-0.070\,\textcolor{BrickRed}{\chi_B})}
+3.337\,N_d
}{
0.102/(1.501\,\textcolor{BrickRed}{\chi_B})
}\\
&\quad
+
\frac{3.764\,\textcolor{BrickRed}{Q_A}}
{-0.063\,\textcolor{blue}{\rho}+
\dfrac{(-0.838\,\textcolor{BrickRed}{\chi_B})/(-0.084\,\textcolor{BrickRed}{\chi_B})}
      {0.651\,\textcolor{blue}{\rho}}}
\end{aligned}
}
& 0.095
& 0.062
& 28.378 \\

\reprowsep

\pysr
& \exprcell{
\displaystyle
0.016\,\textcolor{blue}{\rho}
+
4.001\,\textcolor{BrickRed}{\mu}
}
& 0.121
& 0.068
& 158.251 \\

\reprowsep

\bms
& \exprcell{
\displaystyle
\begin{aligned}[t]
&\frac{
\sqrt{|-\textcolor{BrickRed}{\mu}|}
}{
\sqrt{\left|
(a_0+R_A)
\left[
-\textcolor{BrickRed}{\chi_B}+\sqrt{|\textcolor{RoyalBlue}{\mathcal A_3}|}
\right]
\right|}
}
+\textcolor{BrickRed}{\mu}
+\sqrt{|a_0|},\\[2pt]
\textcolor{RoyalBlue}{\mathcal A_3}
&=
\sqrt{\left|
\sqrt{\left|
\sqrt{|\textcolor{blue}{\rho}|}
+
\left(
\textcolor{blue}{\rho}+\dfrac{\textcolor{BrickRed}{\chi_B}}{R_A/N_d}
\right)
\right|}
\right|} 
\left(\frac{\textcolor{BrickRed}{t}/\textcolor{BrickRed}{\mu}}
{\textcolor{blue}{\rho}+a_0}
+\textcolor{BrickRed}{t}
+
\frac{1}
{\textcolor{BrickRed}{\mu}\textcolor{blue}{\rho}+\textcolor{BrickRed}{\mu}} \right)^{-1}
\end{aligned}
}
& 0.096
& 0.065
& 504.364 \\

\reprowsep

\bsr
& \exprcell{
\displaystyle
\begin{aligned}[t]
&1.748
+0.132\,\textcolor{BrickRed}{Q_A}\textcolor{BrickRed}{\mu}+\frac{0.028}
{\textcolor{blue}{\rho}\textcolor{BrickRed}{t}-\textcolor{BrickRed}{\chi_B}}
+0.003(\textcolor{blue}{\rho}+\textcolor{BrickRed}{Q_A})N_d-\frac{0.042}{R_A}
-\frac{0.025}
{(1/\textcolor{BrickRed}{t})R_A}\\
&+\frac{0.0002}
{1/\textcolor{BrickRed}{t}-\textcolor{BrickRed}{\chi_A}}
-0.241\left(
R_A\textcolor{BrickRed}{\chi_B}+\textcolor{BrickRed}{\mu}-\textcolor{BrickRed}{\chi_A}
\right)
+\frac{1.179}{N_d}
\end{aligned}
}
& 0.124
& 0.074
& 509.069 \\
\end{longtable}
\endgroup


\begingroup
\tiny
\setlength{\tabcolsep}{2.8pt}
\renewcommand{\arraystretch}{1.20}

\begin{longtable}{@{}
>{\centering\arraybackslash}p{0.10\textwidth}
>{\raggedright\arraybackslash}p{0.56\textwidth}
>{\centering\arraybackslash}p{0.085\textwidth}
>{\centering\arraybackslash}p{0.085\textwidth}
>{\centering\arraybackslash}p{0.085\textwidth}
@{}}
\caption{Results for split 4 of the oxide perovskite dataset. Test \texttt{RMSE} and \texttt{MAE} are computed on the corresponding $10\%$ held-out test set. Descriptors highlighted in \textcolor{BrickRed}{\textbf{---}} were identified as particularly important for \texttt{OER} activity~\citep[Table 2]{Weng2020SimpleDescriptor}. For \bms, $a_0$ denotes the fitted constant. For \hierbosss, the effective symbolic forest size is denoted by \textcolor{RoyalBlue}{$K_{\mathrm{eff}}$}.}
\label{tab:perovskite-results-split-4}\\

\toprule
\toprule
Method
& Learned descriptor expression
& Test \texttt{RMSE} (eV)
& Test \texttt{MAE} (eV)
& Runtime (s) \\
\midrule
\endfirsthead

\caption[]{\emph{(Continued)}. Results for split 4 of the oxide perovskite dataset.}\\
\toprule
\toprule
Method
& Learned descriptor expression
& Test \texttt{RMSE} (eV)
& Test \texttt{MAE} (eV)
& Runtime (s) \\
\midrule
\endhead

\midrule
\multicolumn{5}{r}{\emph{Continued on next page}}\\
\endfoot

\bottomrule
\bottomrule
\endlastfoot

\rowcolor{ExprBack}\hierbosss
& \exprcell{\exprbox{\finaltag}{
\begin{aligned}[t]
& {0.720}\textcolor{BrickRed}{\frac{1}{t}}
+\frac{0.509}{\textcolor{BrickRed}{Q_A}}
+0.468\sqrt{\textcolor{BrickRed}{Q_A}+\textcolor{blue}{\rho}}
-0.169\sqrt{\textcolor{blue}{\rho}}
+{0.238\textcolor{blue}{\rho}}\textcolor{BrickRed}{\frac{1}{\chi_B}}\\
&
+0.035N_d\textcolor{blue}{\rho}
-0.148\sqrt{N_d}\textcolor{blue}{\rho}
\end{aligned}
}}
& 0.099
& 0.062
& 68.618 \\

& \exprcell{\exprbox{\textcolor{RoyalBlue}{$K_{\mathrm{eff}}$}}{
7
}}
&
&
& \\

\reprowsep

\dsr
& \exprcell{
\displaystyle
\sqrt{R_A}\,
\frac{
\textcolor{BrickRed}{\mu}
}{
\sqrt{\textcolor{BrickRed}{\mu}}
}
\left[
\sqrt{
\sqrt{
\sqrt{
\sqrt{\textcolor{BrickRed}{t}}-\textcolor{blue}{\rho}
}
}
}
+
\textcolor{BrickRed}{\chi_A}
\right]
}
& 0.142
& 0.079
& 35.367 \\

\reprowsep

\qlattice
& \exprcell{
\displaystyle
\begin{aligned}[t]
&3.233\,\textcolor{BrickRed}{\mu}
-0.078\,
(0.180\,\textcolor{blue}{\rho}-1.021)\\
&\times
\Big[
0.122\,N_d
+(2.519-0.328\,N_d)
(2.080\,\textcolor{BrickRed}{\chi_B}-3.561)
(0.119\,\textcolor{blue}{\rho}+0.938)-2.143
\Big]\\
&
+0.416
\end{aligned}
}
& 0.101
& 0.061
& 84.629 \\

\reprowsep

\sisso$++$
& \exprcell{
\displaystyle
\begin{aligned}[t]
&1.070
-0.088
\Big[
\textcolor{blue}{\rho}\,\textcolor{BrickRed}{t}
-(\textcolor{blue}{\rho}-\textcolor{BrickRed}{\chi_A})
\Big]
+2.061
\frac{\textcolor{BrickRed}{\mu}N_d}
{N_d+R_A}
+0.006
\frac{\textcolor{blue}{\rho}\,\textcolor{BrickRed}{\chi_A}}{\textcolor{BrickRed}{\chi_B}-R_A}
\end{aligned}
}
& 0.099
& 0.061
& 5.706 \\

\reprowsep

\gplearn
& \exprcell{
\displaystyle
\sqrt{
\sqrt{
\sqrt{
\textcolor{BrickRed}{\chi_B}-\textcolor{BrickRed}{Q_A}
-\frac{\textcolor{blue}{\rho}}{R_A}
}
+
\textcolor{BrickRed}{t}
}
}
+
\textcolor{BrickRed}{\mu}
}
& 0.143
& 0.075
& 43.619 \\

\reprowsep

\operon
& \exprcell{
\displaystyle
\begin{aligned}[t]
&
1.947+0.253\textcolor{RoyalBlue}{\mathcal A_4},\\[2pt]
\textcolor{RoyalBlue}{\mathcal A_4}
&=
\frac{
(0.219\,\textcolor{BrickRed}{Q_A}+0.179\,\textcolor{blue}{\rho})
(0.257\,N_d)
}{
(0.147\,\textcolor{BrickRed}{\chi_B}-0.345\,\textcolor{BrickRed}{\mu})
(0.318\,\textcolor{BrickRed}{\chi_B})
(5.830\,\textcolor{BrickRed}{\chi_B})
(0.434\,\textcolor{BrickRed}{\chi_B})
} + \frac{10.141}{3.138\,\textcolor{BrickRed}{\chi_B}},\\
&
-\frac{
\dfrac{-0.909\,\textcolor{BrickRed}{\mu}/0.743}
{-1.065+0.179\,\textcolor{blue}{\rho}-2.692\,\textcolor{BrickRed}{t}}
-0.693\,\textcolor{BrickRed}{Q_A}
}{
3.138\,\textcolor{BrickRed}{\chi_B}
}\\
&-
\frac{
-0.910\,\textcolor{BrickRed}{\mu}^{2}
}{
(1.901\,\textcolor{BrickRed}{\mu}
 +0.179\,\textcolor{blue}{\rho}
 -2.692\,\textcolor{BrickRed}{t})
(3.138\,\textcolor{BrickRed}{\chi_B})
}
\end{aligned}
}
& 0.099
& 0.065
& 26.507 \\

\reprowsep

\pysr
& \exprcell{
\displaystyle
0.016\,\textcolor{blue}{\rho}
+
4.001\,\textcolor{BrickRed}{\mu}
}
& 0.119
& 0.066
& 157.644 \\

\reprowsep

\bms
& \exprcell{
\displaystyle
\begin{aligned}[t]
&
-\textcolor{BrickRed}{\mu}
\left[
-\sqrt{|a_0|}
+
\frac{\textcolor{blue}{\rho}/\textcolor{RoyalBlue}{\mathcal A_4}}{-N_d}
+
\frac{\textcolor{BrickRed}{\mu}}{a_0}
\right],\\[2pt]
\textcolor{RoyalBlue}{\mathcal A_4}
&=
\frac{\textcolor{BrickRed}{\chi_B}}{\textcolor{BrickRed}{\chi_A}}
\sqrt{|\textcolor{BrickRed}{\chi_B}|} 
\sqrt{\left|
\sqrt{\left|
\sqrt{\left|
\sqrt{\left|
(-\textcolor{BrickRed}{\chi_A})
\frac{a_0}{\textcolor{blue}{\rho}}
\sqrt{|R_A|}
\right|}
\,
\frac{
\textcolor{BrickRed}{\chi_A}+\textcolor{BrickRed}{Q_A}
}{
R_A+\textcolor{BrickRed}{t}
}+\sqrt{|a_0\textcolor{blue}{\rho}\textcolor{BrickRed}{\chi_B}|}
\right|}
\right|}
\right|}
\end{aligned}
}
& 0.104
& 0.067
& 415.098 \\

\reprowsep

\bsr
& \exprcell{
\displaystyle
\begin{aligned}[t]
&1.788\times10^{-20}
+1.430\times10^{-11}(2\textcolor{blue}{\rho})+7.232\times10^{-15}
\textcolor{BrickRed}{\mu}
(\textcolor{BrickRed}{\mu}+R_A-R_A)\\
&-2.356\times10^{-14}(\textcolor{BrickRed}{\chi_B}-R_A)
-7.117\times10^{-12}(\textcolor{BrickRed}{\chi_A}-\textcolor{blue}{\rho})+\frac{1.230\times10^{-14}}{R_A}\\
&+\frac{1.788\times10^{-10}}
{(\textcolor{BrickRed}{\chi_B}-\textcolor{BrickRed}{\chi_B})N_d}-1.593\times10^{-14}
(\textcolor{BrickRed}{t}
+\textcolor{BrickRed}{t}
-\textcolor{BrickRed}{t})-\frac{2.263\times10^{-14}}
{\textcolor{BrickRed}{Q_A}\textcolor{BrickRed}{t}}
\end{aligned}
}
& 0.190
& 0.119
& 508.556 \\
\end{longtable}
\endgroup


\begingroup
\tiny
\setlength{\tabcolsep}{2.8pt}
\renewcommand{\arraystretch}{1.20}

\begin{longtable}{@{}
>{\centering\arraybackslash}p{0.10\textwidth}
>{\raggedright\arraybackslash}p{0.56\textwidth}
>{\centering\arraybackslash}p{0.085\textwidth}
>{\centering\arraybackslash}p{0.085\textwidth}
>{\centering\arraybackslash}p{0.085\textwidth}
@{}}
\caption{Results for split 5 of the oxide perovskite dataset. Test \texttt{RMSE} and \texttt{MAE} are computed on the corresponding $10\%$ held-out test set. Descriptors highlighted in \textcolor{BrickRed}{\textbf{---}} were identified as particularly important for \texttt{OER} activity~\citep[Table 2]{Weng2020SimpleDescriptor}. For \bms, $a_0$ denotes the fitted constant. For \hierbosss, the effective symbolic forest size is denoted by \textcolor{RoyalBlue}{$K_{\mathrm{eff}}$}.}
\label{tab:perovskite-results-split-5}\\

\toprule
\toprule
Method
& Learned descriptor expression
& Test \texttt{RMSE} (eV)
& Test \texttt{MAE} (eV)
& Runtime (s) \\
\midrule
\endfirsthead

\caption[]{\emph{(Continued)}. Results for split 5 of the oxide perovskite dataset.}\\
\toprule
\toprule
Method
& Learned descriptor expression
& Test \texttt{RMSE} (eV)
& Test \texttt{MAE} (eV)
& Runtime (s) \\
\midrule
\endhead

\midrule
\multicolumn{5}{r}{\emph{Continued on next page}}\\
\endfoot

\bottomrule
\bottomrule
\endlastfoot

\rowcolor{ExprBack}\hierbosss
& \exprcell{\exprbox{\finaltag}{
\begin{aligned}[t]
& 0.371\textcolor{BrickRed}{t}
+{1.288}\textcolor{BrickRed}{\frac{1}{t}}
+\frac{0.334\textcolor{blue}{\rho}}{N_d}
+0.014N_d\textcolor{blue}{\rho}
-0.069\textcolor{BrickRed}{\chi_B}\textcolor{blue}{\rho}
\end{aligned}
}}
& 0.108
& 0.072
& 94.344 \\

& \exprcell{\exprbox{\textcolor{RoyalBlue}{$K_{\mathrm{eff}}$}}{
5
}}
&
&
& \\

\reprowsep

\dsr
& \exprcell{
\displaystyle
\sqrt{
\textcolor{BrickRed}{\mu}+
\frac{
\sqrt{
\sqrt{\textcolor{BrickRed}{Q_A}}
+\textcolor{BrickRed}{Q_A}
+2R_A+\textcolor{BrickRed}{\chi_A}+\textcolor{blue}{\rho}
}
+
\sqrt{R_A}
}{
\textcolor{BrickRed}{\chi_B}
}
}
}
& 0.128
& 0.086
& 31.566 \\

\reprowsep

\qlattice
& \exprcell{
\displaystyle
\begin{aligned}[t]
&0.011\,\textcolor{blue}{\rho}
+3.347\,\textcolor{BrickRed}{\mu}
-0.083\,
(1.095-0.289\,\textcolor{blue}{\rho})
(19.800-18.681\,\textcolor{BrickRed}{t})\\
&\times
(22.870-12.463\,\textcolor{BrickRed}{\chi_B})
(23.077-23.015\,\textcolor{BrickRed}{t})+0.302
\end{aligned}
}
& 0.106
& 0.070
& 90.842 \\

\reprowsep

\sisso$++$
& \exprcell{
\displaystyle
\begin{aligned}[t]
&1.546
+0.001\,
\textcolor{blue}{\rho} N_d(N_d-\textcolor{BrickRed}{Q_A})
+0.067
\frac{\textcolor{blue}{\rho}+\textcolor{BrickRed}{\chi_A}}{\textcolor{BrickRed}{\chi_B}-\textcolor{BrickRed}{\chi_A}}
-0.034
\frac{\textcolor{blue}{\rho}\,\textcolor{BrickRed}{\chi_B}}{\textcolor{BrickRed}{\chi_B}-\textcolor{BrickRed}{\chi_A}}
\end{aligned}
}
& 0.108
& 0.072
& 5.764 \\

\reprowsep

\gplearn
& \exprcell{
\displaystyle
\sqrt{
\sqrt{
\sqrt{
\frac{\textcolor{blue}{\rho}}{\textcolor{BrickRed}{t}}
+
2\textcolor{BrickRed}{Q_A}
-
R_A
}
}
}
+
\textcolor{BrickRed}{\mu}
}
& 0.154
& 0.083
& 41.193 \\

\reprowsep

\operon
& \exprcell{
\displaystyle
\begin{aligned}[t]
&
1.639+0.001\,\textcolor{RoyalBlue}{\mathcal A_5},\\[2pt]
\textcolor{RoyalBlue}{\mathcal A_5}
&=
\frac{-0.579\,\textcolor{BrickRed}{\chi_B}+0.602\,N_d}
     {-0.435+0.853\,\textcolor{BrickRed}{\mu}}
-
\frac{
(0.825-0.175\,\textcolor{BrickRed}{Q_A})
}{
(-0.435)(-0.459)
(-0.435+0.853\,\textcolor{BrickRed}{\mu})
}\\
&\quad
-
\frac{
1.187\,\textcolor{blue}{\rho}
}{
(4.182\,\textcolor{BrickRed}{\chi_B})
(1.293\,\textcolor{BrickRed}{\mu})
(-0.435+0.872\,\textcolor{BrickRed}{\mu})
(-0.421\,\textcolor{BrickRed}{\chi_B})
}\\
&\quad
-
\frac{
0.602\,\textcolor{BrickRed}{t}
-0.175\,\textcolor{BrickRed}{Q_A}
}{
\dfrac{1.247\,\textcolor{blue}{\rho}}{4.970\,\textcolor{BrickRed}{\chi_B}}
\bigg/
\dfrac{
0.602\,\textcolor{BrickRed}{t}
-0.175\,\textcolor{BrickRed}{Q_A}
}{
-0.298\,R_A
}
}
\frac{1}{
-0.579\,\textcolor{BrickRed}{\chi_B}
+1.293\,\textcolor{BrickRed}{\mu}
}
\end{aligned}
}
& 0.107
& 0.073
& 26.964 \\

\reprowsep

\pysr
& \exprcell{
\displaystyle
0.016\,\textcolor{blue}{\rho}
+
4.005\,\textcolor{BrickRed}{\mu}
}
& 0.137
& 0.083
& 157.306 \\

\reprowsep

\bms
& \exprcell{
\displaystyle
\begin{aligned}[t]
&
\sqrt{\left|
\sqrt{\left|
\frac{
\sqrt{\left|
\textcolor{BrickRed}{Q_A}/a_0
\right|}
}{
-\textcolor{BrickRed}{\chi_B}+\textcolor{RoyalBlue}{\mathcal A_5}
}
\right|}
\right|}
+
\sqrt{|\textcolor{BrickRed}{\mu}|},\\[2pt]
\textcolor{RoyalBlue}{\mathcal A_5}
&=
\frac{
\sqrt{\left|
\textcolor{blue}{\rho}+\textcolor{BrickRed}{\chi_B}\sqrt{|-\textcolor{blue}{\rho}|}
\right|}
}{
\textcolor{BrickRed}{\chi_B}(a_0+R_A)
}\,a_0
\end{aligned}
}
& 0.108
& 0.074
& 391.923 \\

\reprowsep

\bsr
& \exprcell{
\displaystyle
\begin{aligned}[t]
&1.231
-0.001\left[
\textcolor{BrickRed}{\chi_A}-
(\textcolor{BrickRed}{t}-\textcolor{blue}{\rho}+R_A)N_d
\right]-0.0003\,\textcolor{BrickRed}{\chi_A}\textcolor{blue}{\rho}
+0.061\frac{\textcolor{blue}{\rho}}{\textcolor{BrickRed}{Q_A}}\\
&+0.047\left[
\textcolor{BrickRed}{Q_A}
\left(
\textcolor{BrickRed}{t}+R_A+\frac{1}{\textcolor{BrickRed}{\chi_B}}
\right)
-R_A+\textcolor{BrickRed}{t}
\right]+0\left[
\frac{\textcolor{BrickRed}{\chi_A}-\textcolor{BrickRed}{\chi_A}}{\textcolor{BrickRed}{t}}
\right]\\
&
-\frac{0.824}{N_d+R_A}+\frac{1.429}{N_d\textcolor{BrickRed}{t}}
+0.170\left(
\textcolor{blue}{\rho}-\textcolor{BrickRed}{t}\textcolor{blue}{\rho}
\right)
\end{aligned}
}
& 0.126
& 0.082
& 517.036 \\

\end{longtable}
\endgroup

\newpage
\section{Occam's Window Sets \texorpdfstring{($\mathcal J_r, r=10$)}{Jr} of Descriptors Learned by \texorpdfstring{\hierbosss}{HierBOSSS} for the Oxide Perovskite Dataset}
\label{sec:Occams-window-set-perovskite-BayeSymX}

\begingroup
\scriptsize
\setlength{\tabcolsep}{2.0pt}
\renewcommand{\arraystretch}{1.22}

\begin{longtable}{@{}H c
>{\raggedright\arraybackslash}p{0.73\textwidth}
c c c@{}}
\caption{The Occam's window set ($\mathcal J_r, r=10$) of descriptors ranked by $\mathrm{JMP}$ learned by \hierbosss\ for the oxide perovskite dataset across various $90/10$ train/test splits.}
\label{tab:BayeSymX-occams-window}\\

\toprule
\toprule
Split
& Rank
& Learned descriptor expression
& Test \texttt{RMSE} (eV)
& Test \texttt{MAE} (eV)
& $K_{\mathrm{eff}}$\\
\midrule
\endfirsthead

\caption[]{\emph{(Continued)}. The Occam's window set ($\mathcal J_r, r=10$) of symbolic expressions ranked by $\mathrm{JMP}$ learned by \hierbosss\ for the oxide perovskite dataset across various $90/10$ train/test splits.}\\
\toprule
\toprule
Split
& Rank
& Learned descriptor expression
& Test \texttt{RMSE} (eV)
& Test \texttt{MAE} (eV)
& $K_{\mathrm{eff}}$\\
\midrule
\endhead

\midrule
\multicolumn{6}{r}{\emph{Continued on next page}}\\
\endfoot

\bottomrule
\bottomrule
\endlastfoot

\addlinespace[3pt]
\multicolumn{6}{@{}l}{
\textcolor{BrickRed}{\textit{\textbf{Split 1}}}
}\\
\addlinespace[2pt]

1 & 1--8 &
\exprcell{\exprbox{\finaltag}{
\begin{aligned}[t]
& \frac{0.429}{\textcolor{BrickRed}{\mu}}
-0.004\textcolor{blue}{\rho}^{2}
-\frac{0.811}{\textcolor{BrickRed}{\sqrt{\chi_A}}-\sqrt{\textcolor{blue}{\rho}}}
-0.454\sqrt{\textcolor{blue}{\rho}}
+0.006N_d\textcolor{blue}{\rho}
+{0.339\textcolor{blue}{\rho}}\textcolor{BrickRed}{\frac{1}{\chi_B}}\\
&
+\frac{0.206\textcolor{BrickRed}{\mu}}{\textcolor{blue}{\rho}}
\end{aligned}
}}
& 0.084
& 0.058
& 7 \\
\reprowsep

1 & 9--10
& 
\exprcell{\exprbox{\finaltag}{
\begin{aligned}[t]
& 0.719\textcolor{BrickRed}{\sqrt{\mu}}
+\frac{0.421}{\textcolor{BrickRed}{\mu}}
-\frac{0.524}{\textcolor{BrickRed}{\chi_A}-\sqrt{\textcolor{blue}{\rho}}}
-0.004\textcolor{blue}{\rho}^{2}
-0.520\sqrt{\textcolor{blue}{\rho}}
+0.006N_d\textcolor{blue}{\rho}
\\
&+{0.318\textcolor{blue}{\rho}}\textcolor{BrickRed}{\frac{1}{\chi_B}}
+\frac{0.108\textcolor{BrickRed}{\mu}}{\textcolor{blue}{\rho}}
\end{aligned}
}}
& 0.084
& 0.057
& 8 \\

\reprowsep

\addlinespace[3pt]
\midrule
\addlinespace[3pt]
\multicolumn{6}{@{}l}{
\textcolor{BrickRed}{\textit{\textbf{Split 2}}}
}\\
\addlinespace[2pt]

2 & 1--3
& 
\exprcell{\exprbox{\finaltag}{
\begin{aligned}[t]
& 1.158
+0.292\textcolor{BrickRed}{\sqrt{Q_A}}
+\frac{0.530\textcolor{blue}{\rho}}{N_d}
+0.095\sqrt{N_d}\textcolor{blue}{\rho}
-0.068\textcolor{BrickRed}{\chi_B}\textcolor{blue}{\rho}
\\
&-0.073R_A\textcolor{blue}{\rho}
-0.073\textcolor{BrickRed}{\chi_A}\textcolor{blue}{\rho}
\end{aligned}
}}
& 0.087
& 0.064
& 5 \\

\reprowsep

2 & 4--5
& 
\exprcell{\exprbox{\finaltag}{
\begin{aligned}[t]
& 0.496R_A
+0.896\textcolor{BrickRed}{\chi_A}
+\frac{0.596\textcolor{blue}{\rho}}{N_d}
+0.108\sqrt{N_d}\textcolor{blue}{\rho}
-0.070\textcolor{BrickRed}{\chi_B}\textcolor{blue}{\rho}
\\
&-0.089R_A\textcolor{blue}{\rho}
-0.089\textcolor{BrickRed}{\chi_A}\textcolor{blue}{\rho}
\end{aligned}
}}
& 0.087
& 0.063
& 6 \\

\reprowsep

2 & 6--8
& 
\exprcell{\exprbox{\finaltag}{
\begin{aligned}[t]
& 1.158
+0.292\textcolor{BrickRed}{\sqrt{Q_A}}
+\frac{0.530\textcolor{blue}{\rho}}{N_d}
+0.095\sqrt{N_d}\textcolor{blue}{\rho}
-0.068\textcolor{BrickRed}{\chi_B}\textcolor{blue}{\rho}
\\
&-0.073R_A\textcolor{blue}{\rho}
-0.073\textcolor{BrickRed}{\chi_A}\textcolor{blue}{\rho}
\end{aligned}
}}
& 0.087
& 0.064
& 5 \\

\reprowsep

2 & 9--10
&
\exprcell{\exprbox{\finaltag}{
\begin{aligned}[t]
& {1.181}\textcolor{BrickRed}{\frac{1}{t}}
+0.475\textcolor{BrickRed}{\sqrt{t}}
+\frac{0.327\textcolor{blue}{\rho}}{N_d}
+0.014N_d\textcolor{blue}{\rho}
-0.066\textcolor{BrickRed}{\chi_B}\textcolor{blue}{\rho}
\end{aligned}
}}
& 0.086
& 0.063
& 5 \\

\reprowsep

\addlinespace[3pt]
\midrule
\addlinespace[3pt]
\multicolumn{6}{@{}l}{
\textcolor{BrickRed}{\textit{\textbf{Split 3}}}
}\\
\addlinespace[2pt]

3 & 1--3
& 
\exprcell{\exprbox{\finaltag}{
\begin{aligned}[t]
& 2.103\textcolor{BrickRed}{\sqrt{\mu}}
+0.282\textcolor{BrickRed}{\chi_A}
+{0.254\textcolor{blue}{\rho}}\textcolor{BrickRed}{\frac{1}{\chi_B}}\textcolor{BrickRed}{\frac{1}{\chi_B}}
+0.005N_d\textcolor{blue}{\rho}
-0.214\textcolor{BrickRed}{\mu}\textcolor{blue}{\rho}
\end{aligned}
}}
& 0.094
& 0.060
& 5 \\

\reprowsep

3 & 4--5
& 
\exprcell{\exprbox{\finaltag}{
\begin{aligned}[t]
& 0.753
+0.645\textcolor{BrickRed}{\mu}
+0.608\textcolor{BrickRed}{\sqrt{\chi_A}}
+0.006N_d\textcolor{blue}{\rho}
+{0.294\textcolor{blue}{\rho}}\textcolor{BrickRed}{\frac{1}{\chi_B}}\textcolor{BrickRed}{\frac{1}{\chi_B}}
-0.260\textcolor{BrickRed}{\mu}\textcolor{blue}{\rho}
\end{aligned}
}}
& 0.094
& 0.059
& 5 \\

\reprowsep

3 & 6--7
&
\exprcell{\exprbox{\finaltag}{
\begin{aligned}[t]
& 0.565\textcolor{BrickRed}{\sqrt{\chi_A}}
+\frac{1.030}{N_d}
+0.077N_d
+{0.820}\textcolor{BrickRed}{\frac{1}{\chi_B}}
-0.080\textcolor{blue}{\rho}
+{0.173\textcolor{blue}{\rho}}\textcolor{BrickRed}{\frac{1}{\chi_B}}
\end{aligned}
}}
& 0.096
& 0.063
& 6 \\

\reprowsep

3 & 8--9
&
\exprcell{\exprbox{\finaltag}{
\begin{aligned}[t]
& 1.034
+0.310\textcolor{BrickRed}{\chi_A}
+0.684\textcolor{BrickRed}{\mu}
+0.006N_d\textcolor{blue}{\rho}
+{0.291\textcolor{blue}{\rho}}\textcolor{BrickRed}{\frac{1}{\chi_B}}\textcolor{BrickRed}{\frac{1}{\chi_B}}
-0.257\textcolor{BrickRed}{\mu}\textcolor{blue}{\rho}
\end{aligned}
}}
& 0.094
& 0.059
& 5 \\

\reprowsep

3 & 10
&
\exprcell{\exprbox{\finaltag}{
\begin{aligned}[t]
& 0.069N_d
+\frac{1.303}{N_d}
+{1.038}\textcolor{BrickRed}{\frac{1}{t}}
-0.094\textcolor{blue}{\rho}
+{0.200\textcolor{blue}{\rho}}\textcolor{BrickRed}{\frac{1}{\chi_B}}
\end{aligned}
}}
& 0.096
& 0.063
& 5 \\

\reprowsep

\addlinespace[3pt]
\midrule
\addlinespace[3pt]
\multicolumn{6}{@{}l}{
\textcolor{BrickRed}{\textit{\textbf{Split 4}}}
}\\
\addlinespace[2pt]

4 & 1--10
& 
\exprcell{\exprbox{\finaltag}{
\begin{aligned}[t]
& {0.720}\textcolor{BrickRed}{\frac{1}{t}}
+\frac{0.509}{\textcolor{BrickRed}{Q_A}}
+0.468\sqrt{\textcolor{BrickRed}{Q_A}+\textcolor{blue}{\rho}}
-0.169\sqrt{\textcolor{blue}{\rho}}
+{0.238\textcolor{blue}{\rho}}\textcolor{BrickRed}{\frac{1}{\chi_B}}
+0.035N_d\textcolor{blue}{\rho}
\\
&-0.148\sqrt{N_d}\textcolor{blue}{\rho}
\end{aligned}
}}
& 0.099
& 0.062
& 7 \\

\reprowsep

\addlinespace[3pt]
\midrule
\addlinespace[3pt]
\multicolumn{6}{@{}l}{
\textcolor{BrickRed}{\textit{\textbf{Split 5}}}
}\\
\addlinespace[2pt]

5 & 1
&
\exprcell{\exprbox{\finaltag}{
\begin{aligned}[t]
& 0.371\textcolor{BrickRed}{t}
+{1.288}\textcolor{BrickRed}{\frac{1}{t}}
+\frac{0.334\textcolor{blue}{\rho}}{N_d}
+0.014N_d\textcolor{blue}{\rho}
-0.069\textcolor{BrickRed}{\chi_B}\textcolor{blue}{\rho}
\end{aligned}
}}
& 0.108
& 0.072
& 5 \\

\reprowsep

5 & 2
& 
\exprcell{\exprbox{\finaltag}{
\begin{aligned}[t]
& 0.506\textcolor{BrickRed}{\sqrt{t}}
+{1.153}\textcolor{BrickRed}{\frac{1}{t}}
+\frac{0.335\textcolor{blue}{\rho}}{N_d}
+0.014N_d\textcolor{blue}{\rho}
-0.069\textcolor{BrickRed}{\chi_B}\textcolor{blue}{\rho}
\end{aligned}
}}
& 0.108
& 0.072
& 5 \\

\reprowsep

5 & 3--6
& 
\exprcell{\exprbox{\finaltag}{
\begin{aligned}[t]
& 0.371\textcolor{BrickRed}{t}
+{1.288}\textcolor{BrickRed}{\frac{1}{t}}
+\frac{0.334\textcolor{blue}{\rho}}{N_d}
+0.014N_d\textcolor{blue}{\rho}
-0.069\textcolor{BrickRed}{\chi_B}\textcolor{blue}{\rho}
\end{aligned}
}}
& 0.108
& 0.072
& 5 \\

\reprowsep

5 & 7
&
\exprcell{\exprbox{\finaltag}{
\begin{aligned}[t]
& {1.454}\textcolor{BrickRed}{\frac{1}{t}}
+{0.228}\frac{1}{\textcolor{BrickRed}{\chi_A}}
+\frac{0.331\textcolor{blue}{\rho}}{N_d}
+0.014N_d\textcolor{blue}{\rho}
-0.067\textcolor{BrickRed}{\chi_B}\textcolor{blue}{\rho}
\end{aligned}
}}
& 0.107
& 0.072
& 5 \\

\reprowsep

5 & 8--9
&
\exprcell{\exprbox{\finaltag}{
\begin{aligned}[t]
& {1.230}\textcolor{BrickRed}{\frac{1}{t}}
+0.370\sqrt{R_A}
+\frac{0.328\textcolor{blue}{\rho}}{N_d}
+0.014N_d\textcolor{blue}{\rho}
-0.067\textcolor{BrickRed}{\chi_B}\textcolor{blue}{\rho}
\end{aligned}
}}
& 0.108
& 0.072
& 5 \\

\reprowsep

5 & 10
&
\exprcell{\exprbox{\finaltag}{
\begin{aligned}[t]
& 0.736
+{0.924}\textcolor{BrickRed}{\frac{1}{t}}
+\frac{0.334\textcolor{blue}{\rho}}{N_d}
+0.014N_d\rho
-0.068\textcolor{BrickRed}{\chi_B}\textcolor{blue}{\rho}
\end{aligned}
}}
& 0.107
& 0.072
& 4 \\

\end{longtable}
\endgroup

\newpage
\section{Accuracy-Complexity Trade-off of Learned Descriptors for the Oxide Perovskite Dataset}
\label{sec:accuracy-complexity-tradeoff-perovskites}

We compare the predictive accuracy and symbolic complexity of the descriptors learned by \hierbosss\ and competing \sr\ methods on the oxide perovskite dataset. In light of our observations from the experiments on Feynman equations in~\hyperref[superdiv:feynman]{Part \ref{superdiv:feynman}}, predictive accuracy alone does not fully characterize the scientific interpretability of a learned descriptor, since a low-error expression may still contain excessive nesting, redundant transformations, or many constants and operators. We therefore evaluate held-out test \texttt{MAE} jointly with symbolic model size~\citep{Imai-SRBench, LaCava-NIPS} across the $5$ independent $90/10$ train/test splits of the dataset. Experimental configurations for \hierbosss\ and the competing methods are kept the same as those described in~\hyperref[sec:experimental-settings-perovskite]{Appendix \ref{sec:experimental-settings-perovskite}}.

For each learned symbolic expression representing a descriptor, symbolic model size is obtained similarly as in the case of Feynman equations; refer to \hyperref[subsec:expression-complexity-Feynman]{Appendix \ref{subsec:expression-complexity-Feynman}}. For \hierbosss, complexity is computed from the \finaltag\ descriptor expression after post-\texttt{MCMC} symbolic model refinement in \hyperref[alg:post-mcmc-symbolic-model-refinement-single]{Algorithm \ref{alg:post-mcmc-symbolic-model-refinement-single}}; for each competing method, it is computed from the descriptor expression returned by the corresponding implementation.

\begin{figure}[H]
\centering
\includegraphics[width=\textwidth]{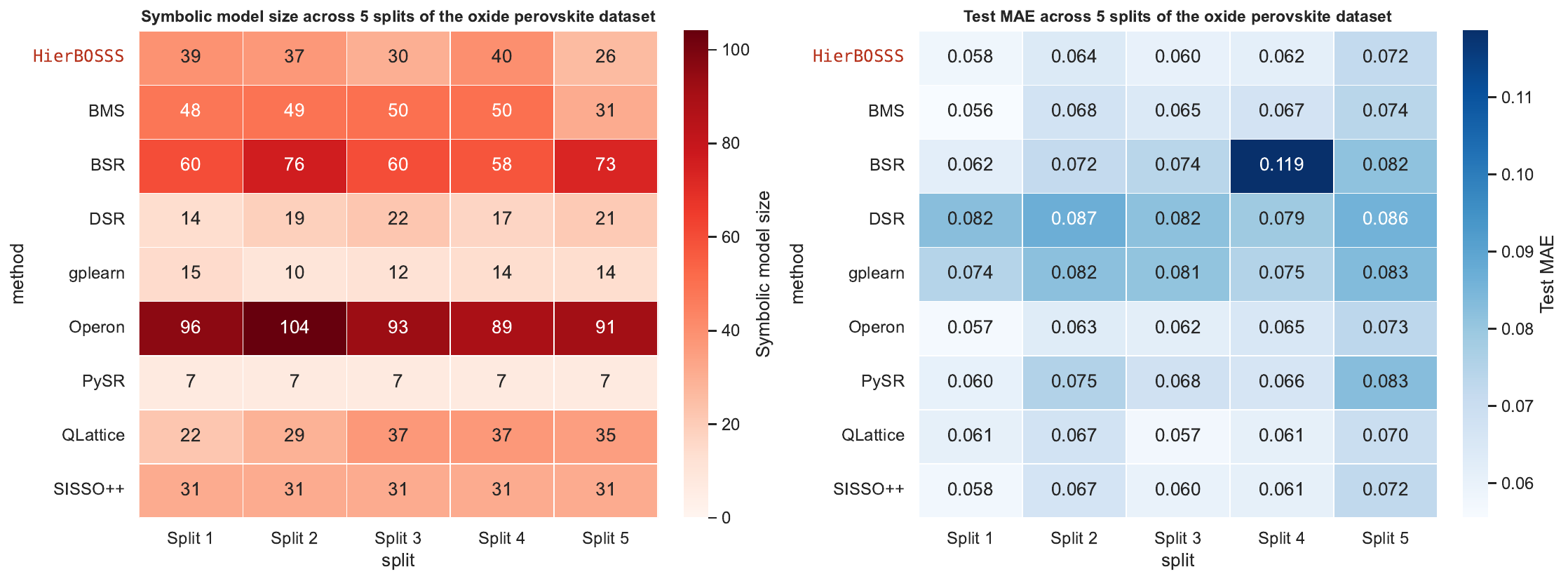}
\caption{Symbolic model size and test \texttt{MAE} of learned descriptor expressions across $5$ independent train/test splits of the oxide perovskite dataset.}
\label{fig:perovskite-complexity-mae-heatmaps}
\end{figure}

\hyperref[fig:perovskite-complexity-mae-heatmaps]{Figure~\ref{fig:perovskite-complexity-mae-heatmaps}} shows that low predictive error does not necessarily imply symbolic parsimony. \operon\ and \bsr\ often attain competitive test \texttt{MAE} through substantially larger descriptor expressions, while \pysr\ and \gplearn\ return very compact symbolic models but generally exhibit weaker predictive performance. \bms, \qlattice, and \sisso$++$ occupy intermediate regimes, with moderate-to-large descriptor expressions and competitive accuracy on selected data splits.

\begin{figure}[!htp]
\centering
\includegraphics[width=0.9\textwidth]{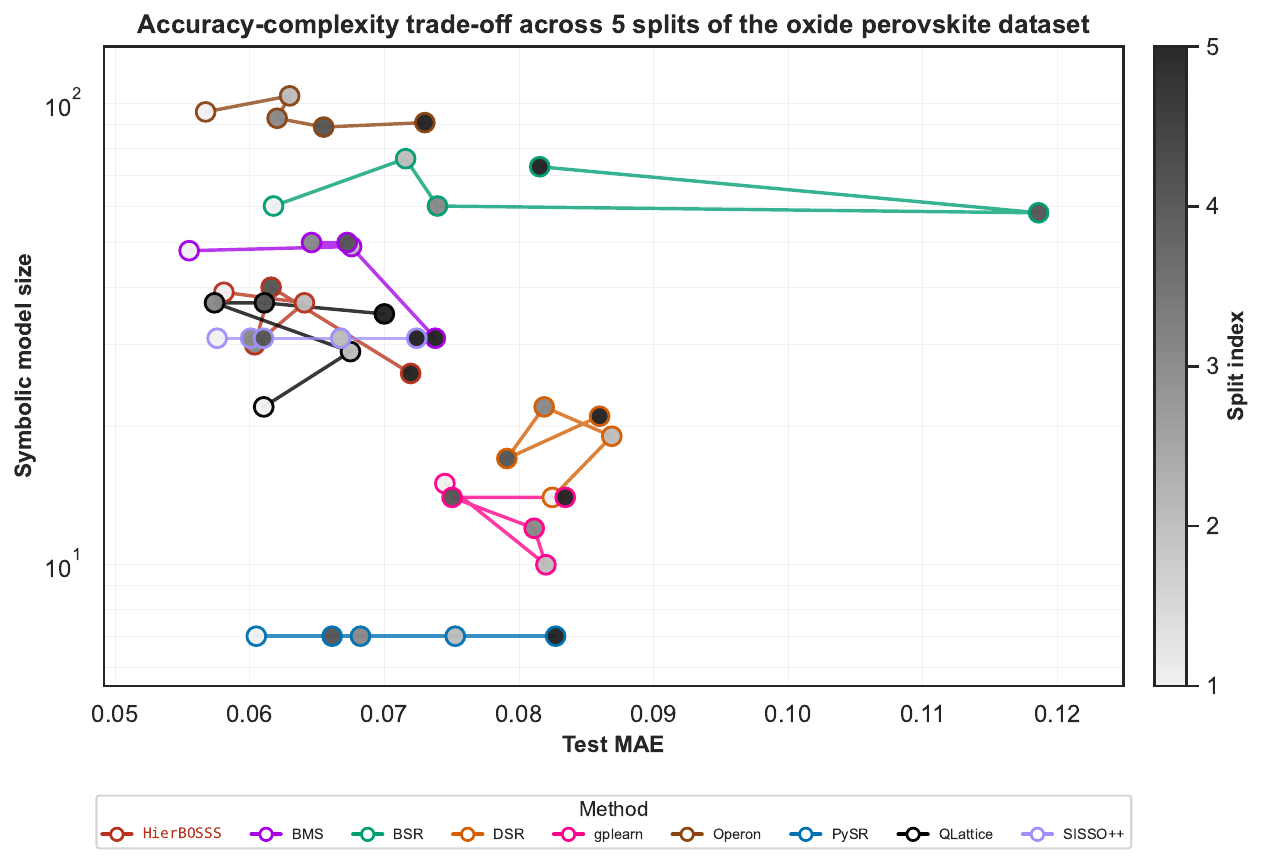}
\caption{Accuracy-complexity trade-off across $5$ independent train/test splits of the oxide perovskite dataset. Each point represents one method on one split, with test \texttt{MAE} on the horizontal axis and symbolic model size on the vertical axis. Point shading indicates the split index.}
\label{fig:perovskite-accuracy-complexity-tradeoff}

\end{figure}

The joint accuracy-complexity plot in~\hyperref[fig:perovskite-accuracy-complexity-tradeoff]{Figure~\ref{fig:perovskite-accuracy-complexity-tradeoff}} provides a clearer view of this trade-off. Across the $5$ data splits, \hierbosss\ consistently attains low test \texttt{MAE} with substantially smaller symbolic models than several high-performing competitors. Its \finaltag\ descriptor expressions contain only $26$-$40$ nodes, while descriptor expressions from several competitors often require $50$-$100$ nodes to achieve comparable predictive accuracy. Moreover, the learned expressions repeatedly recover descriptors such as \textcolor{BrickRed}{$t$}, \textcolor{BrickRed}{$1/t$}, \textcolor{BrickRed}{$\mu$}, \textcolor{BrickRed}{$\sqrt{\mu}$}, \textcolor{BrickRed}{$\sqrt{Q_A}$}, \textcolor{BrickRed}{$\chi_A$}, \textcolor{BrickRed}{$\sqrt{\chi_A}$}, \textcolor{BrickRed}{$\chi_B$}, and \textcolor{BrickRed}{$1 / \chi_B$} which were also identified as scientifically important for \texttt{OER} activity~\citep[Table 2]{Weng2020SimpleDescriptor}. This favorable behavior is consistent with the Feynman experiments and reflects the combined effect of the Occam's razor-supported depth regularizing symbolic tree prior~\citep{Occams-Razor-1} and the post-\texttt{MCMC} symbolic model refinement applied to the Occam's window-based model selection using $\mathrm{JMP}(\mathcal T)$~\citep{madigan1994model}, which preserves high-posterior, scientifically meaningful descriptor structures while removing redundant symbolic components. Overall, \hierbosss\ achieves a strong balance among predictive accuracy, symbolic parsimony, and scientific interpretability.

\newpage
\section{Runtime Comparison of \texorpdfstring{\hierbosss}{HierBOSSS} with the Bayesian \texorpdfstring{\sr}{SR} Methods}
\label{sec:runtime-complexity}

\begin{figure}[H]
    \centering
    \includegraphics[width=0.7\linewidth]{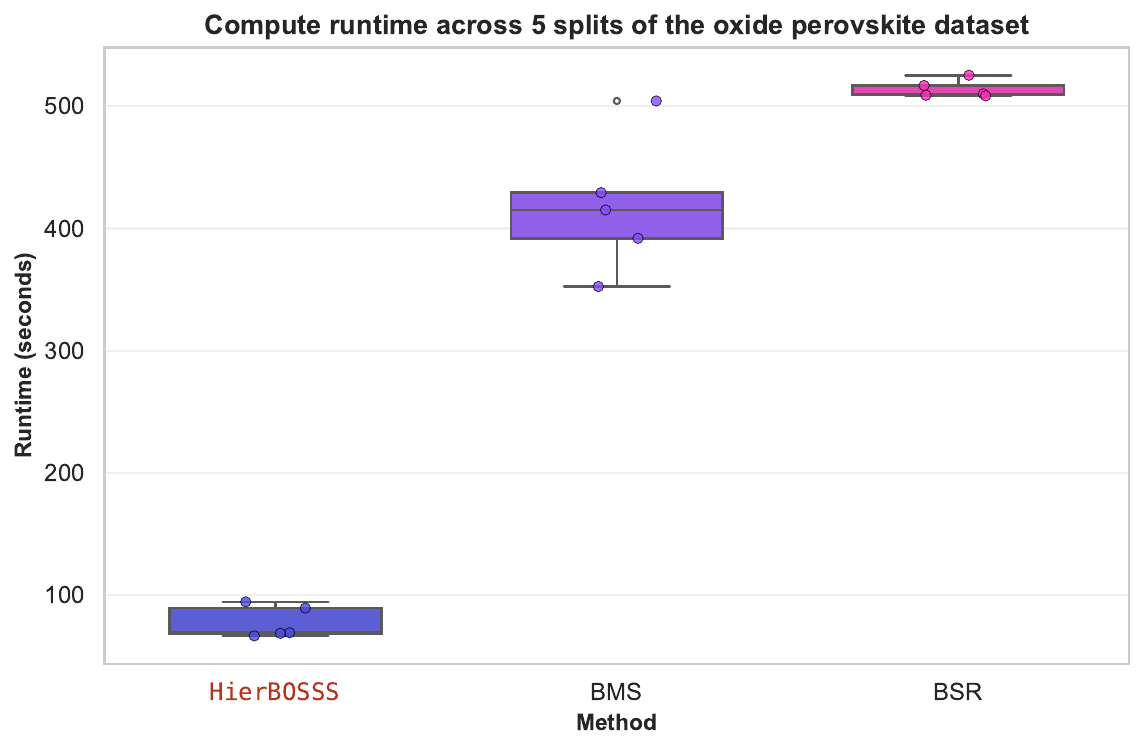}
    \caption{Compute runtimes of \hierbosss, \bms\ ($5$ parallel \texttt{MCMC} chains), and \bsr\ ($5$ restarts) across $5$ splits of the oxide perovskite dataset. Implementation details of \hierbosss\ and the Bayesian competitors follow from~\hyperref[subsec:bayesymx-experimental-settings-perovskite]{Appendices~\ref{subsec:bayesymx-experimental-settings-perovskite}},~\ref{subsec:bms-experimental-settings-perovskite}, and~\ref{subsec:bsr-experimental-settings-perovskite}.}
    \label{fig:runtime-BayeSymX-BMS-BSR-perovskite}
\end{figure}

\newpage
\section{Posterior Diagnostics of the Bayesian \texorpdfstring{\sr}{SR} Methods for the Oxide Perovskite Dataset}
\label{sec:trace-plots-perovskites}

We provide the trace plots to assess the stochastic search and posterior convergence behavior of \hierbosss, \bms, and \bsr\ while learning symbolic descriptor expressions for the oxide perovskite dataset. For each method, diagnostics are reported across the $5$ independent train/test splits of the dataset under the corresponding experimental configurations described in \hyperref[subsec:bayesymx-experimental-settings-perovskite]{Appendices \ref{subsec:bayesymx-experimental-settings-perovskite}}, \hyperref[subsec:bms-experimental-settings-perovskite]{\ref{subsec:bms-experimental-settings-perovskite}}, and \hyperref[subsec:bsr-experimental-settings-perovskite]{\ref{subsec:bsr-experimental-settings-perovskite}}, respectively. As in~\hyperref[sec:trace-plots-Feynman]{Appendix \ref{sec:trace-plots-Feynman}}, the plotted trace quantities are method-specific and should therefore be interpreted as within-method diagnostics of posterior exploration and stability.

\subsection{Posterior Diagnostics for \texorpdfstring{\hierbosss}{HierBOSSS}}
\begin{figure}[H]
\centering
\begin{subfigure}[t]{0.45\textwidth}
    \centering
    \includegraphics[width=\linewidth]{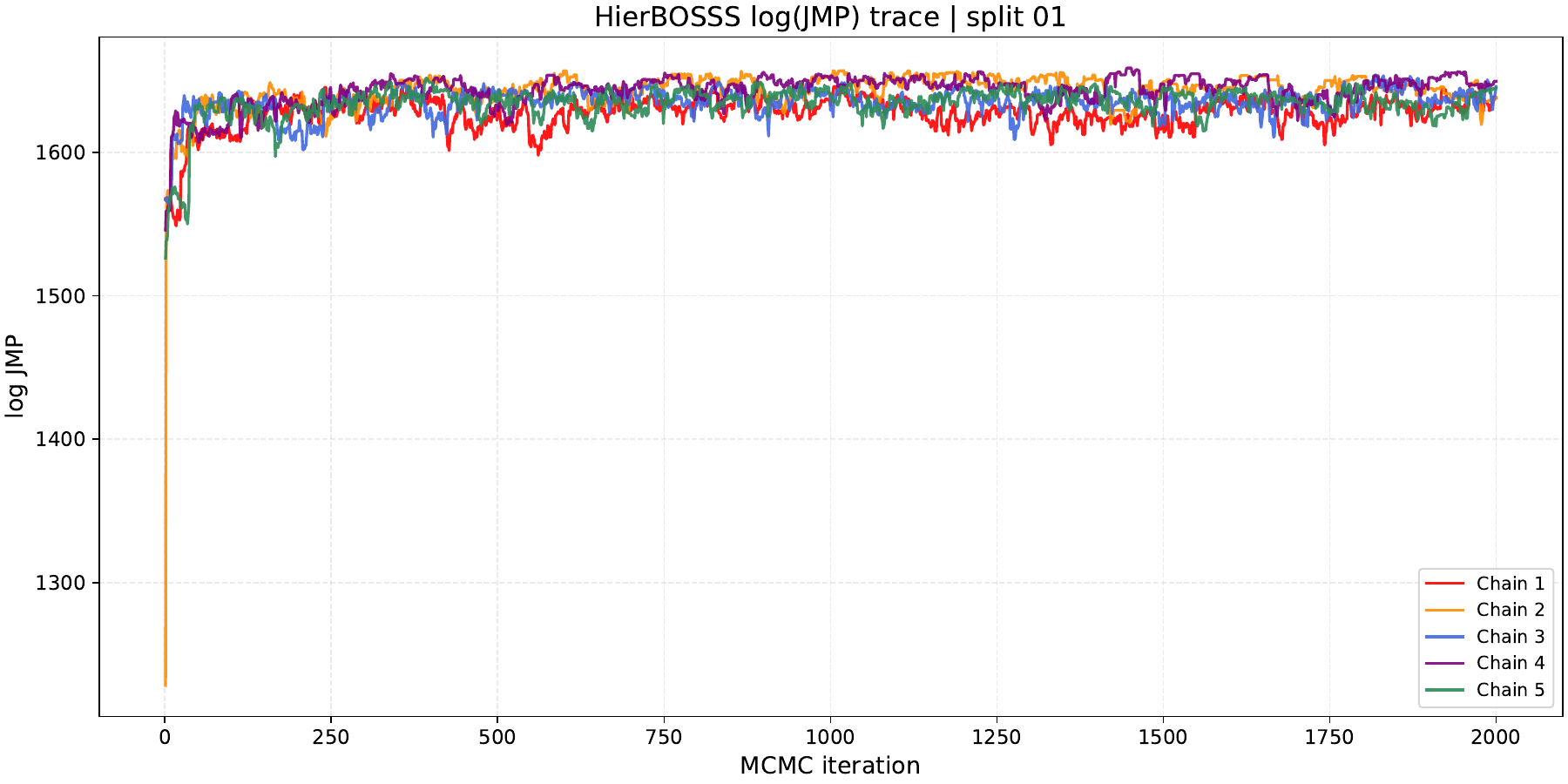}
    \caption{Split 1}
    \label{fig:bayesymx-trace-plot-perovskite-split-1}
\end{subfigure}
\hfill
\begin{subfigure}[t]{0.45\textwidth}
    \centering
    \includegraphics[width=\linewidth]{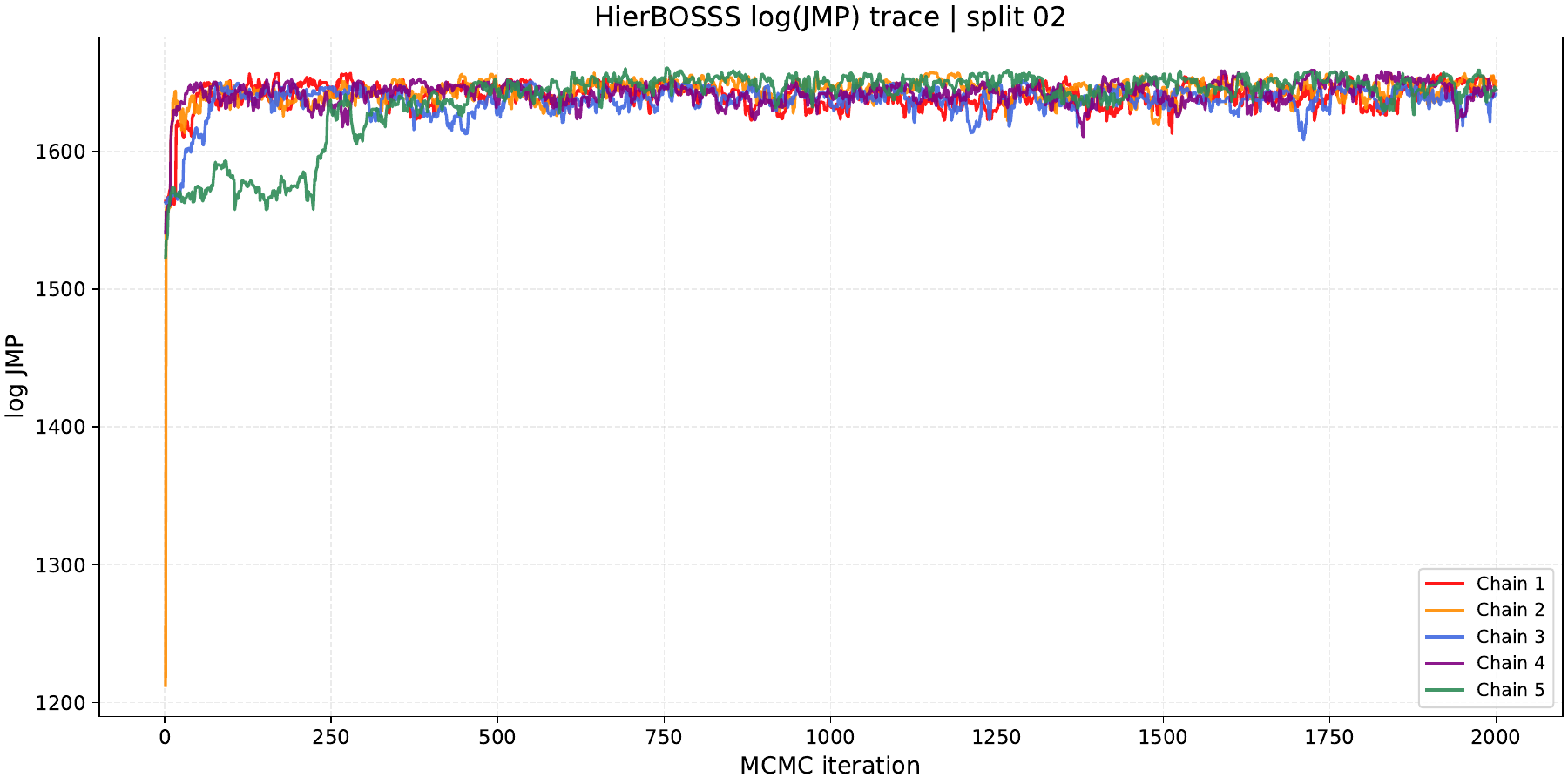}
    \caption{Split 2}
    \label{fig:bayesymx-trace-plot-perovskite-split-2}
\end{subfigure}
\hfill
\begin{subfigure}[t]{0.45\textwidth}
    \centering
    \includegraphics[width=\linewidth]{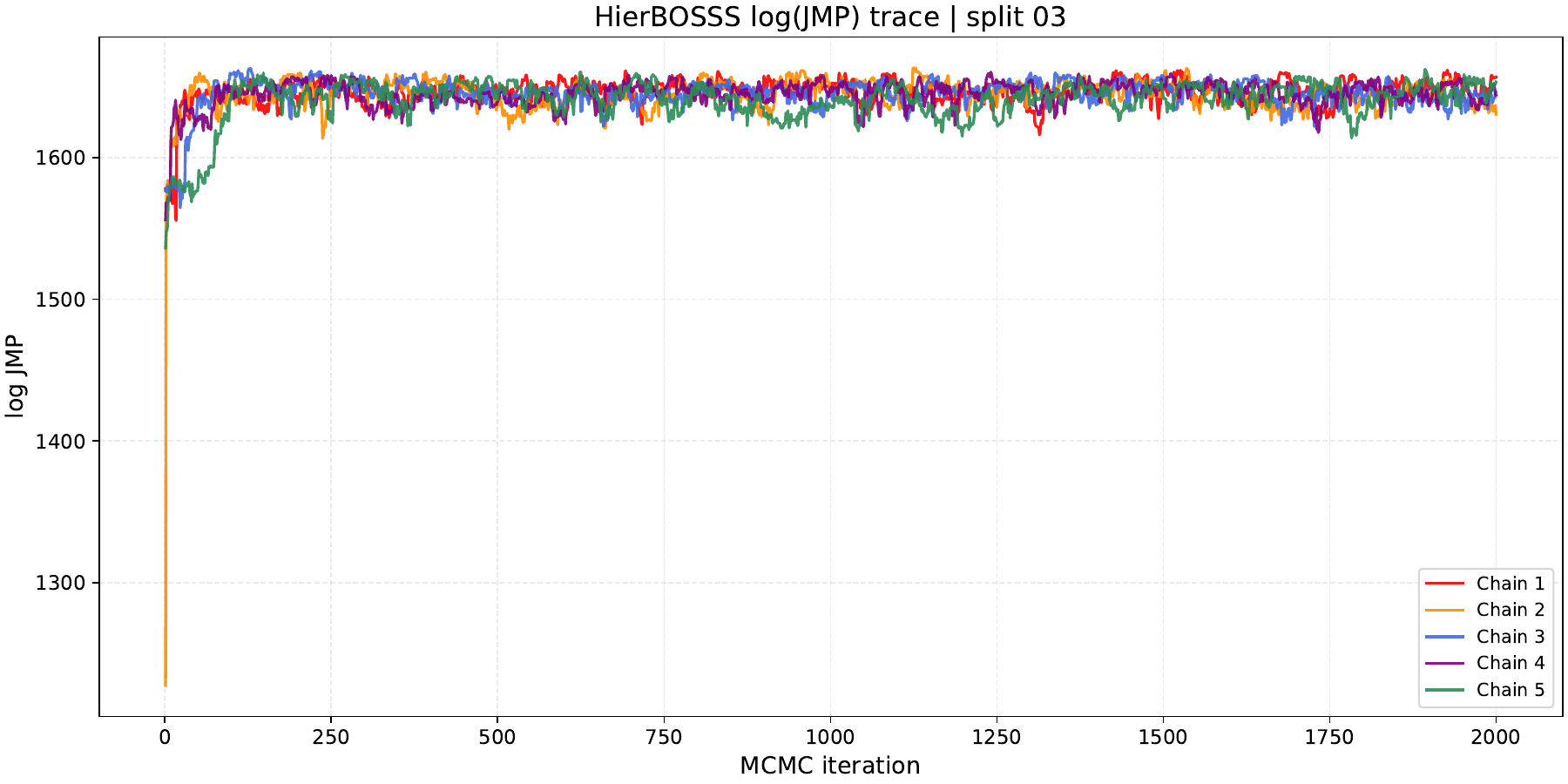}
    \caption{Split 3}
    \label{fig:bayesymx-trace-plot-perovskite-split-3}
\end{subfigure}

\begin{subfigure}[t]{0.45\textwidth}
    \centering
    \includegraphics[width=\linewidth]{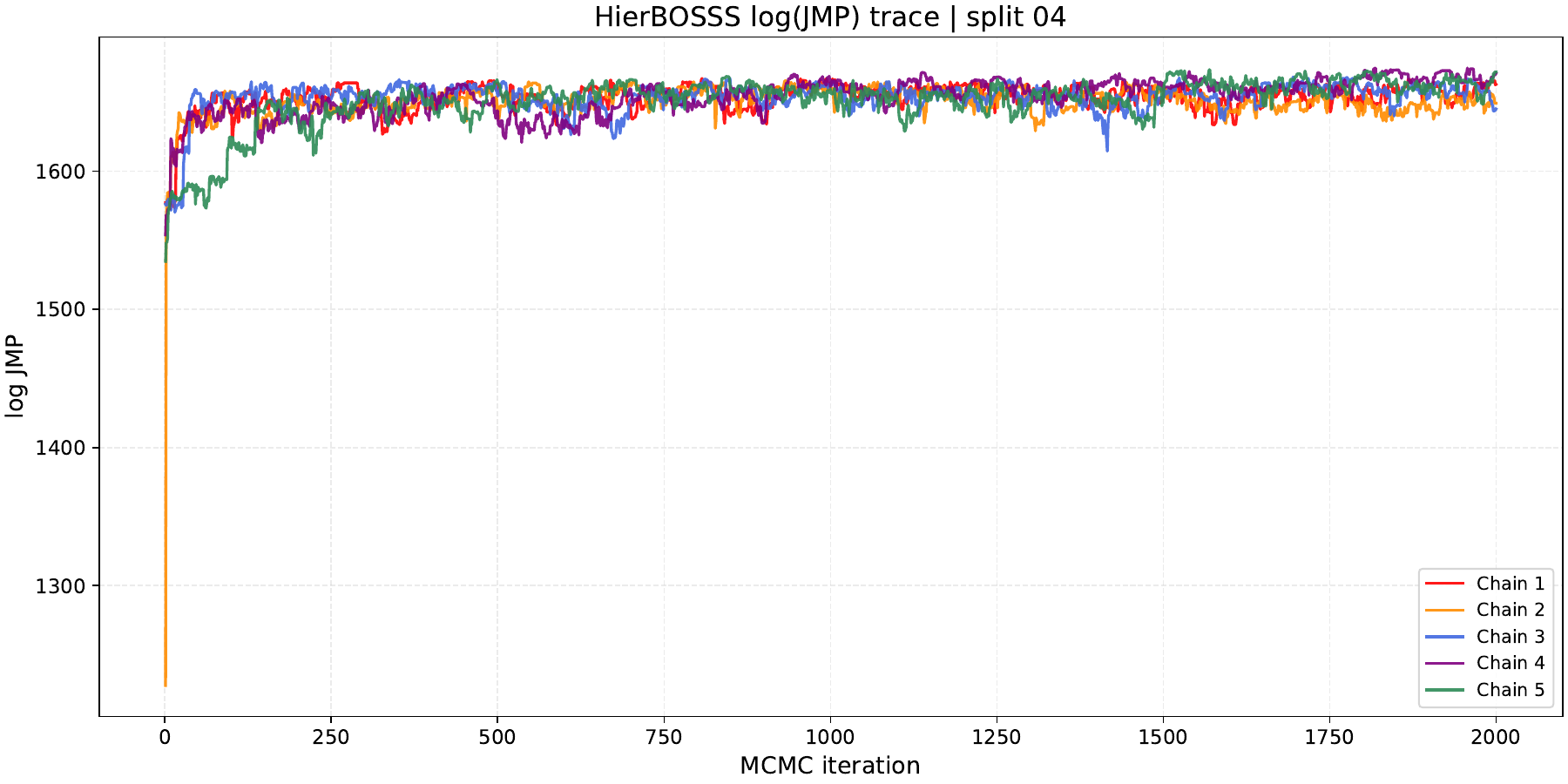}
    \caption{Split 4}
    \label{fig:bayesymx-trace-plot-perovskite-split-4}
\end{subfigure}
\hspace{0.06\textwidth}
\begin{subfigure}[t]{0.45\textwidth}
    \centering
    \includegraphics[width=\linewidth]{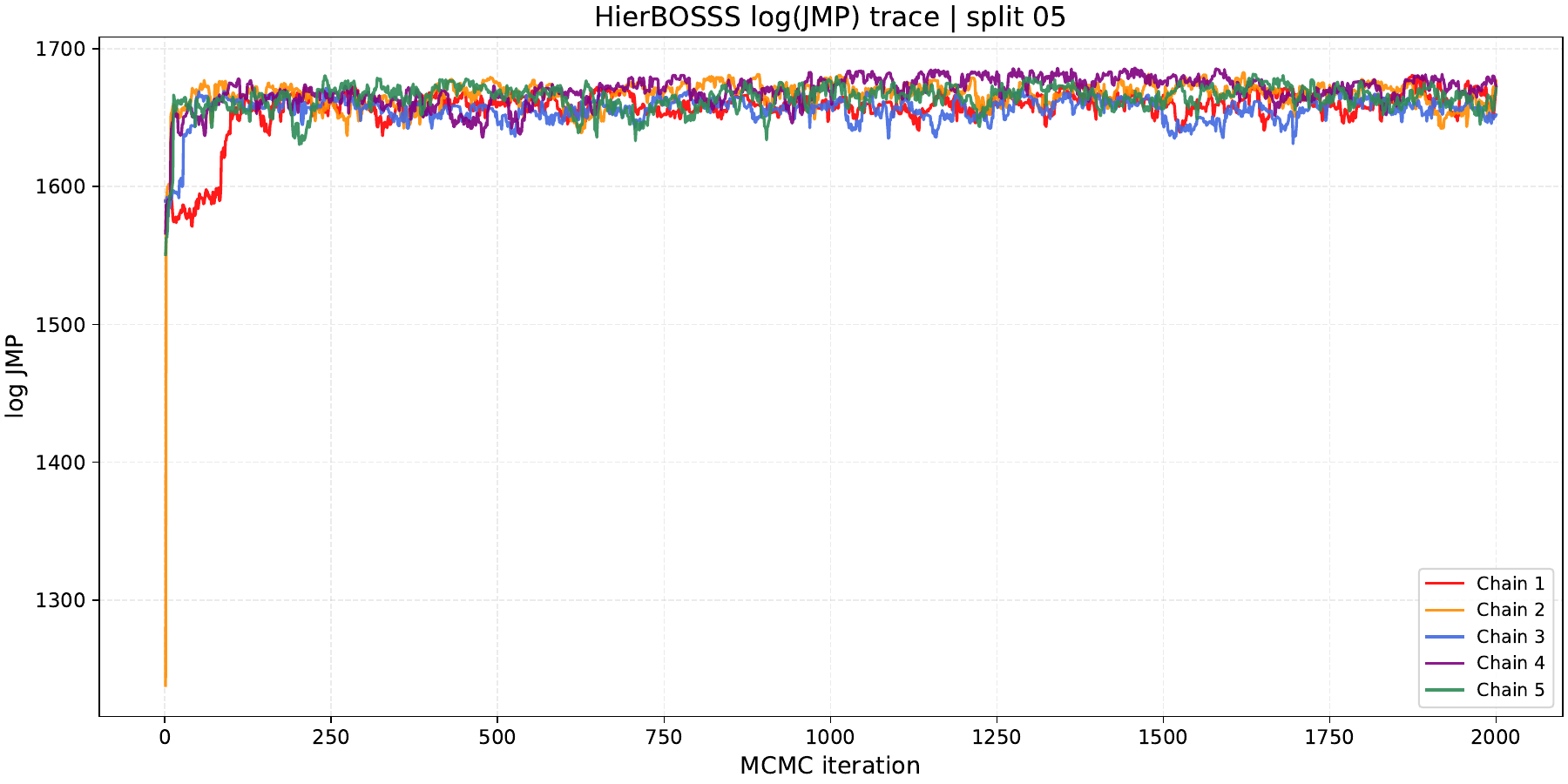}
    \caption{Split 5}
    \label{fig:bayesymx-trace-plot-perovskite-split-5}
\end{subfigure}

\caption{Trace plots of \hierbosss\ across the $5$ splits of the oxide perovskite dataset.}
\label{fig:bayesymx-perovskite-trace-plots}
\end{figure}

\newpage
\subsection{Posterior Diagnostics for \texorpdfstring{\bms}{BMS}}

\begin{figure}[H]
\centering
\begin{subfigure}[t]{0.45\textwidth}
    \centering
    \includegraphics[width=\linewidth]{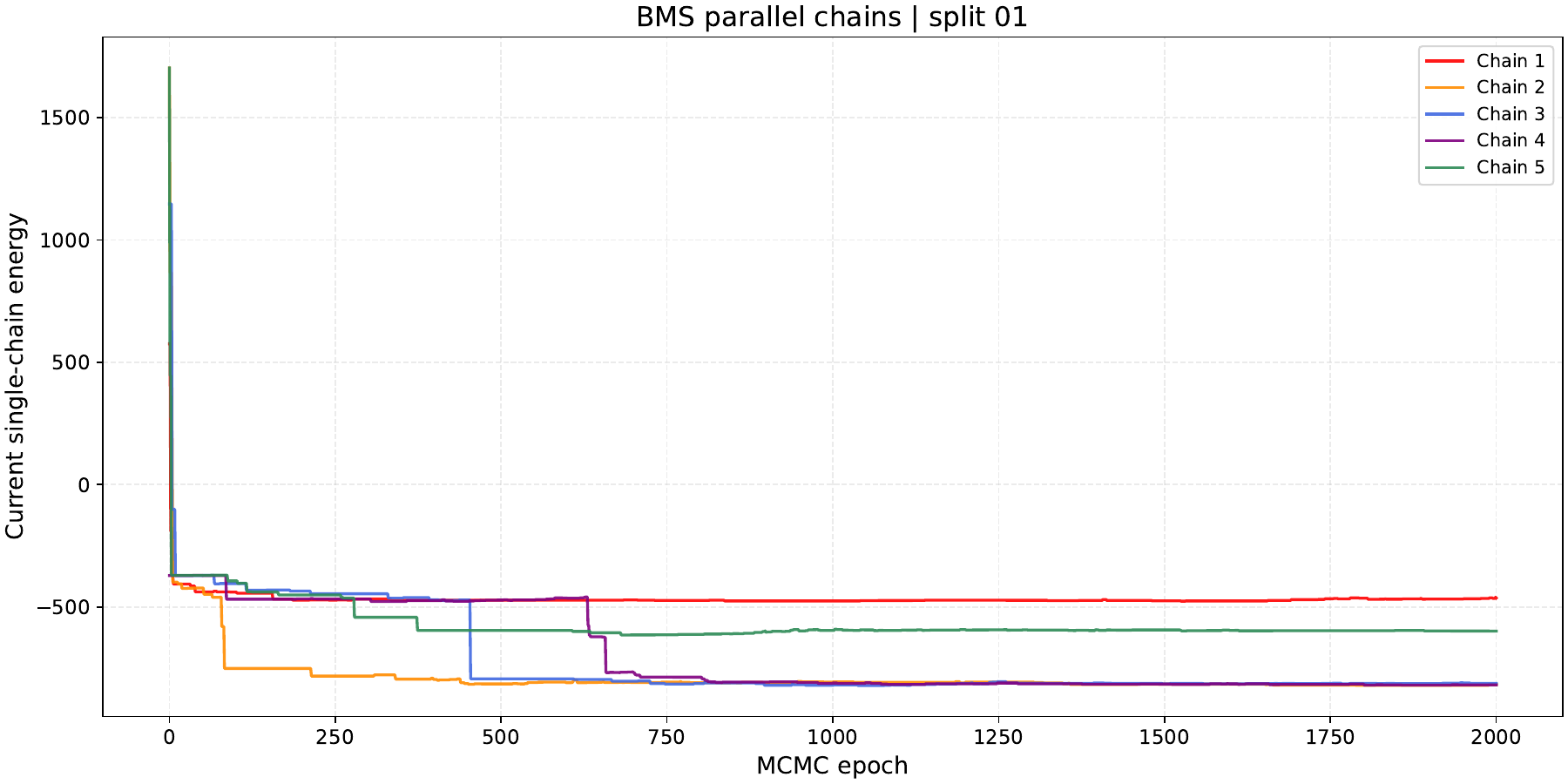}
    \caption{Split 1}
    \label{fig:bms-trace-plot-perovskite-split-1}
\end{subfigure}
\hfill
\begin{subfigure}[t]{0.45\textwidth}
    \centering
    \includegraphics[width=\linewidth]{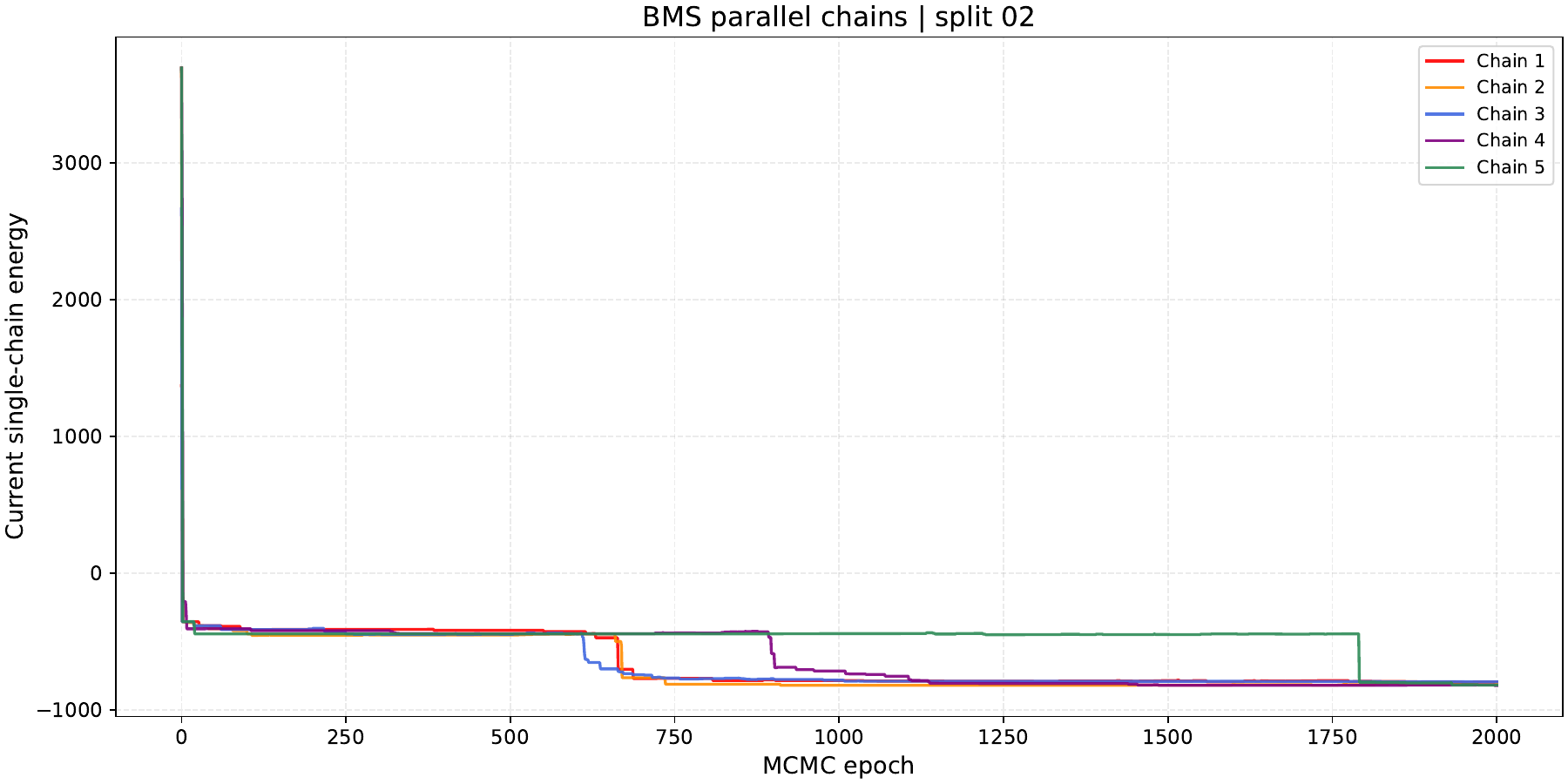}
    \caption{Split 2}
    \label{fig:bms-trace-plot-perovskite-split-2}
\end{subfigure}
\hfill
\begin{subfigure}[t]{0.45\textwidth}
    \centering
    \includegraphics[width=\linewidth]{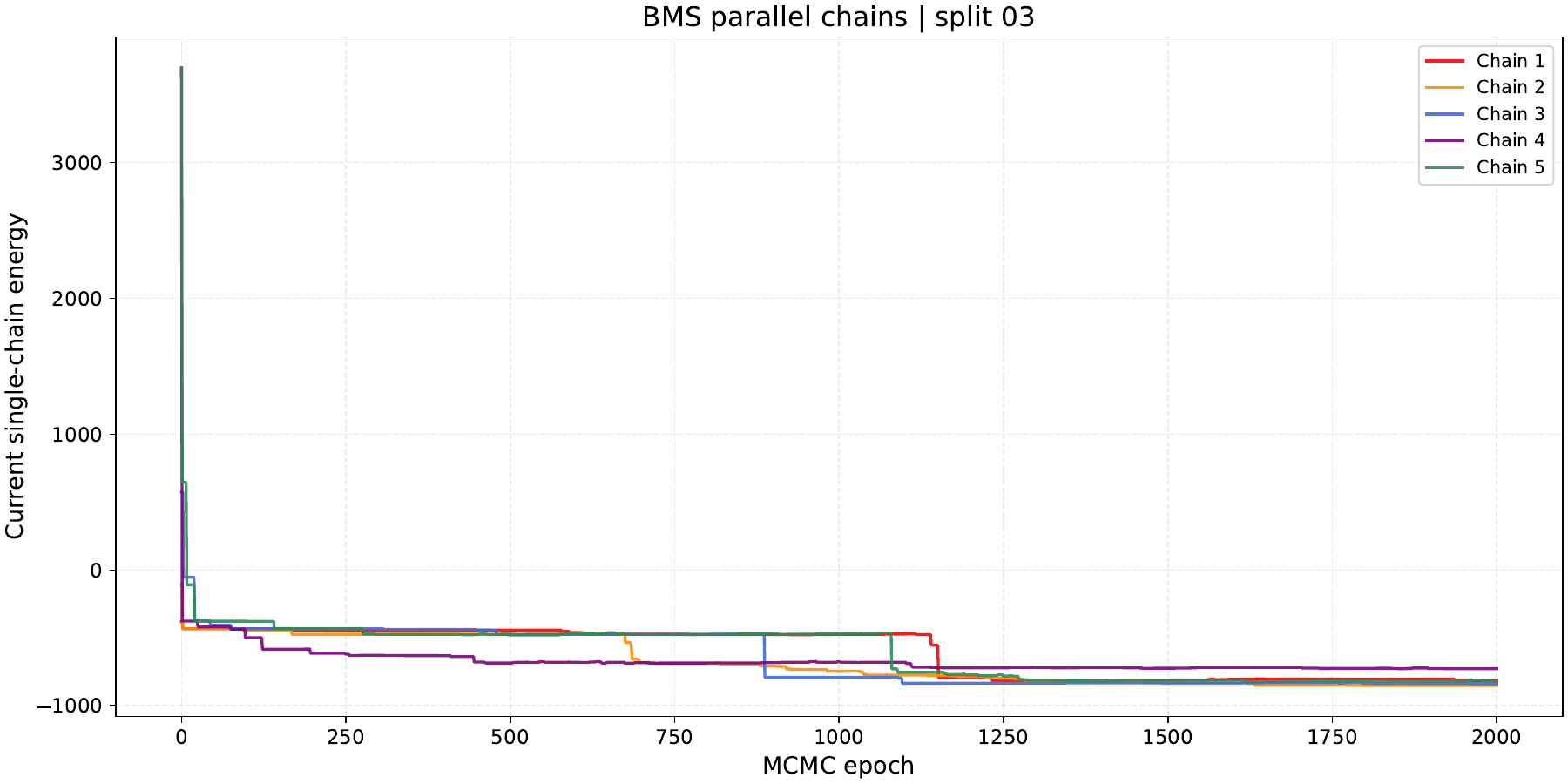}
    \caption{Split 3}
    \label{fig:bms-trace-plot-perovskite-split-3}
\end{subfigure}

\begin{subfigure}[t]{0.45\textwidth}
    \centering
    \includegraphics[width=\linewidth]{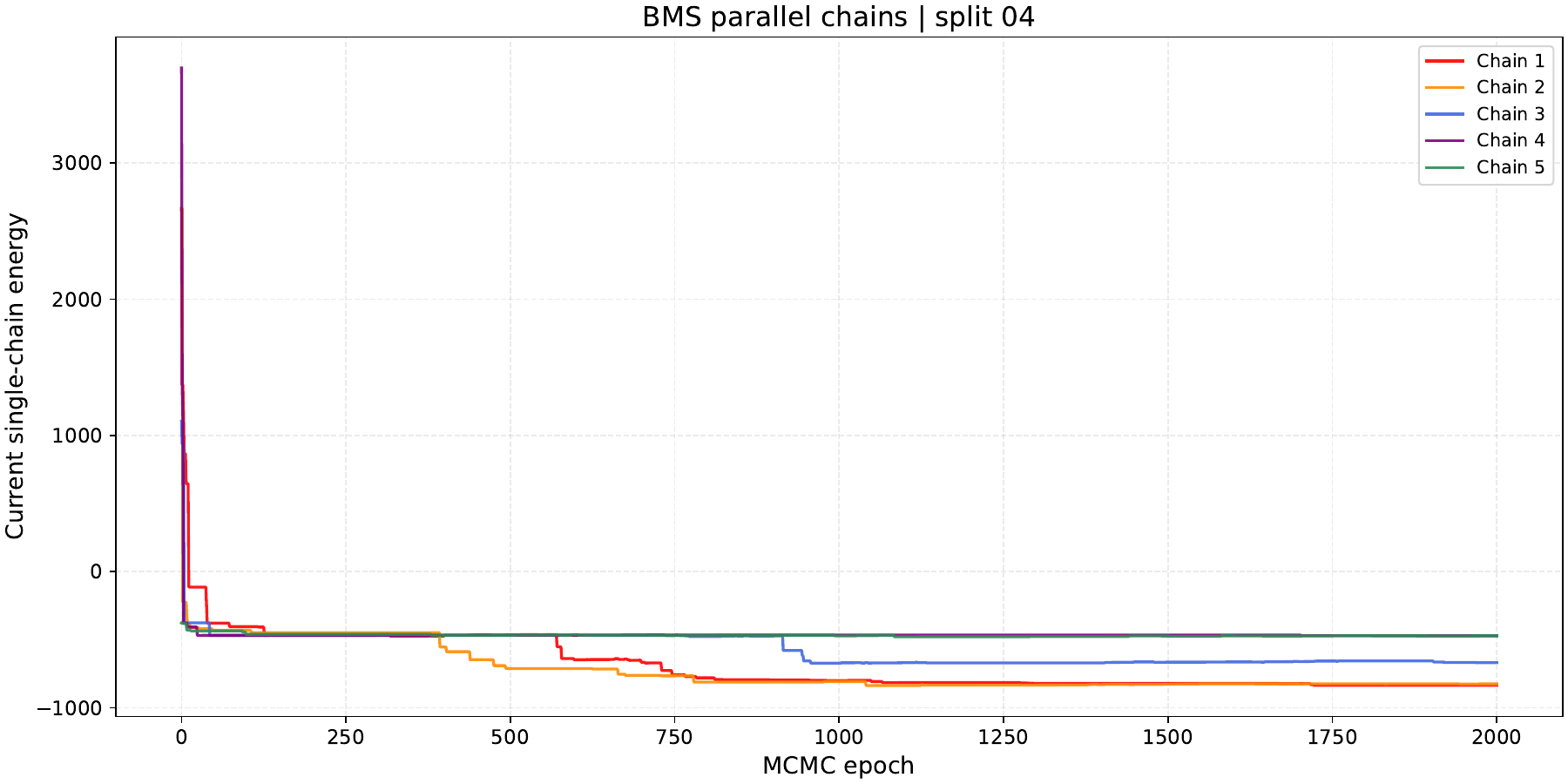}
    \caption{Split 4}
    \label{fig:bms-trace-plot-perovskite-split-4}
\end{subfigure}
\hspace{0.06\textwidth}
\begin{subfigure}[t]{0.45\textwidth}
    \centering
    \includegraphics[width=\linewidth]{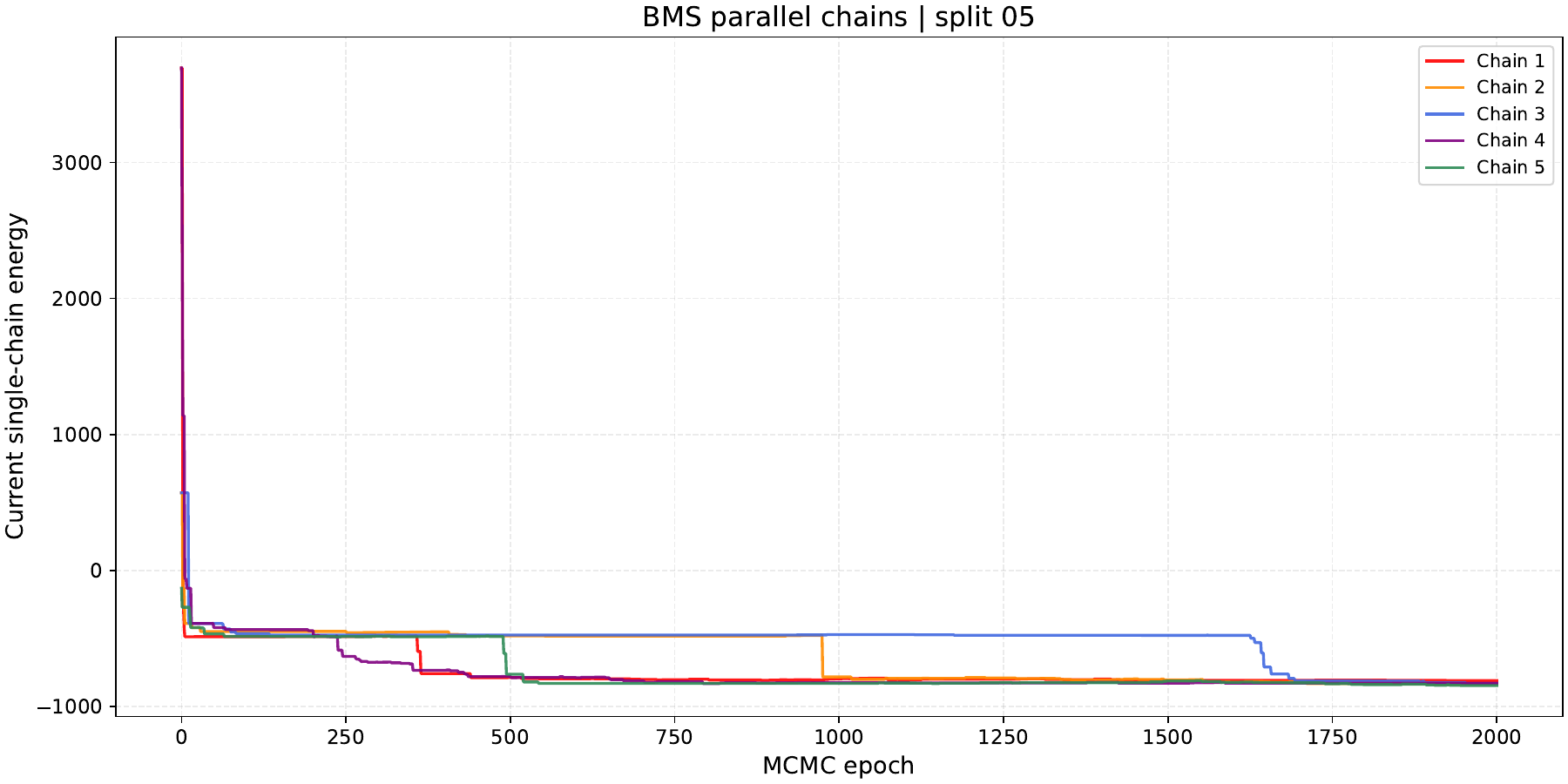}
    \caption{Split 5}
    \label{fig:bms-trace-plot-perovskite-split-5}
\end{subfigure}

\caption{Trace plots of \bms\ across the $5$ splits of the oxide perovskite dataset.}
\label{fig:bms-perovskite-trace-plots}
\end{figure}

\newpage
\subsection{Posterior Diagnostics for \texorpdfstring{\bsr}{BSR}}

\begin{figure}[H]
\centering

\begin{subfigure}[t]{0.45\textwidth}
    \centering
    \includegraphics[width=\linewidth]{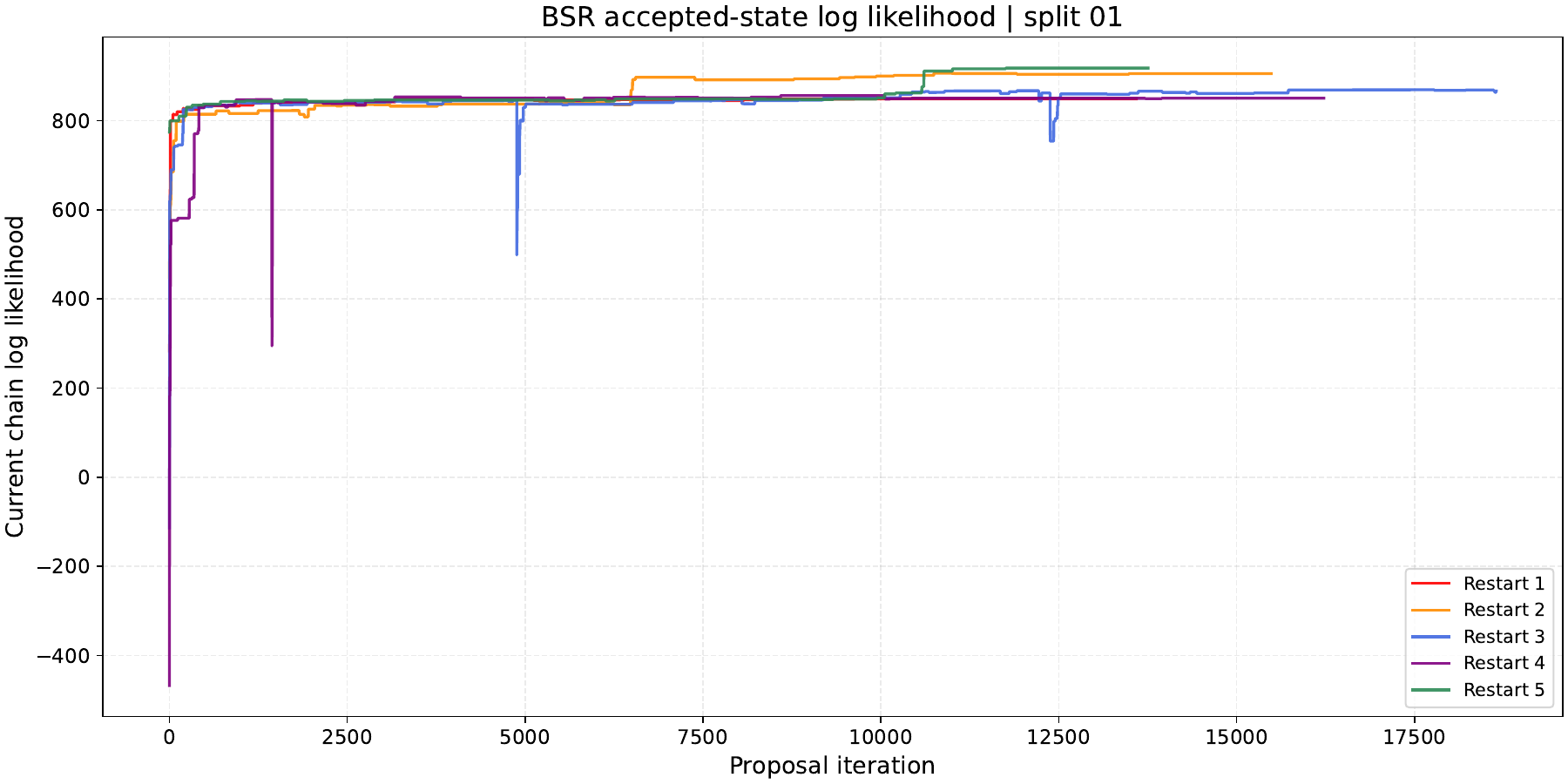}
    \caption{Split 1}
    \label{fig:bsr-trace-plot-perovskite-split-1}
\end{subfigure}
\hfill
\begin{subfigure}[t]{0.45\textwidth}
    \centering
    \includegraphics[width=\linewidth]{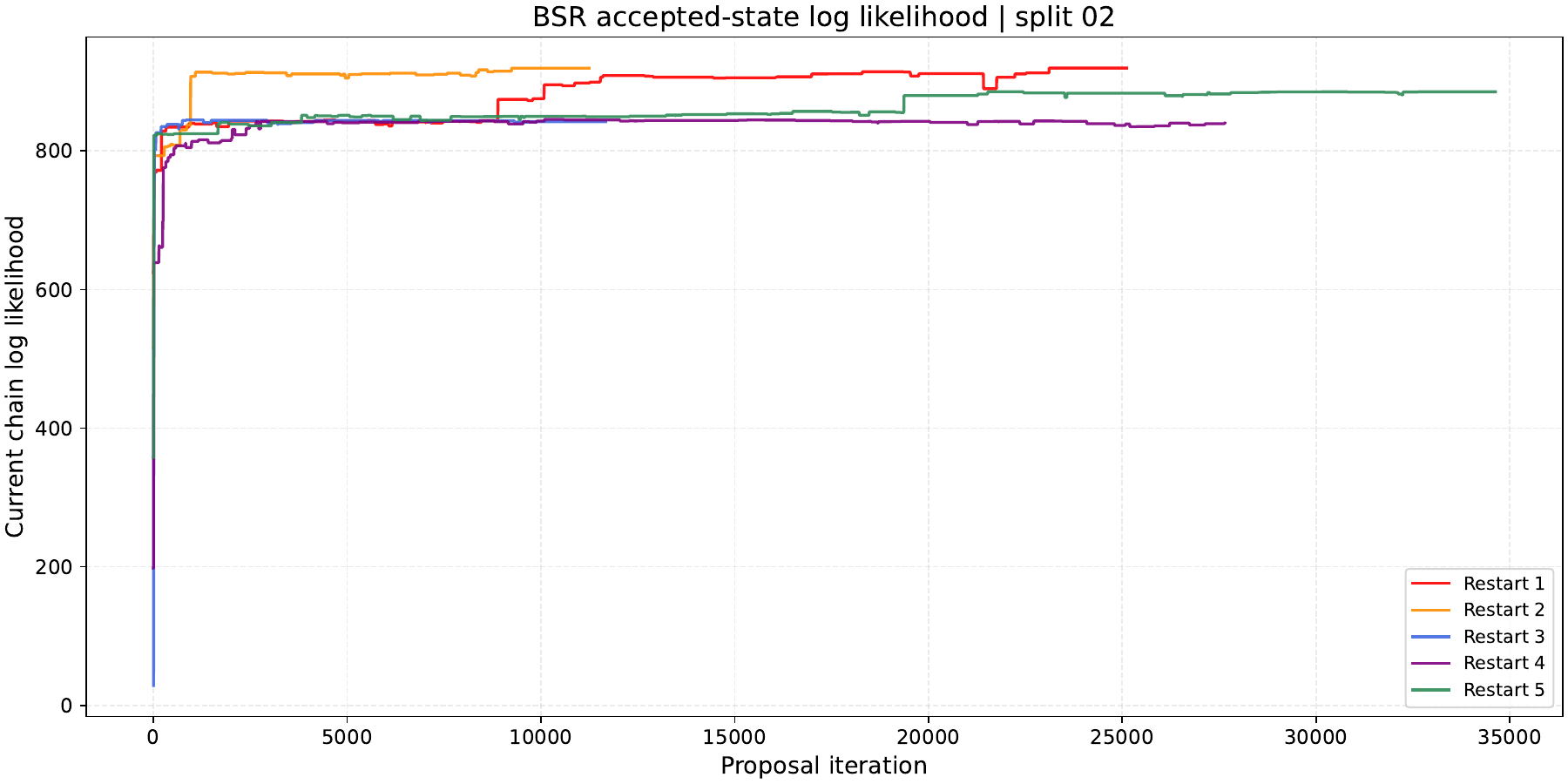}
    \caption{Split 2}
    \label{fig:bsr-trace-plot-perovskite-split-2}
\end{subfigure}
\hfill
\begin{subfigure}[t]{0.45\textwidth}
    \centering
    \includegraphics[width=\linewidth]{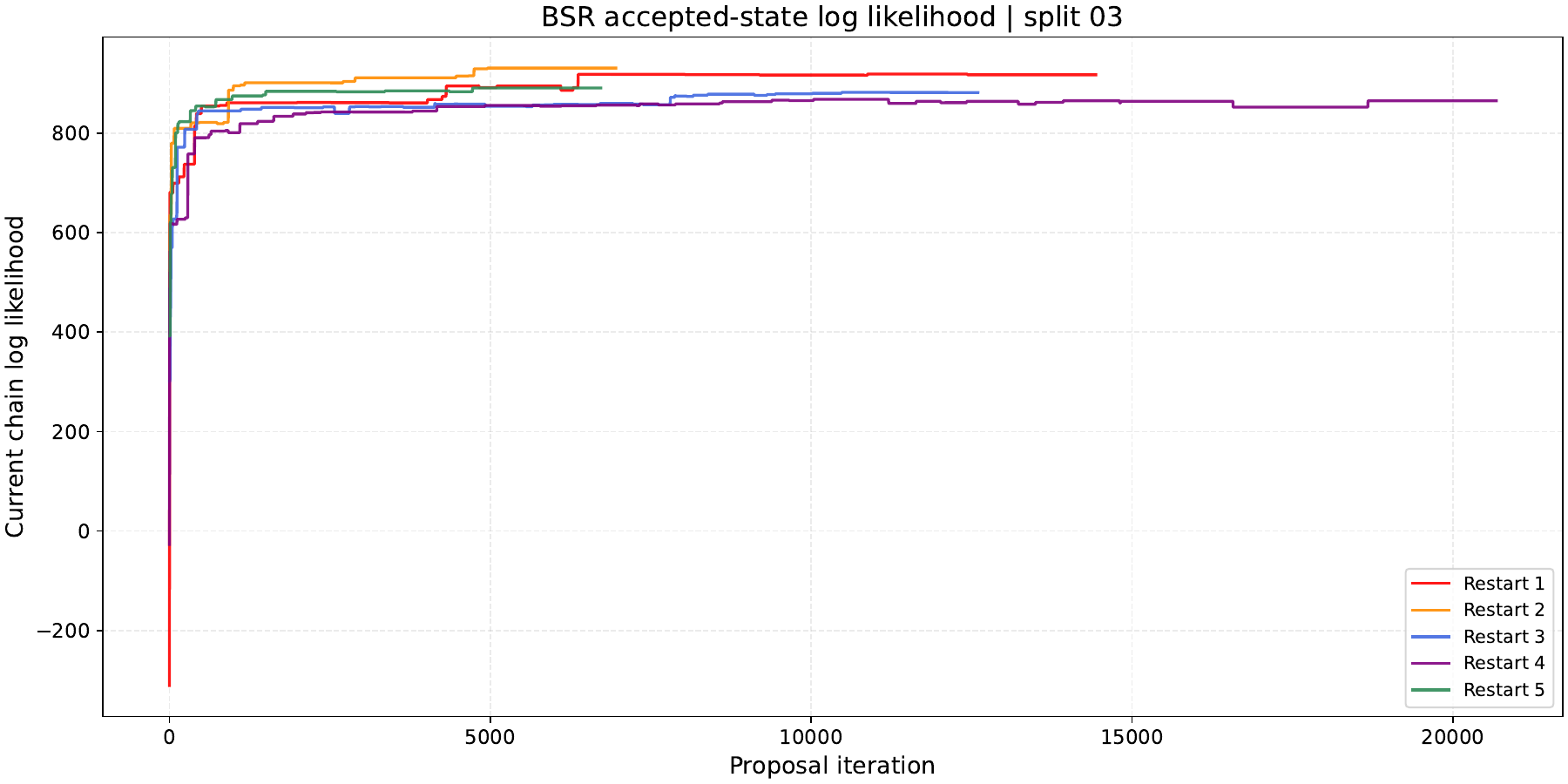}
    \caption{Split 3}
    \label{fig:bsr-trace-plot-perovskite-split-3}
\end{subfigure}

\begin{subfigure}[t]{0.45\textwidth}
    \centering
    \includegraphics[width=\linewidth]{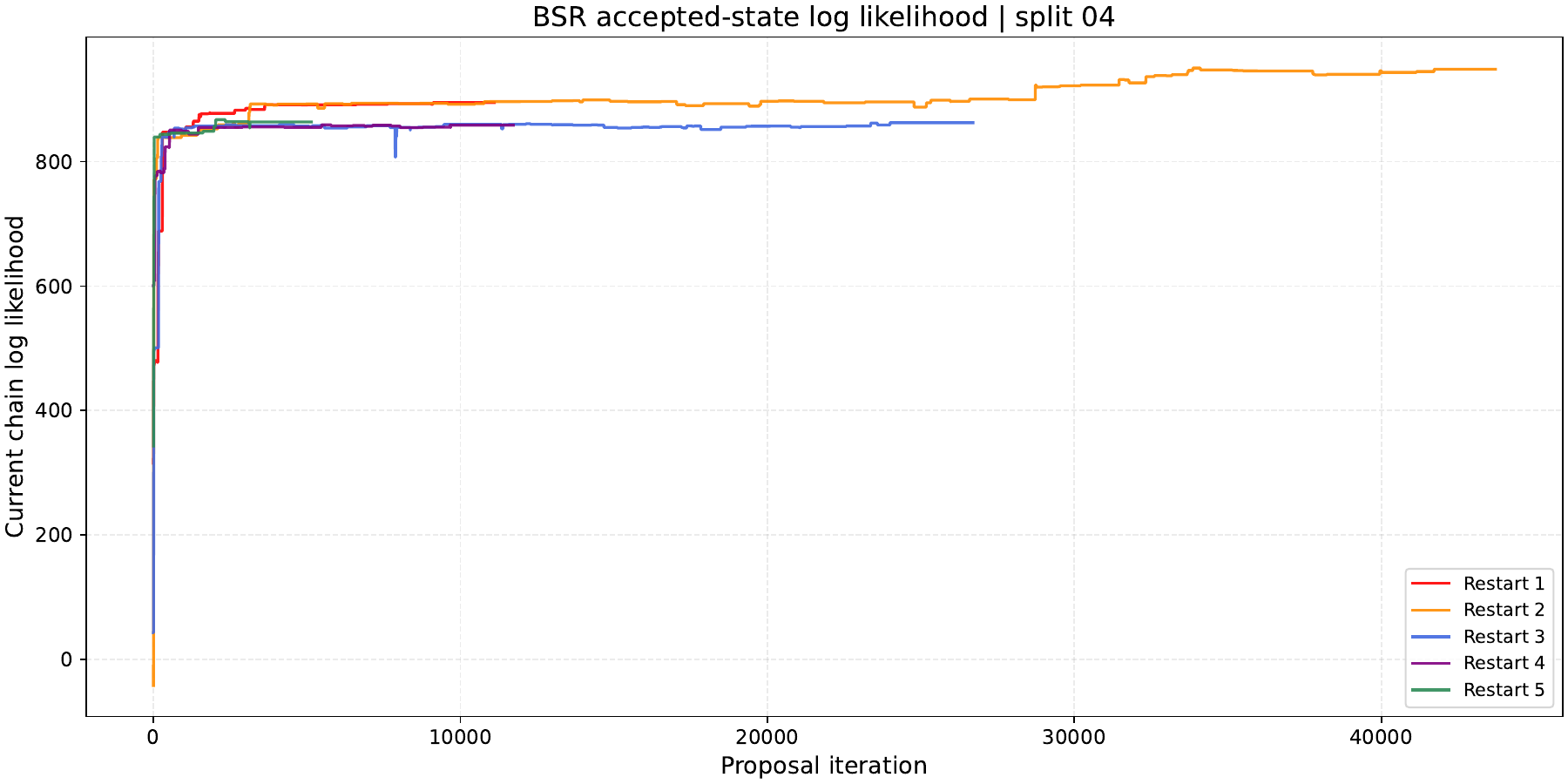}
    \caption{Split 4}
    \label{fig:bsr-trace-plot-perovskite-split-4}
\end{subfigure}
\hspace{0.06\textwidth}
\begin{subfigure}[t]{0.45\textwidth}
    \centering
    \includegraphics[width=\linewidth]{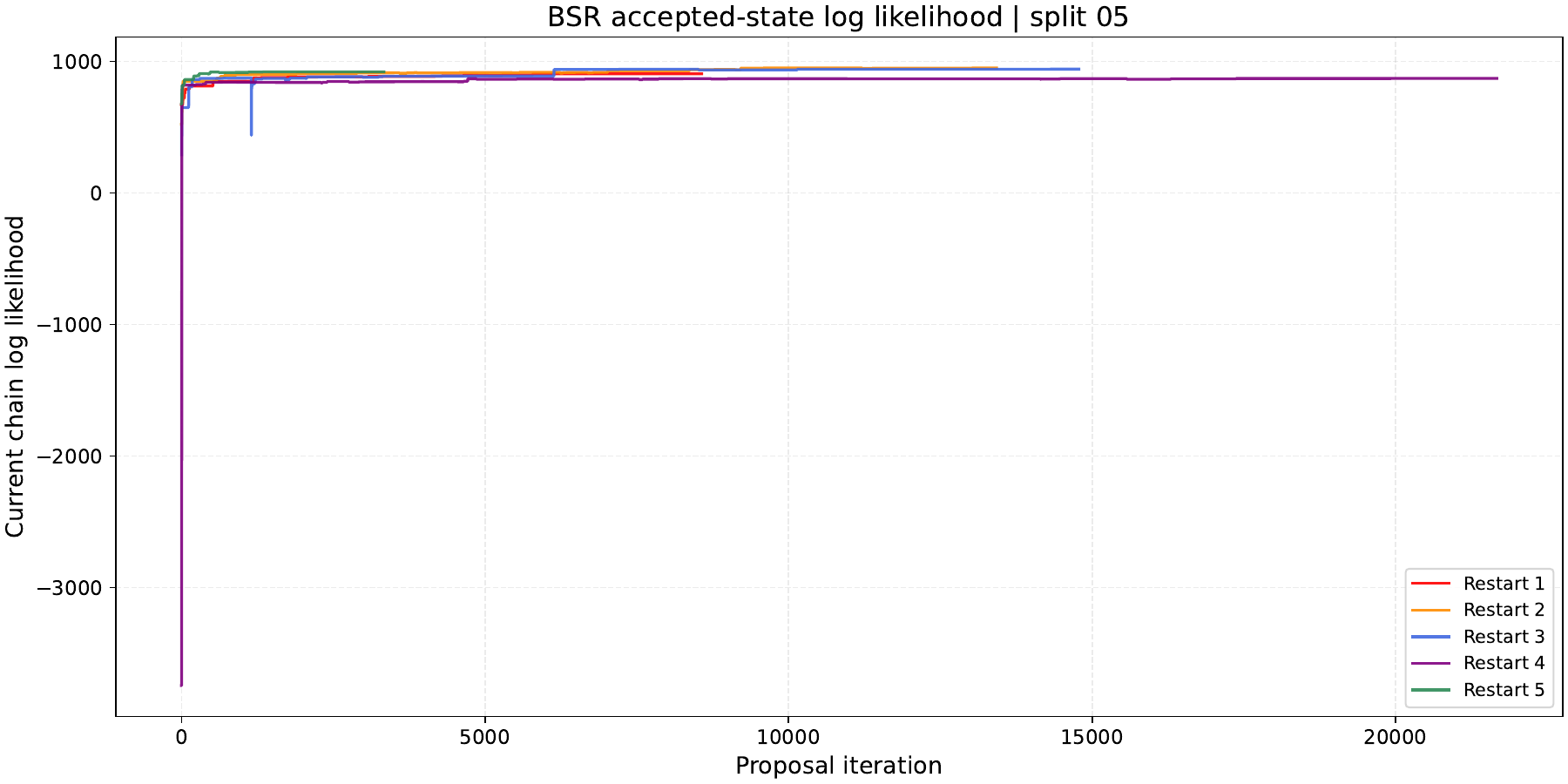}
    \caption{Split 5}
    \label{fig:bsr-trace-plot-perovskite-split-5}
\end{subfigure}

\caption{Trace plots of \bsr\ across the $5$ splits of the oxide perovskite dataset.}
\label{fig:bsr-perovskite-trace-plots}
\end{figure}

\clearpage
\bibliography{references}

\end{document}